\pdfoutput=1
\documentclass[12pt,pdftex]{book}

\usepackage{a4wide,setspace}
\usepackage[dvipsnames,usenames]{color}
\usepackage{amsmath,amssymb,graphicx,feynarts,enumerate}
\usepackage{hyperref}
\allowdisplaybreaks
\newcommand{\ord}{{\cal O}}
\newcommand{\Q}{\mathcal Q}
\newcommand{\beq}{\begin{equation}}
\newcommand{\eeq}{\end{equation}}
\newcommand{\ie}{{\it i.e.}}
\newcommand{\eg}{{\it e.g.\ }}
\newcommand{\etc}{{\it etc.}}
\newcommand{\CF}{C_F} 
\newcommand{\TF}{T_F} 

\newcommand{\Nc}{N_c} 
\newcommand{\AFBt}{A_{\rm FB}^{t}}
\newcommand{\Atc}{A_c^{t}}
\newcommand{\sigtot}{\sigma_{t\bar{t}}}
\newcommand{\dsig}{d\sigma_{t\bar{t}}/dM_{t\bar{t}}}
\newcommand{\Mtt}{M_{t\bar{t}}}
\newcommand{\ff}{f\hspace{-0.4em}f}
\newcommand{\sws}{s_w^2}

\newcommand{\cws}{c_w^2}

\def\bm#1{\mbox{\boldmath$#1$\unboldmath}}
\def\sgn{\mbox{sgn}}
\def\Mkk{M_{\rm KK}}
\def\delslash{\rlap{\hspace{0.02cm}/}{\partial}}
\def\Dslash{\rlap{\hspace{0.08cm}/}{D}}

\def\Gslash{\rlap{\hspace{0.08cm}/}{G}}
\def\etal {\em et al.}
\def\pslash{\rlap{\hspace{0.02cm}/}{p}}
\def\kslash{\rlap{\hspace{0.02cm}/}{k}}

\def\new#1{{\color{black}#1}}
\definecolor{BLUE}{RGB}{0,0,1}

\begin{document}

\pagenumbering{roman}

\begin{titlepage}

\vspace*{1.8cm}
\begin{center}
  {\LARGE\bf\begin{spacing}{1.1}
  Warped Extra Dimensions:\\
  Flavor, Precision Tests and Higgs Physics
  \end{spacing}}
\end{center}
\onehalfspacing
\vspace{2.8cm}
\begin{center}
 {\large \bf Dissertation}\\
 {\large zur Erlangung des Grades
 ``Doktor der Naturwissenschaften''\\
 am Fachbereich Physik, Mathematik und Informatik\\
 der Johannes Gutenberg-Universit\"at\\ 
 in Mainz}
\end{center}
\vspace{3cm}
\begin{center}
 {\large{\bf Florian Goertz}\\
geboren in Bad Kreuznach}
\end{center}
\vspace{2cm}
\begin{center}
Mainz, 2011
\end{center}

\end{titlepage}
\singlespacing
\allowdisplaybreaks

\vspace*{\fill}
{\noindent \it erster Berichterstatter}: Prof. Dr. Matthias Neubert\\
{\it zweiter Berichterstatter}: Prof. Dr. Hubert Spiesberger\\
{\it dritter Berichterstatter}: Prof. Dr. Andrzej Buras\\[1mm]
{\it Datum der m\"undlichen Pr\"ufung}: 15. Dezember 2011\\[1mm]
Dissertation an der Universit\"at Mainz (D77)\\[3mm]
MZ-TH/11-T5

\newpage

\begin{center}
\bf{Abstract}
\end{center}

\noindent In this thesis, the phenomenology of the Randall-Sundrum setup is investigated. 
In this context models with and without an enlarged $SU(2)_L\times SU(2)_R\times U(1)_X\times P_{LR}$ gauge symmetry, which removes 
corrections to the $T$ parameter and to
the $Z b_L\bar b_L$ coupling, are compared with each other.
The Kaluza-Klein decomposition is formulated within the mass basis, which allows for a 
clear understanding of various model-specific features. A complete discussion of tree-level flavor-changing effects is presented.
Exact expressions for five dimensional propagators are derived,
including Yukawa interactions that mediate flavor-off-diagonal transitions.

The symmetry that reduces the corrections to the left-handed $Z b \bar b$ coupling is analyzed in detail.
In the literature, Randall-Sundrum models have been used to address the measured anomaly in the $t\bar t$ 
forward-backward asymmetry. However, it will be shown that this is not possible
within a natural approach to flavor.
The rare decays $t \to cZ$ and $t \to ch$ are investigated,
where in particular the latter could be observed at the LHC.
A calculation of $\Gamma_{12}^{B_s}$  in the presence of new physics
is presented. It is shown that the Randall-Sundrum setup allows for an improved agreement 
with measurements of $A_{\rm SL}^s$, $S_{\psi\phi}$, and $\Delta\Gamma_s$. 
For the first time, a complete one-loop calculation
of all relevant Higgs-boson production and decay channels in the custodial Randall-Sundrum setup
is performed, revealing a sensitivity to large new-physics scales at the LHC.
\vspace{3mm}

\begin{center}
\bf{Zusammenfassung}
\end{center}

\noindent Gegenstand dieser Arbeit ist die Untersuchung der Ph\"anomenologie von Randall-Sundrum-Modellen.
Es werden Modelle, bei denen Beitr\"age zum $T$-Parameter und zur $Z b_L\bar b_L$-Kopplung 
durch eine erweiterte $SU(2)_L\times SU(2)_R\times U(1)_X\times P_{LR}$ Eich-Symmetrie verboten sind und solche ohne diese Symmetrie 
gegen\-\"ubergestellt. Die Kaluza-Klein-Zer\-legung
wird direkt in der Massenbasis vorgenommen, wo\-durch ein vertieftes Verst\"andnis verschiedener
Eigenschaften der Modelle erm\"oglicht wird. 
Eine vollst\"andige Analyse flavor-\"andernder Prozesse auf Born-Niveau wird
pr\"asentiert.
Es werden exakte Ausdr\"ucke f\"ur f\"unf-dimensionale Propagatoren hergeleitet, 
unter Ber\"ucksichtigung flavor-\"andernder Yukawa Kopplungen.

Die Unterdr\"uckung der Korrekturen zum $Z b_L \bar b_L$-Vertex wird detailliert analysiert.
Randall-Sundrum-Modelle wurden in der Literatur herangezogen, um die gemessene Erh\"ohung in der 
$t\bar t$ Vorw\"arts-R\"uckw\"arts-Asymmetrie zu erkl\"aren.
Es wird gezeigt, dass dies in einem nat\"urlichen Ansatz bez\"uglich
der Flavor-Struktur nicht m\"oglich ist.
Die seltenen Zerf\"alle $t \to cZ$ und $t\to ch$ werden untersucht, wobei
insbesondere der letz\-tere eine Gr\"o\ss enordnung erreichen kann, die am LHC messbar ist.
Es wird eine Berechnung von $\Gamma_{12}^{B_s}$
unter Ber\"ucksichtigung von Beitr\"agen neuer Physik durchgef\"uhrt
und gezeigt, dass in Randall-Sundrum-Modellen eine bessere \"Uber\-ein\-stimmung mit den gemessenen Werten
f\"ur $A_{\rm SL}^s$, $S_{\psi\phi}$ und $\Delta\Gamma_s$ erreicht werden kann. 
In dieser Arbeit wird die erste vollst\"andige Ein-Schleifen-Rechnung  f\"ur alle relevanten
Produktions- und Zerfallskan\"ale des Higgs-Bosons in Randall-Sundrum-Modellen pr\"asentiert.
Dabei wird eine Sensitivit\"at auf gro\ss e Skalen, jenseits der direkten Reichweite des LHC, festgestellt.

\thispagestyle{empty}
\cleardoublepage

\thispagestyle{empty}
\tableofcontents

\newpage

\section*{Preface}
\allowdisplaybreaks
The Randall-Sundrum (RS) setup provides promising possibilities to address several puzzles in particle physics.
The aim of this thesis is to study the phenomenology of RS models with bulk gauge and matter fields
in an anarchic approach to flavor. 
We formulate the Kaluza-Klein (KK) decomposition of the theory directly in the mass basis. 
The couplings to a brane-localized Higgs sector are included in an exact way via boundary conditions. This avoids the truncation of the KK 
towers, which is in contrast to the usual approach of treating these couplings as a perturbation.
In consequence, a clear and analytical understanding of important model-specific features is possible. 
For gauge fields, we present the decomposition in a covariant $R_\xi$ gauge. In the case of the fermion sector, the flavor mixing is
included in a completely general way. We show how the hierarchies of the quark sector are generated naturally in warped models,
starting from anarchical 5D Yukawa matrices (featuring neither symmetries nor hierarchies). To this end, the analogy to the Froggatt-Nielsen mechanism is demonstrated.
We present a detailed discussion of the interactions of the model, including a complete survey of tree-level flavor-changing effects.

A first look at electroweak precision observables reveals that the minimal RS model generically induces sizable contributions to 
the $T$ parameter. However, we show that a heavy Higgs boson $m_h\lesssim 1\,$~TeV cancels part of the corrections. This
allows for new-physics scales as low as a few TeV. At the same time, this option potentially improves the agreement between theory and experiment in the 
$Z\to b\bar b$ pseudo observables. Nevertheless, the minimal RS model features sizable corrections to these observables.
Extending the gauge group to the custodial group $SU(2)_L\times SU(2)_R\times U(1)_X \times P_{LR}$ and
including an appropriate embedding of the fermions, allows to remove the problematic contributions. In addition, the $T$ parameter is 
then protected by a {\it symmetry}. We explore the custodial RS model in detail. The approach of treating electroweak symmetry breaking
exactly allows for a clear analysis of the model specific features and protection mechanisms.
As we formulate the KK decomposition of the fermion sector of the minimal model in a general way, allowing for additional 
fermion representations, the application to the extended fermion sector of the custodial model can be performed conveniently.
In turn, it is straightforward to address questions about the model-dependence of the resulting gauge- as well as Higgs-boson
interactions with the Standard-Model fermions.
We demonstrate explicitly the protection of the $T$ parameter and of the left-handed $Z b \bar b$ couplings.
In particular, we work out, which contributions are protected and which terms inevitably escape protection. We identify them 
with the irreducible sources of custodial symmetry breaking.

Due to the presence of the towers of KK excitation one often encounters infinite sums over gauge boson profiles. We show how to 
perform these complete sums for towers with massless as well as massive zero modes, by using completeness relations for the profiles.
As a first application, these results are used to study effective four-quark interactions in both RS variants.
We pay special attention to the interactions with the Higgs sector and to the correct inclusion of Yukawa interactions that couple $Z_2$-odd fermions.
In the perturbative approach these contributions would be naively lost. 
This formal part of the thesis, already including some 
phenomenological considerations is presented in Chapter~\ref{sec:WED}.

A promising method to treat fields in five dimensions that appear as internal states in Feynman diagrams is to use five 
dimensional propagators. This avoids the KK decomposition and the need to sum up the contributions afterwards again.
Especially for the case of sums involving several fermion generations, performing the KK sum turns out to be impractical. 
In Chapter~\ref{sec:5Dprop}, we obtain exact expressions for the
five dimensional propagators of massive gauge bosons. Beyond that, analytic results for five-dimensional fermion-propagators, including
a generic flavor structure with off-diagonal Yukawa couplings, are presented for the first time. These expressions are useful for studying
loop mediated flavor-changing neutral currents, like $B\to X_s \gamma$, in warped extra dimensions.

The phenomenological survey of this thesis in Chapter~\ref{sec:Pheno} starts with an analysis of precision tests.
After exploring the modification of Standard Model parameters in the presence of warped extra dimensions, aspects of the electroweak 
precision parameters will be treated again. Since the anarchic RS setup predicts modified couplings especially for the third generation 
of quarks, the focus will be on this sector. A detailed analysis of the
$Z\to b \bar b$ pseudo observables is presented. To a great extend, we will pay attention to the comparison of the results in the custodial RS model 
with those of the minimal model. The $t\bar t$ forward-backward asymmetry, measured at Tevatron, shows a deviation from the Standard
Model prediction of about 2\,$\sigma$. We analyze if this discrepancy could be caused by the Randall-Sundrum setup calculating
the corresponding prediction at next-to-leading order in the strong coupling constant. The results presented are valid for
a broader class of new physics models. Then we study the flavor (changing) sector of the model.
After an analysis of the non-unitarity of the CKM matrix, the flavor-changing rare decays $t\to c Z$ and $t\to ch$ will be examined. 
\new{We will apply the formalism of five dimensional propagators to study the anomalous magnetic moment of the muon
in the RS setup.}
Another focus is on CP violating observables in the $B_s^0$-system, where recently some anomalies have been reported.
We perform a calculation of the absorptive part of the $B_s^0$--$\bar B_s^0$ mixing amplitude in the presence of new physics
and explore, if warped extra dimensions can lead to an improved agreement between theory and experiment in several 
observables. 

One of the main questions which the CERN Large Hadron Collider (LHC) is supposed to answer is
how the electroweak symmetry is broken in nature. 
In consequence, one focus of the phenomenological part is a detailed discussion of Higgs
physics at Tevatron and the LHC. For the first time, we present a complete one-loop calculation
for all relevant Higgs-boson production and decay channels in the Randall-Sundrum setup, 
incorporating the effects stemming from the extended electroweak gauge-boson and fermion sectors. 
We discuss the impact on physics at the LHC and demonstrate \new{that the Higgs sector could be very
viable for finding} physics beyond the Standard Model.

As this thesis is about extensions of the Standard Model of particle physics, it is mandatory to explore the motivations
to go beyond this successful model. Therefore, important aspects of the Standard Model are reviewed in Chapter~\ref{sec:IntroSM},
starting from an effective field theory approach. 
The fact that this minimal model works extremely well at low energies poses
tight challenges on theories that are meant to complete it. In this context, it is important to explore
the peculiarities of the Standard Model that lead to its successful agreement with experiment.
Such considerations gave rise to several extensions of the original Randall-Sundrum proposal, which finally
brought up the custodial Randall-Sundrum model with bulk-gauge and -matter fields.
In the first chapter we also review the evidence that causes the notion that the Standard Model will have to be replaced by 
another theory above a certain scale. This discussion will lead to the gauge hierarchy problem, which is 
finally examined, together with the puzzle of fermion hierarchies.
Models with (warped) extra dimensions that can address these problems are introduced in Chapter~\ref{sec:XD} 
and at the beginning of Chapter~\ref{sec:WED}.

The chapters~\ref{sec:WED}-\ref{sec:Pheno} contain the main results of this thesis. Chapter~\ref{sec:WED}, besides the introductory part
at the very beginning, and Chapter~\ref{sec:Pheno} 
are based on my publications \cite{Casagrande:2008hr,Casagrande:2010si,Bauer:2010iq,Goertz:2011nx}. 
Chapter~\ref{sec:5Dprop} on propagators of massive gauge and fermion fields in five dimensions
as well as Section~\ref{sec:AMM} on the anomalous magnetic moment of the muon
contain unpublished results.

A conclusion is presented in Chapter~\ref{sec:concl}.
Supplementary calculations and useful formulae can be found in the appendices.

This eprint version of the thesis contains minor modifications compared to the original version
(correction of typing errors, \etc). The version submitted to the University of Mainz can be retrieved 
from the university library.

\chapter{Introduction: The Standard Model and Hierarchies in Nature}

\chaptermark{Introduction: The SM and Hierarchies in Nature}

\pagenumbering{arabic}
\label{sec:IntroSM}
\vspace{-6mm} 
The first chapter is meant to introduce aspects of the Standard Model (SM) of particle physics and of 
effective field theories, with a focus on those topics which will be important for the following chapters. 
We will present the merits of the SM, which suggest that it is 
the right theory to describe nature up to (at least) the weak scale. Beyond that, we will illustrate~its shortcomings which 
bring us to the conclusion that it is, however, not the final theory 
of~nature and we will expound our expectations on the theory to complete it.
In particular, we will discuss several puzzling hierarchies that we observe in particle physics. These represent a 
convincing motivation to study warped extensions of the SM, which will be the focus of~this~thesis.

\section{Aspects of the SM and Effective Field Theories}

As the SM has been extremely successful up to present collider 
energies of several hundred GeV, it is a good starting point for studying new ideas in particle physics.
Its extensions should always have a SM-like theory as a low energy limit. Coming from an effective field theory approach, we will 
go into the main building blocks and tests of the model. 
When exploring warped setups, we will start from and refer to the structure of the SM, which represents the low energy
tail of the KK decomposition of these setups.
Since the phenomenological part of this thesis is mainly related to the Higgs and the flavor sector, as well as to precision tests, we 
will focus on these fields. 
During this introduction we will meet problems of the SM that provide a motivation to go beyond
it. However, we will also see that the SM leads to a multitude of non-trivial well-tested predictions.
Many of those will be jeopardized, when extending the minimal model of particle physics. 
For a successful model building it is thus important to be aware of the features that have lead to these predictions.
\nopagebreak
\subsection{The SM of Particle Physics and Symmetries of Nature}
\nopagebreak
\label{sec:SM1}
\nopagebreak
The dream of having a consistent, anomaly free, and renormalizable theory that describes nature down to length scales of $\sim(1-10)$\, TeV$^{-1}$,
{\it beyond the weak scale}, has become true with the advent of the Standard Model of Particle Physics.
These scales correspond to the energy frontier currently probed directly by mankind at large collider 
experiments. The success story of the SM started in the 1960s with the Glashow-Weinberg-Salam (GWS) theory of 
electroweak unification \cite{Glashow:1961tr,Weinberg:1967tq,Salam:1968rm
} 
(Nobel Prize in Physics 1979). 
Other milestones were the inclusion of the theory of strong interactions, as a gauge theory of elementary particles 
carrying color charge, the discovery of asymptotic freedom by Gross, Politzer, and Wilcszek 
\cite{Gross:1973id,Politzer:1973fx,Gross:1973ju} (Nobel Prize in Physics 2004), and the prove of renormalizability 
of the SM by 't Hooft and Veltman \cite{'tHooft:1972fi} (Nobel Prize in Physics 1999). The SM incarnates the
idea of combining special relativity and quantum mechanics into a quantum field theory that describes the world 
around us.\footnote{\label{fn:G}Note that the SM does not incorporate a (quantum) theory of gravity, which is expected to become 
important at the Planck scale $M_{\rm Pl}=\sqrt{\hbar c/G_N}\approx 1.22 \cdot 10^{19}$ GeV/$c^2$ 
\cite{Nakamura:2010zzi} at the 
latest. Here $G_N$ is the gravitational constant, $\hbar$ is the reduced Planck constant and $c$ is the speed of light 
in the vacuum. From now on we will work in natural units, setting $\hbar=c=1$. For the discussion of the first chapter
we will not consider gravitational interactions explicitly.} It has been tested at the quantum level 
in a plethora of experiments and so far resisted every attempt to refute its validity.
Moreover, it lead to many successful predictions of phenomena and particles, before their 
experimental discovery, like the existence of neutral currents and of the massive $W^\pm$ and $Z$ gauge bosons, including the 
corresponding mass ratio. The experimental observation of jets in the mid-1970s, in particular three-jet events at 
PETRA in 1979 \cite{Brandelik:1979bd}, provided a striking confirmation of quantum chromodynamics (QCD), the theory of strong 
interactions. The experimental evidence for quarks, some years before, confirmed that the quark model, introduced by 
Gell-Mann and Zweig, is not only a mathematical model \cite{GellMann:1964nj,Zweig:1981pd}, but is realized in nature. 
Beyond that, the prediction of the tau neutrino, the charm quark, and the third generation of quarks, before their 
experimental discovery (in particular the indirect determination of the approximate top mass), is to be assigned to the SM.   

Before looking at the model in more detail, let us recapitulate a general guiding principle for constructing 
theories, that will finally lead us to the SM. The most fundamental properties of nature, and thus also of a theory 
that shall describe it, are probably symmetries. From Noether's theorem, we know that continuous symmetries 
lead to conservation laws. Without symmetries, the universe (if it would exist) 
would behave much more randomly
and it would be very hard to derive meaningful predictions. The dynamics of a system in quantum field theory (QFT), which we 
will assume to be the right framework to describe particle physics, is given by 
its Lagrangian (including a quantization rule). Thus, one 
should write down the most general Lagrangian that corresponds to an experimentally given particle content and exhibits the 
symmetries of the system. Every combination of fields of the theory, whose absence can not be explained 
in terms of a symmetry, should appear. Certainly, a well established symmetry of nature is Poincar\'{e} invariance.
The particles of a theory should correspond to irreducible representations of the Poincar\'{e} group. Beyond this space-time 
symmetry, internal {\it local} symmetries seem to be important, so-called (local) {\it gauge symmetries}. 

The known matter fields of the universe 
(leptons and quarks) transform according to fundamental representations of a given gauge-symmetry group, if they are charged with 
respect to that group, and thus their appearance is related to these symmetries. Global invariance under such transformations corresponds to the freedom of  redefining multiplets of matter fields with the same transformation parameter at every space-time point
\beq
\label{eq:mattergau}
\Psi \to e^{i\,\alpha^a T^a} \Psi\,.
\eeq
Here, $\alpha^a \in \mathbb{R}$ and $T^a$ are the generators of the gauge group, in generalization of $U(1)$ phase rotations, where $T^a=Q$\ $(a=1)$
is the charge. For $SU(2)$, for example, the Pauli matrices $T^a=\sigma^a/2$\ $(a=1,\dots,3)$ can be used and 
$\Psi=(\Psi_1(x),\Psi_2(x))^T$, whereas the Gell-Mann matrices are generators of $SU(3)$, $T^a=\lambda^a/2$\ $(a=1,\dots,8)$ , 
see Appendix~\ref{app:PDG}. Throughout this thesis, a summation over repeated indices is understood. 
If the symmetry transformations (\ref{eq:mattergau}) shall be local, \ie, depending on space-time coordinates 
\beq
\alpha^a\to \alpha^a(x)\,,
\eeq
one has to introduce corresponding {\it gauge fields} in order for the kinetic terms of the matter fields, 
containing derivatives, to be invariant. Without kinetic terms there would not be any physical propagating fields.
Explicitly, these terms become invariant by replacing the ordinary derivative with the {\it covariant derivative}
\beq
\label{eq:covd1}
\partial_\mu \to D_\mu=\partial_\mu-ig A_\mu^a T^a\,,
\eeq
where $A_\mu^a$ are (gauge) vector fields, belonging to the generators $T^a$ of the corresponding gauge group with coupling constant $g$.
These fields have to transform according to the adjoint representation of the gauge group. In summary, the local symmetry transformations
read
\beq
\begin{split}
\label{eq:gau2}
\Psi(x) \to V(x)\,\Psi(x)\,,\quad V(x)=e^{i\,\alpha^a(x) T^a}\,,\\
A_\mu^a(x) T^a \to V(x)\left( A_\mu^a(x) T^a + \frac{i}{g}\,\partial_\mu \right)V^\dagger(x)\,.
\end{split}
\eeq
Also from a geometrical perspective, the need for local gauge invariance of the 
kinetic terms can be understood, if one wants to introduce local gauge transformations for the matter fields.
One requires a connection to define the derivative properly, which compares fields at different space-time point, 
where they transform differently. This gauge connection 
is provided by the gauge fields, together with their transformation properties, which are interpreted as physical, 
propagating degrees of freedom. These bosonic fields then act as force carriers. For more details see \eg \cite{Peskin:1995ev}.
Local invariance under a certain gauge group induces and restricts 
the interactions of the fields (via gauge bosons) just how they are discovered in nature and prevents the theory from 
being possibly trivial. It is a defining
criterion for a theory and, as we will see below, also essential from a formal point of view for 
constructing a consistent, sensible theory containing vector bosons. There is compelling experimental evidence (see Section~\ref{sec:Higgs})
that the gauge group of nature contains the structure
\beq
\label{eq:GSM}
G_{SM}=SU(3)_c \times SU(2)_L \times U(1)_Y\,.
\eeq
The corresponding gauge bosons, induced by local gauge invariance, are the gluons, mediating the strong interactions, and the $W^a$ and $B$ bosons, 
responsible for the electroweak interactions, see below. Having identified this gauge invariance as an appropriate
symmetry to describe interactions of matter fields, it should be imposed on the whole Lagrangian (in agreement 
with observation). Note however that a part of (\ref{eq:GSM}) will have to be hidden in a certain way, in order to obtain 
massive gauge bosons. This ``encryption'' of the symmetries also corresponds to a fundamental property of nature 
which has to be accounted for. The fact that we do not see the full gauge group in the form (\ref{eq:GSM})
at low energies, made it a demanding task to find the correct gauge theory. After the gauge group of the theory
has been identified, the matter content can be assigned to certain representations of the gauge group, according
to experimental observations. These representations determine the interactions between matter and gauge fields,
as explained above. For the known matter fields, the representations will be given in Table~\ref{tab:SMQN} below.  

In addition to these symmetry constraints, remember that a Lagrangian should be hermitian in order to 
lead to a unitary scattering matrix (S-matrix) that conserves probabilities. It should preserve causality, which is achieved by including 
proper anti-particle operators in addition to particle operators and brings about the necessity of {\it field} operators
\cite{Peskin:1995ev}. 
Furthermore, to obtain a stable theory with a well-defined ground state, the spectrum of the corresponding 
Hamiltonian has to be bounded from below. With these conditions in mind, it seems sensible to write down the 
most general, translation-invariant, Lorentz scalar which contains spinor fields for the observed matter representations and is 
(locally) invariant with respect to the gauge group $G_{SM}$ (\ref{eq:GSM}) - inducing the gauge fields  - 
and call it the Standard Model. If one counts the number of terms that are allowed under 
these restrictions, one ends up with the rather large number of infinity, which would spoil the predictivity of the 
model. Luckily, it turns out that the situation is not as hopeless as it seems. 

It is instructive to sort the terms in the Lagrangian according to the mass dimensions of the products of field 
operators that they contain. The naive mass dimension of a field operator can be
read off from its standard kinetic term of its free theory (bilinear in the fields), 
by using the fact that a derivative has the mass dimension of a momentum ($D=1$) and that the action has to be dimensionless in natural 
units. As in four space-time dimensions the 
Lagrangian needs to have mass dimension $D=4$, terms that have a mass dimension smaller (bigger) than four have to appear 
with (inverse) powers of some yet unspecified mass scale\footnote{A more rigorous treatment of the
idea of effective field theories (EFTs), used here, is given in Appendix~\ref{app:EFT}.} $M$
\beq
\label{eq:Lexp}
{\cal L}={\cal L}_{\leq4}+\frac{{\cal L}_5 }{M}+\frac{{\cal L}_6}{M^2}+\cdots\, .
\eeq
The subscripts in (\ref{eq:Lexp}) denote the mass dimension of the corresponding sub-Lagrangians. Note that ${\cal L}_{\leq4}$
contains all terms with mass dimensions equal to or less than four, including the possibly dimensionful coefficients.
These are expected to scale like powers of $M$, if there is no second scale generated in the theory. The terms with 
$D<4$ are generically problematic, see sections \ref{sec:SMProblems} and \ref{sec:HP}. Accordingly, all dimensionless 
coefficients present in the Lagrangian are expected to be of $\ord(1)$, if there are no symmetries
present, that would explain other values. 
For the moment, let us not assume that it is possible to construct the final theory of nature, valid up 
to arbitrary high energy scales, with just the particle content observed below the TeV scale. Let us rather
assume that we see the low energy tail of a more complete theory. However,
without direct hints for new particles, we would like to set up a theory, which is based on the observed particle content 
and symmetries, and is valid up to some fixed energy scale. This is assumed to 
lie above the scales at which we have already looked for new degrees of freedom. Considering the Lagrangian above as 
belonging to such a theory, this cutoff scale can now be identified with 
the scale $M$, see Appendix~\ref{app:EFT}.
For example, $M$ could be the mass of a new heavy particle, not present in our low energy theory. If we are interested in 
energies $E \ll M$, a description without this heavy particle as a propagating degree of freedom is perfectly fine.

Crucially, the terms $\propto {\cal L}_i$, with $i>4$, will be suppressed by powers of $E/M \ll 1$. As a consequence the most important part 
of the Lagrangian at low energies is given by ${\cal L}_{\leq 4}$, while the additional terms can be neglected to more or less good approximation 
(depending on the ratio $E/M$). Thus, one ends up with a finite number of terms, depending on the truncation of 
(\ref{eq:Lexp}). The part with $D \leq 4$ is called the Standard Model of Particle Physics (with one ingredient
related to the masses of the particles still missing, see below). Experiments can give a handle on the cutoff scale $M$ by measuring effects of higher dimensional terms. 
As mentioned, all this holds, if we assume to observe the low energy part of a more fundamental theory with a scale $M$ (the cutoff 
of our theory). The existence of a final ultraviolet (UV) completion guarantees the suppression of higher dimensional ($D>4$) terms just by this 
new physics (NP) scale. This is the modern point of view of seeing the SM, see Section~\ref{sec:SMProblems}. 
Historically, the SM was not introduced as part of a more general Lagrangian, but rather 
as {\it the} Lagrangian of nature. The concept that singled out the terms with $D \leq 4$ will be introduced 
in Section~\ref{sec:Renorm}. 
From the SM Lagrangian one obtains the corresponding SM-Feynman rules to compute 
S-matrix elements perturbatively (as a sum over all possible Feynman diagrams) via standard textbook techniques  \cite{Peskin:1995ev, Cheng:1985bj}. 

From the previous discussion one expects ${\cal L}_{\leq4}$ to be explored first in particle collisions. Having identified
gauge invariance as a crucial concept, it is thus no wonder that the SM works so well (up to a certain energy). It just corresponds to the most important 
part of the most general gauge invariant Lagrangian that one can write down, given the observed particle content.\footnote{By 
gauge invariance we mean invariance with respect to the {\it local} symmetry transformations corresponding 
to (\ref{eq:GSM}), unless stated otherwise.} It does not mean that there is no NP, it rather means 
that the NP should be rather heavy. Nevertheless, the construction of the SM as a gauge theory with the appropriate (gauge) symmetries 
was a tremendous success. This is even more the case, given the fact that when the GWS model was constructed, not all 
gauge particles of the SM had been observed, yet. Their existence has been deduced from measuring fermion interactions.
Along the lines discussed above, NP should already show up below being directly produced in terms of new particles,
due to suppressed terms with $D>4$. Measuring deviations from SM expectations can give a handle on the scale $M$, as mentioned before. 
It will turn out that in some sectors the SM works so well that the expected NP scale is much higher than currently
accessible scales.
Note that it is in principle not excluded that there exist unobserved particles which are very light, perhaps with masses much below our 
designated cutoff for the SM. These particles then have to posses very suppressed couplings to the known particles 
in order to have escaped detection. As an example, consider the so called dark photon, which appears as a force mediator 
in some models of Dark Matter. This could be rather light, \eg $m_{\gamma^\prime}\sim 1$\,GeV, 
and have couplings to the SM via kinetic mixings with a strongly suppressed coefficient $\epsilon \ll 1$, see \eg \cite{ArkaniHamed:2008qn}.

Let us finally write down the SM Lagrangian 
\beq
{\cal L}_{\rm SM}\equiv{\cal L}_{\leq 4}. 
\eeq
It reads\footnote{For a review on QFT and the SM see \eg \cite{Peskin:1995ev,Cheng:1985bj,Weinberg:1995mt,Weinberg:1996kr,Zee:2003mt}.} 
\beq
   \label{eq:LSM}
	{\cal L}_{\rm SM}={\cal L}_{\rm ferm}+{\cal L}_{\rm W,B,G}+{\cal L}_{\rm Higgs}+{\cal L}_{\rm Yukawa}
	+{\cal L}_{\rm GF}+{\cal L}_{\rm FP}\, .
\eeq
This is the Lagrangian that we will study in the next pages, and, for the time being, we will not 
consider the possible higher dimensional terms. The first term in (\ref{eq:LSM}) contains kinetic terms for 
the matter fields (fermions) that we have observed in nature so far. The different fermion flavors present in the 
SM are grouped into three-component vectors (each component corresponding to a quark or lepton family) of 
up-type quark flavors $u\equiv(u,c,t)^T$, down-type quarks $d\equiv(d,s,b)^T$, charged leptons $e\equiv(e,\mu,\tau)^T$, 
and neutral leptons (neutrinos) $\nu\equiv(\nu_e,\nu_\mu,\nu_\tau)^T$. We end up with
\beq
\label{eq:fermkin}
{\cal L}_{\rm ferm}=\bar Q_L i\Dslash Q_L + \bar u_R i\Dslash u_R + \bar d_R i\Dslash d_R + \bar E_L i\Dslash E_L 
										+ e_R i\Dslash e_R\, ,
\eeq
where $\bar F \equiv F^\dagger \gamma^0$, $F=Q,u,d,E,e$. The left-handed and right-handed components of the field operators belong 
to the two irreducible spin-1/2 representations of the Poincar\'{e} group in four dimensions (4D), respectively,
and are projected out as $F_{L,R}=P_{L,R}\,F$, where $P_{L,R}=(1\mp \gamma^5)/2$. The Dirac-gamma matrices are defined in Appendx~\ref{app:PDG}.
 In order for the kinetic terms of the matter 
fields to be gauge invariant, the spatial derivative $\partial_\mu$ has been replaced by the covariant derivative 
(in analogy to (\ref{eq:covd1})) corresponding to the full SM gauge group (\ref{eq:GSM})
\beq
\begin{split}
\label{eq:covD1}
\Dslash :=& \gamma^\mu D_\mu\\
D_\mu=&\partial_\mu -i g^\prime\, Y B_\mu -i g\, T^i W^i_\mu -i g_s\, t^a G^a_\mu.
\end{split}
\eeq
This induces local interactions, \ie, terms which contain more than two fields at the same space-time point, with 
the (spin-1) $B,W^i$, and $G^a$ gauge fields, in agreement with observation. Here, $Y$, $T^i=\sigma^i/2$ 
$(i=1,\dots,3)$, and $t^a=\lambda^a/2$ ($a=1,\dots,8$) correspond to the standard generators of $U(1)_Y$ (hypercharge), $SU(2)_L$ 
(weak isospin), and $SU(3)_c$ (color), as introduced before. The corresponding coupling constants are denoted by $g^\prime$, $g$, 
and $g_s$. Through quantum (loop) corrections, these couplings are running couplings, depending on the renormalization scale $\mu$, see 
\eg \cite{Peskin:1995ev}. Their values at the $Z$-pole, 
$\mu=m_Z\approx91.2$\,GeV, read \cite{Nakamura:2010zzi}
\beq
\label{eq:coupZ}
g^\prime(m_Z)\approx0.36\,,\quad g(m_Z)\approx0.65\,, \quad g_s(m_Z)\approx1.22\,.
\eeq
Note that the non-abelian $SU(3)_c$ gauge theory of QCD becomes strongly coupled at low energies 
$\mu\sim 1$\,GeV, explicitly $\alpha_s(\Lambda_{QCD})\equiv g_s^2(\Lambda_{QCD})/(4\pi)\gtrsim \ord(1)$,
where $\Lambda_{QCD}\approx 200$\,MeV.
This calls for non-perturbative methods. On the other hand, it becomes asymptotically free for high energies (which holds for less than 17 participating flavors) \cite{Gross:1973id,Politzer:1973fx,Gross:1973ju}. 
 
Experimentally, it turns out that the left-handed fermions form doublets under $SU(2)_L$
\beq
	Q_L^i=
\left( \begin{array}[h!]{c}
	u_L^i \\
	d_L^i
\end{array} \right)\, , \quad
	E_L^i=
\left( \begin{array}[h!]{c}
	\nu_L^i \\
	e_L^i
\end{array} \right)\,,
\eeq
while all quarks are triplets under $SU(3)_c$, which is not made explicit in (\ref{eq:fermkin}). The representations 
of the SM fields with respect to the SM gauge group, including the hypercharge quantum numbers, \ie, eigenvalues of the generator $Y$, 
are summarized in Table~\ref{tab:SMQN}, which provides an overview of the particle content of the SM. Here, singlets (which do not transform under the corresponding gauge transformation) are denoted by $\bm{1}$, doublets by $\bm{2}$, \etc
\begin{table}
\begin{center}
\begin{tabular}{|c|c|c|}
\hline
spin & field & representation $(SU(3)_c,SU(2)_L)_{U(1)_Y}$  \\
\hline
      & $Q_L^i$ &\ $(\bm{3},\bm{2})_{1/6}$   \\
      & $E_L^i$ &\quad $(\bm{1},\bm{2})_{-1/2}$  \\
  1/2 & $u_R^i$ &\ $(\bm{3},\bm{1})_{2/3}$  \\
      & $d_R^i$ &\quad $(\bm{3},\bm{1})_{-1/3}$  \\
      & $e_R^i$ &\ $(\bm{1},\bm{1})_{-1}$  \\
 \hline
      & $B_\mu$ & $(\bm{1},\bm{1})_{0}$   \\
   1  & $W_\mu^i$ & $(\bm{1},\bm{3})_{0}$  \\
      & $G_\mu^a$ & $(\bm{8},\bm{1})_{0}$  \\
  \hline
   0  & $\Phi$ &\quad  $(\bm{1},\bm{2})_{1/2}$\\
  \hline
\end{tabular} \quad 
\end{center}
\vspace{-3.5mm}
\parbox{15.5cm}{\caption{\label{tab:SMQN} Representations of the different fields in the SM. Right handed neutrinos have not been included,
as they are not present in the original formulation of the SM. They are singlets under the SM gauge group, \ie, transform as $(\bm{1},\bm{1})_0$.
The Higgs field $\Phi$ will be introduced further below.}}
\end{table}
This field content, together with the corresponding representations/quantum numbers, guarantees that the SM is anomaly free, \ie, that
no gauge symmetries of the classical Lagrangian are broken at the quantum level.

As gauge invariance called for the existence of gauge fields, we should now look for further gauge invariant 
Lorentz scalars that can be constructed out of them. In addition to the interaction terms above, we end up with kinetic 
terms for the gauge fields, which allow them to propagate. They read 
\beq
\label{eq:LWBGSM}
   {\cal L}_{\rm W,B,G}= -\frac 1 4 B_{\mu\nu} B^{\mu\nu} -\frac 1 4 W^i_{\mu\nu} W_i^{\mu\nu}
   -\frac 1 4 G^a_{\mu\nu} G_a^{\mu\nu}
\eeq
and contain the field-strength tensors, defined as
\beq
\label{eq:SMFS}
\begin{split}
  B_{\mu\nu}=& \partial_\mu B_\nu - \partial_\nu B_\mu \\
	W_{\mu\nu}^i=& \partial_\mu W_\nu^i - \partial_\nu W_\mu^i + g\, \epsilon^{ijk}\, W_\mu^j W_\nu^k \\
	G_{\mu\nu}^a=& \partial_\mu G_\nu^a - \partial_\nu G_\mu^a + g_s\, f^{abc}\, G_\mu^b G_\nu^c\, .
\end{split}
\eeq
Here, $\epsilon^{ijk}$ and $f^{abc}$ are the structure constants of $SU(2)$ and  $SU(3)$, respectively,
where
\beq
\left[T^i,T^j\right]=i \epsilon^{ijk} T^k
\eeq
for the generators of $SU(2)$ and a similar relation holds for $SU(3)$.
Before detailing the remaining parts of the Lagrangian, several comments are in order. 

In its original formulation, 
the SM does not contain right handed neutrinos which correspond to the only possibility to account for massive neutrinos 
at the level of gauge invariant $D\leq4$ terms. We know that neutrinos have non-zero masses 
from the observation of neutrino oscillations. Neutrino masses could be interpreted as the first evidence for physics 
beyond the SM (BSM). However, it is straightforward to extend the SM by adding right handed neutrinos (though without being 
able to explain the tininess of neutrino masses, see Section~\ref{sec:SMProblems}).
Furthermore, the allowed term $\theta_{QCD}\, g_s^2/(64\pi^2)\epsilon^{\mu\nu\rho\sigma}G_{\mu\nu}^a G_{\rho\sigma}^a$ is not 
included in the SM, as experiments tell us that it seems not to be present. This term would 
violate the combined charge-conjugation and parity (CP) invariance, \ie, the invariance with respect to the substitution 
of a particle by its anti-particle in combination with a reflection at the origin $\vec x \to -\vec x$, in the sector of strong interactions (see \eg \cite{Peskin:1995ev}). Its unexplained absence or the extreme and unnatural 
smallness of its coefficient $\theta_{QCD}$ is called the {\it strong CP problem}. A possible solution is provided by 
the Peccei-Quinn theory, introducing a new particle, called the axion \cite{Peccei:1977hh}. Note that analogous 
terms for the $U(1)$ and $SU(2)$ field strengths have no observable effects \cite{Peskin:1995ev}. Most important, with 
just the fields introduced so far, corresponding to particles which been observed in nature, all fields of the SM would be massless, 
which is clearly in conflict with observation. The particles introduced so far thus still do not quite match those we 
observe in experiments. Given just these fields and no additional structure, it is simply not 
possible to write down gauge invariant mass terms for the (non-abelian) vector bosons and chiral fermions.
However, no additional particles have been observed yet, so the mechanism that gives masses to the SM particles is still 
experimentally unobserved. We certainly do not want to abandon gauge 
invariance as a concept, as it was a key in constructing the SM with the correct interactions and results in non trivial 
relations between couplings, which are well tested. We will discuss some of these relations in Section~\ref{sec:Higgs}. 
Moreover, note that breaking gauge invariance leads to big problems in the high energy behavior of the theory. This is already the case,
if it is just broken in mass terms for the observed massive gauge bosons, while keeping the gauge connection for the fermion fields as 
well as internal gauge interactions as they are (motivated by phenomenology). Although gauge invariance in the massless gauge sector 
can be seen as a redundancy in the description of a spin-1 particle with two degrees of freedom by a four-component Lorentz vector, 
it guarantees that the SM is renormalizable and does not lose its validity at the TeV scale, see below. If we want to construct a decent 
theory for the TeV scale, gauge invariance should better not be broken in an uncontrolled way.

\subsection{The Higgs Sector, Custodial Symmetry and Precision Tests}
\label{sec:Higgs}
The most famous possibility to break the electroweak gauge symmetry in order to obtain massive SM particles while keeping
the fundamental Lagrangian gauge invariant is the Higgs mechanism (which is a part of the SM) due to Brout, Englert, Higgs, Guralnik, Hagen and Kibble 
\cite{Englert:1964et,Higgs:1964pj,Guralnik:1964eu}. 
We will study the phenomenology of the corresponding sector in the context of RS models in detail in Section~\ref{sec:RSHiggs}.
The mechanism works as follows. A complex scalar doublet under $SU(2)_L$ (with hypercharge $Y=1/2$) 
\beq
\label{eq:Hdoub+}
   \Phi(x)=\left( \begin{array}{c}
    \phi^+(x) \\
    \phi^0(x)
   \end{array} \right)\,,
\eeq 
is included in the SM, with a potential, that results in a vacuum expectation value (VEV) for this doublet.
{\nopagebreak Given the aforementioned appropriate charges for the Higgs field, the VEV 
\beq
\label{eq:VEV}
\langle \Phi \rangle = \frac{1}{\sqrt2}
   \left( \begin{array}{c}
    0 \\
    v
   \end{array} \right)\, 
\eeq
will break the electroweak part of the gauge symmetry 
{\it spontaneously} (\ie, in the ground state of the theory, not at the fundamental Lagrangian level), according to the pattern}
\beq
\label{eq:break}
SU(2)_L \times U(1)_Y \longrightarrow U(1)_{EM}\,,
\eeq
which we observe in nature, and thus will generate masses for three electroweak gauge bosons. The photon, corresponding 
to the unbroken $U(1)_{EM}$ symmetry of quantum electrodynamics (QED) will remain massless, as the gluon does. 
The (gauge invariant) Higgs Lagrangian reads 
\beq
\label{eq:LHiggs}
   {\cal L}_{\rm Higgs} = (D_\mu\Phi)^\dagger\,(D^\mu\Phi) - V(\Phi) ,
   \qquad V(\Phi) = - \mu^2\Phi^\dagger\Phi + \frac{\lambda}{2} \left( \Phi^\dagger\Phi \right)^2\,.
\eeq
Note the negative sign of the mass term ($\mu^2>0$) in the Higgs potential. This sign is responsible for the
so-called mexican hat form of the potential, which allows for a nontrivial VEV $\langle |\Phi| \rangle >0$.
The Higgs quartic coupling $\lambda>0$ has to be positive in order for the Higgs potential $V(\Phi)$
to be bounded from below and thus for the vacuum to be stable, see below.
After electroweak symmetry breaking (EWSB), \ie, after acquiring the VEV (\ref{eq:VEV}), 
we parametrize the Higgs doublet in terms of fluctuations around that VEV via the four real scalar 
fields $\varphi^i$, $i=1,2,3$ and $h$ as
\beq
\label{eq:Hdoubl}
   \Phi(x) = \frac{1}{\sqrt2}
   \left( \begin{array}{c}
    -i\sqrt2\,\varphi^+(x) \\
    v + h(x) + i\varphi^3(x)
   \end{array} \right)\, .
\eeq
Here, we have defined $\varphi^\pm=(\varphi^1\mp i\varphi^2)/\sqrt2$. 
The field $h$ corresponds to the famous Higgs particle. It is the last missing particle that is needed to complete 
the SM, after the discovery of the predicted $SU(2)_L$ partner of the tau, the tau neutrino, through the DONUT experiment at 
Fermilab in 2000 \cite{Kodama:2000mp}. It corresponds to the residual physical scalar degree of freedom after EWSB and parametrizes massive fluctuations 
around $v$, whereas the unphysical Goldstone bosons $\varphi^\pm,\varphi^3$ correspond to flat directions in the potential. 
According to the Goldstone theorem, every spontaneously broken generator of a continuous symmetry results in a 
Goldstone boson. The breaking pattern
(\ref{eq:break}) thus results in three Goldstone bosons. By a gauge transformation leading to the unitary gauge, they can be 
formally removed from the theory. Their degrees of freedom are absorbed by the electroweak gauge bosons, which now are massive 
and thus need a third polarization degree of freedom. 

Let us have a look at how these masses are generated. After EWSB, (\ref{eq:LHiggs}) contains 
- besides interaction terms, kinetic terms, and the Higgs potential - the following quadratic terms in the fields 
\beq
\label{eq:SMWBmas}
{\cal L}_{\rm Higgs} \supset v^2\,\frac{g^2 + g^{\prime 2}}{8}\,Z_\mu Z^\mu + v^2\,\frac{g^2}{4}\,W_\mu^+ W^{-\mu}\,,
\eeq
where we have used the charges of the Higgs doublet given in Table \ref{tab:SMQN} and the definition of the covariant derivative 
(\ref{eq:covD1}). Moreover, we have already performed a transformation to the mass basis, \ie, the basis in which the gauge boson 
mass matrix is diagonal, via the orthogonal transformation
\beq \label{eq:ZASM}
  \left( \begin{array}{c}
      Z_\mu\\
      A_\mu
    \end{array} \right)= 
  \left( \begin{array}{cr}
      c_w & - s_w \\
      s_w & c_w
    \end{array} \right)
  \left( \begin{array}{c}
      W_\mu^3\\
      B_\mu
    \end{array} \right)\,. 
\eeq
Here, the sine and cosine of the weak mixing angle, $s_w$ and $c_w$, are given by
\beq
\label{eq:weakmixingSM}
  s_w\equiv\sin\theta_w=\frac{g^\prime}{\sqrt{g^2+g^{\prime 2}}} \,, \qquad
  c_w\equiv\cos\theta_w=\frac{g}{\sqrt{g^2+g^{\prime 2}}} \,,
\eeq
and
\beq
	W_\mu^\pm = \frac{1}{\sqrt2} \left( W_\mu^1\mp i W_\mu^2 \right)\,. 
\eeq
Thus, the physical $W^\pm$ and $Z$ bosons get masses
\beq
\label{eq:MWSM}
	m_W=v\, \frac g 2\, , \qquad m_Z=v\, \frac{\sqrt{g^2 + g^{\prime 2}}}{2} \, ,
\eeq
while the photon $A$ corresponds to a combination of symmetry generators which is left unbroken and remains 
massless.  
Furthermore the Higgs boson receives a mass $m_h=\sqrt{2}\mu=\sqrt{\lambda} v$.
In the mass basis the covariant derivative (\ref{eq:covD1}) becomes
\beq
\label{eq:covder}
   D_\mu=\partial_\mu -i \frac{g}{\cos\theta_w}\left(T^3-\sin^2\theta_w Q\right) Z_\mu -i e\,Q A_\mu 
   - i \frac{g}{\sqrt{2}} \left(T^+ W^+_\mu + T^- W^-_\mu\right) -i g_s t^a G^a_\mu\,,
\eeq
where we have defined the generators
\beq
	T^\pm=(T^1\pm i\,T^2)\,,
\eeq
as well as the $U(1)_{EM}$ generator of electromagnetism and the electric charge
\beq
\label{eq:QT}
	Q=T^3+Y\,, \quad e=g \sin\theta_w\,.
\eeq
Due to the gauge invariant mechanism of giving masses to the electroweak gauge bosons
via the Higgs doublet, we arrive 
from (\ref{eq:MWSM}) at the SM (tree-level) prediction
\beq
\label{eq:mWmZ}
	m_W=m_Z \cos\theta_w\,.
\eeq
Thus, at tree level, the properties of electroweak gauge bosons can be fully specified by the three parameters 
$e,\theta_w$, and $m_Z$. Clearly, one can also trade one of these inputs for another quantity which provides the same 
information like \eg the {\it Fermi coupling} $G_F\equiv g^2/(4\sqrt{2} m_W^2)$ (see Appendix~\ref{app:EFT}). From measuring 
the gauge-boson masses one can deduce $v = 246$\,GeV as the scale of EWSB. The experimental results for these
masses can be found in Appendix~\ref{app:ref}.

\begin{figure}[!t]
\begin{center}
\includegraphics[height=2.55in]{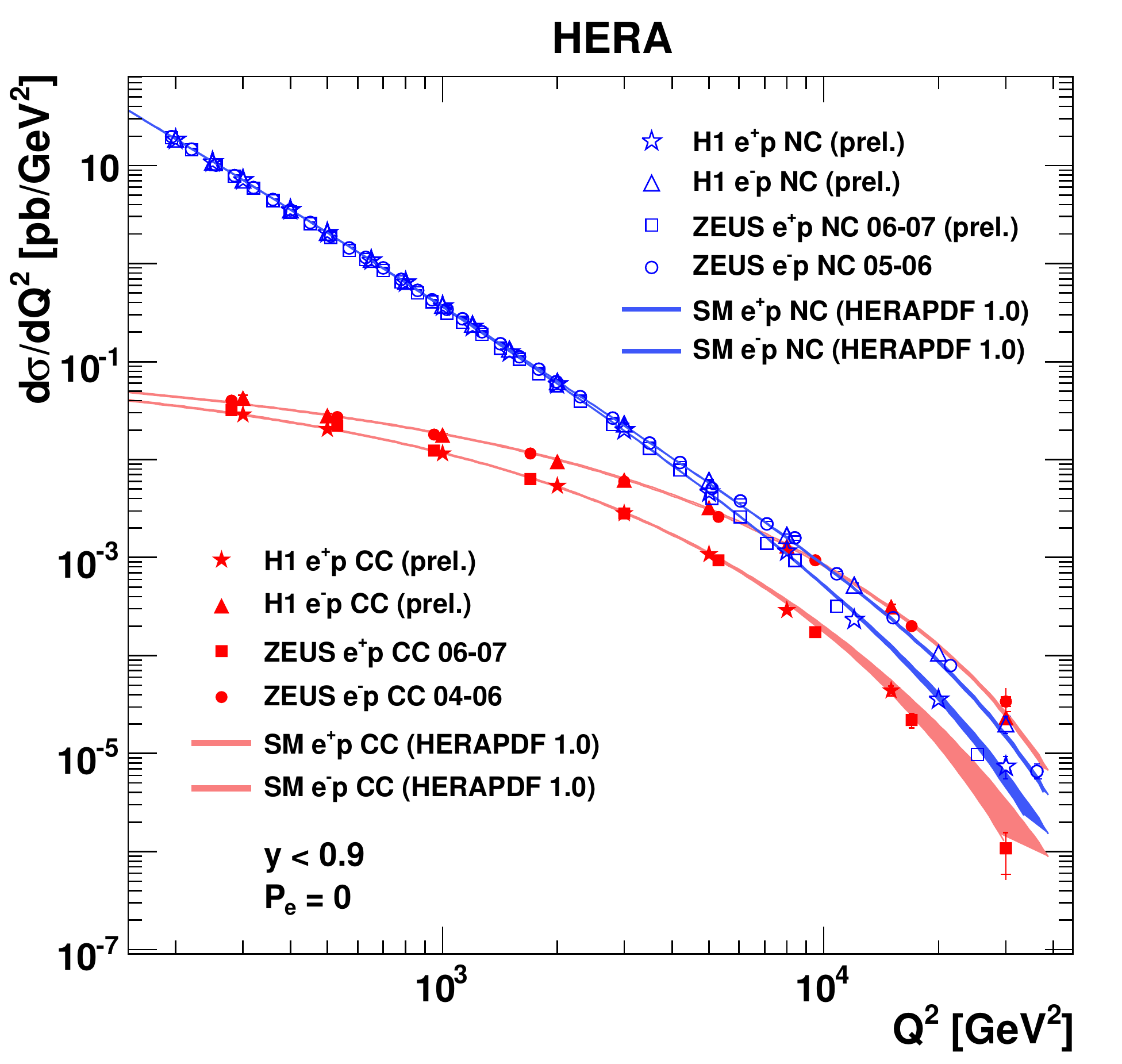}
\vspace{-4mm}
\parbox{15.5cm}{\caption{\label{fig:EWuni}Neutral current (blue) and charged current (red) differential cross 
sections in $e^\pm p$ collisions with respect to the virtuality $Q^2$. Figure from \cite{zeus} (with permission). Electroweak unification 
happens above $M_{\rm EW}\sim 100$ GeV,  
see text for details.}}
\end{center}
\end{figure}

The here described {\it GWS model} merged the electromagnetic and weak interactions together in the framework of a compact 
gauge theory. This was a big step towards having a more and more unified and simple picture of nature. It was developed
some time before the direct experimental discovery of several of its ingredients. Due to the chosen gauge symmetry group 
(\ref{eq:GSM}), which is broken according to the pattern (\ref{eq:break}), it predicted the existence 
of neutral weak currents, in addition to the well known charged currents. These neutral currents have been discovered by 
the Gargamelle experiment some years later in 1973. The profound knowledge about the charged current interactions, like 
the maximally P violating structure and Cabibbo universality of the couplings, was very helpful in finding the correct theory. 
The apparent weakness of the weak force with respect to the electromagnetic force at low momentum transfer is not due to a 
small coupling constant (see (\ref{eq:coupZ})) but due to 
the suppression of the corresponding propagators by the large $W^\pm$- and $Z$-boson masses. This suppression is lifted above the electroweak scale 
$M_{\rm EW}\sim 100$ GeV ($\sim m_W \sim m_Z$). This feature of electroweak unification has been tested successfully at HERA. In Figure \ref{fig:EWuni}
the measured $e^\pm p$ neutral-current and charged-current differential cross sections with respect to the virtuality $Q^2$ of the exchanged 
gauge boson are shown in blue and red, respectively.
At low $Q^2$, the neutral-current cross section is dominated by photon exchange and the charged-current cross section is
strongly suppressed by the large $W^\pm$-boson mass. At high $Q^2\sim m_W^2$, however, the exchange of massive gauge bosons becomes 
comparably important and the neutral and charged cross sections become similar, indicating electroweak unification.

The non-trivial relations of the GWS model discussed above, related to the special way of breaking electroweak symmetry spontaneously (without 
destroying the gauge invariance of the Lagrangian explicitly), diminish the number of parameters in the gauge sector and provide 
the possibility to test the given mechanism of symmetry breaking. 
For example, the SM prediction for the parameter 
\beq
\label{eq:rhopar}
\varrho\equiv m_W^2/(m_Z^2 \cos^2\theta_w)\,,
\eeq 
encoding the relation between gauge couplings and masses ({\it c.f.} (\ref{eq:mWmZ})), is $\varrho_{SM}=1$, at the tree level. Quantum corrections within the SM will alter this 
relation, as will NP contributions to the electroweak gauge boson masses. A measurement of the $\varrho$ parameter can thus 
test the SM and constrain BSM physics, see Section~\ref{sec:mod}. Note that the massive $W^\pm$ and $Z$ bosons, predicted 
within the SM, have only been discovered experimentally in 1983 by the UA1 \cite{Arnison:1983rp,Arnison:1983mk} and UA2 
\cite{Banner:1983jy,Bagnaia:1983zx} collaborations at the CERN Super Proton Synchrotron, which had been upgraded to a two-beam 
collider in order to reach sufficient center of mass energies (Nobel Prize in Physics 1984 for Rubbia and van der Meer). 
Their masses were found to lie just in 
the region suggested by the relation (\ref{eq:mWmZ}), which provides a nontrivial experimental confirmation that some mechanism 
like described above is at work. The reference values compiled in Appendix~\ref{app:ref} lead to $\varrho_{exp}\approx 1.01$, where
the one per cent deviation could be due to loop corrections to the relation (\ref{eq:mWmZ}). The fact that a single value of 
the weak mixing angle accounts for all kinds of different observables of the electroweak theory (including interactions with different 
fermions), provides strong evidence for the existence of an underlying, spontaneously broken, gauge symmetry in nature \cite{Peskin:1995ev}.

\begin{figure}[!t]
\begin{center}
\includegraphics[height=2.6in]{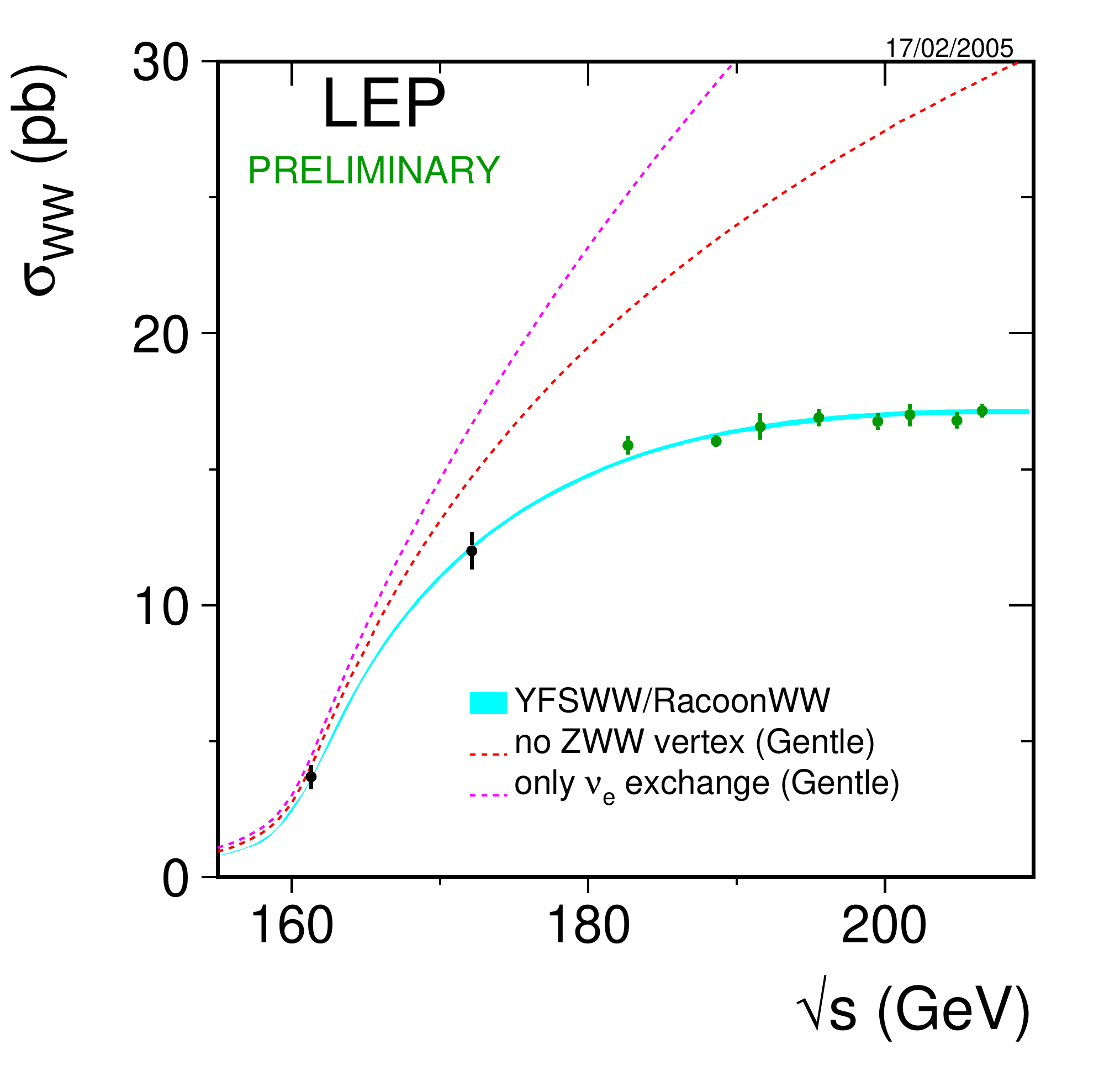}
\vspace{-8mm}
\parbox{15.5cm}{\caption{\label{fig:ZWW}Cross section $\sigma(e^+e^-\to W^+W^-)$, measured at LEP2.
The $ZWW$-vertex contribution, with exactly the relative strength as dictated by gauge invariance, is needed in order to preserve perturbative unitarity. 
Figure from \cite{Alcaraz:2006mx} (with permission). See text for details.}}
\vspace{2mm}
\end{center}
\end{figure}

The $SU(2)_L\times U(1)_Y$ gauge structure of the SM has been successfully tested at the Large Electron-Positron Collider (LEP) 
in trilinear gauge-boson couplings as well as in couplings to fermions in $e^+e^-\to W^+W^-$. Several Feynman graphs
contributing to this process (neutrino exchange, $Z$-boson exchange, photon exchange) have to add up in a certain way, 
not to lead to a cross section that raises proportional to the squared center of mass energy $s$ and violates (perturbative) unitarity 
at around $\sqrt{s}=1$\,TeV. In the SM, the necessary cancellation is guaranteed by Ward identities, following just from gauge invariance 
with respect to the SM gauge group. The resulting, well behaved, SM cross section agrees very well with experiment, see Figure \ref{fig:ZWW}.
In fact, the only theories of massive vector bosons that do not have a violent high-energy behavior are those arising from a spontaneously 
broken gauge symmetry \cite{Cornwall:1974km}.

\subsubsection*{Custodial Symmetry and the $W^\pm$-Boson Mass}

In summary, we can be quite certain that the $W^\pm$ and $Z$ bosons arise from a spontaneously broken $SU(2) \times U(1)$ 
gauge symmetry with masses fulfilling the relation (\ref{eq:mWmZ}) to good accuracy.
One can ask the question, if this already provides evidence for the existence of the SM Higgs mechanism, \ie, if this 
relation is unique to the symmetry breaking by a scalar SM-Higgs doublet, or
if other mechanisms are possible that feature the same breaking pattern including the relation (\ref{eq:mWmZ}). 
Being well tested, any model of NP should reproduce this mass relation to reasonable accuracy.

As a consequence, we will have a closer look on how this relation comes about and how much it depends on the mechanism 
that breaks $SU(2)_L \times U(1)_Y$.
For that purpose, we construct the most general mass matrix for the electroweak gauge bosons, consistent with the breaking 
pattern $SU(2)_L \times U(1)_Y \to U(1)_{EM}$. In the basis corresponding to $(W_\mu^1,W_\mu^2,W_\mu^3,B_\mu)$, we
make the ansatz
\beq
\label{eq:custoM}
   M^2 =
   \left( \begin{array}{cccc}
    m_W^2 & 0 & 0 & 0 \\
     0 & m_W^2  & 0 & 0 \\
      0 & 0 & m_3^2 & m^2 \\
       0 & 0 & m^2 & m_0^2 \\
   \end{array} \right)\,.
\eeq
Note that due to the $U(1)_{EM}$ gauge invariance, the coefficients of the quadratic terms $W_\mu^1 W^{\mu\,1}$ and $W_\mu^2 W^{\mu\,2}$ 
must be identical $m_1^2=m_2^2\equiv m_W^2$. Electromagnetic gauge invariance together with hermiticity is also responsible for the zero entries in 
(\ref{eq:custoM}). The lower-right block represents the most general symmetric (squared) mass matrix for the corresponding
two-component system. Its entries can be further specified by demanding one eigenvalue to be zero, corresponding to the massless photon,
and a non-vanishing eigenvalue, denoted by $m_Z^2$. 
It turns out that, in order to arrive at the sought mass relation (\ref{eq:mWmZ}), one needs $m_3^2=m_W^2$, see Appendix~\ref{app:cus}. 
Thus, the mass matrix (\ref{eq:custoM}) has to fulfill
\beq
\label{eq:customa}
m_1^2=m_2^2=m_3^2=m_W^2\,.
\eeq
This relation holds for the SM Higgs mechanism due to the fact that the Higgs boson is a scalar $SU(2)_L$ doublet.
It is related to an accidental global $SU(2)_L\times SU(2)_R$ symmetry in the Higgs Lagrangian, broken by the Higgs VEV 
according to
\beq
\label{eq:LRbreaka}
SU(2)_L\times SU(2)_R  \longrightarrow SU(2)_V \,,
\eeq
which leaves its imprint in the gauge boson mass matrix, see Appendix~\ref{app:cus}.
The residual $SU(2)_V$ symmetry guarantees that the mass relation (\ref{eq:customa}) is fulfilled.
{\it Any} sector that breaks the SM gauge group according to (\ref{eq:break}) and exhibits 
this so-called {\it custodial symmetry} \cite{Sikivie:1980hm} thus has the potential to lead to a viable phenomenology in the 
gauge boson sector. For a more detailed discussion see Appendix~\ref{app:cus} and \cite{Peskin:1995ev,Grojean:2005ud}.
In Section~\ref{sec:WED}, a version of this symmetry will be used to protect 
the RS setup from large corrections to the $T$ parameter, related to a breaking of the relation (\ref{eq:mWmZ}) (see
the end of this section). 
Expressing this relation through a symmetry has the additional advantage that one can easily identify corrections to it, 
arising \eg through symmetry breaking interactions at the quantum level. In the SM, the custodial symmetry is broken at the
loop level through Yukawa interactions and interactions with the $U(1)_Y$ gauge boson and thus $\varrho_{SM}\neq 1$ beyond
the tree level, as mentioned before. Analysing these interactions can give a handle on the size of the effect. On the other hand, 
measuring $\varrho$ very accuratly could also reveal (small) deviations from the SM prediction.
\begin{figure}[!t]
	\begin{center}
	\mbox{\includegraphics[height=2.45in]{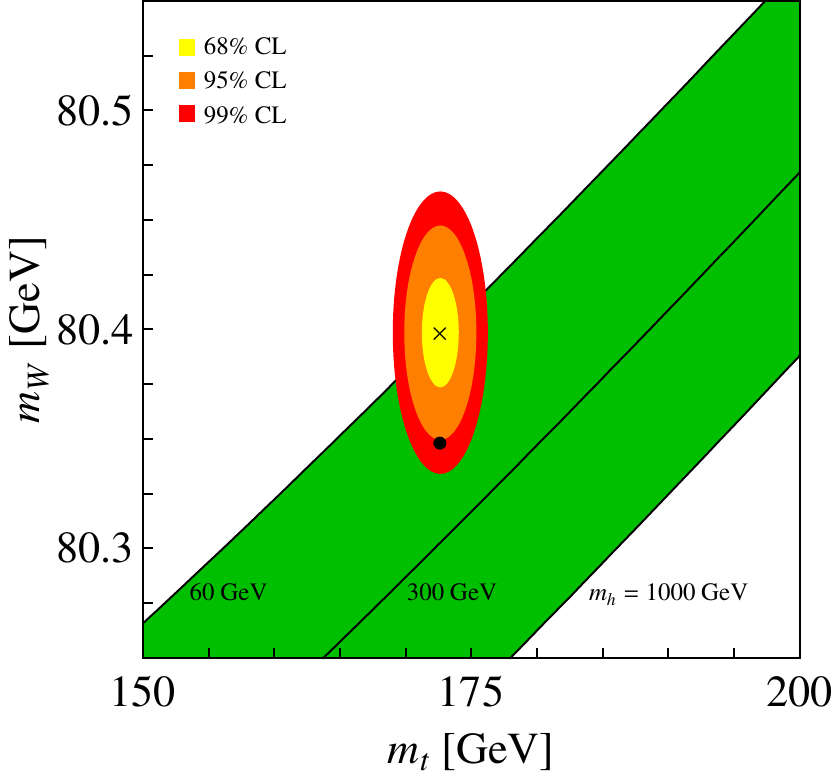}}
	\vspace{-2mm}
\parbox{15.5cm}{\caption{\label{fig:MWind}
Regions in the $m_t$--$m_W$ plane following from the direct and indirect determination of $m_W$. The colored ellipses 
correspond to 68\%, 95\%, and 99\% probability, and are obtained from the direct measurements of $m_W$ and $m_t$ at LEP2 and 
the Tevatron. The black dot shows the SM prediction based on $G_F$ for our reference SM input, 
whereas the green (gray) shaded band shows the SM expectation for values of the Higgs-boson mass $m_h\in [60,1000]$
\,GeV. See text for details.}}
\end{center}
\end{figure}

Another possibility to test the SM is to extract $m_W$ {\it indirectly} from a measurement of $e, \theta_w$ and the Fermi coupling 
$G_F$, measured in muon decay. According to the SM relations (assuming no BSM physics in muon decay) one arrives at
\beq
\label{eq:Mindr}
	(m_W^2)_{\rm indirect}\equiv \frac{e^2}{4 \sqrt{2} \sin^2\theta_w\, G_F}.
\eeq 
This extraction can be compared to the direct reconstruction of the $W^\pm$-boson mass. The results are shown in Figure
\ref{fig:MWind} in the $m_W - m_t$ plane, where $m_t$ is the mass of the top quark. 
The colored ellipses indicate the regions of 68\%, 95\%, and 99\% confidence level (CL), following from the direct 
measurements at LEP2 \cite{LEPEWWG:2005ema} and the Tevatron \cite{Nakamura:2010zzi}. 
The results obtained from (\ref{eq:Mindr}) depend, due to loop effect, significantly on the Higgs mass and are shown by the green (gray) band for 
$m_h\in [60,1000]$\,GeV.
The central values and 1\,$\sigma$ errors of the direct and indirect $W^\pm$-mass determinations are given by
\beq\label{eq:mwmasses}
   m_W = (80.399\pm 0.023) \, \mbox{GeV} \,, \qquad 
   (m_W)_{\rm indirect} = (80.348\pm 0.015) \, \mbox{GeV} \,.
\eeq
The value of $(m_W)_{\rm indirect}$ has been derived with the help of {\tt{ZFITTER}} \cite{Bardin:1999yd, 
Arbuzov:2005ma} and we have used the SM reference values for $\Delta\alpha^{(5)}_{\rm had}(m_Z)$, $\alpha_s(m_Z)$, $m_Z$, 
and $m_t$, as well as $m_h=150$\,GeV, as given in Appendix~\ref{app:ref}. 
The $2\,\sigma$ shift of around 50\,MeV between the two measurements, which are expected to coincide within the SM, could
be just a statistical fluctuation or already a hint for NP, like \eg for the RS model, as we will discuss in Section
\ref{sec:mod}.

\subsubsection*{Fermion Masses and Theoretical Constraints on the Higgs-Boson Mass}
\label{sec:theoH}

Besides giving masses to gauge bosons, the Higgs mechanism can also provide gauge invariant mass terms for the fermion
fields via Yukawa couplings. The corresponding Lagrangian reads
\beq
\label{eq:Lyuk}
	\mathcal{L}_{\rm Yukawa}= - \bar{Q}_L \Phi^c\,\bm{Y}_u u_R -  \bar{Q}_L \Phi\,\bm{Y}_d\,d_R 
	-  \bar{E}_L \Phi\,\bm{Y}_e\,e_R  + {\rm h.c.}\,,
\eeq
where the Yukawa couplings $\bm{Y}_{u,d,e}$ are $3 \times 3$ matrices in flavor space. The left-handed fermion-doublets and the 
Higgs doublet combine to $SU(2)$ singlets, which makes ${\cal L}_{\rm Yukawa}$ gauge invariant. After EWSB, (\ref{eq:Lyuk})
will contain mass terms for the charged SM fermions, see Section~\ref{sec:SMflavor}. Note that in the SM it
is possible to give masses to up- and down- type quarks with a single Higgs doublet by defining $\Phi^c \equiv i 
\sigma_2 \Phi^* $, whereas this is not the case for models of supersymmetry (SUSY), which will be briefly 
reviewed later. Here, at least two $SU(2)$ doublet scalars are needed. 
Let us stress again that the SM Lagrangian including ${\cal L}_{\rm Higgs}$ and ${\cal L}_{\rm Yukawa}$ is still gauge invariant under the 
full gauge group (\ref{eq:GSM}). The gauge invariance of the full theory is just encrypted, looking at the ground 
state. So it would be more correct to say that the symmetry is hidden, rather than broken.
While symmetries are very interesting, spontaneously broken symmetries are even more exciting, as they allow for a 
richer phenomenology, while still possessing an underlying symmetry, as we have seen above.
For additional details on the Higgs mechanism see \eg the literature cited at the beginning of this section and \cite{Djouadi:2005gi}
as well the one on QFT and the SM given in the last footnote. The construction of such a mechanism was quite remarkable, since what 
we observe in nature are solutions following from a Lagrangian in the broken phase and not the Lagrangian itself. 
Since the SM Higgs mechanism of giving masses to gauge bosons and fermions seems to be quite promising, we should try to discover 
the Higgs boson and to test this sector of the SM directly. This is one of the main tasks of the LHC which started 
operation in 2008. However, already at LEP and at the Tevatron, searches for the Higgs boson have been performed, excluding
several ranges for the mass of a SM Higgs, see below. 
Before introducing briefly the main production channels for Tevatron and the LHC, as well as the relevant decay channels, 
let us have a look at theoretical bounds on the Higgs-boson mass within the SM (see \eg \cite{Djouadi:2005gi} and references therein). 
To this end, we start with constraints from {\it unitarity} of longitudinal $W^\pm$-boson scattering $W_L^+ W_L^- \to W_L^+ W_L^-$. 
As indicated before, it turns out that the presence of a Higgs sector, keeping the SM with {\it massive} gauge 
bosons gauge invariant, is essential for an appropriate high energy behavior of the theory. Without such (or a similar) sector, perturbative
unitarity will also be violated in $W_L^+ W_L^- \to W_L^+ W_L^-$ around the TeV scale, due to a growth of the scattering amplitude with the center 
of mass energy $\sqrt{s}$. This process puts an upper bound on the mass of the physical scalar of the Higgs sector, which contributes to this process
as a virtual particle, meaning that it has to complete the SM below a certain maximal scale \cite{Lee:1977eg}. 
Performing a partial wave analysis of the scattering amplitude and using the optical theorem, see Appendix~\ref{app:HB}, one arrives at
\beq
\label{eq:HB}
m_h\lesssim870\,{\rm GeV}\,.
\eeq
This limit becomes slightly stronger if additional scattering channels of massive bosons are included,
decreasing to a value slightly above 700\,GeV \cite{Djouadi:2005gi} .
Note that the tree-level derivation presented in Appendix~\ref{app:HB} relies on perturbativity of the theory. For large Higgs masses the Higgs self 
coupling $\lambda$ becomes strong and higher order corrections could in principle restore unitarity for a Higgs mass
slightly above the TeV scale, at the cost of ending up with a strongly coupled theory and loss of the predictive power
of the SM.

Additional constraints on the Higgs mass can be obtained by studying the Higgs potential in (\ref{eq:LHiggs}) directly.
The quartic coupling $\lambda$ depends due to quantum corrections on the energy scale $\mu$, see Appendix~\ref{app:HB}. 
The requirement for $\lambda$ not to exhibit a Landau pole nor to become negative below a scale $\mu_c$ leads to 
the constraint \cite{Djouadi:2005gi,Hambye:1996wb} 
\beq
\label{eq:theomhb}
\begin{split}
\mu_c\gtrsim10^{16}\, {\rm GeV}\ \Rightarrow 130\,{\rm GeV} \lesssim m_h \lesssim 180\,{\rm GeV}\\
\mu_c\gtrsim1\,{\rm TeV}\ \Rightarrow 50\,{\rm GeV} \lesssim m_h \lesssim 800\,{\rm GeV}\,,
\end{split}
\eeq
where the idea of Grand Unification set the first cutoff scale above (see below). The corresponding leading order (LO) calculations 
are performed explicitly in Appendix~\ref{app:HB}. 
\begin{figure}[!t]
\vspace{-4mm}
	\begin{center}
	\mbox{\includegraphics[height=2.4in]{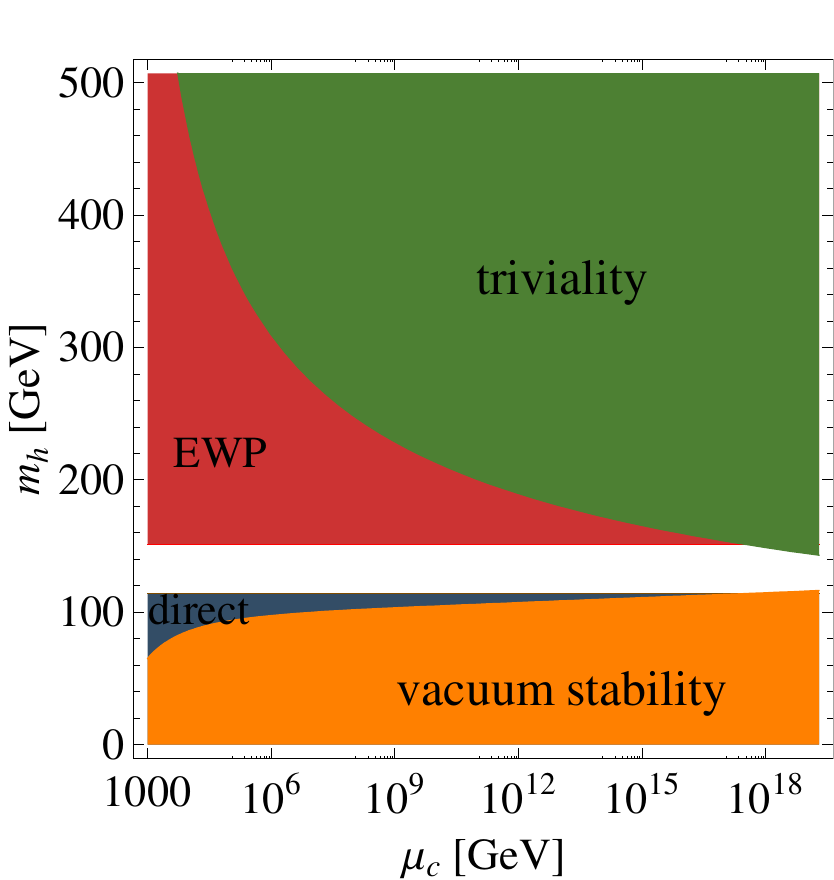}}
	\vspace{-3mm}
\parbox{15.5cm}{\caption{\label{fig:trivstab} Bounds on the Higgs-boson mass from triviality and vacuum stability, as well as
from electroweak precision tests (EWP) and the direct searches at LEP (@\,~95\%\,~CL). The colored regions are excluded. 
Recent bounds from Tevatron and the LHC are not shown for presentation purposes. See text for details.}}
\vspace{-2mm}
\end{center}
\end{figure}
The upper bound on the Higgs mass in (\ref{eq:theomhb}), resulting from the requirement to avoid a Landau pole,
is called {\it triviality bound}. If this pole would appear in the range of validity of
the theory, one would have to set $\lambda\equiv0$ in order to avoid this pole, rendering the Higgs sector non-interacting
and thus trivial. The lower bound, coming from demanding $\lambda>0$, is called {\it vacuum stability bound} since it has 
to be fulfilled in order for the Higgs potential to be bounded from below, as needed for the vacuum to be stable, see (\ref{eq:LHiggs}).
For more details see Appendix~\ref{app:HB}. Requiring the Higgs sector of the SM to be well behaved for the whole possible 
energy range of the theory, up to the Planck scale $\mu_c\sim M_{\rm Pl}$, cuts the allowed region for $m_h$ to 
a small stripe, corresponding to a light Higgs boson. However, allowing for a low Landau pole at around a TeV, meaning that the SM only 
makes sensible predictions below that scale, allows for a heavy Higgs boson $m_h \sim 800$\,GeV.

Again, this has to be taken with caution, since if the quartic coupling becomes large, higher order terms can alter the
running and possibly avoid the emergence of a Landau pole. However, non-perturbative calculations on the lattice also lead to 
a bound of $m_h<640$\, GeV \cite{Hasen} in a pure scalar theory (assuming a SM cutoff at $2\pi\, m_h$), which shows a reasonable
agreement with the perturbative estimations.
Further note that the bounds from vacuum stability can be relaxed if the vacuum turns out to be metastable \cite{Anderson:1990aa}.
The emerging picture is presented in Figure \ref{fig:trivstab} which shows the constraints from triviality and vacuum stability 
considerations, following from the calculations presented in Appendix~\ref{app:HB}, together with those from electroweak
precision measurements and the direct searches at LEP, see below. 
It is mandatory to emphasize that all the constraints derived above hold for the SM. Within BSM models, they can be altered. For 
example in RS models with a SM-type Higgs sector, studied in the main part of this thesis, it could be possible that unitarity can 
be preserved for higher values of $m_h$ than 870\,GeV due to an exchange of light Kaluza Klein gravitons \cite{Grzadkowski:2006nx}. 
Moreover, NP could alter the running of the Higgs quartic coupling and thus the corresponding triviality and vacuum stability bounds within BSM 
models can be different. Studying theoretical constraints on the Higgs mass within the RS setup (with bulk fields) in more detail,
beyond a rough estimation of an upper bound of $m_h\lesssim 1$\, TeV, would be interesting but is beyond the scope of this thesis. Nevertheless, in 
the light of the discussion of bounds in dependence on the cutoff of the theory, one should already mention that the cutoff of 
this model for processes involving couplings to the Higgs sector will be $\Lambda_{UV}\sim \ord$(some TeV), see (\ref{eq:RSIRcut}).

In summary, if we do {\it not} find a Higgs in the mass range of $130\,{\rm GeV} \lesssim m_h \lesssim 180\,{\rm GeV}$ (\ref{eq:theomhb}) this means that
the SM is probably not valid up to the scale of Grand Unification and that at least {\it new phenomena} are expected to appear well below 
that scale. Also if we find an Higgs in the mass range suggested by the SM, it is still important to test, whether the scalar sector really
behaves exactly like the SM Higgs sector. For that purpose, one has to determine further properties of the Higgs boson, like \eg its branching fractions,
or look at the production cross section. One main purpose of this thesis is to study these observables in the context of RS
models, to have an idea what to expect in this BSM scenario. We will find that {\it large corrections} with respect to the SM are possible even 
for high NP scales, see Section \ref{sec:RSHiggs}. These results help in discriminating between different models, should one see deviations from the SM expectations in the Higgs sector, and exhibit an (indirect) sensitivity to large scales within this sector.
Furthermore they show that, although if one does not find a Higgs boson with some years of LHC data, this does not mean that there is
no standard-type Higgs sector but merely could indicate that there is BSM physics that alters this sector such that the Higgs boson is much
more difficult to discover.
\begin{figure}[!t]
	\begin{center}
	\mbox{\includegraphics[height=2.55cm]{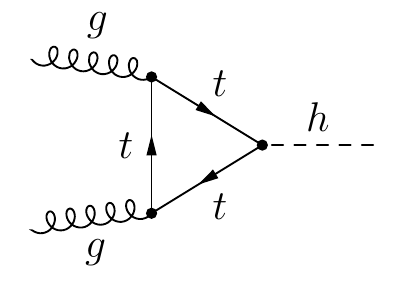}}
	\vspace{-1cm}
\parbox{15.5cm}{\caption{\label{fig:glufus}Leading-order contribution to Higgs-boson production via gluon fusion.}}
\end{center}
\vspace{0.3cm}
\end{figure}

\subsubsection*{Higgs-Boson Production and Decay in the SM}

The couplings of the physical Higgs boson $h$ to gauge bosons and fermions can easily be obtained from the covariant derivative (\ref{eq:covder})
and the Yukawa couplings (\ref{eq:Lyuk}). We will work out the Higgs couplings in more detail in the context of RS models in 
Section~\ref{sec:HcouplingsRS}.
Nevertheless, we already want to give a brief overview of the main production cross sections and branching fractions in the SM at this point.
Due to the particular mechanism of giving masses to the SM fields, the couplings of $h$ to these fields within the SM are directly 
proportional to their {\it masses}. This means also that there are no flavor-changing tree-level couplings to the Higgs boson - 
a fact which, however, can change, if additional contributions to the fermion masses arise (see Section~\ref{sec:HcouplingsRS}). 
In the SM, Higgs decays to heavy particles are generically enhanced. 
For the same reason, the direct production due to couplings to the light quark content of protons or due to electrons in the initial 
state are suppressed. The main production mechanisms at hadron colliders such as the Tevatron 
and the LHC is gluon fusion, which receives the main contribution from a heavy top-quark loop, see Figure \ref{fig:glufus}. It can be 
described by the $D=5$ operator $h/v\, G_{\mu\nu}^a G^{a\,\mu\nu}$ (see Section~\ref{sec:RSHiggs} and Appendix~\ref{app:EFT}), which 
interestingly remains valid for $m_h>m_t$, see \eg \cite{Plehn:2009nd}. Subleading production channels
are associated $W^\pm$-boson production, $q \bar q^{\hspace{0.25mm} \prime} \to W^\ast \to Wh$, which is the only channel that 
could allow for a Higgs discovery at the Tevatron, as well as weak gauge-boson fusion, $q q^{(\prime)} \to qq^{(\prime)} V^{\ast}
V^{\ast} \to qq^{(\prime)} h$ with $V = W,Z$, which is known to be quite useful for a potential Higgs discovery at the LHC.
An overview of the relevant formulae to compute the corresponding cross sections and the Higgs branching fractions within
the SM can be found in \cite{Djouadi:2005gi}. 

\begin{figure}[!t]
\begin{center} 
\hspace{-2mm}
\mbox{\includegraphics[height=2.5in]{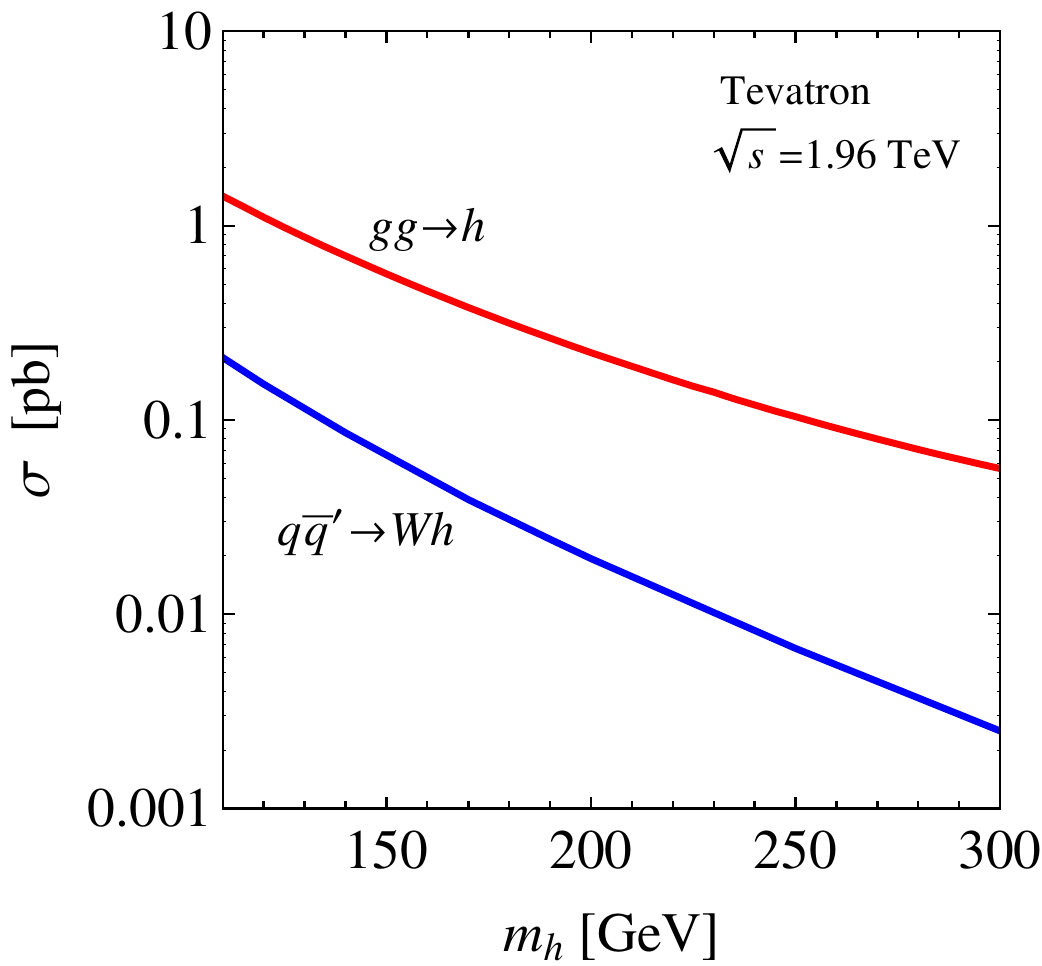}} 
\hspace{4mm}
\mbox{\includegraphics[height=2.5in]{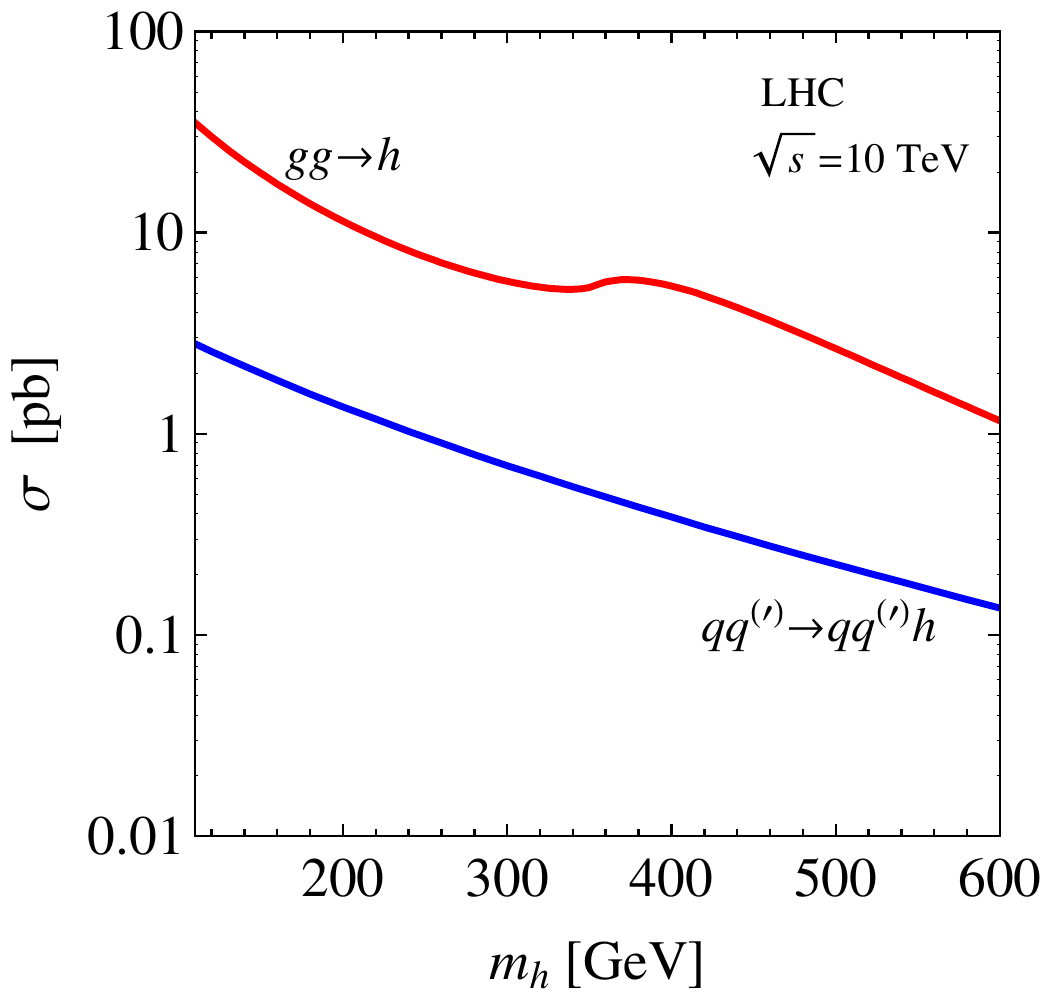}}
\parbox{15.5cm}{\caption{\label{fig:HprodSM} Main Higgs-boson
    production cross sections at the Tevatron (left) and the LHC
    (right) for center-of-mass energies of $\sqrt{s} = 1.96 \, {\rm
      TeV}$ and $\sqrt{s} = 10 \, {\rm TeV}$, respectively. For the
    Tevatron the plot displays gluon-gluon fusion (red) and associated
    $W^\pm$-boson production (blue), while for the LHC gluon-gluon (red) 
    and weak gauge-boson fusion (blue) are shown.}}
\end{center}
\end{figure}

\begin{figure}[!t]
\vspace{-2mm}
\begin{center}
\mbox{\includegraphics[width=11cm]{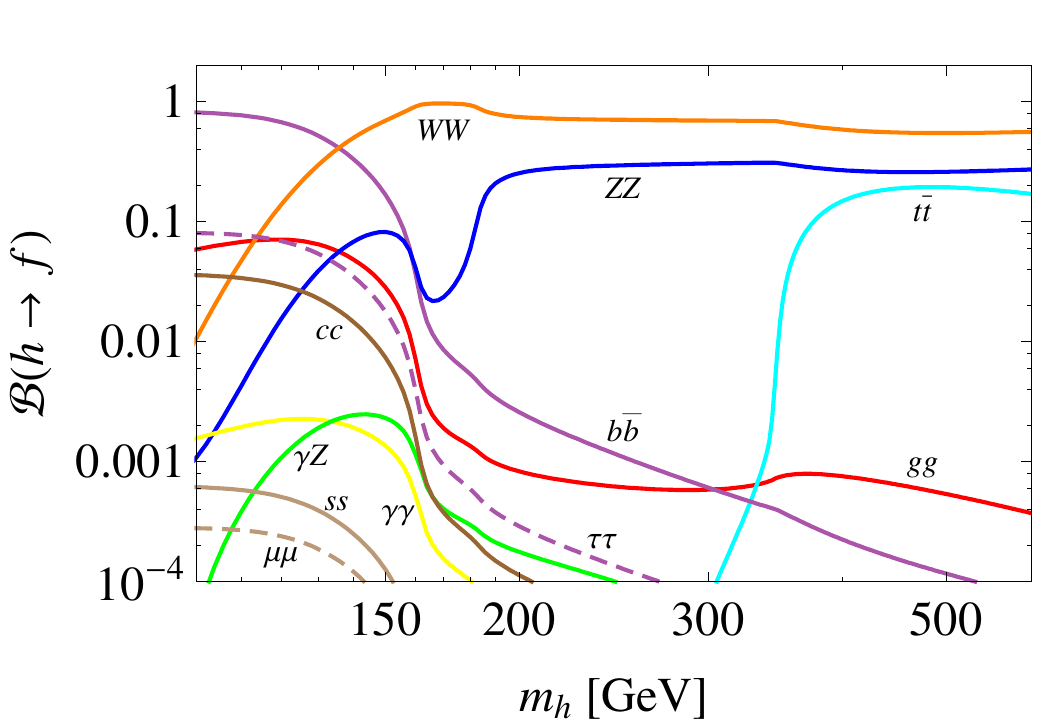}}
\parbox{15.5cm}{\caption{\label{fig:SMele} Branching fractions for $h \to
    f$ as functions of the Higgs-boson mass. Channels with fractions of
    less than $10^{-4}$ are not shown.}}
\end{center}
\end{figure}

The SM results for the Higgs-boson production cross sections at
the Tevatron and LHC for center-of-mass energies of $\sqrt{s} = 1.96 \,{\rm TeV}$ and $\sqrt{s} = 10 \, {\rm TeV}$ are shown
in Figure \ref{fig:HprodSM}. For the gluon-fusion channel, results from \cite{Ahrens:2008nc} are used, which combines the 
next-to-next-to-leading order (NNLO) corrections \cite{Harlander:2002wh, Anastasiou:2002yz, Ravindran:2003um} in fixed order
perturbation theory with a resummation of both threshold logarithms from soft-gluon emission \cite{Moch:2005ky, Laenen:2005uz, 
Idilbi:2005ni, Ravindran:2006cg,Idilbi:2006dg} 
and terms of the form $(N_c \pi \alpha_s )^n$ \cite{Ahrens:2008qu}, where $N_c$ is the number of colors. Moreover, the {\tt MRST2006NNLO} parton distribution functions \cite{Martin:2007bv} 
and the associated normalization $\alpha_s(m_Z) = 0.1191$ for the strong coupling constant have been employed.
The SM predictions for the subleading production channels have been obtained from \cite{Aglietti:2006ne}.
The branching fractions of the Higgs boson within the SM are depicted in Figure \ref{fig:SMele}.
The plot confirms the expectation that couplings to heavy particles are most important (for $h\to gg$ corresponding
to the top quark in the loop). The shown results have been 
calculated with the help of {\tt HDECAY} \cite{Djouadi:1997yw}.\footnote{Expect for the parameters listed in
Appendix~\ref{app:ref}, the original input file of {\tt HDECAY} version 3.51 has been used.} We will compare these results to
the predictions of the RS setup in Section~\ref{sec:RSHiggs}.

\begin{figure}[!t]
\begin{center}
\mbox{\includegraphics[width=10.2cm]{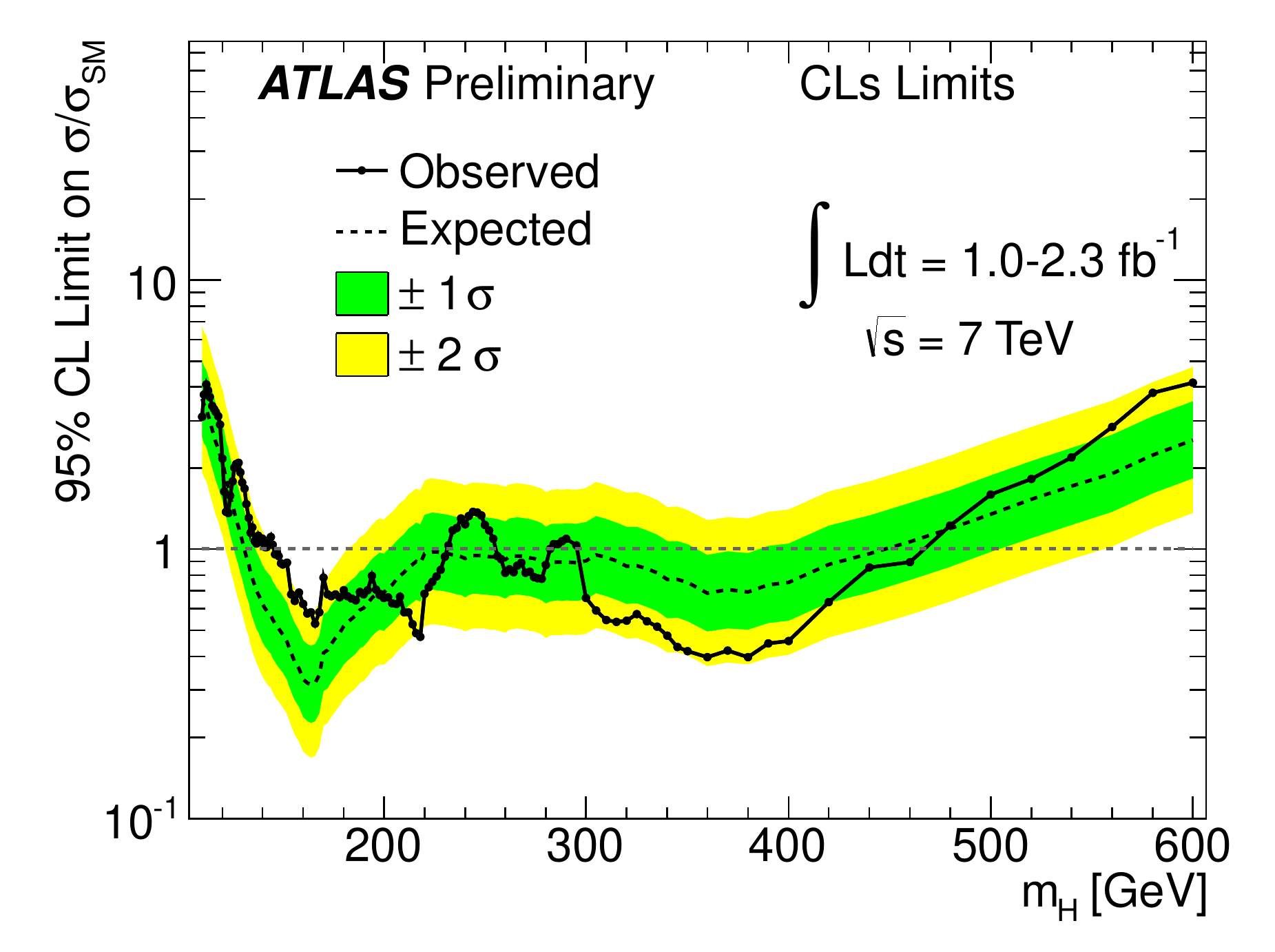}}
\vspace{-3mm}
\parbox{15.5cm}{\caption{\label{fig:LHCHiggs} Higgs search at the LHC. The solid line shows the combined upper limit on the SM Higgs boson 
production cross section divided the expected cross section in the SM as a function of $m_h$. The limit corresponds to $95\%$\, CL. 
The median expected limit in the absence of a signal is shown by the dotted line, while the green and yellow bands indicate the corresponding 
68\% and 95\% regions. Figure from \cite{AHiggs}.}}
\end{center}
\end{figure}

The Higgs boson has been searched in various channels at the Tevatron and LEP.
Currently the LHC is the main machine trying to discover the Higgs.
A direct lower bound on $m_h$ has been obtained from the LEP searches. It reads \cite{Barate:2003sz}
\beq
m_h>114.4\,{\rm GeV}\ @\, 95\%\, {\rm CL}\, .
\eeq
Lately, Tevatron excluded a region \cite{CDF:2011cb}
\beq
m_h\notin[156,177]\, {\rm GeV}\ @\, 95\%\, {\rm CL}\, .
\eeq
A very recent exclusion window has been given by the LHC. The ATLAS results, obtained from $\sim 2 {\rm fb}^{-1}$ of $pp$ collisions at 
$\sqrt{s}=7$\,TeV, are shown in Figure \ref{fig:LHCHiggs}. In that plot, the ratio of the limit on the Higgs boson production cross section 
to the expected SM cross section is plotted in dependence on the Higgs-boson mass. Values below 1 indicate that a SM Higgs boson is 
excluded in the corresponding mass region at $95\%$\,CL. However, for all these limits it is important to recall that, if the 
production cross sections and/or branching fractions are changed with respect to the SM, the corresponding 
exclusion plots will also change. The ATLAS exclusion for the SM reads \cite{AHiggs}
\beq
m_h\notin [149,222] \cup [276,470]\, {\rm GeV}\ @\, 95\%\, {\rm CL}\, .
\eeq
Beyond that, the plot shows a $2\,\sigma$ excess in the low mass region $m_h\in[120,140]$\,GeV. However, the significance is not big enough
to already speak of an evidence for the Higgs boson. More statistics will be needed to clarify the situation.

\subsubsection*{Electroweak Precision Tests}
\begin{figure}[!t]
\vspace{-4mm}
\begin{center}
\mbox{\includegraphics[width=7.6cm]{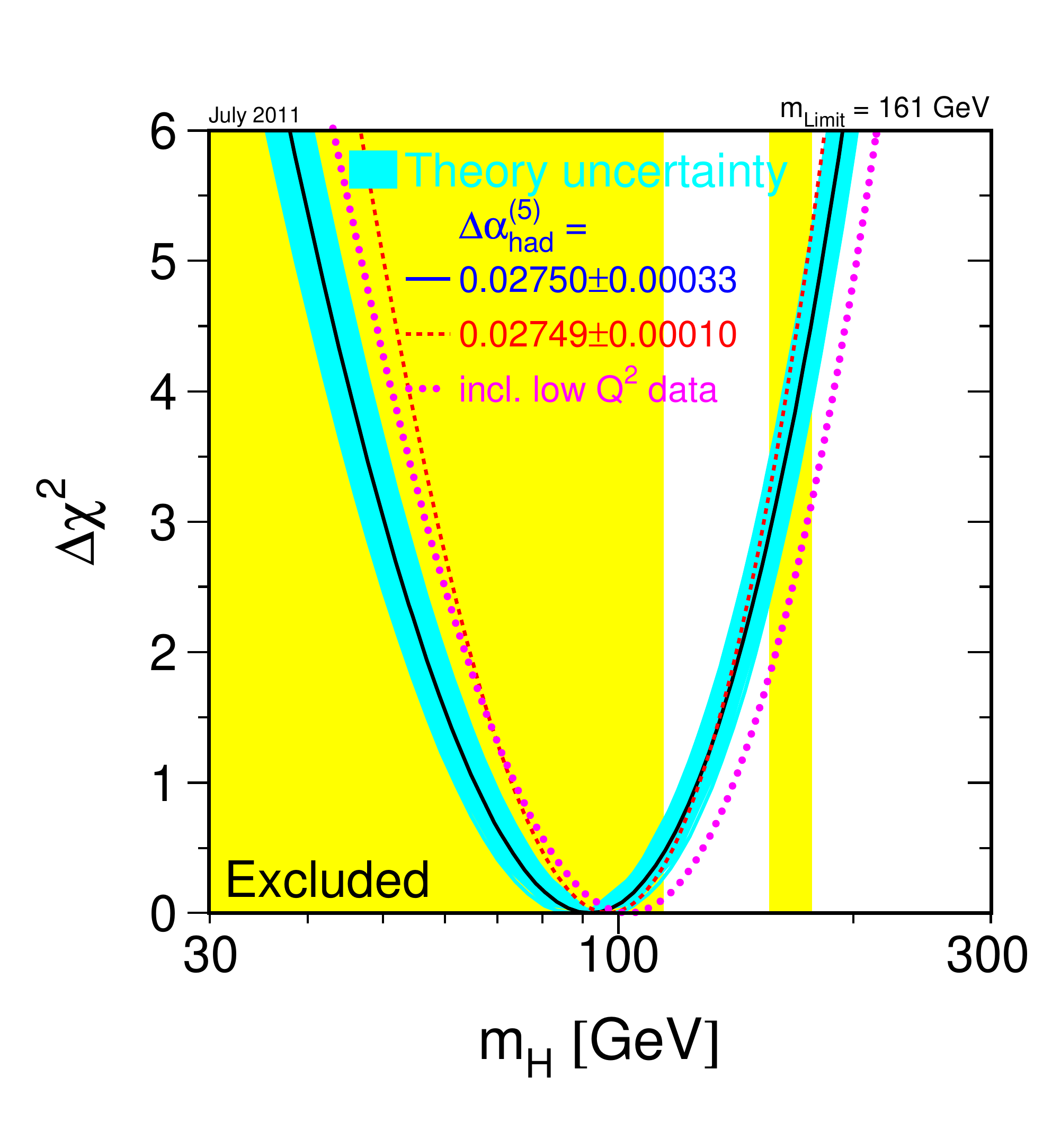}}
\end{center}
\begin{center}
\vspace{-0.7cm}
\parbox{15.5cm}{\caption{\label{fig:blueband} $\Delta \chi^2$ of the SM-fit, obtained from
precision electroweak measurements, as a function of the Higgs-boson mass. The direct exclusion limits from LEP and the Tevatron are indicated by 
the yellow regions. Figure from \cite{:2010vi} (with permission)~. See text for details.}}
\end{center}
\vspace{-3mm}
\end{figure}

In addition to these direct limits, one can get a handle on the Higgs-boson mass indirectly from precision measurements. The Higgs boson (if it exists) 
will enter many processes as a virtual particle and thus potentially leaves an imprint in various observables, which thus depend on $m_h$ (see \eg 
Figure \ref{fig:MWind}). The situation is summarized in Figure \ref{fig:blueband}, which shows the $\Delta \chi^2$ of the SM-fit, obtained from
high-$Q^2$ precision electroweak measurements, performed at LEP and by SLD, CDF, and D{\O}, as a function of the Higgs-boson mass \cite{LEPEWWG:2005ema,
:2010vi}. 
The best fit ($\Delta \chi^2=0$) and the corresponding 68\%/,CL region are given by $m_h=92^{+34}_{-26}\,{\rm GeV}$, without taking into 
account the theoretical uncertainty. Moreover, alternative fits for a different value of $\Delta\alpha^{(5)}_{\rm had}$ as well as for an 
inclusion of low-$Q^2$ data are shown. The maximal possible value of the Higgs-boson mass in the SM, suggested by electroweak precision measurements, reads
\beq
m_h<161\,{\rm GeV}\ @ 95\%\, {\rm CL}\,,
\eeq
without taking into account direct exclusions.
Thus, all considerations within the SM (theoretical as well as experimental) suggest that the Higgs boson is light, just around
the LEP limit. Note that this picture will change in the minimal RS model, studied in chapters \ref{sec:WED}-\ref{sec:Pheno}.
Seen from another point of view, electroweak precision measurements together with the LEP exclusion hint to the fact that the SM could be valid
up to very high energies, as reflected in Figure~\ref{fig:trivstab} by the fact that the experimentally allowed region matches with the closing throat
of theoretical constraints at large cutoff. This perception is supported by the fact that the SM is a renormalizable theory.

Let us finally mention that there are several alternatives to the SM Higgs mechanism.
The simplest extension is just a model of two Higgs doublets, one giving masses to the up type and one to the down type quarks,
like used in models of SUSY. Furthermore, there are various ways of realizing a Higgs boson. It does not even 
have to correspond to an elementary particle, but could also be composite in the spirit of technicolor models, see Section~\ref{sec:solHP}
and Chapter~\ref{sec:WED}. Another possibility to arrive at massive SM particles is to break the gauge invariance by boundary 
conditions in Higgsless models, see Chapter~\ref{sec:WED}.

\begin{figure}[!t]
\vspace{-4.5mm}
\begin{center}
\mbox{\includegraphics[width=9cm]{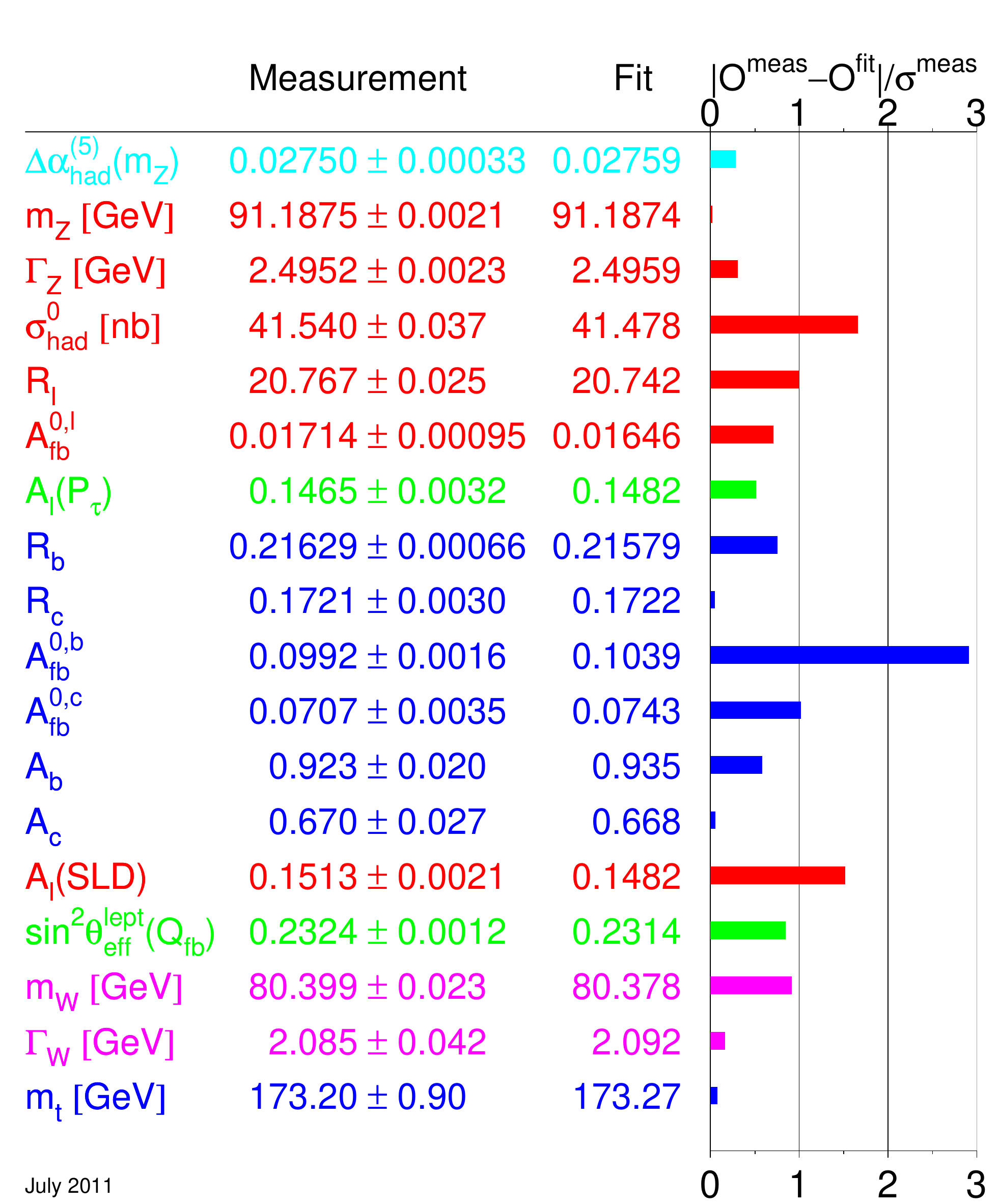}}
\vspace{-5.5mm}
\parbox{15.5cm}{\caption{\label{fig:Zobs} Comparison of a direct measurement of the $Z$-pole observables with the values obtained 
from the SM-fit. Figure from \cite{:2010vi} (with permission). See text for details.}}
\end{center}
\end{figure}

In order to get a feeling for the overall agreement of the SM with data at the mass scale of the weak gauge bosons, 
we will have a detailed look at the fit of the $Z$-pole (pseudo) observables, studied at LEP. For a detailed 
descriptions of these variables see \cite{LEPEWWG:2005ema}. 
These depend through quantum corrections on the masses of the top quark and the Higgs-boson which were 
not accessible directly at LEP with a center of mass energy of $\sqrt{s} \approx 91$\, for LEP1. Including loop 
corrections, the SM input parameters relevant for the $Z$-pole observables are the electromagnetic and the strong coupling 
constants $\alpha(m_Z)\equiv e^2(m_Z)/(4\pi)$ and $\alpha_s(m_Z)$, evaluated at the scale $m_Z$, 
as well as the masses of the top-quark, the $Z$-boson and the Higgs-boson. Note that the masses of all fermions, besides 
the top quark, are small compared to $m_Z$ and their values are known to 
sufficient accuracy, such that their impact on $Z$-pole observables can be neglected. They are assumed to be fixed, 
which also holds for $G_F=1.16637(1)\times 10^{-5}$\,GeV$^{-2}$ which is known to excellent accuracy through muon decay. 
The values of the input parameters are now fitted to agree best with the measurements of the $Z$-pole observables together with the
(five-flavor) hadronic vacuum polarization $\Delta \alpha_{\rm had}^{(5)}(m_Z)$. The resulting predictions for 
these observables are shown in Figure \ref{fig:Zobs}.
 
In total, the SM predictions agree well with the direct measurements of the corresponding observables. This means 
that the input parameters, introduced above, are able to describe the $Z$-pole consistently, which provides
a test of the SM at the quantum level. The only discrepancy well 
above the level of $2\,\sigma$ is found in the forward-backward asymmetry for bottom quarks $A_{FB}^{0,b}$.
While this could still be a statistical fluctuation, it might also be interpreted
as a hint for BSM physics. We will study this possibility in Section \ref{sec:bpseudo}.
Note that, before its direct discovery at Tevatron, the electroweak precision fit at the $Z$-pole {\it predicted} the
mass of the top quark to be approximately equal to 170\,GeV (see \cite{Peskin:1995ev}), in remarkable agreement with the direct measurement, see
Appendix~\ref{app:ref}. Moreover, from measuring the width of the $Z$-boson it was possible 
to exclude the existence of a fourth neutrino with standard-model couplings and a mass below $m_Z/2$. Comparing the 
measured invisible width with the SM prediction, depending on the number of light neutrinos $N_\nu$, one concludes 
$N_\nu=2.9840 \pm 0.0082$ \cite{LEPEWWG:2005ema}, in agreement with the three leptonic doublets of the SM. This provides another 
remarkable confirmation of the consistency of the SM. 

Another important test of the SM is provided by the electroweak precision observables $S$, $T$, and $U$, the
Peskin-Takeuchi parameters. They parametrize (universal) deviations from the expected electroweak 
radiative corrections in the SM, entering through vacuum polarization diagrams \cite{Peskin:1995ev,Peskin:1990zt,Peskin:1991sw,Golden:1990ig,Holdom:1990tc,Altarelli:1990zd,Altarelli:1993sz,Altarelli:1993bh}.
Thus, at a SM reference point, they are normalized to zero. In terms of vacuum polarization amplitudes involving electroweak gauge bosons and their derivatives,
evaluated at zero momentum, they read
\begin{equation} \label{eq:STUdef}
  \begin{split}
    S & = \frac{16 \hspace{0.3mm} \pi s^2_w c^2_w}{e^2} \left[ \,
      \Pi_{ZZ}^{\hspace{0.25mm} \prime}(0)+ \frac{s^2_w-c^2_w}{s_w
        c_w} \, \Pi_{ZA}^{\hspace{0.25mm}
        \prime}(0)-\Pi_{AA}^{\hspace{0.25mm}
        \prime}(0) \, \right] ,\\
    T & = \frac{4 \hspace{0.3mm} \pi}{e^2 c^2_w m_Z^2} \, \Big
    [\Pi_{WW}(0)-c^2_w \, \Pi_{ZZ}(0) -2 \, s_w c_w \, \Pi_{ZA}(0)-
    s^2_w \, \Pi_{AA}(0) \Big]\,,\\
    U & = \frac{16 \hspace{0.3mm} \pi s^2_w}{e^2} \, \Big
    [\Pi_{WW}^{\hspace{0.25mm} \prime}(0)-c^2_w \,
    \Pi_{ZZ}^{\hspace{0.25mm} \prime}(0) -2 \, s_w c_w \,
    \Pi_{ZA}^{\hspace{0.25mm} \prime}(0) -s^2_w \,
    \Pi_{AA}^{\hspace{0.25mm} \prime}(0) \Big ] \,.
  \end{split}
\end{equation}
The parameter $T$ is sensitive to the difference between the corrections to the $W^\pm$- and $Z$-boson 
vacuum-polarization amplitudes and thus measures isospin violation. It is thus directly related to the $\varrho$ parameter
introduced before. The experimental central values and 68\% CL bounds on the $S$ and $T$ parameters, corrected to the 
world average of the top-quark mass \cite{Group:2008nq}, as well as their correlation matrix are given by \cite{LEPEWWG:2005ema}
\beq\label{eq:STexp}
   \begin{array}{l}
    S = 0.07 \pm 0.10 \,, \\[0.5mm] 
    T = 0.16 \pm 0.10 \,,
   \end{array} \qquad 
   \rho = \begin{pmatrix} 1.00~ & 0.85 \\ 
                          0.85~ & 1.00 
          \end{pmatrix}\, .
\eeq
In the global fit to the SLC and LEP measurements, the parameter $U$ is set to zero.
The subtraction point corresponds to the SM reference values as given in Appendix~\ref{app:ref}.
The experimental values in (\ref{eq:STexp}) are in rather good agreement with zero, indicating that there are no unexpectedly 
large electroweak corrections from BSM physics.
The corresponding regions of 68\%, 95\%, and 99\% probability in the $S$--$T$ plane are shown in Figure \ref{fig:STgLR} in Section 
\ref{sec:pheno1}, where we will discuss corrections to the electroweak precision parameters in the RS setup.
Finally, an important test of the QED part of the SM deeply at the quantum level ($\geq 4$ loops) is provided by studies of 
the anomalous magnetic moment of the electron $a_e$. The experimental result agrees with the SM calculation at the level of $10^{-9}$ 
(one part per billion) \cite{Hanneke:2008tm,2011PhRvL.106h0801B}!
\beq
\begin{split}
a_e^{theo} = 0.001 159 652 181 13(84),\\
a_e^{theo} = 0.001 159 652 180 73(28).
\end{split}
\eeq
This agreement between theory and experiment in one of the most precisely measured quantities in whole science
is an impressing confirmation of QED.

\subsection{Flavor and CP Violation}
\label{sec:SMflavor}

The Yukawa Lagrangian of the SM gives rise to a rich phenomenology in the flavor sector, \ie, the sector that 
differentiates between different fermion flavors. For an introduction see \eg \cite{Nir:2001ge,
Grossman:2010gw,Buras:2005xt}. 
After EWSB, the Lagrangian (\ref{eq:Lyuk}) leads to (non-diagonal) mass terms for the SM fermions which are 
proportional to the Higgs VEV $v$ and entries of the Yukawa matrices 
\beq
\bm{m}_f=\frac{1}{\sqrt2}\,v\,\bm{Y}_f\,,\ f=u,d,e\,.
\eeq
These are the only terms in the SM that distinguish between different fermion flavors of the same 
electromagnetic charge and break the $SU(3)_Q\times SU(3)_u\times SU(3)_d\times SU(3)_E\times SU(3)_e$ flavor symmetry that 
the SM would have otherwise. In order to end up in the physical basis of propagating mass eigenstates, we diagonalize 
these terms by bi-unitary transformations
\beq
\label{eq:SVDSM}
	\bm{m}_u =\bm{U}_u\, {\rm diag}\left(m_u,m_c,m_t\right) \bm{W}_u^\dagger
\eeq
and analogously for down-type quarks and charged leptons. The fermion mass eigenstates are thus obtained from the
flavor eigenstates as $u_L^{\rm mass} = \bm{U}_u^\dagger u_L,\, u_R^{\rm mass} = \bm{W}_u^\dagger u_R$, \etc\
Experimental results for the quark and lepton masses can be found in \cite{Nakamura:2010zzi}. Besides the Yukawa terms and
those involving charged current interactions with the $W^\pm$-bosons, all remaining terms in the Lagrangian are 
invariant under this change of basis. Furthermore, as neutrinos are massless within the SM, it is always possible 
to go to a basis in which the charged current interactions in the lepton sector, as well as the mass matrix of 
the charged leptons are diagonal. 
The charged current interactions of quarks however become 
\beq
	{\cal L}_{\rm ferm} \ni  \frac{g}{\sqrt{2}} \gamma^\mu\, \bar u_L^{\rm mass} W^+_\mu \bm{V}_{\rm \hspace{-1mm} CKM}\, d_L^{\rm mass} + {\rm h.c.},
\eeq
where 
\beq
\bm{V}_{\rm \hspace{-1mm} CKM}\equiv \bm{U}_u^\dagger \bm{U}_d\,,
\eeq
and the superscript of the quark fields signals that they are now in the 
mass basis. In the following, we will omit this superscript, unless it is necessary for clarity. Due to
the misalignment of the up-type and down-type Yukawa matrices, encoded in the off-diagonal entries of the 
Cabibbo-Kobayashi-Maskawa (CKM) matrix 
\beq
\bm{V}_{\rm \hspace{-1mm} CKM}\equiv
\left(\begin{array}{ccc}
V_{ud} & V_{us} & V_{ub}  \\
V_{cd} & V_{cs} & V_{cb} \\
V_{td} & V_{ts} & V_{tb}
\end{array}\right)\,,
\eeq 
there are family-changing interactions at tree level 
in the charged current sector of the SM. However, due to the observed hierarchies in the CKM matrix, these are 
suppressed compared to interactions within the same family, see Section~\ref{sec:SMhi}. Moreover, due to the unitarity 
of $\bm{U}_f$ and $\bm{W}_f$, there are {\it no flavor-changing neutral currents} (FCNCs) at tree 
level in the SM. This agrees nicely with the observed tininess of FCNCs in nature, which on the other hand leads to tight constraints 
on BSM physics, see Section~\ref{sec:SMProblems}. 
Due to this unitarity (leading to a unitary CKM matrix) FCNCs in the SM are also suppressed on the loop level. In fact, the sum over 
internal fermions in loop mediated FCNCs leads to combinations of CKM-matrix elements such as
\beq
\label{eq:UT}
V_{ud}V_{ub}^* + V_{cd}V_{cb}^* + V_{td}V_{tb}^* = 0\,,
\eeq
which are zero due to the unitarity of $\bm{V}_{\rm \hspace{-1mm} CKM}$. Such relations, corresponding to the off-diagonal entries
in $\bm{V}_{\rm \hspace{-1mm} CKM}\bm{V}_{\rm \hspace{-1mm} CKM}^\dagger=\bm{1}_{3\times 3}$, can be interpreted 
as triangle equations in the complex plane, corresponding to so-called {\it unitarity triangles}. 
However, if the fermion masses are not degenerate,
there are also terms, where each factor in (\ref{eq:UT}) is multiplied by a corresponding (different) quark mass,
spoiling the cancellation. If the masses are small compared to the electroweak scale (or the mass differences are small), 
the effect will be suppressed. Only the large top quark mass will lead to significant corrections, see \eg \cite{Buras:1998raa}.
This suppression of FCNCs within the SM is known under the name of Glashow-Iliopoulos-Maiani (GIM) mechanism \cite{Glashow:1970gm}.
Note that these authors used this mechanism to explain the smallness of the branching ratio of the 
decay $K^0\to\pi^+\pi^-$. In this context, the prediction of the charm quark is usually assigned to them. Back in 1970, this quark 
had not been discovered yet. However, this fourth quark was needed in order to explain the smallness of the 
FCNC-Kaon-decay via the (2 generation) GIM mechanism, in addition to the need of an equal numbers of quarks and 
leptons for anomaly cancellation \cite{Bjorken:1964gz}. This shows how (precise) measurements in flavor physics 
can be sensitive to new physics. Since FCNCs are in general rather constrained by experiment, having a type 
of suppression as described above also in BSM models would be very useful. For up-type quark decays the GIM mechanism 
is generally extremely effective (since not broken by internal top quarks), leading to the tiny SM prediction for the 
branching fraction of a top quark decaying into a charm quark and a $Z$-boson 
\cite{Eilam:1990zc,AguilarSaavedra:2004wm}
\beq
{\cal B}(t\to c Z) \simeq 1\times10^{-14}.
\eeq
A suppression of phenomenological dangerous flavor changing processes for generic models can be achieved by assuming
the only source of flavor violation, \ie, breaking of the $SU(3)^5$ global flavor symmetry of the Lagrangian, 
to be the SM-Yukawa matrices. This assumption is called {\it minimal flavor violation} (MFV) \cite{Buras:2000dm,D'Ambrosio:2002ex}.

The CKM matrix, being a unitary $3 \times 3$ matrix, is described by nine parameters, out of which 
six are phases. However, by redefining the phases of the fermion fields, five of these six phases can be eliminated (corresponding 
to five independent phase differences between up- and down-type quark fields). So we end up with three mixing angles and
one physical phase, which results in CP violation in weak interaction. It is the only source of CP violation in the SM. 
As a general unitary $n \times n$ matrix has $n(n+1)/2$ phases, out of which $2n-1$ can be eliminated, $n=3$ is the 
minimum number of generations that allow for a CP-violating phase in the SM, however, see Section~\ref{sec:hierarchies}. 
As CP was known to be violated in Kaon decays, Kobayashi and Maskawa conjectured the existence of a third generation of 
quarks before their experimental discovery, which was another triumph of the SM \cite{Kobayashi:1973fv}. In 2008 they 
received the Nobel Prize in Physics ``for the discovery of the 
origin of the broken symmetry which predicts the existence of at least three families of quarks in nature''.
Another requirement for having a CP violating phase in the SM are Yukawa couplings with a non-vanishing imaginary 
part. More precisely, the necessary and sufficient condition for obtaining CP violation in the SM reads 
\cite{Jarlskog:1985ht}
\beq
	{\rm Im}\{\rm{det}[Y_u Y_u^\dagger,Y_d Y_d^\dagger]\} \neq 0\,.
\eeq
The fact that {\it complex} Yukawa couplings in the Lagrangian lead to CP violation can be understood intuitively 
as follows. For example, if we look at the down-type part of the Yukawa Lagrangian, and write down the hermitian 
conjugation explicitly, we end up with
\beq
	\mathcal{L}_{\rm Yukawa} \ni - \left(Y_d\right)_{ij}\,\bar{Q}_{Li} \Phi d_{Rj} - \left(Y_d\right)_{ij}^*\,\bar{d}_{Rj} \Phi^\dagger Q_{Li},
\eeq
where $i$ and $j$ are flavor indices. A CP transformation on this Lagrangian interchanges the operators
$\bar{Q}_{Li} \Phi d_{Rj}$ and $\bar{d}_{Rj} \Phi^\dagger Q_{Li}$, but leaves their coefficients unchanged. Thus, only 
if the Yukawa matrices above are complex, it is in general not CP-invariant.
CP violation has been experimentally well established in the neutral Kaon and $B_d^0$-meson systems, see \eg \cite{Nir:2001ge}.

Since CP violation is one of the Sakharov criteria \cite{Sakharov:1967dj} (together with C violation, baryon-number 
non-conservation and interactions out of thermal equilibrium) it is necessary for baryogenesis. 
While the SM could in principle fulfill the additional Sakharov criteria - baryon-number violation via 
unsuppressed non-perturbative (sphaleron) effects at high temperatures and a departure from thermal equilibrium 
during phase transitions in the expanding universe - the CP violation in the SM is orders of magnitude too small to 
account for the observed baryon asymmetry. Thus, CP violation {\it beyond the SM} seems to be necessary for baryogenisis 
to work. 
Furthermore, within the SM, the LEP bound $m_h>114.4$\, GeV seems to exclude the first order phase transition needed 
for a viable model of electroweak baryogenesis. For a review on the subject see \eg \cite{Bernreuther:2002uj}.
We have seen that the number of physical parameters in the flavor sector turned out to be smaller than the number of parameters 
we started with. Some of them could be rotated to zero in a certain basis. Let us now recall a general strategy which is useful
for determining the number of physical parameters due to the flavor sector of a theory. We will use this method in
Section~\ref{sec:hierarchies} to count the parameters of the RS setup.

Consider a theory with gauge invariant kinetic terms. In addition to the gauge symmetry, the theory in general also
has a certain additional global symmetry group $G$ with $N_G$ generators. When adding a gauge invariant potential with $N_{\rm general}$ parameters (in 
a general basis) to the theory, the global symmetry might be broken down to a smaller symmetry $H$ with $N_H$ generators. Due to this breaking 
there is a freedom of rotating away {\it unphysical parameters}, \ie, choosing a certain ``direction'', which leaves the theory invariant
under the residual global symmetry, see \eg \cite{Grossman:2010gw}. Denoting the difference in the number of symmetry generators
of $G$ and $H$ by $N_{\rm broken}=N_G-N_H$, the number of physical parameters, affecting measurements, is given by
\beq
N_{\rm phys}=N_{\rm general}-N_{\rm broken}\,,
\eeq
which holds separately for real and imaginary parameters.
Let us apply this rule to the flavor sector of the SM. The kinetic terms of the quarks have the global symmetry
\beq
G=U(3)_Q\times U(3)_u \times U(3)_d,
\eeq
which corresponds to $N_G=(9,18)$ real and imaginary generators. The Yukawa matrices $\bm{Y}_{u,d}$, being general 
$3 \times 3$ complex matrices, possess $N_{\rm general}=N_Y=(18,18)$ real moduli and CP-odd phases (in a general basis). 
They break down the global symmetry $G$ down to
\beq
H=U(1)_B\,,
\eeq
see Section~\ref{sec:FL}, which has $N_H=(0,1)$ generators, resulting in $N_{\rm broken}=(9,17)$. The number of
physical parameters, which cannot be rotated to zero, is thus given by
\beq
N_{\rm phys}=(18,18)-(9,17)=(9,1)\,.
\eeq
These parameters can be identified with the six quark masses and the three mixing angles and the CP violating phase of the
CKM matrix.

A useful parametrization of the CKM matrix has been given to Wolfenstein. In terms of the four 
parameters $(\lambda,A,\bar\rho,\bar\eta)$, it can be written as
\beq
\label{eq:CKMWolf}
\bm{V}_{\rm \hspace{-1mm} CKM}=
\left(\begin{array}{ccc}
1-\frac{\lambda^2}{2} & \lambda & A\lambda^3 (\bar\rho-i\bar\eta)  \\
-\lambda & 1-\frac{\lambda^2}{2} & A\lambda^2 \\
A\lambda^3 (1-\bar\rho-i\bar\eta) & -A\lambda^2 & 1
\end{array}\right)+O(\lambda^4)\,,
\eeq
where $\lambda\approx0.23\ll1$ plays the role of an expansion parameter and exhibits the diagonal-dominant, hierarchical, structure
of the matrix. The Wolfenstein parameters are related to the entries of the CKM matrix as
\beq\label{eq:lAre}
   \lambda 
   = \frac{\left|V_{us}\right|}%
          {\sqrt{\left|V_{ud}\right|^2 + \left|V_{us}\right|^2}} ,
    \qquad
   A = \frac{1}{\lambda} \left| \frac{V_{cb}}{V_{us}} \right| ,
    \qquad
   \bar\rho - i\bar\eta 
   = - \frac{V_{ud}^* V_{ub}}{V_{cd}^* V_{cb}} \,.
\eeq
Their experimental values are given in Appendix~\ref{app:ref}.
A phase-convention invariant measure (independent of phase redefinitions of the fermion fields) of CP violation
in $\bm{V}_{\rm \hspace{-1mm} CKM}$ is given by the Jarlskog invariant $J_{\rm CKM}$ \cite{Jarlskog:1985ht}, which is defined via
\beq
   {\rm Im}\left[(V_{\rm CKM})_{ij}(V_{\rm CKM})_{kl}(V_{\rm CKM}^*)_{il}(V_{\rm CKM}^*)_{kj}\right] =J_{\rm CKM} \sum_{m,n=1}^3
   \epsilon_{ikm}\epsilon_{jln}\quad (i,j,k,l=1,2,3)\,,
\eeq
where $\epsilon_{ikm}$ is the Levi-Civita tensor.
It corresponds to two times the area of any of the unitarity triangles. 
Note that $J_{\rm CKM}$ is not small due to a small CP violating phase but due to small mixing angles in the CKM matrix 
The fact that it is not $\lesssim \ord(1)$, but
\beq
   J_{\rm CKM}\simeq \lambda^6 A^2 \bar\eta \sim 10^{-5}
\eeq
quantifies the statement that CP violation is small within the SM.
As mentioned before, flavor physics is sensitive (indirectly) to undiscovered particles. 
Beyond being known approximately from electroweak precision tests, the top-quark mass
could also be determined from studies of $B$ mesons, since the rates for rare FCNC processes 
such as $B\to X_s \gamma$ and those for $B$--$\bar B$-mixing are strongly sensitive on it. 
Moreover, many other properties of the top quark, like its flavor-changing couplings and CP-violating 
interactions, are known from Kaon and B-meson physics \cite{Neubert:2005mu}.
This approach of probing new physics {\it indirectly} is complementary to direct searches and often sensitive to much higher scales.
It will be applied in many of the phenomenological studies in Chapter~\ref{sec:Pheno}.

An important task in flavor physics is to overconstrain the flavor parameters (\eg in terms of an unitarity triangle)
by various measurements, in order to test the CKM mechanism. In particular, the fact that CP violation in the SM is 
induced by a single CP violating phase results in relations between different CP violating observables, which provide 
stringent tests for the model. Despite some tensions, the experimental situation provides evidence that the CKM mechanism 
is the dominant source of flavor violation as well as of CP violation in flavor-changing processes at low energies 
\cite{Nir:2001ge,Grossman:2010gw,Bevan:2010gi}. We will come back to CP violation and the corresponding observables as well as 
related tensions in Section~\ref{sec:asl}. BSM physics generically introduces new sources of CP violation.
In the latter section we will study in detail the impact of the RS proposal on certain CP violating observables.
\vspace{-0.4cm}

\subsection{Symmetries and Parameters of the SM Lagrangian}

\label{sec:FL}

After having discussed aspects of the SM (and beyond), a final look at its complete Lagrangian, 
its symmetries and the parameters contained is instructive. 
So far we have discussed all ingredients of the SM Lagrangian (\ref{eq:LSM}),
besides one sector which shall be shortly touched on before continuing. In order to eliminate unphysical degrees 
of freedom, related to the gauge redundancies in the description of the spin-1 gauge bosons by 4D Lorentz 
vectors, we have to introduce a gauge fixing Lagrangian ${\cal L}_{\rm GF}$.
Otherwise the equivalence classes of physical undistinguishable fields will lead to singularities in the quantized theory. 
However, we do not want to destroy symmetries, but merely rewrite the theory in a way to avoid these singularities. 
Thus, in the course of fixing the gauge, we also end up with a Lagrangian of Faddeev-Popov ghost fields ${\cal L}_{\rm FP}$ 
(which are necessary to preserve unitarity). The rigorous way of quantizing a (non-abelian) gauge theory along the lines 
discussed above is called BRST quantization \cite{Becchi:1975nq,Iofa:1976je}. The final Lagrangian is invariant under BRST transformations 
which generalize the concept of gauge transformations. We choose the (SM) gauge fixing 
\beq
\label{eq:SMGF}
\begin{split}
   {\cal L}_{\rm GF}
   &= - \frac{1}{2\xi} \left( \partial^\mu A_\mu \right)^2 
    - \frac{1}{2\xi} \left( \partial^\mu Z_\mu
    - \xi m_Z \varphi_Z \right)^2 \\
   &\quad\mbox{}- \frac{1}{\xi} \left( \partial^\mu W_\mu^+
    - \xi m_W\varphi_W^+ \right)
    \left( \partial^\mu W_\mu^-
    - \xi m_W\varphi_W^- \right)\, ,
\end{split}
\eeq
which also has the convenient feature of eliminating mixed terms between gauge fields and Goldstone bosons in the 
Lagrangian. A later generalization of ${\cal L}_{\rm GF}$ to the RS model will be straightforward. The parameter $\xi$ interpolates 
between different so-called $R_\xi$ (renormalizable) gauges and drops out in physical observables. The Faddeev-Popov Lagrangian 
is not needed for the calculations performed throughout this thesis. We refer the reader to the 
literature for more details \cite{Peskin:1995ev}.
Now we have collected all terms of the SM Lagrangian which are necessary to calculate scattering amplitudes and cross 
sections within the SM by means of standard techniques \cite{Peskin:1995ev, Cheng:1985bj}. 

The Lagrangian (\ref{eq:LSM}) exhibits several symmetries, of which two - BRST invariance 
and Poincar\'{e} invariance - have already been mentioned. Furthermore, the Lagrangian has {\it accidental} global symmetries. It is 
invariant under a global phase rotation of all quark fields $U(1)_B$ (see Section~\ref{sec:SMflavor}) which corresponds via Noether's theorem 
to the conservation of baryon number ($B$). As the SM Lagrangian does not include mass terms for the neutrino fields, it is furthermore 
invariant under separate phase rotations of the three lepton families $U(1)_e\times U(1)_\mu\times U(1)_\tau$, which 
corresponds to a conservation of the lepton-family number. If one includes Dirac mass terms for the neutrinos, this symmetry
breaks down to the total lepton number ($L$). The conservation of lepton number can further be broken by including Majorana 
mass terms for the neutrinos, see Section~\ref{sec:SMProblems}. These symmetries are called 
accidental, since they do not have to be put in by hand, but rather follow from the particle content of the gauge theory
and the restriction to $D\leq4$ terms in the SM Lagrangian. This matches nicely with the experimental non-observation 
of the $B$-violating proton decay ($\tau_{p\to e^+ \pi^0}>8.2 \times 10^{33}$\,yr \cite{:2009gd}). 

The SM is not only in good agreement with experiment at the places, where it was designed to be,
but also in many other accidental details. On the other hand, the non-observation of proton decay poses 
a problem for the {\it simplest Grand Unified Theories} (GUTs) \cite{Georgi:1974sy}, which allow for a unification of 
coupling constants but generically induce $B$ violation (\eg $p\to e^+ \pi^0$) at a too large rate \cite{Langacker:1994vf}. 
At the quantum level however, $L$ and $B$ are broken inevitably already in the SM, due to the 
chiral anomaly. Non-perturbative sphaleron processes violate $B+L$ (but respect $B-L$). 
These processes are just important for very high temperatures $T>100$\,GeV, \eg during the early universe.
However, if a symmetry shall be fundamental, it seems that it has to be realized locally ({\it c.f.} general relativity).
Further discrete symmetries that are for example present in electromagnetism, are violated in the SM. P is violated 
maximally in charged-current interactions which just couple {\it left-handed} fermions. It is 
furthermore violated in neutral-current interactions with the $Z$-boson. Moreover, the Yukawa Lagrangian 
(\ref{eq:Lyuk}) leads to a violation of CP, as discussed in Section \ref{sec:SMflavor}. Due to the small 
(non-maximal) CP violation, also C is violated within the weak sector of the SM. However, CPT is a symmetry of the 
SM, as it is for every local and Lorentz invariant QFT. At the end it is our task to take a model, determine the 
parameters present in the Lagrangian and see if this Lagrangian consistently describes nature. Therefore we have 
to perform more independent measurements than there are independent parameters in the theory in order to overconstrain 
and thus test the model. If we find tensions, we should think about a modification of our Lagrangian, see 
Section~\ref{sec:SMProblems}.  

The SM has 18 parameters, which can be chosen to be
\begin{itemize}
	\item{3 gauge couplings: $g, g^\prime, g_s$}
	\item{15 parameters in the Higgs/flavor sector:}
		\begin{itemize}
			\item $m_h$, $v$
			\item 6 quark masses: $(m_u,m_c,m_t),\,(m_d,m_s,m_b)$ 
			\item 3 lepton masses: $(m_e,m_\mu,m_\tau)$
			\item 3 mixing angles and one phase of ${\bm V}_{\rm \hspace{-1mm} CKM}$\,.
		\end{itemize}			
\end{itemize}
Note that we did not count $\theta_{QCD}=0$ as a free parameter of the SM. Furthermore, 
allowing for neutrino masses increases the number of parameters, in dependence on the possible assumption of lepton 
number conservation (Dirac- vs. Majorana-mass terms). Note also that the choice of the parameters above is not
unique. As discussed in Section~\ref{sec:Higgs}, we can \eg trade $g$ and $g^\prime$ for the experimentally
more common $e$ and $\sin\theta_w$. In contrast to the gauge sector of the SM, which, due to gauge invariance 
has only a few parameters, there are many free parameters in the flavor sector. This suggests that we have not yet completely 
understood the fundamental characteristics of matter. Moreover, in contrast to $\ord(1)$ gauge couplings, the flavor sector
exhibits non understood hierarchies. We will come back to this point in Section~\ref{sec:SMhi}.

\subsection{Effective Field Theories, Renormalizability and the TeV Scale}

\label{sec:Renorm}

We have so far introduced the SM as the leading ${\cal L}_{\leq 4}$ term in (\ref{eq:Lexp}). Thus we assume, that it is the right theory at low 
energies - the theory we have studied in the last sections - but at the same time we allow for more than this 
theory, encoded in the terms with higher mass dimensions and suppressed by powers of the cutoff. That is the 
modern way of thinking about the SM. 

Historically, its genesis was different. The SM was constructed on the 
purpose of providing an elegant and quite minimalistic gauge theory in order to describe electrodynamics as well as charged current 
interactions without limitations, which lead to the GWS theory, and later also included QCD. One of the 
paradigms was renormalizability. If a theory involves couplings 
with negative mass dimensions, it is not (power-counting) renormalizable (see \eg 
\cite{Cheng:1985bj}). Calculating Feynman diagrams containing closed loops of virtual particles often leads
to divergent results, since one has to integrate over undetermined internal four-momenta. These results can be 
regularized by \eg introducing a cutoff $\Lambda$ in the (euclidean absolute) momentum integration. 
A theory is renormalizable, if all cutoff-dependent terms that appear can be canceled by introducing a finite number of 
counterterms in the Lagrangian. This corresponds to absorbing the terms which would diverge in the limit of removing
the regularization ($\Lambda\rightarrow\infty$) into a redefinition of the (finite number of) fundamental parameters
of the theory and is called renormalization. However, if a theory involves couplings with negative mass dimension, 
such a procedure is not possible. In the course of renormalization one would end up with an infinite number of counterterms 
of all possible mass dimensions, increasing with the order of the perturbative series. These can no longer be absorbed into 
redefinitions of a finite number of parameters of the theory. This means that the theory (foreseen as a theory valid up to arbitrary scales)
would lose its predictivity as it is impossible to measure an infinite number of unknown couplings. As a consequence, it was 
argued that no terms in the Lagrangian should be allowed to have mass dimensions bigger than four, 
as otherwise their coefficients (coupling constants) would need to have a negative mass dimension. Renormalizability was the 
concept that singled out ${\cal L}_{\leq4}$ out of (\ref{eq:Lexp}) and is a useful method 
of constraining the possible form of a theory. Note that for a theory to be renormalizable it is also important that the
counterterms are compatible with the defining symmetries of the model. In the early days of the SM, theories 
that contained terms with mass dimensions bigger than four were considered unsatisfactory and unpredictive at the level of quantum corrections. 
The SM was defined as the most general, {\it renormalizable} (\ie\,$D\leq 4$) theory, 
consistent with the symmetries of nature, which exactly corresponds to the term 
${\cal L}_{\leq4}$. Higher dimensional operators have not been included. Complemented by a concept of 
treating gravity, like asymptotic safe general relativity \cite{WeinbAS}, 
the renormalizable SM including the Higgs mechanism could {\it in principle} be a UV complete theory of 
nature. This makes the SM very attractive and outstanding with respect to ideas 
like the (non renormalizable) Fermi theory of weak interactions, 
which is known to break down at scales of the order of $m_W$, see Appendix~\ref{app:EFT}. The fact that the electroweak 
gauge bosons get their masses through {\it spontaneous} symmetry breaking is essential for the renormalizability 
of the SM. It guarantees a gentle high energy behavior of processes involving massive gauge bosons (as we have seen before), 
which would otherwise violate the line of reasoning with naive mass dimensions above \cite{'tHooft:1972fi}. As a 
consequence, the SM can make {\it exact} predictions at a certain level of perturbation theory, with no unknown cutoff 
effects. Importantly, the SM is anomaly free which means that its gauge symmetries also hold at the quantum level. 

However, although the SM is very successful up to energies 
of $\ord(100\,{\rm GeV})$, this does not mean that it has to be at higher scales and indeed there is a bunch of evidence that the SM is not the final 
answer. For the time being, let us just mention the lack of a Dark Matter candidate and of unification of 
the coupling constants as well as the absence of a quantum theory of gravity. We will discuss more 
shortcomings of the SM in Section~\ref{sec:SMProblems}. Today, theories with higher dimensional operators are 
considered perfectly fine as {\it effective} field theories, valid up to a cutoff.
If the expansion of the Lagrangian is stopped at a certain level of precision (\ie, mass dimension), there is by definition 
just a finite number of parameters present in the theory which thus is automatically renormalizable (in the ``modern'' sense)
and predictive up to suppressed terms. Indeed, also the RS proposal will turn out to be an EFT.
With the model not being valid until arbitrary high energy scales, there is no reason not to include higher 
dimensional operators. In Appendix~\ref{app:EFT} we will treat this concept, which we have already sketched at the beginning of this chapter. 
The content is assumed to be known in the following. It also demonstrates how our theory (\ref{eq:Lexp}) with the particle 
content of the SM (\ie, the SM including higher dimensional operators) 
could arise from a more complete, and possibly renormalizable, theory. This low energy EFT is the theory we see today and, 
being the most general one, should work with. The effects of its UV completion will be present in the coupling constants
of the EFT - the {\it Wilson coefficients}. As suggested by the shortcomings of the SM, we 
should construct a theory that completes the SM at high energies. Considering the SM as an EFT, \ie, allowing 
for higher dimensional operators and trying to determine their coefficients, is a first step on this way. 
The final theory, valid up to arbitrary high energies (if it 
exists and is a QFT), should however be renormalizable in the sense that it just contains operators with mass 
dimensions less or equal than four. 

In Appendix~\ref{app:EFT} we review an important virtue of the concept of EFTs. 
If a theory involves several widely separated scales, \eg two scales $m\ll M$, 
loop calculations will generically feature large logarithms of these scales which can spoil the perturbative expansion. 
By integrating out physics at the larger scale, \ie, by removing the corresponding dynamical degrees of freedom 
and constructing a theory just valid below the scale $M$, one can circumvent this problem and it is possible to
{\it resum} the large logarithms. This is done with the help of renormalization group (RG) equations, which evolve
the Wilson coefficients between one scale and another. Effective theories provide the possibility to separate high energy physics 
(residing in the Wilson coefficients) from low energy physics (residing in the EFT matrix elements).
We will use these techniques in sections \ref{sec:afbt} and \ref{sec:asl}. 
EFTs are often used for so-called model-independent studies (see Section~\ref{sec:afbt}). This corresponds just to 
writing down a general low-energy Lagrangian, including higher dimensional operators (respecting appropriate symmetries)
and deriving predictions in terms of the undetermined coefficients of this Lagrangian. 
It avoids specifying a certain BSM model and resembles just the way in which we have introduced the SM in (\ref{eq:Lexp}).

We have seen that the Higgs sector is important for the renormalizability of the SM. Let us stress
that in the EFT picture of the SM, the Higgs boson remains relevant for the high energy behavior of the theory. It guarantees at least that 
the SM does not necessarily break down already at the TeV scale, due to uncontrolled high energy behavior such as unitarity 
violation in longitudinal $W^\pm$-boson scattering, but could 
have a much higher cutoff. One can see the Higgs sector as the UV completion of the SM 
without the Higgs, which provides the longitudinal components of the massive gauge bosons in a renormalizable way. 
This (or another) sector {\it has to} complete the SM at scales $\lesssim 1$\, TeV, which is a convincing argument to build 
a collider just for that energy range. If there is no Higgs-boson significantly below a TeV, the SM breaks down at a TeV. 

Moreover, the mass of such a Higgs boson should also be stabilized with respect to quantum corrections
which might result in further interesting physics around the TeV scale, see Section~\ref{sec:HP}. In the following, we
will assume the existence of a Higgs-like scalar sector that provides masses 
for the electroweak gauge bosons and fermions, given all its successful properties, and the virtues due to the resulting gauge invariance. 
We assume that the possible new physics will complete the SM without making a (sub) TeV scale Higgs 
sector unnecessary for the proper behavior of the theory.
It is mandatory trying to explore this important sector of particle physics. As mentioned, it is possible 
that BSM physics relies on a standard Higgs sector, but changes its properties quite dramatically. Thus, it is very 
important to study Higgs physics also in BSM scenarios, to be prepared for
those changes.

\subsection{Problems and Open Questions in the SM - the SM as an EFT}
\label{sec:SMProblems}

Up to now we have discussed many successful predictions of the SM and the gauge principle. It describes nature extremely well up 
to current collider energies. Despite all these agreements, there are compelling arguments suggesting 
that there is physics beyond the SM. Some of these we have already sketched before. First of all,
the SM does not include a quantum theory of gravity, which is expected to become important at the Planck scale 
$M_{\rm Pl}\sim 10^{19}\,$GeV, and thus seems not to be valid in the very early universe. However, if \eg a scenario of 
fixed point gravity leads to a harmless high energy behavior of gravity, the SM, augmented with such a theory of 
gravity, could be in principle UV complete without the need of new particles (if the Higgs boson will be found in
the appropriate mass range, see Section~\ref{sec:theoH}).

A striking argument for the 
existence of BSM physics is given by cosmology. There is evidence for the existence of non baryonic {\it Dark Matter} 
which cannot be explained within the SM and triggered a lot of model building, see \eg \cite{ArkaniHamed:2008qn}. Moreover, the 
running of the coupling constants implies that the SM gauge group could emerge from a single larger gauge group which is broken down to $G_{SM}$ 
below the scale of Grand Unification $M_{\rm GUT} \sim 10^{15}$\,GeV. However, in the SM the coupling constants 
slightly miss each other at that scale. The mere fact that they come so close to each other suggests nevertheless 
the existence of such a GUT. In consequence, NP below $M_{\rm GUT}$ is needed for an appropriate 
renormalization group running for the couplings to match. In general, the many parameters in the SM $(\geq18)$, 
especially within the flavor sector, call for a more unified description. In addition to the already mentioned strong CP problem there 
is another even more severe problem with a parameter (of gravity) being unnaturally close to zero, which cannot be understood within 
the framework of a combined theory of general relativity and the SM. The so-called cosmological constant problem is 
caused by vacuum fluctuation corrections in quantum field theory, which drive the coefficient of the allowed renormalizable 
$D=0$ operator in ${\cal L}_{\rm SM}$ to values of at least $\rho_\Lambda \equiv g_0 \sim m_t^4$, about 55 orders of magnitude above the 
experimental value, which is in the range of some ${\rm meV}^4$.\footnote{Assuming contributions of Planck scale physics makes the problem 
even worse and amounts to a difference of about 120 orders of magnitude.}
Furthermore, as stated in Section~\ref{sec:SMflavor}, the SM is not able to account for the right amount of CP violation in order 
for baryogenesis to work. 
The isotropy, homogenity and flatness of the visible universe also calls for an explanation, like inflation, 
triggered by a BSM field - the inflaton. For a review see \cite{Lyth:1998xn}.
However, there are also ideas to generate inflation with the help of the SM Higgs boson \cite{Bezrukov:2007ep}, restricting its possible mass to $m_h\in[126,194]$\,GeV \cite{Bezrukov:2009db}.
Let us finish this first collection of theoretical and cosmological arguments with some fundamental questions, not answered within the SM.\\ 
What causes the representations of the fields to be like given in the SM?\\ Why are there three generations of fermions, just distinguishable 
by their masses?\\  Why is the gauge symmetry group given by $G_{SM}$ and broken by a scalar $SU(2)_L$ doublet?\\ 
Why are the charges quantized in a certain way (see in this context \cite{Nowakowski:1992ff})?\\
What causes the mass parameter $- \mu^2$ in the Higgs Potential in (\ref{eq:LHiggs}) to become negative?\\
There exist many more questions like that and even more fundamental ones, \eg about space time. There is a class of theories 
that has the ambition to solve all these puzzles, see Section~\ref{sec:string}. 
However, these theories have little predictive power at low energies. 
The existence of non-zero neutrino masses can already be seen as BSM physics, 
but, as mentioned before, can easily be included into the SM, see also below.
Two important issues related to hierarchies and not addressed in the SM will be discussed in more detail below in
Section~\ref{sec:hiera} - the large and radiatively unstable discrepancy between the electroweak and the Planck scale (the 
gauge hierarchy problem, which is not a problem {\it of} the SM) as well as the non-understood hierarchies 
within the fermion masses and mixings. Especially the first issue caused a big portion of the model building activity 
in the last decades. Both of them can be addressed in RS models, which provides a convincing motivation for studying 
the phenomenology of these models in detail. In addition, it is also possible to address further of the aforementioned puzzles in 
this setup, as we will comment on later.

In the sector of (precision) measurements there are several $\gtrsim 1 \sigma$ effects which do not come unexpected 
when many observables are measured. However, worth mentioning are deviations in CP violating observables, 
like the discrepancy of nearly $4\sigma$ in the like-sign dimuon charge asymmetry in semileptonic b-hadron decays, $A_{\rm SL}^b$ 
\cite{Abazov:2011yk} or the deviation of around $2\sigma$ in the 
combined CDF and D{\O} measurement in the $(\beta_s^{J/\psi\phi},\Delta\Gamma_s)$--plane \cite{pubn,pubn2}, see Section~\ref{sec:asl}.\footnote{Note
that the brand new LHCb measurement does not seem to confirm this deviation \cite{LHCb:2011}.}
Moreover, a deviation between theory and experiment of about $3\sigma$ has been observed in the branching fraction 
for $B^\pm\to \tau^\pm \nu$ decays \cite{Bevan:2010gi}. 
A tension of about $2\sigma$ has been found between exclusive and inclusive determinations of $V_{ub}$ \cite{Mannel:2010zz} 
(see also \cite{Crivellin:2009sd}). In addition, the forward-backward asymmetry in top-quark pair-production $(A_{FB}^{t})^{p\bar p}$, measured 
at Tevatron \cite{CDFbrandnew,Aaltonen:2011kc,note10398}, shows a deviation of similar size, see Section~\ref{sec:afbt}. 
Finally, there are the already stated discrepancy of nearly $3\sigma$ in $A_{FB}^{0,b}$ at the Z pole, see Section~\ref{sec:bpseudo} 
and the ($3$-$4)\,\sigma$ deviation in the anomalous magnetic moment of the muon $a_\mu$ \cite{Passera:2010ev},
see Section~\ref{sec:AMM}. 
It is interesting that the largest effects seem to be present in observables involving the heavier fermion generations ($a_\tau$ is only 
poorly measured). A framework in which such a feature arises naturally, is the RS setup. 
We will discuss many of the aforementioned observables in these models and address 
the question if they can account for some of the measured deviations {\it quantitatively}, while still being in 
agreement with other important precision tests. Let us finally 
mention the recent anomaly in the invariant mass distribution of two-jet final states at the Tevatron, produced in association 
with a $W^\pm$-boson. The excess at ($120$-$160$)GeV, seen by CDF, resulted in many attempts to explain this bump in BSM 
scenarios, but it could also be explained by SM physics, due to the treatment of the single-top-quark background \cite{Sullivan:2011hu,Plehn:2011nx}.

Let us stress again that, despite these tensions and the theoretical arguments for NP given before, the SM still 
works extremely well up to at least the electroweak scale $M_{\rm EW}$. In the light of the plethora of its 
successful predictions, the SM, with its peculiarities, should at least be seen as (part of) the low energy limit
of the theory that will replace it. Such a theory most probably exists, given the sum of all the striking hints and shortcomings of the 
SM discussed before. Note that many arguments hint at BSM physics around the TeV scale, most notably the
gauge hierarchy problem, but also for example the so called WIMP miracle, \ie, the fact that a {\it weakly} interacting
massive particle with a mass around the TeV scale just leads to the correct relic abundance to account for the Dark Matter
in the universe, see \eg \cite{Jungman:1995df}.

When going beyond the SM, one has to give up some assumptions that lead to its construction. One of the easiest things to do
is just to add further matter content. One can also change the (gauge-)symmetry group, add internal global symmetries, or add 
scalars, always taking care that at low energies the agreement with experiment is not spoiled. This is generally achieved by assuming 
the new particles to be heavy or having small couplings to the SM particles. There is a plethora of those models, like 
four-generation models, see-saw models featuring \eg right handed neutrinos with large Majorana masses,
$Z^\prime$ models, $SU(2)_L \times SU(2)_R$-symmetric models, GUTs, Peccei-Quinn models, ``dark forces'', 
two-Higgs-doublet models, and many more. Most of them try to solve the one or the other problem 
of the SM or aim on being theoretically more appealing. Some are just possibilities which are not excluded yet. When thinking about 
further options, it does not seem sensible to abandon the very successful gauge principle - its merits and
the resulting problems in giving up this principle have already been examined before. 
It is possible to think about changes related to space-time symmetries. However, as special relativity is very well tested, we 
want to keep Poincar\'{e} invariance. One could anyhow try to extend the given Lie-algebra of symmetry generators, keeping those 
of the Poincar\'{e} group as well as internal generators, by involving new Lie-algebra generators related to space-time symmetry 
in a non-trivial way. Due to the {\it Theorem of Coleman and Mandula} \cite{Coleman:1967ad} this is (under very general assumptions, 
like analyticity and non-triviality of the S-matrix as well as the presence of a mass gap) however not possible. Nevertheless, the theorem 
leaves a little ``loophole'' since it only talks about Lie-algebras (with bosonic generators). Giving up this assumption, there is a single 
class of extensions of the trivial direct product of [space-time(=Poincar\'{e})]$\otimes$[internal symmetries] by employing so-called 
Lie-superalgebras (with fermionic symmetry generators which fulfill {\it anti}-commutation relations). Due to the theorem of Haag, 
Lopuszanski and Sohnius (HLS) \cite{Haag:1974qh} the most general (super)algebra of symmetry generators, that can 
combine states with different spin, is the so-called supersymmetry algebra, see Section~\ref{sec:solHP}.
Another radical possibility would be to give up the concept of fundamental point-like constituents, see Section~\ref{sec:string}.
This leads to UV finite results without the need of regularization, a very attractive feature 
for a UV completion of the SM. Other approaches just assume that some of the SM particles, like the Higgs boson, 
are composite states of elementary particles (see Section~\ref{sec:solHP}). Others extend the concept of particles by 
introducing so-called unparticles. These are states that arise as low energy degrees of freedom from a conformal sector weakly 
coupled to the SM \cite{Georgi:2007ek}. They have non-integer scaling dimensions, leading to a spectral density depending like a fractional
power on momentum. However, it was shown that theories like QCD can produce spectral densities for quarks and gluons
that are virtually indistinguishable from those of unparticles \cite{Neubert:2007kh}.
We will stop the list here, however many more (more or less exotic) ideas like that are on the market.

From the phenomenology side, we have already explored several {\it constraints on the NP}. It is expected to have a highly non-generic 
flavor structure and new CP violating effects. In addition it ought to possess a Dark Matter candidate. Beyond that, it should address the 
gauge hierarchy problem as well as the fermion hierarchies, see Section~\ref{sec:hiera}. Let us mention that this (rough and not complete) list of requirements can be fulfilled in the RS framework, 
as we will see below.
From now on we will consider the SM, including higher dimensional operators, as the low energy EFT of a 
more fundamental theory. At low energies, we consider the whole Lagrangian (\ref{eq:Lexp}), truncated at a certain operator-dimension, which we now write 
more explicitly as
\beq
\label{eq:Lef}
{\cal L}_{\rm eff} ={\cal L}_{\rm SM}+ \sum_{i}\frac{C_i^{(5)}}{M}\,{\cal Q}_i^{(5)} + \sum_{i}\frac{C_i^{(6)}}{M^2}\,{\cal Q}_i^{(6)}+\cdots\,.
\eeq
Here, $C_i^{(D)}$ are the (running) Wilson coefficients of the operators ${\cal Q}_i^{(D)}$ of mass dimension $D$.
They contain the information about the high energy physics that has been ``integrated out'', as introduced in Appendix~\ref{app:EFT}.
Let us have a first look at possible higher dimensional operators \cite{Weinberg:1979sa} and see how they give a handle on BSM physics.
At the level of $D=5$ there is just one possible gauge invariant term that can be constructed out of the SM fields.
It reads 
\beq
\label{eq:D5n}
{\cal L}_{\rm eff} \supset \frac{\lambda_{ij}}{\Lambda_L} (E_i\Phi)^T(E_j\Phi)\,,
\eeq
where the couplings $\lambda_{ij}$ are assumed to be of $\ord(1)$. This term leads (after EWSB) to a Majorana mass term 
for the left handed neutrinos with $m_\nu \sim v^2/\Lambda_L$. It breaks $L$ by 2 units.
Note that the scale of neutrino masses $m_\nu\approx 0.1$\,eV indicates a fundamental scale $\Lambda_L\sim10^{15}\, {\rm GeV} \sim M_{\rm GUT}$,
around the GUT scale. Neutrino masses could be a first hint to a new energy scale in physics, below $M_{\rm Pl}$.
The SM as an EFT can thus account for these masses without introducing new fields at low energies. At the same time, 
assuming a large cutoff scale around $M_{\rm GUT}$, they are naturally tiny. This would have to be put
in by hand for the case of pure Dirac mass terms with a right handed neutrino by assuming an unnaturally small $y_\nu\sim10^{-12}$.
Note that the $D=5$ term (\ref{eq:D5n}) could arise from a UV completion of the SM via the see-saw mechanism, for
a review see \eg \cite{Mohapatra:2006gs}.
At the level of $D=6$ operators, proton decay is induced by $B$ and $L$ violating terms like
\beq
\label{eq:D6}
{\cal L}_{\rm eff} \supset \frac{1}{\Lambda_B^2} \epsilon_{def}((\bar u_R^d)^c d_R^e)((\bar u_R^f)^c e_R)\,,
\eeq
where $d,e,f$ are $SU(3)_c$ indices. These operators (which could be induced \eg by a GUT) mediate processes like $p\to e^+ \pi^0$ which leads to the bound
from Super-Kamiokande (see Section~\ref{sec:FL} above) of $\Lambda_B\gtrsim (10^{15}-10^{16})$\,GeV.
The SM as an EFT with a large cutoff allows for interesting effects, \eg the breaking of its global symmetries, and at the same time explains the
tininess of these effects, which is needed to be not in conflict with observation. However, lowering the cutoff leads to problems, because then generically 
$B$- and $L$-violating operators will not be sufficiently suppressed to be in agreement with observation, given there is no 
symmetry which forbids them, see sections \ref{sec:LED} and \ref{sec:SMinB} in the context of BSM models.

For the time being, let us assume that the NP that will be possibly found well below the GUT scale does not induce the operators examined above.
Still, measurements in flavor physics performed at low energies can constrain the scale or structure of BSM physics quite significantly.
Remember that the flavor sector of the SM works impressively well. The facts that charged current interactions are universal (the CKM matrix is unitary)
and that FCNCs are highly suppressed due to the GIM mechanism agrees with results from experiments.
However, this flavor structure is special and NP models will generically not reproduce the same structure.
To estimate the expected effects, consider $D=6$ four-quark operators like
\beq
{\cal L}_{\rm eff} \supset \frac{q_1\bar q_2\,q_3\bar q_4}{\Lambda_{\rm flavor}^2}\,,
\eeq
where $q_i$ can be arbitrary quarks, which however have to feature charges that sum up to zero.
Measurements of meson mixing and CP violation can strongly constrain the scale $\Lambda_{\rm flavor}$ for generic $\ord(1)$
dimensionless couplings and lead to the estimate  (see \eg \cite{Nir:cargese})
\beq
\Lambda_{\rm flavor} \geq 10^4\,{\rm TeV}\,.
\eeq
Thus, if there are no symmetries that account for a special flavor structure, from this estimate we expect new physics
not to be lighter than $10^4$\,TeV\,! However, as we will go into in Section~\ref{sec:HP}, the gauge hierarchy problem indicates NP
at a scale $\Lambda_{\rm UV}\sim 1$\,TeV.
This suggests that the NP cannot have a generic flavor structure which leads to assumptions like \eg MFV (which however does not provide
an {\it explanation} of this structure). This problem is called the {\it new physics flavor problem}. Note that a very strong constraint
comes from the CP violating parameter $\epsilon_K$ in the Kaon system, see Section~\ref{sec:asl}. We will come back to
these issues later in the context of RS models. 

\section{Hierarchies as a Motivation for Physics Beyond the Standard Model}
\sectionmark{Hierarchies as a Motivation for BSM Physics}
\label{sec:hiera}

Many models given above address the one or the other unsolved problem, like \eg the Peccei-Quinn theory solving 
the strong CP problem, but no testable theory known so far addresses all open questions. Some theories solve several problems 
at once, making them suitable candidates for completing the SM in a more unified way. In the following we focus on two important
problems mentioned before, related to hierarchies, and briefly introduce models that are able to address at least one of these issues. 
The RS model can address both and will be introduced in detail in Chapter~\ref{sec:WED}.

\subsection{The Gauge Hierarchy Problem and Possible Solutions}
\label{sec:HP}

The infamous gauge hierarchy problem has caused lots of attempts to construct models which are able to resolve it. 
This problem, in its modern formulation, arises, if we consider the SM as an EFT, UV completed
by some NP below $M_{\rm Pl}$ (\eg by a GUT). In the last section we have shown that there are many good reasons for such an assumption. 
The gauge hierarchy problem is caused by the enormous hierarchy of 17 orders of magnitude between the electroweak and the Planck scale, 
see the Figure on the left of this page. It is difficult to understand why the weak scale should be so much lower than the Planck scale, $M_{\rm EW}\ll M_{\rm Pl}$, or
equivalently, why gravity is so much weaker than the electroweak force. After studying the details of the problem, we 
will introduce the most famous models that try to address it.

\subsubsection{The Problem}
\label{sec:HPP}

\begin{tabular}{cc}
\vspace{-1.2cm}
	\includegraphics[height=10cm]{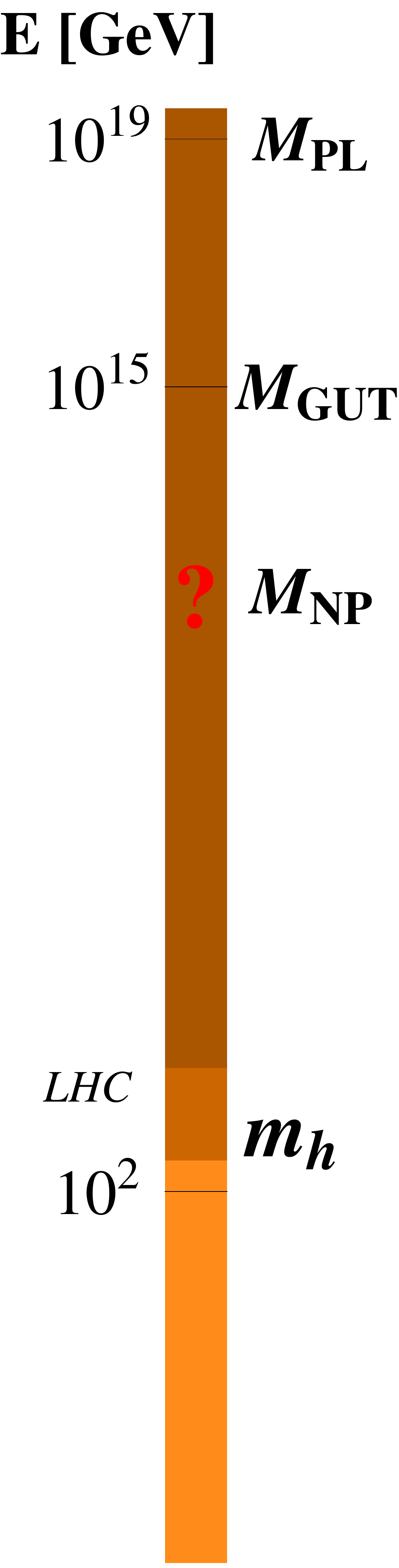}
&
\raisebox{4.5cm}{\parbox{12.99cm}{
From the arguments of Section~\ref{sec:Higgs} we expect the mass of the Higgs boson, setting the electroweak scale, to be very light
compared to the Planck scale
\beq
m_h \sim 100\,{\rm GeV} \ll M_{\rm Pl}\,.
\eeq
It is not difficult to account for such a mass on the tree level by choosing the Lagrangian parameters accordingly. However,
the Higgs boson, being a fundamental scalar particle, has a priori no protection mechanism for its mass via symmetries. From an EFT point
of view, one would naively expect the super-renormalizable (UV sensitive) $D=2$ operator $h^2$ in ${\cal L}_{\rm SM}$ 
to have a coefficient of the order of the cutoff of the QFT, say 
\beq
\label{eq:HPL}
m_h^2 \sim M_{\rm Pl}^2\,.
\eeq
It is now difficult to understand why the Higgs boson should be light, like indirect experiments and theory tell us. The only possibility to avoid this 
puzzle would be to introduce a mechanism which saves the Higgs mass from corrections above a certain scale.}}
\end{tabular}

\begin{figure}[!t]
\begin{center}
\mbox{\includegraphics[height=3.8cm]{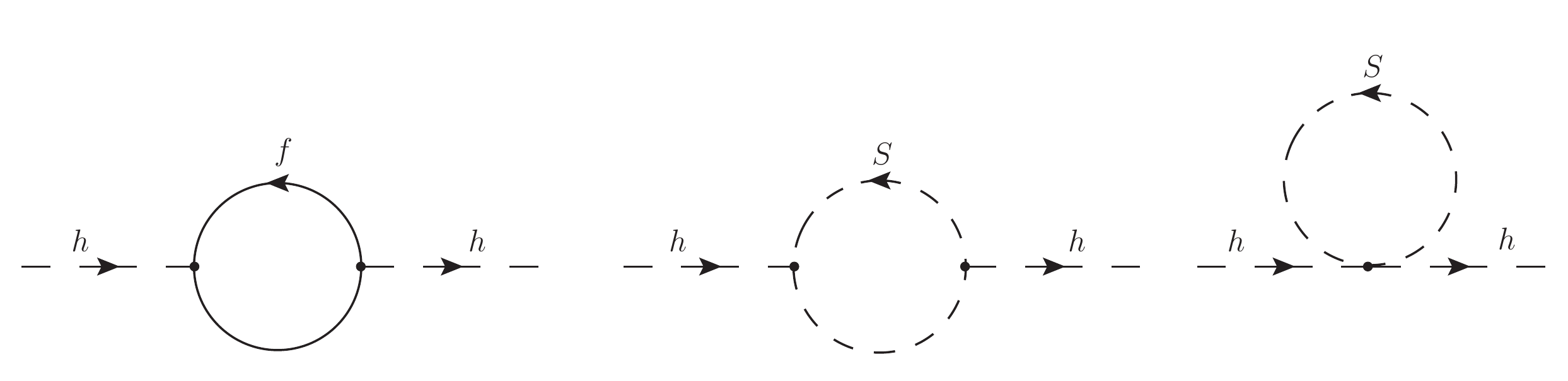}}
\end{center}
\vspace{-1cm}
\begin{center}
  \parbox{15.5cm}{\caption{\label{fig:Hpdiag}One-loop contribution to the Higgs-boson mass-squared from fermion loops (left) and scalar loops (middle, right).}}
\end{center}
\vspace{-0.8cm}
\end{figure}

\newpage
That such large corrections to the mass are indeed created at the quantum level can be seen by studying the loop 
diagrams shown in Figure \ref{fig:Hpdiag}.
The diagram on the left results in a quadratically divergent correction $\sim \Lambda_{\rm UV}^2$ 
(in cutoff regularization) to the Higgs-mass squared (see \eg \cite{Djouadi:2005gj}) of
\beq
\label{eq:Dmhf}
\Delta m_h^2= \frac{|y_f|^2}{8\pi^2}\left[-\Lambda_{\rm UV}^2+6\,m_f^2\, {\rm ln}\frac{\Lambda_{\rm UV}}{m_f}-2\,m_f^2\right]+\ord(\Lambda_{\rm UV}^{-2})\,,
\eeq
for every fermion with mass $m_f$ and Yukawa coupling $y_f$ running in the loop. For simplicity, we have assumed the fermion to be heavy, so that we can 
neglect the external Higgs momentum. The diagrams on the right, corresponding to scalar contributions (with mass $m_S$, trilinear coupling $v \lambda_S$, and quartic coupling $\lambda_S$) in the loop, also lead to quadratically divergent corrections of
\beq
\label{eq:Dmhs}
\Delta m_h^2= \frac{\lambda_s}{16\pi^2}\left[-\Lambda_{\rm UV}^2+2\,m_S^2\, {\rm ln}\frac{\Lambda_{\rm UV}}{m_S}\right]
-\frac{\lambda_s^2}{16\pi^2}v^2\left[-1+2\,{\rm ln}\frac{\Lambda_{\rm UV}}{m_S}\right]+\ord(\Lambda_{\rm UV}^{-2})\,.
\eeq
Note that there are in addition contributions of the gauge bosons $\Delta m_h^2\sim (3g^2+g^{\prime\,2})\,\Lambda_{\rm UV}^2$.
\footnote{An alternative method to quickly obtain the quadratically divergent corrections to the scalar potential, without calculating Feynman
diagrams, is to compute the Coleman-Weinberg potential \cite{Coleman:1973jx}, see also \cite{Grojean:2005ud}.}  
It is in principle possible to renormalize the theory (by adding counterterms to cancel the UV sensitive terms
that would diverge in the limit $\Lambda_{\rm UV}\to \infty$) in such a way as to arrive 
at a Higgs mass around the electroweak scale, despite the quadratic UV sensitivity. 
However, assuming $M_{\rm Pl}$ as the scale of the corrections, this would need incredible fine-tuning of the parameters at the level of 1 in 
$10^{34}$, and seems very unnatural. 

Note that in order to arrive at this conclusion, one does not have to assign a significance to the 
quadratic divergences, which could be related to non understood high energy behavior of the theory. The crucial point is that all the heavy 
particles in the QFT, running in the loop, give a contribution to $m_h^2$ proportional to their mass squared and not coming with $\Lambda_{\rm UV}^2$, see (\ref{eq:Dmhf}) and (\ref{eq:Dmhs}). In consequence one still has a physical correction to the Higgs mass, which is set by the heaviest particle 
running in the loop. In the following we assume the existence of heavy BSM physics, indicated by $M_{\rm NP}$ and $M_{\rm GUT}$ in the Figure on the left
of the last page, for which we have collected very good arguments. Thus, if one believes for example in a GUT, one would expect the Higgs mass 
to reside at the scale
\beq
m_h^2 \sim M_{\rm GUT}^2\,,
\eeq
and not around the mass scale of the electroweak gauge bosons. One would expect that in an effective Lagrangian like (\ref{eq:Lef}) 
\beq
C_h^{(2)}=\ord(1)\,, \quad  {\rm with\ } M^2=M_{\rm GUT}^2,
\eeq
and not $C_h^{(2)} \ll 1$. 
Arriving at a Higgs mass around the electroweak scale would require a cancellation between the bare Lagrangian Higgs-mass parameter 
and the loop corrections to it, which are now both expected to be around $10^{30}\,{\rm GeV}^2$, at a level of 1 in $10^{26}$. 
Let us stress that, seen in this way, the gauge hierarchy problem is not a problem of the SM but rather a problem that 
comes about when extending the SM by new heavy particles - an issue that every extension of the 
SM has to face at a more or less severe level. Although not logically excluded, a desert without NP between the electroweak and the Planck scale seems not very likely, given all the arguments collected in Section~\ref{sec:SMProblems}. Also in such a scenario one still would have an unexplained 
large (though radiatively stable) discrepancy between two fundamental scales.
In contrast to fundamental scalars, gauge bosons or chiral fermions do not have a problem in keeping a low mass. Without 
spontaneous symmetry breaking the chiral fermions of the SM as well as the gauge bosons would be massless. Before EWSB, they do not
have mass terms, corresponding to relevant operators. Those operators are forbidden by the gauge symmetry. After EWSB, the (hidden) symmetry 
saves those particles from corrections to their masses above the scale of EWSB of $v \sim \ord(100\,{\rm GeV})$. 
Thus it seems natural that the massive gauge bosons have masses just around that value. This argument can also be related to the 
fact that a massless spin-1 particle has two degrees of freedom and there is no continuous transformation to a massive boson with three 
degrees of freedom. In the Higgs mechanism, this additional longitudinal polarization is provided by a goldstone boson. However, without 
such a mechanism, massless gauge bosons will not receive a mass at the loop level. The same is true for massless (chiral) fermions, which only 
have one helicity. Note that corrections to the fermion masses are proportional to their own masses and only logarithmically divergent 
$\Delta m_f\, \sim m_f\, {\rm ln}(\Lambda_{\rm UV}/m_f)$. However, for the spin-0 Higgs boson, there is a priori no such argument, which leads to the 
expectation (\ref{eq:HPL}).  

The gauge hierarchy problem hints strongly at NP at the TeV scale ($\Lambda_{\rm UV} \lesssim 4\pi M_{\rm EW} \sim 1$\,TeV) that solves this issue. 
On the other hand, the good agreement of the SM with experiment in many details seems not to suggest new physics at such low scales. For example,
electroweak precision tests rather indicate that generic new physics is not to be found below several times this scale ($\Lambda_{\rm UV} 
\gtrsim$ a few TeV), needless to mention constraints from Kaon mixing. As with a cutoff for the SM at such a scale, one ends up with a fine 
tuning of some per cent, a ``little hierarchy problem'' could remain after a possible solution to the large gauge hierarchy problem. 
As we will see later, much effort is made to address this issue in BSM models and to allow for lower NP scales.
Excitingly the LHC just starts to approach the TeV scale and the gauge hierarchy problem lets us expect further discoveries, in addition 
to that of the sector of EWSB. The LHC will hopefully also reveal the sector that stabilizes the mass 
of the fundamental scalar Higgs boson (if present) against quantum fluctuations.

\subsubsection{Possible Solutions: From SUSY to Classicalization}
\label{sec:solHP}

\begin{figure}[!t]
\begin{center}
\vspace{0.2cm}
\mbox{\includegraphics[width=11cm]{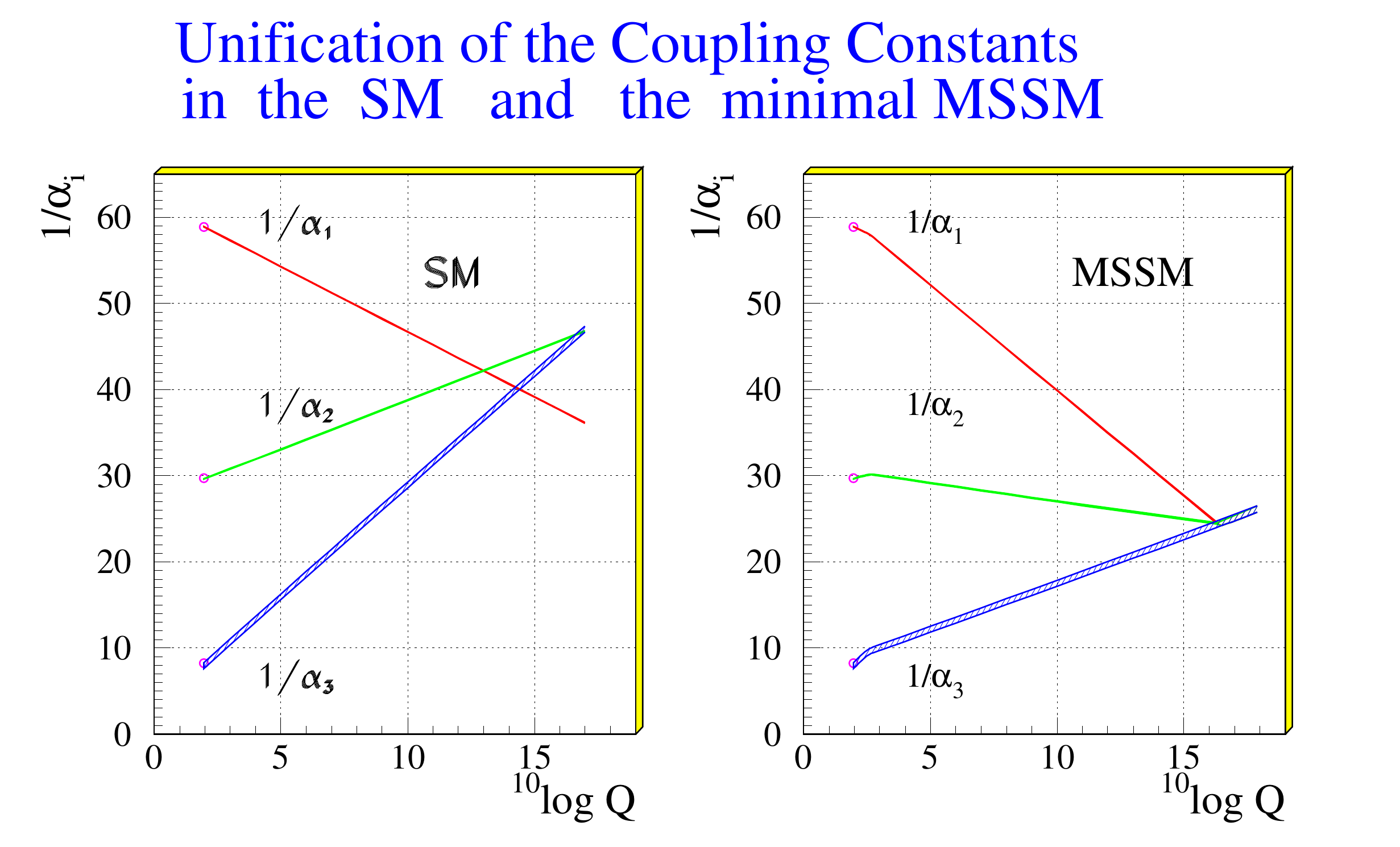}}
\end{center}
\vspace{-1cm}
\begin{center}
  \parbox{15.5cm}{\caption{\label{fig:unif} Renormalization group evolution of the gauge couplings in the SM as well as in the MSSM.
  Figure from \cite{deBoer:2003xm}, with permission from the author. See text for details.}}
\end{center}
\end{figure}

The most popular ideas to address the gauge hierarchy problem are to invoke a new symmetry (to protect
the Higgs mass),
to introduce a new strongly coupled sector (with the scalar EWSB sector consisting of bound states
of this new interaction) or to lower the cutoff of the theory (by extending space-time). Some models relate several of these
classes to each other. Let us start with recalling the problematic contributions to the Higgs mass due to fermion and boson loops,
see (\ref{eq:Dmhf}) and (\ref{eq:Dmhs}). Given the absence of a symmetry to protect the Higgs mass,
these terms contribute to the gauge hierarchy problem.
Consider the corrections of $N_f$ fermions and those due to $N_S$ scalar particles.
If we now {\it assume} $y_f^2=-\lambda_S$ and $N_S=2N_f$ and add up both contributions (of particles with different 
spin-statistics) we arrive at \cite{Djouadi:2005gj}
\beq
\Delta m_h^2= \frac{y_f^2 N_f}{4\pi^2}\left[(m_f^2-m_S^2)\,{\rm ln}\frac{\Lambda_{\rm UV}}{m_S}+3\,m_f^2\,{\rm ln}
\frac{m_S}{m_f}\right]+\ord(\Lambda_{\rm UV}^{-2}).
\eeq
We see that the problematic quadratic divergences have vanished, seemingly ameliorating the gauge hierarchy problem.
However the corrections to the Higgs mass are still proportional to the mass of the heaviest particle running in the loop.
If in addition $m_f=m_S$ holds, the correction to the Higgs mass vanishes completely (note that all that is not possible 
within the SM).
Thus, if there would be a symmetry that would lead to $y_f^2=-\lambda_S$, $N_S=2N_f$, and $m_f=m_S$, the gauge hierarchy problem
would be solved (given a similar cancellation takes place for {\it all} contributions to the Higgs mass, including 
gauge boson corrections). Such a symmetry exists indeed and is just the so-called {\bf supersymmetry}, introduced
in Section~\ref{sec:SMProblems}, which allows to extend the spin-considerations of Section~\ref{sec:HPP} to scalar particles. 
Interestingly, this new symmetry, relating bosonic to fermionic degrees of freedom, is the only possibility to generalize the direct product 
of [space-time]$\otimes$[internal symmetries], along the lines of the HLS theorem
as discussed in \ref{sec:SMProblems}. For reviews on SUSY and a list of references see \eg \cite{Dine:2007zp,Martin:1997ns,Signer:2009dx,Wess:1992cp}.

A SUSY Lagrangian is invariant (up to a total derivative) under   
SUSY transformations, mediated by fermionic sets of generators $Q_\alpha^i$. These transform fermions into bosons 
(with the same internal quantum numbers) and vice versa. 
Here, $\alpha$ is a spinor index, and the $N$ possible sets of generators are labeled by $i$. In the following we will consider $N=1$ SUSY, 
corresponding to a single set of SUSY generators. Schematically, one has
\beq
Q_\alpha |{\rm fermion}\rangle^\alpha=|{\rm boson}\rangle\,,\quad Q_\alpha |{\rm boson}\rangle=|{\rm fermion}\rangle_\alpha\,.
\eeq
The generators of the $N=1$ super Poincar\'{e} algebra fulfill the relations
\beq
\begin{split}
\left[Q_\alpha,P^\rho\right]&=0\\
\left\{Q_\alpha,\bar Q_{\dot\beta}\right\}&=2\left(\sigma^\rho\right)_{\alpha\dot\beta}P_\rho\\
\left[M^{\rho\sigma},Q_\alpha\right]&=-i\left(\sigma^{\rho\sigma}\right)_\alpha^{\hspace{1mm}\beta} Q_\beta\\
\left\{Q_\alpha,Q_\beta\right\}&=\left\{\bar Q_{\dot\alpha},\bar Q_{\dot\beta}\right\}=0\,,
\end{split}
\eeq
where $Q_{\alpha}$ and $\bar Q^{\dot\alpha}=(Q^\alpha)^\dagger$ are Weyl spinors, corresponding to the upper and lower components of a four-component 
Dirac spinor. The indices $\alpha,\dot\alpha$ can be raised and lowered with the totally antisymmetric tensor 
$\epsilon_{\alpha\beta}$ \cite{Signer:2009dx} and $\sigma^{\mu\nu}=i/2[\gamma^\mu,\gamma^\nu]$. The generators $P^\rho$ and $M^{\rho\sigma}$ correspond to the four-momentum and the generalized angular momentum.
The fields of a supersymmetric theory are grouped into irreducible representations of the SUSY algebra, so-called {\it supermultiplets}.
If SUSY was an exact symmetry, as an immediate consequence the SUSY partners would have exactly
the same mass as the corresponding SM fields.  As such a scenario is clearly excluded by experiment, SUSY has to be broken. This is in general 
achieved by introducing super-renormalizable (soft) SUSY breaking terms. However, the mass scale of the SUSY partners of the SM particles 
$M_{SUSY}$ should not be much larger than the TEV scale as otherwise the gauge hierarchy problem would be introduced again to a certain amount.
In consequence, if SUSY solves the gauge hierarchy problem, we expect to find it at the LHC. In the 
following, we explore briefly some phenomenological consequences of the minimal supersymmetric extension of the SM (MSSM). 
This minimal extension features a fermionic partner for each SM gauge boson, as well as a bosonic partner for each SM (Weyl) fermion,
such that in the end, as for every SUSY model, the number of fermionic and bosonic degrees of freedom is equal.

As mentioned before, SUSY models require the 
existence of more than one Higgs doublet. The MSSM features two Higgs doublets with 5 physical Higgs bosons (after EWSB) - two neutral CP even 
$h,H$, a CP odd one $A$, and two charged $H^\pm$. At the tree level the MSSM leads to the constraint
\beq
m_h<m_Z|\cos(2\beta)|
\eeq
for the mass of the lightest neutral Higgs boson, where $\tan\beta=v_u/v_d$ and $v_u$, $v_d$ are the VEVs belonging to the Higgs doublets 
giving masses to up and down type quarks, respectively. Although in the MSSM the LEP exclusion limit can be weakend (depending on $\tan\beta$) \cite{Djouadi:2005gj}, this constraint would put the MSSM into some trouble. However, loop corrections can lift the limit to
\beq
m_h\lesssim 135\,{\rm GeV}\,, 
\eeq
assuming that all sparticles that can contribute to $m_h$ have masses below a TeV \cite{Djouadi:2005gj}.
Note that the properties of the lightest Higgs boson of the MSSM are similar to the SM Higgs in the decoupling regime (where $m_A \sim \ord$(TeV)), 
however, outside this regime important differences can arise. For example, the couplings to bottom quarks can be strongly
enhanced in the case of large $\tan\beta \gg 1$.
There is still a puzzle in the Higgs sector of the MSSM, the so-called $\mu-$problem. This corresponds to the fact that
it is difficult to understand why the magnitude of the SUSY preserving squared-mass in the Higgs potential should be in the region of 
$(100-1000)$\,GeV, which coincides with the scale of the (unrelated) soft breaking terms, as required by phenomenology, 
and not for example at the Planck scale. Some extensions of the minimal setup try to solve this problem by introducing the 
$\mu$ term as the VEV of a new field.
Moreover, the number of parameters in the MSSM is significantly increased with respect to the SM, especially due to the SUSY 
breaking sector. In total, the MSSM possesses $\sim 120$ new parameters with respect to the SM.

Note that, beyond solving the gauge hierarchy problem, the MSSM solves several of the problems dealt with in Section~\ref{sec:SMProblems}.
First of all, let us mention that the MSSM features a so-called $R$-parity in order to prevent the presence of $B$ and $L$ violating 
couplings in the Lagrangian, where $R=(-1)^{2s+{3B+L}}$ and $s$ is the spin of the particle the operator acts on.
This parity leads to the fact that SUSY partners of the SM particles can only be produced in pairs. In particular, the lightest
supersymmetric partner (LSP) will be stable. In many cases, the LSP corresponds to the lightest of the so-called neutralinos, 
and is massive, weakly interacting and electrically neutral. This provides a good candidate for the {\it Dark Matter} of the universe
within the MSSM.
Moreover, the modification in the running of the gauge couplings due to the SUSY partners provides the possibility for them to meet
each other to good accuracy around the scale $M_{\rm GUT}$, allowing for a {\it grand unification} within the MSSM \cite{Amaldi:1991cn}. 
This fact is illustrated in Figure \ref{fig:unif}, where the evolution of the gauge couplings $\alpha_1\equiv 5/3\,\, g^{\prime\,2}/(4\pi)\, ,\alpha_2\equiv g^2/(4\pi)\, ,\alpha_3\equiv g_s^2/(4\pi)$ is shown. Above the mass scale of the SUSY partners $M_{\rm SUSY}\sim 1$\, TeV, which has been fitted 
such that unification is possible, the running is modified with respect to the SM. The factor 5/3 is needed for the correct normalization at 
the unification point. In addition, the MSSM, allows for {\it radiative electroweak symmetry breaking}. In the SM, the Higgs mechanism is only
a description of EWSB but offers no explanation why it occurs. In contrast to the SM, where one has 
to put in the negative squared mass for the Higgs doublet by hand, the MSSM can generate the necessary conditions for a potential
to result in electroweak symmetry breaking radiatively, at a certain scale, via renormalization group evolution. 
It is interesting that the MSSM solves all these puzzles, without being invented for that purpose. It was rather built as the 
most general extension of the Poincar\'{e} algebra, as explained before.
Finally note that, if SUSY is introduced as a local symmetry, it naturally incorporates gravity ({\it supergravity}).
For more details, \eg on the SUSY breaking mechanism, related flavor issues or collider signatures see the literature cited before. 
If the SUSY-breaking mechanism is not flavor blind, SUSY models will generically have problems with flavor constraints, requiring 
an additional concept like MFV, see Section~\ref{sec:SMProblems}. For details on Higgs physics in SUSY see \cite{Djouadi:2005gj}.

The next type of models that can address the gauge hierarchy problem are those that do not feature an elementary
scalar and thus avoid the fine tuning problem.
An example of this class of models is {\bf technicolor} \cite{Weinberg:1979bn,Susskind:1978ms}.
Here, the scalar degrees of freedom responsible for the masses of the weak gauge bosons are realized as bound states of new massless fermions
(techni-fermions), that are charged under a new QCD-like $SU(N)$ gauge group as well as under the electroweak gauge group of the SM. The new strong 
interaction is asymptotically free at high energies, but assumed to be 
confining at $M_{\rm EW}$. In analogy to QCD, chiral symmetry breaking is triggered and the resulting Goldstone bosons (techni-pions),
corresponding just to bound states of the techni-fermions, provide the longitudinal degrees of freedom for the electroweak gauge bosons.
In consequence, EWSB is realized dynamically. Importantly, it is possible to reproduce the relation (\ref{eq:mWmZ}) to leading order in technicolor models.
Note however, that the simplest models are ruled out by electroweak precision tests, due to large 
corrections to the $S$ parameter \cite{Peskin:1991sw}.
For a review, also about newer developments in the sector, see \cite{Hill:2002ap}.

Similar ideas are used in {\bf Little Higgs} theories \cite{Georgi:1974yw,Georgi:1975tz,ArkaniHamed:2001nc, ArkaniHamed:2002qy}, where the Higgs boson 
itself is realized as a (pseudo-)Goldstone boson of
a spontaneously broken global symmetry. This symmetry is also assumed to be broken explicitly (making the Higgs boson massive), 
but only by the interplay of at least two coupling constants. This {\it collective symmetry breaking} leads to a vanishing of the 
quadratic divergences in the Higgs mass at one loop. If one of these couplings does not contribute, the Higgs boson mass
is zero due to the {\it shift symmetry} of the Goldstone boson. Little Higgs theories are weakly coupled effective field theories with a 
cutoff scale of the order of $\Lambda_{\rm UV}\sim 10$\, TeV. They stabilize the hierarchy between 
this cutoff (which could be identified with a scale suggested by electroweak precision tests) and the electroweak scale 
$M_{\rm EW}$. Above the cutoff, these theories have to be UV completed. 

Such a completion could 
feature a breaking of the global symmetry by strong dynamics, resulting in a {\bf composite Higgs} boson \cite{Katz:2003sn}.
In this general class of theories the Higgs-boson mass is protected by the fact that at a certain scale it will not behave as 
an elementary particle anymore and a form factor will regularize the UV behavior. 
For interesting relations with the models studied in this thesis see \cite{Contino:2003ve,Agashe:2004rs,Thaler:2005en}.

Another recent approach that provides (amongst other interesting possible consequences) an idea how the gauge hierarchy problem could be solved, 
assumes the existence of {\bf classicalization} \cite{Dvali:2010jz} at large energies. In this approach, the collission of two particles with 
a momentum transfer larger than the characteristic scale $M_*$ of the (non-renormalizable) theory will cause the formation of an extended classical 
configuration of a classicalizer field, with radius $r_*$, in analogy to a black hole of gravity, UV completing the theory. As a consequence,
the short distance behavior does not have to be specified to know the high energy behavior. The radius $r^*$ grows with the energy localized 
in the system. If the Higgs scalar is sourced 
by energy-momentum of the SM particles, \eg by an operator
\beq
\Phi^\dagger\Phi\, T^\mu_\mu\,,
\eeq 
where $ T^\mu_\mu$ corresponds to the trace of the energy-momentum tensor of the SM fields,
it could just act as such a classicalizer \cite{Dvali:2010jz}. It would develope a classical configuration above a scale $M_*$, which could solve the gauge hierarchy problem.

When trying to construct solutions to the gauge hierarchy problem, we have so far only talked about changing the SM along 
the lines discussed in \ref{sec:SMProblems}. We have not explored another possibility yet, which would change quite drastically 
the picture of the universe we have so far. When looking at the world around us and feeling time passing by, we seem to know, that
we are living in four space-time dimensions, an implicit assumption we have made before. When talking about 
Poincar\'{e} invariance, we were thinking about four-dimensional Minkowski space-time. However, already from 
high-school physics we know that our eyes or the human senses are not sufficient as measuring instruments. In the next chapters 
we will introduce models that involve {\it extra dimensions} and that are not in immediate conflict with observation. Some of these 
models, in particular RS models, offer an alternative possibility to solve the gauge hierarchy problem. Beyond that,
they have further very interesting features.

\subsection{Hierarchies in the Fermion Sector}
\label{sec:SMhi}

In addition to the large hierarchy between the electroweak and the Planck scale, there are also non-understood hierarchies in the flavor 
sector of the SM. The masses and mixings of the SM fermions exhibit large ratios, which can be accounted for within the SM but which cannot
be explained by the model. The tiny neutrino masses have already been discussed. However, there are also sizable hierarchies in between
the quark masses as well as the charged lepton masses. Moreover, the mixing matrix in the quark sector also is strictly hierarchical, 
while the one of the lepton sector is not. Let us have a look at the hierarchies in the quark sector more quantitatively. 

\begin{figure}[!t]
\begin{center}
\mbox{\includegraphics[width=7.2cm]{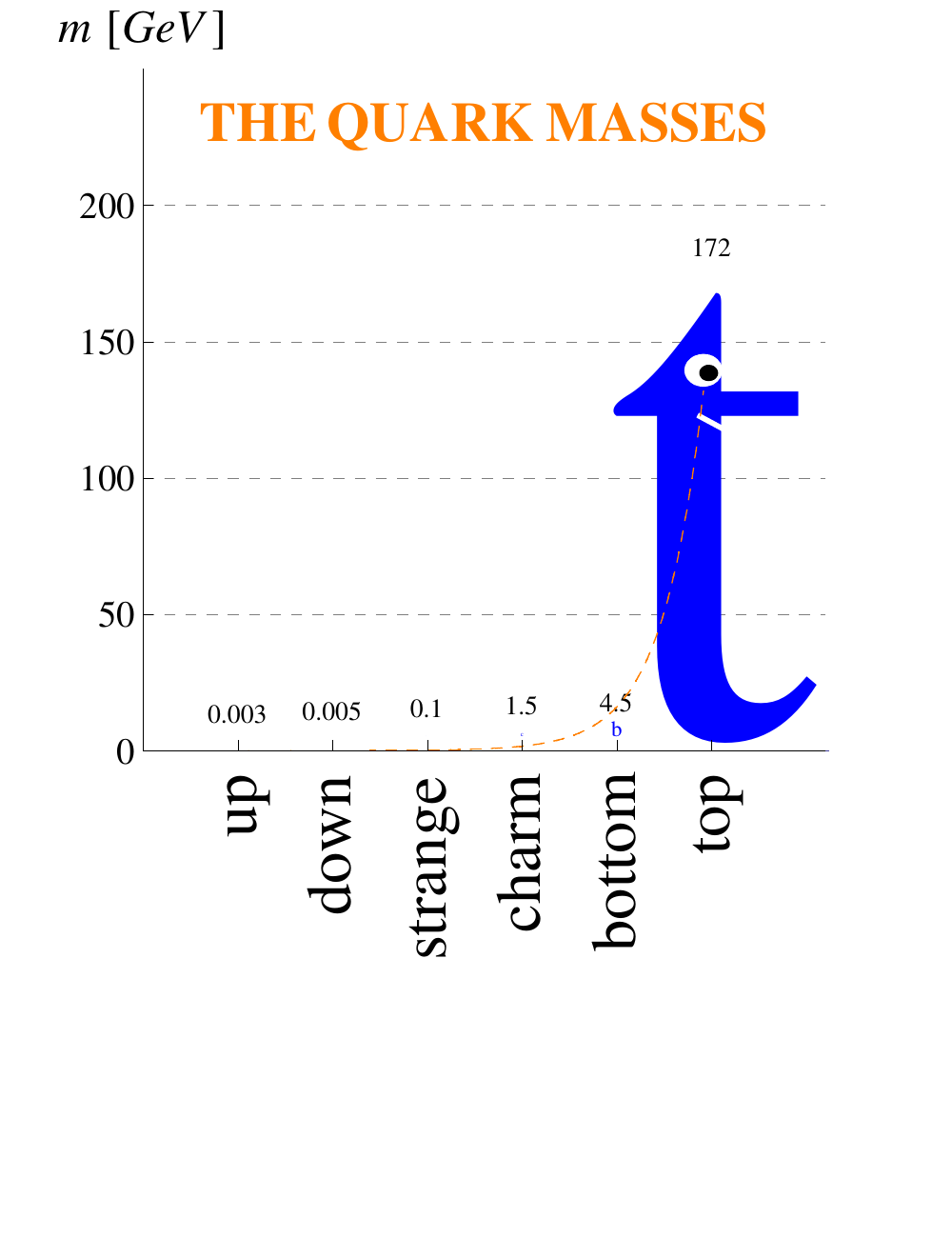}}
\end{center}
\vspace{-3.15cm}
\begin{center}
  \parbox{15.5cm}{\caption{\label{fig:top} Hierarchies in the quark masses.}}
\end{center}
\end{figure}
The quarks of the same charge in the SM are identical in all their properties besides their mass. Here, they exhibit very large ratios of roughly 
(see Appendix~\ref{app:ref}) \footnote{As we are just interested in rough hierarchies, we do not have to worry about the issues in defining the quark masses.}
\beq
\label{eq:quarkhier}
m_t/m_c/m_u\approx100000/500/1\,,\quad m_b/m_s/m_d\approx1000/20/1\,,
\eeq
see Figure \ref{fig:top}.
Although the (running) quark masses depend on the energy scale, the hierarchies quoted before are to good accuracy stable with 
respect to scale variations \cite{Froggatt:1978nt}. The natural value for the scale of 
the quark masses would be $m \sim v/\sqrt{2}$, corresponding to an $\ord(1)$ Yukawa coupling. This expectation is fulfilled only for 
the top quark. Although the hierarchies (\ref{eq:quarkhier}) can be put into the SM, it offers no {\it explanation} for 
the question why the masses of its fundamental constituents in the matter sector differ so much from each other. 

Also the CKM matrix exhibits strong hierarchies, which are reflected in the expansion in the small Wolfenstein parameter $\lambda\approx 0.23$ (\ref{eq:CKMWolf}).
To LO in each entry
\beq
\label{eq:CKMWmag}
\left|\bm{V}_{\rm \hspace{-1mm} CKM}\right|\sim
\left(\begin{array}{ccc}
1 & \lambda & \lambda^3 \\
\lambda & 1& \lambda^2 \\
\lambda^3 & \lambda^2 & 1
\end{array}\right)\,.
\eeq
The fact that the diagonalizations of the up-type and down-type mass matrices lead, to first order, to the same rotation matrices $\bm{U}_u\approx \bm{U}_d$
can not be explained within the SM and calls for elucidation.

A model which can address both of these fermion hierarchies, has been presented by Froggatt and Nielsen \cite{Froggatt:1978nt}.
Starting from anarchic Yukawa matrices, \ie, matrices with random $\ord(1)$ entries with arbitrary phases, the observed hierarchies 
are generated by assuming the left-handed and right-handed components of the quark fields to have different values for
an almost conserved quantum number with respect to an abelian symmetry group $U(1)_F$. This model will be introduced in 
Section~\ref{sec:quark} and we will directly apply its results to the fermion sector of RS models, which, as we will show,
feature a similar mechanism. Here, the Froggat-Nielsen mechanism arises automatically, 
if one puts fermions in the bulk, and no additional structure is needed. So, beyond solving the gauge hierarchy problem, the fermion hierarchies 
can also be addressed in these models, without the setup having been invented for that purpose.

\chapter{Extending Space-Time to Address Hierarchies}
\label{sec:XD}
\vspace{-6mm} 

In this chapter we will give a survey of how the idea that nature could feature more than four space-time dimensions finally
lead to an option to address the gauge hierarchy problem. After a short historical overview, we will review the 
Arkani-Hamed$-$Dimopoulos$-$Dvali model in some detail. This model can be seen as the forerunner of the RS model.
Although the approach to address the large hierarchy between the electroweak and the Planck scale is different, it
certainly influenced the following developments quite strongly.

\section{Introduction: Extra Dimensions before ADD}
\label{sec:string}

\begin{figure}[!t]
	\centering
		\includegraphics[width=5.5cm]{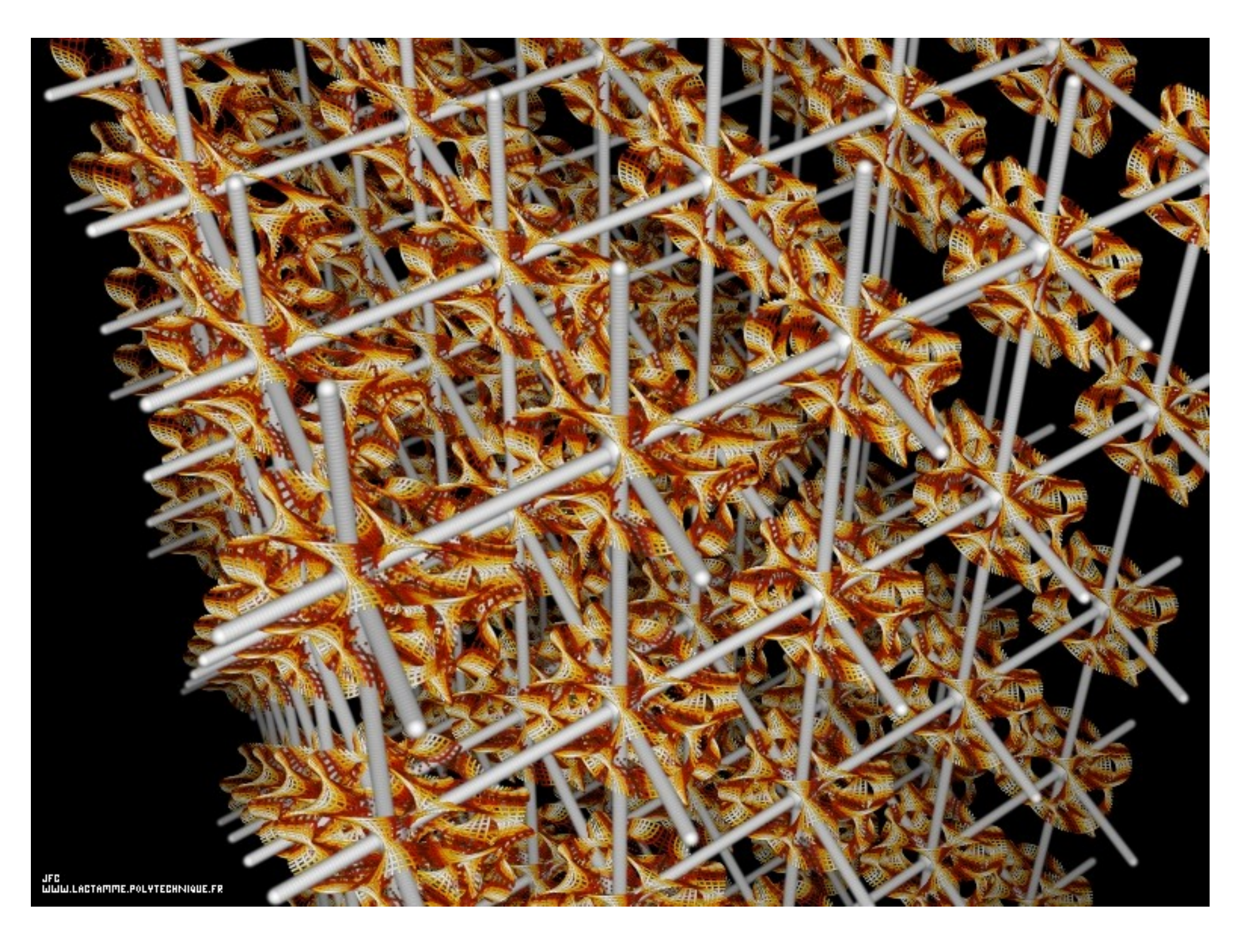}
		\vspace{-2mm}
	\caption{	\label{fig:Col} Illustration of a compactified higher-dimensional space-time. A Calabi-Yau manifold is attached 
	to every point of the usual 4D space-time. Courtesy of J.-F. Colonna \cite{Col}.}
\end{figure}

The first serious appearance of the idea that space-time could feature more than four dimensions was due to {\it Kaluza} 
\cite{Kaluza:1921tu}  and {\it Klein} \cite{Klein:1926tv} in the 1920s. They tried to unify gravity and electromagnetism,
the only fundamental forces known at that time, by merging the photon vector field together with the 4D Minkowski metric 
into a $5\times5$ metric. However, their idea turned out to be not a correct description of nature, in particular after
the discovery of the additional forces and the rise of the electroweak standard model. Moreover, already before, the attempt
to quantize the theory lead to serious problems. The ideas of Kaluza and Klein were revived in the 1980s by the 
advent of {\it string theory} in its modern form (for a review and further references see \eg \cite{Dine:2007zp,Ooguri:1996ik,Green:1987sp,Green:1987mn}).
The roots of string theory go back to the 1960s, when Gabriele Veneziano \cite{Veneziano:1968yb}
and others tried to use strings to describe the strong interaction. During the first 
superstring revolution (1984-1989) it was realized that string theory could serve as a fundamental description of nature 
if it is formulated in 9+1 space-time dimensions. The number of dimensions was set by the need for anomaly cancellation. 
In the framework of superstring theory, the fundamental ingredients of nature are not point-like particles but one dimensional 
strings of Planck-length size, characterized by their tension. This leads to a unification of the various different
particles of the SM into one type of fundamental ``particle'', the string. 
How is it possible that the extra dimensions have not been discovered so far? For not being in conflict with observation
these dimensions have to be hidden in some way. This is possible by compactifying them, \eg on a higher dimensional torus (or on a more complicated manifold) 
with a natural length scale of the order of $M_{\rm Pl}$. One can imagine this additional geometry sitting at every space-time point of 4D Minkowski space-time, as illustrated
in Figure \ref{fig:Col}
 
The second superstring revolution took place in the mid 1990s and was initiated by Edward Witten. He discovered that the 
different formulations of string theory known at that time (as well as 11D supergravity) are related to each other 
by dualities and could all be manifestations of the same theory, called M-theory. With the emergence of the 
concept of branes another option to hide the extra dimensions became available. Open strings, corresponding to SM particles 
in the low energy limit, have to end on their higher dimensional cousins, the branes. It could thus be possible that the particles we know 
are confined to live on a 4D sub-manifold of space-time, a 3-brane, see below.
It seems to be possible to model the SM spectrum by an appropriate geometry of the extra dimensions. Beyond that, these 
theories have in principle the ambition to {\it explain} the presence of a certain gauge group and particle content present 
in nature and to address all other theoretical questions stated in \ref{sec:SMProblems}. In particular they are UV finite 
due to the non vanishing extension of their fundamental constituents and contain a quantum theory of gravity. In this sense, 
they can be called theory of everything (TOE) (although it is in principle not possible to prove that a real TOE has been found). 
Despite their many promising features, string theories also have serious drawbacks. An important issue is that in the original spirit of 
residing at the Planck scale of $M_{\rm Pl}\sim10^{19}$\,GeV, string theories make (currently) no unique predictions for energies 
accessible in the near future. The expected excitations of the strings would have masses around $M_{\rm Pl}$. The mechanism of compactifying 
the extra dimensions is not yet understood sufficiently. As mentioned before, the geometry of space-time is related to the low energy 
spectrum. However, it seems that the incredible number of $10^{500}$ different vacua are possible and thus, if there is 
no mechanism to chose a special vacuum, the geometry cannot be predicted and neither can the low energy spectrum. It is however possible 
to interpret a discovery of supersymmetry as a hint for string theory, since it can only be formulated in a supersymmetrized way. This 
still does not make the model really testable at current collider energies. Moreover, note that string theory is still at the level of quantum mechanics 
(no second quantization). 

This pessimistic picture changed after it was realized that extra (spatial) dimensions could help in solving the gauge hierarchy 
problem in the late 1990s by Nima Arkani-Hamed, Savas Dimopoulos and Gia Dvali (ADD) \cite{ArkaniHamed:1998rs}.
To see how this works, let us recall that the gauge hierarchy problem is due to the fact that the Planck scale is so much bigger 
than the electroweak scale. An important point is that one deduces the gigantic size of the Planck scale from the enormous 
weakness of gravity (see footnote~\ref{fn:G}) $G_N=M_{\rm Pl}^{-2}$. In terms of the fundamental Planck mass 
$M_{\rm Pl}^{(d)}$, the gravitational potential between two test masses $m_1$ and $m_2$ in $d$ dimensions, has the form
\beq
\label{eq:VG}
V^{d}(r) \sim \frac{m_1 m_2}{[M_{\rm Pl}^{(d)}]^{d-2}}\ \frac{1}{r^{d-3}}\,.
\eeq
From measuring the gravitational force one deduces $d=4$ and determines the Planck scale $M_{\rm Pl}=
M_{\rm Pl}^{(4)}\sim 10^{19}\,$GeV. However, it is important to note that Newtons gravitational law has so far only be tested down to 
length scales of about $50\,\mu$m $\sim 250\,$eV$^{-1}$ \cite{Kapner:2006si}. 

\section{From Large Extra Dimensions to Warped Extra Dimensions}

The fact that Newton's law, (\ref{eq:VG}) with $d=4$, has by far not been tested up to the same energy scales as the other interactions of nature, 
not speaking of up to the Planck scale, lead ADD to the idea that the gravitational potential could be changed at small distances $r$, with however 
$r\gg M_{\rm EW}^{-1}$. This allows for the existence of {\it large extra dimensions}, if only gravity is allowed to propagate into them \cite{ArkaniHamed:1998rs}. 
How does this work in detail? 

\subsection{Large Extra Dimensions to Address the Gauge Hierarchy Problem}
\label{sec:LED}

Imagine $n$ extra dimensions, all of size $\sim R$ for simplicity, compactified on a corresponding manifold 
$M_n$ with volume $\sim R^n$.
For distances $r\ll R$, Gauss's law in $d$ dimensions leads just to the gravitational potential (\ref{eq:VG}) with $d=n+4$
\beq
    V^{4+n}(r) \sim \frac{m_1 m_2}{[M_{\rm Pl}^{(4+n)}]^{n+2}}\ \frac{1}{r^{n+1}}\qquad (r \ll R)\,.
\eeq
However, for distances $r\gg R$ the field lines of gravity will not resolve the extra dimensions, leading to the usual $1/r$ 
behavior we observe
\beq
		V^{4+n}(r) \sim \frac{m_1 m_2}{[M_{\rm Pl}^{(4+n)}]^{n+2}} \frac{1}{R^{n}}\ \frac{1}{r}\qquad (r \gg R)\,.
\eeq
An interesting thing has happened here. If we do not assume the existence of extra dimensions, we would just identify the whole 
expression
\beq
\label{eq:MPLeff}
M_{\rm Pl}^{\rm eff} \equiv [M_{\rm Pl}^{(4+n)}]^{\frac{n+2}{2}} R^{\frac{n}{2}}
\eeq
with the Planck scale (the cutoff of the SM) $M_{\rm Pl}=M_{\rm Pl}^{eff}\sim10^{19}\,$GeV, without noticing the
presence of additional dimensions. Thus, if only gravity is allowed to propagate into the extra dimensions they could be possibly 
as large as $50\,\mu$m, since below that scale the gravitational interaction has not been tested and above we would not notice the presence of the additional dimensions.
The only consequence at large distances would be the fact that we would have determined the effective Planck scale (\ref{eq:MPLeff})
and not the fundamental one $M_{\rm Pl}^{(4+n)}$.
It is not digressive that (\ref{eq:VG}) does not hold for the whole range from macroscopic distances down to the Planck length 
$l_p\sim10^{-35}$ m. We know many examples for theories in physics that are replaced
by another one at a certain scale, see Appendix~\ref{app:EFT}. The law of gravity that we measure at macroscopic distances does not have to 
be the final answer. Furthermore, the fact that only gravity might propagate into the additional dimensions can be motivated by string 
theory. Here, the corresponding force mediators correspond to closed strings, allowed to propagate into the whole space-time, while the 
other particles would correspond to open strings, attached to D-branes \cite{Antoniadis:1998ig}. Moreover, due to general relativity, gravity 
is directly related to
space-time, whereas the other forces are not. In this picture, the weakness of gravity would be just a dynamically generated effective 
weakness, seen for large distances, due to the dilution of gravity propagating into extra dimensions. 
As the other fundamental interactions have been probed up to the electroweak scale, extra dimensions would have to be smaller
than $\sim 1$\,fm to have escaped detection, if the SM would be allowed to enter them. However this would still allow
for additional dimensions much larger than the (4D) Planck scale. The SM could be localized on a sub-manifold of the whole space-time, with 
TeV$^{-1}$ ``thickness'', for example due to topological defects. In order not to dilute the SM interactions, the thickness should not
be bigger than the fundamental Planck scale.

Looking again at (\ref{eq:MPLeff}), we see how the gauge hierarchy problem could be addressed in such a setup of gravity propagating into
large extra dimensions. The Planck scale would just appear to be huge due to the presence of additional dimensions that we do not
resolve. We would just think that we measure the fundamental Planck scale, but in reality we would measure $M_{\rm Pl}^{\rm eff}$ (\ref{eq:MPLeff}),
which is enhanced due to the volume of the extra dimensions, given that
\beq
R>\left(M_{\rm Pl}^{(4+n)}\right)^{-1}.
\eeq 
The {\it fundamental} Planck scale could well be of the order of
\beq
\label{eq:effPLTeV}
M_{\rm Pl}^{(4+n)} \sim M_{\rm EW}\,,
\eeq
just as large that we would not have already noticed it, say around a TeV. Now this scale would set the cutoff for the model and for
quantum corrections to the Higgs mass. Above this energy, a quantum theory of gravity like string theory could set in, cutting away 
potentially large radiative corrections at a TeV. The natural scale for the Higgs-boson mass would then be around this fundamental Planck 
scale and the fine-tuning problem as discussed in \ref{sec:HP} would nearly completely vanish. The theory would have only one fundamental 
scale, the electroweak scale. When starting to probe the extra dimension, gravity would be of weak scale strength in $4+n$ dimensions 
and not suppressed by the huge effective Planck mass. From the requirement (\ref{eq:effPLTeV}) one can derive the corresponding size 
of the extra dimensions $R$, in dependence on their number $n$. With the help of (\ref{eq:MPLeff}) we arrive at
\beq
\label{eq:RMPL}
R\sim \left[M_{\rm Pl}^{eff}\right]^{\frac 2 n} \left[M_{\rm EW}\right]^{-\frac{2}{n}-1}\sim 2 \cdot 10^{\frac{32}{n}-19} {\rm m} \times \left[\frac{\rm TeV}{M_{\rm EW}}\right]^{\frac 2 n +1},
\eeq
where we have used the relation 200 MeV\,fm $\approx$ 1.

For $\underline{n=1}$ we get $R\sim 10^{13}\,$m, setting $M_{\rm EW}\sim1\,$TeV for this discussion. This implies that if there is exactly one large extra dimension, it 
would have to be larger than the extension of the solar system, which is around $7 \cdot 10^{12}\,$m . Such a scenario is certainly excluded. 

For $\underline{n=2}$ we end up with $R\sim 1\,$mm, which is excluded by the experimental constraint $R\leq44\,\mu$m \cite{Kapner:2006si}. 
However, if we are willing to accept some amount of tuning, we can relax the requirement (\ref{eq:effPLTeV}) and use the experimental 
result to set a lower limit on the fundamental Planck scale. We arrive at $M_{\rm Pl}^{(6)}>7\,$TeV which would again introduce a little 
hierarchy, but still would be a big improvement compared to the full gauge hierarchy problem. However, there are more stringent 
constraints for $n=2$ from astrophysics, see below. 

For $\underline{n\geq3}$ we get $R\leq10^{-8}\,$m, which is not directly testable in the near future.

A weakness of the ADD model is that one could also see it as just rephrasing the gauge hierarchy problem. Instead of asking why $ M_{\rm EW}\ll M_{\rm Pl}$
one can now ask the question why the extra dimensions should be so much bigger than their natural size of $M_{\rm EW}^{-1}$
\beq
\frac{R}{(M_{\rm EW})^{-1}} \sim \left[ \frac{M_{\rm Pl}}{M_{\rm EW}} \right]^{\frac 2 n}.
\eeq
Nevertheless, the problem of the radiative instability of the large separation between two fundamental scales is not present anymore. 
A possibility to circumvent the necessity to justify the size of the extra dimensions is to take the limit $n\rightarrow \infty$. In 
this setup, the extra dimensions just have to be infinitesimally larger than the
fundamental Planck scale $M_{\rm Pl}^{(4+n)} \sim M_{\rm EW}$ in order to generate an arbitrary high effective Planck scale (\ref{eq:MPLeff}). 
So, all input parameters could be of natural size $\sim \ord(M_{\rm EW})$ and also the SM could propagate nearly 
into the whole higher-dimensional space-time, in the spirit of {\it universal extra dimensions}. 

In conclusion, low energy strings {\it can} arise, if compactified extra dimensions are present, being much larger 
than their natural length of $1/M_{\rm Pl}$ or if there are many extra dimensions (or if the geometry is warped, 
see below). This ``low-energy'' (higher dimensional) setup will be UV completed by a 10 or 11 dimensional string theory at energies around $M_{\rm EW}$. 
Such a UV completion should explain the emergence of (large) extra dimensions and their topology, which is not addressed within 
the ADD model. With a fundamental Planck scale of $\ord(M_{\rm EW})$, gravity will already become strongly coupled at LHC energies and it 
could in principle be possible to create micro black holes at the LHC. However, these black holes are predicted to 
be very short-lived, evaporating quickly via Hawking radiation. Beyond that, potentially stable micro black holes would
have already shown up in cosmic rays \cite{Ellis:2008hg}.  

A generic problem of theories with such a low cutoff is that higher dimensional operators mediating for example proton decay or
FCNCs are not sufficiently suppressed, see Section~\ref{sec:SMProblems}. A possible 
solution to this issue is to split the fermions by
localizing them differently within the extra dimensions, suppressing these higher dimensional operators by small overlaps \cite{ArkaniHamed:1999dc}. 
This scenario also provides the possibility to address hierarchies within the fermion masses geometrically, by means of different 
overlaps with the Higgs sector. Note furthermore that, within the ADD model, a potential Grand Unification of couplings would already 
have to happen at a much lower scale than $M_{\rm GUT}$, see \eg \cite{Dienes:1998vh}. Moreover, the mechanism of generating tiny neutrino masses 
via the strongly suppressed $D=5$ operator introduced in Section~\ref{sec:SMProblems} does not work any longer (however, see \cite{ArkaniHamed:1998vp}).

\subsection{Constraints on Large/Flat Extra Dimensions}
\label{sec:ADDconstr}
\vspace{-0.1cm}
Before introducing models of warped extra dimensions, which can solve the gauge hierarchy problem without unnatural parameters and
with only one additional dimension, let us have a short look on some of the tightest constraints on large extra dimensions, coming
from astrophysics. 
Gravity propagating into compactified extra dimensions leads, from a 4D perspective (after integrating out the additional dimensions 
in the action) to the emergence of an infinite tower of massive KK excitations. These so-called KK gravitons, couple to matter suppressed by $1/M_{\rm Pl}$.  
This can be seen in analogy to putting a particle into a box (corresponding to the compactified dimensions), which can be described by 
an infinite set of eigenfunctions, corresponding to different, quantized, energy levels. From the 4D perspective, the quantized components 
of the momentum in the compactified extra dimension correspond to a mass. For details on this method of performing a {\it KK decomposition} 
(in the context of warped extra dimensions) see Section~ \ref{sec:KKgauge}. Note that in the following 
we only consider the spin-2 part $G_{\mu\nu},\, \mu,\nu=0,\dots,3$ of the higher dimensional metric tensor. 
The KK gravitons have masses $\sim 1/R$, explicitly 
\beq
m_k\sim k/R
\eeq 
for one extra dimension ($d=5$), where $k=0,\dots,\infty$ denotes the KK level and the {\it zero mode} ($k=0$) corresponds to the standard massless (4D) graviton. Thus, with (\ref{eq:RMPL}), we arrive at mass splittings of
\beq
\label{eq:Dm}
\Delta m \sim \frac 1 R \sim 10^{19-\frac{32}{n}} {\rm m^{-1}} \sim 10^{12- \frac{32}{n}}\, {\rm eV}\,,
\eeq
This means that from a high energy 
point of view one could produce a nearly continuous spectrum of real massive KK gravitons, given the number of extra 
dimensions is not too large.
 
Before reviewing the constraints from astrophysics, let us estimate the production cross section for arbitrary spin-2 
(KK) gravitons at high energy colliders. After noting that the possible number of graviton final 
states for an available energy of $\Delta E$ is ${\left( \frac{\Delta E}{\Delta m} \right)}^n \sim {\left(\Delta E\,R\right)}^n$, we arrive at
\beq
\sigma \sim \frac{1}{M_{\rm Pl}^2} {\left(\Delta E\,R \right)}^n \sim \frac{\Delta E^{\,n}}{{[M_{\rm Pl}^{(4+n)}]}^{n+2}}\,.
\eeq
For energies available to the graviton of $\Delta E\lesssim M_{\rm EW}$ and $M_{\rm Pl}^{(4+n)}\sim M_{\rm EW}$ one thus 
ends up with cross sections that could become as large as
\beq
\sigma \lesssim \frac{1}{{\rm TeV}^2}\,.
\eeq
Thus, graviton production can become important, if the available energies start to approach the weak scale region.
This agrees with the statements of strongly coupled gravity at the electroweak scale made before.
From the 4D point of view, the size of the cross section is due to the large multiplicity of possible graviton final states.
Due to the small interaction cross section of a {\it single} graviton, they would be seen as missing energy at collider experiments.
For example, at the LHC or the Tevatron one would have the characteristic signal of a single jet in association with missing energy,
whereas at electron positron colliders like LEP one would have for example a photon plus missing energy
\beq
\begin{split}
	p p \rightarrow j + E_{\rm miss}\,,\\
	e^+ e^- \rightarrow \gamma + E_{\rm miss}\,.
\end{split}
\eeq
For more details, see \eg \cite{Mirabelli:1998rt}.

For $n=2,3$\,, the most stringent constraints however still come from astrophysics and are hard to challenge in current collider 
experiments. (KK) Gravitons can carry away energy in supernova collapses, thus leading to a faster cooling.
From analyzing the data of the supernova SN1987A, one can set an upper limit on the rate of graviton emission. 
According to the standard theory of type-II supernovae, most of the gravitational binding energy released during the core collapse
of a supernova, is carried away via neutrinos. Since the measured neutrino flux agreed well with the theory value for the corresponding
energy, one can put tight constraints on the emission of other particles like axions or 
KK gravitons. The analysis of \cite{Cullen:1999hc} considered ``gravi-strahlung''
\begin{equation}
	N + N \rightarrow N + N + G
\end{equation}
within the ADD setup as the dominant graviton-emission process. Here, $G$ stands for any (KK) graviton. The nucleon-nucleon interactions
are modelled by one-pion exchange. Due to the low temperature of the core ($30-70$\,MeV), we expect strong limits only for $n\leq4$.
For higher $n$, the mass splittings (\ref{eq:Dm}) are approaching the available energy, which leads to a vanishing of the high multiplicity 
enhancement in graviton final states. The authors derive the following (conservative) lower bounds on the fundamental Planck scale
\begin{align}
	M_{\rm Pl}^{(6)} & \geq 50\ TeV\,,\quad n=2\,, \nonumber
	\\M_{\rm Pl}^{(7)} & \geq 4\ TeV\,,\quad n=3\,,
	\\M_{\rm Pl}^{(8)} & \geq 1\ TeV\,,\quad n=4\,. \nonumber
\end{align}
For n $\geq 4$ the LHC should be able to improve the limits significantly.
The biggest uncertainty in the supernova analysis is due to the unknown temperature of the core. The authors are conservative and assume
$T=30$ MeV. With more realistic values tighter bounds would be possible. The supernova data clearly exclude the possibility to address 
the large hierarchy between the Planck scale and the electroweak scale within the ADD setup for 2 extra dimensions.

Now we move on to a model which completely avoids the appearance of different scales, without the necessity 
of introducing infinitely many extra dimensions. The RS model is able to solve the gauge hierarchy problem with 
only one additional spatial dimension with a large curvature and all fundamental parameters of the 
order of the Planck scale $M_{\rm Pl}$. When exploring possibilities to extend the SM in Section~\ref{sec:SMProblems},
we were thinking about an extension of the Poincar\'{e} group, which was finally realized in SUSY models and 
extra dimensions. However, we were still assuming space-time to feature a Minkowsk metric in the vaccum.
The RS model will even modify this assumption in a way, that is not in conflict with observation.

 It is also able to address naturally the hierarchies in the fermion sector, by allowing the quarks and 
leptons to propagate into the warped extra dimension. Furthermore, it features an automatic suppression mechanism for FCNCs and allows 
for gauge coupling unification as well as for the construction of a Dark Matter candidate, see below. It offers a rich 
phenomenology in the reach of the LHC, especially within the Higgs sector. In the next chapter, we will discuss the theoretical setup of 
the model and its further virtues and properties in detail. 

The following chapters will be based on the publications \cite{Casagrande:2008hr,Casagrande:2010si,Bauer:2010iq,Goertz:2011nx} 
(besides the first, introductory pages), which occurred in the context of preparing this thesis. Beyond that, they will
also include updated analyses and several new aspects, in particular unpublished work on five-dimensional propagators
in warped space, as well as on the anomalous magnetic moment of the muon, see Chapter~\ref{sec:5Dprop} and Section~\ref{sec:AMM}.

\chapter[Warped Extra Dimensions: Theoretical Aspects\\Hierarchies - Interactions - Custodial Extension - Summing KK Towers]
{Warped Extra Dimensions: Theoretical Aspects\\{\normalsize Hierarchies - Interactions - Custodial Extension - Summing KK Towers}}
\chaptermark{Warped Extra Dimensions: Theoretical Aspects}
\label{sec:WED}
\vspace{-6mm} 

In this chapter, we will first introduce the basic concepts of the original RS model, including the derivation of the 
RS metric as a solution to Einstein's equations. We will then review, how the model is able to solve the gauge hierarchy problem. 
After that introductory part, we will study in detail gauge fields as well as fermions in the bulk, coupled to a brane localized 
Higgs sector within the minimal RS setup. In contrast to the perturbative approach usually 
used in the literature, where the couplings to the Higgs sector are introduced as a small correction after having obtained the bulk solutions,
we perform the KK decomposition directly in the mass basis. Electroweak symmetry breaking 
will be included exactly via boundary conditions (BCs), avoiding the truncation of the KK tower \cite{Goertz:2008vr}. In this way 
we are able to derive simple analytic results for the profiles and masses of the fields as well as for the interactions, which 
allow for a clear understanding of important effects of the model.
For example, we will see that the summation over the entire KK tower of intermediate gauge bosons leads to a characteristic 
dependence on the coordinates in the extra dimension, reflecting the full 5D structure, which is lost through truncation 
\cite{Bauer:2008xb}.
For fermions, the mixings between different generations are included in a completely
general way. The hierarchies observed in the fermion masses and CKM mixing angles are explained by wave-function overlaps with the 
Higgs sector, starting from anarchic fundamental Yukawa matrices. Corresponding analytical expressions are given by demonstrating and 
exploring an analogy to the Froggatt-Nielsen mechanism. The special localization pattern of fermions, resulting from the RS
setup, leads to the possibility to make generic predictions about the size of effects for the different quark generations. 
Although the main phenomenological part is devoted to Chapter~\ref{sec:Pheno}, we will already discuss some 
aspects of electroweak precision tests as a motivation to extend the gauge group of the minimal model, in
order to achieve a custodial protection. In this context
we will also explore the possibility of having a heavy Higgs boson $m_h\lesssim 1$  TeV in the RS setup.
The custodial extension to the gauge group $SU(2)_L\times SU(2)_R \times U(1)_X \times P_{LR}$ will then be discussed 
in detail. In particular, analytic formulae for the Peskin-Takeuchi parameters as well as $Zb\bar b$ couplings will be 
given that show explicitly which terms can be protected and which inevitably escape protection. Beyond that, simple 
and exact expressions for general interactions between gauge bosons and fermions as well as for the couplings to the 
Higgs boson will be presented and compared for both models. These allow for a complete discussion of tree-level flavor-changing effects in the RS setup. 
Moreover, we will show how to perform sums over infinite towers of gauge-boson profiles in the minimal as well as 
in the custodial model.
\vspace{0.1cm}

\section{Introduction and Solution to the Gauge Hierarchy Problem}
\label{sec:RSIntro}

\subsection{The General Setup}

\begin{figure}[!t] 
	\centering
		\includegraphics[width=13.5cm]{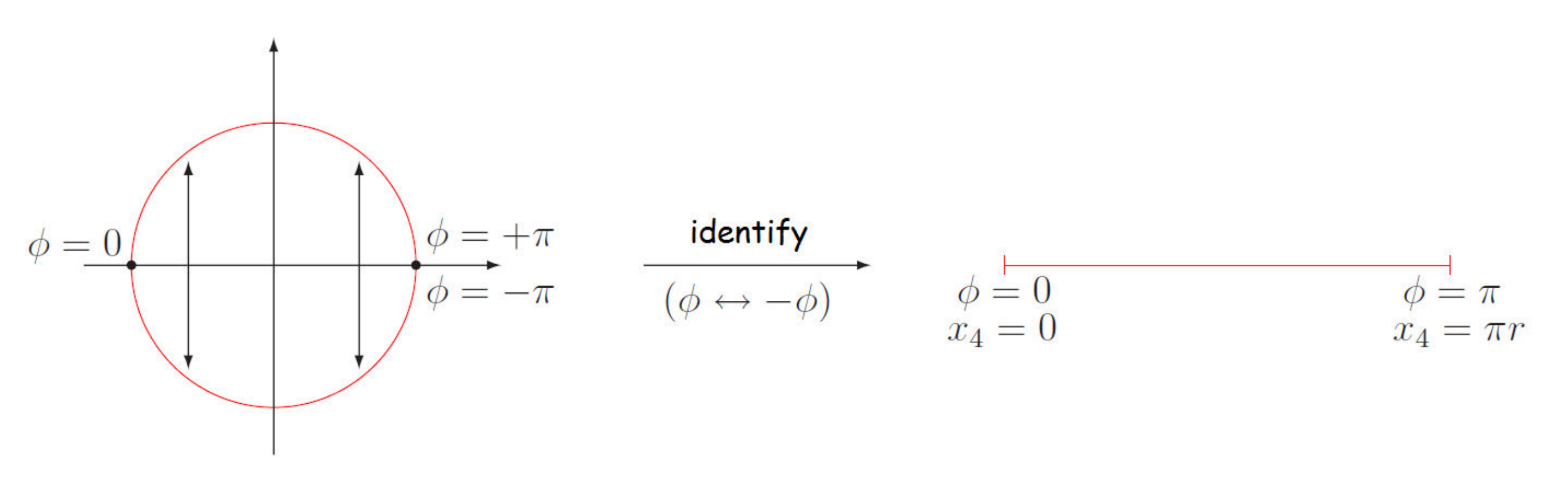}
		\vspace{-0.65cm}
	\caption{\label{fig:Orbi} The Orbifold.}
	\vspace{-0.6cm}
\end{figure}

The RS model \cite{Randall:1999ee} offers an elegant possibility to address the large hierarchy 
between the Planck scale and the electroweak scale by means of a non-trivial geometry in a five dimensional (5D)
anti-de Sitter (AdS$_5$) space. The non-factorizable RS metric
\beq
\label{eq:RSmetric}
ds^2=e^{-2 \sigma(\phi)}\eta_{\mu\nu}\,dx^\mu dx^\nu-r^2d\phi^2\,,
\eeq
is constructed such that length scales within the usual 4D space-time of constant $\phi$, labeled by coordinates $x^\mu$ ($\mu=0\dots3$), 
are rescaled via an exponential {\it warp factor}, depending on the position $\phi \in [-\pi,\pi]$ in the extra dimension.
In this thesis we will use the west coast convention for the Minkowski metric $\eta_{\mu\nu}=\,diag(1,-1,-1,-1)$. 
The exponential factor will turn out to be responsible for the solution to the gauge hierarchy problem and will 
be specified later. Importantly, the metric respects 4D Poincar\'{e} invariance. The fifth dimension is compactified on an orbifold 
$S_1/Z_2$, \ie, a circle with radius $r$ and with points identified, that 
are related to each other by a $Z_2$ (symmetry) transformation
\beq
(x^\mu,\phi)\leftrightarrow (x^\mu,-\phi)\,,
\eeq
see figure \ref{fig:Orbi}. The radius is assumed to be not much larger than $\ord(M_{\rm Pl})$, however, due to the warping, the model 
will have observable consequences down to the TeV scale. 
\setlength{\unitlength}{1mm}
We assume the action to be invariant under $Z_2$ transformations $Z$. However, the fields do not 
have to be identical at points which are identified by the orbifold structure and can differ by 
a symmetry transformation which leaves the action invariant. Since $Z^2=id$, $Z$ has eigenvalues 
$\pm 1$, corresponding to even and odd functions on the orbifold, respectively
\beq
\Phi(x^{\mu},-\phi)=Z \Phi(x^{\mu},\phi)= \pm \Phi(x^{\mu},\phi)\,.
\eeq
The $Z_2$ parity, gives rise to orbifold fixed points at $\phi=0,\pi$, providing support for 3-branes, \ie, 
sub-manifolds with three spatial and one time dimension, the so-called UV (or Planck) 
and IR (or TeV) branes. The reason for the names will become clear later. 
On these fixed points 4D field theories can be defined. The introduction of an orbifold also helps for
phenomenological reasons, in particular for obtaining the correct low energy spectrum of the SM 
within the RS model (with the SM in the bulk), see below.
Due to compactifying the extra dimension on an orbifold, one effectively ends up with an interval 
$\phi \in [0,\pi]$, bounded by the branes, which together with the $Z_2$ eigenvalues of the fields 
contain the whole information. The region in between the branes is called the ``bulk''. We impose 
Neumann or Dirichlet BCs for the fields at the branes, dictated by the low energy 
phenomenology, \ie, the requirement to arrive at the SM spectrum in the low energy limit. Furthermore, 
the orbifold structure leads to periodic boundary conditions
\beq
\Phi(x^{\mu},\phi)=\Phi(x^{\mu},\phi+2\pi)\,.
\eeq

In the original RS proposal only gravity was allowed to propagate into the extra dimensions, which is sufficient
to solve the gauge hierarchy problem. The SM was assumed to be confined to the IR brane. Thus we consider 
the (for the time being, classical) action of gravity propagating into the compactified fifth dimensions\footnote{Note the difference in the signature of the metric with 
respect to \cite{Randall:1999ee}, which results in a negative sign for the Ricci-scalar $R$.}
\beq
\label{eq:RSaction}
\begin{split} 
  S & = S_{\rm bulk} + S_{\rm UV} + S_{\rm IR}\,, \\
  S_{\rm bulk} & = \int{d^4x}\int_{-\pi}^{\pi}{d\phi \sqrt{G} \left\{-\Lambda - 2 M^3 R  \right\}}\,, \\
  S_{\rm UV}  & = \int{d^4x \sqrt{-g^{UV}}\left\{{\cal L}_{\rm UV} - V_{\rm UV} \right\}}\,, \\
  S_{\rm IR}  & = \int{d^4x \sqrt{-g^{IR}}\left\{{\cal L}_{\rm IR} - V_{\rm IR} \right\}}\,.
\end{split}
\eeq
where $G_{MN}(x^{\mu},\phi)$ is the 5D metric (not to be confused with the field strength tensor of the gluon), defined via (\ref{eq:RSmetric}) in the vacuum.
Note that, due to the curved space-time, the square root of the determinant 
of the metric $\sqrt{G}$ has to be considered, which is not longer the identity and is needed to obtain an invariant integration measure. 
Latin (greek) indices $M,N=0, \dots,3,\phi$\, ($\mu,\nu=0,\dots,3$) correspond to the 5D (4D) space-time. The 4D metric on 
the 3-branes is given by evaluating the 4D part of the bulk metric at the corresponding positions
\beq
	g_{\mu\nu}^{UV}(x^\mu) \equiv G_{\mu\nu}(x^\mu,\phi=0)\,,\quad g_{\mu\nu}^{IR}(x^\mu)
	\equiv G_{\mu\nu}(x^\mu,\phi=\pi)\,.
\eeq
Above, $M \sim \bar M_{\rm Pl}^{(5)}$ is the (reduced) fundamental scale of the theory and $\Lambda$ is the 5D cosmological 
constant. In the most simple setup of an IR-brane SM and no UV localized fields we have ${\cal L}_{\rm IR} = {\cal L}_{\rm SM}$ from (\ref{eq:LSM})
and ${\cal L}_{\rm UV}=0$. However, due to the different metric, the fields in ${\cal L}_{\rm SM}$ might have to be 
rescaled in order to obtain a canonical normalization and the numerical values of the input parameters might 
differ from their values in the ``standard'' SM (\ref{eq:LSM}) to match phenomenology. Non-vanishing 
``vacuum energies'' $V_{\rm UV,IR}$ on the branes, so-called brane tensions, are needed to allow for a non-trivial 
metric, responsible for the solution to the gauge hierarchy problem.

\subsection{Derivation of the Warp Factor}
We now specify the form of the exponential warp factor $e^{-\sigma(\phi)}$. The metric (\ref{eq:RSmetric}) 
has to be a solution to Einstein's equations. In the form given, it corresponds to a general ansatz, respecting 4D 
Poincar\'{e} invariance. Since such an ansatz is not forbidden by any known principle, it is the most natural
choice not to restrict it to the case of a vanishing warp factor. In the vacuum, Einstein's equations
for the setup (\ref {eq:RSaction}) read
\beq
\label{eq:Eins1}
\begin{split}
	\sqrt{G} \left( R_{MN}-\frac 1 2 G_{MN}R \right) & =  \frac{1}{4 M^3} [ \Lambda \sqrt{G}\, G_{MN} \\ 
	& + V_{\rm IR} \sqrt{-g^{IR}}\, g_{\mu\nu}^{IR}\, \delta^\mu_M \delta^\nu_N\, \delta(\phi-\pi) \\ 
	& + V_{\rm UV} \sqrt{-g^{UV}}\, g_{\mu\nu}^{UV}\, \delta^\mu_M \delta^\nu_N\, \delta(\phi) ]\,.
\end{split}
\eeq
Since the metric (\ref{eq:RSmetric}) is diagonal and just depends on $\phi$, it is not intricate 
to calculate the corresponding Christoffel symbols. We obtain the Ricci-tensor
\beq
\begin{split}
   R_{\mu\nu} & = \frac{e^{-2\sigma(\phi)}}{r^2} \left(-\sigma^{\prime\prime}(\phi) 
   + 4\left[\sigma^\prime(\phi) \right]^2\right) \eta_{\mu\nu}\,,\\
   R_{\phi\phi} & = 4\left(\sigma^{\prime\prime}(\phi)-\left[\sigma^\prime(\phi) \right]^2\right)\,,\\
   R_{\phi\nu} & =R_{\mu \phi}=0\,,
\end{split}
\eeq
as well as the Ricci-scalar
\beq
R = \frac{4}{r^2} \left(-2 \sigma^{\prime\prime}(\phi) + 5 \left[\sigma^\prime(\phi) \right]^2\right)\,.
\eeq
Inserting these results into (\ref{eq:Eins1}) we arrive at
\beq
\label{eq:Eins2}
\begin{split}
6 \left[\sigma^\prime(\phi)\right]^2 &= - \Lambda \frac{r^2}{4M^3}\,,\\
3 \sigma^{\prime\prime}(\phi) &= \frac{r}{4M^3}\left[V_{\rm IR}\, \delta(\phi-\pi) + V_{\rm UV}\, \delta(\phi)\right]\,.
\end{split}
\eeq
From the first equation in (\ref{eq:Eins2}) one deduces directly that a non-trivial solution $\sigma\neq {\rm const.}$
is just possible for a non-vanishing 5D cosmological constant $\Lambda\neq 0$. The general solution which is consistent 
with the orbifold symmetry reads\footnote{Note that we have not included an unobservable additive constant in the 
solution, which could be absorbed into a redefinition of the 4D coordinates $x^\mu$, moreover the ambiguity in 
the sign does not change the physical setup, since it just corresponds to interchanging the branes.}
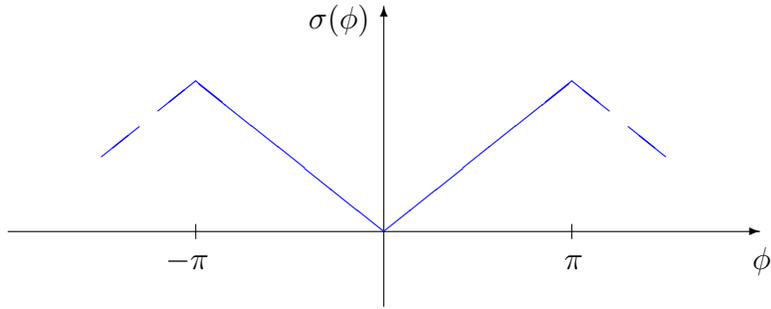
\begin{figure}[!t]
	\centering
	\begin{picture}(120,50)
		\put(10,15){\vector(1,0){100}}
		\put(60,5){\vector(0,1){40}}
		\put(35,14){\line(0,1){2}}
		\put(85,14){\line(0,1){2}}	
		\put(31,10){$-\pi$}
		\put(84,10){$\pi$}
		\put(109,10){$\phi$}
		\put(50,42){$\sigma(\phi)$}
		\put(60,15){\color{blue}\line(5,4){25}}
		\put(60,15){\color{blue}\line(-5,4){25}}
		\put(85,35){\color{blue}\line(5,-4){5}}
		\put(92.5,29){\color{blue}\line(5,-4){5}}
		\put(35,35){\color{blue}\line(-5,-4){5}}
		\put(27.5,29){\color{blue}\line(-5,-4){5}}
	\end{picture}
	\caption{\label{fig:per} Periodic solution for the exponent of the warp factor (schematically).}
\end{figure}
\beq
\sigma(\phi) =\sqrt{\frac{-\Lambda}{24 M^3}}\ r\left|\phi\right|\,.
\eeq
This function, featuring periodic BCs $\sigma(\phi)=\sigma(\phi+2\pi)$, has kinks
at $\phi=0,\pi$, see Figure \ref{fig:per}, leading to $\delta$-distributions in the
second derivative. These contributions have to be compensated by the brane vacuum energies, when inserting 
the solution into the second equation in (\ref{eq:Eins2}), which requires
\beq
	V_{\rm UV}=-V_{\rm IR}=24 M^3 k\,.
\eeq
Here, we have defined the RS-curvature
\beq
	k \equiv \sqrt{\frac{-\Lambda}{24M^3}}\,,\quad \Lambda=-24 M^3 k^2\,.
\eeq

\begin{figure}[!t]
	\centering
		\includegraphics[width=10.5cm]{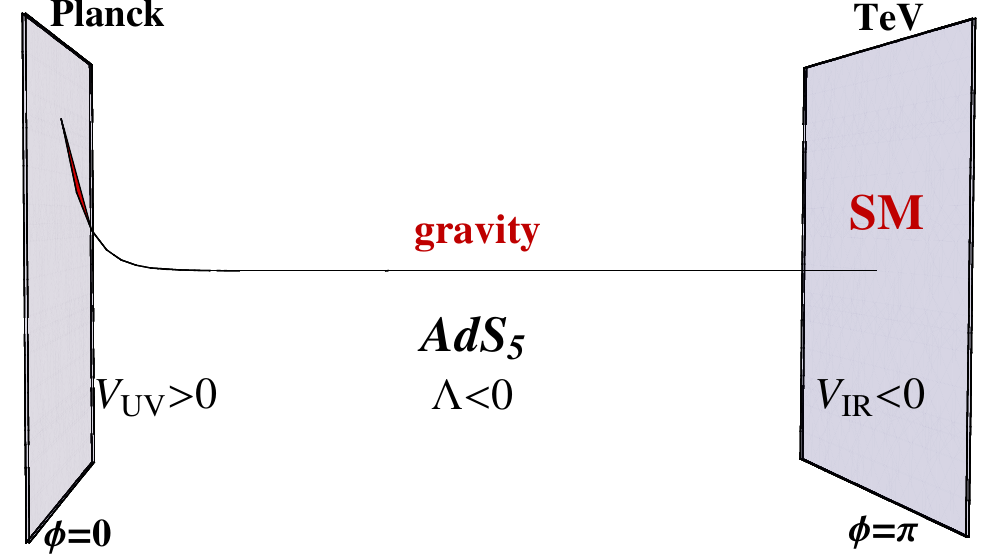}
		\put(-82,40){\footnotesize $ ds^2=e^{-2kr \left|\phi\right|}\eta_{\mu\nu}\,dx^\mu dx^\nu-r^2d\phi^2$}
	\caption{	\label{fig:AdSSM} The Randall-Sundrum setup corresponds to a five dimensional anti-de Sitter space with a non-factorizable metric, bounded by two four 
	dimensional sub-manifolds, the Planck brane and the TeV brane. The vacuum energies are adjusted such that the effective four dimensional 
	cosmological constant vanishes. See text for details.}
\vspace{-0.3cm}
\end{figure}

The brane tensions and the 5D cosmological constant are now linked in such a way, that the resulting 4D 
effective cosmological constant vanishes. This makes the 4D universe static and flat, consistent with 
the ansatz respecting 4D Poincar\'{e} invariance. In the RS model, the 
cosmological constant problem can be reformulated into the question why the brane tensions exactly 
cancel the 5D cosmological constant, a fact that has been put in by hand when choosing the ansatz. Remember 
that without the brane tensions, just the trivial solution $\sigma={\rm const.}$ would be consistent with a 
vanishing 4D cosmological constant, which would not allow for a solution to the gauge hierarchy problem. The explicit 
RS metric is now given by (\ref{eq:RSmetric}) with
\beq
   \sigma(\phi)=k r \left|\phi\right|.
\eeq
Due to the non-vanishing 5D cosmological constant, the extra dimension has a finite curvature within the vacuum. 
Therefore RS models are also called {\it warped extra dimensions}. The setup is shown in Figure \ref{fig:AdSSM}.
The factor $e^{-kr \left|\phi\right|}$, which describes the 
change of length scales when moving along the extra dimension is accordingly called {\it warp factor}. It is just the large
curvature, which will allow for a solution to the gauge hierarchy problem. For the consistency of the solution, 
the 5D cosmological constant has to be negative, resulting in an AdS$_5$ space. Furthermore, note that the 
curvature has to be smaller than the fundamental Planck scale of the setup. At this point, the theory 
possesses three fundamental scales ($M,k,r$) which, for naturalness reasons, should all be of the same order. Indeed, this 
will turn out to be possible.

\subsection{The Planck Scale}
\label{sec:MPLRS}
Let us now explore how the effective (4D) Planck scale $M_{\rm Pl}$, that we observe, arises from the fundamental scale 
$M$ of the RS model. Since we assume the extra dimension to be unresolvable directly by gravity experiments in the intermediate
future, we will use an 4D EFT description. We extend the metric (\ref{eq:RSmetric}) by 
{\it massless} fluctuations around the vacuum solution, which will provide the right degrees of freedom to 
describe gravity at low energies \cite{Randall:1999ee}
\beq
\label{eq:metricfluc}
	ds^2=e^{-2k\bar{b}(x)\left|\phi\right|} \left[ \eta_{\mu\nu}+\bar{h}_{\mu\nu}(x)\right]dx^\mu dx^\nu - \bar{b}^2(x) d\phi^2\,.
\eeq
Here, $\bar{h}_{\mu\nu}(x)$ corresponds to the massless zero mode in the KK decomposition of the (spin-2) graviton and
is the graviton of the 4D effective theory with a wave function $\psi(\phi)\sim e^{-2kr\left|\phi\right|}$ in the extra dimension, 
see also \cite{Davoudiasl:1999jd}. The effects of the massive KK excitations are suppressed 
at low energies by powers of the NP scale (see Appendix~\ref{app:EFT}) which will be above the TeV scale, see (\ref{eq:Mkk}) below. 
Locally, (\ref{eq:metricfluc}) is identical to the vacuum solution (\ref{eq:RSmetric}),
given that $\bar{h}_{\mu\nu}(x)$ and $\bar{b}$ are smooth functions. 
We did not include a term $A_\mu dx^\mu d\phi$, corresponding to vector fluctuations, since these off-diagonal fluctuations
will be heavy and thus not part of the EFT \cite{Randall:1999ee}. $G_{\phi\phi}(x) \equiv b^2(x)$ corresponds to the {\it radion}, 
its zero mode $\bar{b}(x)$ determines the radius of the extra dimension. A priori, it is massless, since the radius, 
which we have fixed so far just at $r$, does not possess a potential in the first place. However, from the phenomenology 
side, this would not be acceptable, since a massless radion would lead to a measurable violation of the equivalence principle. 
It is essential that $\bar{b}(x)$ can be stabilized at a VEV of $\left\langle \bar{b} \right\rangle =r$
and obtain a mass of at least $10^{-4}$\,eV \cite{Randall:1999ee}. This can be achieved due to the Goldberger-Wise mechanism 
\cite{Goldberger:1999uk} by introducing a new scalar field in the bulk. A radius which is in agreement with the solution to the gauge 
hierarchy problem can be reached without much fine tuning of parameters. 
In the following we assume that a mechanism a la Goldberger and Wise is at
work and replace $\bar{b}$ by its VEV $r$.

Inserting (\ref{eq:metricfluc}) into the action (\ref{eq:RSaction}) one can, on the one hand, explicitly verify that
the effective four dimensional cosmological constant vanishes $\Lambda_4=0$ and that the graviton zero mode has no effective
potential. On the other hand, one deduces from the curvature term, written in terms of the effective 4D metric
\beq
   \bar g_{\mu\nu}(x)\equiv \eta_{\mu\nu}+\bar h_{\mu\nu}(x)\,,
\eeq 
how the reduced effective (4D) Planck scale $\bar M_{\rm Pl}\approx2\times 10^{18}$\,GeV 
emerges from the (reduced) fundamental scale $M$ of the theory \cite{Randall:1999ee}. \footnote{Note that this relation can also be obtained 
by studying the interaction of the gravity zero mode with matter.} 
\beq
\label{eq:MPLRS}
	\bar M_{\rm Pl}^2=\frac{M^3}{k} \left[1-e^{-2kr\pi}\right]\,.
\eeq
For $L \equiv kr\pi \gg 1$, which will be necessary to solve the gauge hierarchy problem, the second term becomes unimportant
and we observe that $\bar M_{\rm Pl}$ only depends very weekly on the compactification radius $r$, in contrast to the ADD setup. 
Not wanting to produce large hierarchies between the input parameters of the theory, one sets
\beq
r^{-1}\lesssim k \sim M\,,
\eeq
which leads to an effective Planck mass of the order of the fundamental scale of the RS model
\beq
\bar M_{\rm Pl}\sim M\,.
\eeq
In consequence, this scale is expected to be $M\sim 10^{18}$\,GeV and not around the electroweak scale like in the ADD scenario. In the next section 
we discuss how the gauge hierarchy problem can be solved in such a completely natural setup. 
From (\ref{eq:MPLRS}) it follows that $\bar M_{\rm Pl}$ stays finite even in the limit $r \rightarrow \infty$! This setup has 
been worked out in \cite{Randall:1999vf}.

\subsection{Solution to the Gauge Hierarchy Problem}

So far the RS model just contains fundamental scales which reside at $M_{\rm Pl}$. Let us now see how, under this assumption,
a Higgs-boson mass around the scale $M_{\rm EW}$ can arise and can be radiatively stable. Therefore, consider a fundamental Higgs
scalar on the IR brane ($\phi=\pi$). The corresponding action reads
\beq
\label{eq:Ha1}
	S_{\rm IR} \supset \int d^4x\,r\int_{-\pi}^\pi\!d\phi\, {\cal L}_{\rm Higgs}\,, 
\eeq
where, in analogy to (\ref{eq:LHiggs})
\beq
\label{eq:LHiggs5D}
   {\cal L}_{\rm Higgs} = \delta(|\phi|-\pi) \frac{\sqrt G}{r^2}
   \left[G^{\mu\nu} (D_\mu\Phi)^\dagger\,(D_\nu\Phi) - V(\Phi) \right]\, ,
    \qquad
   V(\Phi) = - \mu_5^2\Phi^\dagger\Phi 
    + \frac{\lambda_5}{2} \left( \Phi^\dagger\Phi \right)^2\,.
\eeq
Here, $\sqrt G$ denotes the square root of the determinant of the complete 5D metric. Note that 
$G_{\mu\nu}(x^\mu,\phi=\pi)\equiv \bar g_{\mu\nu}^{IR}(x^\mu)$. 
Moreover, we have already switched to a notation which suggests that in principle the fields of the theory are 5D bulk fields, making 
the IR localization of the Higgs sector explicit via the $\delta$-distribution. The form of the Higgs doublet $\Phi(x)$
is identical with the one given in (\ref{eq:Hdoubl}) with however $v$ replaced by $v_5$. Here, and above, the subscripts $5$ denote that the 
input parameters are those of the fundamental (5D) theory with the metric (\ref{eq:metricfluc}) and before integrating out the extra 
dimension. Due to naturalness arguments, we assume $v_5\lesssim M_{\rm Pl}$.
First we rewrite the Higgs potential in terms of the VEV $v_5=\sqrt{2\mu_5^2/\lambda_5}$ as
\beq
   V(\Phi) = \frac{\lambda_5}{2} \left( \Phi^\dagger\Phi-\frac{v_5^2}{2} \right)^2\,,
\eeq
where we have dropped a, for our discussion irrelevant, constant term. Evaluating (\ref{eq:Ha1}) with the RS metric (\ref{eq:metricfluc})
and performing the integral over the extra dimension, we arrive at
\beq
   S_{\rm IR} \supset \int d^4x \sqrt{-\bar g}\, e^{-4kr\pi} \left\{e^{2kr\pi} \bar g^{\mu\nu} (D_\mu\Phi)^\dagger\,(D_\nu\Phi) - \frac{\lambda_5}{2} \left(\Phi^\dagger\Phi-\frac{v_5^2}{2} \right)^2\right\}\,.
\eeq
In order to obtain a canonically normalized 4D Higgs doublet, we have to perform a wave-function renormalization 
$\Phi \rightarrow e^{kr\pi} \Phi$, which leads to the 4D action
\beq
   S_{eff} \supset \int d^4x \sqrt{-\bar g} \left\{ \bar g^{\mu\nu} (D_\mu\Phi)^\dagger\,(D_\nu\Phi) - \frac{\lambda_5}{2} 
   \left(\Phi^\dagger\Phi - e^{-2kr\pi}\,\frac{v_5^2}{2} \right)^2\right\}\,.
\eeq
Looking at the last term, we see that a remarkable thing has happened. The fundamental Higgs VEV $v_5$ gets rescaled
by the (quadratic) warp factor $e^{-2kr\pi}$, such that the effective (4D) VEV, which sets the mass scale of the 4D theory, reads
\beq
\label{eq:vresc}
	v=e^{-kr\pi} v_5\,.
\eeq
Such a relation holds for {\it any} fundamental mass parameter $m_5$ present in the 5D theory. Measured on the IR brane, with
the metric of the effective 4D theory $\bar g^{\mu\nu}$, it will correspond to a physical mass of
$m=e^{-kr\pi} m_5$.
Note that also $\mu_5$ gets rescaled according to $\mu=e^{-kr\pi} \mu_5$, whereas dimensionless parameters will
not receive such a rescaling $\lambda_5=\lambda$.
In consequence, starting from a fundamental Higgs VEV of $v_5 \lesssim k \lesssim M_{\rm Pl}$, slightly
below its natural Planck-scale value, one arrives at an effective 4D VEV of $v \sim M_{\rm EW}$ by choosing
\beq
L\equiv kr\pi\approx 37\,.
\eeq
The same holds true for the Higgs mass, where ${m_h}_5 \lesssim k \lesssim M_{\rm Pl}$ results in a (4D) mass for the Higgs boson of
\beq
\label{eq:mhwarped}
	m_h = e^{-L}\, {m_h}_5 \sim M_{\rm EW}\,.
\eeq
Note that the fundamental Higgs mass should be slightly below the Planck scale, because otherwise the Higgs boson would be 
heavier than the KK excitations of the model, see below.
The question about the huge size of the hierarchy between $M_{\rm Pl}$ and $M_{\rm EW}$ which was reformulated into the question why
$R\gg M_{\rm Pl}^{-1}$ in the ADD setup has now been tamed to the question why the size of the extra dimension, in units of the inverse curvature,
is moderately larger than one ($L>$ a few). 

In summary, one can choose the radius of the extra 
dimension to be not far above the inverse RS curvature $r/k^{-1}\sim \ord(10)$, in order to generate the huge hierarchy between
the Planck scale and the electroweak scale, due to the exponential warp factor
\beq
   \epsilon \equiv e^{-L} \approx 10^{-16} \approx \frac{M_{\rm EW}}{M_{\rm Pl}}\,,
\eeq
which arises naturally when solving Einstein's equations. The gauge hierarchy problem is thus solved in the AdS$_5$ 
background by gravitational red shifting, see Figure \ref{fig:AdSH}. The corresponding brane separation can be stabilized by a Goldberger-Wise mechanism, 
as discussed before. Thus all the dimensionful scales of the RS model are within 1-2 orders of magnitude 
\beq
v_5 \sim M \sim 1/r \sim k \sim M_{\rm Pl}\,.
\eeq
Note that many observables in warped extra dimensions are enhanced by the ``volume factor'' $L$.
It is important that the Higgs mass (\ref{eq:mhwarped}) does not receive large corrections at the quantum level. 
For the following discussion, let us already generalize the RS setup, and allow also gauge bosons and fermions to propagate into the bulk.
We will see that this does not spoil the solution to the gauge hierarchy problem.

The absence of large corrections to the Higgs-boson mass is automatically secured in the RS setup, since, from a 5D point of view as well 
as from a 4D point of view, this mass is just below the cutoff of the RS model. Note that the RS model {\it has} to be defined with a cutoff (also in the
presence of gauge couplings only) since, being a QFT in more than four dimensions, it possesses gauge couplings with negative mass 
dimension, see (\ref{eq:g4def}). Thus, it is {\it not} expected to be the final theory of nature. It makes only sense as an EFT, 
see sections \ref{sec:SMProblems} and \ref{app:EFT}, and needs a UV completion like string theory. The 5D cutoff of the RS model is given by a scale 
$\Lambda_5\sim M_{\rm Pl}$. This translates into a cutoff for the 4D theory, depending on the position in the extra dimension, of 
\beq
\label{eq:poscut}
\Lambda_{\rm UV}(\phi)=e^{-kr\left|\phi\right|} \Lambda_5\,,
\eeq
where we have used the generic suppression of fundamental mass scales by the warp factor, seen above. 
In loop diagrams containing general bulk fields, the 4D momentum flow in a propagator has to be cut off at a scale, defined by the 
position-dependent cutoff $\Lambda_{\rm UV}(\phi)$ evaluated at the adjacent vertices. In addition, energy-momentum 
conservation has to be considered. A sensible prescription to choose one of the two possible cutoff scales for a propagator in the bulk thus is to use
$\phi$=max$(\phi_1,\phi_2)$, where $\phi_{1,2}$ are the coordinates in the extra dimension, belonging to the adjacent vertices. 
For further details, including the renormalization of the Green's functions, see \cite{Randall:2001gb}. For the case of an IR-brane 
localized Higgs sector, these considerations lead to a cutoff for the leading corrections to the Higgs-boson mass of 
\beq
\label{eq:RSIRcut}
\Lambda_{\rm UV}(\pi)=\epsilon\,\Lambda_5 = \ord({\rm some\ TeV})
\eeq
in the 4D theory, solving the large gauge hierarchy problem (however still leaving a little hierarchy problem for the case of a light Higgs boson, 
see below). This holds independently of the localization of the other fields. Possible corrections of heavier particles will be effectively 
cut off at this scale. $\Lambda_{\rm UV}(\pi)$ is the scale around which gravity becomes strongly coupled for an 4D observer on the IR 
brane and at which the RS effective 
field theory is expected to be replaced by something else like string/M-theory. This theory would then have to
answer the question how the warped space comes about. In this context, it is interesting to know that higher dimensional spaces with 
warp factors arise naturally in flux compactifications of string theory \cite{Klebanov:2000hb,Giddings:2001yu,
Kachru:2003aw,Brummer:2005sh}, see also \cite{Lukas:1998yy,Verlinde:1999fy}.
The effective mass scales on the UV and IR branes, resulting from fundamental scales of $\ord(M_{\rm Pl})$, also explain their names - the Planck 
brane and the TeV brane. Since the masses of the fermions and gauge bosons are protected by symmetries, as discussed before, a large cutoff
will not lead to large masses in these sectors.
As we will explain below, the solution to the gauge hierarchy problem in RS, due to a higher-dimensional space-time with a non-trivial metric, is
related to the technicolor/strong-coupling ideas of Section~\ref{sec:solHP} via the AdS/CFT correspondence, with the Higgs-boson 
being composite.

\subsection{The Standard Model Propagating into the Bulk}
\label{sec:SMinB}

\begin{figure}[!t]
	\centering
	\vspace{1mm}
		\includegraphics[width=10.5cm]{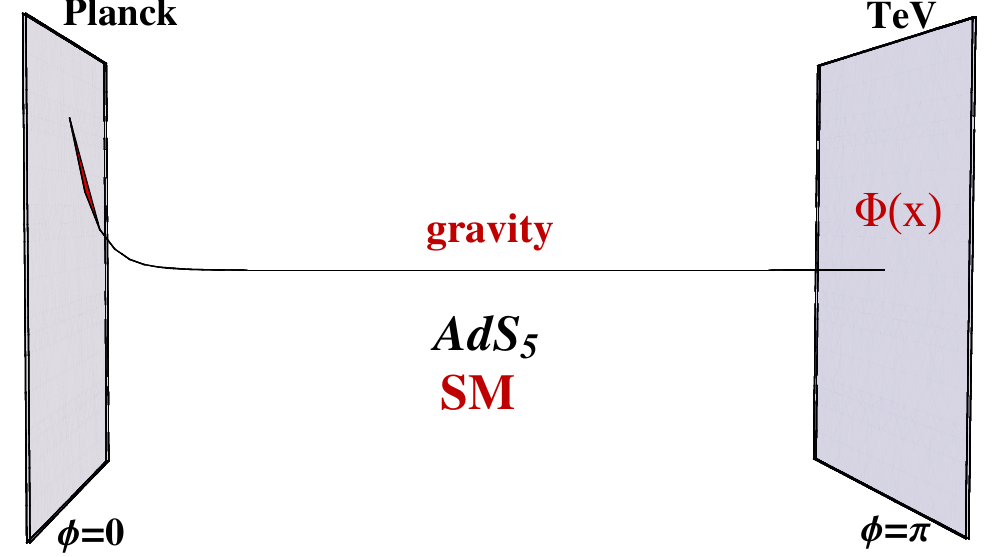}
		\put(-100.5,21){\small $\bm{m_5}$}
		\put(-18,21){\small $\bm{e^{-L} m_5}$}
	\caption{\label{fig:AdSH}Setup of the minimal Randall-Sundrum model. The gauge hierarchy problem is solved since mass scales on the TeV brane, where
	the Higgs sector is localized, are suppressed by the warp factor. The warping is described by the profile of the massless graviton,
	depicted by the exponential line.}
\end{figure}

While in the original RS setup the whole SM was assumed to be localized on the TeV brane, we have seen that the solution
to the gauge hierarchy problem will not be spoiled if gauge bosons \cite{Davoudiasl:1999tf,Pomarol:1999ad,Chang:1999nh,Gherghetta:2000qt} 
and matter fields \cite{Gherghetta:2000qt,Grossman:1999ra} are allowed to propagate into the extra dimension, see Figure \ref{fig:AdSH}. 
In particular, this makes constraints from higher dimensional operators, like those mediating proton decay or FCNCs, less problematic.
As the RS model is an EFT, there is no reason not to consider such operators. While the 
cutoff for an IR-brane localized Higgs sector $\Lambda_{\rm UV}(\pi)$ will be in the sought TeV range, operators of fields that propagate 
into the extra dimension will have a larger suppression factor $\Lambda_{\rm UV}$, that depends on their localization. The more UV localized the 
fields are, the higher is the generic cutoff. This naturally results in different suppressions for different operators, \eg for those leading to proton 
decay and those mediating meson mixing, a pattern which could well be realized in nature, see Section~\ref{sec:SMProblems}. 
Bulk fermions further offer the interesting possibility to address the observed hierarchies within the flavor sector via geometrical
sequestering \cite{ArkaniHamed:1999dc}. Starting from anarchical fundamental Yukawa couplings, as defined in Section~\ref{sec:SMhi},
the large mass hierarchies of the SM fermions are generated naturally by localizing them differently in the fifth dimension, without the 
need to build in hierarchies in the input parameters by hand \cite{Grossman:1999ra,Gherghetta:2000qt,Huber:2000ie,Huber:2003tu}. These are 
generated due to exponentially enhanced differences in the fermion-Higgs couplings, due to the warp factor. In consequence, small mixing 
angles in the CKM matrix are generated automatically \cite{Huber:2003tu}, since the scenario is in complete analogy
to the Froggatt Nielsen mechanism, see Section~\ref{sec:quark}. As a by-product, this way of generating fermion 
hierarchies implies an explicit suppression of dangerous FCNCs within the RS model for processes involving {\it light} quarks. The UV 
localization of these quarks results in small couplings to KK modes which mediate FCNCs already at the tree level \cite{Gherghetta:2000qt}.
This suppression is called RS-GIM mechanism \cite{Agashe:2004ay,Agashe:2004cp}, in analogy to the GIM mechanism within the SM, which both share 
some features, see below. Importantly, bulk fermions mitigate significantly the RS corrections to the Peskin-Takeuchi $S$ parameter, that 
arise through de-localized $W^\pm$ and $Z$ bosons \cite{Gherghetta:2000qt,Huber:2000fh,Davoudiasl:2000wi,Huber:2001gw}.
The remaining sizable corrections to the $T$ parameter will be studied in more detail below in Section~\ref{sec:pheno1}.
This RS setup, with the SM in the bulk, is the model that we will study in this thesis. We will call this setup, featuring the SM gauge group, 
the {\it minimal RS model}, in discrimination from the extension to the {\it custodial RS model}, which will be introduced later.
The Higgs boson may only leave the region of the IR brane if one wants to give up the idea of solving the gauge 
hierarchy problem in RS or if one includes another mechanism to stabilize the weak scale, see Section \ref{sec:solHP}.
 
With the SM (besides the Higgs sector) in the bulk, also the gauge bosons and fermions will develop 
KK excitations. These will turn out to be localized close to the TeV brane, see Figure \ref{fig:AdSBos} below. 
The mass scale for the lightest of these excitations (as well as for KK gravitons), will also be set by the warp factor and is called the 
``KK scale'' 
\beq
\label{eq:Mkk}
   \Mkk\equiv k\epsilon = k\,e^{-L} 
   = \ord(\mbox{TeV}) \,,
\eeq
see sections \ref{sec:mingaugeprofiles} and \ref{sec:minfermionprofiles}. The zero modes of the KK towers will 
correspond to the ``SM-fields'' we observe at low energies.\footnote{Although they can receive non-vanishing masses via couplings to the 
Higgs sector, we will still call the 
lightest modes with masses $m_n \ll \Mkk$ zero modes.} The KK excitations however,
will correspond to BSM physics and their mass scale $\Mkk$ will set the NP scale.\footnote{Note that, after going to the 
mass basis, also the couplings of the zero mode sector will be changed with respect to the SM, due to mixings with the KK excitations.} 
For instance, the masses of the first KK photon and gluon will turn out to be approximately $2.45\,\Mkk$. 
In contrast to the ADD model, the masses of the excitations are thus not given by the inverse compactification radius.
This makes these modes in principle observable at colliders like the LHC, a big advantage compared to usual string theory,
where the compactification radius as well as the mass of the string excitations are {\it both} expected to be around the Planck scale. 
The explicit form of the profiles and masses of the SM fields and their KK excitations in the RS setup will be given in the next sections.
Note that the possibility to generate (KK) particle masses due to a compactified extra dimension without a Higgs sector also lead to the construction 
of Higgsless models of EWSB \cite{Csaki:2003dt,Csaki:2003zu,Nomura:2003du,Barbieri:2003pr,Csaki:2003sh}. 
The breakdown of unitarity in longitudinal $W^\pm$ scattering in these models is raised by the exchange
of KK towers of gauge bosons.
Moreover, the presence of (4D) scalar degrees of freedom in higher dimensional gauge fields offers the possibility
to achieve EWSB by supplying these components with a VEV \cite{Manton:1979kb,Csaki:2002ur,Scrucca:2003ra,Scrucca:2003ut}, 
see also \cite{Agashe:2004rs}.
The gauge symmetry then protects the mass of the Higgs component of the gauge field.
This corresponds to an alternative option to extend the Poincar\'{e} group trivially by adding dimensions, in order to protect the Higgs sector
The scalars and the (4D) gauge fields are then linked by the 5D Poincar\'{e} group ({\it c.f.} SUSY).

After having introduced the basic properties of KK modes within the RS setup, let us again study gravity. The weakness of gravity 
in the RS setup is reflected by the exponential localization of the graviton zero mode close to the UV brane (\ref{eq:metricfluc}), see 
Figure \ref{fig:AdSH}, which leads to dimensionful couplings, suppressed by the Planck scale $M_{\rm Pl}\gtrsim M$. 
The KK excitations of the graviton will however be localized near the IR brane and thus their couplings will be enhanced 
by the warp factor, which, after solving the gauge hierarchy problem, will lead to a coupling, 
suppressed by the weak scale only \cite{Davoudiasl:1999jd}. 
As a consequence, from the 4D point of view of an observer sitting on the IR brane, gravity 
will become strong already at the weak scale. 
This reason for gravity becoming strong at the TeV scale - unsuppressed couplings of single IR localized KK gravitons with TeV masses 
 - should be contrasted with the huge multiplicity of Planck-scale suppressed KK-graviton couplings in the ADD model. The 
fact that KK gravitons in the RS setup have masses above $\Mkk$ removes nearly completely the constraints from cosmology and astrophysics, 
see Section~\ref{sec:ADDconstr}. Furthermore their stronger coupling makes them, in principle, directly observable as decaying spin-2 
resonances at colliders like the LHC. 

Let us finally have another look at the corrections to the Higgs-boson mass in the 4D theory. From naive dimensional 
analysis we conclude that the residual corrections in the RS model scale like 
\beq
   \delta m_h^2\sim \frac{\Lambda_{\rm UV}^2(\pi)}{16\pi^2}\,\frac{\Lambda_{\rm UV}^2(\pi)}{\Mkk^2} \,.
\eeq
These corrections grow like the fourth power of the UV cutoff, not like the second one as in 4D. Thus, the ``little hierarchy problem'' 
in RS is in general more severe than in standard 4D extensions of the SM. These considerations favor a heavy Higgs boson, not too 
far below the cutoff. Indeed, it will turn out that electroweak precision tests within the minimal RS model, studied at the tree level, also 
prefer a heavy Higgs boson $m_h\gtrsim 500$\, GeV, see Section~\ref{sec:pheno1}. However, since we have not seen strong hints for KK excitations 
below the TeV scale so far, the resulting UV cutoff of at least some TeV still results in a fine tuning at the per cent level for a heavy 
Higgs-boson and even worse for a light Higgs of $m_h \sim \ord(100)$ GeV. 

The discussion above can also be summarized from another point of view. By virtue of the AdS/CFT correspondence \cite{Maldacena:1997re,Gubser:1998bc,Witten:1998qj}, a 5D gravitational theory in anti~de-Sitter space is dual to a 
strongly coupled 4D conformal field theory (CFT). For the case of the RS setup, with the conformal symmetry being broken on the IR brane, 
implications of this correspondence have been studied \eg in \cite{Verlinde:1999fy,ArkaniHamed:2000ds,Rattazzi:2000hs,Contino:2004vy}. A 
recent review can be found in \cite{Gherghetta:2010cj}.\footnote{Note that by applying the AdS/CFT duality (approximative) to low energy QCD, one can still use 
warped 5D models as mathematical tools to calculate corresponding observables in a weakly coupled theory, even if nature would turn out 
to be four dimensional. For a review on these aspects, see \cite{Erlich:2008en}.} Holography implies that 5D fields living near the IR brane correspond to 
composite objects in the CFT, whereas fields living near the UV brane correspond to elementary particles. In the RS setup, where the 
Higgs sector is localized on (or close to) the IR brane, the Higgs boson thus can be thought of as a composite object 
of the strongly coupled sector \cite{Kaplan:1983fs,Kaplan:1983sm}, whose compositeness scale and mass is naturally of the order of the scale of the IR brane 
$\Lambda_{\rm UV}(\pi)\gtrsim \Mkk$.
This may be different in models of gauge-Higgs unification, which we 
will not study in this thesis. In this sense, a light Higgs boson with $m_h\ll\Mkk$ would be unnatural in the RS models studied here. 
Along the same lines, the SM-fields propagating into the bulk correspond to elementary particles up to much higher scales. In this language,
a bulk SM has the advantage that many observables can be calculated within the weakly coupled RS setup with less sensitivity on the details of the 
compositeness (\ie, UV completion of the RS EFT).

\section{The Minimal Randall-Sundrum Model: The SM in a Slice of AdS$_5$}
\sectionmark{The Minimal RS Model: The SM in a Slice of AdS$_5$}
After this general overview we will now study in detail the theory of the SM field content propagating into a warped extra dimension, in the presence of a 
brane-localized Higgs sector. First, we will discuss gauge fields in the bulk. After that, we will move on to bulk fermions and study 
the generation of the hierarchies in the fermion sector as discovered in nature. 
By constructing solutions to the bulk EOMs, subject to boundary conditions given by the couplings to the scalar sector, we obtain exact 
results for the masses and profiles of the SM fields and their KK excitations. 
This will pay off in the following discussion of interactions between fermions and gauge 
bosons as well as in studying interactions with the Higgs boson, where in all cases we will give simple analytic formulae for the couplings.
Proceeding in this way we can then clearly distinguish between leading and subleading terms. At the end of this section, we will have a first 
look at the phenomenology of the RS model. Rather large corrections to electroweak precision observables will serve as a motivation to extend 
the gauge group of the minimal RS model. The following discussion is based on \cite{Casagrande:2008hr}.

\subsection{The Gauge Sector}
In this section we study the SM gauge group in the warped RS background, incorporating the gauge fixing in 
a covariant $R_\xi$-gauge. We will focus on the 
electroweak sector as the generalization to the $SU(3)_c$ gauge group is straightforward. 
The starting point for the following 
discussion is the electroweak gauge sector of the SM, as given in Section~\ref{sec:SM1}.

\subsubsection{Action of the 5D Theory}

When generalizing the SM gauge sector to a five dimensional theory in a slice of AdS$_5$, we still want to
keep the low energy sector of this theory similar to the SM version. Therefore we start from the SM action
corresponding to the symmetry group $SU(2)_L\times U(1)_Y$ (\ref{eq:LWBGSM}) (for the time being, without the gluon part), 
account for the additional dimension in the space-time integral and the non-trivial metric, and simply replace the 
gauge fields by five dimensional versions, \ie, five-component vector fields
\beq
\begin{split}
B_\mu \rightarrow B_M\,,\\
W_\mu^a\rightarrow W_M^a\,.
\end{split}
\eeq
As we later want to study a 4D theory, we will decompose the 5D vectors above
into representations of the {\it 4D Lorentz group}, the vector $B_\mu$ and the scalar $B_\phi$, which do not mix under
4D Lorentz transformations, and similar for the other gauge fields. In order to arrive at a low energy spectrum that 
contains just SM-like fields and will be to leading order identical to the SM, we have to take care that the 
$\phi$-components of the 5D gauge fields do not enter the low energy theory. We thus choose the scalar components 
to be odd under the $Z_2$ orbifold symmetry, and supply them with Dirichlet BCs, so that they 
will not possess light (or massless) modes, see below.\footnote{In the models 
of gauge-Higgs unification mentioned before, it is just such scalar components of gauge fields that are taken as candidates 
for a Higgs field.} The vector components $W_\mu^a$ and $B_\mu$ however have to be even under $Z_2$
and obey Neuman BCs.
We arrive at
\beq
\label{eq:Sgauge}
   S_{\rm gauge} = \int d^4x\,r\int_{-\pi}^\pi\!d\phi\, 
   \Big( {\cal L}_{\rm B,W} + {\cal L}_{\rm Higgs}
   + {\cal L}_{\rm GF} + {\cal L}_{\rm FP} \Big) \,,
\eeq
where the Lagrangian of the 5D gauge theory reads
\beq\label{Lgauge}
   {\cal L}_{\rm B,W} = \frac{\sqrt{G}}{r}\,G^{KM} G^{LN}
   \left(- \frac14\,B_{KL} B_{MN} - \frac14\,W_{KL}^a W_{MN}^a 
   \right)\,.
\eeq
The field-strength tensors are defined in analogy to the usual 4D definitions (\ref{eq:SMFS}) by replacing
four-component indices with five-component ones. The action (\ref{eq:Sgauge}) is invariant with respect to 
5D gauge transformations, which are also in complete analogy to the 4D gauge transformations introduced in (\ref{eq:gau2}),
see also \cite{Randall:2001gb}. Note that, due to the higher dimensional space-time, the gauge fields now have mass 
dimension $D=3/2$.

In order to achieve EWSB, we couple these fields to a Higgs sector, which is localized on the IR brane. 
We know that this sector has to be localized close to the IR brane in order to solve the gauge hierarchy 
problem, so we assume it to be defined directly on the brane for simplicity. This enables us to obtain exact 
solutions by accounting for the Higgs couplings via BCs.
The corresponding Lagrangian ${\cal L}_{\rm Higgs}$ has already been given in (\ref{eq:LHiggs5D}), before evaluating 
the RS metric and performing the wave-function renormalization of the Higgs Field. After these operations and
replacing fundamental parameters with 4D parameters according to (\ref{eq:vresc}) and below, we obtain
\beq
\label{eq:LHiggs54}
   {\cal L}_{\rm Higgs} = \frac{\delta(|\phi|-\pi)}{r}
   \left[ (D_\mu\Phi)^\dagger\,(D^\mu\Phi) - V(\Phi) \right]\,,
    \qquad
   V(\Phi) = - \mu^2\Phi^\dagger\Phi 
    + \frac{\lambda}{2} \left( \Phi^\dagger\Phi \right)^2\,.
\eeq
After EWSB, the Higgs doublet can be decomposed in terms of the 
four real scalar fields $\varphi^i$, $i=1,2,3$ and $h$, exactly like in (\ref{eq:Hdoubl}).
Note that, although the parameters in the RS model have the same names as the corresponding SM ones, the exact values
might differ. For example the relation between the Higgs VEV and the $W^\pm$-boson or $Z$-boson (zero mode) mass 
will get corrections, suppressed by the NP scale $\Mkk$, and thus the extracted 
Higgs VEV will differ from the SM value. While for $\Mkk\gtrsim 3$\, TeV, the correction is about $\lesssim 1\%$, 
for lower scales it can become important and exceeds 5\% for $\Mkk=1$\,TeV. Anyway, as the effect appears already at 
$\ord(v^2/\Mkk^2)$ one should include it \cite{Bouchart:2009vq}, see below.

In order to diagonalize the 5D mass terms resulting from (\ref{eq:LHiggs54}), we perform the usual redefinitions of the gauge fields
\begin{eqnarray}
   W_M^\pm &=& \frac{1}{\sqrt2} \left( W_M^1\mp i W_M^2 \right)\,, 
    \nonumber\\
   Z_M &=& \frac{1}{\sqrt{g_5^2+g_5^{\prime 2}}} 
    \left( g_5 W_M^3 - g_5' B_M \right)\,, \\
   A_M &=& \frac{1}{\sqrt{g_5^2+g_5^{\prime 2}}} 
    \left( g_5' W_M^3 + g_5 B_M \right)\,, \nonumber
\end{eqnarray}
where $g_5$ and $g_5'$ are the 5D gauge couplings of $SU(2)_L$ and $U(1)_Y$, respectively. These have mass 
dimension $D[g_5^{(\prime)}]=-1/2$ since interaction operators between two fermion fields and a gauge-boson
field have mass dimension $D=11/2$ in five dimensional space-time. Introducing 4D gauge couplings (see Section~\ref{sec:gaugecouplings})
\beq\label{eq:g4def}
   g = \frac{g_5}{\sqrt{2\pi r}} \,,
\eeq
and similar for the other gauge couplings \cite{Davoudiasl:1999tf}, we see that these field redefinitions
are just like in the SM (\ref{eq:ZASM}). Consequently, the definition of the weak mixing angle between the 5D fields,
\beq
\begin{split}
   \label{eq:weinbRS}
  \sin\theta_w=\frac{g_5^\prime}{\sqrt{g_5^2+g_5^{\prime 2}}}=\frac{g^\prime}{\sqrt{g^2+g^{\prime 2}}}\,\\
  \cos\theta_w=\frac{g_5}{\sqrt{g_5^2+g_5^{\prime 2}}}=\frac{g}{\sqrt{g^2+g^{\prime 2}}}\,,
\end{split}
\eeq
agrees with the one of the SM (\ref{eq:weakmixingSM}).
In analogy to (\ref{eq:covder}), the 5D covariant derivative in this mass basis, including QCD 
for completeness, reads
\beq
\label{eq:covderRS}
   D_M=\partial_M -i \frac{g_5}{\cos\theta_w}\left(T^3-\sin^2\theta_w Q\right) Z_M -i e_5 Q A_M 
   - i \frac{g_5}{\sqrt{2}} \left(T^+ W^+_M + T^- W^-_M\right) -i g_{s,5}\, t^a G^a_M\,,
\eeq
with $Q$ defined as in (\ref{eq:QT}).
Evaluating the four-vector part of this expression, acting on the Higgs doublets in (\ref{eq:LHiggs54}), we see that the 
bulk gauge fields $W^\pm$ and $Z$ get ``masses''
\beq
   M_W = \frac{v g_5}{2} \,, \qquad
   M_Z = \frac{v\sqrt{g_5^2+g_5^{\prime 2}}}{2} \,,
\eeq
while the photon remains massless ($M_A=0$).
This is again in complete analogy to the mechanism of EWSB in the SM with the symmetry breaking sector now localized 
at the boundary of an extra dimension. However, note that the mass parameters $M_W$ and $M_Z$ now have mass 
dimension $D=1/2$.

The kinetic term for the Higgs field contains mixed terms of the gauge bosons and the scalar fields 
$\varphi^\pm$ and $\varphi^3$. They can be read off from
\beq
   D_\mu\Phi = \frac{1}{\sqrt2} \left( \begin{array}{c}
    -i\sqrt2 \left( \partial_\mu\varphi^+ + M_W\,W_\mu^+ \right) \\
    \partial_\mu h 
     + i \left( \partial_\mu\varphi^3 + M_Z\,Z_\mu \right)
   \end{array} \right) 
   + \mbox{terms bi-linear in fields \,.}
\eeq
Moreover, the kinetic terms for the gauge fields in (\ref{Lgauge}) comprise mixed terms consisting of 
the gauge boson vector fields and their scalar components $W_\phi^\pm$, $Z_\phi$, and $A_\phi$. We remove all of 
these mixed terms with a suitable choice of the gauge-fixing Lagrangian ({\it cf.} (\ref{eq:SMGF}))
\beq\label{Sgf}
\begin{split}
   {\cal L}_{\rm GF}
   &= - \frac{1}{2\xi} \left( \partial^\mu A_\mu - \xi \left[ 
    \frac{\partial_\phi\,e^{-2\sigma(\phi)} A_\phi}{r^2} \right] 
    \right)^2 \\
   &\quad\mbox{}- \frac{1}{2\xi} 
    \left( \partial^\mu Z_\mu - \xi \left[ 
    \frac{\delta(|\phi|-\pi)}{r}\,M_Z\,\varphi^3 
    + \frac{\partial_\phi\,e^{-2\sigma(\phi)} Z_\phi}{r^2} \right] 
    \right)^2 \\
   &\quad\mbox{}- \frac{1}{\xi} 
    \left( \partial^\mu W_\mu^+ - \xi \left[ 
    \frac{\delta(|\phi|-\pi)}{r}\,M_W\,\varphi^+
    + \frac{\partial_\phi\,e^{-2\sigma(\phi)} W_\phi^+}{r^2} \right] 
    \right) \\
   &\qquad\times
    \left( \partial^\mu W_\mu^- - \xi \left[ 
    \frac{\delta(|\phi|-\pi)}{r}\,M_W\,\varphi^-
    + \frac{\partial_\phi\,e^{-2\sigma(\phi)} W_\phi^-}{r^2} \right] 
    \right).
\end{split}
\eeq
Note that each term above could be written with a different gauge-fixing parameter 
$\xi_i$, however, without loss of generality, we will choose them to be equal. Moreover, there is no problem 
in squaring the $\delta$-distributions in the expression above. As we 
will see below, the derivatives of the scalar components of the gauge fields $W_\phi^\pm$ and $Z_\phi$ also 
contain $\delta$-distributions, which exactly cancel the $\delta$-distributions from the Higgs sector. 
In consequence, we do not have to introduce separate gauge-fixing Lagrangians in the bulk and on the IR 
brane as done in \cite{Csaki:2005vy}.

After integration by parts, the quadratic terms in the action finally become
\begin{eqnarray}\label{Sgauge2new}
   &&S_{\rm gauge,2} 
    = \int d^4x\,r\int_{-\pi}^\pi\!d\phi\,\bigg\{
    - \frac14\,F_{\mu\nu} F^{\mu\nu} 
    - \frac{1}{2\xi} \left( \partial^\mu A_\mu \right)^2 
    \nonumber\\
   &&\quad\mbox{}+ \frac{e^{-2\sigma(\phi)}}{2r^2} \left[ 
    \partial_\mu A_\phi\partial^\mu A_\phi
    + \partial_\phi A_\mu\partial_\phi A^\mu \right]
    - \frac{\xi}{2} \left[
    \frac{\partial_\phi\,e^{-2\sigma(\phi)} A_\phi}{r^2} \right]^2 
    \nonumber\\
   &&\quad\mbox{}- \frac14\,Z_{\mu\nu} Z^{\mu\nu} 
    - \frac{1}{2\xi} \left( \partial^\mu Z_\mu \right)^2
    + \frac{e^{-2\sigma(\phi)}}{2r^2} \left[ 
    \partial_\mu Z_\phi\partial^\mu Z_\phi
    + \partial_\phi Z_\mu\partial_\phi Z^\mu \right] 
    \nonumber\\
   &&\quad\mbox{}- \frac12\,W_{\mu\nu}^+ W^{-\mu\nu} 
    - \frac{1}{\xi}\,\partial^\mu W_\mu^+\,\partial^\mu W_\mu^-
    + \frac{e^{-2\sigma(\phi)}}{r^2} \left[ 
    \partial_\mu W_\phi^+\partial^\mu W_\phi^-
    + \partial_\phi W_\mu^+\partial_\phi W^{-\mu} \right] 
    \\
   &&\quad\mbox{}+ \frac{\delta(|\phi|-\pi)}{r} \left[
    \frac12\partial_\mu h\partial^\mu h - \lambda v^2 h^2
    + \partial_\mu\varphi^+\partial^\mu\varphi^- 
    + \frac12\partial_\mu\varphi^3\partial^\mu\varphi^3 
    + \frac{M_Z^2}{2}\,Z_\mu Z^\mu + M_W^2\,W_\mu^+ W^{-\mu}
    \right] 
    \nonumber\\
   &&\quad\mbox{}- \frac{\xi}{2} \left[ 
    \frac{\delta(|\phi|-\pi)}{r}\,M_Z\,\varphi^3 
    + \frac{\partial_\phi\,e^{-2\sigma(\phi)} Z_\phi}{r^2} 
    \right]^2 
    \nonumber\\
   &&\quad\mbox{}- \xi \left[ 
    \frac{\delta(|\phi|-\pi)}{r}\,M_W\,\varphi^+
    + \frac{\partial_\phi\,e^{-2\sigma(\phi)} W_\phi^+}{r^2} \right] 
    \left[ \frac{\delta(|\phi|-\pi)}{r}\,M_W\,\varphi^-
    + \frac{\partial_\phi\,e^{-2\sigma(\phi)} W_\phi^-}{r^2} \right]
    + {\cal L}_{\rm FP} \bigg\} \,. \nonumber
\end{eqnarray}
The Faddeev-Popov ghost Lagrangian ${\cal L}_{\rm FP}$ will not be needed for the following derivations. Its 
form will however be discussed after the KK decomposition.

\begin{figure}[!t]
\begin{center}
\includegraphics[height=1.6in]{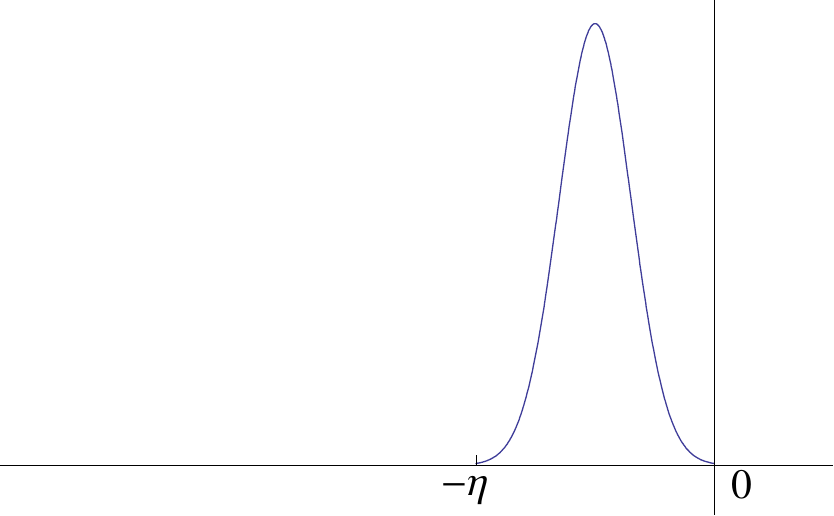}
\vspace{-4mm}
\parbox{15.5cm}{\caption{\label{fig:del} Regularized $\delta$-distribution (schematically). See \cite{Casagrande:2010si} and text for details.}}
\end{center}
\end{figure}

Before proceeding, we have to explain the precise definition of the $\delta$-distributions above.
For the consistency of the 5D gauge theory, it is important that we can integrate by parts 
in the action without encountering boundary terms. Otherwise the Lagrangian is not hermitian
as required by unitarity, see Section~\ref{sec:SM1}. However, the presence of 
$\delta$-distribution terms at the IR boundary gives rise to discontinuities of some of the fields 
at $|\phi|=\pi$, which seems to jeopardize this important feature. 
In order to avoid this problem, we will always regularize the $\delta$-distributions by moving them
infinitesimally into the bulk. When treating fermions, we will also have to furnish them with a finite width. We 
will thus view the $\delta$-distribution as the limit $\eta\to 0$ of a sequence of regularized functions $\delta^\eta(x)$ 
with support on the interval $x \in [-\eta, 0]$, see Figure \ref{fig:del}. 
This limit is understood in the weak sense such that
\begin{equation} \label{eq:deltadis}
  \lim_{\eta \to 0^+} \int_{-\infty}^{+\infty} dx \; \delta^{\eta} (x) 
  \hspace{0.25mm} f(x) = f(0) \,,
\end{equation}
for all test functions $f(x)$, \ie, smooth functions having compact support.

It turns out that for the gauge boson sector, we can take the limit of the width approaching zero
from the beginning, nevertheless still keeping the $\delta$-distribution infinitesimally away from the brane. 
The orbifold symmetry thus leads to the definition
\beq
\label{eq:delIBP}
   \delta(|\phi|-\pi)\equiv \lim_{\eta\to 0^+}\,
   \frac12\,\Big[ \delta(\phi-\pi+\eta) + \delta(\phi+\pi-\eta)
   \Big] \,.
\eeq
As a consequence, the discontinuities are moved into the bulk and we can assign standard BCs to the 
fields on the branes, being consistent with integration by parts (without boundary terms). All
calculations are performed at small but finite $\eta$ and in the end the (smooth) limit $\eta\to 0^+$ is taken, giving rise
to well-defined jump conditions for $Z_2$-odd functions on the IR brane. We will use the notation 
$f(\pi^-)\equiv\lim_{\eta\to 0^+}\,f(\pi-\eta)$ to specify the value of a function which is discontinuous at $|\phi|=\pi$. 
For bulk fermions, however, it will turn out that we have to work with a finite width in the course of calculating the 
spectrum. The reason is, that in this sector, the $\delta$-distributions multiply functions which feature 
discontinuities (in the limit $\eta \rightarrow 0$) within the support of the $\delta$-distributions, see Section
\ref{sec:fermions}.

\subsubsection{Kaluza-Klein Decomposition}
\label{sec:KKgauge}

We now perform a KK decomposition of the various 5D gauge fields. Not being able to probe directly the extra dimension,
it is sensible to choose a 4D description by decomposing the fields into components, just depending on the non-compact 4D 
space-time, supplied with profile functions, depending on $\phi$. Thus, after performing the integral over the fifth dimension, 
we will end up with an effective 4D theory containing infinite towers of (massive) 4D fields, as discussed in Section~\ref{sec:ADDconstr}. 
The zero mode sector has to correspond (to LO) to the SM, as the low-energy tail
of the NP model. 

In the spirit of 
performing a Fourier decomposition, we decompose the 4D-Lorentz vector (scalar) components of the 5D boson fields with respect 
to the complete set of (derivatives of) even functions on the orbifold $\chi_n^a(\phi)$, $a=A,Z,W$, which can be taken to obey the 
orthonormality conditions
\beq\label{chinorm}
   \int_{-\pi}^\pi\!d\phi\,\chi_m^a(\phi)\,\chi_n^a(\phi)
   = \delta_{mn} \,.
\eeq
The Fourier coefficients, still depending on $x^\mu$, will then correspond to standard 4D gauge fields. We write
\beq
\begin{aligned}
\label{eq:KKdecbos}
   A_\mu(x,\phi) 
   &= \frac{1}{\sqrt r} \sum_n A_\mu^{(n)}(x)\,\chi_n^A(\phi) \,,
    &\qquad 
   A_\phi(x,\phi) 
   &= \frac{1}{\sqrt r} \sum_n a_n^A\,\varphi_A^{(n)}(x)\,
    \partial_\phi\,\chi_n^A(\phi) \,, \\
   Z_\mu(x,\phi) 
   &= \frac{1}{\sqrt r} \sum_n Z_\mu^{(n)}(x)\,\chi_n^Z(\phi) \,,
    &\qquad
   Z_\phi(x,\phi) 
   &= \frac{1}{\sqrt r} \sum_n a_n^Z\,\varphi_Z^{(n)}(x)\,
    \partial_\phi\,\chi_n^Z(\phi) \,, \\
   W_\mu^\pm(x,\phi) 
   &= \frac{1}{\sqrt r} \sum_n W_\mu^{\pm(n)}(x)\,\chi_n^W(\phi)
    \,, &\qquad
   W_\phi^\pm(x,\phi) 
   &= \frac{1}{\sqrt r} \sum_n a_n^W\,\varphi_W^{\pm(n)}(x)\,
    \partial_\phi\,\chi_n^W(\phi) \,,
\end{aligned}
\eeq
where the {\it Kaluza Klein modes} $A_\mu^{(n)}$ \etc\ are already the correct 4D mass eigenstates and thus no further transformation 
to the mass basis is necessary.

The 4D scalar fields can also be expanded in the basis of mass eigenstates as
\beq
\label{eq:gol}
   \varphi^\pm(x) = \sum_n b_n^W\,\varphi_W^{\pm(n)}(x) \,, \qquad 
   \varphi^3(x) = \sum_n b_n^Z\,\varphi_Z^{(n)}(x) \,.
\eeq
The masses of the 4D vector fields will be denoted by $m_n^a\ge 0$ (with $a=A,Z,W$) and those of the scalar 
fields $\varphi_a^{(n)}$ are related to them by gauge invariance.
Just as the scalar Goldstone bosons of the Higgs sector provide the longitudinal degrees of freedom for the massive gauge bosons
(see Section~\ref{sec:Higgs}), the scalar components of the 5D gauge fields provide the degrees of freedom for the 4D vector KK fields
to become massive. After EWSB those fields will mix. The similar role that they play is reflected in the gauge 
fixing Lagrangian (\ref{Sgf}).

The EOMs which determine the form of the profiles $\chi_n^a$ can be obtained \eg by inserting the decompositions 
(\ref{eq:KKdecbos}) into the action and demanding that the 4D theory, after performing the integral over the unresolvable fifth 
dimension, looks like a standard 4D theory of (massive) gauge bosons (and scalar ``Goldstone'' bosons). 
We find \cite{Davoudiasl:1999tf,Pomarol:1999ad}
\beq\label{eq:gaugeeom}
   - \frac{1}{r^2}\,\partial_\phi\,e^{-2\sigma(\phi)}\,
   \partial_\phi\,\chi_n^a(\phi) = (m_n^a)^2\,\chi_n^a(\phi) 
   - \frac{\delta(|\phi|-\pi)}{r}\,M_a^2\,\chi_n^a(\phi) \,.
\eeq
As we work with a compact extra dimension with boundaries, we have to specify the BCs for our profiles.
Integrating (\ref{eq:gaugeeom}) over an infinitesimal interval across the orbifold fixed points and keeping in 
mind that the functions $\chi_n^a$ have even $Z_2$ parity one arrives at
\beq\label{eq:bcs}
\begin{aligned}
   \partial_\phi\,\chi_n^a(0) &= 0 
   &&\mbox{(UV brane)}\,, \\
   \partial_\phi\,\chi_n^a(\pi^-) 
   &= - \frac{r M_a^2}{2\epsilon^2}\,\chi_n^a(\pi) 
   &\quad &\mbox{(IR brane)}\,.
\end{aligned}
\eeq
The IR BCs determine the mass eigenvalues $m_n^a$ of the gauge-boson mass-eigenstates. Already the lightest
modes of the $W^\pm$ and $Z$ bosons receive finite masses $m_n^a\sim v$, due to the couplings to the Higgs sector, reflected
by $M_a>0,\, a=W,Z $. Note that $Z_2$-odd profiles, obeying canonical Dirichlet BCs, do not posses zero modes, 
\ie, they have no solution with $m_n^a=0$ (or $m_n^a\sim v$ after EWSB). Their lightest modes will
develop masses of the order of $\ord(\Mkk)$. This holds in particular for the scalar components of the gauge 
fields, that have profiles proportional to the derivative of the vector components. However, they can still mix with light modes
like the Goldstone bosons from the Higgs sector.

In the end the bilinear gauge-boson action takes the desired form
\beq\label{gaugefinal}
\begin{split}
   S_{\rm gauge,2} 
   &= \sum_n \int d^4x\,\bigg\{
    - \frac14\,F_{\mu\nu}^{(n)} F^{\mu\nu(n)}
    - \frac{1}{2\xi} \left( \partial^\mu A_\mu^{(n)} \right)^2 
    + \frac{(m_n^A)^2}{2}\,A_\mu^{(n)} A^{\mu(n)} \\
   &\quad\mbox{}- \frac14\,Z_{\mu\nu}^{(n)} Z^{\mu\nu(n)}
    - \frac{1}{2\xi} \left( \partial^\mu Z_\mu^{(n)} \right)^2
    + \frac{(m_n^Z)^2}{2}\,Z_\mu^{(n)} Z^{\mu(n)} \\
   &\quad\mbox{}- \frac12\,W_{\mu\nu}^{+(n)} W^{-\mu\nu(n)}
    - \frac{1}{\xi}\,\partial^\mu W_\mu^{+(n)}\,
    \partial^\mu W_\mu^{-(n)} 
    + (m_n^W)^2\,W_\mu^{+(n)} W^{-\mu(n)} \\
   &\quad\mbox{}
    + \frac12\partial_\mu\varphi_A^{(n)}
    \partial^\mu\varphi_A^{(n)} 
    - \frac{\xi (m_n^A)^2}{2}\,\varphi_A^{(n)}\varphi_A^{(n)} 
    + \frac12\partial_\mu\varphi_Z^{(n)}
    \partial^\mu\varphi_Z^{(n)} 
    - \frac{\xi (m_n^Z)^2}{2}\,\varphi_Z^{(n)}\varphi_Z^{(n)} \\
   &\quad\mbox{}+ \partial_\mu\varphi_W^{+(n)}
    \partial^\mu\varphi_W^{-(n)} 
    - \xi (m_n^W)^2\,\varphi_W^{+(n)}\varphi_W^{-(n)} 
    \bigg\} \\
   &\quad\mbox{}+ \int d^4x \left( 
    \frac12\partial_\mu h\partial^\mu h - \lambda v^2 h^2 \right) 
    + \sum_n \int d^4x\,{\cal L}_{\rm FP}^{(n)} \,,
\end{split}
\eeq
if and only if, in addition to the EOMs (\ref{eq:gaugeeom}), the relations
\beq\label{coefs}
   a_n^a = - \frac{1}{m_n^a} \,, \qquad
   b_n^a = \frac{M_a}{\sqrt r}\,\frac{\chi_n^a(\pi^-)}{m_n^a}
\eeq
hold.
Looking at (\ref{gaugefinal}) we see that, as expected, we end up with an infinite number of copies of the 4D SM gauge sector,
to be distinguished by the mass of the fields. We observe towers of massive gauge bosons with masses $m_n^a$, accompanied by towers of 
massive scalars with masses $\sqrt{\xi}\,m_n^a$, as well as the Higgs field $h$ with mass $\sqrt{2\lambda} v$.
The low energy tail of the gauge-boson spectrum, \ie, the zero modes, can be interpreted as the SM gauge bosons we observe in nature. 
However, due to $v^2/\Mkk^2$ suppressed mixing effects with the KK modes (already included in our KK decomposition), the massive gauge 
bosons will couple differently compared to the SM, see below. Measuring these deviations, as well as deviations due to the virtual 
exchange of KK excitations, provides a possibility to test the RS scenario, in addition to detecting directly the heavy KK modes with 
masses $m_1^a \sim \Mkk$.

Note that, after applying (\ref{coefs}), the 4D gauge-fixing Lagrangian derived from (\ref{Sgf}) can be written as
\beq
   r\int_{-\pi}^\pi\!d\phi\,{\cal L}_{\rm GF}
   = \sum_n\,{\cal L}_{\rm GF}^{(n)} \,,
\eeq
where
\beq
\begin{split}
   {\cal L}_{\rm GF}^{(n)}
   &= - \frac{1}{2\xi} \left( \partial^\mu A_\mu^{(n)}
    - \xi m_n^A\varphi_A^{(n)} \right)^2 
    - \frac{1}{2\xi} \left( \partial^\mu Z_\mu^{(n)}
    - \xi m_n^Z\varphi_Z^{(n)} \right)^2 \\
   &\quad\mbox{}- \frac{1}{\xi} \left( \partial^\mu W_\mu^{+(n)}
    - \xi m_n^W\varphi_W^{+(n)} \right)
    \left( \partial^\mu W_\mu^{-(n)}
    - \xi m_n^W\varphi_W^{-(n)} \right) .
\end{split}
\eeq
For each KK mode these expressions resemble those of the SM. 
As a consequence, the Faddeev-Popov ghost Lagrangians ${\cal L}_{\rm FP}^{(n)}$ in (\ref{gaugefinal}) are 
completely analogous to the one of the SM. The only generalization being, that a ghost field is required for every 
KK mode. 

\subsubsection{Bulk Profiles}
\label{sec:mingaugeprofiles}

In this section we derive the explicit form of the gauge-boson profiles $\chi_n$ which are solutions to the EOMs 
(\ref{eq:gaugeeom}). From now on, we omit the superscript $a$, denoting the gauge-boson type, unless it is needed for clarity. 
The corresponding expressions were first obtained in \cite{Davoudiasl:1999tf,Pomarol:1999ad} for the case 
of an unbroken gauge symmetry. As the EOMs in the bulk are the same, the structure of this solution remains valid 
also for a spontaneously broken symmetry. However, we have to take into account the modified BCs on 
the IR brane due to the couplings to the Higgs sector (\ref{eq:bcs}). 

In the following it will turn out to be convenient to switch to the variable $t=\epsilon\,e^\sigma$ to describe the orbifold 
\cite{Grossman:1999ra}, which will take values between $t=\epsilon$ (UV brane) and $t=1$ (IR brane), corresponding to the 
interval $\phi\in[0,\pi]$. The reflection at $\phi=0$ will be defined via the $Z_2$ symmetry. Integrals over the orbifold can 
be obtained (for an $Z_2$ even integrand) using
\beq
   \int_{-\pi}^\pi\!d\phi 
   \to \frac{2\pi}{L} \int_\epsilon^1\!\frac{dt}{t} \,, 
    \qquad
   \int_{-\pi}^\pi\!d\phi\,e^{\sigma(\phi)}
   \to \frac{2\pi}{L\epsilon} \int_\epsilon^1\!dt \,,
    \qquad \etc
\eeq
After this transformation of coordinates, the EOMs (\ref{eq:gaugeeom}) become
\footnote{Note that another useful formulation of the RS background is given by the conformally flat metric \cite{Csaki:2002gy} 
$$
   ds^2 = \left( \frac{R}{z} \right)^2 
   \left( \eta_{\mu\nu}\,dx^\mu dx^\nu - dz^2 \right) ,
$$
restricted to the interval $z\in[R,R']$, where $z=\frac{t}{\Mkk}$. Here, $R = \frac{1}{k}$ and $R'= \frac{1}{\Mkk}$ denote the 
positions of the UV brane and the IR brane, respectively.}
\beq
   t\,\partial_t\,t^{-1}
   \partial_t\,\chi_n(t) = -x_n^2\,\chi_n(t) 
   +\delta(t-1)\frac{M^2}{2 k \epsilon^2}\,\chi_n(t) \,.
\eeq
The parameters
\beq
   x_n\equiv \frac{m_n}{\Mkk}
\eeq
are the dimensionless versions of the masses of the gauge bosons and their KK excitations in the 4D theory. 
The ansatz
\beq\label{chin}
   \chi_n(t) = N_n\sqrt{\frac{L}{\pi}}\,t\,c_n^+(t) \,,
\eeq
finally leads to the Bessel equation
\beq
x_n^2 t^2\, c_n^{+\,\prime\prime}(x_n t)+x_n t\, c_n^{+\,\prime}(x_n t)+(x_n^2 t^2-1) c_n^{+}(x_n t)=0
\eeq
in the bulk $(t\ne \epsilon,1)$. This equation is solved by
\beq\label{cpl}
   c_n^+(t) = Y_0(x_n\epsilon)\,J_1(x_n t)
    - J_0(x_n\epsilon)\,Y_1(x_n t)\,,
\eeq
where the coefficients have been determined by the UV BCs (\ref{eq:bcs}).
We furthermore define
\beq\label{cmi}
   c_n^-(t) = \frac{1}{x_n t}\,
    \frac{d}{dt} \left[ t\,c_n^+(t) \right]
    = Y_0(x_n\epsilon)\,J_0(x_n t)
    - J_0(x_n\epsilon)\,Y_0(x_n t) \,.
\eeq
The normalization constants $N_n$ are now fixed by the orthonormality condition (\ref{chinorm}) to obey
\beq\label{Nngauge}
   N_n^{-2} = \left[ c_n^+(1) \right]^2 + \left[ c_n^-(1) \right]^2
   - \frac{2}{x_n}\,c_n^+(1)\,c_n^-(1)
   - \epsilon^2 \left[ c_n^+(\epsilon) \right]^2\,.
\eeq
Since $c_n^-(\epsilon)=0$, we easily see that (\ref{chin}) satisfies the BCs 
$\partial_t\chi_n(\epsilon)=0$ on the UV brane. 

The BCs (\ref{eq:bcs}) on the IR brane lead 
to the relation
\beq\label{condi}
   x_n\,c_n^-(1) 
   = - \frac{g^2 v^2}{4\Mkk^2}\,L\,c_n^+(1) \,,
\eeq
from which the mass eigenvalues $x_n$ can be derived. Note that the condition (\ref{condi}) as stated above 
holds for the profiles of the $W^\pm$ bosons and their KK partners. For the case of the $Z$ boson, the 
$SU(2)$-coupling $g^2$ has to be replaced by the combination $(g^2+g'^2)$. In the absence of EWSB ($v=0$), which corresponds to the case of the photon and 
the gluon which do not couple to the Higgs sector, the right hand side becomes zero. In this case the spectrum 
contains a massless zero mode ($m_0=0$) with a flat profile 
\beq\label{eq:chi0flat}
   \chi_{\gamma,g}(t) = \frac{1}{\sqrt{2\pi}} \,.
\eeq
These massless modes are identified with the SM photon and gluon. The masses of their KK excitations will simply correspond 
to zeros of a combination of Bessel functions $c_n^-(1)=0$, which, to good approximation corresponds to zeros of the Bessel 
function $J_0(x_n)$ . This leads to $x_1\approx 2.45$ and the heavier modes follow in spacings of about $\pi$. Due to the 
flat profile, the photon and the gluon zero modes will couple to bulk fermions exactly like in the SM, which will be important 
for the discussions of FCNCs in Section~\ref{sec:gaugecouplings}. 

The results for the massive SM gauge bosons, which correspond to (light) modes with $m_0\ne0$, can be simplified, since $x_0\ll 1$. 
Expanding (\ref{condi}) in powers of $x_0$ we arrive at
\beq\label{eq:m02}
   m_W^2 = \frac{g^2 v^2}{4} \left[
   1 - \frac{g^2 v^2}{8\Mkk^2} 
   \left( L - 1 + \frac{1-\epsilon^2}{2L} \right) 
   + \ord\left( \frac{v^4}{\Mkk^4} \right) \right]\,.
\eeq
An analogous relation with $g^2$ replaced by $(g^2+g'^2)$ holds for the $Z$-boson mass. 
Comparing these results with the SM relations (\ref{eq:MWSM}), one sees immediately 
that the tree-level ratio of the $W^\pm$-boson and $Z$-boson mass in RS models deviates from the SM 
prediction.\footnote{Note that, here and in the following, by $Z$ boson ($W^\pm$ boson), 
we denote the {\it zero modes} of the $Z$ boson ($W^\pm$ bosons).} This is due to the mixings with the KK 
modes, \ie, the contributions to the zero-mode masses from the compactification of the extra dimension. 
Moreover, the Higgs VEV, extracted from (\ref{eq:m02}) will get corrections with respect to 
the SM value of $v_{\rm SM}=246\,$ GeV. We will come back to both of these issues later in Section~\ref{sec:mod}.
It will also turn out to be useful to derive an approximate expression for the ground-state 
profile $\chi_0$ for $x_0\ll 1$. We obtain, in agreement with
\cite{Csaki:2002gy},
\beq\label{eq:chi0WZ}
   \chi_{W,Z}(t) = \frac{1}{\sqrt{2\pi}} \left[
   1 + \frac{m_{W,Z}^2}{4\Mkk^2} \left( 1 - \frac{1-\epsilon^2}{L} 
   + t^2 \left( 1 - 2L - 2\ln t\right) \right) 
   + \ord\left( \frac{m_{W,Z}^4}{\Mkk^4} \right) \right]\,.
\eeq
\begin{figure}[!t]
	\centering
		\includegraphics[width=10cm]{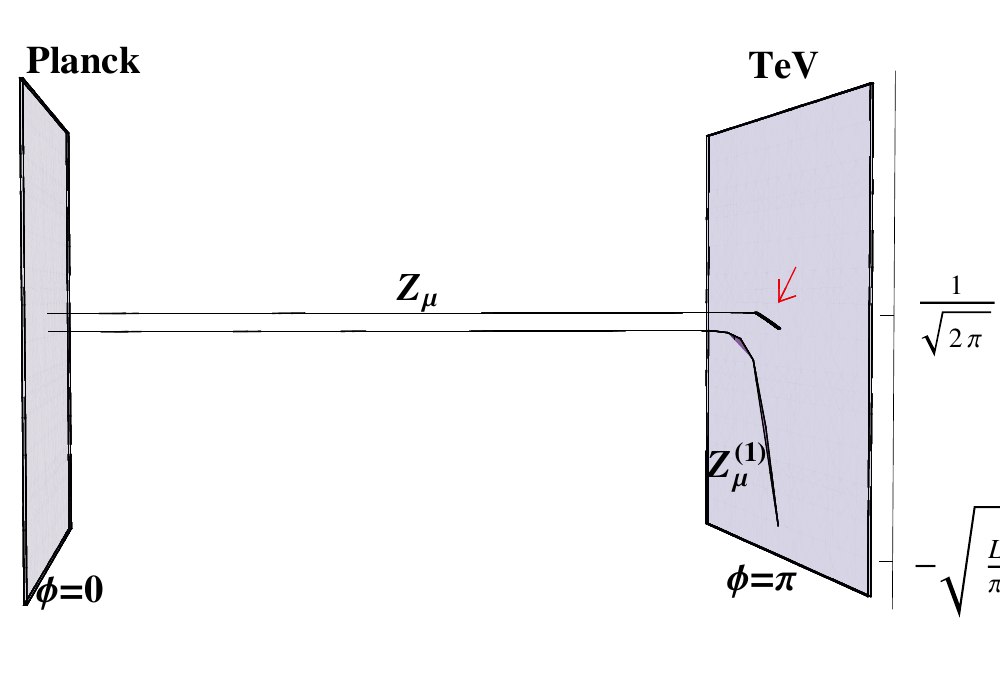}
	\caption{Zero mode and first KK excitation of the $Z$ boson in a slice of AdS$_5$.}
	\label{fig:AdSBos}
\end{figure}
The deviations from the flat zero mode solution without EWSB (\ref{eq:chi0flat}) can be understood from a perturbative point of view
as the mixings of the massive KK modes, being localized close to the IR brane, into the zero mode. This will lead to a dip
in the zero mode profiles at this boundary, proportional to the mixing parameter $m_{W,Z}/\Mkk^2$, see Figure \ref{fig:AdSBos}. The absolute 
value of the profiles of generic KK modes on the IR brane is to good approximation given by
\beq
|\chi_n(1)|\approx\sqrt{\frac L \pi}\,, n\geq1\,.
\eeq
The accuracy of this approximation increases with $n$.

\subsection{Bulk Matter and Flavor Mixing}
\label{sec:fermions}

Now we move to the SM fermion content, propagating into the extra dimension. The warped background provides interesting possibilities
for model building in this sector. After discussing the setup, we will perform the KK decomposition in the 
mass basis, incorporating flavor mixing in a completely general way. In this way we are again able to derive 
exact expressions from which the masses and mixings can be obtained. We will concentrate on the quark sector. The 
generalization to include the lepton sector, which will not be studied in this thesis, is straightforward.

In the subsequent section we will then show how hierarchies in the quark spectrum, as well as in the CKM mixing angles, 
arise naturally within the RS model. By performing an expansion in small parameters, we will work out explicit analytic 
expressions for the masses of the SM quarks (corresponding to the light ``zero modes'' in the RS model), as well as the CKM parameters, 
in terms of the input parameters of the RS model. To this end, we demonstrate and apply an analogy between the generation of 
flavor hierarchies in warped extra dimensions and in the Froggatt-Nielsen model.

\subsubsection{Action of the 5D Theory}

As we consider the SM field content, we start from (\ref{eq:fermkin}), with the quantum numbers of the fermion fields still 
as given in Table \ref{tab:SMQN}, and generalize the corresponding action to 5D. However, in an odd number of space-time dimensions we have to 
face a subtlety when trying to define chiral fermions. To describe matter fields in five dimensions, we need a representation
of the 5D-Clifford Algebra
\beq
	\left\{\Gamma_M,\Gamma_N\right\}=2\eta_{MN}\,.
\eeq
An appropriate four dimensional representation in 5D is provided by taking the usual 4D representation, consisting of the Dirac matrices $\gamma_\mu$
(see Appendix~\ref{app:PDG}), 
and subjoining
$\gamma_5$
\beq
\label{eq:5DDM}
	\Gamma_M=\left(\gamma_\mu, i \gamma_5 \right)\,.
\eeq
Note that the vector of Dirac matrices above is contracted with Lorentz vectors in curved space-time via the vielbein formalism, see \eg 
\cite{Grossman:1999ra}. 

In an even number of dimensions, the Dirac-spinor representation is reducible. For example, in 4D it can be decomposed into 
the well known (left-handed and right-handed) Weyl-spinor representations that do not mix under Lorentz transformations. 
The corresponding projections are just provided by the projectors $P_{L,R}\equiv(1\mp\gamma^5)/2$, defined with the help 
of the chirality matrix $\gamma^5$, see Section~\ref{sec:SM1}. However, it turns out that in an odd number of dimensions, the Dirac-spinor representation 
is irreducible. Although we can formally still define projection operators in 5D out of $\gamma^5$, the spectrum will a priori have to consist of 
``vector-like'' four-component Dirac fermions, \ie, for every left handed fermion a right handed counterpart will appear with the same quantum numbers, and vice versa.
Yet, as discussed before, nature is based on chiral matter fields. The left-handed fermions we observe are doublets under 
$SU(2)_L$, whereas the right-handed ones are singlets under this gauge group. At first sight, it seems impossible to generate such a 
setup in 5D. However, the obrifold will again furnish a way out and offer a possibility to arrive at chiral fermions in the low
energy spectrum of the theory, after KK decomposition \cite{Grossman:1999ra}. The $Z_2$ symmetry of the action results in opposite $Z_2$ parities for 
the left-handed and right-handed components of a Dirac fermion.
This suggests to start with two sets of Dirac fermions in 5D, one containing doublets under $SU(2)_L$ and one containing the corresponding 
singlets, and assign the $Z_2$-parities/BCs properly, such that the doublets just have left-handed (4D) zero modes, while the singlets 
possess only right-hand zero modes. In 
consequence, the low energy spectrum will contain just the chiralities that we observe in Nature (in the limit $v \to 0$). At each 
level of KK excitations however, two full (vector-like) Dirac-fermions, a singlet and a doublet, will be present and thus twice the amount of degrees
of freedom with respect to the zero mode level. Note that after EWSB, the doublets and singlets will mix, for example the right handed zero
mode will receive a small doublet admixture from the KK excitations, leading to a (suppressed) ``right handed'' CKM matrix, see below. 

So, let us consider $N$ generations of 5D (SM-like) quarks in the bulk, denoting components of $SU(2)_L$ doublets by $u\,,\,d$ and singlets by 
$u^{\,c}\,,\,d^{\,c}$. As we will discuss more complicated fermion contents in Section~\ref{sec:custo}, we will give the action 
already in a notation which is appropriate for a richer fermion structure. The bilinear part of the 5D fermion action can be 
written as ({\it cf.} \cite{Grossman:1999ra,Gherghetta:2000qt})
\beq\label{eq:Sferm2}
\begin{split} 
   S_{\rm ferm,2} & = \int d^4x\,r\int_{-\pi}^\pi\!d\phi\;\Bigg\{
    \sum_{q=U,u,D,d} \Bigg( e^{-3\sigma(\phi)}\,
    \bar{\vec q}\,i\delslash\,\vec q -
    e^{-4\sigma(\phi)}\,\sgn(\phi)\,
    \bar{\vec q}\,\bm{M}_{\vec q}\,\vec q\\
    &\quad\mbox{}- \frac{1}{2r}\, \bigg[ \, \bar{\vec
      q}_L\,e^{-2\sigma(\phi)} \,
    {\stackrel{\leftrightarrow}{\partial}}_\phi \,
    e^{-2\sigma(\phi)}\,\vec q_R + \mbox{h.c.}\,
    \bigg] \Bigg)\\
   &\quad\mbox{}- \delta(|\phi|-\pi)\,e^{-3\sigma(\phi)} 
    \frac{v}{\sqrt2 r} \sum_{(Q,q)=(U,u),(D,d)} \left[ \bar{\vec Q}_L\,\bm{Y}_{\vec q}^{\rm (5D)C}\,\vec q_R
    + \bar{\vec Q}_R\,\bm{Y}_{\vec q}^{\rm (5D)S}\,\vec q_L+ \mbox{h.c.} \right] 
    \bigg\} \,,
\end{split}
\eeq
where $\stackrel{\leftrightarrow}{\partial}_\phi\,\equiv\,\stackrel{\rightarrow}{\partial}_\phi
-\stackrel{\leftarrow}{\partial}_\phi$, and we have already evaluated explicitly the $\mu$ and $\phi$ components of 
the vielbein, entering the kinetic terms \cite{Grossman:1999ra}. Moreover, note that the spin connection gives no 
contribution to (\ref{eq:Sferm2}). For the SM fermion content (Table \ref{tab:SMQN}), the vectors of 
fermions introduced here correspond merely to $N$-component vectors in flavor space (\eg $u = (u,c,t)^T$ for the up-type 
quarks of the SM). We identify for the {\it minimal RS model}
\beq
\label{eq:SMembedding}
\begin{split}
\vec U \equiv u\,,\ \vec D \equiv d\,,\quad \vec u \equiv u^c\,,\ \vec d \equiv d^c\,,\\
\end{split}
\eeq
and thus end up with the standard SM notation, as used in \cite{Casagrande:2008hr}. Note that for the extended model studied in
Section~\ref{sec:custo}, the fermion vectors will contain several types of quarks.
In the spirit of writing down every term which is not forbidden by a symmetry, we include explicit gauge-invariant
5D Dirac mass terms for the vector-like quarks, with diagonal and real $N\times N$ bulk mass matrices 
\beq
\bm{M}_{\vec U}=\bm{M}_{\vec D} \equiv\bm{M}_Q
\eeq
for the $SU(2)_L$ doublets as well as $\bm{M}_{\vec q}\equiv \bm{M}_q\,,\, q=u,d$ for the singlets. Since the corresponding 
combination of field operators is always $Z_2$ odd, the mass matrices have to be multiplied by a $Z_2$ odd function in order to arrive 
at a $Z_2$ invariant action. Starting from generic hermitian bulk mass matrices, it is easy to see that going to the {\it bulk 
mass basis} in which those matrices are real and diagonal does not spoil the canonical and diagonal form of the kinetic terms. 
Note however that the bulk masses can be positive as well as negative. In contrast to 4D theories, the sign of the Dirac mass 
term cannot be reversed by a field redefinition.
In order to arrive at the correct low energy spectrum, as discussed before, the left-handed (right-handed) components 
of the $SU(2)_L$ doublet $Q \equiv(u,d)^T$ are taken to be even (odd) under the $Z_2$ orbifold symmetry, with appropriate
canonical BCs. Likewise, the right-handed (left-handed) components of the singlets $u^c$ and $d^c$ are chosen even (odd), 
such that the zero modes of the even fields correspond to the SM particles. After EWSB on the IR brane, the quark fields 
are coupled by 5D $N\times N$ Yukawa matrices 
$\bm{Y}_{\vec q}^{\rm (5D)C}$, connecting the $Z_2$ even components of the $SU(2)_L$ 
doublets with those of the singlets, as well as corresponding matrices $\bm{Y}_{\vec q}^{\rm (5D)S}$, 
connecting $Z_2$-odd fields, where $\,q=u,d$. If not stated otherwise, those matrices correspond to the bulk mass basis of real and diagonal 
matrices $\bm{M}_{Q,q}$. Note that the second type of Yukawa couplings has not been taken into account in \cite{Casagrande:2008hr}. 
While such a setup is consistent and it is possible to generate the fermion masses without those couplings, 
it is not the most general or natural possibility, and we will include these couplings. Furthermore, if one considers the 
brane-localized Higgs-sector as the limit of a Higgs, propagating into the bulk, both types of Yukawa matrices have to be 
identical $\bm{Y}_{\vec q}^{\rm (5D)C}=\bm{Y}_{\vec q}^{\rm (5D)S}\equiv\bm{Y}_{\vec q}^{\rm (5D)}$, which we will 
assume in the following. This is, again, due to the fact that in 5D the Dirac-spinor representation is irreducible. In a perturbative approach, 
the Yukawa terms containing $Z_2$ odd fields do not seem to contribute, as the corresponding profiles vanish on the 
IR brane. However, in the end this naive point of view will turn out to be wrong, 
as we will see below \cite{Azatov:2009na}. 
For the KK decomposition it will be convenient to define dimensionless 4D Yukawa matrices via 
\beq\label{eq:Y4Ddef}
   \bm{Y}_{\vec q}^{\rm (5D)}\equiv \frac{2\bm{Y}_{\vec q}}{k} \,,
   \qquad q=u,d \,,
\eeq
where again in the {\it minimal RS model} the vector notation is not necessary $\bm{Y}_{\vec q}\equiv\bm{Y}_{q}$. The chiral components of the spinor field are defined as usual as $\vec q=\vec q_L+\vec q_R\,$, \etc\ As discussed before, without the 
presence of the Yukawa interactions, each 5D fermion would induce a massless chiral fermion in the 4D theory, accompanied by a tower 
of massive, vector-like, KK excitations \cite{Grossman:1999ra}. After EWSB, the Yukawa couplings replace the massless 
modes by light modes (compared to the KK scale), corresponding to the SM fermions.

\subsubsection{Kaluza-Klein Decomposition}
\label{sec:KKferm}

We want to arrive at a 4D theory of massive Dirac fermions $q^{(n)}=q_L^{(n)}+q_R^{(n)}$ with masses $m_n>0$. 
To this end we perform a KK decomposition of the 5D fields in the mass basis and write
\begin{equation}\label{eq:KKdecferm}
  \begin{split}
    \vec Q_L(x,\phi) &= \frac{e^{2\sigma(\phi)}}{\sqrt r} \sum_n
    \bm{C}_n^{Q}(\phi)\,\vec a_n^{\hspace{0.5mm} Q}\,q_L^{(n)}(x) \,,
    \qquad \vec Q_R(x,\phi) = \frac{e^{2\sigma(\phi)}}{\sqrt r} \sum_n
    \bm{S}_n^{Q}(\phi)\, \vec a_n^{\hspace{0.5mm} Q}\,q_R^{(n)}(x) \,,
    \hspace{4mm} \\
    \vec q_L(x,\phi) &= \frac{e^{2\sigma(\phi)}}{\sqrt r} \sum_n
    \bm{S}_n^{\hspace{0.25mm} q}(\phi)\, \vec a_n^{\hspace{0.25mm}
      q}\,q_L^{(n)}(x) \,, \hspace{1.2cm} \vec q_R(x,\phi) =
    \frac{e^{2\sigma(\phi)}}{\sqrt r} \sum_n \bm{C}_n^{\hspace{0.25mm}
      q}(\phi)\,\vec a_n^{\hspace{0.5mm} q}\,q_R^{(n)}(x) \,,
  \end{split}
\end{equation}
where $Q=U,D$ and correspondingly $q=u,d$. For the SM fermion content studied in this section, we have with 
(\ref{eq:SMembedding}) $\vec U=u$ and $\vec u=u^c$ and similar for down type quarks. In that case, the spinor fields on the left 
hand side of the equations in (\ref{eq:KKdecferm}), as well as the objects $\vec a_n^{Q,q}$, describing flavor mixing, 
are $N$ component vectors in flavor space. The $Z_2$ even profiles $\bm{C}_n^{Q,q}$ and odd profiles $\bm{S}_n^{Q,q}$, 
however, are diagonal $N\times N$ matrices, where each entry corresponds to a different bulk mass parameter (in the bulk mass 
basis).\footnote{Note that we have already exploited the fact that $\vec Q_{L,R} (x,\phi)$ ($\vec q_{L,R} (x,\phi)$) can be 
expanded in terms of the same vector $\vec a_n^{\hspace{0.5mm} Q}$ ($\vec a_n^{\hspace{0.25mm} q}$). With this choice, 
the profiles ${\bm C}^Q_n(\phi)$ and ${\bm S}^Q_n (\phi)$ (${\bm C}^q_n (\phi)$ and ${\bm S}^q_n (\phi)$) will be normalized 
in the same way.} The 4D spinors $q^{(n)}$, on the right hand side of the equations in (\ref{eq:KKdecferm}) are single spinor 
fields in the mass basis. The index $n$ labels the mass eigenstates of masses $m_n$. For the case of up-type quarks these are
$m_1=m_u,m_2= m_c,m_3= m_t$, as well as the higher KK masses. Note that, due to gauge invariance, the components of the 
$SU(2)_L$ doublet $Q$ have the same profiles 
\beq
\label{eq:doublpr}
\bm{C}_n^{U}=\bm{C}_n^{D}\equiv\bm{C}_n^{Q}\,,\quad \bm{S}_n^{U}=\bm{S}_n^{D}\equiv\bm{S}_n^{Q}.
\eeq
The associated vectors $\vec a_n^{U}$ and $\vec a_n^{D}$, nevertheless differ for the different components 
of $Q$.

Inserting the decompositions (\ref{eq:KKdecferm}) into the action (\ref{eq:Sferm2}), one derives the equations of motion 
\begin{equation} \label{eq:fermEOM}
  \begin{split}
    \left( - \frac{1}{r}\,\partial_\phi - \bm{M}_{\vec Q}\,\sgn(\phi)
    \right) \bm{S}_n^Q(\phi)\, \vec a_n^{\hspace{0.25mm} Q} &= -
    m_n\,e^{\sigma(\phi)}\,\bm{C}_n^Q(\phi)\, \vec
    a_n^{\hspace{0.25mm} Q} \\ & \phantom{=} + \delta (|\phi|-\pi) \,
    e^{\sigma(\phi)} \, \frac{\sqrt{2} \, v}{kr} \, {\bm Y}_{\vec q}
    \hspace{1mm} \bm{C}_n^{\hspace{0.25mm} q}(\phi) \, \vec
    a_n^{\hspace{0.5mm} q} \,, \\[1mm]
    \left( \frac{1}{r}\,\partial_\phi - \bm{M}_{\vec q}\,\sgn(\phi)
    \right) \bm{S}_n^{\hspace{0.25mm} q}(\phi)\, \vec
    a_n^{\hspace{0.5mm} q} &= -m_n\,e^{\sigma(\phi)}\,
    \bm{C}_n^{\hspace{0.25mm} q}(\phi)\, \vec a_n^{\hspace{0.5mm} q} \\
    & \phantom{=} + \delta (|\phi|-\pi) \, e^{\sigma(\phi)} \,
    \frac{\sqrt{2} \, v}{kr} \, {\bm Y}_{\vec q}^\dagger \hspace{1mm}
    \bm{C}_n^{Q}(\phi) \, \vec a_n^{\hspace{0.25mm} Q} \,, \\[1mm]
    \left( \frac{1}{r}\,\partial_\phi - \bm{M}_{\vec Q}\,\sgn(\phi)
    \right) \bm{C}_n^Q(\phi)\,\vec a_n^{\hspace{0.25mm} Q} &= -
    m_n\,e^{\sigma(\phi)}\,\bm{S}_n^Q(\phi)\,\vec a_n^{\hspace{0.5mm}
      Q} \\
    & \phantom{=} + \delta (|\phi|-\pi) \, e^{\sigma(\phi)} \,
    \frac{\sqrt{2} \, v}{kr} \, {\bm Y}_{\vec q} \hspace{1mm}
    \bm{S}_n^{\hspace{0.25mm} q}(\phi) \, \vec a_n^{\hspace{0.5mm} q}
    \,, \\[1mm]
    \left( - \frac{1}{r}\,\partial_\phi - \bm{M}_{\vec q}\,\sgn(\phi)
    \right) \bm{C}_n^{\hspace{0.25mm} q}(\phi)\,\vec a_n^{\hspace{0.5mm} q} &= -
    m_n\,e^{\sigma(\phi)}\,\bm{S}_n^{\hspace{0.25mm}
      q}(\phi)\,\vec a_n^{\hspace{0.5mm} q} \\
    & \phantom{=} + \delta (|\phi|-\pi) \, e^{\sigma(\phi)} \,
    \frac{\sqrt{2} \, v}{kr} \, {\bm Y}_{\vec q}^\dagger \hspace{1mm}
    \bm{S}_n^{Q}(\phi) \, \vec a_n^{\hspace{0.25mm} Q} \,,
  \end{split}
\end{equation} 
where $(Q,q)=(U,u)\,,\,(D,d)$.
In the bulk, \ie, for $\phi \neq 0, \pm \pi$, these equations reduce to the relations first obtained in \cite{Grossman:1999ra}.
The corresponding general solutions can be written as linear combinations of Bessel functions, (see Section~\ref{sec:minfermionprofiles}). 
The presence of the source terms on the IR brane, however, modifies the boundary behavior of the fields and causes both the $Z_2$-even 
and $Z_2$-odd profiles to become discontinuous on the IR brane with $\bm{C}_n^{Q,q} (\pm \pi) \neq \bm{C}_n^{Q,q} (\pm \pi^-)$ and 
$\bm{S}_n^{Q,q} (\pm \pi) = 0$ but $\bm{S}_n^{Q,q}(\pm \pi^-) \neq 0$ \cite{Bagger:2001qi}. While the UV BCs remain canonical,
\beq\label{bcUV}
   \bm{S}_n^{Q,q}(0) = 0\,,
\eeq
finding the correct IR BCs requires a proper finite-width regularization of the $\delta$-distributions appearing in (\ref{eq:fermEOM}). 
Such a regularization is needed since the $\delta$-distributions appear together with functions that are discontinuous within their 
support. We will use the regularization as introduced in (\ref{eq:deltadis}).

After switching to $t$ coordinates, we study the EOMs (\ref{eq:fermEOM}) in an infinitesimal interval $t\in[1-\eta,1]$,
keeping in mind that at the end we will take the limit $\eta\rightarrow 0$. We regularize the 
$\delta$-distributions and integrate these equations from $t\in[1-\eta,1]$ to $1$, taking into account that the odd fermion
profiles vanish identically on the IR brane, \ie, ${\bm S}^{Q,q}_n (1)= 0$ and dropping all terms $\propto \eta$. In this 
way we find (for finite $n$)
\begin{equation} \label{eq:BCsintegrated}
  \begin{split}
    \bm{S}_n^Q(t)\, \vec a_n^Q & = \frac{v}{\sqrt{2} M_{\rm KK}} \,
    {\bm Y}_{\vec{q}} \int_t^1 dt^\prime \, \big [ \delta^\eta
    (t^\prime - 1 ) \, \bm{C}_n^{\hspace{0.25mm} q}(t^\prime) \big ]
    \vec a_n^{\hspace{0.25mm} q} \,, \\[1mm]
    \bm{S}_n^{\hspace{0.25mm} q}(t)\, \vec a_n^{\hspace{0.25mm} q} & =
    -\frac{v}{\sqrt{2} M_{\rm KK}} \, {\bm Y}_{\vec{q}}^\dagger
    \int_t^1 dt^\prime \, \big [ \delta^\eta (t^\prime - 1 ) \,
    \bm{C}_n^{Q}(t^\prime) \big ] \vec a_n^{\hspace{0.25mm}Q} \,, \\[1mm]
    \bm{C}_n^Q(t)\, \vec a_n^Q & = \bm{C}_n^Q(1)\, \vec a_n^Q
    -\frac{v}{\sqrt{2} M_{\rm KK}} \, {\bm Y}_{\vec{q}} \int_t^1
    dt^\prime \, \big [ \delta^\eta (t^\prime - 1 ) \,
    \bm{S}_n^{\hspace{0.25mm} q}(t^\prime) \big ] \vec
    a_n^{\hspace{0.25mm} q} \,, \\[1mm]
    \bm{C}_n^{\hspace{0.25mm} q}(t)\, \vec a_n^{\hspace{0.25mm} q} & =
    \bm{C}_n^{\hspace{0.25mm} q}(1)\, \vec a_n^{\hspace{0.25mm} q}
    +\frac{v}{\sqrt{2} M_{\rm KK}} \, {\bm Y}_{\vec{q}}^\dagger-
    \int_t^1 dt^\prime \, \big [ \delta^\eta (t^\prime - 1 ) \,
    \bm{S}_n^{Q}(t^\prime) \big ] \vec a_n^{\hspace{0.25mm}Q} \,,
  \end{split}
\end{equation}
indicating that the form of the profiles ${\bm C}^{Q, q}_n (t)$ and ${\bm S}^{Q,q}_n (t)$ becomes independent of the mass 
terms entering the EOMs.

In order to solve (\ref{eq:BCsintegrated}), we introduce the regularized Heaviside 
function
\begin{equation}
  \bar \theta^\eta (x) \equiv 1 - \int_{-\infty}^x dy \; \delta^\eta (y) \,,
\end{equation}
which obeys 
\begin{equation}
  \bar \theta^\eta (0) = 0 \,, \qquad 
   \bar \theta^\eta (-\eta) = 1 \,, \qquad 
  \partial_x \hspace{0.25mm} \bar \theta^\eta (x) = - \delta^\eta (x) \,.
\end{equation}
Using these properties it is easy to show that
\begin{equation} \label{eq:delthe}
  \int_t^1 dt^\prime \, \delta^\eta (t^\prime - 1) \, \big [ \bar 
  \theta^\eta (t^\prime - 1) \big ]^n = \frac{1}{n+1} \, \big [ \bar 
  \theta^\eta (t - 1) \big ]^{n+1} \,.
\end{equation}
This relation allows us to integrate hyperbolic functions of regularized
$\theta$-functions with regularized $\delta$-distributions.
For an arbitrary invertible matrix ${\bm A}$ we obtain
\begin{equation} \label{eq:magic1}
  \begin{split}
    \int_t^1 dt^\prime \, \delta^\eta (t^\prime - 1) \, \sinh
    \hspace{0.25mm} \big ( \bar \theta^\eta (t^\prime - 1)
    \hspace{0.25mm} {\bm A} \big ) & = \left [ \cosh \hspace{0.25mm}
      \big ( \bar \theta^\eta (t - 1) \hspace{0.25mm} {\bm A} \big ) -
      {\bm 1} \right ] {\bm A}^{-1} \,, \\
    \int_t^1 dt^\prime \, \delta^\eta (t^\prime - 1) \, \cosh
    \hspace{0.25mm} \big ( \bar \theta^\eta (t^\prime - 1)
    \hspace{0.25mm} {\bm A} \big ) & =\sinh \hspace{0.25mm} \big (
    \bar \theta^\eta (t - 1) \hspace{0.25mm} {\bm A} \big ) \, {\bm
      A}^{-1} \,,
  \end{split}
\end{equation} 
where the hyperbolic sine and cosine of a matrix are defined via their Taylor expansions. 
With the latter relations it is now easy to determine the solutions
to (\ref{eq:BCsintegrated}). We find 
\begin{equation} \label{eq:BCssolution}
  \begin{split}
    \bm{S}_n^Q(t)\, \vec a_n^Q & = {\bm Y}_{\vec{q}} \left (
      \sqrt{{\bm Y}_{\vec{q}}^\dagger \hspace{0.25mm} {\bm
          Y}_{\vec{q}}} \right )^{-1} \, \sinh \left (
      \frac{v}{\sqrt{2} M_{\rm KK}} \, \bar \theta^\eta (t - 1) \,
      \sqrt{{\bm Y}_{\vec{q}}^\dagger \hspace{0.25mm} {\bm
          Y}_{\vec{q}}}\right ) \bm{C}_n^{\hspace{0.25mm} q}(1) \,
    \vec a_n^{\hspace{0.25mm} q} \,, \\[1mm]
    \bm{S}_n^{\hspace{0.25mm} q}(t)\, \vec a_n^{\hspace{0.25mm} q} & =
    - {\bm Y}_{\vec{q}}^\dagger \left ( \sqrt{{\bm Y}_{\vec{q}}
        \hspace{0.25mm} {\bm Y}_{\vec{q}}^\dagger} \right )^{-1} \,
    \sinh \left ( \frac{v}{\sqrt{2} M_{\rm KK}} \, \bar \theta^\eta (t
      - 1) \, \sqrt{{\bm Y}_{\vec{q}} \hspace{0.25mm} {\bm
          Y}_{\vec{q}}^\dagger} \right ) \bm{C}_n^{Q}(1) \, \vec
    a_n^{Q} \,, \\[1mm]
    \bm{C}_n^Q(t)\, \vec a_n^Q & = \cosh \left ( \frac{v}{\sqrt{2}
        M_{\rm KK}} \, \bar \theta^\eta (t - 1) \, \sqrt{{\bm
          Y}_{\vec{q}} \hspace{0.25mm} {\bm Y}_{\vec{q}}^\dagger}
    \right ) \bm{C}_n^Q(1)\, \vec a_n^Q
    \,, \\[1mm]
    \bm{C}_n^{\hspace{0.25mm} q}(t)\, \vec a_n^{\hspace{0.25mm} q} & =
    \cosh \left ( \frac{v}{\sqrt{2} M_{\rm KK}} \, \bar \theta^\eta (t
      - 1) \, \sqrt{{\bm Y}_{\vec{q}}^\dagger \hspace{0.25mm} {\bm
          Y}_{\vec{q}}}\right ) \bm{C}_n^{\hspace{0.25mm} q}(1)\, \vec
    a_n^{\hspace{0.25mm} q} \,.
  \end{split}
\end{equation}

Since no $t$-integration is left, we can now safely take the limit $\eta \to 0^+$ and use the second pair of equations to
trade the brane values $\bm{C}^{Q,q}_n (1)$ in the first equations for the bulk values $\bm{C}^{Q,q}_n (1^-)$, obtained from the solutions to
(\ref{eq:fermEOM}) by a limiting procedure. Introducing the rescaled Yukawa matrices\footnote{Generalizing this 
result to the case where $Z_2$-even and -odd fermion fields couple via different Yukawa matrices to the brane-localized 
Higgs sector requires to perform the replacements $\bm{Y}_{\vec q} \to \bm{Y}_{\vec q}^{C}$ and $\bm{Y}_{\vec q}^\dagger
\to \bm{Y}_{\vec q}^{S \, \dagger}$. The same replacement rules also apply for (\ref{eq:gtil1}) to (\ref{eq:bmh}).}
\begin{equation} \label{eq:Yukresc}
  \bm{\tilde Y}_{\vec q} \equiv \bm{f} \left (
    \frac{v}{\sqrt{2} M_{\rm KK}} \, \sqrt{{\bm Y}_{\vec{q}}
      \hspace{0.25mm} {\bm Y}_{\vec{q}}^\dagger} \right ) \bm{
    Y}_{\vec q} \,, \qquad \bm{f} (\bm{A}) =  
    \tanh \left (\bm{A} \right ) \bm{A}^{-1} \,,
\end{equation}
it is then easy to show that the resulting IR BCs are manifestly regularization independent\footnote{As a cross check, I have shown that 
different explicit regularization functions - rectangular, triangular, and bulk-Higgs motivated - all lead to the same result. However, these straightforward calculations will not be given here.}  
and can be written in $\phi$ coordinates as
\begin{equation} \label{eq:bcIRrescaled}
  \begin{split}
    \bm{S}_n^Q(\pi^-)\, \vec a_n^Q &= \frac{v}{\sqrt2\Mkk}\,\bm{\tilde
      Y}_{\vec q}\;\bm{C}_n^{\hspace{0.25mm}
      q}(\pi^-)\, \vec a_n^{\hspace{0.5mm} q} \,, \\
    - \bm{S}_n^{\hspace{0.25mm} q}(\pi^-)\, \vec a_n^{\hspace{0.5mm}
      q} &= \frac{v}{\sqrt2\Mkk}\,\bm{\tilde Y}_{\vec q}^{
      \dagger}\;\bm{C}_n^Q(\pi^-)\, \vec a_n^Q \,.
  \end{split}
\end{equation}
They hence take precisely the same form as the BCs derived in \cite{Casagrande:2008hr}, with the original Yukawa 
couplings replaced by the rescaled ones, as defined in (\ref{eq:Yukresc}). These coincide at LO in $v^2/\Mkk^2$, \ie, 
$\bm{\tilde Y}_{\vec q} = \bm{Y}_{\vec q} + {\cal O} (v^2/\Mkk^2)$. In practice, the combinations of quark profiles and Yukawa 
matrices are chosen such that the masses of the zero mode fermions, as well as the CKM parameters, reproduce those determined by 
experiment. Thus, the rescaling of the Yukawa matrices, as described above, has no observable effect on the fermion spectrum and the 
mixings. However, as we will explain in detail in Section~\ref{sec:HcouplingsRS}, 
the inclusion of the Yukawa couplings between $Z_2$-odd fermions alters the misalignment between the masses and Yukawa couplings. 
This leads to a change in the tree-level interactions of the Higgs-boson with fermions with respect to the
results derived in \cite{Casagrande:2008hr}.

Without the brane-localized Yukawa couplings, the profiles $\bm{C}_n^{Q,q}$ and $\bm{S}_n^{Q,q}$ form complete sets of even and 
odd functions on the orbifold, respectively, which can be chosen to obey separate orthonormality conditions with respect to the 
measure $d\phi\,e^\sigma$ \cite{Grossman:1999ra}. However, the $\delta$-distribution terms in the equations of motion are inconsistent 
with these orthonormality relations. We thus make the general ansatz
\beq\label{eq:orthonorm}
\begin{split}
   \int_{-\pi}^\pi\!d\phi\,e^{\sigma(\phi)}\,
   \bm{C}_m^{Q,q}(\phi)\,\bm{C}_n^{Q,q}(\phi)
   &= \delta_{mn}\,\bm{1} + \bm{\Delta C}_{mn}^{Q,q} \,, \\
   \int_{-\pi}^\pi\!d\phi\,e^{\sigma(\phi)}\,
   \bm{S}_m^{Q,q}(\phi)\,\bm{S}_n^{Q,q}(\phi) 
   &= \delta_{mn}\,\bm{1} + \bm{\Delta S}_{mn}^{Q,q} \,.
\end{split}
\eeq

We then find that the 4D action reduces to the desired canonical form
\beq
   S_{\rm ferm,2} = \sum_{q=u,d} \sum_n \int d^4x \left[ 
   \bar q^{(n)}(x)\,i\delslash\,q^{(n)}(x)
   - m_n\,\bar q^{(n)}(x)\,q^{(n)}(x) \right] ,
\eeq
if and only if, in addition to the BCs, the relation
\begin{equation} \label{eq:CS}
  \vec a_m^{Q,q\, \dagger} \, \big ( \delta_{mn} \bm 1 + \bm{\Delta
    C}_{mn}^{Q,q} \big ) \; \vec a_n^{Q,q} + \vec a_m^{\hspace{0.5mm}
    q,Q\, \dagger}\,\big (\delta_{mn} \bm 1 + \bm{\Delta
    S}_{mn}^{\hspace{0.25mm} q,Q} \big ) \; \vec a_n^{\hspace{0.5mm} q,Q}
  = \delta_{mn}
\end{equation}
is fulfilled.
It is also straightforward to show that the equations of motion imply 
\beq
   m_m\,\bm{\Delta C}_{mn}^{Q,q} - m_n\,\bm{\Delta S}_{mn}^{Q,q} 
   = \pm\frac{2}{r}\,\bm{C}_n^{Q,q}(\pi^-)\,
   \bm{S}_m^{Q,q}(\pi^-) \,.
\eeq
Since an overall normalization can always be reshuffled between the
profiles ${\bm{ C}}^{Q,q}_n (\phi)$ and ${\bm{S}}^{Q,q}_n (\phi)$ and
the eigenvectors $\vec a_n^{\hspace{0.5mm} Q,q}$, we can choose the sum 
$\bm{\Delta C}_{nn}^{Q,q} + \bm{\Delta S}_{nn}^{Q,q}$ freely, without
changing the physical result. The option $\bm{\Delta C}_{nn}^{Q,q} +
\bm{\Delta S}_{nn}^{Q,q}=0$ turns out to be particular convenient and
thus will be applied hereafter. With this choice (\ref{eq:CS}) splits
up into
\begin{equation} \label{eq:abrel} 
  \vec a_n^{Q\, \dagger}\,\vec a_n^Q + \vec a_n^{\hspace{0.5mm} q\,
    \dagger}\,\vec a_n^{\hspace{0.5mm} q} = 1 \,,
\end{equation} 
and 
\begin{equation} \label{eq:magicCS} 
  \vec a_m^{Q,q\, \dagger}\,\bm{\Delta C}_{mn}^{Q,q}\; \vec a_n^{Q,q} +
  \vec a_m^{\hspace{0.5mm} q,Q\, \dagger}\,\bm{\Delta
    S}_{mn}^{\hspace{0.25mm} q,Q}\; \vec a_n^{\hspace{0.5mm} q,Q} = 0\,.
\end{equation}
Thus, although the odd/even profiles of a certain KK-mode level $n$ (with a definite chirality) 
alone do not fulfill standard orthonormalization conditions, taken together, in combination with the corresponding
vectors $a_n^{Q,q}$, they are orthonormal in the sense of (\ref{eq:CS}).
The profiles alone do not obey such a relation due to the fact that (after EWSB) a certain 4D spinor receives contributions 
from doublets as well as singlets from the 5D theory. A part of the normalization is missing if one of these contributions 
is not present. This doublet-singlet mixing will become important when discussing FCNC couplings to the $Z$ boson in 
Section~\ref{sec:gaugecouplings} and thereafter.

Using the symmetry of the relations (\ref{eq:orthonorm}) in $m$ and $n$,
we finally obtain for $m \ne n$ the explicit expressions
\beq 
\begin{split}   
   \bm{\Delta C}_{mn}^{Q,q} 
   &= \pm\frac{2}{r}\,
    \frac{m_m\,\bm{C}_n^{Q,q}(\pi^-)\,\bm{S}_m^{Q,q}(\pi^-) 
          - m_n\,\bm{C}_m^{Q,q}(\pi^-)\,\bm{S}_n^{Q,q}(\pi^-)}%
         {m_m^2-m_n^2} \,, \\
   \bm{\Delta S}_{mn}^{Q,q} &= \mp\frac{2}{r}\,
    \frac{m_m\,\bm{C}_m^{Q,q}(\pi^-)\,\bm{S}_n^{Q,q}(\pi^-)
          - m_n\,\bm{C}_n^{Q,q}(\pi^-)\,\bm{S}_m^{Q,q}(\pi^-)}%
         {m_m^2-m_n^2} \,,
\end{split}
\eeq
whereas for $m = n$ we get
\beq
\label{eq:DCnn}
   \bm{\Delta C}_{nn}^{Q,q} = - \bm{\Delta S}_{nn}^{Q,q} 
   = \pm\frac{1}{r m_n}\,\bm{C}_n^{Q,q}(\pi^-)\,
   \bm{S}_n^{Q,q}(\pi^-) \,.
\eeq
Since they are proportional to $\bm{S}_n^{Q,q}(\pi^-)$ (which vanishes in the limit $v\rightarrow 0$), one would naively 
expect that the extra terms in the generalized orthonormality conditions (\ref{eq:orthonorm}) are small corrections of
order $v/\Mkk$. However, for the light SM fields, these terms are in fact of $\ord(1)$.

Finally, we want to determine the mass eigenvalues $m_n$ of the 4D Dirac fermions. They can be obtained by studying
the system of $2N$ linear equations for the components of the vectors $\vec a_n^{Q,q}$, following from 
(\ref{eq:bcIRrescaled}). It is straightforward to show that the mass eigenvalues $m_n$ correspond to the solutions of 
the equation
\beq \label{eq:fermeigenvals} 
  \det\left( \bm{1} + \frac{v^2}{2\Mkk^2} \, \bm{\tilde Y}_{\vec q}\,
    \bm{C}_n^{\hspace{0.25mm} q}(\pi^-) \left [ \bm{S}_n^{\hspace{0.25mm}
        q}(\pi^-) \right]^{-1} \bm{ \tilde Y}_{\vec
      q}^\dagger\,\bm{C}_n^Q(\pi^-) \left [ \bm{S}_n^Q(\pi^-) \right]^{-1}
  \right) = 0 \,.
\eeq
Once these are known, the eigenvectors $\vec a_n^{Q,q}$ can be derived from (\ref{eq:bcIRrescaled}). Note that, 
while it is always possible to work with real and diagonal profile matrices $\bm{C}_n^{Q,q}(\phi)$ and $\bm{S}_n^{Q,q}
(\phi)$, the vectors $\vec a_n^{Q,q}$ are, in general, complex-valued.

\subsubsection{Bulk Profiles}
\label{sec:minfermionprofiles}

In this section we derive the explicit form of the profiles $(C_n^{Q,q})_i$ and $(S_n^{Q,q})_i$, associated with bulk 
mass parameters $M_{Q_i,q_i}$ (with $q=u,d$), in the presence of EWSB. The structure of the solutions is similar to the case 
of unbroken electroweak symmetry, as the EOMs within the bulk are the same for both cases. Combining the first and the third equation
in (\ref{eq:fermEOM}), as well as the second and fourth one, we arrive again at Bessel equations. In terms of $t=\epsilon
\,e^\sigma$, these have the general solutions \cite{Grossman:1999ra,Gherghetta:2000qt}\footnote{Since, up to the BCs which in the
end will determine the mass eigenvalues $m_n$ and the flavor mixing vectors in (\ref{eq:KKdecferm}), the problem 
decomposes into independent and similar equations for the different flavors, we will drop the index $i$ in the following.}
\beq\label{eq:fermprofiles}
\begin{split}
   C_n^{Q,q}(t) 
   &= {\cal N}_n(c_{Q,q})\,\sqrt{\frac{L\epsilon t}{\pi}}\,
    f_n^+(t,c_{Q,q}) \,, \\
   S_n^{Q,q}(t) 
   &= \pm {\cal N}_n(c_{Q,q})\,\sgn(\phi)\,
    \sqrt{\frac{L\epsilon t}{\pi}}\,f_n^-(t,c_{Q,q}) \,,
\end{split}
\eeq
where $c_{Q,q}\equiv\pm M_{Q,q}/k$ are dimensionless parameters, derived from the bulk mass terms, and
\beq\label{fplmi}
   f_n^\pm(t,c) 
   = J_{-\frac12-c}(x_n\epsilon)\,J_{\mp\frac12+c}(x_n t) 
   \pm J_{\frac12+c}(x_n\epsilon)\,J_{\pm\frac12-c}(x_n t) \,.
\eeq
Remember that, as before, $x_n=m_n/\Mkk$. The orthonormality relations (\ref{eq:orthonorm}) lead to the normalization conditions
\beq
   2\int_\epsilon^1\!dt\,t \left[ f_n^\pm(t,c) \right]^2
   = \frac{1}{{\cal N}_n^2(c)} 
   \pm \frac{f_n^+(1,c)\,f_n^-(1,c)}{x_n} \,.
\eeq
From these we derive
\beq\label{eq:fermnorm}
   {\cal N}_n^{-2}(c) 
   = \left[ f_n^+(1,c) \right]^2 + \left[ f_n^-(1,c) \right]^2 
    - \frac{2c}{x_n}\,f_n^+(1,c)\,f_n^-(1,c) 
    - \epsilon^2 \left[ f_n^+(\epsilon,c) \right]^2 \,.
\eeq
For the special cases where $c+1/2$ is an integer, the profiles have to be obtained from the above 
expressions by a limiting procedure. 

Since external fermions in low energy processes will always correspond to zero modes of the RS setup ($n=1,2,3$), it will be 
useful to derive simple analytic expressions for the corresponding profiles. For those modes, it is a very good 
approximation to expand the above results in the limit $x_n\ll 1$, as even the top-quark is much lighter than the 
KK scale. In this way we find, dropping an phenomenological irrelevant term \cite{Casagrande:2008hr},
\beq\label{eq:SMfermions}
\begin{split}
   C_n^{Q,q}(\phi) 
   &\approx \sqrt{\frac{L\epsilon}{\pi}}\,
    F(c_{Q,q})\, t^{c_{Q,q}}, \\
   S_n^{Q,q}(\phi) 
   &\approx \pm\sgn(\phi)\,\sqrt{\frac{L\epsilon}{\pi}}\,
    x_n F(c_{Q,q})\,
    \frac{t^{1+c_{Q,q}} - \epsilon^{1+2c_{Q,q}}\,t^{-c_{Q,q}}}%
         {1+2c_{Q,q}} \,.
\end{split}
\eeq
Here, we have introduced the ``zero-mode profile'' \cite{Grossman:1999ra,Gherghetta:2000qt}
\beq\label{eq:Fdef}
   F(c) \equiv \sgn[\cos(\pi c)]\,
   \sqrt{\frac{1+2c}{1-\epsilon^{1+2c}}} \,,
\eeq
which corresponds to the approximate value of the profile on the IR brane, divided by $\sqrt{L\epsilon/\pi}$.
Note that the sign factor in (\ref{eq:Fdef}) is chosen such that the signs in (\ref{eq:SMfermions}) 
agree with those derived from the exact profiles (\ref{eq:fermprofiles}). 
The zeroth order in the expansion for the zero mode profiles performed above is often 
called (zeroth order) ``zero-mode approximation'' (ZMA). At this order, the even profile matrices do not depend on the level 
$n$ and the odd profiles vanish. These results correspond to the starting point in the perturbative approach, in which
one first solves for the fermion bulk profiles without the Yukawa couplings and then treats these 
couplings as a perturbation \cite{Grossman:1999ra,Gherghetta:2000qt,Huber:2000ie}.  
As we will see later in Section~\ref{sec:gaugecouplings}, it will sometimes be necessary to consider higher orders in the
expansion to derive consistent non-trivial results in the ZMA. This can be done comfortably in our approach, as we just have to include 
the next order in the well-defined expansion of the exact profiles in powers of $v^2/\Mkk^2$ (note that $x_n\,,\,n=1,2,3$  
scales like $v/\Mkk$), and do not have to diagonalize (and truncate) infinite dimensional mass matrices.

The quantity $F(c)$ strongly depends on the value of $c$. 
One obtains to excellent approximation 
\begin{equation}
  F(c) \approx \begin{cases} -\sqrt{-1-2c}\,\, \epsilon^{-c-\frac12} \,,
    & -3/2<c<-1/2 \,, \\[4mm] \sqrt{1+2c} \,, & -1/2<c<1/2 \,. \end{cases}
\end{equation}
This behavior signals that for $c<-1/2$ the zero-modes are localized close to the UV brane, while for $c>-1/2$ 
they have the biggest weight close to the IR brane. Thus, it turns out that rather than generating 4D masses, 
the bulk mass parameters control the localization of the 4D fermions in the extra dimension. The fact that for UV-localized 
fermions the zero-mode profile is exponentially small, while it is of ${\cal O}(1)$ for IR-localized fields can be used to 
generate large hierarchies in the fermion spectrum by means of different overlaps with the IR-brane Higgs sector via 
$\ord(1)$ input parameters. This feature will be discussed in more detail in the following section.

\subsection[Fermion Hierarchies: Anarchic RS Model as a Predictive Model of Flavor]{Fermion Hierarchies: The Anarchic RS Model as a Predictive Model of Flavor}
\label{sec:hierarchies}

Due to the possibility of additional gauge invariant bulk mass terms, with hermitian mass matrices $\bm{M}_{Q,u,d}$, the minimal RS model has 
27 new parameters in the flavor sector compared to the SM (for $N=3$ generations). In the light of this significant number of 
new parameters, entering the model as exponents of the warp factor, see (\ref{eq:Fdef}), one should address the question of the predictivity 
of the model. If one allows \eg for arbitrary values for the bulk mass parameters $c_{Q_i,u_i,d_i}$ (addressing the fermion spectrum by hierarchical 
Yukawa matrices), a very broad range of effects is possible within the RS setup. This renders the 
model not very predictive at the first place. However, as mentioned before, a big virtue of the model is the 
possibility to address the hierarchies within the flavor sector, which cannot be understood within the SM. This approach will constrain the 
large parameter space and allow for more generic predictions. So we will understand the RS model in this way, \ie, we do not build in hierarchies 
by hand but rather want to generate the SM flavor structure out of $\ord(1)$ input parameters. To this end, 
we assume anarchical Yukawa matrices and generate the hierarchies via different values for the dimensionless bulk mass parameters. 
These input parameters are then constrained by the requirement to end up with the correct spectrum and CKM mixing angles. 
Due to the warping, as explained above, they can be of $\ord(1)$, while still leading to hierarchies in the fermion spectrum. 
These naturalness considerations do not fix the input parameters completely, however, they significantly decrease 
the spread in the parameter space and thus also in the predictions of the model. While the precise values of the input 
parameters in the flavor sector will not be determined,
their hierarchical structure will. This often allows for generic estimations for observables in the RS model, without an uncertainty in the prediction
over many orders of magnitude (assuming $\Mkk$ to be fixed). Such an uncertainty would be generated when reshuffling too large contributions between the 
localization parameters and the Yukawa matrices, leaving the spectrum invariant. This is prevented by the anarchic approach
(however, some amount of reshuffling is possible, see below). 

To get 
an analytic handle on the relations between input parameters and the resulting spectrum and mixings, we will demonstrate and explore 
similarities between the RS setup of generating hierarchies by different fermion localizations in a slice of AdS$_5$ and the Froggatt-Nielsen mechanism. To this end, it is useful to perform an expansion to the first non-vanishing order in $v^2/\Mkk^2$ 
and to use the approximate formulae for the bulk profiles of the SM fermions, obtained in (\ref{eq:SMfermions}). For the purpose of 
discussing the fermion masses and mixings, we need these profiles evaluated at $t=1$, where they couple to the Higgs sector. We arrive 
at the simple expressions
\beq
   C_n^{Q,q}(1^-)\to \sqrt{\frac{L\epsilon}{\pi}}\,F(c_{Q,q}) \,,
    \qquad
   S_n^{Q,q}(1^-)\to \pm\sqrt{\frac{L\epsilon}{\pi}}\,
    \frac{x_n}{F(c_{Q,q})} \,.
\eeq
With their help, the BCs (\ref{eq:bcIRrescaled}) can be recast to LO in $v^2/\Mkk^2$ into the simple form
\beq\label{eq:BCZMA}
   \frac{\sqrt2\,m_n}{v}\,\hat a_n^{q} 
   = \bm{Y}_q^{\rm eff}\,\hat a_n^{q^c} \,, \qquad 
   \frac{\sqrt2\,m_n}{v}\,\hat a_n^{q^c} 
   = (\bm{Y}_q^{\rm eff})^\dagger\,\hat a_n^{q} \,.
\eeq
Here
\beq\label{eq:Yueff}
   \left( Y_q^{\rm eff} \right)_{ij} 
   \equiv F(c_{Q_i}) \left(Y_{\vec q} \right)_{ij} F(c_{q_j})
\eeq
are effective Yukawa matrices, and the rescaled flavor vectors $\hat a_n^{q}\equiv\sqrt{2}\,\vec a_n^{\hspace{0.25mm}Q}\,,\,
\hat a_n^{q^c}\equiv\sqrt{2}\,\vec a_n^{\hspace{0.25mm}q}$ obey the normalization conditions
\beq\label{eq:ahatnorm}
   \hat a_n^{q\dagger}\,\hat a_n^{q} 
   = \hat a_n^{q^c\dagger}\,\hat a_n^{q^c} = 1 \,.
\eeq
Moreover, we obtain from (\ref{eq:BCZMA}) the equalities
\beq\label{eq:mnZMA}
   \left( m_n^2\,\bm{1} - \frac{v^2}{2}\,
    \bm{Y}_q^{\rm eff}\,(\bm{Y}_q^{\rm eff})^\dagger \right) 
    \hat a_n^q = 0 \,, \qquad 
   \left( m_n^2\,\bm{1} - \frac{v^2}{2}\,
    (\bm{Y}_q^{\rm eff})^\dagger\,\bm{Y}_q^{\rm eff} \right) 
    \hat a_n^{q^c} = 0 \,,
\eeq
and thus the mass eigenvalues can be obtained from the simple equation 
\beq\label{eq:minm}
   \det\left( m_n^2\,\bm{1} - \frac{v^2}{2}\, 
   \bm{Y}_q^{\rm eff}\,(\bm{Y}_q^{\rm eff})^\dagger \right) = 0 \,.
\eeq
These expressions hold to LO in $v^2/\Mkk^2$.  
At this order, but not in general, the vectors $\vec a_n^{Q,q}$ that belong to different $n$ are orthogonal 
on each other. 

The eigenvectors $\hat{a}_n^{q}$ and $\hat{a}_n^{q^c}$ of the Yukawa matrices ${\bm Y}_q^{\rm eff} 
\left( {\bm Y}_q^{\rm eff} \right)^\dagger$ and $\left( {\bm Y}_q^{\rm eff} \right)^\dagger 
{\bm Y}_q^{\rm eff}$ (with $n=1,2,3$ and $q=u,d$) form the columns of the unitary matrices ${\bm U}_q$ and ${\bm W}_q$ that
appear in the singular-value decomposition
\beq\label{eq:singular}
   \bm{Y}_q^{\rm eff} 
   = \bm{U}_q\,\bm{\lambda}_q\,\bm{W}_q^\dagger \,,
\eeq
respectively, where
\beq\label{eq:lambdaud}
   \bm{\lambda}_u = \frac{\sqrt{2}}{v}\,
    \mbox{diag}(m_u,m_c,m_t) \,, \qquad 
   \bm{\lambda}_d = \frac{\sqrt{2}}{v}\,
    \mbox{diag}(m_d,m_s,m_b) \,.
\eeq
To LO in $v^2/\Mkk^2$ the relations between the fundamental 5D fields and the SM mass eigenstates thus involve the matrices 
$\bm{U}_q$ and $\bm{W}_q$. In particular, the CKM matrix in the RS model is to LO given by
\beq\label{eq:VCKM}
   \bm{V}_{\rm \hspace{-1mm} CKM} = \bm{U}_u^\dagger\,\bm{U}_d \,.
\eeq
Before applying the Froggatt-Nielsen analysis to the RS setup 
we will count the number of physical parameters in the flavor sector.

\subsubsection{Parameter Counting in the RS Model}

Applying the method reviewed in Section~\ref{sec:SMflavor}, the number of physical parameters in the flavor sector of the RS model is derived as follows.
To be completely general, we assume $N$ quark generations and thus start with $N_Y=2\,(N^2,N^2)$ real moduli and CP-odd phases for the 5D Yukawa matrices $\bm{Y}_{u,d}^{({\rm 5D})}$, as well as $N_c=3\,(N(N+1)/2,N(N-1)/2)$ parameters for the hermitian bulk mass matrices $\bm{c}_{Q,u,d}$.
Thus 
\beq
N_{\rm general}=N_Y+N_c=(N(7N+3)/2,N(7N-3)/2)\,.
\eeq 
The Yukawa matrices break the global bulk flavor symmetry $G=U(N)_Q\times U(N)_u\times U(N)_d$ with $N_G=3\,(N(N-1)/2,N(N+1)/2)$ parameters down to the 
subgroup $H=U(1)_B$ with $N_H=(0,1)$ parameters, resulting in 
\beq
N_{\rm broken}=N_G-N_H=(3N(N-1)/2,3N(N+1)/2-1)\,.
\eeq 
One thus arrives at 
\beq
N_{\rm phys}=N_{\rm general}-N_{\rm broken}=(N(2N+3),(N-1)(2N-1)) 
\eeq
physical parameters. 
For $N=3$ quark generations we end up with 27 moduli and ten phases \cite{Agashe:2004cp}. In the ZMA we can identify the physical real parameters with the 
six quark masses, twelve mixing angles appearing in the mixing matrices ${\bm U}_{u,d}$ and ${\bm W}_{u,d}$, and the nine zero-mode profiles ${\bm F}_{Q,u,d}$. One of the ten phases can be identified with the phase of the CKM matrix. Moreover, nine new phases enter, in different combinations, 
the various matrices in (\ref{eq:overlapints1}) and (\ref{defdelta}), that describe the flavor-changing interactions of the RS model. 

Interestingly, for $N=2$ quark generations there still remain 14 moduli and three phases. Since, as we have seen in Section~\ref{sec:SMflavor},
the CKM matrix for two-generations can be made real by phase redefinitions, the phases can all be chosen to reside in the new mixing matrices. 
As a consequence, the RS model allows for CP-violating effects which do not involve all three fermion generations. 
Thus, CP violation in the RS model is much less suppressed than in the SM, see also Section~\ref{sec:asl}. 
It would be interesting to work out the consequences of this fact for baryogenesis. Beyond that, note that the new 
CP-violating phases induce electric dipole moments for the electron and the neutron at the one-loop level. The stringent 
experimental bound on the neutron electric dipole moment leads
to a lower bound on the masses of the KK excitations of $\sim 10$\,TeV \cite{Agashe:2004cp}. There are several proposals to address this 
``CP problem'' (and to allow for lighter KK excitations) \cite{Santiago:2008vq,Fitzpatrick:2007sa,Cheung:2007bu}, which postpone the 
contributions to the neutron electric dipole moment to the two-loop level by reducing the number of CP-odd phases.
Neglecting the lepton sector (as well as gravity), the minimal RS model thus has, besides the ``SM'' parameters described in Section~\ref{sec:FL}, 
as additional parameters the KK scale $\Mkk$, the RS volume $L$, and the 27 new parameters in the flavor sector discussed above. 

\subsubsection{Warped-Space Froggatt-Nielsen Mechanism}			
\label{sec:quark}

\begin{figure}[!t]
\begin{center}
\mbox{\includegraphics[width=12cm]{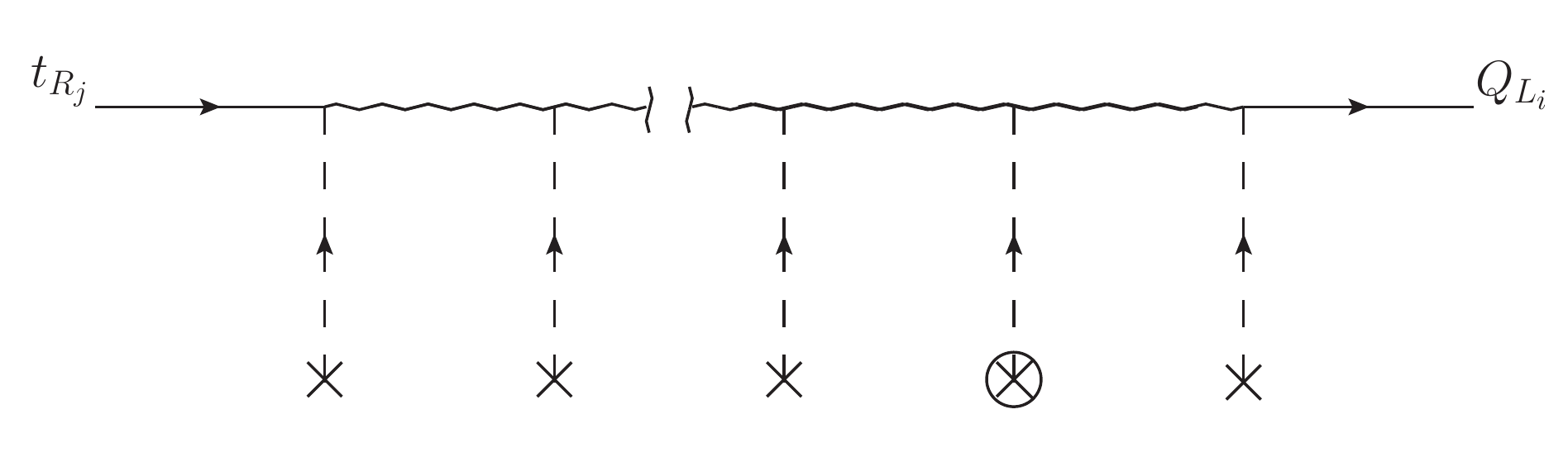}}
\end{center}
\vspace{-0.5cm}
\begin{center}
  \parbox{15.5cm}{\caption{\label{fig:FN} Contribution to the element $(m_u)_{ij}$ of the up-type quark mass matrix. 
  Full lines correspond to SM quarks, whereas wavy lines represent ultra-heavy fermions. 
  Insertions of the VEV $\langle\Phi_1\rangle$ are depicted by crosses, while insertions of $\langle\Phi\rangle$ correspond to crossed 
  circles.}}
\end{center}
\vspace{-0.6cm}
\end{figure}

We will now introduce the Froggatt-Nielsen mechanism and apply the corresponding analysis in the context of warped extra dimensions to show how 
fermion hierarchies arise naturally in these models.
Starting from an anarchical Yukawa matrix, the Froggatt-Nielsen mechanism generates the observed hierarchies in the quark sector
by assuming the left-handed and right-handed components of the SM-quark fields to have different values for
an almost conserved quantum number $R$, belonging to an abelian symmetry group $U(1)_F$. 
Quark-mass terms only arise via interactions with new ultra-heavy fermions and Higgs-scalars that have various different values of $R$, which is integrally quantized.
Explicitly, the model introduces three Higgs fields $\Phi,\Phi_1,\Phi_2$, where the VEV $\langle\Phi_2\rangle>0$ of the neutral 
($R=0$) Higgs scalar $\Phi_2$ generates mass terms for the new fermions. The Higgs scalar $\Phi_1$ is a gauge singlet, but has $R=1$, and thus can mediate between fermions with a different $R$-charge.
Its VEV $\langle\Phi_1\rangle>0$ breaks the $U(1)_F$ symmetry spontaneously and allows for mass terms of the light (SM) quarks via the mechanism
depicted in Figure \ref{fig:FN}.
Depending on the difference in $R$-quantum numbers between the left-handed and right-handed SM quark-fields, a certain number of 
insertions of transitions mediated by the the VEV $\langle\Phi_1\rangle$ is needed to account for this difference
via intermediate ultra-heavy fermions, in order to generate a mass term for the SM quarks. Moreover, an insertion of the VEV of 
the SM Higgs $\langle\Phi\rangle=v/\sqrt 2$ is necessary to balance the different $SU(2)_L$ quantum numbers of left-handed and right-handed
SM quarks. Every $\Delta |R|=1$ transition together with the adjacent propagating ultra-heavy fermion leads to a suppression by the small symmetry breaking parameter 
\beq
   \epsilon\equiv\frac{y(\mu_q)}{y(\mu_0)}\frac{\langle\Phi_1\rangle}{\langle\Phi_2\rangle}\,.
\eeq
The prefactor corresponds to the ratio of generic $\Phi_1$-Yukawa couplings at the scale of the light fermions $\mu_q$ and the fundamental scale 
$\mu_0$ of the model, where the Yukawa couplings are expected to be of $\ord(1)$. Concentrating on the up-type sector,
denoting the quantum numbers of the left-handed SM-doublets and right-handed SM-singlets by $R_{Q_i}=a_i$ and $R_{t_i}=b_i$, and assuming $a_i>0$ and $b_i\leq0$,
the mass matrix arising from diagrams like the one in Figure \ref{fig:FN} reads
\beq
\label{eq:FNmm}
(m_u)_{ij}=\frac{v}{\sqrt2}  \left(G_u\right)_{ij} \epsilon^{a_i-b_j}\,.
\eeq 
Note that this matrix is order-of-magnitude factorizing. In the spirit of the model, all dimensionless couplings should be of $\ord(1)$, which results in the complex
coefficients $\left(G_u\right)_{ij}$ also being of $\ord(1)$ times a factor due to evolving the $\Phi$ Yukawa couplings down to the scale $\langle\Phi\rangle$.
One can now consistently evaluate the mass eigenvalues as well as the CKM matrix to LO in the small parameter $\epsilon$.
First, it can be shown that the product of the $n$ largest eigenvalues of $\bm{m}_u$ reads to LO \cite{Froggatt:1978nt}
\beq
\label{eq:FNprod}
\prod_{i=1}^n m_i= \left(\frac{v}{\sqrt2}\right)^n\,\det(\bm{G}^{(n)}_u)\ \epsilon^{\sum_{i=1}^n (a_i-b_i)}\,,
\eeq
where $\bm{G}^{(n)}_u$ is obtained from $\bm{G}_u\equiv ({(G_u)}_{ij})$ by removing rows and columns with $i>n$ and $j>n$, respectively.
Here, the quarks are ordered such that $a_i>a_{i-1}$ and $b_i<b_{i-1}$. In the case of degenerate values  $a_i=a_{i-1}$ or $b_i=b_{i-1}$ 
the derivations are only order-of-magnitude wise correct.
From (\ref{eq:FNprod}) one obtains
\beq
\label{eq:FNm}
m_n=\frac{v}{\sqrt{2}}\, \frac{\left|\det \bm{G}^{(n)}_u\right|}{\left|\det \bm{G}^{(n-1)}_u\right|}\, \epsilon^{a_n-b_n} 
\eeq
for the $n^{\rm th}$ heaviest quark which leads to the hierarchical structure
\beq
\frac{m_i}{m_j}\sim \epsilon^{a_i-a_j-b_i+b_j}\,.
\eeq
Due to the exponential dependence, this allows to reproduce the hierarchies (\ref{eq:quarkhier}) with a moderately small symmetry breaking parameter $\epsilon$, meaning that
$\langle\Phi_1\rangle$ and $\langle\Phi_2\rangle$ do not have to differ by orders of magnitude, and moderate different values for the 
various quantum numbers $R_{Q_i}$ and $R_{t_i}$. In consequence, hierarchical and suppressed masses are a natural prediction of the model.

A nice feature of the Froggatt-Nielsen mechanism is that, given that the mass hierarchies are generated as described above with 
different quantum numbers for the various quark fields in the right-handed as well as in the left-handed sector, the CKM matrix is predicted
to have a structure similar to (\ref{eq:CKMWmag}).
This can be seen from the form of the diagonalization matrices $\bm{U}_u$ and $\bm{U}_d$, obtained from the singular value decomposition of
the matrix $\bm{m}_u$, see (\ref{eq:SVDSM}), and the corresponding equation for the down-type sector. To LO in $\epsilon$ and for the
case of three generations, these matrices 
read \cite{Froggatt:1978nt}
\beq
\bm{U}_q=(u_q)_{ij}\,\epsilon^{|a_i-a_j|}\,,\ q=u,d\,,
\eeq
where
\beq
   \bm{u}_q = \begin{pmatrix}
    1 & \frac{\displaystyle \left( M_q \right)_{21}}%
             {\displaystyle \left( M_q \right)_{11}} &
    \frac{\displaystyle \left( G_q \right)_{13}}%
         {\displaystyle \left( G_q \right)_{33}} \\ \\[-2mm]
    - \frac{\displaystyle \left( M_q \right)_{21}^\ast}
          {\displaystyle \left( M_q \right)_{11}^\ast} & 1 &
    \frac{\displaystyle \left( G_q \right)_{23}}%
         {\displaystyle \left( G_q \right)_{33}} \\ \\[-2mm]
    \frac{\displaystyle \left( M_q \right)_{31}^\ast}%
         {\displaystyle \left( M_q \right)_{11}^\ast} & 
    -\frac{\displaystyle \left( G_q \right)_{23}^\ast}%
          {\displaystyle \left( G_q \right)_{33}^\ast} & 1
    \end{pmatrix}\,.
\eeq
Here, $(M_q)_{ij}$ denote the minors of $\bm{G}_q$, \ie, the determinants of the matrices obtained by removing the $i^{\rm th}$
row and the $j^{\rm th}$ column of $\bm{G}_q$. Moreover, we have used the invariance of the singular-value decomposition with
respect to phase rotations, to make the diagonal elements $\left(U_q\right)_{ii}$ real.
One obtains, as a prediction, the hierarchical structure of the CKM matrix
\beq
\left(V_{\rm CKM}\right)_{ik}=\sum_{j=1}^3\left(U_u^\dagger\right)_{ij} \left(U_d\right)_{jk} \sim \epsilon^{|a_i-a_k|}\,.
\eeq
This is very similar to (\ref{eq:CKMWmag}) (for appropriate values of $a_i$ and $\epsilon$), without having to put in the structure by hand.
The diagonal dominant form of $\bm{V}_{\rm \hspace{-1mm} CKM}$ is a robust prediction of the model.
In the Froggatt-Nielsen model, both hierarchies, those within the masses as well as those of the CKM matrix, can be traced back 
to the same origin. Note, however that in this approach, the CP violating phase is not suppressed by any small parameter
and thus it is difficult to explain why the measured CP violation is so small.

\begin{table}[!t]
\begin{center}
\begin{tabular}{|c|c|c|}
\hline
     Froggatt-Nielsen Mechanism & Warped Extra Dimensions  \\
  \hline
     symmetry breaking parameter $\epsilon\sim\frac{\langle\Phi_1\rangle}{\langle\Phi_2\rangle}$
     & warp factor $\epsilon=e^{-L}$   \\
     quantum numbers $(a_i,b_j)$ & localization parameters $(-c_{Q_i}-\frac12,c_{q_j}+\frac12)$  \\
     combination of couplings $\bm{G}_q$ & Yukawa matrix $\bm{\tilde Y}_{\vec q}$\\
  \hline
     $\epsilon^{a_i}$ & $F(c_{Q_i})$\\
     $\epsilon^{b_j}$ & $F(c_{q_j})^{-1}$\\
  \hline
\end{tabular} 
\end{center}
\vspace{-0.45cm}
\parbox{15.5cm}{\caption{\label{tab:AnalFN}
Analogy between parameters of the Froggatt-Nielsen mechanism and those of the RS model. The second row holds just for $-3/2<c_{Q_i,q_j}<-1/2$, whereas the range of validity of the last two rows is not restricted.}}
\vspace{-0.6cm}
\end{table}

We will now apply the results presented here to the fermion sector of the Randall-Sundrum model.
As we will see immediately, the RS model with fermions in the bulk automatically provides a complete analogy to the Froggatt-Nielsen mechanism, without 
the need to introduce a new $U(1)_F$-symmetry, new fermion representations, or additional Higgs fields. In that context, we will also determine 
the hierarchies for the RS analogon to the $U(1)_F$ charges, in dependence on the observed quark 
masses and CKM mixings
Due to the form of the effective Yukawa matrices in (\ref{eq:Yueff}), a {\it hierarchical} and order-of-magnitude {\it factorizing} mass matrix 
of exactly the same structure as in (\ref{eq:FNmm}) is generated in the RS setup by assuming a hierarchical structure of the zero-mode profiles
\beq\label{eq:hierarchy}
   |F(c_{A_1})| < |F(c_{A_2})| < |F(c_{A_3})| \,,\quad A=Q,u,d\,.
\eeq
Importantly, in warped extra dimensions, such a hierarchy is induced naturally, since it only requires small differences 
in the bulk mass parameters $c_{A_i}$. Note that we assume the quarks to be ordered such that the relation (\ref{eq:hierarchy}) 
holds. In the case of degenerate profiles $|F(c_{A_i})|=|F(c_{A_{i+1}})|$, the following discussion is again only order-of-magnitude wise correct.
The localization parameters $c_{Q_i,q_i}$ of the minimal RS model play the role of the left-handed and right-handed
$U(1)_F$ charges $a_i,b_i$ of the Froggatt-Nielsen mechanism, however they are not integrally quantized. Explicitly, the product 
of functions $F(c_{Q_i})F(c_{q_j})$ in (\ref{eq:Yueff}) corresponds to the exponential factor $\epsilon^{a_i-b_j}$ of the symmetry 
breaking parameter $\epsilon \sim \langle\Phi_1\rangle$ in (\ref{eq:FNmm}). This correspondence can be split up as in Table \ref{tab:AnalFN}, 
which gives an overview of the analogies of the models. Note that the direct relation between the localization parameters and the 
Froggatt-Nielsen charges, given in the second row, only holds for $-3/2<c_{Q_i,q_j}<-1/2$. However, the general analogy, as well
as the last three exact relations, hold independently of this restriction.

Thus, the hierarchies of fermion masses and mixings in a warped background result without further assumptions from the 
Froggatt-Nielsen mechanism. In the following we perform the corresponding analysis explicitly,
starting from the relation (\ref{eq:Yueff}). 
To LO, the products of up- and down-type quark masses in the RS model are given by, see (\ref{eq:FNprod}),
\beq\label{eq:detdu}
\begin{split}
   m_u\,m_c\,m_t 
   &= \frac{v^3}{2\sqrt2} \left| \det\left( \bm{Y}_u \right ) \right|
    \prod_{i=1,2,3} \left| F(c_{Q_i})\,F(c_{u_i}) \right| , \\
   m_d\,m_s\,m_b 
   &= \frac{v^3}{2\sqrt2} \left| \det\left( \bm{Y}_d \right ) \right|
    \prod_{i=1,2,3} \left| F(c_{Q_i})\,F(c_{d_i}) \right|\,.
\end{split}
\eeq
Using $|F(c_{A_i})|<|F(c_{A_{i+1}})|$, one can evaluate
all the mass eigenvalues to LO in hierarchies. We find, see (\ref{eq:FNm}),
\beq\label{eq:quarkmasses} 
\begin{aligned}
   m_u &= \frac{v}{\sqrt2}\,
    \frac{|\det(\bm{Y}_u)|}{|(M_u)_{11}|}\,
    |F(c_{Q_1}) F(c_{u_1})| \,, & \qquad
   m_d &= \frac{v}{\sqrt2}\,
    \frac{|\det(\bm{Y}_d)|}{|(M_d)_{11}|}\,
    |F(c_{Q_1}) F(c_{d_1})| \,, \\
   m_c & = \frac{v}{\sqrt2}\,
    \frac{|(M_u)_{11}|}{|(Y_u)_{33}|}\,
    |F(c_{Q_2}) F(c_{u_2})| \,, & \qquad
   m_s & = \frac{v}{\sqrt2}\,
    \frac{|(M_d)_{11}|}{|(Y_d)_{33}|}\,
    |F(c_{Q_2}) F(c_{d_2})| \,, \\
   m_t & = \frac{v}{\sqrt2}\,|(Y_u)_{33}|\,
    |F(c_{Q_3}) F(c_{u_3})| \,, & \qquad 
   m_b & = \frac{v}{\sqrt2}\,|(Y_d)_{33}|\,
    |F(c_{Q_3}) F(c_{d_3})| \,,
\end{aligned}
\eeq
where $(M_q)_{ij}$ denote the minors of $\bm{Y}_q$, as defined before.

The elements of the rotation matrices $\bm{U}_q$ and $\bm{W}_q$ are given, to
LO in hierarchies, by
\beq\label{eq:UWq}
   (U_q)_{ij} = (u_q)_{ij}
   \begin{cases} 
    \frac{\displaystyle F(c_{Q_i})}{\displaystyle F(c_{Q_j})} \,,
    & i\le j \,, \\[5mm]
    \frac{\displaystyle F(c_{Q_j})}{\displaystyle F(c_{Q_i})} \,, 
    & i>j \,, 
   \end{cases} \qquad 
   (W_q)_{ij} = (w_q)_{ij}\;e^{i\phi_j}
   \begin{cases} 
    \frac{\displaystyle F(c_{q_i})}{\displaystyle F(c_{q_j})} \,, 
    & i\le j \,, \\[5mm]
    \frac{\displaystyle F(c_{q_j})}{\displaystyle F(c_{q_i})} \,, 
    & i>j \,,
   \end{cases}
\eeq
where the coefficient matrices $\bm{u}_q$ and $\bm{w}_q$ can be expressed through the elements $(Y_q)_{ij}$ of the original 
Yukawa matrices and their minors $(M_q)_{ij}$ and read
\beq\label{eq:uwq} 
   \bm{u}_q = \begin{pmatrix}
    1 & \frac{\displaystyle \left( M_q \right)_{21}}%
             {\displaystyle \left( M_q \right)_{11}} &
    \frac{\displaystyle \left( Y_q \right)_{13}}%
         {\displaystyle \left( Y_q \right)_{33}} \\ \\[-2mm]
    - \frac{\displaystyle \left( M_q \right)_{21}^\ast}
          {\displaystyle \left( M_q \right)_{11}^\ast} & 1 &
    \frac{\displaystyle \left( Y_q \right)_{23}}%
         {\displaystyle \left( Y_q \right)_{33}} \\ \\[-2mm]
    \frac{\displaystyle \left( M_q \right)_{31}^\ast}%
         {\displaystyle \left( M_q \right)_{11}^\ast} & 
    -\frac{\displaystyle \left( Y_q \right)_{23}^\ast}%
          {\displaystyle \left( Y_q \right)_{33}^\ast} & 1
    \end{pmatrix} , \qquad
   \bm{w}_q = \begin{pmatrix}
    1 & \frac{\displaystyle \left( M_q \right)_{12}^\ast}%
             {\displaystyle \left( M_q \right)_{11}^\ast} &
    \frac{\displaystyle \left( Y_q \right)_{31}^*}%
         {\displaystyle \left( Y_q \right)_{33}^*} \\ \\[-2mm]
    - \frac{\displaystyle \left( M_q \right)_{12}}%
           {\displaystyle \left( M_q \right)_{11}} & 1 & 
    \frac{\displaystyle \left( Y_q \right)_{32}^\ast}%
         {\displaystyle \left( Y_q \right)_{33}^\ast} \\ \\[-2mm]
    \frac{\displaystyle \left( M_q \right)_{13}}%
         {\displaystyle \left( M_q \right)_{11}} & 
    - \frac{\displaystyle \left( Y_q \right)_{32}}%
          {\displaystyle \left( Y_q \right)_{33}} & 1
    \end{pmatrix} .
\eeq
Invariance of the singular-value decomposition (\ref{eq:singular}) under field redefinitions allows to make 
either the diagonal elements $(U_q)_{ii}$ or $(W_q)_{ii}$ real. In (\ref{eq:UWq}) we have used that freedom to choose $(U_q)_{ii}$ to be real, 
so that all phase factors $e^{i\phi_j}$ appear in the elements $(W_q)_{ij}$. 
They are given by
\beq\label{eq:expphij}
   e^{i\phi_j} = \sgn\big[ F(c_{Q_j})\,F(c_{q_j})\big]\, 
   e^{-i(\rho_j-\rho_{j+1})} \,,
\eeq
where 
\beq
   \rho_1 = \arg\left( \det(\bm{Y}_q) \right) , \qquad 
   \rho_2 = \arg\left( (M_q)_{11} \right) , \qquad 
   \rho_3 = \arg\left( (Y_q)_{33} \right) , 
\eeq
and $\rho_4=0$. 
We observe that, to LO, the matrices $\bm{U}_q$ and thus also the CKM matrix do not depend on the right-handed 
profiles $F(c_{q_i})$. This has already been pointed out in \cite{Huber:2003tu}. 

It is now straightforward to derive the LO expressions for the Wolfenstein parameters of the CKM matrix 
$\lambda$, $A$, $\bar\rho$, and $\bar\eta$, defined in (\ref{eq:lAre}). From (\ref{eq:VCKM}), (\ref{eq:UWq}), and (\ref{eq:uwq}), we obtain
\beq\label{eq:wolfenstein}
\begin{split}
   \lambda = \frac{|F(c_{Q_1})|}{|F(c_{Q_2})|} 
    \left| \frac{\left( M_d \right)_{21}}{\left( M_d \right)_{11}}
    - \frac{\left( M_u \right)_{21}}{\left( M_u \right)_{11}} \right|
    \,, \qquad 
   A = \frac{|F(c_{Q_2})|^3}{|F(c_{Q_1})|^2\,|F(c_{Q_3})|}
    \left| \frac{%
     \frac{\displaystyle \left( Y_d \right)_{23}}%
          {\displaystyle \left( Y_d \right)_{33}}
     - \frac{\displaystyle \left( Y_u \right)_{23}}%
            {\displaystyle \left(Y_u \right)_{33}}}%
     {\left[ \frac{\displaystyle \left( M_d \right)_{21}}%
                  {\displaystyle \left( M_d \right)_{11}} 
     - \frac{\displaystyle \left( M_u \right)_{21}}%
            {\displaystyle \left( M_u \right)_{11}} \right]^2}
     \right| , \\
   \bar\rho - i\bar\eta
   = \frac{\left( Y_d \right)_{33} \left( M_u \right)_{31}
           - \left( Y_d \right)_{23} \left( M_u \right)_{21} 
           + \left( Y_d \right)_{13} \left( M_u \right)_{11}}%
          {\left( Y_d \right)_{33} \left( M_u \right)_{11} 
           \left[ \frac{\displaystyle \left( Y_d \right)_{23}}%
                       {\displaystyle \left( Y_d \right)_{33}} 
           - \frac{\displaystyle \left( Y_u \right)_{23}}%
                  {\displaystyle \left( Y_u \right)_{33}} \right]
           \left[ \frac{\displaystyle \left( M_d \right)_{21}}%
                       {\displaystyle \left( M_d \right)_{11}} 
           - \frac{\displaystyle \left( M_u \right)_{21}}%
                  {\displaystyle \left( M_u \right)_{11}} \right]}
   \,. \hspace{2.0cm}
\end{split}
\eeq
Notice that, like in the case of the Froggatt-Nielsen mechanism, $\bar{\rho}$ and $\bar{\eta}$ are not suppressed by any small 
parameters \cite{Froggatt:1978nt}. They are to first order independent of the zero-mode profiles $F(c_{Q_i,q_i})$. The RS setup 
thus predicts that these parameters are of $\ord(1)$, while the precise values remain unexplained. 

Let us now try to determine the bulk mass parameters from the measurable quantities discussed here. 
The relations given in (\ref{eq:quarkmasses}) and (\ref{eq:wolfenstein}) do not allow to determine all zero-mode profiles solely in terms of the quark 
masses and Wolfenstein parameters (as well as $\ord(1)$ Yukawa couplings). One profile remains as a free parameter. Choosing $F(c_{Q_2})$ to 
be that parameter and expressing the other profiles in terms of its value, we find for the left-handed quark profiles 
\beq\label{eq:Qwave}
   |F(c_{Q_1})| 
   = \frac{\lambda}{\left| 
    \frac{\displaystyle \left( M_d \right)_{21}}%
         {\displaystyle \left( M_d \right)_{11}} 
    - \frac{\displaystyle \left( M_u \right)_{21}}%
           {\displaystyle \left( M_u \right)_{11}}\right|}\,
    |F(c_{Q_2})| \,, \qquad 
   |F(c_{Q_3})| 
   = \frac{\left| 
    \frac{\displaystyle \left( Y_d \right)_{23}}%
         {\displaystyle \left( Y_d \right)_{33}} 
    - \frac{\displaystyle \left( Y_u \right)_{23}}%
           {\displaystyle \left( Y_u \right)_{33}}\right|}%
          {A\lambda^2}\,|F(c_{Q_2})| \,.
\eeq
For the right-handed profiles, we arrive at
\beq\label{eq:uwave}
\begin{split}
   |F(c_{u_1})| &= \frac{\sqrt{2} m_u}{v}\,
    \frac{\left| \left( M_u \right)_{11} \right| 
          \left| \frac{\displaystyle \left( M_d \right)_{21}}%
                      {\displaystyle \left( M_d \right)_{11}} 
          - \frac{\displaystyle \left( M_u \right)_{21}}%
                 {\displaystyle \left( M_u \right)_{11}} \right|}%
         {\lambda\left| \det(\bm{Y}_u) \right|}\,
    \frac{1}{|F(c_{Q_2})|} \,, \\
   |F(c_{u_2})| &= \frac{\sqrt{2} m_c}{v}\,
    \frac{\left| \left( Y_u \right)_{33} \right|}%
         {\left| \left( M_u \right)_{11} \right|}\,
    \frac{1}{|F(c_{Q_2})|} \,, \\
   |F(c_{u_3})| &= \frac{\sqrt{2} m_t}{v}\,
    \frac{A\lambda^2}%
         {\left| \left( Y_u \right)_{33} \right|
          \left| \frac{\displaystyle \left( Y_d \right)_{23}}%
                      {\displaystyle \left( Y_d \right)_{33}} 
          - \frac{\displaystyle \left( Y_u \right)_{23}}%
                 {\displaystyle \left( Y_u \right)_{33}}\right|}\,
    \frac{1}{|F(c_{Q_2})|} \,,
\end{split}
\eeq
and 
\beq
\begin{split}
   |F(c_{d_1})| &= \frac{\sqrt{2} m_d}{v}\,
    \frac{\left| \left( M_d \right)_{11} \right| 
          \left| \frac{\displaystyle \left( M_u \right)_{21}}%
                      {\displaystyle \left( M_u \right)_{11}} 
          - \frac{\displaystyle \left( M_d \right)_{21}}%
                 {\displaystyle \left( M_d \right)_{11}} \right|}%
         {\lambda\left| \det(\bm{Y}_d) \right|}\,
    \frac{1}{|F(c_{Q_2})|} \,, \\
   |F(c_{d_2})| &= \frac{\sqrt{2} m_s}{v}\,
    \frac{\left| \left( Y_d \right)_{33} \right|}%
         {\left| \left( M_d \right)_{11} \right|}\,
    \frac{1}{|F(c_{Q_2})|} \,, \\
   |F(c_{d_3})| &= \frac{\sqrt{2} m_b}{v}\,
    \frac{A\lambda^2}%
         {\left| \left( Y_d \right)_{33} \right|
          \left| \frac{\displaystyle \left( Y_u \right)_{23}}%
                      {\displaystyle \left( Y_u \right)_{33}} 
          - \frac{\displaystyle \left( Y_d \right)_{23}}%
                 {\displaystyle \left( Y_d \right)_{33}}\right|}\,
    \frac{1}{|F(c_{Q_2})|} \,.
\end{split}
\eeq

As expected, these relations call for a hierarchical structure among the quark profiles,
which is naturally accommodated within warped extra dimensions.
For the left-handed quark profiles we find
\beq\label{eq:WFLhierrarchy}
   \frac{|F(c_{Q_1})|}{|F(c_{Q_2})|} \sim \lambda \,, \qquad
   \frac{|F(c_{Q_2})|}{|F(c_{Q_3})|} \sim \lambda^2 \,, \qquad 
   \frac{|F(c_{Q_1})|}{|F(c_{Q_3})|} \sim \lambda^3 \,.
\eeq
The hierarchies of the right-handed profiles can then be fixed by the observed quark masses 
\beq\label{eq:Fvalues}
\begin{split}
   \frac{|F(c_{u_1})|}{|F(c_{u_3})|} 
    \sim \frac{m_u}{m_t}\,\frac{1}{\lambda^3} \,, \qquad 
   \frac{|F(c_{u_2})|}{|F(c_{u_3})|}
    \sim \frac{m_c}{m_t}\,\frac{1}{\lambda^2} \,, \hspace{1.7cm} \\ 
   \frac{|F(c_{d_1})|}{|F(c_{u_3})|}
    \sim \frac{m_d}{m_t}\,\frac{1}{\lambda^3} \,, \qquad 
   \frac{|F(c_{d_2})|}{|F(c_{u_3})|}
    \sim \frac{m_s}{m_t}\,\frac{1}{\lambda^2} \,, \qquad 
   \frac{|F(c_{d_3})|}{|F(c_{u_3})|}
    \sim \frac{m_b}{m_t} \,.
\end{split}
\eeq
\begin{figure}[!t]
	\centering
		\includegraphics[width=11.5cm]{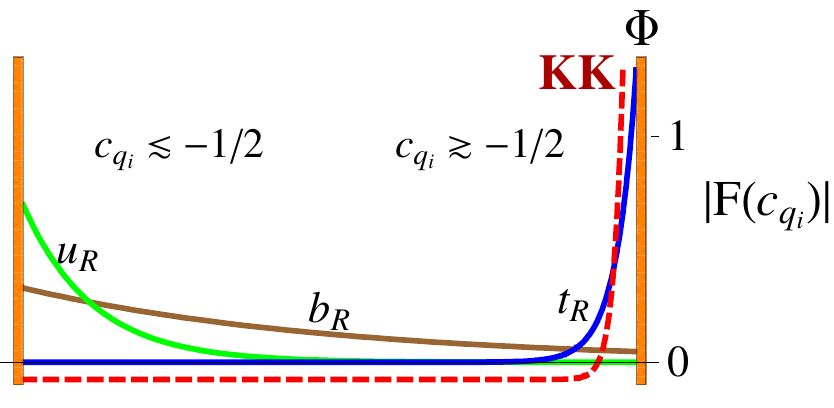}
	\caption{\label{fig:ferms} Fermion localizations in the anarchic approach to flavor in warped extra dimensions. Hierarchies are generated
	from $\ord(1)$ input parameters $c_{q_i}$ due to the warped geometry. See \cite{Casagrande:2008hr} and text for details.}
\vspace{-1.2cm}
\end{figure}

These relations can be used to determine the hierarchical structure of the flavor mixing matrices $\bm{U}_q$ and $\bm{W}_q$ in (\ref{eq:UWq}). 
The rotation matrices in the left-handed quark sector have the same structure as the CKM matrix. Their hierarchies are given by
\beq\label{eq:Uud} 
   \bm{U}_{u,d} \sim \bm{V}_{\rm \hspace{-1mm} CKM} \sim 
   \begin{pmatrix} 1 & ~\lambda~ & \lambda^3 \\ 
                        \lambda & 1 & \lambda^2 \\ 
                        \lambda^3 & \lambda^2 & 1 
   \end{pmatrix} \sim
   \begin{pmatrix} 1 & ~0.23~ & 0.01 \\ 
                   0.23 & 1 & 0.05 \\ 
                  0.01 & 0.05 & 1 
   \end{pmatrix}\, , 
\eeq
whereas in the right-handed quark sector we obtain
\beq\label{eq:WudVR}
\begin{split}
   \bm{W}_u &\sim 
    \begin{pmatrix} 1 & 
     \frac{\displaystyle m_u}{\displaystyle m_c}\,
      \frac{\displaystyle 1}{\displaystyle\lambda} & 
     \frac{\displaystyle m_u}{\displaystyle m_t}\,
      \frac{\displaystyle 1}{\displaystyle \lambda^3} \\ \\[-2mm]
     \frac{\displaystyle m_u}{\displaystyle m_c}\,
      \frac{\displaystyle 1}{\displaystyle \lambda} & 1 &
     \frac{\displaystyle m_c}{\displaystyle m_t}\,
      \frac{\displaystyle 1}{\displaystyle \lambda^2} \\ \\[-2mm]
     \frac{\displaystyle m_u}{\displaystyle m_t}\,
      \frac{\displaystyle 1}{\displaystyle \lambda^3} & 
     \frac{\displaystyle m_c}{\displaystyle m_t}\,
      \frac{\displaystyle 1}{\displaystyle \lambda^2} & 1
    \end{pmatrix} \sim 
    \begin{pmatrix} 1 & ~0.012~ & 0.001 \\ 
                    0.012 & 1 & 0.077 \\ 
                    0.001 & ~0.077~ & 1 
    \end{pmatrix} , \\[2mm]
   \bm{W}_d & \sim 
    \begin{pmatrix} 1 & 
     \frac{\displaystyle m_d}{\displaystyle m_s}\,
      \frac{\displaystyle 1}{\displaystyle \lambda} & 
     \frac{\displaystyle m_d}{\displaystyle m_b}\,
      \frac{\displaystyle 1}{\displaystyle \lambda^3} \\ \\[-2mm]
     \frac{\displaystyle m_d}{\displaystyle m_s}\,
      \frac{\displaystyle 1}{\displaystyle \lambda} & 1 & 
     \frac{\displaystyle m_s}{\displaystyle m_b}\,
      \frac{\displaystyle 1}{\displaystyle \lambda^2} \\ \\[-2mm]
     \frac{\displaystyle m_d}{\displaystyle m_b}\,
      \frac{\displaystyle 1}{\displaystyle \lambda^3} & 
     \frac{\displaystyle m_s}{\displaystyle m_b}\,
      \frac{\displaystyle 1}{\displaystyle \lambda^2} & 1
    \end{pmatrix} \sim 
    \begin{pmatrix} 1 & 0.26 & 0.12 \\ 
                    0.26 & 1 & 0.44 \\ 
                    0.12 & ~0.44~ & 1 
    \end{pmatrix}\,.
\end{split}
\eeq
The numerical values quoted here have been obtained by using the input parameters as compiled in Appendix~\ref{app:ref}.
Note that, while the fermion profiles are expected to be strongly hierarchical in the left-handed quark sector as well as in the right-handed up-type sector, 
this is not the case for the right-handed down-quark sector. Here, the hierarchies are much weaker
\beq
   \frac{|F(c_{d_1})|}{|F(c_{d_2})|}
    \sim \frac{m_d}{m_s}\,\frac{1}{\lambda} \sim 0.3 \,, 
    \qquad 
   \frac{|F(c_{d_2})|}{|F(c_{d_3})|}
    \sim \frac{m_s}{m_b}\,\frac{1}{\lambda^2} \sim 0.4 \,.
\eeq
In this sector, and only there, it is thus a viable possibility to assume equal quark profiles $F(c_{d_i})$ by imposing a
$U(3)$ flavor symmetry and to explain the required 
moderate splittings in terms of $\ord(1)$ variations in the fundamental Yukawa couplings that break this symmetry. Although such a choice might seem 
to be {\it ad hoc}, it has the virtue of strongly suppressing dangerous tree-level FCNC contributions to $K$--$\bar K$ mixing \cite{Santiago:2008vq}.

To summarize the findings of this section, it is possible to address the hierarchies in the fermion sector,
and to generate small masses out of $\ord(1)$ fundamental parameters within warped extra dimensions.
Those models thus provide also a solution to the second problem introduced in Section~\ref{sec:hiera}. At the same time,
the anarchic approach to flavor narrows down many of the new parameters entering the RS setup, making the model more predictive. The 
relations (\ref{eq:WFLhierrarchy}) and (\ref{eq:Fvalues}) demonstrate how the (relative) localizations of 
the different quarks, expressed through profile functions $F(c)$, are given in terms of observable quantities. These new flavor quantities of 
the RS model can be fixed to first approximation by the observed quark masses and CKM parameters.
The emerging picture is shown in Figure \ref{fig:ferms}. The (right handed component of the) top quark, being the heaviest quark of the SM,
should be localized closest to the IR brane, where the Higgs sector is localized. The light quarks, however, reside more closely 
to the UV brane. The fact that the top quark is localized next
to the IR brane, where also KK modes live, results in potentially interesting effects in top physics.
The same holds true for the Higgs sector, being localized directly on the IR brane, see Chapter~\ref{sec:Pheno}.
Note that the localization of single profiles can only be fixed to certain regions, which depend on each other. Within these regions, 
the profiles can be shifted due to reparametrization invariances, leaving the quark masses and CKM elements unchanged, see below. 
The boundaries of these intervals are defined by naturalness, \ie, we do not want to abandon the assumption of $\ord(1)$ localization parameters, 
as well as of $\ord(1)$ fundamental Yukawa couplings. 

\begin{figure}[!t]
\begin{center} 
\hspace{-2mm}
\mbox{\includegraphics[height=2.8in]{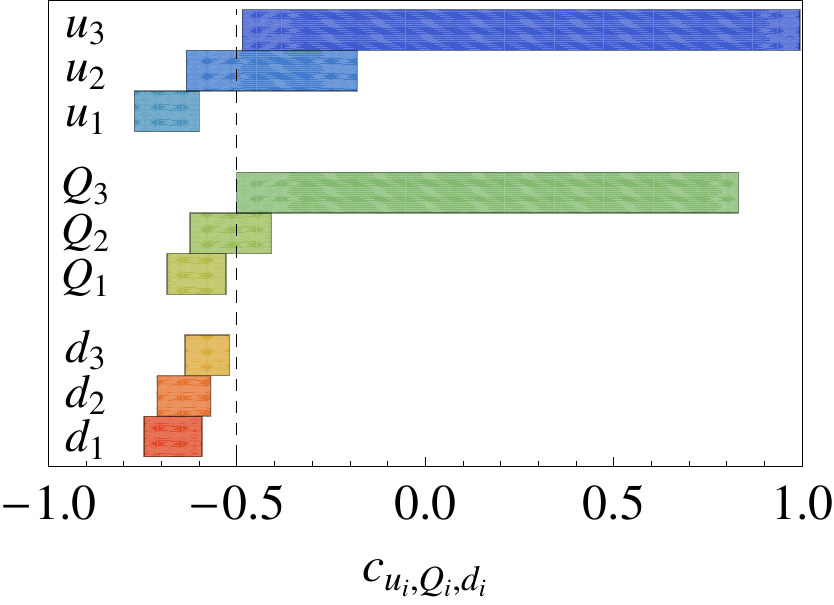}} 
\vspace{-2mm}
\parbox{15.5cm}{\caption{\label{fig:cs}
Localization parameters of quarks in the extra dimension. Values of $c_{Q_i,q_i}<-1/2$ lead to an UV localization,
whereas IR-localization corresponds to $c_{Q_i,q_i}>-1/2$. The shown regions contain 95\% of the points in the parameter space.
See \cite{Casagrande:2008hr} and text for details.}}
\end{center}
\end{figure}

The regions in parameter space which are ocupied by $95\%$ of our RS parameter points (see Chapter~\ref{sec:Pheno}) are plotted in 
Figure~\ref{fig:cs}.
This plot confirms that indeed the quark spectrum and the mixing parameters constrain the localization of the quark fields
and thus the generic RS predictions for observables.
As expected, light quarks are predicted to have $c_{Q_i,q_i}<-1/2$, $i=1,2$,
whereas the right-handed top quark and, to a lesser extend, the third generation doublet feature $c_{Q_3,u_3}>-1/2$.
Importantly, the generic statement about the flavor structure, to expect the largest effects in the sector of third generation quarks, is quite robust.
In particular, the relative localizations within the $SU(2)_L$ doublet as well as within the singlet sector are more or less fixed
due to the observed hierarchies. However, also the absolute localizations are constrained, due to the requirement that {\it none} of the (correlated) parameters leaves its natural range. In that context note that the bulk masses should not significantly exceed the RS curvature $c_{Q_i,q_i}\lesssim 1$.
To improve these order of magnitude estimations and to determine the precise values of the RS parameters, one has to perform significantly 
more measurements than within the SM. In the phenomenological studies presented in this thesis, we will scan over parameter sets, with the 
localization of the quarks determined just as described in this section. Starting from random anarchical Yukawa matrices we will find that, 
despite the remaining uncertainty in the localization of the fermion fields, generic predictions are possible in the RS setup without 
additional input, see sections \ref{sec:rare}, \ref{sec:afbt}, and \ref{sec:RSHiggs}.

What concerns leptons, note that it is also possible to address the tiny neutrino masses in the RS setup, without a see-saw mechanism,
by means of wave function overlaps \cite{Grossman:1999ra,Agashe:2008fe,Carena:2009yt}, see also \cite{Huber:2001ug,Burdman:2002gr,Huber:2003sf}.

\vspace{-0.4cm}

\subsubsection{Reparametrization Invariance}
\vspace{-0.15cm}
As sketched before, the results for the quark masses and mixings in the ZMA are invariant 
under a set of reparametrization transformations, which change the input parameters of the theory. We will now give the explicit
form of the corresponding transformations. The first type of reparametrization invariance (RPI-1) corresponds to a simultaneous 
rescaling of the profiles for the $SU(2)_L$ doublet and singlet fermions by opposite factors, while keeping the fundamental 
Yukawa couplings invariant 
\beq\label{RPI1}
   F(c_{Q_i}) \to e^{-\xi}\,F(c_{Q_i}) \,, \qquad
   F(c_{q_i}) \to e^{+\xi}\,F(c_{q_i}) \,, \qquad
   \bm{Y}_q \to  \bm{Y}_q \,.
   \qquad \mbox{(RPI-1)}
\eeq
For UV localized quarks, with $c_i<-1/2$, these transformations correspond to good approximation 
to the shifts $c_{Q_i}\to c_{Q_i}-\xi/L$ and $c_{q_i}\to c_{q_i}+\xi/L$ of the bulk mass parameters.

A second type of reparametrization invariance (RPI-2) corresponds to a simultaneous rescaling of all fermion profiles by 
a common factor, while the fundamental Yukawa couplings are rescaled with an opposite factor (however still keeping them of $\ord(1)$)
\beq\label{RPI2}
   F(c_{Q_i}) \to \eta\,F(c_{Q_i}) \,, \qquad
   F(c_{q_i}) \to \eta\,F(c_{q_i}) \,, \qquad
   \bm{Y}_q\to \frac{1}{\eta^2}\,\bm{Y}_q \,.
   \qquad \mbox{(RPI-2)}
\eeq
For $c_i<-1/2$, the corresponding shifts of the bulk mass parameters are approximately universal and read
$c_{Q_i}\to c_{Q_i}+L^{-1}\ln\eta$ and $c_{q_i}\to c_{q_i}+L^{-1}\ln\eta$. A stronger form of this relation is provided
by allowing for different transformation parameters for up-type and down-type singlets $\eta_u\ne\eta_d$
\beq\label{RPI2a}
   F(c_{Q_i}) \to \eta\,F(c_{Q_i}) \,, \qquad
   F(c_{q_i}) \to \eta_q\,F(c_{q_i}) \,, \qquad
   \bm{Y}_q\to \frac{1}{\eta\eta_q}\,\bm{Y}_q\,.
   \qquad \mbox{(RPI-$2^{\,\prime}$)}
\eeq
Of course the transformations (\ref{RPI1}) and (\ref{RPI2a}) can be combined in arbitrary ways.
While the masses and mixing angles in the ZMA are not affected by these transformations, the fermion bulk profiles do change.
As a consequence also the result for the flavor-changing interactions (apart from the CKM matrix), that will be derived in the upcoming sections,
will change under these reparametrizations. As discussed before, observables in the RS setup thus depend, besides on the NP scale $\Mkk$, more or less strongly 
on additional new parameters, which are not completely fixed by the fermion hierarchies. However, these hierarchies tell us a lot
about these parameters, and fix the relative sizes in the doublet as well as in the singlet sector to a good extent. The demand for fundamental
$\ord(1)$ Yukawa couplings constrains the deviations due to the second type of reparametrization invariance. The remaining freedom 
should then be constrained further by experiment, \eg by $Z$-pole measurements, see Section~\ref{sec:bpseudo}.

\subsection{Gauge-Boson Interactions with Fermions}
\label{sec:gaugecouplings}

With the results obtained so far, we are able to derive all Feynman rules for the minimal RS model in a straightforward way. In particular, we 
do not have to worry
about the transformation to the mass basis anymore, which has already been implemented in our KK decomposition. We start the discussion with
the interactions between fermions and gauge bosons. The interactions involving the Higgs boson will be explored later in Section~\ref{sec:HcouplingsRS}.
The main results that will be presented here are exact to all orders in $v^2/\Mkk$. However, also ZMA expressions will be given, which 
make the dependence on the model parameters more transparent. It will turn out that for the $Z$-boson interactions with SM fermions it is 
necessary to include the next-to-leading order (NLO) in the ZMA to derive a consistent result at $\ord(v^2/\Mkk^2)$. This contribution
had been neglected in the literature before \cite{Casagrande:2008hr}. 

\subsubsection{Fermion Couplings to Massless Gauge Bosons and their KK Excitations}
\label{subsec:gluons}

First we consider general fermion couplings to gauge bosons which do not receive masses from EWSB, \ie, couplings to (KK) gluons and photons.
Those have massless zero modes with a flat profile (\ref{eq:chi0flat}).
The interaction terms follow, like in 4D, from local gauge invariance. They are analogous to the terms resulting from (\ref{eq:fermkin}) but with 
an additional integration over the extra dimension and a non-trivial metric, as well as 5D instead of 4D fields. In particular,
the 4D covariant derivative (\ref{eq:covD1}) is to be replaced by (\ref{eq:covderRS}). Note that the $\phi$-components of the gauge fields
couple to fermion zero modes only at $\ord(v^2/\Mkk^2)$. Moreover, they do not develop a zero mode themselves. Thus, we do not consider them here.

After KK decomposition, the 4D Lagrangian contains the interaction terms 
\beq\label{eq:gluoncouplings}
\begin{split}
   {\cal L}_{\rm 4D} &\ni \sum_{n_1, n_2, n_3} \bigg\{ \left[ 
     \vec a_{n_2}^{U\dagger}\,\bm{I}_{n_1 n_2 n_3}^{C(Q)}\,\vec a_{n_3}^{U}
     + \vec a_{n_2}^{u\dagger}\,\bm{I}_{n_1 n_2 n_3}^{S(u)}\,
     \vec a_{n_3}^{u} \right]
    \bar u_L^{(n_2)} g_s\Gslash^{(n_1)a}\,t^a\,u_L^{(n_3)} \\
   &\hspace{1.6cm}\mbox{}+ \left[ 
    \vec a_{n_2}^{u\dagger}\,\bm{I}_{n_1 n_2 n_3}^{C(u)}\,\vec a_{n_3}^{u} 
    + \vec a_{n_2}^{U\dagger}\,\bm{I}_{n_1 n_2 n_3}^{S(Q)}\,
    \vec a_{n_3}^{U} \right]
    \bar u_R^{(n_2)} g_s\Gslash^{(n_1)a}\,t^a\,u_R^{(n_3)} \bigg\} \,,
\end{split}
\eeq
corresponding to couplings of gluons to up-type quarks. Analogous expressions hold for down-type quarks. 
Here, $g_s=g_{s,5}/\sqrt{2\pi r}$ is the 4D gauge coupling of QCD and we have defined the overlap integrals
\beq\label{eq:overlap}
   \bm{I}_{n_1 n_2 n_3}^{C(A)} 
   = \int_{-\pi}^\pi\!d\phi\,\sqrt{2\pi}\,\chi_{n_1}(\phi)\,
   e^{\sigma(\phi)}\,\bm{C}_{n_2}^{A}(\phi)\,
   \bm{C}_{n_3}^{A}(\phi) \,,
   \qquad A = Q, u, d \,.
\eeq
The overlaps $\bm{I}_{n_1 n_2 n_3}^{S(A)}$ are defined similarly in terms of integrals over $\bm{S}_n^{(A)}$ profiles. For fixed $n_1,n_2,n_3$ these integrals 
are $N\times N$ diagonal matrices in generation space. For the gluon zero mode, we obtain using (\ref{eq:chi0flat}) and (\ref{eq:orthonorm})
\beq
   \bm{I}_{0 n_2 n_3}^{C(A)} 
   = \delta_{n_2 n_3}\,\bm{1} + \Delta \bm{C}_{n_2 n_3}^{A} \,, 
    \qquad
   \bm{I}_{0 n_2 n_3}^{S(A)} 
   = \delta_{n_2 n_3}\,\bm{1} + \Delta \bm{S}_{n_2 n_3}^{A} \,.
\eeq
Thus, the relation (\ref{eq:CS}) implies that the couplings of massless gauge boson modes are flavor diagonal and take the 
same form as in the SM, \ie,
\beq\label{eq115}
   {\cal L}_{\rm 4D}\ni \sum_n 
   \bar u^{(n)} g_s\Gslash^{(0)a}\,t^a\,u^{(n)} \,,
   \vspace{-1mm}
\eeq
where $u^{(n)}=u_L^{(n)}+u_R^{(n)}$. The couplings of KK gluons are however not flavor diagonal and must 
be worked out from (\ref{eq:gluoncouplings}). 
Due to the structure of the overlap integrals (\ref{eq:overlap}), the effective couplings between heavy fermions and KK gluons, which are both 
localized close to the IR brane is not $g_s$ but $\sqrt{L}\,g_s$ \cite{Davoudiasl:1999tf,Pomarol:1999ad}.
The interactions between photons and fermions can be obtained from the relations above by the replacement $g_s t^a\to e Q_f$, where $f$ can be any  
fermion species charged under electromagnetism.

\subsubsection[Fermion Couplings to the $Z$ Boson]{Fermion Couplings to the $\bm{Z}$ Boson}

The weak interactions of the minimal RS model can be worked out in a similar way. Nevertheless, some differences arise in this sector. 
In principle, the emerging overlap integrals have the same form as in (\ref{eq:gluoncouplings}). But, since the weak gauge bosons couple 
differently to $SU(2)_L$-doublet and -singlet fermions, the integrals appear in different combinations compared to (\ref{eq:gluoncouplings}). 
Moreover, already the zero modes of the weak gauge bosons, corresponding to the $W^\pm$ and $Z$ bosons of the SM, are massive and thus have 
a non-trivial profile in the extra dimension. These facts result in modified weak gauge-boson couplings to fermions with respect to the SM.
The RS corrections are non-diagonal in flavor space, leading to tree-level FCNCs due to interactions with the $Z$-boson.
Interestingly, these corrections are parametrically enhanced by the ``volume factor'' $L$ (measuring the radius of the extra dimension in 
units of the inverse curvature divided by $\pi$) with respect to the FCNC couplings of KK gauge bosons \cite{Bauer:2009cf},
see Section~\ref{sec:KKsum}.

Technically, the emergence of tree-level FCNC-couplings to the $Z$-boson in the RS model has two reasons. 
First, the fermion profiles do not fulfill standard orthonormality relations but rather fulfill (\ref{eq:orthonorm}). As discussed above, 
this reflects the fact, that after EWSB the $SU(2)_L$ doublets and singlets mix. Thus, only the sum of the $SU(2)_L$ doublet and 
singlet contribution to a certain chirality leads to diagonal couplings due to the combined 
orthonormality relation (\ref{eq:CS}), as is the case for the massless SM gauge bosons. If one term is coming with another prefactor, 
like in the $Z$-boson couplings, being different for singlets and doublets, an off-diagonal contribution to the interaction will remain, 
due to fermion mixing. Second, due to the non-trivial dependence of the $Z$-boson profile on $\phi$, the interaction integrals can not 
collapse to the (combined) orthonormality relation (\ref{eq:CS}), which would lead to diagonal couplings, even if singlets and doublets
would couple similarly. This source of tree FCNCs would remain even without the presence of doublet-singlet mixing, \ie, when the odd profiles 
of the zero modes vanish and the even profiles obey standard orthonormality relations with $\Delta \bm{C}_{mn}^{A}\to \delta_{mn}\, \bm{1}$. From a 
perturbative point of view, this can be understood through the fact that before going from the flavor basis to the fermion mass basis, the 
$Z$-couplings are flavor (and KK mode) diagonal but non universal due to the different overlaps. This will induce off-diagonal entries in 
the interaction matrix, after going to the fermion-mass basis. In the same way, one can understand the tree-level FCNCs due to fermion 
mixing, since here the $SU(2)_L$ doublets and singlets, that are contributing to the same chirality (due to the presence of vector-like 
KK excitations), couple differently \cite{Huber:2003tu, delAguila:2000kb, Hewett:2002fe}, see also \cite{delAguila:2000rc}.
This source of tree-level FCNCs would remain even for couplings to {\it massless} gauge bosons that couple differently to $SU(2)_L$ doublets and singlets.
Before, this effect has usually been neglected in the literature since it is proportional
to the masses of the light SM fermions. However, we will show that it is parametrical as well as numerical as important as the other
source of FCNCs. In \cite{Casagrande:2008hr}, we have given for the first time exact expressions for the corresponding contributions and 
presented compact analytical results, valid at first non-trivial order in the ZMA.

Including RS corrections to $\ord(m_Z^2/\Mkk^2)$, the couplings of the $Z$-boson to SM-fermions and 
their KK excitations can be written as
\beq\label{eq:Zff}
\begin{split}
   {\cal L}_{\rm 4D}
   &\ni \frac{g}{\cos\theta_w}
    \left[ 1 + \frac{m_Z^2}{4\Mkk^2} \left( 1 - \frac{1}{L} \right) 
    \right] Z_\mu \\
   &\quad\times \sum_{f,m,n}
    \left[ \big( g_L^f \big)_{mn}\,
    \bar f_{L,m}\gamma^\mu f_{L,n}
    + \big( g_R^f \big)_{mn}\,\bar f_{R,m}\gamma^\mu f_{R,n} 
    \right] ,
\end{split}
\eeq  
where $g$ and $\cos\theta_w$ are the 4D weak $SU(2)_L$ gauge coupling and the cosine of the weak mixing angle as defined in (\ref{eq:g4def}) 
and (\ref{eq:weinbRS}). The prefactor in brackets in the first line above corresponds to a universal correction due to the constant terms 
in the bulk profile (\ref{eq:chi0WZ}). The left- and right-handed couplings $\bm{g}_{L,R}^f$ are infinite-dimensional matrices in the space 
of fermion modes, depending on the fermion type $f=u,d,\nu,e$.\footnote{The discussion of the $Z$ couplings in this section refers to quarks 
as well as leptons. In the rest of this thesis we will, however, focus on the quark sector.}  They can be parametrized as
\beq\label{eq:gLR}
\begin{split}
   \bm{g}_L^f 
   &= \left( T_3^f - \sin^2\theta_w\,Q_f \right)
    \left[ \bm{1} - \frac{m_Z^2}{2\Mkk^2}
    \left( L\,\bm{\Delta}_F - \bm{\Delta}'_F \right) \right] 
    - T_3^f \left[ \bm{\delta}_F - \frac{m_Z^2}{2\Mkk^2} 
    \left( L\,\bm{\varepsilon}_F - \bm{\varepsilon}'_F \right)
    \right] , \hspace{6mm} \\
   \bm{g}_R^f 
   &= - \sin^2\theta_w\,Q_f
    \left[ \bm{1} - \frac{m_Z^2}{2\Mkk^2}
    \left( L\,\bm{\Delta}_f - \bm{\Delta}'_f \right) \right] 
    + T_3^f \left[ \bm{\delta}_f - \frac{m_Z^2}{2\Mkk^2} 
    \left( L\,\bm{\varepsilon}_f - \bm{\varepsilon}'_f \right)
    \right] ,
\end{split}
\eeq
where $T_3^f$ and $Q_f$ denote the weak isospin and the electric charge (in units of $e$) of the fermion $f$.
A subscript $F$ on the matrices $\bm{\Delta}$, $\bm{\Delta}'$, $\bm{\delta}$, $\bm{\varepsilon}$, and $\bm{\varepsilon}'$, refers to a 
fermion of an $SU(2)_L$ doublet ($F=U,D$ in the quark sector, and $F=\nu,E$ in the lepton sector), while a subscript $f$ refers to a singlet ($f=u,d$ or $f=\nu_R,e$). The matrices $\bm{\Delta}^{(\prime)}$ and $\bm{\varepsilon}^{(\prime)}$ arise due to the 
$t$-dependent terms in the gauge-boson profile (\ref{eq:chi0WZ}) and mediate FCNCs. The elements of the former read
\begin{eqnarray}\label{eq:overlapints1}
\begin{split}
   \left( \Delta_F \right)_{mn}
   &= \frac{2\pi}{L\epsilon} \int_\epsilon^1\!dt\,t^2
    \left[ \vec a_m^{F\dagger}\,\bm{C}_m^{F}(\phi)\, 
    \bm{C}_n^{F}(\phi)\,\vec a_n^{F} 
    + \vec a_m^{f\dagger}\,\bm{S}_m^{f}(\phi)\, 
    \bm{S}_n^{f}(\phi)\,\vec a_n^{f} \right] , \\
   \left( \Delta_f \right)_{mn}
   &= \frac{2\pi}{L\epsilon} \int_\epsilon^1\!dt\,t^2
    \left[ \vec a_m^{f\dagger}\,\bm{C}_m^{f}(\phi)\, 
    \bm{C}_n^{f}(\phi)\,\vec a_n^{f} 
    + \vec a_m^{F\dagger}\,\bm{S}_m^{F}(\phi)\, 
    \bm{S}_n^{F}(\phi)\,\vec a_n^{F} \right] , \\
   \left( \Delta'_F \right)_{mn}
   &= \frac{2\pi}{L\epsilon} \int_\epsilon^1\!dt\,t^2
    \left( \frac12 - \ln t \right)
    \left[ \vec a_m^{F\dagger}\,\bm{C}_m^{F}(\phi)\, 
    \bm{C}_n^{F}(\phi)\,\vec a_n^{F} 
    + \vec a_m^{f\dagger}\,\bm{S}_m^{f}(\phi)\, 
    \bm{S}_n^{f}(\phi)\,\vec a_n^{f} \right] , \\
   \left( \Delta'_f \right)_{mn}
   &= \frac{2\pi}{L\epsilon} \int_\epsilon^1\!dt\,t^2
    \left( \frac12 - \ln t \right)
    \left[ \vec a_m^{f\dagger}\,\bm{C}_m^{f}(\phi)\, 
    \bm{C}_n^{f}(\phi)\,\vec a_n^{f} 
    + \vec a_m^{F\dagger}\,\bm{S}_m^{F}(\phi)\, 
    \bm{S}_n^{F}(\phi)\,\vec a_n^{F} \right] . \hspace{4mm}
\end{split}
\end{eqnarray}
Keep in mind that for the profiles of $SU(2)_L$ doublet fermions no distinction is to be made between $F=U$ and $F=D$ 
(\ref{eq:doublpr}), as well as between  $F=\nu$ and $F=E$. For the minimal RS model, the matrices 
$\bm{\varepsilon}^{(\prime)}$ are identical to those given above, just with the even profiles $\bm{C}_n^{F,f}$ omitted. 
Finally, there are also the contributions from the fact that the fermion profiles are not orthonormal on each other, \ie, from
doublet-singlet mixing. They are parametrized by the matrices $\bm{\delta}$, defined as 
\beq\label{defdelta}
   \left( \delta_F \right)_{mn} 
   = \vec a_m^{f\dagger} \left( \delta_{mn}
    + \Delta\bm{S}_{mn}^{f} \right) \vec a_n^{f} \,, \qquad
   \left( \delta_f \right)_{mn}
   = \vec a_m^{F\dagger} \left( \delta_{mn} 
    + \Delta\bm{S}_{mn}^{F} \right) \vec a_n^{F} . 
\eeq
We will now have a closer look on the couplings of the $Z$-boson to the light (SM) fermions, which will be particular relevant for 
the following discussions. These are described by the upper-left $3\times 3$ blocks of the matrices $\bm{g}_{L,R}^f$. 
First, note that the weight factor $t^2$, present in the overlap integrals (\ref{eq:overlapints1}), emphasizes the region 
close the IR brane, where $t=\ord(1)$. However, the SM fermions are, with the exception of the heavy top quark and, to
a lesser extend, the left-handed component of the bottom quark, all localized near the UV brane (see Figure \ref{fig:ferms} and the 
related discussion). However, in that region the weight factor is exponentially suppressed $t^2\to\ord(\epsilon^2)$. This leads to a strong 
suppression of FCNCs containing light quarks, which is known as the RS-GIM mechanism \cite{Agashe:2004ay,Agashe:2004cp,Agashe:2005hk}.
This is a very important feature of RS models, as without such a mechanism, bounds from flavor physics would immediately drive the 
KK scale way beyond the reach of the LHC, see the discussion of the {\it new physics flavor problem} in Section~\ref{sec:SMProblems}.
Like the GIM mechanism of the SM, RS-GIM is broken by the large top-quark mass, see Section~\ref{sec:SMflavor}, which leads to an IR 
localization of the corresponding singlet profile as well as (less strongly) of the corresponding doublet profile. However, its dynamical
origin is different from the GIM mechanism of the SM.

It will be very useful to have approximate analytical expressions for the exact overlap integrals at hand, from which the size of 
their effects as well as the dependence on the input parameters of the model will become more transparent. To this end, we employ 
the zeroth order in the ZMA to the fermion profiles entering the overlap integrals of (\ref{eq:overlapints1}). This
is justified, since in (\ref{eq:Zff}) 
the profiles already feature a coefficient scaling like $v^2/\Mkk^2$ and higher orders are neglected anyway. We perform 
the replacements
\beq\label{eq:ZMA}
   \bm{C}_n^{Q,u}(\phi)\,\vec a_n^{U,u}
   \to \sqrt{\frac{L\epsilon}{2\pi}}\,
   \mbox{diag}( F(c_{Q_i,u_i})\,
    t^{c_{Q_i,u_i}} )\,\hat a_n^{u,u^c} \,, \qquad
   \bm{S}_n^{Q,u}(\phi)\,\vec a_n^{U,u} \to 0 \,,
\eeq
and similar for down-type quarks,
meaning that the even profiles reduce to the zero mode profiles obtained without EWSB \cite{Grossman:1999ra}, while the odd 
profiles are suppressed by an extra power of $x_n=m_n/\Mkk$ and thus vanish to the order considered. Analogous expressions
hold for the down-quark sector.

With these replacements we find \cite{Burdman:2002gr,Agashe:2003zs,Delgado:2007ne}
\beq\label{ZMA1}
\begin{split}
   \bm{\Delta}_F
   &\to \bm{U}_f^\dagger\,\,\mbox{diag} \left[ 
    \frac{F^2(c_{F_i})}{3+2c_{F_i}} \right] \bm{U}_f \,, \\
   \bm{\Delta}_f
   &\to \bm{W}_f^\dagger\,\,\mbox{diag} \left[ 
    \frac{F^2(c_{f_i})}{3+2c_{f_i}} \right] \bm{W}_f \,, \\
   \bm{\Delta}'_F
   &\to \bm{U}_f^\dagger\,\,\mbox{diag} \left[ 
    \frac{5+2c_{F_i}}{2(3+2c_{F_i})^2}\,F^2(c_{F_i}) \right] 
    \bm{U}_f \,, \\
   \bm{\Delta}'_f
   &\to \bm{W}_f^\dagger\,\,\mbox{diag} \left[ 
    \frac{5+2c_{f_i}}{2(3+2c_{f_i})^2}\,F^2(c_{f_i}) \right] 
    \bm{W}_f \,.
\end{split}
\eeq
All these objects are $3\times 3$ matrices in generation space, and the diagonal matrices contain the elements in the corresponding 
brackets on the diagonal. Note that, with the $c_{F_i,f_i}$ parameters being close to $-1/2$, we have 
$\bm{\Delta}_A\approx\bm{\Delta}'_A$ for $A=F,f$ to good approximation. The matrices $\bm{\varepsilon}_A^{(\prime)}$ vanish at zeroth
order in the ZMA - they are suppressed by an additional factor of $v^2/\Mkk^2$. The same is true for the matrices $\bm{\delta}_A$,
since they can also be written as integrals containing only $\bm{S}_n^{(A)}$ profiles. However, we have to be careful, since these 
contributions come with an unsuppressed $\ord(1)$ coefficient in (\ref{eq:gLR}). Thus we have to include the NLO in the ZMA which
yields
\beq\label{ZMAforS}
   \bm{S}_n^{Q,u}(\phi)\,\vec a_n^{U,u}
   \to \pm\sgn(\phi)\,\sqrt{\frac{L\epsilon}{2\pi}}\,x_n\,
   \mbox{diag}\left( F(c_{Q_i,u_i})\,
   \frac{t^{1+c_{Q_i,u_i}} - \epsilon^{1+2c_{Q_i,u_i}}\,
         t^{-c_{Q_i,u_i}}}%
        {1+2c_{Q_i,u_i}} \right) \hat a_n^{u,u^c} \,,
\eeq
from which we derive
\beq\label{eq:ZMA2}
\begin{split}
   \bm{\delta}_F
   &\to \bm{x}_f\,\bm{W}_f^\dagger\,\,
    \mbox{diag}\left[ \frac{1}{1-2c_{f_i}} 
    \left( \frac{1}{F^2(c_{f_i})} 
    - 1 + \frac{F^2(c_{f_i})}{3+2c_{f_i}} \right) \right] 
    \bm{W}_f\,\bm{x}_f \,, \\
   \bm{\delta}_f
   &\to \bm{x}_f\,\bm{U}_f^\dagger\,\,
    \mbox{diag}\left[ \frac{1}{1-2c_{F_i}} 
    \left( \frac{1}{F^2(c_{F_i})} 
    - 1 + \frac{F^2(c_{F_i})}{3+2c_{F_i}} \right) \right] 
    \bm{U}_f\,\bm{x}_f \,. 
\end{split}
\eeq
Here, the diagonal matrix $\bm{x}_f=\mbox{diag}(m_{f_1},m_{f_2},m_{f_3})/\Mkk$ contains the masses of the SM fermions.
The important expressions (\ref{eq:ZMA2}) have not been presented before \cite{Casagrande:2008hr}. Approximate results for the admixture
of singlet contributions in the wave functions of predominantly $SU(2)_L$ doublet SM fermions due to mixing with their 
KK excitations have been given in \cite{delAguila:2000kb,Hewett:2002fe,Csaki:2008zd,Chang:2008zx}. However, this effect has
not been discussed systematically in the context of flavor-changing processes. By making use of the scaling relations obtained
via the Froggatt-Nielsen analysis in Section~\ref{sec:hierarchies} we arrive at
\beq
\label{eq:FNZMA}
\begin{split}
   \left( \Delta_F \right)_{ij}
   &\sim \left( \Delta'_F \right)_{ij} 
    \sim F(c_{F_i})\,F(c_{F_j}) \,, \\
   \left( \Delta_f \right)_{ij}
   &\sim \left( \Delta'_f \right)_{ij}
    \sim F(c_{f_i})\,F(c_{f_j}) \,, \\
   \left( \delta_F \right)_{ij}
   &\sim \frac{m_{f_i} m_{f_j}}{\Mkk^2}\,
    \frac{1}{F(c_{f_i})\,F(c_{f_j})}
    \sim \frac{v^2\,Y_f^2}{\Mkk^2}\,F(c_{F_i})\,F(c_{F_j}) \,, \\
   \left( \delta_f \right)_{ij}
   &\sim \frac{m_{f_i} m_{f_j}}{\Mkk^2}\,
    \frac{1}{F(c_{F_i})\,F(c_{F_j})}
    \sim \frac{v^2\,Y_f^2}{\Mkk^2}\,F(c_{f_i})\,F(c_{f_j}) \,,
\end{split}
\eeq
valid to LO in hierarchies.
Here, $Y_f$ represents an (combination of) element(s) of the Yukawa matrix $\bm{Y}_f$. The relations (\ref{eq:FNZMA}) 
make the RS-GIM suppression factors due to the fermion zero-mode profiles explicit. Moreover, one observes explicitly that the 
contributions of the (usually neglected) $\bm{\delta}_A$ matrices in (\ref{eq:gLR}), are of the same order as the effects 
proportional to the $\bm{\Delta}^{(\prime)}_A$ matrices. The chiral suppression is lifted by the inverse powers of the 
corresponding fermion zero-mode profiles. Explicit expressions for the mixing matrices, in which the relevant 
combinations of Yukawa matrices are included, can be found in \cite{Casagrande:2008hr}.

Let us finally mention how the flavor-changing couplings discussed here transform under the reparametrizations of 
Section~\ref{sec:hierarchies}, which leave the quark spectrum and CKM parameters (in the ZMA) invariant.
From (\ref{RPI1}), we obtain for a RPI-1 transformation
\beq\label{RPI1Delta}
\begin{aligned}
   \bm{\Delta}_F &\to e^{-2\xi}\,\bm{\Delta}_F \,,
   &\qquad
   \bm{\Delta}_f &\to e^{+2\xi}\,\bm{\Delta}_f \,, \\
   \bm{\delta}_F &\to e^{-2\xi}\,\bm{\delta}_F \,,
   &\qquad
   \bm{\delta}_f &\to e^{+2\xi}\,\bm{\delta}_f \,.
\end{aligned}
\eeq
which redistributes effects between the left- and right-handed sectors. 
For a RPI-$2^{\, \prime}$ transformation, we obtain from (\ref{RPI2a})
\beq\label{RPI2Delta}
\begin{aligned}
   \bm{\Delta}_F &\to \eta^2\,\bm{\Delta}_F \,,
   &\qquad
   \bm{\Delta}_f &\to \eta_q^2\,\bm{\Delta}_f \,, \\
   \bm{\delta}_F &\to \frac{1}{\eta_q^2}\,\bm{\delta}_F \,,
   &\qquad
   \bm{\delta}_f &\to \frac{1}{\eta^2}\,\bm{\delta}_f \,.
\end{aligned}
\eeq
This corresponds to a reshuffling of contributions between the two sources of flavor violations, one arising from 
the non-trivial gauge-boson profiles ($\bm{\Delta}_A$) and one from fermion mixing ($\bm{\delta}_A$). 

\vspace{-0.4cm}
\subsubsection[Fermion Couplings to $W^\pm$ Bosons]{Fermion Couplings to $\bm{W^\pm}$ Bosons}
\label{sec:chargedcurrents}

Similar to the way of deriving the interactions between fermions and the $Z$ boson, we can obtain expressions for the couplings
of the charged $W^\pm$ bosons to fermions. In this sector, of course, flavor-changing effects at the tree-level are already unsuppressed 
within the SM. Focusing on the quark sector and working up to $\ord(m_W^2/\Mkk^2)$, we obtain
\beq\label{Wff}
   {\cal L}_{\rm 4D}\ni \frac{g}{\sqrt 2}\,W_\mu^+ 
   \sum_{n_1,n_2} \left[ \big( \tilde V_L \big)_{n_1 n_2}\,
   \bar u_{L,n1}\gamma^\mu d_{L,n_2}
   + \big( \tilde V_R \big)_{n_1 n_2}\,\bar u_{R,n1}\gamma^\mu d_{R,n_2} 
   \right] + \mbox{h.c.}\,,
\eeq
where
\beq\label{eq:VLVRdef}
   \left( \tilde V_L \right)_{n_1 n_2} 
   = \vec a_{n_1}^{U\dagger}\,
   \bm{I}_{0 n_1 n_2}^{C(Q)}\,\vec a_{n_2}^D \,,
    \qquad
   \left( \tilde V_R \right)_{n_1 n_2} 
   = \vec a_{n_1}^{U\dagger}\,
   \bm{I}_{0 n_1 n_2}^{S(Q)}\,\vec a_{n_2}^D \,
\eeq
and the corresponding overlap integrals have been defined in (\ref{eq:overlap}).

The couplings to the SM fermions are again encoded in the upper-left $3\times 3$ blocks of these matrices. In the leading
order of the ZMA we obtain
\beq
   \bm{\tilde V}_L\to \bm{U}_u^\dagger\,\bm{U}_d = \bm{V}_{\rm \hspace{-1mm} CKM} \,,
    \qquad
   \bm{\tilde V}_R\to 0 \,,
\eeq
where the tilde symbol indicates that the charged-current interaction-matrices differ from the quantities which would be experimentally 
identified with the CKM matrix (for the left handed interactions) as well as the corresponding object in the right handed sector (which arises
due to doublet-singlet mixing). Here, the definition corresponds to the single $W u_L^i d_L^j$ and $W u_R^i d_R^j$ vertices, whereas 
a more physical definition includes the exchange of the whole tower of $W^\pm$-boson KK modes.
We will give such a definition, based on four-fermion interactions, after having discussed the necessary formalism to sum up
KK towers in Section~\ref{sec:KKsum}.
At $\ord(v^2/\Mkk^2)$, corrections to the two matrices in (\ref{eq:VLVRdef}) arise. These lead to a non-unitarity of the matrix
$\bm{\tilde V}_L$ describing $W u_L^i d_L^j$ interactions and to right-handed charged currents. The matrix $\bm{\tilde V}_R$ can be estimated by 
using the first non-trivial order in the ZMA for the $S_n^A$ profiles, as given in (\ref{ZMAforS}). We obtain
\beq\label{VRres}
   \bm{\tilde V}_R\to \bm{x}_u\,\bm{U}_u^\dagger\,\,
   \mbox{diag}\left[ \frac{1}{1-2c_{Q_i}} 
   \left( \frac{1}{F^2(c_{Q_i})} - 1 
   + \frac{F^2(c_{Q_i})}{3+2c_{Q_i}} \right) \right] \bm{U}_d\,
   \bm{x}_d \,.
\eeq
The scaling relations from the Froggatt-Nielsen analysis of Section~\ref{sec:hierarchies} imply
\beq
   \left(\tilde V_R \right)_{ij}
   \sim \frac{v^2}{\Mkk^2}\,F(c_{u_i})\,F(c_{d_j}) 
   \sim \frac{m_{u_i} m_{d_j}}{\Mkk^2}\,
   \frac{1}{F(c_{Q_i})\,F(c_{Q_j})} \,.
\eeq
The non-unitarity of the matrix $\bm{\tilde V}_L$ is determined by the deviations of the $Z_2$-even fermion profiles and the 
corresponding flavor vectors $\vec a_n^{Q,q}$ from their ZMA expressions, \ie, fermion mixing, as well as by the 
$t$-dependent terms in the gauge-boson profile (\ref{eq:chi0WZ}). These effects are easiest to study with the help of the exact results 
for the fermion profiles and eigenvectors derived in Section~\ref{sec:minfermionprofiles}, see \cite{Casagrande:2008hr}. 
However, we will focus on the definition of the CKM matrix via the exchange of the whole KK tower, see Section~\ref{sec:4Fint}.
After that discussion, we will also explore the fermion couplings to the Higgs boson, directly in the context of both RS variants 
studied in this thesis, see Section~\ref{sec:HcouplingsRS}.

\section[$T$ and $Zb\bar b$ Observables as a Motivation to Extend the Gauge Group]{The $\bm{T}$ Parameter and \boldmath $Zb\bar b$ \unboldmath Observables as a Motivation to Extend the Gauge Group}
\sectionmark{Precision Tests as a Motivation to Extend the Gauge Group}
\label{sec:pheno1}

After having examined the interactions of the SM fields propagating in a warped extra dimension in detail in the last sections, we
will have a first look at the phenomenological consequences of this setup, which can be derived with the help of these results.
For the time being, we will concentrate on the electroweak precision parameters $S,T$ and $U$ due to Peskin and Takeuchi, and corrections to $Zb\bar b$ couplings.
It will turn out that the RS setup presented so far generically produces sizable contributions to these quantities. 
This will provide a motivation for the following part in this theory chapter - the introduction of the custodial Randall-Sundrum model, 
which can help improving the electroweak fit with respect to the minimal model. However, we will also advocate another 
option of allowing for low NP scales $\Mkk\sim (2-3)$ TeV, within the minimal RS model (at the tree level).

The $S$, $T$, and $U$ parameters have been introduced in Section~\ref{sec:Higgs}, see (\ref{eq:STUdef}). They measure deviations from the electroweak radiative 
corrections of the SM, due to NP contributions in universal electroweak corrections. Thus, they are defined as zero
for a SM reference point. In the following, we will derive these parameters in the minimal RS model. Here, they 
already get contributions on the born level, due to the tree-level corrections to the $W^\pm$- and $Z$-boson masses, as well as corrections
to the trivial flat profiles. Thus we will not consider loop corrections to the electroweak parameters,
although a complete one-loop calculation, extending the work of \cite{Carena:2006bn, Carena:2007ua,Burdman:2008gm}
would be desirable. However, this is beyond the scope of this thesis.
To calculate the RS contributions to the electroweak precision parameters, we use an effective Lagrangian approach \cite{Csaki:2002gy}.
First note that, due to the RS-GIM mechanism, non universal corrections to the $W^\pm$- and $Z$-boson interactions are strongly suppressed for
the first two generations of SM fermions. In such a case, the corrections are adequately parametrized by the $S$, $T$ and $U$ parameters.
Neglecting non-universal corrections, we first rescale the gauge fields in order to bring the 
interactions with zero-mode fermions to their SM form, working at lowest order in the ZMA. Using (\ref{eq:m02}) and (\ref{eq:chi0WZ}) 
we can then read off (from the kinetic terms and mass terms of the electroweak gauge bosons) the contributions to the correlators in the RS model
\beq \label{eq:mincor}
  \begin{split}
  \allowdisplaybreaks
    \Pi_{WW}(0)&=-\frac{g^4 v^4}{32 \Mkk^2}
    \left(L-\frac 1{2L}\right) ,\\
    \Pi^{\hspace{0.25mm} \prime}_{WW}(0)&=\frac{g^2v^2}{8 \Mkk^2}
    \left(1-\frac 1 L\right) ,
    \end{split}\quad
    \begin{split}
    \Pi_{ZZ}(0)&=-\frac{(g^2+{g^\prime}^2)^2 \, v^4}{32
      \Mkk^2}\left(L-\frac1{2L}\right) ,\\
    \Pi^{\hspace{0.25mm} \prime}_{ZZ}(0)&=\frac{(g^2+{g^\prime}^2) \,
      v^2}{8 \Mkk^2} \left(1-\frac 1 L\right) .
  \end{split}
\eeq
Gauge invariance ensures that $\Pi_{AA}(0)=0$ to all orders in perturbation theory. Further, working at tree-level results in
$\Pi_{ZA}(0)=\Pi_{ZA}^{\hspace{0.25mm} \prime}(0)=0$.

Inserting the results (\ref{eq:mincor}) into the definitions of the Peskin-Takeuchi parameters (\ref{eq:STUdef}), we find,
in agreement with \cite{Delgado:2007ne,Carena:2003fx}, the positive corrections
\beq\label{eq:STURS}
   S = \frac{2\pi v^2}{\Mkk^2} \left( 1 - \frac{1}{L} \right) ,
    \qquad 
   T = \frac{\pi v^2}{2\cos^2\theta_w\,\Mkk^2}
    \left( L - \frac{1}{2L} \right) ,
\eeq
while $U$ vanishes. As mentioned in Section~\ref{sec:SMinB}, allowing the fermion fields to propagate into the bulk significantly diminishes
the corrections to $S$, with respect to a scenario with bulk gauge fields and brane fermions, for which $S, T\sim -L\pi v^2/\Mkk^2$ are both 
large and negative \cite{Csaki:2002gy}. The expressions above show the expected behavior of decreasing with an increasing mass scale $\Mkk$ of the NP.
\begin{figure}[!t]
\begin{center} 
\hspace{-2mm}
\mbox{\includegraphics[height=2.85in]{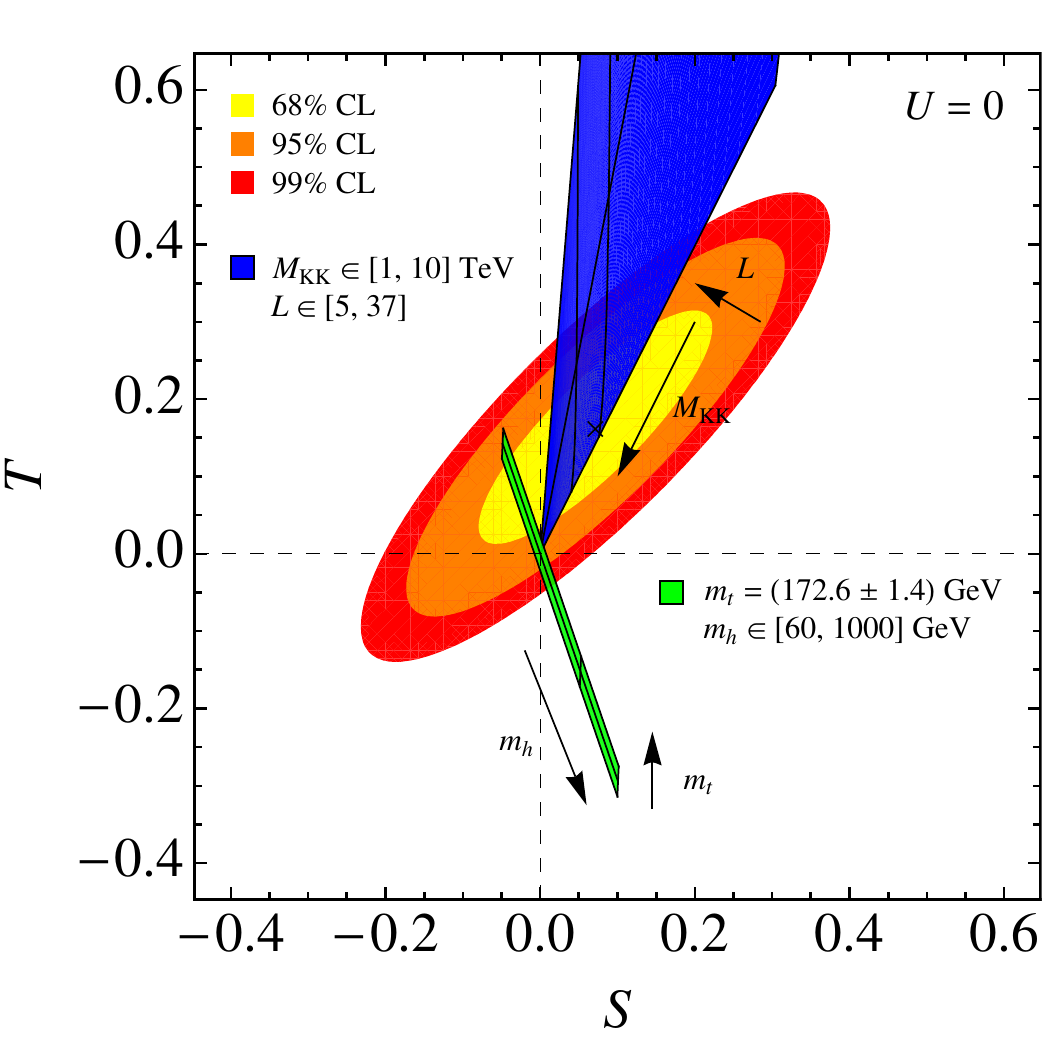}} 
\qquad 
\mbox{\includegraphics[height=2.85in]{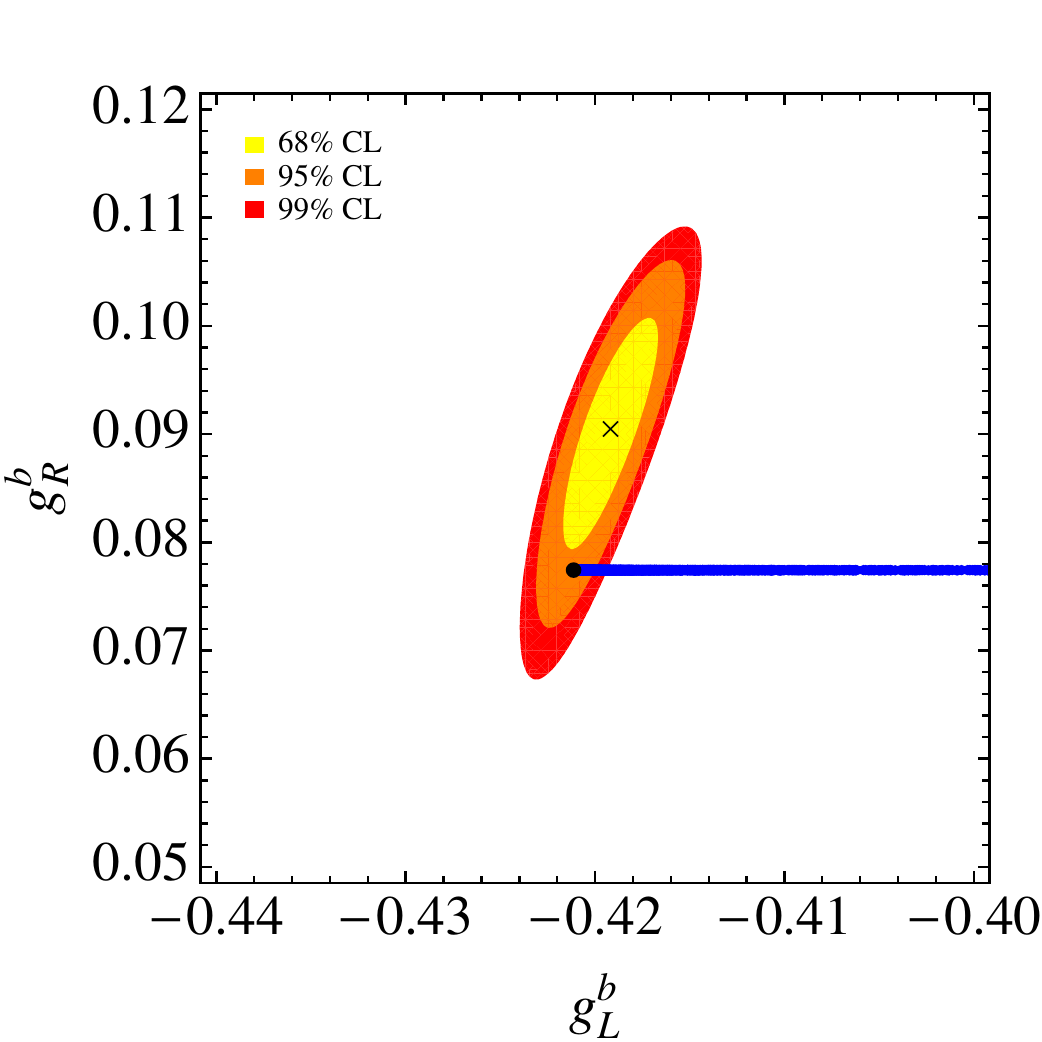}}
\vspace{-2mm}
\parbox{15.5cm}{\caption{\label{fig:STgLR}
Left: Regions of 68\%, 95\%, and 99\% probability in the $S$--$T$ plane. The green (light-shaded) shaded stripe 
shows the SM predictions for $m_t=(172.6\pm 1.4)$\,GeV and $m_h\in [60,1000]$\,GeV. The blue (dark-shaded) area indicates the 
RS corrections for $\Mkk\in [1,10]$\,TeV and $L\in [5,37]$.
Right: Same regions of probability in the $g_L^b$--$g_R^b$ plane. The horizontal stripe consists of a large number 
of points in the RS parameter space, while the black dot corresponds to the SM prediction. See \cite{Casagrande:2008hr} and text for details.}}
\end{center}
\end{figure}
Explicit values for the experimental 68\% CL bounds on the $S$ and $T$ parameters and their correlation matrix are given in (\ref{eq:STexp}).
The corresponding regions of 68\%, 95\%, and 99\% probability in the $S$--$T$ plane are depicted in the left panel of Figure~\ref{fig:STgLR}. 
The SM predictions for different values of $m_h$ and $m_t$ are shown by the green (light-shaded) stripe, whereas the blue (dark-shaded) area
represents the RS corrections for different values of the NP scale $\Mkk$ and the volume factor $L$, which we will keep variable for
this discussion. From requiring the RS corrections to satisfy the experimental bounds from $S$ and $T$ leads, for the SM reference point of
Appendix~\ref{app:ref} and $L=37$, to the constraint on the KK scale
\beq\label{eq:MKKboundslighthiggs}
   \Mkk > 4.0\,\mbox{TeV} \quad (99\%\;{\rm CL}) \,.
\eeq
Recalling that the lightest KK excitations have masses around $2.45\Mkk$, we arrive at masses of the first KK-gauge bosons of at least
about 10\,TeV. This large scale is driven by the sizable corrections to $T$, which is enhanced by the volume factor $L$, and makes the 
KK excitations impossible to be discovered directly at the LHC. 
Moreover, such a high scale increases the fine-tuning problem discussed in Section~\ref{sec:SMinB} since it leads to a higher cutoff for the 
minimal RS model (recall that $\Lambda_{\rm UV}(\pi)>\Mkk$). This sizable ``little hierarchy problem'' 
calls for a cure. Interestingly, it is possible to show that the rather tight constraint form $T$ is present in any 5D warped model with 
the SM gauge group in the bulk, which solves the gauge hierarchy problem by a moderately large volume factor \cite{Delgado:2007ne}.
There are at least {\it five possibilities} to mitigate the strong constraint from the $T$ parameter in models with a warped extra dimension.

Looking at the left panel of Figure \ref{fig:STgLR} it seems to be possible to compensate the large positive corrections (\ref{eq:STURS}) 
to $T$ in the minimal RS model by negative corrections due to a {\it heavy Higgs boson} \cite{Dobrescu:1997nm,Chivukula:1999az,Peskin:2001rw,Choudhury:2001hs,
Barbieri:2006dq}. Explicitly, the leading logarithmic corrections to $S$ and $T$, due to a Higgs-boson mass different from the 
reference value of $m_h^{\rm ref}=150$\,GeV, read \cite{Peskin:1991sw}
\beq\label{eq:SThiggs} 
   \Delta S = \frac{1}{6\pi}\,\ln\frac{m_h}{m_h^{\rm ref}} \,,
    \qquad 
   \Delta T = -\frac{3}{8\pi\cos^2\theta_w}\,
    \ln\frac{m_h}{m_h^{\rm ref}} \,,
\eeq
whereas $U$ remains unchanged. From these relations one concludes that taking for example $m_h=1$\,TeV, can significantly relax the bounds from the $T$ parameter by providing
a negative contribution of $\Delta T\approx -0.30$. Note that a large Higgs-boson mass also induces a positive shift in the parameter $S$. However, 
since $\Delta T/\Delta S\approx -3$ the combined effect will still lead to a less stringent bound on $\Mkk$. Taking $m_h=1$\,TeV we arrive at
\beq\label{eq:MKKboundsheavyhiggs}
   \Mkk > 2.6\,\mbox{TeV} \quad (99\% \; {\rm CL}) \,.
\eeq
Remember that the bound on the Higgs-boson mass from unitarity of longitudinal $W^\pm$-boson scattering, see (\ref{eq:HB}), can be relaxed in RS models \cite{Grzadkowski:2006nx}. Thus we consider $m_h\lesssim 1$\,TeV as a rough upper bound for our analysis.
As we additionally argued that in warped models, with the Higgs sector residing on the IR-brane, we naturally {\it do expect} the mass of the Higgs boson to be of the order of the KK scale and not the electroweak scale, the option to see electroweak precision tests as a hint for a heavy Higgs boson, rather than a high KK scale
of the minimal RS model, should not be directly discarded. Moreover, a smaller value of the top-quark mass can relax the bound further as it provides a negative
contribution to $T$ without changing $S$. The total error on the top-quark mass of $\Delta m_t = 1.4$\,GeV translates into a possible shift 
of $\Delta T\approx\pm 0.02$. A heavy Higgs boson in combination with a slightly lighter top quark might thus allow for first KK gauge-bosons
as light as 6\,TeV, without changing the setup of the minimal RS model.

As the corrections to $T$ are enhanced by the RS volume $L$, a second, obvious, way to diminish them would be to assume a {\it smaller RS volume}.
This removes the possibility of addressing the complete hierarchy between the Planck scale and the electroweak scale within
the RS model. However, as this model is an EFT, it is anyway well possible that it will be replaced by a UV completion well {\it below}
the Planck scale $\Lambda_{\rm UV}(0)\ll M_{\rm Pl}$. In the spirit of ``little Higgs'' models, stabilizing the Higgs mass only up to scales of the order of (10--100)\,TeV \cite{ArkaniHamed:2001nc,ArkaniHamed:2002qy}, see Section~\ref{sec:HP}, the ``little RS'' model could be replaced by a more fundamental theory already at such a low scale. 
As many observables in the RS model are enhanced by $L$, such a possibility seems viable to improve the 
agreement between the model and experiment also in other sectors. However, note that the situation is not that simple. As has been 
shown in \cite{Bauer:2008xb}, a small volume factor leads to stronger constraints from the CP-violating quantity $\epsilon_K$,
due to a softening of the RS-GIM mechanism. Moreover, the true solution to the gauge hierarchy problem is only postponed to larger energies.
We will not consider this possibility in the main phenomenological part of this thesis.
However, choosing for example a volume-truncated background with $L=\ln(10^3)\approx 7$ to address the hierarchy between the electroweak scale 
and $10^3$\,TeV, the RS bound from electroweak precision is weakened to the ``little RS'' bound of
\beq\label{eq:MKKboundsLRS}
   \Mkk > 1.5\,\mbox{TeV} \quad (99\% \; {\rm CL}) \,.
\eeq
The lightest KK modes in such a scenario would have masses of approximately $2.65\,\Mkk$. As a result, $T$ only constrains the masses 
of the lowest-lying KK gauge-boson excitation to be heavier than around 4\,TeV. This bound relaxes further for a larger Higgs-boson mass. 
For example, using $m_h=500$\,GeV instead of the reference value of $m_h=150$\,GeV would allow for light KK gauge bosons of around 3\,TeV.

A third possibility to lower the KK scale while still achieving good agreement with electroweak precision data would be to 
introduce {\it large brane-localized kinetic terms} for the electroweak gauge bosons, allowing for masses of the first KK gauge-boson 
modes of the order of 5\,TeV \cite{Carena:2003fx,Davoudiasl:2002ua,Carena:2002dz}. The appearance of such terms in orbifold
theories is expected on general grounds since they are needed as counterterms to cancel divergences arising at the loop level
\cite{Georgi:2000ks,Cheng:2002iz}. However, in our analysis, we do not want to follow this direction as a possibility to weaken
constraints from electroweak precision measurement. The bare contributions to these terms correspond to unknown UV physics above the 
cutoff. We simply assume these contributions to be small in order to retain the predictivity of the model.
Moreover we ignore possible loop contributions to brane-localized kinetic terms since we concentrate on the leading contributions
to the precision observables. Note that, even if these assumptions would be relaxed, a heavier Higgs boson in RS would still
help in lowering the limit on the KK scale, so that this scenario will still be viable in the presence of brane-localized kinetic 
terms, see also \cite{Carena:2003fx}.

A fourth option is given by promoting the {\it Higgs to a bulk field} by removing it slightly from the IR brane. This will lead to a smaller 
overlap of the Higgs boson with KK modes and in consequence to a slightly reduced contribution to $T$. If one assumes in addition
a small {\it deformation of the geometry} from the AdS$_5$ metric (\ref{eq:RSmetric}) near the IR brane, the constraint on the lightest KK 
gauge-boson masses due to electroweak precision tests can be lowered to about $2$\,TeV \cite{Cabrer:2010si,Cabrer:2011fb,Cabrer:2011vu,
Carmona:2011ib}. However, in the following we will stick to the original RS solution and do not consider additional structure that 
could lead to such a deformation of the geometry.

Before coming to the fifth possibility to evade large corrections to the $T$ parameter, which will be studied detailed 
in the next sections, let us mention another sector of electroweak precision, where sizable 
RS corrections are expected. The left-handed bottom quark, residing in the same $SU(2)_L$ doublet as the top quark, will have a 
rather large overlap with the IR brane (see Figure~\ref{fig:cs}). Therefore, sizable deviations in couplings of this quark
to massive gauge bosons are generated, since those have the largest deviation from a flat profile close to the IR brane, too. Precise 
measurements of different bottom-quark pseudo observables at the $Z$-pole have been performed at LEP. These lead to 
quite stringent constraints on the left-handed and right-handed couplings of bottom quarks to the $Z$-boson, $g_L^b$ and $g_R^b$. 
The corresponding expressions in the minimal RS model have been given in (\ref{eq:gLR}), where $g_c^b \equiv 
(g_c^d)_{33}\,,\,c=L,R$. Note that the universal prefactor in (\ref{eq:Zff}) cancels out in the observables considered. 
A thorough analysis of the $Zb\bar b$ couplings in RS models will be performed in 
Section~\ref{sec:bpseudo}, where we will also specify the observables, mentioned before. The results in the $g_L^b$--$g_R^b$ plane for 
the minimal RS model are shown in the right panel of Figure~\ref{fig:STgLR}. The horizontal stripe corresponds to a scan over 
the RS parameter space, guaranteeing that the quark masses as well as the CKM mixing angles and phase are reproduced within 
the 1\,$\sigma$ range, see Chapter~\ref{sec:Pheno}. The SM prediction for our reference point is depicted by the black dot.
One can see that large positive corrections to $g_L^b$ arise, leading potentially out 
of the $3\,\sigma$ range.\footnote{Note that the tiny corrections in $g_R^b$ are always negative.} Moreover, the anomaly 
in $A_{FB}^{0,b}$ can not be resolved directly by RS corrections, see Section~
\ref{sec:bpseudo}. One might interpret this measurement as a {\it determination} of the RS input 
parameters, constraining their values. Note that \eg for $\Mkk \sim 1.5$\, TeV it is always possible to find flavor parameters
in the anarchic RS setup, such that the constraints from $Z \to b\bar b$ are fulfilled,
see the analysis of Section~\ref{sec:bpseudo}. For example for the RS reference point, 
given in \cite{Casagrande:2008hr}, we arrive at a constraint on the KK scale of $\Mkk>1.6\,$TeV $(@\,99\%\;{\rm CL})\,.$
Thus, after an adjustment of the flavor parameters, the constraint from $Z \to b\bar b$ is in general weaker than the one from
the Peskin-Takeuchi parameters. However, in the light of the generically large corrections to $g_L^b$, it would be interesting 
to find a mechanism to evade these large RS contributions.

A setup that can avoid large corrections to the $T$ parameter as well as to the left-handed $Zb\bar b$ coupling
is provided by the {\it custodial Randall-Sundrum model}. As we have seen in Section~\ref{sec:Higgs} and Appendix~\ref{app:cus}, the custodial symmetry of the Higgs 
sector is responsible for a protection of the $\rho$ parameter (and thus also of the $T$ parameter) in the SM. If we want to have a protection for the $T$ parameter in a slice of AdS$_5$, it seems to be a good idea to explore the possibility
of a similar protection mechanism. First of all, as we are using a SM Higgs sector one can ask the question, why the standard protection
of the $T$ parameter does no longer work in the minimal RS model. This can be understood again from the dual 4D perspective.
To have a protection for the $T$ parameter, we would like to have a custodial symmetry for the CFT of which the minimal Higgs
is a light composite. However, a {\it global symmetry group} on the CFT side (with a weakly gauged subgroup) corresponds to a 
{\it gauge symmetry} in the bulk of the AdS$_5$ space-time \cite{ArkaniHamed:2000ds}. In consequence, we need to gauge the complete
$SU(2)_L\times SU(2)_R$ symmetry responsible for a custodial protection \cite{Agashe:2003zs}, see Section~\ref{sec:Higgs}.
The absence of a $SU(2)_R$ gauge symmetry in the minimal RS model is the reason for its large generic contributions to the $T$ parameter
at the tree-level. Explicitly, the mixing with the heavy KK modes, that get their masses from compactification, destroys the sought relation 
for the $W^\pm$ and $Z$-boson {\it zero modes} in the minimal RS model, in the absence of the gauged custodial symmetry.
In addition to providing such a symmetry, the custodial RS model features a protection for
$Z b_L\bar b_L$ couplings, given an appropriate embedding of the left handed bottom quark (i.e., the bottom quark
which has a left-handed zero mode in the perturbative approach) and invariance under the interchange of 
both $SU(2)$ groups \cite{Agashe:2006at}. This requires an extended fermion sector with respect to the minimal RS model. We
will elaborate on this protection in sections \ref{sec:custodialprotection} and \ref{sec:bpseudo}. 
To summarize this discussion, the gauge group, which leads to a custodial protection of the $T$ parameter in warped extra
dimensions, and which is also appropriate for a protection of the $Z b_L\bar b_L$ vertex, reads
\beq
\label{eq:Gcus}
SU(3)_c\times SU(2)_L\times SU(2)_R\times U(1)_X\times P_{LR}\,,
\eeq
where $P_{LR}$ interchanges the two $SU(2)$ groups.
The RS variant employing this gauge group is called the custodial Randall-Sundrum model (with $P_{LR}$ symmetry).
Note, however, that this symmetry group has to experience a first breaking above the electroweak scale, as we do not see additional 
gauge bosons at low energies. Thus, we will break the extended hypercharge gauge group $SU(2)_R\times U(1)_X$ down to $U(1)_Y$ on the 
UV brane via BCs, so as to arrive at the SM gauge group (\ref{eq:GSM}) in the low energy theory. In the end, it will turn 
out that the extended symmetry results in a vanishing of the {\it leading} RS contribution to the $T$ parameter, which is enhanced by the 
RS volume $L$, see (\ref{eq:STUcu}) below. Thus, the $T$ parameter will become tiny which leads to a lower bound on the KK scale, due 
to the $S$ parameter which remains unchanged, of
\beq\label{eq:MKKboundscustodial}
   \Mkk > 2.4 \,\mbox{TeV} \quad (99\% \; {\rm CL}) \,.
\eeq
This translates into a lower bound on the first KK gauge-boson masses of about 6\,TeV. This bound is marginally better that the 
one (\ref{eq:MKKboundsheavyhiggs}) of the minimal RS setup without custodial symmetry, but with a heavy Higgs boson.
In this context, note that in the case of the RS scenario with extended electroweak sector, the existence of a heavy Higgs boson would 
be rather problematic with respect to the global electroweak fit. The remaining corrections to $T$ are generically too small to 
compensate for the negative shift $\Delta T$ due to a large Higgs mass. 

So far we have only talked about tree-level corrections to the electroweak parameters.
Another virtue of the gauged custodial symmetry is that it makes one-loop corrections to the $T$ parameter finite and thus calculable
within the RS setup \cite{Agashe:2003zs}. 
These can, depending on the realization of the model and the region in parameter space, improve or worsen the agreement 
with experiment. On the other hand, without a custodial gauge symmetry, uncontrolled loop effects might also raise the lower bounds 
(\ref{eq:MKKboundsheavyhiggs}) and (\ref{eq:MKKboundsLRS}) to substantially higher values.
We will study the custodial protection of the $T$ parameter as well as the $Z b_L \bar b_L$ 
couplings in the custodial model in detail in the upcoming sections, where we also will comment more specifically on the loop corrections 
mentioned before. Nevertheless, in the light of the heavy Higgs option, we still do not discard the minimal RS model and will 
study the phenomenology of both variants in Chapter \ref{sec:Pheno}. 

\section{The Custodial Randall-Sundrum Model}
\label{sec:custo}

In this section we will perform a thorough analysis of the structure of the RS proposal featuring custodial protection, due to 
an extended gauge group in the bulk. 
We will again avoid to expand the theory in powers of $v^2/\Mkk^2$ from the very beginning and to truncate the KK tower after one (or a few) 
modes, as done in the literature on the custodial RS model. The approach of performing the KK decomposition directly in the mass basis 
is particularly suited for understanding clearly and analytically important features of the custodial model, like the level 
of protection of the left-handed $Z b\bar b$ couplings. Calculating analytically all terms of order $v^2/\Mkk^2$, we will 
identify (ir)reducible sources of custodial symmetry breaking in different sectors, which remain somewhat hidden if a 
perturbative approach is used. Our exact approach allows to include the mixing of fermions between different generations in a completely general way.
This makes the dependence on the exact realization of the matter sector transparent and it becomes straightforward to study 
the model-dependence of the gauge- and Higgs-boson interactions with the SM fermions. A thorough treatment of the 
perturbative approach featuring truncation after the first mode can be found in \cite{Albrecht:2009xr}.

We start with discussing the KK decomposition of the bulk gauge fields of the custodial gauge group, in the presence of a 
brane-localized Higgs sector, working in a covariant $R_\xi$ gauge. Here we will also study the custodial protection of the 
$T$ parameter. After that we will examine the extended fermion sector of the custodial RS model. The particular fermion representations 
that we choose allow for a protection of the $Z b_L \bar b_L$ couplings \cite{Agashe:2006at}.
It will turn out to be possible to write the KK decomposition in a form which allows to apply directly the general formalism developed
in Section~\ref{sec:fermions}. Then we will study the structure of gauge-boson interactions with SM
fermions in detail. A crucial part is the analysis of the custodial protection mechanism. We will give analytic formulae 
that show, on the one hand, the requirements for achieving a custodial protection of the left-handed $Z$-boson couplings and that expose, 
on the other, the terms that necessarily escape protection. We will distinguish between the protection from 
gauge-boson corrections and from those arising from fermion mixing. Moreover, we will show
explicitly that no protection mechanism is present in the charged-current sector, confirming existing model-independent 
results. The interactions of the Higgs boson with matter will be studied further below, together for both, the 
minimal RS variant and the custodial RS model. The exact dependence of the interactions on the realization of the fermion sector 
will be worked out. The following is based on \cite{Casagrande:2010si}.

\subsection{The Gauge Sector in the Custodial RS Model}
\label{sec:gauge}

In this section we will perform the KK decomposition of the extended gauge 
sector of the custodial RS model in the mass basis. We will derive exact 
solutions for the profiles and masses of the bulk fields, including the 
effects of an IR brane-localized Higgs sector. The formulae derived here 
will build the basis for studying interactions of the model and in particular 
for analyzing the custodial protection mechanism.

\subsubsection{Action of the 5D Theory}
We consider the RS model with custodial protection as proposed in
\cite{Agashe:2003zs} and introduced above, with the bulk gauge symmetry 
$SU(2)_L\times SU(2)_R\times U(1)_X\times P_{LR}$. On the IR brane, the
symmetry-breaking pattern $SU(2)_L\times SU(2)_R\to SU(2)_V$ provides
a custodial symmetry, which protects the $T$ parameter,
whereas the breaking $SU(2)_R\times U(1)_X\to U(1)_Y$ on the UV brane generates the 
SM gauge group, which we observe at low energies. The further breaking
down to $U(1)_{\rm EM}$ is due to an interplay of UV and IR BCs and will become
clear later. The 5D action of the gauge sector takes the form
\begin{equation} \label{eq:Sgaugecus} 
  S_{\rm gauge} = \int \!d^4x\,r\int_{-\pi}^\pi\, d\phi\, \Big( {\cal
    L}_{\rm L,R,X} + {\cal L}_{\rm Higgs} + {\cal L}_{\rm GF}\Big) \,,
\end{equation}
with the gauge-kinetic terms
\begin{equation} \label{eq:Lgauge} 
  {\cal L}_{\rm L,R,X} = \frac{\sqrt{G}}{r}\,G^{KM} G^{LN} \left( -
    \frac14\,L_{KL}^a L_{MN}^a - \frac14\,R_{KL}^a R_{MN}^a -
    \frac14\,X_{KL} X_{MN} \right) ,
\end{equation}
where the assignment of the fields to the gauge groups
should be self-explanatory. We choose the four-vector components of the gauge fields to be 
even under the $Z_2$ parity, while the scalar fifth components are odd, in order to arrive at 
a low-energy spectrum that is consistent with observation. As it is not needed for our
analysis, we ignore the Faddeev-Popov Lagrangian.

The Higgs Lagrangian
\begin{equation} \label{eq:higgslag} 
  {\cal L}_{\rm Higgs} = \frac{\delta(|\phi|-\pi)}{r} \left( \frac12 \,
    {\rm Tr} \left|(D_\mu\Phi)\right|^2 - V(\Phi) \right)
\end{equation} 
is localized on the IR brane. A prescription of 
how to deal with $\delta(|\phi|-\pi)$ has already been
presented before in (\ref{eq:delIBP}). The Higgs is extended to a 
bi-doublet under $SU(2)_L \times SU(2)_R$, and is responsible for
breaking $SU(2)_L \times SU(2)_R$ to the diagonal subgroup $SU(2)_V$
on the IR brane. It transforms as $\left(\bm{2},\bm{2}\right)_0$ and
explicitly reads (see (\ref{eq:HiggsLR}) and \cite{Burdman:2008gm}), 
\begin{equation} \label{eq:Higgsbi}
  \Phi(x) = \frac{1}{\sqrt2} \left( \begin{array}{cc} v + h(x) -
      i\varphi^3(x) & -i\sqrt2\,\varphi^+(x) \\
      -i\sqrt2\,\varphi^-(x) & v + h(x) + i\varphi^3(x)
    \end{array} \right) ,
\end{equation}
with real scalar fields $\varphi^i$ and $h$, $\varphi^\pm = (\varphi^1\mp
i\varphi^2)/\sqrt2$. The VEV of the Higgs fields is again only to first approximation given by the SM
value, $v \approx 246 \, {\rm GeV}$, and will receive corrections at $\ord(v^2/\Mkk^2)$, see Section~\ref{sec:Pheno}.
In the notation above, 
$SU(2)_L$ transformations act from the left on the bi-doublet, while the $SU(2)_R$ transformations act from the right. 
The covariant derivative acting on the Higgs sector reads
\begin{equation} \label{eq:D}
  D_\mu\Phi = \partial_\mu\Phi - i {g_L}_5\, L_\mu^a \hspace{0.25mm} 
  T_L^a\; \Phi + i {g_R}_5\, \Phi\, R_\mu^a \hspace{0.25mm} T_R^a \,,
\end{equation}
with $T^a_{L,R}=\sigma^a/2$. An explicit calculation leads to
\begin{align} \label{eq:cohigg}
    D_\mu\Phi & = \frac{1}{\sqrt2} \left( \begin{array}{c c}
        \partial_\mu \left(h - i\varphi^3 \right) -\displaystyle  i\, \frac v2
        \left({g_L}_5\, L_\mu^3 - {g_R}_5\, R_\mu^3 \right) & \phantom{i}
        -\partial_\mu i \sqrt 2 \varphi^+ - \displaystyle  i\,  \frac v2 \left({g_L}_5\,
          L_\mu^+ - {g_R}_5\, R_\mu^+ \right) \vspace{2mm} \\ 
        -\partial_\mu i \sqrt 2 \varphi^- - \displaystyle  i\,  
        \frac v2 \left({g_L}_5\,
          L_\mu^- - {g_R}_5\, R_\mu^- \right) & \phantom{i} 
        \partial_\mu \left(h +
          i\varphi^3 \right) + \displaystyle  i\, 
        \frac v2 \left({g_L}_5\, L_\mu^3 -
          {g_R}_5\, R_\mu^3 \right)
      \end{array} \right)  \hspace{4mm} \nonumber \\[-3mm] \\[-3mm] 
    & \phantom{xx} + \text{terms bi-linear in fields} \,, \nonumber 
\end{align}
where we have introduced
\begin{equation} \label{eq:LRmu}
  L_\mu^\pm = \frac{1}{\sqrt2} \left( L_\mu^1\mp i L_\mu^2 \right)
  ,\qquad R_\mu^\pm = \frac{1}{\sqrt2} \left( R_\mu^1\mp i R_\mu^2
  \right)\,,
\end{equation}
in analogy to the charged $W^\pm$ bosons of the SM.
The structure of (\ref{eq:cohigg}) motivates us to define the new
fields 
\begin{equation}
  \left( \begin{array}{c}
      \tilde A_M\\
      V_M
    \end{array} \right)= 
  \frac{1}{\sqrt{g_L^2 + g_R^2}} \,
  \left( \begin{array}{cr}
      g_L & - g_R \\
      g_R & g_L
    \end{array} \right)
  \left( \begin{array}{c}
      L_M\\
      R_M 
    \end{array} \right) ,
\end{equation}
which result in a diagonal mass matrix. The 4D gauge
couplings are related to 5D couplings as introduced in (\ref{eq:g4def}). The rotations above are in analogy to
the usual definitions of the $Z$ boson and photon fields, like in the SM,
which are themselves postponed to (\ref{eq:ZA}). Note that the $P_{LR}$ symmetry
forces $g_L=g_R$. However, for the time being, we will keep both couplings separately, in order to
be able to identify the dependence of observables on the presence of the extra symmetry. 
Finally, the mass term adopts the form
\begin{equation} \label{eq:Lm} 
  {\cal L}_{\rm Higgs} \supset \frac{\delta(|\phi|-\pi)}{r} \,
  \frac{\left({g_L^2}_5+{g_R^2}_5\right) v^2}{8} \, \tilde A_\mu^a
  \tilde A^{\mu\, a} \equiv \frac{\delta(|\phi|-\pi)}{r} \, \frac 1 2 \,
  M_{\tilde A}^2 \, \tilde A_\mu^a \tilde A^{\mu\, a} \,,
\end{equation}
and reveals the breaking pattern ({\it c.f.} (\ref{eq:LRbreak}))
\begin{equation} \label{eq:IRbreaking}
  SU(2)_L \times SU(2)_R \xrightarrow{\rm IR} SU(2)_V \,,
\end{equation}
induced by the Higgs VEV $\left \langle \Phi\right \rangle= v/\sqrt{2}
\; \bm{1}$. Appropriate BCs break the extended electroweak gauge group
down to the SM gauge group on the UV boundary
\begin{equation} \label{eq:UVbreaking}
  SU(2)_R \times U(1)_X \xrightarrow{\rm UV} U(1)_Y \,.
\end{equation}
Explicitly, this is achieved by introducing the new fields
\begin{equation} \label{eq:Zp}
  \left( \begin{array}{c}
      Z_M^\prime\\
      B_M^Y 
    \end{array} \right)= 
  \frac{1}{\sqrt{g_R^2 + g_X^2}} \, 
  \left( \begin{array}{cr}
      g_R & - g_X \\
      g_X & g_R
    \end{array} \right)
  \left( \begin{array}{c}
      R_M^3\\
      X_M
    \end{array} \right) ,
\end{equation}
and giving Dirichlet BCs to $Z_\mu^\prime$ and $R_\mu^{1,2}$ on the UV
brane, which prevent the emergence of corresponding zero modes. The $U(1)_Y$ hypercharge coupling is related to the $SU(2)_R
\times U(1)_X$ couplings by
\begin{equation} \label{eq:gY}
  g_Y= \frac{g_R\,g_X}{\sqrt{g_R^2+g_X^2}}
\end{equation}
and the SM-like neutral electroweak gauge bosons are defined in the
standard way through
\begin{equation} \label{eq:ZA}
  \left( \begin{array}{c}
      Z_M\\
      A_M
    \end{array} \right)= 
  \frac{1}{\sqrt{g_L^2 + g_Y^2}} \,
  \left( \begin{array}{cr}
      g_L & - g_Y \\
      g_Y & g_L
    \end{array} \right)
  \left( \begin{array}{c}
      L_M^3\\
      B_M^Y
    \end{array} \right) .
\end{equation}
It follows that the definitions of the sine and cosine of the
weak-mixing angle,
\begin{equation} \label{eq:weakmixing}
  \sin\theta_w=\frac{g_Y}{\sqrt{g_L^2+g_Y^2}} \,, \qquad
  \cos\theta_w=\frac{g_L}{\sqrt{g_L^2+g_Y^2}} \,,
\end{equation}
agree again formally with those in the SM, see (\ref{eq:weinbRS}) and (\ref{eq:weakmixingSM}). Note that the fields $V_M^3$ and
$X_M$ can be rotated to the photon field $A_M$ and a state $Z_M^H$ via
\begin{equation}
  \left( \begin{array}{c}
      Z_M^H\\
      A_M
    \end{array} \right)= 
  \frac{1}{g_{LRX}^2} \,
  \left( \begin{array}{cc}
      g_L\,g_R & -g_X \sqrt{g_L^2+g_R^2} \\
      g_X \sqrt{g_L^2+g_R^2}& g_L\,g_R
    \end{array} \right)
  \left( \begin{array}{c}
      V_M^3\\
      X_M
    \end{array} \right) ,
\end{equation}
where 
\begin{equation} \label{eq:gLRX2}
  g_{LRX}^2= \sqrt{g_L^2\,g_R^2+g_L^2\,g_X^2+g_R^2\,g_X^2} \,.
\end{equation}
Moreover, we write $\tilde Z_M \equiv \tilde A_M^3$, as we will see that it is a linear
combination of $Z_M$ and $Z_M^\prime$, which is orthogonal to $Z_M^H$.

\begin{figure}[!t]
\begin{center}
\vspace{6mm}
\mbox{\includegraphics[width=12cm]{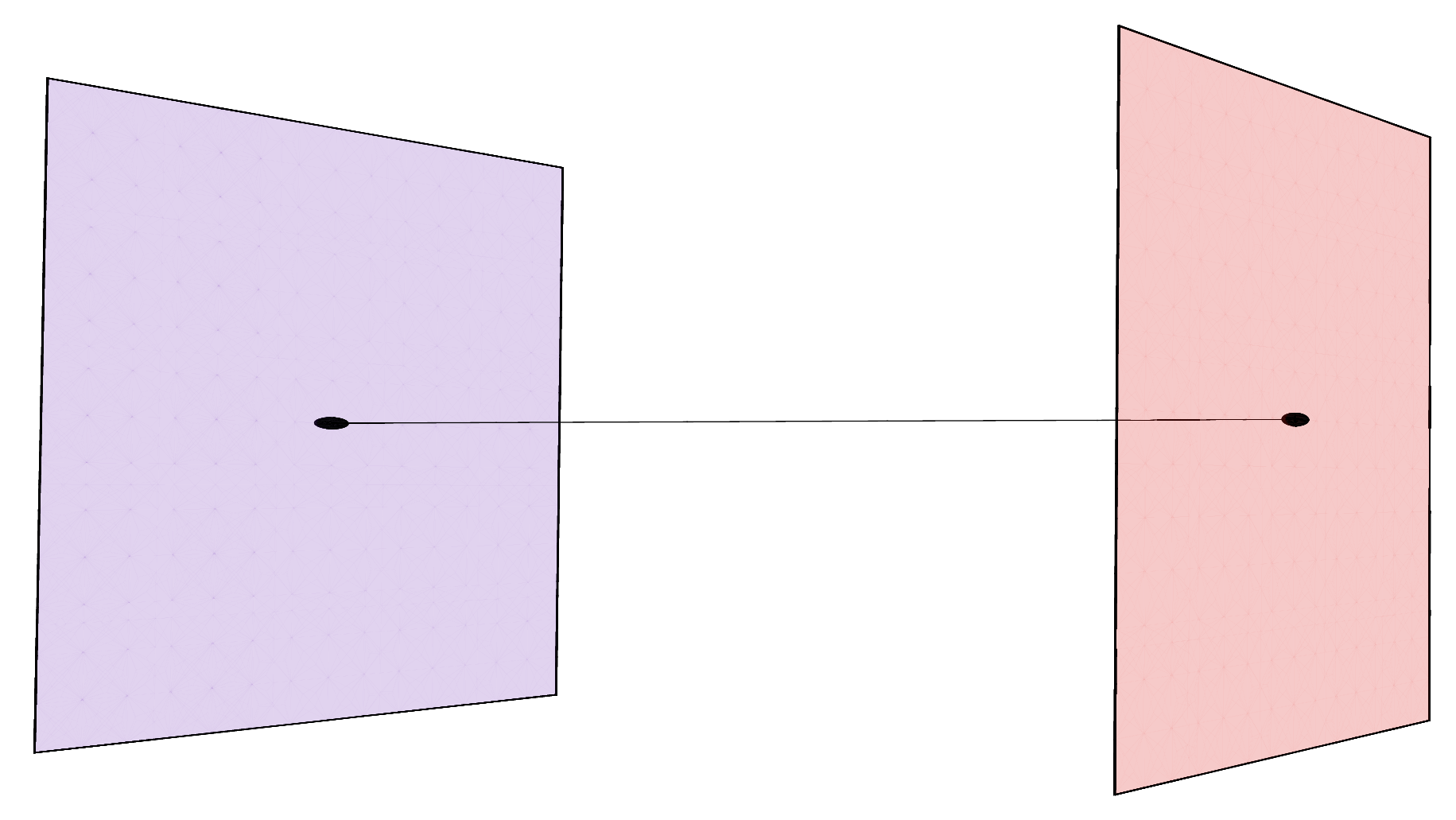}}
\begin{picture}(0,0)(0,0) 
\put(-103,3){\rotatebox{8}{UV brane}}
\put(-120,64){\rotatebox{-8.75}{$SU(2)_R \times U(1)_X \to U(1)_Y$}}
\put(-102,20){\rotatebox{7}{$Z^\prime_M, R_M^\pm$}}
\put(-105,45){\rotatebox{-7.25}{$A_M, Z_M, L_M^\pm$}}
\put(-22,-1){\rotatebox{17}{IR brane}}
\put(-40,72){\rotatebox{-18.5}{$SU(2)_L \times SU(2)_R \to SU(2)_V$}}
\put(-23,20){\rotatebox{9.25}{$Z^H_M, V_M^\pm$}}
\put(-28,47){\rotatebox{-9.5}{$A_M, \tilde{Z}_M, \tilde{A}_M^\pm$}}
\end{picture}
\vspace{6mm}

\parbox{15.5cm}{\caption{\label{fig:bases} UV and IR basis, \ie, gauge fields
  with individual BCs on the corresponding branes. The fields in the
  first (second) row on the UV brane do (do not) possess a zero
  mode. The symmetry-breaking pattern on the UV and IR brane is also
  indicated. See \cite{Casagrande:2010si} and text for details.}}
\end{center}
\end{figure}
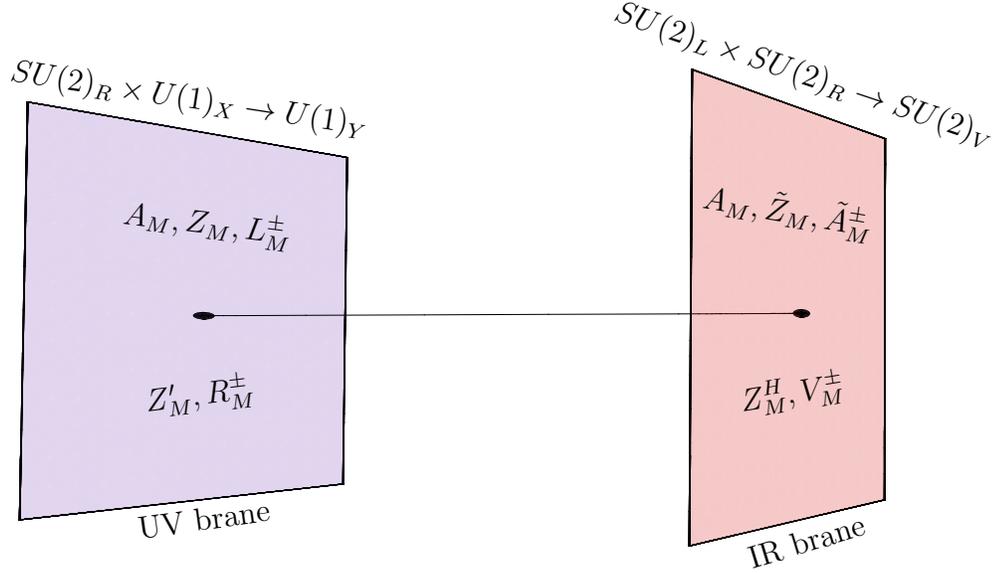

\begin{table}
  \begin{equation}
    \begin{tabular}{|c|c|}
      \hline
      $\partial_\phi L_\mu^\pm(x,0)=0$ & $L_5^\pm(x,0)=0$\\
      $R_\mu^\pm(x,0)=0$ & $R_5^\pm(x,0)=0$\\
      $\partial_\phi Z_\mu(x,0)=0$ & $Z_5(x,0)=0$\\
      $Z_\mu^\prime(x,0)=0$ & $Z_5^\prime(x,0)=0$\\
      $\partial_\phi A_\mu(x,0)=0$ & $A_5(x,0)=0$\\
      \hline
    \end{tabular}
    \quad
    \begin{tabular}{|c|c|}
      \hline
      $\partial_\phi \tilde A_\mu^\pm(x,\pi^-)=-\frac{r}{2 \epsilon ^2} 
      M_{\tilde A}^2 \tilde A_\mu^\pm(x,\pi)$
      & $\tilde A_5^\pm(x,\pi)=0$\\
      $\partial_\phi V_\mu^\pm(x,\pi)= 0$ & $V_5^\pm(x,\pi)=0$\\
      $\partial_\phi \tilde Z_\mu(x,\pi^-)=-\frac{r}{2 \epsilon ^2} 
      M_{\tilde A}^2 \tilde Z_\mu(x,\pi)$
      & $\tilde Z_5(x,\pi)=0$\\
      $\partial_\phi Z_\mu^H(x,\pi)=0$ & $Z_5^H(x,\pi)=0$\\
      $\partial_\phi A_\mu(x,\pi)=0$ & $A_5(x,\pi)=0$\\
      \hline
    \end{tabular}\nonumber
  \end{equation}
  \parbox{15.5cm}{\caption{\label{tab:BCs} UV (left) and IR (right) BCs.}}
\end{table}
In Table~\ref{tab:BCs} we collect the BCs that we choose for the
fields in order to obtain the correct SM spectrum. They are given in terms 
of fields with individual BCs on the
corresponding branes. In the following we will refer to these sets of
fields as the UV and the IR basis, respectively. The situation is
summarized in Figure~\ref{fig:bases}, where we also recall the
symmetry-breaking patterns on the different branes. The BCs can easily
be transformed to another basis at the expense of obtaining
expressions that mix different fields. The photon $A_{\mu}$ has
individual and source-free Neumann BCs on both branes, and therefore
its zero mode remains massless. Note that there is just one mass
parameter $M_{\tilde A}$ entering the IR BCs, in contrast to the two
parameters $M_Z$ and $M_W$ appearing in the minimal model. In the
custodial model, the different masses for the lightest electroweak
gauge bosons are accomplished through the mixed UV BCs of the gauge
fields in the IR basis (see (\ref{eq:rots}) below). The fact that
there is just one fundamental mass parameter is crucial for the
custodial protection of the $T$ parameter. We will elaborate on this
later in this section.

The action of the theory contains again mixed terms between gauge
fields and scalars, which can be removed by an appropriate
gauge-fixing Lagrangian. As the Higgs sector is localized on the IR
brane, it is natural to work in the IR basis for that purpose. For
this reason, we define the 5D theory in the IR basis. The concrete
form of the gauge fixing will be given below in (\ref{eq:Sgf}).

Before discussing the KK decomposition, we summarize the relations
between the UV (right) and the IR basis (left). They read
\begin{equation} \label{eq:rots}
  \begin{split}
    \left( \begin{array}{c}
        \tilde Z_M\\
        Z_M^H
      \end{array} \right) & = 
    \left( \begin{array}{cr}
        \cos\theta_Z & -\sin\theta_Z \\
        \sin\theta_Z & \cos\theta_Z
      \end{array} \right)
    \left( \begin{array}{c}
        Z_M\\
        Z_M^\prime 
      \end{array} \right)
    \equiv {\bm R}_Z \left( \begin{array}{c}
        Z_M\\
        Z_M^\prime 
      \end{array} \right) , \\[1mm]
    \left( \begin{array}{c}
        \tilde A_M^\pm\\
        V_M^\pm 
      \end{array} \right) & = 
    \left( \begin{array}{cr}
        \cos\theta_W & -\sin\theta_W \\
        \sin\theta_W & \cos\theta_W
      \end{array} \right)
    \left( \begin{array}{c}
        L_M^\pm\\
        R_M^\pm 
      \end{array} \right)
    \equiv {\bm R}_W \left( \begin{array}{c}
        L_M^\pm\\
        R_M^\pm 
      \end{array} \right) ,
  \end{split}
\end{equation}
where 
\begin{align} \label{eq:thetas}
    \sin\theta_Z & =\frac{g_R^2}{\sqrt{(g_L^2+g_R^2)(g_R^2+g_X^2)}}\,,
    & \qquad \cos\theta_Z &
    =\frac{g_{LRX}^2}{\sqrt{(g_L^2+g_R^2)(g_R^2+g_X^2)}} \, , \nonumber \\[1mm]
    \sin\theta_W & =\frac{g_R}{\sqrt{g_L^2+g_R^2}} \,, & \qquad
    \cos\theta_W & =\frac{g_L}{\sqrt{g_L^2+g_R^2}} \,,
\end{align}
and $g_{LRX}^2$ has been defined in (\ref{eq:gLRX2}). In order to
shorten the notation we will hereafter employ the abbreviations $s_a
\equiv \sin \theta_a$ and $c_a \equiv \cos \theta_a$ for $a = w, Z,
W$.

\subsubsection{Kaluza-Klein Decomposition}
\label{sec:KKdec}

We now perform the KK decomposition of the fields of the 5D theory, which we formulated in
the IR basis. However, it is convenient to work with profiles that obey definite Neumann ($+$) or Dirichlet
($-$) BCs on the UV brane. Therefore we include a rotation to the UV
basis, \ie, the basis in which the UV BCs decouple, in our
decomposition. Furthermore, as different UV fields get mixed by the IR
BCs, these fields should be expressed through the same 4D basis. We
consequently introduce the vectors $\vec Z_M=(\tilde Z_M, Z_M^H)^T$
and $\vec W_M^\pm=(\tilde A_M^\pm, V_M^\pm)^T$ and write
\begin{align} \label{eq:KKdec}
    A_\mu(x,\phi) &= \frac{1}{\sqrt r} \sum_n \chi_n^{(+)}(\phi)\,
    A_\mu^{(n)}(x)\,, & A_\phi(x,\phi) &= \frac{1}{\sqrt r} \sum_n
    \partial_\phi\chi_n^{(+)}(\phi)\,a_n^A\, \varphi_A^{(n)}(x)\,, \nonumber\\ 
     \vec Z_\mu(x,\phi) &= \frac{{\bm R}_Z}{\sqrt r} \sum_n {\bm
      \chi}_n^+(\phi)\,\vec A_n^{\, Z}\, Z_\mu^{(n)}(x)\,, & \vec
    Z_\phi(x,\phi) &= \frac{{\bm R}_Z}{\sqrt r} \sum_n \partial_\phi
    {\bm \chi}_n^+(\phi)\,\vec A_n^{\, Z} a_n^Z\,\varphi_Z^{(n)}(x)\,,\nonumber\\
    \vec W_\mu^\pm(x,\phi) &= \frac{{\bm R}_W}{\sqrt r} \sum_n {\bm
      \chi}_n^+(\phi)\,\vec A_n^{\, W}\, W_\mu^{\pm(n)}(x)\,, &  \vec
   W_\phi^\pm(x,\phi) &= \frac{{\bm R}_W}{\sqrt r} \sum_n
    \partial_\phi {\bm \chi}_n^+(\phi)\,\vec A_n^{\, W}
    a_n^W\,\varphi_W^{\pm(n)}(x)\,,
\end{align}
where the sums run over $n = 0, \ldots, \infty$. Note that
$A_\mu^{(n)}(x)$ \etc \ are 4D mass eigenstates and the lightest modes
are identified with the SM gauge bosons, which we observe at low energies.
\newpage 
The matrices ${\bm R}_{Z,W}$ have been defined in (\ref{eq:rots}) and we have 
introduced the diagonal profile matrices

\begin{equation} \label{eq:chimatrix} 
  {\bm \chi}_n^+(\phi)=
  \left( \begin{array}{cc}
      \chi_n^{(+)}(\phi) & 0 \\
      0 & \chi_n^{(-)}(\phi)
    \end{array} \right) ,
\end{equation}
as well as two-component vectors $\vec A_n^a$, with $a=Z,W$,
representing the mixings between the different gauge fields and their
KK excitations.  These vectors are normalized according to
\begin{equation}
 {(\vec A_n^a)}^T \vec A_n^a =1 \,.
\end{equation}

Notice that the matrices ${\bm \chi}_n^+ (\phi)$ should in principle
also carry a superscript $a$, indicating the field to which they
belongs, but we will not show it, as the correct index should be
always clear from the context. The superscripts $(+)$ and $(-)$ label
the type of BC we impose on the profiles on the UV brane, \ie, they
indicate untwisted and twisted even functions\footnote{We use the term
twisted even functions for profiles with even $Z_2$-parity, which obey
Dirichlet BC on the UV brane and are thus not smooth at this orbifold
fix point. These fields are sometimes called odd, as they look like an
odd function (at the UV brane) if one just considers half of the orbifold.  Untwisted
even functions correspond to ordinary profiles with Neumann UV BCs.}
on the orbifold. Remember from Table \ref{tab:BCs} that both types of profiles
satisfy Neumann BCs on the IR boundary, which we do not indicate
explicitly by a superscript $(+)$ to avoid unnecessary clutter of
notation. Let us also introduce the shorthand notations
\begin{equation} \label{eq:ZWvecs}
  \vec{\chi}_n^{\, Z} (\phi) =\left( \begin{array}{c} \chi_n^Z(\phi)\\
      \chi_n^{Z^\prime}(\phi)
    \end{array} \right)={\bm \chi}_n^+(\phi)\,\vec A_n^{\, Z} \,,\qquad
  \vec{\chi}_n^{\, W} (\phi) =\left( \begin{array}{c} \chi_n^L(\phi)\\
      \chi_n^R(\phi)
    \end{array} \right)={\bm \chi}_n^+(\phi)\,\vec A_n^{\, W} \,,
\end{equation}
for the profiles of the UV fields. In analogy to the fermion profiles
of the minimal model (\ref{eq:orthonorm}), the profiles ${\bm
\chi}_n^+(\phi)$ do not obey exact orthonormality conditions. This
fact is related to the decomposition of different fields into the same 
4D states (\ie, mixing). The complete vectors $\vec{\chi}_n^{\hspace{0.5mm} a} 
(\phi)$ with $a = Z, W$ are however orthonormal on each other,
\begin{equation} \label{eq:ortho}
  \int_{-\pi}^\pi\!d\phi\,{\vec \chi}_m^{\, a \, T}(\phi)\,{\vec
    \chi}_n^{\, a} (\phi) = \delta_{mn} \,.
\end{equation}
Note also that the photon obeys the standard orthonormality
condition (\ref{chinorm}), as before. As in (\ref{eq:gol}), we also expand the 4D Goldstone bosons in the basis of mass
eigenstates $\varphi_Z^{(n)}(x)$ and $\varphi_W^{ \pm(n)}(x)$ by
writing
\begin{equation} \label{eq:phivecs}
  \vec \varphi^{\, 3}(x) = \sum_n \vec b_n^{\, Z}\,\varphi_Z^{(n)}(x)
  \,, \qquad \vec \varphi^{\, \pm}(x) = \sum_n \vec b_n^{\,
    W}\,\varphi_W^{\pm(n)}(x) \,.
\end{equation}
Employing the notation introduced in this section, the gauge-fixing
Lagrangian takes the form 
\beq
\begin{split} \label{eq:Sgf}
    {\cal L}_{\rm GF} &= - \frac{1}{2\xi}
    \left( \partial^\mu A_\mu - \xi \left[
        \frac{\partial_\phi\,e^{-2\sigma(\phi)}}{r^2} \, A_\phi
      \right] \right)^2\\
    &\quad\mbox{}- \frac{1}{2\xi} \left( \partial^\mu \vec Z_\mu - \xi
      \left[ \frac{\delta(|\phi|-\pi)}{r}\,M_{\tilde
          A}\,\vec\varphi^{\, 3} +
        \frac{\partial_\phi\,e^{-2\sigma(\phi)}}{r^2}\, \vec
        Z_\phi \right] \right)^2 \\
    &\quad\mbox{}- \frac{1}{\xi} \left( \partial^\mu \vec W_\mu^+ -
      \xi \left[ \frac{\delta(|\phi|-\pi)}{r}\,M_{\tilde
          A}\,\vec\varphi^{\, +} +
        \frac{\partial_\phi\,e^{-2\sigma(\phi)}}{r^2} \, \vec
        W_\phi^+ \right] \right)^T \\
    &\qquad\times \left( \partial^\mu \vec W_\mu^- - \xi \left[
        \frac{\delta(|\phi|-\pi)}{r}\,M_{\tilde A}\,\vec\varphi^{\, -}
        + \frac{\partial_\phi\,e^{-2\sigma(\phi)}}{r^2} \, \vec
        W_\phi^- \right] \right).
\end{split}
\eeq

Inserting the decompositions (\ref{eq:KKdec}) into the action and
defining the projectors ${\bm P}_{(+)}={\rm diag}(1,0)$ and ${\bm
P}_{(-)}={\rm diag}(0,1)$, we derive the EOMs
\cite{Casagrande:2008hr, Davoudiasl:1999tf, Pomarol:1999ad}
\begin{equation} \label{eq:gaugeeomcust} 
 -\frac{1}{r^2}\,\partial_\phi\,e^{-2\sigma(\phi)}\,
  \partial_\phi {\bm R}_a\,{\bm \chi}_n^+(\phi) \vec A_n^{\, a} =
  (m_n^a)^2\, {\bm R}_a\,{\bm \chi}_n^+(\phi) \vec A_n^{\, a} -
  \frac{\delta(|\phi|-\pi)}{r}\,M_a^2\, {\bm P}_{(+)} {\bm R}_a\,{\bm
    \chi}_n^+(\phi) \vec A_n^{\, a} \,,
\end{equation} 
where $a = Z, W, A$ with $M_Z=M_W=M_{\tilde A}$ and $M_A=0$, as well
as ${\bm R}_A={\bm 1}$ and $\vec A_n^{\, A}=(1,0)^T$. In order to
avoid boundary terms due to integration by parts, we move the
$\delta$-distribution by an infinitesimal amount into the bulk,
as explained before (\ref{eq:delIBP}). We will again indicate values 
obtained by a limiting procedure by a superscript in the argument, \eg 
by writing ${\bm\chi}_n^+(\pi^-)$. The appropriate IR BCs for the profiles can be
obtained by integrating the EOMs (\ref{eq:gaugeeomcust}) over an
infinitesimal interval around $|\phi|=\pi$. At the 5D level the BCs have
already been presented in Table \ref{tab:BCs}. However, note that the 
discontinuities of the scalar components, whose profiles are proportional to 
the $\phi$-derivative of the vector profiles, have not been given yet. We arrive at
\begin{equation} \label{eq:IRBC2} 
  \frac{m_n^a}{\Mkk} \, {\bm R}_a\, {\bm \chi}_n^-(\pi^-) \vec A_n^{\,
    a} =- X^2\hspace{0.25mm} L\hspace{0.25mm} {\bm P}_{(+)} {\bm R}_a\,
  {\bm \chi}_n^+(\pi) \vec A_n^{\, a} \,,
\end{equation} 
where
\begin{equation} \label{eq:chiminus} 
  {\bm \chi}_n^- (\phi) \equiv \frac{1}{m_n^a r }\,
  e^{-\sigma(\phi)}\partial_\phi \, {\bm \chi}_n^+ (\phi) \,, \qquad X^2
  \equiv \frac{ (g_L^2+g_R^2) \, v^2}{4 \Mkk^2} \,.
\end{equation} 
For the photon the right-hand side in (\ref{eq:IRBC2}) is
equal to zero.
After applying the EOMs and the orthonormality condition
(\ref{eq:ortho}), we observe that the 4D action takes the desired
canonical form, given that
\begin{equation} \label{eq:coeffs} 
  a_n^a = - \frac{1}{m_n^a} \,, \qquad \vec b_n^{\, a} =
  \frac{M_a}{\sqrt r\, m_n^a} {\bm P}_{(+)} {\bm R}_a\,{\bm
    \chi}_n^+(\pi^-)\vec A_n^{\, a} \,.
\end{equation}

Thus, we finally end up again with a low energy theory that corresponds
to the SM gauge fields, with quadratic terms in analogy to (\ref{gaugefinal}).
However, due to the extended gauge group we started with, the theory now contains
a custodial protection mechanism for electroweak precision observables, as discussed
before. Due to the need to reproduce the SM at low energies, this protection
cannot be perfect. It will be broken by the BCs that we chose
in order to match to the particle content that we observe in nature. 
The spectrum of the theory is determined by the IR BCs
(\ref{eq:IRBC2}). The dimensionless eigenvalues $x_n^a \equiv m_n^a/M_{\rm KK}$
are thus solutions of
\begin{equation} \label{eq:IRBC3}
  {\rm det}\left[x_n^a \, {\bm \chi}_n^-(\pi^-) + L \hspace{0.25mm} X^2
    \hspace{0.25mm} {\bm D}_a {\bm \chi}_n^+(\pi)\right]=0\,,
\end{equation} 
with 
\begin{equation} \label{eq:Dmatrix} 
  {\bm D}_a = {\bm R}_a^{-1} {\bm P}_{(+)} {\bm R}_a =
  \left( \begin{array}{cc}
      c_a^2 & -s_a c_a \\
      -s_a c_a & s_a^2
    \end{array} \right) .
\end{equation}
Once the eigenvalues are known, the eigenvectors $\vec A_n^{\, a}$ are
determined from (\ref{eq:IRBC2}).

\vspace{-0.6cm}
\subsubsection{Bulk Profiles}

We now derive expressions for the profiles $\chi_n^{(\pm)} (\phi)$.
In order to obtain the EOMs for the UV-basis profiles, we multiply
(\ref{eq:gaugeeomcust}) by ${\bm R}_a^T$ from the left. We write 
the solutions as
\begin{equation} \label{eq:sol} 
  \chi_n^{(+)}(t) = N_n^{(+)}\sqrt{\frac{L}{\pi}}\,t\,c_n^{(+)+}(t) \,,
  \qquad \chi_n^{(-)}(t) = N_n^{(-)}\sqrt{\frac{L}{\pi}}\,
  t\,c_n^{(-)+}(t)\,,
\end{equation}
where 
\begin{equation}
  \begin{split}
    c_n^{(+)+}(t) &= Y_0(x_n \epsilon) J_1(x_n t)
    -J_0(x_n \epsilon) \,Y_1(x_n t) \,, \\[2mm]
    c_n^{(-)+}(t) &= Y_1(x_n \epsilon) J_1(x_n t)
    -J_1(x_n \epsilon) \,Y_1(x_n t) \,, \\[1mm]
    c_n^{(+)-}(t) &= \frac{1}{x_n t}\, \frac{d}{dt} \Big(
    t\,c_n^{(+)+}(t) \Big) =Y_0(x_n \epsilon) \,J_0(x_n t)
    -J_0(x_n \epsilon) \,Y_0(x_n t) \,, \\
    c_n^{(-)-}(t) &= \frac{1}{x_n t}\, \frac{d}{dt} \Big(
    t\,c_n^{(-)+}(t) \Big) =Y_1(x_n \epsilon) \,J_0(x_n t) -J_1(x_n
    \epsilon) \,Y_0(x_n t) \,.
  \end{split}
\end{equation}
The normalized masses $x_n$ are
determined by the IR BCs as explained above. From the latter
expressions, it is obvious that the profiles fulfill the sought UV BCs, since
$c_n^{(+)-}(\epsilon) = c_n^{(-)+}(\epsilon) =0$. The normalization
constants $N_n^{(\pm)}$ are determined from the orthonormality
condition (\ref{eq:ortho}). With respect to the minimal model
(\ref{Nngauge}), they contain additional terms
due to the different UV BCs. We obtain
\begin{equation} \label{eq:Nngaugecus}
  \begin{split}
    \big ( N_n^{(\pm)} \big )^{-2} = &\left[ c_n^{(\pm)+}(1) \right]^2
    + \left[ c_n^{(\pm)-}(1) \right]^2 - \frac{2}{x_n} \Big
    (\,c_n^{(\pm)+}(1)\,c_n^{(\pm)-}(1)-
    \epsilon \,c_n^{(\pm)+}(\epsilon)\,c_n^{(\pm)-}(\epsilon) \Big)\\
    &- \epsilon^2 \left( \left[ c_n^{(\pm)+}(\epsilon) \right]^2 +
      \left[ c_n^{(\pm)-}(\epsilon) \right]^2\right) .
  \end{split}
\end{equation} 
Note that, depending on the type of the UV BCs, some of the terms
in (\ref{eq:Nngaugecus}) vanish identically.

Again, it will be useful to have simple analytical expressions
for the masses and profiles of the lightest gauge-boson modes. Expanding
(\ref{eq:IRBC3}) in powers of $v^2/\Mkk^2$ and inserting the
definitions of the mixing angles (\ref{eq:thetas}), which connect the
UV and IR bases, we arrive at analytic expressions for the masses of
the $W^\pm$ and $Z$ bosons. They read
\begin{align} \label{eq:mwmz}
    m_W^2 & = \frac{g_L^2 v^2}{4} \left[ 1 - \frac{g_L^2 v^2}{8\Mkk^2}
      \left( L - 1 + \frac{1}{2L}\right) - \frac{g_R^2 v^2}{8\Mkk^2}
      \,L
      + \ord\left( \frac{v^4}{\Mkk^4} \right) \right] , \nonumber \\
    m_Z^2 & = \frac{(g_L^2+g_Y^2) \, v^2}{4} \left[ 1 -
      \frac{(g_L^2+g_Y^2) \, v^2}{8\Mkk^2} \left( L - 1 +
        \frac{1}{2L}\right) - \frac{(g_R^2-g_Y^2) \, v^2}{8\Mkk^2} \,
      L + \ord\left( \frac{v^4}{\Mkk^4} \right) \right] , \quad 
  \end{align}
where the last terms inside the square brackets are new compared to
the minimal model, respectively, {\it c.f.} (\ref{eq:m02}).
Interestingly, the latter terms in (\ref{eq:mwmz}) will be responsible for the custodial
protection of the Peskin-Takeuchi parameter $T$, which is sensitive to the difference between the
corrections to the $W^\pm$-boson and $Z$-boson vacuum-polarization functions.
The RS corrections lead again to a shift in the Higgs VEV compared to the SM value, see Section~\ref{sec:mod}.

The zero-mode profiles, expanded in $v^2/\Mkk^2$, read
\begin{equation} \label{eq:expprof}
  \begin{split}
    \chi_0^{(+)}(t) & = \frac{1}{\sqrt{2\pi}} \left[ \, 1 +
      \frac{x_{a}^2}{4} \left( 1 - \frac{1}{L} + t^2 \, \big( 1 - 2L -
        2\ln t\big) \right)
      + \ord\left( x_{a}^4\right) \right] ,\\[1mm]
    \chi_0^{(-)}(t) & = \sqrt{\frac{L}{2\pi}}\; t^2 \left[ \, -2 +
      \frac{x_{a}^2}{4} \left(t^2-\frac 2 3\right) + \ord\left( x_a^4
      \right) \right] ,
  \end{split}
\end{equation} 
for $a = W, Z$. Here $x_a^2 \equiv (m_0^a)^2/M_{\rm KK}^2$ denotes the
corresponding zero-mode solution, given in (\ref{eq:mwmz}). The profiles
$\chi_0^{(+)}(t)$, featuring Neumann IR BCs are identical to those appearing
in the minimal model, while the profiles $\chi_0^{(-)} (t)$, satisfying
Dirichlet IR BCs, are new and scale like $\sqrt{L}$, which reflects the 
localization of KK modes close to the IR boundary.  Notice that (\ref{eq:expprof})
contains, besides the $t$-independent terms, also $t$-dependent contributions that will in
general lead to non-universal vertex corrections. While these
corrections do modify the interactions of the SM fermions with the $W^\pm$
and $Z$ bosons, they turn out to be negligibly small for light
fermions localized near the UV brane. This is the case for the first
two generations of SM fermions, and it helps to avoid excessive
contributions to FCNCs, as discussed before.

Finally, we can also expand the mixing vectors $\vec A_0^a$. Including corrections up to $v^2/\Mkk^2$, we find
\begin{equation} \label{eq:vecA0a}
  \vec A_0^{\, a} = \left( \begin{array}{c}
      1\\
      \displaystyle -s_a c_a\, \frac{X^2}{4}\, \sqrt L
    \end{array} \right),
\end{equation}
where the second component parametrizes the admixture of $\chi_0^{(-)}
(t)$ in the zero mode. As we will see below in Section~\ref{sec:custodialprotection}, 
the results (\ref{eq:expprof})
  and (\ref{eq:vecA0a}) play a crucial role in the custodial protection
mechanism of the $Z b_L \bar b_L$ vertex and its flavor-changing
counterparts.

\subsection{Bulk Matter in the Custodial RS Model}
\label{sec:cusfermions}

We will now present our explicit realization of the quark sector in
the custodial RS model. Then we will turn to the KK
decomposition and derive the bulk profiles for the corresponding
fields. As we want to have a custodial protection of the $Z b_L \bar
b_L$ vertex, we impose the aforementioned discrete $P_{LR}$ symmetry
\cite{Agashe:2006at} on (part of) the Lagrangian and take the left-handed bottom 
quark to reside in a $SU(2)_L \times SU(2)_R$ bi-doublet, with isospin quantum 
numbers $T_L^3 =-T_R^3 =-1/2$. This will turn out to lead to the sought protection, 
see Section~\ref{sec:custodialprotection}, and fixes the
quantum numbers of the other quark-fields uniquely, implying the following
multiplet structure for the fields with even $Z_2$ parity:
\beq \label{eq:multiplets}
 \begin{split}
    & \quad Q_L \equiv \left(\begin{array}{cc}
        {u_L^{(+)}}_{\frac 23} & {\lambda_L^{(-)}}_{\frac 53}\\
        {d_L^{\hspace{0.25mm} (+)}}_{-\frac 13} & {u_L^{\prime \,
            (-)}}_{\frac 23}
      \end{array}\right)_{\frac 23}\, ,\qquad \qquad 
    u_R^c \equiv \left ( {u_R^{c\, (+)}}_{\frac 23}
    \right )_{\frac 23} \, ,\\[1mm]
    {\cal T}_R & \equiv {\cal T}_{1R}\oplus{\cal
      T}_{2R}\equiv\left(\begin{array}{c}
        {\Lambda_R^{\prime\, (-)}}_{\frac 53}\\
        {U_R^{\prime\, (-)}}_{\frac 23}\\
        {D_R^{\prime\, (-)}}_{-\frac 13}
      \end{array}\right)_{\frac 23}
    \oplus\left(\begin{array}{ccc} {D_R^{(+)}}_{-\frac 13}\
        {U_R^{(-)}}_{\frac 23}\ {\Lambda_R^{(-)}}_{\frac 53}
      \end{array}\right)_{\frac 23}\,.
  \end{split}
\eeq
Here, the superscripts $(+)$ and $(-)$ of the chiral fields specify the
type of BC on the UV boundary, and as before we have not explicitly
shown the BCs on the IR brane, which are understood to be of Neumann
type in all cases.

The choice of the parities/BCs is motivated by the
constraint to arrive at a low-energy spectrum of the theory that is
consistent with observations. Fields that feature a Dirichlet boundary condition 
will not develop a zero mode. The inner (outer) subscripts correspond to the
$U(1)_{\rm EM}$ ($U(1)_X$) charges, which are
connected through the relations $Q=T_L^3+Y$ and $Y=-T_R^3+Q_X$ . For
completeness and future reference, we summarize the quantum numbers of
the quark fields in Table~\ref{tab:charges}. The right-handed
down-type quarks have to be embedded in a $SU(2)_R$ triplet in order
to arrive at an $U(1)_X$-invariant Yukawa coupling. Note that we have
chosen the same $SU(2)_L \times SU(2)_R$ representations for all three
generations, which is necessary, if one wants to
incorporate consistently quark mixing in the fully anarchic approach to flavor in
warped extra dimensions. The chosen representations play a
crucial role in the suppression of flavor-changing left-handed
$Z$-boson couplings \cite{Blanke:2008zb}. Altogether they feature 15
different quark fields in the up-type and nine in the down-type
sector (for the case of three generations). Due to the BCs, there will be three light modes in each sector
to be identified with the SM quarks. These are accompanied by KK
towers, which consist of groups of 15 and nine modes of similar masses
in each {\it KK level} in the up- and down-type quark sector, respectively. Moreover one
also faces a KK tower of exotic fermion fields of electric charge
$5/3$, which exhibits nine excitations with small mass splitting in
each level and no light modes with $m_n\ll \Mkk$. In addition to (\ref{eq:multiplets}), 
there is a second set of multiplets, belonging to the components of opposite chirality. 
The corresponding states have opposite BCs. In particular, they all obey
Dirichlet BCs on the IR brane. Thus one ends up with the same particle content as in
the minimal RS model at low energies. Remember that the $SU(2)_L$
transformations act vertically, while the $SU(2)_R$ transformations
act horizontally on the multiplets.

\begin{table}
\begin{center}
\begin{tabular}{|c|c|c|c|c|c|}
\hline
& $Q$ & $Q_X$ & $Y$ & $T_L^3$ & $T_R^3$ \\
\hline
$u_L^{(+)}$ & $\phantom{-} 2/3$ & $2/3$ & $1/6$ & $\phantom{-} 1/2$ 
& $\phantom{-} 1/2$ \\
$d_L^{\hspace{0.25mm} (+)}$ & $-1/3$ & $2/3$ & $1/6$ & 
$-1/2$ & $\phantom{-} 1/2$ \\
$\lambda_L^{(-)}$ & $\phantom{-} 5/3$ & $2/3$ & $7/6$ & 
$\phantom{-} 1/2$ & $-1/2$ \\
$u_L^{\prime \, (-)}$ & $\phantom{-} 2/3$ & $2/3$ & $7/6$ & 
$-1/2$ & $-1/2$ \\ \hline
\end{tabular} \quad 
\begin{tabular}{|c|c|c|c|c|c|} 
\hline
& $Q$ & $Q_X$ & $Y$ & $T_L^3$ & $T_R^3$ \\
\hline
$u_R^{c \, {(+)}}$ & $\phantom{-} 2/3$ & $2/3$ & $2/3$ & 
$\phantom{-} 0$ & $0$ \\
$\Lambda_R^{\prime \, (-)}$ & $\phantom{-} 5/3$ & $2/3$ & $2/3$ & 
$\phantom{-} 1$ & $ 0$ \\
$U_R^{\prime \, {(-)}}$ & $\phantom{-} 2/3$ & $2/3$ & $2/3$ & 
$\phantom{-} 0$ & $0$ \\
$D_R^{\prime \, {(-)}}$ & $-1/3$ & $2/3$ & $2/3$ & 
$-1$ & $ 0$ \\ \hline
\end{tabular}
\end{center}
\begin{center}
\begin{tabular}{|c|c|c|c|c|c|} 
\hline
& $Q$ & $Q_X$ & $Y$ & $T_L^3$ & $T_R^3$ \\
\hline
$D_R^{(+)}$ & $-1/3$ & $2/3$ & $-1/3$ & 
$0$ 
& $\phantom{-} 1$ \\
$U_R^{(-)}$ & $\phantom{-} 2/3$ & $2/3$ & $\phantom{-} 2/3$ & $0$ & 
$\phantom{-} 0$ \\
$\Lambda_R^{(-)}$ & $\phantom{-} 5/3$ & $2/3$ & $\phantom{-} 5/3$ & 
$0$ & $-1$ \\ \hline
\end{tabular}
\end{center}
\parbox{15.5cm}{\caption{\label{tab:charges} Charge assignments for the different quark fields in the 
  custodial RS model.}}
  \vspace{-0.4cm}
\end{table}

\subsubsection{Fermionic Action and Yukawa Couplings}

The structure of the 5D action of the quark fields has already been
given in (\ref{eq:Sferm2}). It is straightforward to
generalize this action to the custodial model \cite{Albrecht:2009xr}. 
The only non-trivial part is due to the Yukawa couplings, where the possible
gauge-invariant terms take the form
\begin{align} \label{eq:Sfermyuk}
    S_{\rm Yukawa} & = -\int d^4x\,r \int_{-\pi}^\pi\!d\phi\;
    \delta(|\phi|-\pi)\; \frac{e^{-3\sigma(\phi)}}{r} \, \Bigg [
    \Big \{ \! \left(\bar
      Q_L\right)_{a\alpha}\,\bm{Y}_{u}^{\rm (5D)C}\, u_R^{c} + \left(\bar
      Q_R\right)_{a\alpha}\,\bm{Y}_{u}^{\rm (5D)S}\, u_L^{c}
    \Big \} \, \Phi_{a \alpha} \hspace{4mm} \nonumber\\
    & + \frac{1}{\sqrt{2}}\, \bigg\{ \Big [ \left(\bar
      Q_L\right)_{a\alpha} \bm{Y}_{d}^{\rm (5D)C}\, \left({\cal T}_{1R}\right)^c +
    \left(\bar Q_R\right)_{a\alpha}\bm{Y}_{d}^{\rm (5D)S}\, \left({\cal
        T}_{1L}\right)^c\Big ] \left(\sigma^c\right)_{ab} \Phi_{b
      \alpha} \nonumber\\
    & +\Big [ \left(\bar
      Q_L\right)_{a\alpha} \bm{Y}_{d}^{\rm (5D)C}\, \left({\cal T}_{2R}\right)^\gamma +
    \left(\bar Q_R\right)_{a\alpha} \bm{Y}_{d}^{\rm (5D)S}\, \left({\cal
        T}_{2L}\right)^\gamma \Big ]
    \left(\sigma^\gamma\right)_{\alpha\beta} \Phi_{a \beta} \bigg\}
    +{\rm h.c.} \, \Bigg]\,.
\end{align}
Here, $\Phi$ is the Higgs bi-doublet introduced in (\ref{eq:Higgsbi}),
$\sigma^{c, \gamma}$ are the Pauli matrices,
and repeated indices are understood to be summed over. 
The Latin (Greek) letters from the beginning of
the alphabet refer to $SU(2)_L$ ($SU(2)_R$) indices. Moreover, for the case of 3 generations, 
all components of the quark multiplets are three-vectors in flavor space. 
Note that the components of the triplets in the expression above refer to the representations
\begin{equation} 
  {\cal T}_{1R}=\left(\begin{array}{c} \frac{1}{\sqrt
        2}\left({D_R^{\prime\, (-)}}_{-\frac 13}+ {\Lambda_R^{\prime\,
            (-)}}_{\frac 53}\right)\\
      \frac{i}{\sqrt 2}\left({D_R^{\prime\, (-)}}_{-\frac 13}-
        {\Lambda_R^{\prime\, (-)}}_{\frac 53}\right)\\
      {U_R^{\prime\, (-)}}_{\frac 23} \end{array}\right) , \qquad {\cal
    T}_{2R}=\left(\begin{array}{c} \frac{1}{\sqrt
        2}\left({D_R^{(+)}}_{-\frac 13}+ {\Lambda_R^{(-)}}_{\frac 53}\right)\\
      \frac{i}{\sqrt 2}\left({-D_R^{(+)}}_{-\frac 13}+
        {\Lambda_R^{(-)}}_{\frac 53}\right)\\
    {U_R^{(-)}}_{\frac 23} \end{array}\right)^T \!,
\end{equation}
which ensures that one ends up in the desired mass basis. 
After electroweak
symmetry breaking, the Yukawa couplings (\ref{eq:Sfermyuk}) give rise to
mass terms which mix different 5D fields with the same $U(1)_{\rm EM}$
charge. Similar as for the case of the gauge bosons in the custodial model,
as well as of the fermions in the minimal model, the fields that mix will
be decomposed into the same 4D fields. It is sensible to collect them into the
vectors 
\begin{equation} \label{eq:fermvec}
  \vec U \equiv
  \left(\begin{array}{c}
      u\\
      u^\prime
    \end{array}\right)\,,\quad 
  \vec u \equiv \left(\begin{array}{c}
      u^c\\
      U^\prime\\
      U
    \end{array}\right)\,,\quad 
  \vec D \equiv d\,,\quad 
  \vec d \equiv \left(\begin{array}{c}
      D\\
      D^\prime
    \end{array}\right)\,,\quad 
  \vec \Lambda \equiv \lambda\,,\quad 
  \vec \lambda \equiv
  \left(\begin{array}{c}
      \Lambda^\prime\\
      \Lambda
    \end{array}\right)\, ,
\end{equation}
which leads to a formal one-to-one correspondence between the analysis of
fermions in the minimal model and the one presented here. The resulting 
action will have exactly the same form as in the minimal model (\ref{eq:Sferm2}),
with the only differences of the extension of the sums  
to account for the charge-5/3 $\lambda$-quarks
\beq
S_{\rm ferm,2}=(\ref{eq:Sferm2})\quad {\rm with}\  \sum_{q=U,u,D,d} \to   \sum_{q=U,u,D,d,\Lambda,\lambda} \,, \quad \sum_{(Q,q)=(U,u),(D,d)} \to \sum_{(Q,q)=(U,u),(D,d),(\Lambda,\lambda)}\,
\eeq
and the higher dimension of the fermion structure, given by (\ref{eq:fermvec})
and to be compared with (\ref{eq:SMembedding}).
The Yukawa matrices appearing in the action for the custodial model are higher-dimensional
with respect to those of the minimal model and read
\begin{equation} \label{eq:yukawam}
  \bm{Y}_{\vec u}^{\rm (5D)} \equiv
  \left(\begin{array}{ccc} \bm{Y}_u^{\rm (5D)}& \frac{1}{\sqrt
        2}\bm{Y}_d^{\rm (5D)}&
      \, \frac{1}{\sqrt 2}\bm{Y}_d^{\rm (5D)}\\
      \bm{Y}_u^{\rm (5D)}& - \frac{1}{\sqrt 2}\bm{Y}_d^{\rm (5D)}& -
      \, \frac{1}{\sqrt 2}\bm{Y}_d^{\rm (5D)}
    \end{array}\right) , \qquad
  \bm{Y}_{\vec d}^{\rm (5D)} \equiv \bm{Y}_{\vec \lambda}^{\rm (5D)}
  \equiv \left(\begin{array}{cc} 
      \hspace{0.5mm} \bm{Y}_d^{\rm (5D)}&
      \hspace{0.5mm} \bm{Y}_d^{\rm (5D)}
    \end{array}\right) .
\end{equation}
Here, we have already chosen the same Yukawa matrices for the
couplings of both chirality structures ($LR$ and $RL$), setting $\bm{Y}_{\vec q}^{\rm (5D)C}=\bm{Y}_{\vec q}^{\rm (5D)S}\equiv\bm{Y}_{\vec q}^{\rm (5D)}$. 
Like for the minimal model, this should be regarded as the limit of a set-up 
with a bulk Higgs, approaching the IR brane. The generalization 
to the case of different Yukawa matrices is straightforward. The relation between 5D and
4D Yukawa matrices is still given by (\ref{eq:Y4Ddef}).
In the case of three generations, each entry of (\ref {eq:yukawam}) is a
$3\times 3$ matrix. The generalized bulk mass matrices ${\bm M}_{\vec q}$ take the
form
\begin{equation}
\label{eq:bulkmcus}
  \begin{array}{ccc}
    \bm{M}_{\vec U} \equiv \left(\begin{array}{cc}
        \bm{M}_Q&0\\
        0&\bm{M}_Q
      \end{array}\right) ,&
    \bm{M}_{\vec D} \equiv \bm{M}_Q\, , &
    \bm{M}_{\vec \Lambda} \equiv \bm{M}_Q\, ,\\[8mm]
    \bm{M}_{\vec u} \equiv \left(\begin{array}{ccc}
        \bm{M}_{u^c}&0&0\\
        0&\bm{M}_{{\cal T}_1}&0\\
        0&0&\bm{M}_{{\cal T}_2}
      \end{array}\right) ,&\quad 
    \bm{M}_{\vec d} \equiv \left(\begin{array}{cc}
        \bm{M}_{{\cal T}_2}&0\\
        0&\bm{M}_{{\cal T}_1}
      \end{array}\right) , &\quad 
    \bm{M}_{\vec \lambda} \equiv \left(\begin{array}{cc}
        \bm{M}_{{\cal T}_1}&0\\
        0&\bm{M}_{{\cal T}_2}
      \end{array}\right) ,
  \end{array}
\end{equation}
where $\bm{M}_A$ are the $3 \times 3$ bulk mass matrices of the
corresponding multiplets $A = Q,u^c,{\cal T}_1,{\cal T}_2$. We turn now to the KK decomposition
of the matter sector introduced in this Section.

\subsubsection{Kaluza-Klein Decomposition}

Due to the effort of formulating the fermion sector in the minimal
RS model in a completely general way, the KK decomposition performed there 
can be applied to the extended matter sector of the custodial model straightforwardly. 
The main formulae are identical to the ones presented in Section~\ref{sec:fermions}. 
However, the contained flavor structure is now given by (\ref{eq:fermvec}), 
(\ref{eq:yukawam}), and (\ref{eq:bulkmcus}), as well as 
(\ref{eq:custoprof}) and (\ref{eq:custoavecs}) below,
which replace the expressions of Section~\ref{sec:fermions}
for the custodial model.
Explicitly, 
\begin{itemize}
\item{the KK decomposition is given by (\ref{eq:KKdecferm}), with $(Q,q)=(U,u),(D,d),(\Lambda,\lambda)$.}
\item{the EOMs are given by (\ref{eq:fermEOM}), where $(Q,q)=(U,u),(D,d),(\Lambda,\lambda)$.}
\item{the BCs are given by (\ref{eq:bcIRrescaled}).}
\item{the orthonormality relations are given by (\ref{eq:orthonorm}), together with (\ref{eq:CS})-(\ref{eq:DCnn}).}
\newpage
\item{the determinant, determining the masses of the 4D states, is given by (\ref{eq:fermeigenvals}).}
\vspace{-0.4cm}
\item{all fermion fields, Yukawa matrices, bulk-mass matrices, profiles, and flavor vectors are to be replaced by (\ref{eq:fermvec}), 
(\ref{eq:yukawam}),(\ref{eq:bulkmcus}), (\ref{eq:custoprof}), and (\ref{eq:custoavecs}), respectively.}
\end{itemize}

The profiles in the custodial model are given by
\begin{equation}\label{eq:custoprof}
  \begin{aligned}
    \bm{C}_n^{\hspace{0.25mm} U} &\equiv{\rm diag} \big
    (\bm{C}_n^{\hspace{0.25mm} Q(+)}, \bm{C}_n^{\hspace{0.25mm} Q(-)}
    \big )\, , & \qquad \bm{C}_n^{\hspace{0.25mm} u} &\equiv{\rm diag}
    \big (\bm{C}_n^{\hspace{0.25mm} u^c(+)}, \bm{C}_n^{\hspace{0.25mm}
      {\cal T}_1(-)},
    \bm{C}_n^{\hspace{0.25mm} {\cal T}_2(-)} \big )\, ,\\[1mm]
    \bm{S}_n^{\hspace{0.25mm} U} &\equiv{\rm diag} \big
    (\bm{S}_n^{\hspace{0.25mm} Q(+)}, \bm{S}_n^{\hspace{0.25mm} Q(-)}
    \big )\, , & \qquad \bm{S}_n^{\hspace{0.25mm} u} &\equiv{\rm diag}
    \big (\bm{S}_n^{\hspace{0.25mm}u^c(+)},
    \bm{S}_n^{\hspace{0.25mm}{\cal T}_1(-)},
    \bm{S}_n^{\hspace{0.25mm} {\cal T}_2(-)} \big )\, ,\, \\[4mm]
    \bm{C}_n^{\hspace{0.25mm} D} &\equiv \bm{C}_n^{\hspace{0.25mm}
      Q(+)}\, ,& \bm{C}_n^{\hspace{0.25mm} d} &\equiv{\rm diag} \big
    (\bm{C}_n^{\hspace{0.25mm} {\cal T}_2(+)},
    \bm{C}_n^{{\cal T}_1(-)} \big )\, ,\\[1mm]
    \bm{S}_n^{\hspace{0.25mm} D}&\equiv \bm{S}_n^{\hspace{0.25mm}
      Q(+)}\, ,& \bm{S}_n^{\hspace{0.25mm} d} &\equiv{\rm diag}\big
    (\bm{S}_n^{\hspace{0.25mm} {\cal T}_2 (+)},
    \bm{S}_n^{\hspace{0.25mm} {\cal T}_1 (-)} \big )\, , \, \\[4mm]
    \bm{C}_n^{\hspace{0.25mm} \Lambda}&\equiv
    \bm{C}_n^{\hspace{0.25mm} Q(-)}\, ,& \bm{C}_n^{\hspace{0.25mm}
      \lambda}&\equiv{\rm diag} \big (\bm{C}_n^{\hspace{0.25mm} {\cal
        T}_1(-)},
    \bm{C}_n^{\hspace{0.25mm} {\cal T}_2(-)})\, ,\\[1mm]
    \bm{S}_n^{\hspace{0.25mm} \Lambda}&\equiv
    \bm{S}_n^{\hspace{0.25mm} Q(-)}\, ,& \bm{S}_n^{\hspace{0.25mm}
      \lambda}&\equiv{\rm diag} \big (\bm{S}_n^{\hspace{0.25mm} {\cal
        T}_1(-)}, \bm{S}_n^{\hspace{0.25mm} {\cal T}_2(-)} \big )\,,
  \end{aligned}
\end{equation} 
while the flavor vectors read
\begin{eqnarray} \label{eq:custoavecs}
  \begin{array}{cccccc}
    \vec a_n^{\hspace{0.35mm} U} \equiv 
    \left(\begin{array}{c} a_n^{\hspace{0.25mm} u}\\
        a_n^{\hspace{0.25mm} u^\prime}\end{array}\right) , & 
    ~~ \vec a_n^{\hspace{0.5mm} u} \equiv 
    \left(\begin{array}{c} a_n^{\hspace{0.25mm} u^c}\\ 
        a_n^{\hspace{0.25mm} U^\prime}\\ 
        a_n^{\hspace{0.25mm} U}\end{array}\right) , & ~~ 
    \vec a_n^{\hspace{0.25mm} D} \equiv a_n^{\hspace{0.25mm} d}\, , & 
    ~~ \vec a_n^{\hspace{0.35mm} d} \equiv 
    \left(\begin{array}{c}a_n^{\hspace{0.25mm} D}\\
        a_n^{\hspace{0.25mm} D^\prime}\end{array}\right) , &
    ~~ \vec a_n^{\hspace{0.25mm} \Lambda} \equiv 
    a_n^{\hspace{0.25mm} \lambda}\, , & ~~ 
    \vec a_n^{\hspace{0.5mm} \lambda} 
    \equiv \left(\begin{array}{c}a_n^{\hspace{0.25mm} \Lambda^\prime}\\
        a_n^{\hspace{0.25mm} \Lambda}\end{array}\right). \hspace{11mm}
  \end{array}
\end{eqnarray}
The $3 \times 3$ matrices $\bm{C}_n^{A(\pm)}(\phi)$ ($\bm{S}_n^{A(\pm)}(\phi)$) with
$A=Q,u^c,{\cal T}_1,{\cal T}_2$ correspond to even (odd) profiles on
the orbifold, and the superscript $(\pm)$ indicates the type of BC on
the UV brane. With some abuse of notation, the superscripts of the odd
profiles refer to the UV BCs of the associated even profiles. The
quarks present already in the minimal RS model hence all carry a $(+)$
superscript. Labels for the IR BCs are again omitted to simplify the
notation. The flavor structure is encoded in the three-component
vectors $a_n^{A}$ with $A=u, u^\prime, u^c, U^\prime, U, d, D^\prime,
D, \lambda, \Lambda^\prime, \Lambda$, which are then combined into the
larger flavor vectors defined above. As stated below (\ref{eq:KKdecferm}), 
the spinors on the right hand side of the KK decomposition $q_{L}^{(n)}(x)$ 
and $q_{R}^{(n)}(x)$ are still 4D spinors in the mass basis, and the index $n$ labels 
the different mass eigenstates with masses $m_n$, \ie, $m_1=m_u,\ m_2=m_c, \ m_3 = m_t$, \etc \ 
in the case of up-type quarks.

\subsubsection{Bulk Profiles}
\label{sec:fermionprofilescusto}

The form of the solutions $\big (C_n^{A(+)} (\phi) \big )_i$
and $ \big (S_n^{A(+)} (\phi) \big )_i$ associated with bulk mass
parameters $M_{A_i}$, has already been given in (\ref{eq:fermprofiles}). The functions 
$\big (C_n^{A(-)} (\phi) \big )_i$ and $\big (S_n^{A(-)} (\phi) \big )_i$ can be 
derived in a similar fashion by requiring a Dirichlet BC for the even mode, 
$\big(C_n^{A(-)}(0) \big )_i = 0$, to account for the additional twist of the
non-SM-like fermions on the UV boundary. In consequence, the treatment is
analogous to that of the odd modes of the SM-like fermions, for which
$\big(S_n^{A(+)} (0) \big )_i = 0$. In the following we will drop the label $A$ 
and the index $i$ since they should be clear from the context. In $t$-notation, 
one finds in the bulk (\ie, for $t \in ]\epsilon, 1[$)
\begin{equation} \label{fermprofiles}
  \begin{split}
    C_n^{(\pm)}(\phi) &= {\cal N}_n^{(\pm)}(c)\,\sqrt{\frac{L\epsilon
        t}{\pi}}\, f_n^{(\pm)+}(t,c) \,, \\
    S_n^{(\pm)}(\phi) &= \pm {\cal N}_n^{(\pm)}(c)\,\sgn(\phi)\,
    \sqrt{\frac{L\epsilon t}{\pi}}\,f_n^{(\pm)-}(t,c) \,,
  \end{split}
\end{equation} 
where the overall ``$+$'' sign entering the $Z_2$-odd profiles holds
if $c = c_Q \equiv +M_{Q}/k$ refers to the bi-doublet, while the
``$-$'' sign applies in the case of $c = c_A \equiv -M_A/k$, where $A
= u^c,{\cal T}_1,{\cal T}_2$. The functions $f^{(\pm)\pm}_n (t,c)$ are 
given by
\begin{equation} \label{eq:fplmi}
  \begin{split}
    f_n^{(+)\pm}(t,c) & =
    J_{-\frac12-c}(x_n\epsilon)\,J_{\mp\frac12+c}(x_n t) \pm
    J_{+\frac12+c}(x_n\epsilon)\,J_{\pm\frac12-c}(x_n t) \,, \\
    f_n^{(-)\pm}(t,c) & =
    J_{+\frac12-c}(x_n\epsilon)\,J_{\mp\frac12+c}(x_n t) \mp
    J_{-\frac12+c}(x_n\epsilon)\,J_{\pm\frac12-c}(x_n t) \,.
  \end{split}
\end{equation} 
They satisfy the relations 
\begin{equation} \label{eq:frel}
  f_n^{(+)+}(t,c)=f_n^{(-)-}(t,-c)\,, \qquad 
  f_n^{(+)-}(t,c)=-f_n^{(-)+}(t,-c)\,.
\end{equation}
The orthonormality conditions (\ref{eq:orthonorm}) imply 
\begin{equation} 
  2\int_\epsilon^1\!dt\,t \left[ f_n^{(a)\pm}(t,c)
  \right]^2 = \frac{1}{\big [ {\cal N}_n^{(a)}(c) \big ]^2} \pm
  \frac{f_n^{(a)+}(1,c)\,f_n^{(a)-}(1,c)}{x_n} \,, 
\end{equation} 
where $a=\pm$, from which we derive
\begin{equation} \label{eq:fermnormcus}
  \begin{split}
    \big[{\cal N}_n^{(a)}(c)\big]^{-2} & = \left[ f_n^{(a)+}(1,c)
    \right]^2 + \left[ f_n^{(a)-}(1,c) \right]^2 \\
    & \phantom{xx} -
    \frac{2c}{x_n}\,f_n^{(a)+}(1,c)\,f_n^{(a)-}(1,c) - \epsilon^2
    \left ( \left[ f_n^{(a)+}(\epsilon,c) \right]^2 + \left[
        f_n^{(a)-}(\epsilon,c) \right]^2 \right ) .
  \end{split}
\end{equation}
This extends the result (\ref{eq:fermnorm}) of Section~\ref{sec:fermions} to the
case of $Z_2$-odd profiles with a non-zero value on the UV boundary. For
the special case where $c+1/2$ is an integer, the profiles must be
again obtained from the above relations by a limiting procedure.

For the SM fermions, it is again a very good approximation to perform
an expansion in $x_n\ll 1$. Using the
results (\ref{eq:SMfermions}), in combination with
(\ref{eq:frel}), we obtain
\begin{equation} \label{eq:profileexp}
\begin{split}  
  C_n^{(+)}(\phi) \approx \sqrt{\frac{L\epsilon}{\pi}}\, F(c) \, t^{c}
  \, , \qquad S_n^{(+)}(\phi) \approx
  \pm\sgn(\phi)\,\sqrt{\frac{L\epsilon}{\pi}}\; x_n F(c)\;
  \frac{t^{1+c} - \epsilon^{1+2c}\,t^{-c}} {1+2c} \,,
  \quad \\[1mm]
  C_n^{(-)}(\phi) \approx -\sqrt{\frac{L\epsilon}{\pi}}\; x_n F(-c)\;
  \frac{t^{1-c} - \epsilon^{1-2c}\,t^{c}} {1-2c} \, , \qquad
  S_n^{(-)}(\phi) \approx \pm\sgn(\phi) \sqrt{\frac{L\epsilon}{\pi}}\,
  F(-c) \, t^{-c} \,, 
\end{split}
\end{equation}
with the zero-mode profile $F(c)$ as defined in (\ref{eq:Fdef}).
Remember that this profile is exponentially small for UV-localized fermions, 
while it is of ${\cal O}(1)$ for IR-localized fields.
Moreover, note that the profiles $C_n^{(+)} (\phi)$ and
$S_n^{(-)} (\phi)$ are of ${\cal O}(1)$, while $C_n^{(-)} (\phi)$ and
$S_n^{(+)} (\phi)$ are of ${\cal O} (v/M_{\rm KK})$. As we will
explain in detail in the next section, this feature will be important for
the (partially) shielding of the $Z b_L \bar b_L$ and $Z d_L^{\hspace{0.25mm} i}
\bar d^{\hspace{0.25mm} j}_L$ vertices from corrections due to mixing
of zero-mode quarks with their KK excitations.

\subsection{Gauge-Boson Interactions with Fermions}
\label{sec:custodialprotection}

In this section we examine explicitly, how a protection of the left-handed
down-type couplings of the $Z$ boson can be achieved.
A protection from both gauge-boson as well as fermion corrections is possible
 by choosing an appropriate embedding of the fermions into the enlarged 
gauge group in the bulk. Later we will also derive the four-fermion charged-current
interactions and show that a custodial protection is not at
work in this case. First, let us give the covariant derivative of
the custodial RS model in the
UV basis. It reads
\begin{align}
\label{eq:covdcus}
  \begin{split}
    D_\mu=\partial_\mu & - i \, \frac{{g_L}_5}{\sqrt 2}
    \left(L_\mu^+\hspace{0.25mm} T_L^++L_\mu^-\hspace{0.25mm}
      T_L^-\right) + i \, \frac{{g_R}_5}{\sqrt 2} \left(R_\mu^+
      \hspace{0.25mm}
      T_R^++R_\mu^- \hspace{0.25mm} T_R^-\right) \\
    & - i \, {g_Z}_5\, Q_Z \hspace{0.25mm} Z_\mu - i \, {g_{Z'}}_5 \,
    Q_{Z^{\, \prime}} \hspace{0.25mm} Z_\mu^\prime - i \, e_5
    \hspace{0.25mm} Q A_\mu \, .
  \end{split}
\end{align}
The fundamental $Z$-boson couplings to fermionic currents are given by
\begin{equation} \label{eq:gQZZp}
  \begin{aligned}
    & g_Z=\sqrt{g_L^2+g_Y^2}\,, && \qquad g_{Z^{\, \prime}} =
    \sqrt{g_R^2+g_X^2}\,, \\
    & Q_Z =T_L^3 - \frac{g_Y^2}{g_Z^2}Q\,, && \qquad Q_{Z^{\,
        \prime}}= -T_R^3 - \frac{g_X^2}{g_{Z^{\, \prime}}^2}Y\,,
  \end{aligned}
\end{equation}
where we have replaced 5D couplings by 4D couplings as in (\ref{eq:g4def}).
At this point we have to specify the form of the $SU(2)_{L,R}$
generators when acting on different fermion representations. Acting on
fermion bi-doublets, the generators $T_{L,R}^i$ with $i=1,2,3$ are
given by the Pauli matrices in the standard convention
times a factor of $1/2$, as defined below (\ref{eq:covD1}). As usual, we define 
$T_{L,R}^\pm=T_{L,R}^1 \pm i \, T_{L,R}^2$. If, on the other hand, 
the generators act on $SU(2)_{
L,R}$ triplets, they read explicitly
\begin{equation}
  T_{L,R}^+=\begin{pmatrix} \,0\, & \sqrt 2 & 0 \\
    \,0\, & 0 & \sqrt 2 \\
    \,0\, & 0 & 0
  \end{pmatrix} ,\qquad
  T_{L,R}^-=\begin{pmatrix} 0 & 0\, & 0\, \\
    \sqrt 2 & 0\, & 0\, \\
    0 & \sqrt 2\, & 0\,
  \end{pmatrix} ,\qquad
  T_{L,R}^3=\begin{pmatrix} \,1\; & 0 & 0 \\
    \,0\; & 0 & 0 \\
    \,0\; & 0 & -1
  \end{pmatrix} .
\end{equation}
Note that the current-operators in interaction terms involve a trace with respect to the
fundamental gauge indices, and that again the $T_R^i$ act to the
left. When they are not needed, we will often drop the subscripts $L,R$ in the following.
For the calculation of the corrections to the
quark-mixing matrices (which will be performed in Section~\ref{sec:4Fint}) it will turn out to be 
useful to introduce the vector of couplings
\begin{equation}
  \vec{g}_Z = \begin{pmatrix} g_Z \hspace{0.25mm} Q_Z \\
    g_{Z^\prime} \hspace{0.25mm} Q_{Z^{\, \prime}} \end{pmatrix} ,
\end{equation}
as well as the charged-current vectors
\begin{align} \label{eq:currents}
    \vec J_{W_Q}^{\, \mu\pm}&=\frac 1{\sqrt 2}\, \big(g_L\,
    \text{Tr}\left[\bar Q\,\gamma^\mu\,T^\pm\,Q\right],
    g_R\, \text{Tr}\left[\bar Q\,\gamma^\mu\,Q\, T^\pm\right]\big)\,,\nonumber\\
    \vec J_{W_{\cal T}}^{\, \mu\pm}& = \frac 1{\sqrt 2}\, \big( g_L
    \bar {\cal T}_1\,\gamma^\mu\,T^\pm\,{\cal T}_1\, ,\, g_R\,
    \text{Tr}\left[\bar {\cal T}_2\,\gamma^\mu\,{\cal T}_2
      \,T^\pm\right] \big)\,,
\end{align}
which will multiply $\left(L_\mu^\pm\ , R_\mu^\pm \right)^T$ from the left.

Before going into the interactions of fermions with massive gauge bosons, 
note that in the custodial model the interactions with gauge bosons that have massless zero modes (which can be worked 
out from (\ref{eq:covdcus})) are in complete analogy to the corresponding expressions presented in Section~\ref{sec:gaugecouplings}.
The fermion structure present in the latter section is to be trivially replaced with the objects of the custodial model, presented in 
Section~\ref{sec:cusfermions}. In addition to the interactions with up- and down-type quarks, there will
also be photon as well as gluon interactions with the charge-5/3 $\lambda$-quarks.

\subsubsection{Custodial Protection: Gauge-Boson Contributions}

Using (\ref{eq:expprof}) and (\ref{eq:vecA0a}), we find that the
coupling of the $Z$ boson to a current of $q$-quarks is proportional to
\begin{align} \label{eq:Zqq}
  \big( \vec{g}_Z^{\, q} \big)^{\mkern-3mu T} \vec{\chi}^{\, Z}_0 (\phi)
  = \frac{g_Z \hspace{0.25mm} Q_Z^{\, q}}{\sqrt{2 \pi}} \left \{ 1 +
    \frac{m_Z^2}{4 M_{\rm KK}^2} \left [ \,- 2 L \, t^2\,
      \omega_Z^q + 1 - \frac{1}{L}  + 2 \, t^2 \!\left( \frac12 - \ln t \right) \right ] \right
  \} + \ord\left(\frac{m_Z^4}{\Mkk^4}\right) , \hspace{6mm}
\end{align}
with 
\begin{equation} \label{eq:omegaZ_definition}
\omega_Z^q = 1 - \frac{s_Z}{c_Z} \, \frac{g_{Z^{\, \prime}}
\hspace{0.25mm} Q^q_{Z^{\, \prime}} }{ g_Z \hspace{0.25mm} Q_Z^q
}\,.
\end{equation}
This is an important result, as it allows us to understand the
custodial protection of the $Z b_L\bar b_L$ vertex. Note that the first
term in the square bracket in (\ref{eq:Zqq}), which is enhanced by the volume factor $L$, gets
modified by the prefactor $\omega_Z^q$, \ie, a combination of the
fundamental charges and couplings. While $\omega_Z^q = 1$ for all
quarks in the minimal RS model, it is possible to arrange for
\beq \label{eq:custodialQ}
  \omega_Z^{b_L} = 0 \quad \Longleftrightarrow \quad g_Z \hspace{0.25mm}
  Q_Z^{b_L} = \frac{s_Z}{c_Z} \, g_{Z^{\, \prime}} \hspace{0.5mm}
  Q^{b_L}_{Z^{\, \prime}} \;,
\eeq
by virtue of the extension of the gauge group in the bulk. It is
interesting to observe that the interplay of neutral gauge bosons can only 
protect leading term in $L$, while no such mechanism is available for the subleading
terms in $L$. Those arise from the fact that the profiles
$\chi_0^{(\pm)} (t)$ obey different BCs, which represents an irreducible 
source of $P_{LR}$ symmetry breaking.
Numerically, the corrections to the $Z b_L \bar b_L$ vertex arising
from the gauge sector are thus suppressed by a factor of $L\approx 37$
in the $SU(2)_L \times SU(2)_R \times P_{LR}$ custodial model relative
to the minimal RS model.

Formula (\ref{eq:omegaZ_definition}) can be recast into the form
\begin{equation} \label{eq:omegaZ}
  \omega_Z^q = \frac{c_w^2}{2 g_L^2} \, \frac{ \, (g_L^2 + g_R^2) \,
    (T_L^{3 \, q} + T_R^{3 \, q}) + (g_L^2 - g_R^2) \, (T_L^{3 \, q} -
    T_R^{3 \, q}) \, } { T_L^{3 \, q} - s_w^2 \hspace{0.25mm} Q_q } \,,
\end{equation}
which allows to read off that the choices 
\begin{equation} \label{PC}
  T_L^{3 \, q} = T_R^{ 3 \, q} = 0 \qquad (P_C \ \text{symmetry}) \,,
\end{equation} 
and 
\begin{equation} \label{PLR}
  g_L = g_R\,, \qquad T_L^{3 \, q} = - T_R^{ 3 \, q} \qquad
  (P_{LR} \ \text{symmetry}) \,,
\end{equation}
are suitable to protect the $Z$-boson vertices from receiving
$L$-enhanced corrections. Since the representation
(\ref{eq:multiplets}) features $T_L^{3\, d_L} = -T_R^{3\, d_L} = -1/2$
and $T_L^{3\, u_R} = T_R^{3\, u_R} = 0$, it is then immediately clear
that the $Z d_L^{\hspace{0.25mm} i} \bar d_L^{\hspace{0.25mm} j}$ and
$Z u_R^{\hspace{0.25mm} i} \bar u_R^{\hspace{0.25mm} j}$ vertices are
protected to leading order in $L$ by the $P_{LR}$ and $P_C$
symmetries, respectively. On the other hand, the $Z
d_R^{\hspace{0.25mm} i} \bar d_R^{\hspace{0.25mm} j}$ and $Z
u_L^{\hspace{0.25mm} i} \bar u_L^{\hspace{0.25mm} j}$ vertices do
receive $L$-enhanced corrections, since the corresponding quantum
numbers are $T_L^{3\, d_R} = 0$, $T_R^{3\, d_R} = 1$ and $T_L^{3\,
u_L} = T_R^{3\, u_L} = 1/2$. We also add that devising the quark
sector as in (\ref{eq:multiplets}) implies $\omega_Z^{b_R} > 0$, so
that the shift in the right-handed $Z$-boson coupling to bottom quarks
arising from the gauge-boson sector is predicted to be strictly
negative. This suggests that the well-known tension in the global fit
to the $Z \to b \bar b$ pseudo observables cannot be softened in the
model under considerations. We will come back to this point in Section
\ref{sec:bpseudo}.

\subsubsection[Fermion Couplings to the $Z$ Boson]{Fermion Couplings to the $\bm{Z}$ Boson}

We will now identify all phenomenologically relevant RS contributions
to weak neutral gauge interactions of quark currents at relative order $v^2/\Mkk^2$ in the custodial 
model. The $Z$-boson
couplings to left- and right-handed quarks can be read off from the
Lagrangian
\begin{equation} \label{eq:Zffcus}
  {\cal L}_{\rm 4D} \, \ni \, \frac{g_L}{c_w} \left[ 1 +
    \frac{m_Z^2}{4\Mkk^2} \left( 1 - \frac{1}{L} \right) \right]  Z_\mu \;\times
  \sum_{q,m,n} \Big[ \big( g_L^{\, q} \big)_{m n} \left ( \bar
    q_{L}^{\hspace{0.25mm} m} \gamma_\mu q_{L}^n \right ) + \big( g_R^{\,
    q} \big)_{mn} \left ( \bar q_{R}^{\hspace{0.25mm} m} \gamma_\mu
    q_{R}^n \right ) \Big]\,,
\end{equation}
where the prefactor accounts for a universal correction due to the
$t$-independent terms in (\ref{eq:Zqq}), identical to (\ref{eq:Zff}).  The left- and right-handed
couplings $\bm{g}_{L,R}^q$ are infinite-dimensional matrices in the
space of quark modes, and can be parametrized as
\begin{eqnarray}\label{gLR}
  \begin{split}
    \bm{g}_L^q &= \left(T_L^{3 \, q_L} - s_w^2 Q_q\right) \left[
      \bm{1} - \frac{m_Z^2}{2\Mkk^2} \left ( \omega_Z^{q_L} L
        \hspace{0.5mm} \bm{\Delta}_Q -\bm{\Delta}'_Q \right )\right]
    -\bm{\delta}_Q + \frac{m_Z^2}{2\Mkk^2} \,\left (
      \frac{c_w^2}{g_L^2} \, L \hspace{0.5mm} \bm{\varepsilon}_Q -
      \bm{\varepsilon}'_Q \right ) ,
    \hspace{6mm} \\
    \bm{g}_R^q &= -s_w^2 Q_q \left[ \bm{1} - \frac{m_Z^2}{2\Mkk^2}
      \left ( \omega_Z^{q_R} L \hspace{0.5mm} \bm{\Delta}_q -
        \bm{\Delta}'_q \right ) \right ] +\bm{\delta}_q
    -\frac{m_Z^2}{2\Mkk^2} \left ( \frac{c_w^2}{g_L^2} \, L
      \hspace{0.5mm} \bm{\varepsilon}_q - \bm{\varepsilon}'_q \right)\,.
  \end{split}
\end{eqnarray}
The labels of the charges appearing in these expressions (as well as above)
indicate that the quantum numbers of the corresponding zero modes are to be employed. 
In $\bm{g}_L^q$ they read $T_L^{3
  \, u_L} (= T_L^{3 \, u}) = T_R^{3 \, u_L} (= T_R^{3 \, u}) = T_R^{3
  \, d_L} (= T_R^{3 \, d}) = 1/2$ and $T_L^{3 \, d_L} (= T_L^{3 \, d})
=-1/2$, whereas for $\bm{g}_R^q$ one has $T_L^{3 \, u_R} (= T_L^{3
  \, u^c}) = T_R^{3 \, u_R} (= T_R^{3 \,u^c}) = T_L^{3 \, d_R} (=
T_L^{3 \, D}) = 0$ and $T_R^{3 \, d_R} (= T_R^{3 \, D}) = 1$. The
quoted numerical values correspond to the choice
(\ref{eq:multiplets}). We do not consider the sector of $\lambda$ and
$\Lambda^{(\prime)}$ quarks at this point, as these fields do not
possess zero modes. The $L$-enhanced term proportional to
$\omega_Z^{q}$ vanishes for the assignments (\ref{PC}) and
(\ref{PLR}), making the custodial protection explicit. Following
(\ref{eq:gLR}), we have split the corrections to the
$Z$-boson couplings into leading contributions in the ZMA, denoted by
$\bm{\Delta}^{(\prime)}_{Q,q}$, and subleading ones, parametrized by
$\bm{\varepsilon}^{(\prime)}_{Q,q}$. The elements of the leading-order
matrices ${\bm \Delta}^{(\prime)}_{Q,q}$ are defined as
\begin{align}\label{overlapintsLO}
    \left( \Delta_Q \right)_{mn} &= \frac{2\pi}{L\epsilon}
    \int_\epsilon^1\!dt\,t^2 \, \Big [ \vec a_m^{\hspace{0.25mm} Q
      \dagger}\,\bm{C}_m^{Q}(t)\, \bm{C}_n^{Q}(t)\,\vec
    a_n^{\hspace{0.25mm} Q} + \vec a_m^{\hspace{0.25mm}
      q\dagger}\,\bm{S}_m^{q}(t)\,
    \bm{S}_n^{q}(t)\,\vec a_n^{\hspace{0.25mm}q} \Big] , \nonumber\\
    \left( \Delta_q \right)_{mn} &= \frac{2\pi}{L\epsilon}
    \int_\epsilon^1\!dt\,t^2 \, \Big[ \vec
    a_m^{\hspace{0.25mm}q\dagger}\,\bm{C}_m^{q}(t)\,
    \bm{C}_n^{q}(t)\,\vec a_n^{\hspace{0.25mm}q} + \vec
    a_m^{\hspace{0.25mm} Q\dagger}\,\bm{S}_m^{Q}(t)\,
    \bm{S}_n^{Q}(t)\,\vec a_n^{\hspace{0.25mm}Q} \Big] , \nonumber\\
    \left( \Delta'_Q \right)_{mn} &= \frac{2\pi}{L\epsilon}
    \int_\epsilon^1\!dt\,t^2 \left( \frac12 - \ln t \right) \Big [
    \vec a_m^{\hspace{0.25mm}Q\dagger}\,\bm{C}_m^{Q}(t)\,
    \bm{C}_n^{Q}(t)\,\vec a_n^{\hspace{0.25mm}Q} + \vec
    a_m^{\hspace{0.25mm}q\dagger}\,\bm{S}_m^{q}(t)\,
    \bm{S}_n^{q}(t)\,\vec a_n^{\hspace{0.25mm}q} \Big] , \nonumber\\
    \left( \Delta'_q \right)_{mn} &= \frac{2\pi}{L\epsilon}
    \int_\epsilon^1\!dt\,t^2 \left( \frac12 - \ln t \right) \Big[ \vec
    a_m^{\hspace{0.25mm} q\dagger}\,\bm{C}_m^{q}(t)\,
    \bm{C}_n^{q}(t)\,\vec a_n^{\hspace{0.25mm} q} + \vec
    a_m^{\hspace{0.25mm} Q\dagger}\,\bm{S}_m^{Q}(t)\,
    \bm{S}_n^{Q}(t)\,\vec a_n^{\hspace{0.25mm} Q} \Big] ,
\end{align} 
while the elements of the matrices ${\bm\varepsilon}^{(\prime)}_{Q,q}$
take the form
\begin{eqnarray}\label{overlapintsNLO}
  \begin{split}
    \left( \varepsilon_Q \right)_{mn} &= \frac{2\pi}{L\epsilon}
    \int_\epsilon^1\!dt\,t^2 \, \Big[ \vec a_m^{\hspace{0.25mm} Q
      \dagger}\,\bm{C}_m^{Q}(t) \, \Big \{g_L^2 \hspace{0.5mm}
    \Big(T_L^{3 \, q_L}{{\bm 1}-{\bm T}_L^{3 \, Q}}\Big)+
    g_R^2\hspace{0.5mm} \Big(T_R^{3 \, q_L}{{\bm 1}-{\bm T}_R^{3 \,
        Q}}\Big)\Big\}\,
    \bm{C}_n^{Q}(t)\, \vec a_n^{\hspace{0.25mm} Q}\\
    &\qquad \qquad \qquad + \vec
    a_m^{\hspace{0.25mm}q\dagger}\,\bm{S}_m^{q}(t)\, \Big \{g_L^2
    \hspace{0.5mm} \Big (T_L^{3 \, q_L}{\bm 1}-{\bm T}_L^{3 \, q}\Big
    )+ g_R^2 \hspace{0.5mm} \Big (T_R^{3 \, q_L}{\bm 1}-{\bm T}_R^{3\,
      q}\Big
    )\Big \} \, \bm{S}_n^{q}(t)\, \vec a_n^{\hspace{0.25mm}q} \Big] , \\
    \left( \varepsilon_q \right)_{mn} &= \frac{2\pi}{L\epsilon}
    \int_\epsilon^1\!dt\,t^2 \, \Big[ \vec a_m^{\hspace{0.25mm} q
      \dagger}\,\bm{C}_m^{q}(t) \, \Big \{g_L^2 \hspace{0.5mm} {\bm
      T}_L^{3 \, q}- g_R^2\hspace{0.5mm} \Big(T_R^{3 \, q_R}{{\bm
        1}-{\bm T}_R^{3 \,
        q}}\Big)\Big\}\, \bm{C}_n^{q}(t)\, \vec a_n^{\hspace{0.25mm} q}\\
    &\qquad \qquad \qquad + \vec
    a_m^{\hspace{0.25mm}Q\dagger}\,\bm{S}_m^{Q}(t)\, \Big \{g_L^2
    \hspace{0.5mm} {\bm T}_L^{3 \, Q}- g_R^2 \hspace{0.5mm} \Big
    (T_R^{3 \, q_R}{\bm 1}-{\bm T}_R^{3\, Q}\Big )\Big \} \,
    \bm{S}_n^{Q}(t)\,
    \vec a_n^{\hspace{0.25mm}Q} \Big] , \hspace{6mm} \\
    \left( \varepsilon'_Q \right)_{mn} &= \frac{2\pi}{L\epsilon}
    \int_\epsilon^1\!dt\,t^2 \left( \frac12 - \ln t \right) \Big[ \vec
    a_m^{\hspace{0.25mm} Q\dagger}\,\bm{C}_m^{Q}(t)\, \Big(T_L^{3 \,
      q_L}{\bm 1}-{\bm T}_L^{3 \, Q}\Big) \bm{C}_n^{Q}(t)\, \vec
    a_n^{\hspace{0.25mm} Q}\\
    &\hspace{46.5mm} + \vec a_m^{\hspace{0.25mm}
      q\dagger}\,\bm{S}_m^{q}(t)\, \Big(T_L^{3\, q_L}{\bm 1}-{\bm
      T}_L^{3\,
      q}\Big)\bm{S}_n^{q}(t)\, \vec a_n^{\hspace{0.25mm}q} \Big] , \\
    \left( \varepsilon'_q \right)_{mn} &= \frac{2\pi}{L\epsilon}
    \int_\epsilon^1\!dt\,t^2 \left( \frac12 - \ln t \right) \Big[ \vec
    a_m^{\hspace{0.25mm}q\dagger}\,\bm{C}_m^{q}(t)\, {\bm T}_L^{3\,
      q}\, \bm{C}_n^{q}(t)\,\vec a_n^{\hspace{0.25mm} q} + \vec
    a_m^{\hspace{0.25mm} Q\dagger}\,\bm{S}_m^{Q}(t)\, {\bm
      T}_L^{3\,Q}\, \bm{S}_n^{Q}(t)\,\vec a_n^{\hspace{0.25mm} Q}
    \Big] \,. \hspace{11mm}
  \end{split}
\end{eqnarray} 
Finally, the elements of the matrices $\bm{\delta}_{Q,q}$, which arise
because of the non-orthonormality of the quark profiles and describe
mixings between the different multiplets, read
\begin{align} \label{eq:delta2}
    \left( \delta_Q \right)_{mn} &= \frac{2\pi}{L\epsilon}
    \int_\epsilon^1\!dt \, \Big[ \vec a_m^{\hspace{0.25mm} Q
      \dagger}\,\bm{C}_m^{Q}(t)\, \Big(T_L^{3 \, q_L}{\bm 1}-{\bm
      T}_L^{3\,
      Q}\Big) \, \bm{C}_n^{Q}(t)\,\vec a_n^{\hspace{0.25mm} Q} \nonumber \\
    & \hspace{2.1cm} + \vec a_m^{\hspace{0.25mm}
      q\dagger}\,\bm{S}_m^{q}(t) \, \, \Big (T_L^{3 \, q_L}{\bm
      1}-{\bm T}_L^{3\, q} \Big) \, \bm{S}_n^{q}(t)\,\vec
    a_n^{\hspace{0.25mm} q}
    \Big ] , \nonumber \\
    \left( \delta_q \right)_{mn} &= \frac{2\pi}{L\epsilon}
    \int_\epsilon^1\!dt \, \Big [ \vec a_m^{\hspace{0.25mm}
      q\dagger}\,\bm{C}_m^{q}(t)\, {\bm T}_L^{3\,q}\,
    \bm{C}_n^{q}(t)\,\vec a_n^{\hspace{0.25mm} q} + \vec
    a_m^{\hspace{0.25mm} Q\dagger}\,\bm{S}_m^{Q}(t)\, {\bm T}_L^{3\,
      Q}\, \bm{S}_n^{\hspace{0.25mm} Q}(t)\,\vec a_n^{\hspace{0.25mm}
      Q} \Big] .
\end{align} 
In the expressions above we have used the charge matrices ${\bm
T}_{L,R}^{3 \, Q,q}$, defined as
\begin{equation}
  \begin{split}
    {\bm T}_{L,R}^{3\, U} & =\left(\begin{array}{cc}
        T_{L,R}^{3 \, u}&0\\
        0&T_{L,R}^{3 \, u^\prime}
      \end{array}\right) , \qquad 
    {\bm T}_{L,R}^{3 \, u}=\left(\begin{array}{ccc}
        T_{L,R}^{3 \, u^c}&0&0\\
        0&T_{L,R}^{3 \, U^\prime}&0\\
        0&0&T_{L,R}^{3 \, U}
      \end{array}\right) , \\
    {\bm T}_{L,R}^{3 \, D} & =T_{L,R}^{3 \, d}\, , \qquad \qquad
    \qquad \phantom{i} {\bm T}_{L,R}^{3 \, d}=\left(\begin{array}{cc}
        T_{L,R}^{3 \, D}&0\\
        0&T_{L,R}^{3 \, D^\prime}
      \end{array}\right) .
  \end{split}
\end{equation}
One can easily check that for our choice (\ref{eq:multiplets}), the
quantities $\bm{\varepsilon}^{(\prime)}_{Q,q}$ are indeed suppressed
by $v^2/\Mkk^2$ with respect to the matrices
$\bm{\Delta}^{(\prime)}_{Q,q}$. Note that with this embedding, 
the matrices ${\bm T}_{L,R}^{3 \, u}$ vanish identically.

\subsubsection{Custodial Protection: Fermionic Contributions}

Finally, we want to have a look at the custodial protection of the $Z
b_L \bar b_L$ vertex from effects arising from quark mixing,
parametrized by the matrices $\bm{\delta}_{Q,q}$. These objects scale in general as
$v^2/\Mkk^2$, but as they appear with an $\ord(1)$ coefficient in
(\ref{gLR}) they contribute at the same order as the matrices
$\bm{\Delta}^{(\prime)}_{Q,q}$. In the case of the left-handed
down-type quark-sector, one has
\begin{equation} \label{deltaD}
  \begin{split}
    \left( \delta_D \right)_{mn} &= \frac{2\pi}{L\epsilon}
    \int_\epsilon^1\!dt \, \Big[a_m^{ D\dagger}\,\bm{S}_m^{{\cal
        T}_2(+)}(t)\,
    \Big(T_L^{3\hspace{0.25mm}d_L}-T_L^{3\hspace{0.25mm}D}\Big)
    \bm{S}_n^{{\cal T}_2(+)}(t)
    \,a_n^{ D}\\
    &\hspace{22mm}+a_m^{D^\prime\dagger}\,\bm{S}_m^{{\cal
        T}_1(-)}(t)\, \Big(T_L^{3 \, d_L}-T_L^{3 \, D^\prime}\Big)
    \bm{S}_n^{{\cal T}_1(-)}(t)
    \,a_n^{D^\prime} \Big]\\
    &=- \frac 1 2 \, \frac{2\pi}{L\epsilon} \int_\epsilon^1\!dt \,
    \Big [a_m^{D\dagger}\,\bm{S}_m^{{\cal T}_2(+)}(t) \,
    \bm{S}_n^{{\cal T}_2(+)}(t)\,a_n^{D}
    -a_m^{D^\prime\dagger}\,\bm{S}_m^{{\cal T}_1(-)}(t)\,
    \bm{S}_n^{{\cal T}_1(-)}(t)\,a_n^{D^\prime} \Big] \,, \hspace{6mm}
  \end{split}
\end{equation} 
where in the second step we have inserted the quantum numbers
corresponding to our choice (\ref{eq:multiplets}) of multiplets. 

The relative sign between the two terms in the second line of
(\ref{deltaD}) suggests that also for the corrections due to quark
mixing, a custodial protection mechanism could be at work. To see analytically if
this is indeed the case, let us derive the ZMA expression for ${\bm
\delta}_D$. Using the approximate expressions (\ref{eq:profileexp}),
the system of equations (\ref{eq:bcIRrescaled}) can be brought into
the form
\begin{equation} \label{IRBCZMA}
  \frac{\sqrt 2\, m_n}{v} \, {\hat a}_n^d = {\bm Y}_d^{\rm eff} \, {\hat
    a}_n^D \,, \qquad \frac{\sqrt 2\, m_n}{v} \, {\hat a}_n^D = \left (
    {\bm Y}_d^{\rm eff} \right )^\dagger {\hat a}_n^d \,,
\end{equation}
and 
\begin{equation} \label{IRBC3}
  \hat a_n^{D^\prime} = x_n \, {\rm diag} \; \Big (
  F^{-1}(c_{{\cal T}_{2 i}}) \hspace{0.5mm} F^{-1}(-c_{{\cal T}_{1 i}})
  \Big ) \, \hat a_n^{D} \,,
\end{equation}
where we have defined the effective Yukawa couplings
\begin{equation}\label{eq:effY}
  ({\bm Y}_d^{\rm eff} )_{ij} \equiv F(c_{Q_i}) \left (Y_d \right )_{ij}
  F(c_{{\cal T}_{2 j}}) \,,
\end{equation}
in analogy to (\ref{eq:Yueff}). Moreover, the rescaled vectors $\hat a_n^A \equiv \sqrt{2} \, a_n^A$ with $A
=d,D,D^\prime$, obey the normalization conditions
\begin{equation}
  \hat a_n^{D \hspace{0.5mm} \dagger} \, \hat a_n^{D} = 1 \, , \qquad
  \hat a_n^{d \hspace{0.5mm} \dagger} \, \hat a_n^{d} + \hat
  a_n^{D^\prime \hspace{0.5mm} \dagger} \, \hat a_n^{D^\prime} = 1 \,.
\end{equation}
We obtain from (\ref{IRBCZMA}) the LO equalities
\begin{equation}\label{eq:EVeq1}
  \Big ( m_n^2\,\bm{1} - \frac{v^2}{2} \, \bm{Y}_d^{\rm eff} \left
    (\bm{Y}_d^{\rm eff} \right )^\dagger \Big ) \, \hat a_n^{d} = 0 \,,
  \qquad \Big ( m_n^2\,\bm{1} - \frac{v^2}{2} \, \left (\bm{Y}_d^{\rm
      eff} \right )^\dagger \bm{Y}_d^{\rm eff} \Big) \, \hat a_n^{D} = 0 \,,
\end{equation}
and the mass eigenvalues are the solutions to the equation 
\begin{equation}\label{eq:EVeq2}
  \det\left( m_n^2\,\bm{1} - \frac{v^2}{2} \, \bm{Y}_d^{\rm eff} \left
      (\bm{Y}_d^{\rm eff} \right )^\dagger \right) = 0 \,.
\end{equation}
This exactly resembles (\ref{eq:minm}) and implies that to LO in $v/M_{\rm KK}$ the values $m_n$
are unaffected by the presence of the $D^\prime$ quarks embedded in 
the multiplet ${\cal T}_1$. In the ZMA, but not in general, the vectors $\hat a_n^{d}$ and
$\hat a_n^{D}$ belonging to different $n$ are orthogonal on each
other.

In complete analogy to the discussion below (\ref{eq:minm}),
the eigenvectors $\hat{a}_n^{d}$ and $\hat{a}_n^{D}$ of
the matrices ${\bm Y}_d^{\rm eff} \left( {\bm Y}_d^{\rm eff}
\right)^\dagger$ and $\left( {\bm Y}_d^{\rm eff} \right)^\dagger {\bm
Y}_d^{\rm eff}$, with $n=1,2,3$, form the columns of the unitary matrices ${\bm U}_d$
and ${\bm W}_d$ appearing in the singular-value decomposition
\begin{equation} \label{eq:singularcus}
  \bm{Y}_d^{\rm eff} = \bm{U}_d\,\bm{\lambda}_d\,\bm{W}_d^\dagger \,,
\end{equation}
with $\bm{\lambda}_d$ as given in (\ref{eq:lambdaud}).
Similar relations hold in the up-type quark
sector and will be given explicitly in Section
\ref{sec:HcouplingsRS}. In the ZMA, the definition of the
CKM matrix is identical to the one of the minimal RS model, given in (\ref{eq:VCKM}).
As the quark profiles, in combination with the Yukawa matrices, are fixed such that
the physical quark masses and CKM parameters are reproduced (and the formulae above are identical
to those of the minimal model), the profiles of the SM-like quarks will be to LO 
identical to those of the minimal RS model. 

With these results at hand, it is a matter of simple algebra to find
the expression for ${\bm \delta}_D$\ in the ZMA. Working to first
order in $v^2/M_{\rm KK}^2$, and using (\ref{eq:profileexp}) and
(\ref{IRBC3}) we arrive at
\beq \label{deltaDZMA}
  {\bm \delta}_D = -\frac{1}{2} \, {\bm x}_d \, {\bm W}_d^\dagger \, \,
  {\rm diag } \left [ \frac{1}{1 - 2 \hspace{0.25mm}c_{{{\cal T}}_{2
          i}}} \left ( \frac{1}{F^2(c_{{{\cal T}}_{2 i}})} \left [ 1 - \frac{1 -
          2 \hspace{0.25mm} c_{{{\cal T}}_{2 i}}}{F^2(-c_{{{\cal T}}_{1 i}})}
      \right ] - 1 + \frac{F^2(c_{{{\cal T}}_{2 i}})}{3 + 2 \hspace{0.25mm}
        c_{{{\cal T}}_{2 i}}} \right ) \right ] {\bm W}_d \, {\bm x}_d\,,
\eeq
where ${\bm x}_d \equiv {\rm diag} (m_d, m_s, m_b)/M_{\rm
  KK}$. Compared to the ZMA result in the minimal RS model
(\ref{eq:ZMA2}), this relation contains an additional term
involving the zero-mode profile $F(-c_{{{\cal T}}_{1 i}})$. It stems
from the admixture of the ${\cal T}_1$ multiplet in the zero mode,
which is parametrized by the value of $\hat a^{D^\prime}_n$. Notice
that although this admixture is suppressed by $v/M_{\rm KK}$, the fact
that the profile ${\bm S}^{{\cal T}_1(-)}_n (t)$ is enhanced with
respect to ${\bm S}^{{\cal T}_2(+)}_n (t)$ by the reciprocal factor
promotes the second term in the last line of (\ref{deltaD}) to a
leading contribution.

Note that the relations (\ref{IRBCZMA}) and (\ref{IRBC3}) are valid to leading
order in $v/M_{\rm KK}$. Beyond that order the first relation in
(\ref{IRBCZMA}) receives corrections from the profiles ${\bm C}^{{\cal
    T}_1(-)}_n (1^-)$ which scale like $x_n/F (-c_{{\cal T}_{1 i}})$
as can be seen from (\ref{eq:profileexp}). Thus in order to avoid
exponentially enhanced terms of the form $v/M_{\rm KK} \,
\epsilon^{1-2 \hspace{0.25mm} c_{{{\cal T}}_{1 i}}}$ in the mass
eigenvalues $m_n$, which, barring accidental cancellations, would make
it impossible to reproduce the observed zero-mode down-type quark
masses, one has to require that all the bulk mass parameters belonging
to the multiplet ${\cal T}_1$ obey the relation $c_{{{\cal T}}_{1 i}}
< 1/2$. In this case the profiles ${\bm C}^{{\cal T}_1(-)}_n (t)$ are
IR localized and one has to an excellent accuracy
\beq \label{eq:deltaD2}
  {\bm \delta}_D = -\frac{1}{2} \, {\bm x}_d \, {\bm W}_d^\dagger \, \,
  {\rm diag } \left [ \frac{1}{1 - 2 \hspace{0.25mm}c_{{{\cal T}}_{2
          i}}} \left ( \frac{1}{F^2(c_{{{\cal T}}_{2 i}})} \left [ 1 - \frac{1 -
          2 \hspace{0.25mm} c_{{{\cal T}}_{2 i}}}{1 - 2 \hspace{0.25mm}
          c_{{{\cal T}}_{1 i}}} \right ] - 1 + \frac{F^2(c_{{{\cal T}}_{2
            i}})}{3 + 2 \hspace{0.25mm} c_{{{\cal T}}_{2 i}}} \right ) \right ] \,
  {\bm W}_d \, {\bm x}_d \hspace{0.5mm} . \hspace{6mm}
\eeq
This result implies that the leading term in ${\bm \delta}_D$, \ie,
the contribution that is inversely proportional to $F^2(c_{{{\cal T}}_{2
i}})\ll 1$, is absent if the bulk-mass parameters $c_{{{\cal T}}_{1
i}}$ satisfy
\begin{equation} \label{eq:cbcb}
  c_{{{\cal T}}_{1 i}} = c_{{{\cal T}}_{2 i}} \,.
\end{equation}

Thus, due to quark mixing, the conditions (\ref{PC}) and (\ref{PLR}) alone 
are not sufficient to entirely shield the $Z b_L \bar b_L$ vertex from the leading
corrections. However, since already for not too different values of $c_{{{{\cal T}}_{2
i}}} \approx -1/2$ and $c_{{{{\cal T}}_{1 i}}} \lesssim 0$ the
first term in brackets in (\ref{eq:deltaD2}) is smaller in magnitude than 1, a
partial protection is in place for a large range of bulk
parameters. In consequence, effects due to quark mixing entering the left-handed
down-type $Z$-boson couplings are generically suppressed in the custodial RS
model relative to the minimal scenario as long as the corresponding 
$Z_2$-odd quark fields are not too far localized in the UV. The subleading terms in
${\bm \delta}_D$ are independent of $c_{{{\cal T}}_{1 i}}$ and
therefore not protected even if $c_{{{\cal T}}_{1 i}} = c_{{{\cal
T}}_{2 i}}$. 

Notice that (\ref{eq:cbcb}) can be enforced by requiring the
action to be invariant under the exchange of the
$D^\prime$ and $D$ quark fields,
\begin{equation} \label{eq:extendedPLR}
P_{LR} (D^\prime) = D \,, 
\qquad (\text{extended} \ P_{LR} \ \text{symmetry})
\end{equation}
which extends the $P_{LR}$ symmetry to the part of the quark sector
that mixes with the left-handed down-type zero modes. This extended symmetry 
will necessarily be broken by the different BCs 
of $D^\prime$ and $D$, which embody irreducible sources of symmetry
breaking and lead to non-vanishing sub-leading terms in (\ref{eq:deltaD2}). 
The symmetry (\ref{eq:extendedPLR}) can also be broken softly by choosing bulk 
masses for $D^\prime$ that differ from those
of $D$, which is a phenomenological viable option as long as
$c_{{{\cal T}}_{1 i}} < 1/2$, because it does not affect the SM
down-type quark masses in an appreciable way.  The protection
mechanism discussed here has also been studied in \cite{Buras:2009ka}
employing a perturbative approach. Our analysis based on the exact
solution of the EOMs (\ref{eq:fermEOM}) including the BCs
(\ref{eq:bcIRrescaled}) goes beyond the latter work in the sense that
it makes the dependence of ${\bm \delta}_D$ on the bulk
masses ${\bm M}_{{\cal T}_{1,2}}$ explicit. It therefore allows for a
clear understanding of the custodial protection mechanism in two
respects. First, it makes transparent what the requirements are that need
to be satisfied to achieve a protection and, second, which the
terms in ${\bm \delta}_D$ are that inevitably escape protection. Compared
to the perturbative approach, the exact solution thus has the
salient advantage that the protection of the $Z d_L^i \bar d_L^j$
vertices from effects due to quark mixing can be clearly deciphered.

\section{Summing over Kaluza-Klein Excitations}
\label{sec:KKsum}
When calculating Feynman diagrams involving tree-level exchange of a SM gauge boson, accompanied by corresponding 
KK excitations, one encounters a combination of propagator and vertex functions, which in the low-energy 
limit, \ie, for small momentum transfer $q^2$, can be expanded as \vspace{-3mm}
\beq\label{expandedprop}
\begin{split}
   \sum_n\,\frac{{\vec \chi}_n^{\, a}(t)\,{\vec\chi}_n^{\,a\, T}(t')}{(m_n^a)^2-q^2}
   =& \sum_{N=1}^\infty \left( q^2 \right)^{N-1}
   \sum_n\,\frac{{\vec \chi}_n^{\, a}(t)\,{\vec\chi}_n^{\,a\, T}(t')}%
                { (m_n^a)^{2N}}\\
   \equiv& \sum_{N=1}^\infty \frac{1}{q^2} \left( \frac{q^2}{\Mkk^2} \right)^N \bm{\Sigma}_a^{(N)}(t,t')\,.
\end{split}
\eeq
The gauge-boson profiles ${\vec\chi}_n^a$ will be integrated with further profiles at each end of the propagator. 
As stated above, the formula holds for gauge bosons in the {\it custodial RS variant} with a massive zero mode, for which we will 
discuss these sums in the following. However, the special case of the minimal RS model will be easy to obtain at the end, where we 
will also study towers featuring a massless zero mode.

The infinite sums over profiles weighted by inverse powers of $(m_n^a)^2$ can be calculated in closed form by generalizing a method developed in \cite{Hirn:2007bb}.
The key to perform these sums is to make use of the fact that the bulk profiles $\vec \chi_n$ form a complete set of orthonormal, 
even functions on the orbifold, subject to the BCs given in (\ref{eq:IRBC2}) and in Table \ref{tab:BCs}. The corresponding completeness relations
read
\beq
\label{eq:gaugcompl}
   \sum_n\,{\vec \chi}_n^{\, a}(\phi)\,{\vec \chi}_n^{\, a}(\phi')
   = \frac12 \left[ \delta(\phi-\phi') + \delta(\phi+\phi') \right] {\bm
    1}\, , 
  \ \sum_n \frac{1}{t} \, {\vec \chi}_n^{\, a} (t) \, {\vec \chi}_n^{\,
    a\, T} (t^\prime) = \frac{L}{2 \pi} \, \delta (t - t^\prime) \, {\bm
    1}\,.
\eeq

Integrating the EOMs (\ref{eq:gaugeeomcust}) twice and accounting for the
BCs on both the UV and the IR brane leads to
\begin{equation}
  \frac{\vec{\chi}_n^{\, a} (t)}{(x_n^a)^2} \, = \, \vec{\cal I}_n^{\,
    a} (t) - \left (t^2 - \epsilon^2 \right ) {\bm X}_a \, \vec{\cal
    I}_n^{\, a} (1) + \left [ \, \bm{1} - \left (t^2 - \epsilon^2 \right )
    {\bm X}_a \, \right ] \! {\bm P}_{(+)} \, \frac{\vec{\chi}_n^{\, a}
    (\epsilon)}{\big ( x_n^a \big)^2} \,,
\end{equation}
where we have defined
\begin{equation}
  {\vec{\cal I}}_n^{\, a} (t) \equiv \int_{\epsilon}^t \! dt^\prime \,
  t^\prime \int_{t^\prime}^{1^-} \! \frac{dt^{\prime \prime}}{t^{\prime
      \prime}} \, \vec{\chi}_n^{\, a} (t^{\prime\prime}) \,, \qquad {\bm
    X}_a \equiv \tilde{X}^2 \, {\bm D}_a \equiv \frac{L X^2}{2 + L X^2 \,
    (1-\epsilon^2)} \, {\bm D}_a \,.
\end{equation}
Using the completeness relation (\ref{eq:gaugcompl}), it is then easy to prove that
\begin{equation}
  \sum_n {\vec{\cal I}}_n^{\, a} (t) \, \vec{\chi}_n^{\, a \, T}
  (\phi^\prime) = \frac{L}{4 \pi} \left (t_<^2 - \epsilon^2 \right )
  {\bm 1} \,,
\end{equation}
where $t_< \equiv {\rm min} (t, t^\prime)$. 
With the help of these results we finally arrive at the result
\begin{equation}
  \begin{split}
    \bm{\Sigma}_a^{(1)} (t, t^\prime) & = \frac{L}{4 \pi} \, \Big [ \left
      (t_<^2 - \epsilon^2 \right ) {\bm 1} + \left (t^2 - \epsilon^2
    \right ) \left
      (t^{\prime \, 2} - \epsilon^2 \right ) {\bm X}_a \Big ] \\[1mm]
    & \phantom{xx} + \left [ \,\bm{1} - \left (t^2 - \epsilon^2 \right
      ) {\bm X}_a \, \right ] {\bm P}_{(+)} \, {\bm \Sigma}_a^{(1)}
    (\epsilon, \epsilon) \, {\bm P}_{(+)} \left [ \, \bm{1} - \left
        (t^{\prime \, 2} - \epsilon^2 \right ) {\bm X}_a \, \right]^T ,
  \end{split}
\end{equation}
for the leading sum, which is exact to all orders in $v^2/{M_{\rm KK}^2}$.

Using the orthonormality relation (\ref{eq:ortho}), the remaining sum over gauge profiles 
evaluated on the UV brane can be written as 
\begin{equation}
  \begin{split}
    {\bm P}_{(+)} \, {\bm \Sigma}_a^{(1)} (\epsilon, \epsilon) \, {\bm
      P}_{(+)} = \frac{L}{2 \pi x_a^2} \, \big ( \vec{\chi}_0^{\, a}
    (\epsilon) \big )_1 \, & \bigg [ \, \int_{\epsilon}^1 \frac{dt}{t} \,
    \Big [ \left ( 1 - c_a^2 \hspace{0.25mm} \tilde X^2 \left ( t^{2} -
        \epsilon^2 \right ) \right ) \big ( \vec{\chi}_0^{\, a} (t) \big )_1
    \\ & \phantom{x} + s_a c_a \hspace{0.25mm} \tilde X^2 \left ( t^{2} -
      \epsilon^2 \right ) \big ( \vec{\chi}_0^{\, a} (t) \big )_2 \Big ] \,
    \bigg ]^{-1} {\bm P}_{(+)}\,,
  \end{split}
\end{equation}
where $\big(\vec{\chi}_0^{\,a}(t)\big)_i$ denotes the $i^{\rm th}$
component of the corresponding zero-mode vector. This formula can 
easily be expanded in powers of $v^2/\Mkk^2$ by employing (\ref{eq:expprof}) and (\ref{eq:vecA0a}), leading to
\begin{equation}
  {\bm P}_{(+)} \, {\bm \Sigma}_a^{(1)} (\epsilon, \epsilon) \, {\bm P}_{(+)}
  = \left ( \frac{1}{2 \pi x_a^2} + \frac{1}{4 \pi} \left [ \, 1 -
      \frac{1}{2 L} - \epsilon^2 \left ( L - \frac{1}{2 L} \right ) \right ]
    + {\cal O} (x_a^2) \right ) {\bm P}_{(+)}\,.
\end{equation}
In the end, we obtain
\begin{equation} \label{eq:Sigmafinal}
  \bm{\Sigma}_a^{(1)} (t, t^\prime) = \frac{L}{4 \pi} \, \Big [ \, t_<^2 \,
  {\bm 1} - {\bm P}_a \, t^2 - {\bm P}_a^{\, T} \, t^{\prime \, 2} \,
  \Big ] + \left [ \frac{1}{2 \pi x_a^2} + \frac{1}{4 \pi} \left ( 1 -
      \frac{1}{2 L} \right ) \right ] {\bm P}_{(+)} + {\cal O} (x_a^2) \,,
\end{equation}
where we have employed that $X^2 = x_a^2/c_a^2 + O(x_a^4)$ and defined
\begin{equation} \label{eq:Pa}
  {\bm P}_a= \left( \begin{array}{cc}
      1~ & 0\\
      -\frac{s_a}{c_a}~ & 0
    \end{array} \right)\, .
\end{equation}
Notice that we have dropped phenomenological irrelevant terms of $\ord(\epsilon)\sim 10^{-16}$,
which we will also do in the following.

It is also useful to have an analytic expression for the zero-mode contribution
to (\ref{expandedprop}) alone
\begin{equation}
  \bm{\Pi}_a (t, t^\prime) \equiv \frac{ \vec{\chi}_0^{\, a} (t) \,
    \vec{\chi}_0^{\, a \, T} (t^\prime)}{ x_a^2 } \,.
\end{equation}
With the help of
(\ref{eq:expprof}) and (\ref{eq:vecA0a}), we obtain
\begin{eqnarray} \label{eq:Pifinal}
  \begin{split}
    \bm{\Pi}_a (t, t^\prime) & = -\frac{L}{4 \pi} \, \Big [ {\bm P}_a
    \, t^2
    + {\bm P}_a^{\hspace{0.25mm} T} \, t^{\prime \, 2} \, \Big ] \\[1mm]
    & \phantom{xx} + \left [ \, \frac{1}{2 \pi x_a^2} + \frac{1}{4
        \pi} \left ( 1 - \frac{1}{L} + t^2 \left ( \frac{1}{2} - \ln t
        \right ) + t^{\prime \, 2} \left ( \frac{1}{2} - \ln t^\prime
        \right ) \right ) \right ] {\bm P}_{(+)} + {\cal O} (x_a^2)
    \,. \hspace{4mm}
  \end{split}
\end{eqnarray}

Comparing (\ref{eq:Sigmafinal}) to (\ref{eq:Pifinal}), we infer that all the
$L$-enhanced terms in $\bm{\Sigma}_a^{(1)} (t, t^\prime)$, besides the 
non-factorizable term proportional to $t_<^2$\,, arise from the
zero-mode contribution $\bm{\Pi}_a (t, t^\prime)$. Factorizable
contributions due to the $W^\pm$- and $Z$-boson zero-modes are thus
enhanced by the logarithm of the inverse warp factor with respect to the
contributions from the tower of KK excitations
\cite{Casagrande:2008hr}. The term $t_<^2$
reflects the full 5D structure of the RS model, which is lost when considering
 only a few low-lying KK modes \cite{Bauer:2008xb}.

To obtain the corresponding results for the weak gauge bosons of the {\it minimal RS model},
one just needs to replace the two component profile vectors to start with in (\ref{expandedprop})
by the simple profiles $\chi_n^a$ of the minimal model, see (\ref{eq:KKdecbos}),
and to use the appropriate relations of the minimal model. However, it is easy to 
see that the corresponding sum is already contained in the upper left element of
$\bm{\Sigma}_a^{(1)}$ (\ref{eq:Sigmafinal})
\beq
\label{eq:minsum}
\Sigma_a^{(1)\,{\rm min}}=\left(\bm{\Sigma}_a^{(1)}\right)_{11}\,.
\eeq
For the case of the photon and gluon, which possess massless zero modes, the ground state with $x_{\gamma,g}=0$ 
must be subtracted from the sum. We find
\beq\label{important2}
   {\sum_n}^\prime\,\frac{\chi_n(t)\,\chi_n(t')}{x_n^2}
   = \frac{1}{4\pi} \left[ L\,t_<^2 
   - t^2 \left( \frac12 - \ln t \right) 
   - t'^2 \left( \frac12 - \ln t' \right) + \frac{1}{2L} \right]\, ,
\eeq
where the prime on the sum indicates that $n$ runs from 1 to $\infty$, \ie, the zero mode
contribution is missing.
Notice that the terms proportional to $t^2$ and $t'^2$ in the ``massive'' sum (\ref{eq:Sigmafinal}) are enhanced by a factor $L$, whereas 
this is not the case for the ``massless'' sum (\ref{important2}). This fact has important consequences for the phenomenology of 
flavor-violating processes (see \cite{Bauer:2009cf}). It implies that in the RS model the NP contributions to $\Delta F=2$ processes, 
such as $K$--$\bar K$ or $B$--$\bar B$ mixing, are dominated by the tree-level exchange of KK gluons, while those to $\Delta F=1$ processes,
such as rare meson decays, arise predominantly from the FCNC couplings of the $Z$-boson zero-mode.

It is possible to extend the procedure presented above in an iterative way to sums with $N>1$ in (\ref{expandedprop}). For example,
for $N=2$ we get
\beq
\begin{split}
   \Sigma_a^{(2)\,{\rm min}}(t,t')
   &= \frac{1}{2\pi x_a^4}
    + \frac{1}{4\pi x_a^2} \left( 1 - \frac{1}{L} \right) 
    - \frac{1}{32\pi} \left( L - 5 + \frac{29}{4L} 
    - \frac{3}{L^2} \right) \\
   &\quad\mbox{}- \left[ \frac{1}{4\pi x_a^2}
    + \frac{1}{8\pi} \left( 1 - \frac{1}{2L} \right) \right] 
    \left[ t^2 \left( L - \frac12 + \ln t \right)
    + t'^2 \left( L - \frac12 + \ln t' \right) \right] \\
   & \quad\mbox{}+ \frac{L}{32\pi} \left[
    t_>^4 + 4 t^2 t'^2 \left( L - \frac12 + \ln t_< \right) 
    \right] + \ord\left( x_a^2 \right)\,,
\end{split}
\eeq
while for the massless case we arrive at
\beq\label{m4rela2}
\begin{split}
   {\sum_n}^\prime\,\frac{\chi_n(t)\,\chi_n(t')}{x_n^4}
   &= \frac{1}{32\pi} \left( \frac{5}{8L} 
    - \frac{1}{L^2} \right) \\
   &\quad\mbox{}- \frac{1}{32\pi} \left[
    t^4 \left( L - \frac54 + \ln t \right)
    + \frac{2t^2}{L} \left( L - \frac12 + \ln t \right) \right] \\
   &\quad\mbox{}- \frac{1}{32\pi} \left[
    t'^4 \left( L - \frac54 + \ln t' \right)
    + \frac{2t'^2}{L} \left( L - \frac12 + \ln t' \right) \right] \\
   &\quad\mbox{}+ \frac{1}{32\pi} \left[ L\,t_>^4 
    + 4 t^2 t'^2 \left[ L\ln t_< 
    - \left( L - \frac12 \right) 
    \left( \ln t t' - \frac12 \right) 
    - \ln t\ln t' \right] \right]\, .
\end{split}
\eeq
The analytic results presented here are phenomenologically quite important,
since they allow for a clear understanding of the structure
of $\Delta F = 1$ and $\Delta F = 2$ FCNC interactions,
mediated by towers of KK modes.

\vspace{-0.9cm}
\section{Four-Fermion Charged-Current Interactions}
\label{sec:4Fint}
\vspace{-0.3cm}

\begin{figure}[!t]
\begin{center} 
\hspace{-2mm}
\mbox{\includegraphics[height=1in]{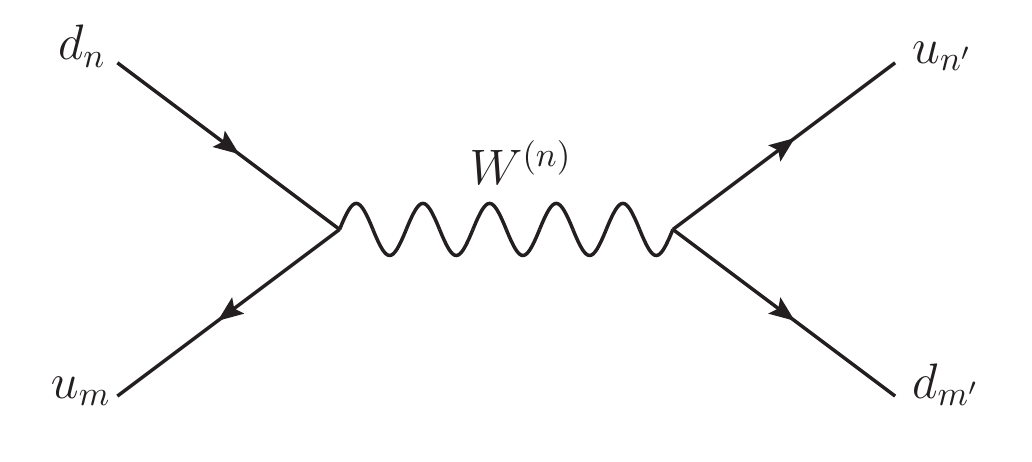}} 
\vspace{-2mm}
\parbox{15.5cm}{\caption{\label{fig:Wtow} Four-fermion interactions due to the exchange of the $W^\pm$-bosons and their KK excitations.}}
\end{center}
\end{figure}
With the help of the low energy expansion derived in the last section, we are able to integrate out the complete
towers of the $W^\pm$-bosons and their KK excitations and to derive the induced effective four-fermion interactions, 
for the minimal as well as for the custodial model, see Figure \ref{fig:Wtow}. 
The effective Hamiltonian reads
\beq \label{eq:HeffW}
\begin{split}
{\cal H}_{\rm eff}^{(W)} =\,2\sqrt 2\, G_F 
&\Big{\lbrace} 
[\,\bar u_{m_L}\gamma_\mu {({\bf V}_L)}_{mn} d_{n_L} + 
\bar u_{m_R}\gamma_\mu {({\bf V}_R)}_{mn} d_{n_R}]\\
&\;\otimes
[\,\bar d_{m'_L}\gamma^\mu {({\bf V}_L^\dagger)}_{m'n'} u_{n'_L} + 
\bar d_{m'_R}\gamma^\mu {({\bf V}_R^\dagger)}_{m'n'} u_{n'_R}] (+ {\rm h.c.})\Big{\rbrace}\,,
\end{split}
\eeq
where $m,n,m',n'\in\lbrace1,2,3\rbrace$ and a summation over the flavor 
indices is understood. The tensor symbol merely indicates that the full analytic result contains 
terms that can not be separated into independent matrix products. This is due to the sum over 
$W^\pm$-boson profiles (\ref{eq:Sigmafinal}), which contains a term $\propto t_<^2$ 
that prevents a factorization into separate vertex factors.

The elements of the mixing matrices $\bm{V}_{\!  \!  L,R}$ are computed with the help of this expression,
as well as the currents (\ref{eq:currents}) and the respective simpler expressions for the minimal model.  
Including corrections up to ${\cal O} (v^2/M_{\rm KK}^2)$, we obtain
\beq
\begin{split}\label{eq:LL}
{({\bf V}_L)}_{mn}\otimes{({\bf V}_L^\dagger)}_{m'n'} =&
    \left(\Delta_{mn}^{+\, Q} + \sqrt 2\, \varepsilon_{mn}^{+\, q} \right) 
    \left( \Delta_{n'm'}^{+\, Q} + \sqrt 2\, \varepsilon_{n'm'}^{+\, q} \right)^\ast\\ 
   -&\frac{m_W^2}{2 \Mkk^2}\,L
    \hspace{0.25mm}\left[ \left(\bar\Delta_{mn}^{+\, Q} + \sqrt 2\, \bar\varepsilon_{mn}^{+\, q} \right)
    \left( \Delta_{n'm'}^{+\, Q} + \sqrt 2\, \varepsilon_{n'm'}^{+\, q} \right)^\ast\right.\\ 
    & \left. + \left(\Delta_{mn}^{+\, Q} + \sqrt 2\, \varepsilon_{mn}^{+\, q} \right)
    \left( \bar\Delta_{n'm'}^{+\, Q} + \sqrt 2\, \bar\varepsilon_{n'm'}^{+\, q} \right)^\ast \right.\\
    & \left. -\left(\Delta_{mn}^{UD}+ \sqrt 2\, \varepsilon_{mn}^{ud}\right)\otimes
    \left(\Delta_{m'n'}^{DU}+ \sqrt 2\, \varepsilon_{m'n'}^{du}\right) \right]\,,
\end{split}
\eeq
where the other combinations are obtained by the replacements
\beq
{\bf V}_L \to {\bf V}_R:\quad \Delta \leftrightarrow \varepsilon\,,
\eeq
and, for the {\it custodial model},
\begin{align} \label{eq:Deltaepsilonplus}
    \Delta_{mn}^{+\, Q,q} & = \frac{2 \pi}{L \epsilon} \,
    \int_\epsilon^1 \! dt \; \vec a_m^{\hspace{0.5mm} U,u \dagger} \,
    \bm{C}_m^{U,u} (t) \, \bm{\Omega}^{Q,q} \, \bm{C}_n^{D,d}(t) \,
    \vec a_n^{D,d} \,, \nonumber\\
    \bar\Delta_{mn}^{+\, Q,q} & = \frac{2 \pi}{L \epsilon} \,
    \int_\epsilon^1 \! dt \; t^2 \, \vec a_m^{\hspace{0.5mm} U,u
      \dagger} \, \bm{C}_m^{U,u} (t) \, \bm{\bar\Omega}^{Q,q} \,
    \bm{C}_n^{D,d}(t) \,
    \vec a_n^{D,d} \,, \nonumber\\
    \Delta_{mn}^{UD}\otimes \Delta_{m'n'}^{du} & = \left(\frac{2 \pi}{L \epsilon}\right)^2
    \int_\epsilon^1 dt \int_\epsilon^1 dt' t_<^2 \, \left(\vec a_m^{\hspace{0.5mm} U\dagger} \,
    \bm{C}_m^{U} (t) \, \bm{\tilde\Omega}^{Q} \, \bm{C}_n^{D}(t) \,
    \vec a_n^{D}\right) \bm{\otimes}
    \left(\vec a_{m'}^{\hspace{0.5mm} d\dagger} \,
    \bm{C}_{m'}^{d} (t') \ \bm{\tilde\Omega}^{q\, \dagger} \ \bm{C}_{n'}^{u}(t') \,
    \vec a_{n'}^{u}\right),\nonumber\\
    & {\rm \etc}
\end{align} 
Note that in the last integrals $(a+b)\,\bm{\otimes}\,(c+d)\equiv ac+bd$ and the corresponding expressions with $\Delta \to \epsilon$ are
obtained by the replacements $\bm{C}_n^A \to \bm{S}_n^A$,\, $A=U,D,u,d$ at the appropriate places.
Moreover
\begin{equation} 
\begin{split}
   \bm{\Omega}^Q&=\begin{pmatrix}
   \bm{1}\\[1mm]
   \bm{0}
 \end{pmatrix} , \qquad \bm{\Omega}^q=
 \begin{pmatrix}
   \bm{0}~ &  \;\bm{0}\; \\
   \bm{0}~ &  \;\bm{1}\;\\
   \bm{0}~ &\, \;\bm{0}\;
  \end{pmatrix} , \qquad
 \bm{\bar\Omega}^Q=\begin{pmatrix}
   \bm{1}\\[1mm]
   - \displaystyle \frac{g_R^2}{g_L^2} \, \bm{1}
 \end{pmatrix} , \qquad \bm{\bar\Omega}^q=
 \begin{pmatrix}
   \bm{0}~ &  \;\bm{0}\; \\
   \bm{0}~ &  \;\bm{1}\;\\[1mm]
   - \displaystyle \frac{g_R^2}{g_L^2}\, \bm{1}~ &\, \;\bm{0}\;
  \end{pmatrix} ,\\
  \bm{\tilde\Omega}^Q&=\begin{pmatrix}
   \bm{1}\\[1mm]
   \displaystyle \frac{g_R}{g_L} \, \bm{1}
 \end{pmatrix} , \qquad 
 \bm{\tilde\Omega}^q=
 \begin{pmatrix}
   \bm{0}~ &  \;\bm{0}\; \\
   \bm{0}~ &  \;\bm{1}\;\\[1mm]
   \displaystyle \frac{g_R}{g_L}\, \bm{1}~ &\, \;\bm{0}\;
  \end{pmatrix}\,.
\end{split}
\end{equation}

Notice that in \cite{Casagrande:2010si},
the simplifying assumption of an interaction with leptons featuring SM-couplings 
at one vertex was made, whereas here we have included deviations from the SM 
at both vertices. 

For the semileptonic case we assume
all the left- and right-handed 5D lepton fields to have the same bulk mass parameters
and to be localized sufficiently close to the UV brane so as not to violate the constraints imposed 
by the electroweak precision tests. Then, the interactions of the lepton zero modes with the $W^\pm$ boson and its KK
excitations are flavor universal and the deviations from the SM are numerically insignificant.
In that case, the right hand side of the tensor structure in (\ref{eq:HeffW}) reduces to a SM leptonic current 
$\sum_l (\bar l_L \gamma^\mu {\nu_l}_L)$ and the remaining (factorizing) quark transition is described by the left- and right-handed
``CKM-matrices''
\begin{equation} \label{eq:VLVR}
  \begin{split}
    \bm{V}_{\! L} = \, \bm{\Delta}^{\! +\, Q} + \sqrt 2\,
    \bm{\varepsilon}^{+\, q} -\frac{m_W^2}{2 \Mkk^2} \, L
    \hspace{0.25mm} \left( \bm{\bar\Delta}^{\! +\, Q} + \sqrt 2\,
      \bm{\bar\varepsilon}^{\, +\, q} \right) , \\
    \bm{V}_{\! R} = \, \sqrt 2\, \bm{\Delta}^{\! +\, q} +
    \bm{\varepsilon}^{+\, Q} - \frac{m_W^2}{2 \Mkk^2} \, L
    \hspace{0.25mm} \left(\sqrt 2\, \bm{\bar\Delta}^{\! +\, q} +
      \bm{\bar\varepsilon}^{\, +\, Q} \right)\,.
  \end{split}
\end{equation} 

The definition of the CKM matrix as $\bm{V}_{\! L}$ accounts for the fact that
in four-fermion interactions one automatically measures the effect of the entire towers of $W^\pm$ bosons and their KK excitations.
Thus the experimentally determined CKM-matrix elements rather correspond to those of the matrix $\bm{V}_{\! L}$, and not 
$\bm{\tilde V}_L$ (\ref{eq:VLVRdef}) (if the RS setup is realized). 
This differs from the definition of the CKM matrix employed in
\cite{Casagrande:2008hr, Buras:2009ka}, which is based on the $W
u^i_L d^j_L$ and $W u^i_R d^j_R$ vertices.
 
Note that above we have absorbed a universal factor
$(1 + m_W^2/(2\Mkk^2) ( 1 - 1/(2 \hspace{0.5mm} L) ))$ into the Fermi constant $G_F$ in (\ref{eq:HeffW}),
due to the normalization to muon decay, from which $G_F$ is extracted, see Section~\ref{sec:mod}.
This factor is independent of the (possibly extended) gauge group. Proceeding in this way renders the
individual factors in the combination $G_F \hspace{0.05mm} \bm{V}_{\!
L,R}$ physically observable. 

From the formulae (\ref{eq:LL}) it is evident that no custodial
protection mechanism is at work in the charged-current sector. 
This is due to the embedding of the up-type quarks in (\ref{eq:multiplets}) and
has already been pointed out in \cite{Agashe:2006at}.  
  The leading contribution to $(V_{L})_{mn}$ stems from $\Delta^{+ \,
    Q}_{mn}$, which is unitary to very good approximation. Corrections
  of order $v^2/\Mkk^2$ arise from the non-universality of KK gauge
  boson couplings encoded in $\bar\Delta_{mn}^{+\, Q}$ as well as the admixture
  from $U^\prime$ and $D^\prime$ quarks described by
  $\epsilon_{mn}^{+\, q}$. Contributions arising from the admixture of
  $U$, $D$, and $u^\prime$ quarks are of order $v^4/\Mkk^4$ and will
  be neglected in the following. 
  The full expression for $\bm{V}_{\!L}$, obtained by employing the ZMA, is given by
\begin{equation} \label{eq:VLZMA}
\begin{aligned}
  \bm{V}_{\! L} &= \bm{U}_u^\dagger \, \Bigg [ \bm{1} - \frac{m_W^2}{2
    \Mkk^2} \, L \; {\rm diag}\bigg(
  \frac{F^2(c_{Q_i})}{3+2c_{Q_i}} \bigg) \\
  &\hspace{1.2cm}\mbox{} + \frac{v^2}{2 \Mkk^2} \; {\rm diag}\, \big (
  F(c_{Q_i}) \big ) \, \bm{Y}_{\! d} \, {\rm diag}\, \big
  (F^{-2}(-c_{{\cal T}_{1i}} ) \big ) \, \bm{Y}_{\! d}^{\dagger} \;
  {\rm diag}\, \big ( F(c_{Q_i}) \big ) \Bigg ] \, \bm{U}_d \,,
\end{aligned}
\end{equation}
which is obviously not unitary. Note that this matrix agrees with the matrix $\bm{\tilde V}_L$,
defined in (\ref{eq:VLVRdef}) only to leading order
\beq
\bm{V}_{\! L}= \bm{\tilde V}_L+\ord(v^2/\Mkk^2)\,.
\eeq
As far as $(V_{R})_{mn}$ is concerned,
the dominant contribution is given by $\epsilon_{mn}^{+\, Q}$, which is
suppressed both by $v^2/M_{\rm KK}^2$ and a chiral factor $m_m^u
m_n^d/v^2$. The chiral suppression present in each of the
terms contributing to ${\bm V}_{\! R}$ reflects the mere fact that
they all originate from quark mixing. As a result, right-handed
charged-current interactions are small in RS models.

In the {\it minimal RS model} the expressions above will be modified. First, the \bm{\varepsilon} matrices in (\ref{eq:LL}) 
will not be present (however the replacement rules below still apply).
This means also that in $\bm{V}_{\! L}$ ($\bm{V}_{\! R}$) in (\ref{eq:VLVR}) the $\bm{\varepsilon}^{\, +\, q} $
and $\bm{\bar\varepsilon}^{\, +\, q}$ ($\bm{\Delta}^{\, +\, q} $
and $\bm{\bar\Delta}^{\, +\, q}$) contributions are not present.
Moreover, the profiles appearing under the integrals (\ref{eq:Deltaepsilonplus}) are now given by those of 
the minimal model (\ref{eq:doublpr}). The same holds true for the flavor-mixing vectors  $\vec a_{n}^{A}$, 
which are now three-component vectors. Finally, the $\bm{\Omega}$ matrices are replaced by the identity $\bm{\Omega},\bm{\bar\Omega},\bm{\tilde\Omega}\to\bm{1}_{3\times3}$. 
As a consequence, the third term in the ZMA expression (\ref{eq:VLZMA}) is not present in the minimal RS model. 
However, identifying the whole expression (\ref{eq:VLZMA}) with the experimentally determined CKM matrix, such a deviation will not be observable in
measuring single CKM matrix elements. In Appendix~\ref{app:charged} we will give ZMA expressions for the charged current
four-fermion interactions, after having identified factorizable contributions with measured CKM matrix elements.
We will study the impact of the mentioned unitarity violation briefly in Section~\ref{sec:CKMPheno}.

The four-fermion interactions resulting from the exchange
of the remaining SM gauge boson and their KK partners can be worked out similarly from the formulae of Section
\ref{sec:KKsum}. For the minimal RS variant they can be found in \cite{Bauer:2009cf}.

\section{Fermion Couplings to the Higgs Boson}
\label{sec:HcouplingsRS}
In the SM, the couplings of matter and gauge fields to the Higgs boson are directly proportional to their masses, due to the mechanism
of EWSB, see Section~\ref{sec:Higgs}. Thus they are flavor diagonal in the mass basis. However, within RS models, this is not true 
anymore \cite{Agashe:2006wa}. As the fields receive masses from couplings to the Higgs sector, as well as from 
compactification, a misalignment between the masses and the Yukawa couplings is present, leading for example to FCNCs at 
tree level. As they will be important for the following analyses, we will now discuss the Higgs-couplings in RS models in detail.

Working in unitary gauge, we first identify the relevant terms in the
4D Yukawa Lagrangian, describing the couplings of the Higgs boson to
quarks. They read
\begin{equation}\label{eq:hff}
  {\cal L}_{\rm 4D} \ni - \sum_{q,m,n}\,(g_h^q)_{m n}\, h\,\bar
  q_L^{\hspace{0.5mm} m}\,q_R^{\hspace{0.25mm} n} + {\rm h.c.} \,,
\end{equation}
where the couplings $(g_h^q)_{mn}$ are given by 
\beq\label{eq:ghn1n2}
  (g_h^q)_{mn} = \frac{\sqrt2\hspace{0.25mm} \pi}{L} \!
  \int_{-\pi}^{\pi} \! d\phi \, \delta (|\phi| - \pi) \,
  e^{\sigma(\phi)} \! \left[ \vec a_{m}^{Q\hspace{0.25mm} \dagger}\,\bm{C}_{m}^{Q}
    (\phi)  \bm{Y}_{\vec q} \hspace{1mm} \bm{C}_{n}^q(\phi)\,\vec
    a_{n}^{\hspace{0.25mm} q} +\vec a_{m}^{\hspace{0.25mm} q
      \hspace{0.5mm} \dagger}\,\bm{S}_{m}^{q} (\phi) \bm{Y}_{\vec q}^\dagger
    \hspace{1mm} \bm{S}_{n}^Q(\phi)\,\vec a_{n}^Q\right] \! . \hspace{9mm}
\eeq
Note that this equation, as well as the following derivations, hold for both the custodial
as well as for the minimal RS variants.
By making use of the EOMs one can eliminate the term bi-linear in the $Z_2$-even
profiles from (\ref{eq:ghn1n2}) and express the tree-level Higgs
FCNCs solely in terms of overlap integrals involving $Z_2$-odd fields.

Defining the misalignment $(\Delta g_h^q)_{mn}$ between the SM masses
and the Yukawa couplings via
\begin{equation} \label{eq:mis1}
  (g_h^q)_{mn} \equiv \delta_{mn}\,\frac{m^q_m}{v} - ( \Delta
  g_h^q)_{mn} \,,
\end{equation}
it is easy to show that
\begin{equation} \label{eq:Higgscorrection} 
  (\Delta g_h^q)_{mn} = \frac{m^q_m}{v}\,(\Phi_q)_{mn} +
  (\Phi_Q)_{mn}\,\frac{m^q_n}{v} + (\Delta \tilde g_h^q)_{mn}\,,
\end{equation}
where in $t$-notation  
\begin{equation} \label{eq:Phi}
  \begin{split}
    \left( \Phi_q \right)_{mn} = \frac{2\pi}{L\epsilon}
    \int_\epsilon^1\!dt\, \vec a_m^{\hspace{0.25mm} Q
      \dagger}\,\bm{S}_m^{Q}(t)\, \bm{S}_n^{Q}(t)\,\vec
    a_n^{\hspace{0.25mm} Q} \,, \qquad \left( \Phi_Q \right)_{mn} =
    \frac{2\pi}{L\epsilon} \int_\epsilon^1\!dt\, \vec
    a_m^{\hspace{0.25mm} q \dagger}\,\bm{S}_m^{q}(t)\,
    \bm{S}_n^{q}(t)\,\vec a_n^{\hspace{0.25mm} q} \,,
  \end{split}
\end{equation} 
and 
\begin{equation} \label{eq:Deltagtilde}
  (\Delta \tilde{g}_h^q )_{mn} = - \sqrt{2} \,
  \frac{2\pi}{L\epsilon}\int_\epsilon^1\!dt \, \delta(t-1) \, \vec
  a_m^{\hspace{0.25mm} q\,\dagger}\, \bm{S}_m^{\hspace{0.25mm} q} (t) \,
  \bm{Y}_{\vec q}^{\dagger} \, \bm{S}_n^Q(t)\, \vec a_n^Q \,.
\end{equation}
The latter contribution is induced only by operators
of the form $(Y_q^{\rm (5D)})_{ij} \left(\bar Q_R^i\right)_{a\alpha}
\hspace{0.25mm} q_L^{c \,j} \, \Phi_{a\alpha}$, containing $Z_2$-odd 
fields evaluated at the boundary of the extra dimension.
In order to evaluate $(\Delta \tilde{g}_h^q )_{mn}$
one has to regularize the $\delta$-distribution appearing in
(\ref{eq:Deltagtilde}) as explained in Section~\ref{sec:fermions}. Employing
\begin{equation} \label{eq:magic2} 
    \int_t^1 dt^\prime \,
    \delta^\eta (t^\prime - 1) \left [ \sinh \hspace{0.25mm} \big (
      \bar \theta^\eta (t^\prime - 1) \hspace{0.25mm} {\bm A} \big )
    \right ]^2 = \frac{1}{2} \left [ \sinh \hspace{0.25mm} \big ( \bar
      \theta^\eta (t - 1) \hspace{0.5mm} 2 {\bm A} \big ) \big (2 {\bm
        A} \big )^{-1} - \bar \theta^\eta (t-1) \, {\bm 1} \right ] ,
\end{equation} 
we obtain the following regularization independent result
\begin{equation} \label{eq:gtil1} 
  (\Delta \tilde g_h^q)_{mn} =
  \frac{1}{\sqrt{2}} \, \frac{2\pi}{L\epsilon} \, \frac{v^2}{3 M_{\rm
      KK}^2} \, \vec a_m^{Q\,\dagger}\, \bm{C}_m^{Q} (1^-) \,
  \bm{Y}_{\vec q} \, \bm{Y}_{\vec q}^{\dagger} \,\, \bm{g} \left (
    \frac{v}{\sqrt{2} M_{\rm KK}} \, \sqrt{{\bm Y}_{\vec{q}}
      \hspace{0.25mm} {\bm Y}_{\vec{q}}^\dagger} \right ) \bm{Y}_{\vec
    q} \, \bm{C}^{\hspace{0.25mm} q}_n (1^-) \,
  \vec{a}_n^{\hspace{0.25mm} q} \,,
\end{equation} 
with  
\begin{equation} \label{eq:bmg} 
  \bm{g} (\bm{A}) = \frac{3}{2} \left [
    \sinh \big (2 \bm{A} \big ) \big (2 \bm{A} \big )^{-1} 
    - \bm{1} \right ] \left ( \cosh \big (\bm{A} \big ) \bm{A} 
  \right )^{-2} \,.
\end{equation}

It is also straightforward to express (\ref{eq:gtil1}) in
  terms of the rescaled Yukawa matrices introduced in
  (\ref{eq:Yukresc}). Using
\begin{equation} \label{eq:YYYtYt}
  \frac{v}{\sqrt{2} \Mkk}\, \sqrt{\bm{Y}_{\vec q} \bm{Y}_{\vec q}^\dagger}\,=\,
  \tanh^{-1}\left ( \frac{v}{\sqrt{2} \Mkk} \, \sqrt{\bm{\tilde Y}_{\vec{q}} 
      \hspace{0.25mm} \bm{\tilde Y}_{\vec{q}}^\dagger} \right ) \,, 
\end{equation}
we obtain
\begin{equation} \label{eq:gtil2} 
  (\Delta \tilde g_h^q)_{mn} =
  \frac{1}{\sqrt{2}} \, \frac{2\pi}{L\epsilon} \, \frac{v^2}{3 M_{\rm
      KK}^2} \; \vec a_m^{Q\,\dagger}\, \bm{C}_m^{Q} (1^-) \,
  \bm{\tilde Y}_{\vec q} \, \bm{\bar Y}_{\vec q}^{\dagger} \,
  {\bm{\tilde Y}}_{\vec{q}} \, \bm{C}^{\hspace{0.25mm} q}_n (1^-) \,
  \vec{a}_n^{\hspace{0.25mm} q} \,,
\end{equation} 
where  
\begin{equation} \label{eq:bmh}
  \bm{\bar Y}_{\vec q}^\dagger \equiv \bm{\tilde Y}_{\vec q}^\dagger 
  \,\, \bm{h} \left ( \frac{v}{\sqrt{2} \Mkk} 
    \, \sqrt{\bm{\tilde Y}_{\vec{q}} \hspace{0.25mm} \bm{
        \tilde Y}_{\vec{q}}^\dagger} \right ) \,, \quad \bm{h} (\bm{A}) =
  \frac{3}{2} \, \Big [\bm{A}^{-2} + \tanh^{-1} \big (\bm{A} \big) \, 
  \bm{A}^{-1} \left (\bm{1} - \bm{A}^{-2}\right) \Big ] \, . 
\end{equation}
The relevant matrix-valued functions can be computed by diagonalizing
the hermitian products $\bm{\tilde Y}_{\vec{q}} \, \bm{\tilde Y}_{\vec{q}}^\dagger$ and
$\bm{\tilde Y}_{\vec{q}}^\dagger \, \bm{\tilde Y}_{\vec{q}}$ with the help of
unitary matrices ${\bm {\cal U}}_{\vec q}$ and ${\bm
  {\cal V}}_{\vec q}$,
\begin{equation} \label{eq:diagonalize} 
  \bm{\tilde Y}_{\vec{q}} \, \bm{\tilde Y}_{\vec{q}}^\dagger = {\bm{\cal
      U}}_{\vec q} \; {\bm{\tilde y}}_{\vec q} \, {\bm{\tilde
      y}}_{\vec q}^{\hspace{0.25mm} T} \; {\bm{\cal U}}_{\vec q}^\dagger \,, 
  \qquad
  \bm{\tilde Y}_{\vec{q}}^\dagger \, \bm{\tilde Y}_{\vec{q}} =
  {\bm{\cal V}}_{\vec q} \; {\bm{\tilde y}}_{\vec q}^{\hspace{0.25mm} T} \,
  {\bm{\tilde y}}_{\vec q} \; {\bm{\cal V}}_{\vec q}^\dagger \,.
\end{equation}
Here ${\bm{\tilde y}}_{\vec q}$ is, depending on the value of the
index $\vec q$, a non-square matrix of dimension $3 \times 6$ or $6 \times 9$
(in the custodial model) containing the non-negative eigenvalues of $\sqrt{\bm{\tilde
    Y}_{\vec{q}} \, \bm{\tilde Y}_{\vec{q}}^\dagger}\,$ on its diagonal.
It follows that
\begin{equation} 
  \bm{\tilde Y}_{\vec q} \, \bm{\bar Y}_{\vec q}^{\dagger} \,
  {\bm{\tilde Y}}_{\vec{q}} \, = \, {\bm{\cal U}}_{\vec q} \; 
  {\bm{\tilde y}}_{\vec q}\; {\bm{\tilde y}}_{\vec q}^{\hspace{0.25mm} T} 
  \; \bm{h} \left ( \frac{v}{\sqrt{2} \Mkk} \;
    \sqrt{{\bm{\tilde y}}_{\vec q} \, {\bm{\tilde y}}_{\vec q}^{\hspace{0.25mm} T} 
    } \right )  \, {\bm{\tilde y}}_{\vec q} \, {\bm{\cal V}}_{\vec q}^\dagger \,.
\end{equation}

Note finally that the Yukawa matrices introduced in
(\ref{eq:Yukresc}) and (\ref{eq:bmh}) satisfy ${\bm{\tilde Y}}_{\vec
  q} = {\bm Y}_{\vec q} + {\cal O} (v^2/M_{\rm KK}^2)$ and ${\bm{ \bar
    Y}}_{\vec q}^\dagger = {\bm Y}_{\vec q}^\dagger + {\cal O}
(v^2/M_{\rm KK}^2)$. In consequence, as long as one is interested in
the ZMA results for $(\Delta \tilde g_h^q)_{mn}$ only, one can 
replace $\bm{\tilde Y}_{\vec q} \, \bm{\bar Y}_{\vec q}^{\dagger} \,
{\bm{\tilde Y}}_{\vec{q}}$ by the combination $\bm{Y}_{\vec q} \,
\bm{Y}_{\vec q}^{\dagger} \, {\bm{Y}}_{\vec{q}}$ of original Yukawa
matrices.

It will be useful to derive ZMA results for the elements $(\Delta
g_h^q)_{mn}$. For this purpose, we still need the ${\cal O}
(v^2/\Mkk^2)$ expressions for the rescaled eigenvectors $\hat a_n^A$
with $A = u, u^\prime, u^c, U^\prime, U$ in the custodial RS model. First note that the
relations (\ref{IRBCZMA}), (\ref{eq:effY}), and (\ref{eq:EVeq1}) to
(\ref{eq:singularcus}) also hold in the up-type quark sector after the
replacements $d\rightarrow u,\, D \rightarrow u^c,\, c_{{{\cal T}}_{2
i}}\rightarrow c_{u^c_i}$ with $\bm{\lambda}_u~=~\sqrt
2/v\,\mbox{diag} \, (m_u,m_c,m_t)$. The remaining $\hat a_n^A$ are
found to satisfy
\begin{equation}\label{eq:upUpU}
  \begin{split}
    \hat a_n^{u^\prime} & = x_n\, {\rm diag} \; \Big (
    F^{-1}(-c_{Q_i})
    \hspace{0.5mm} F^{-1}(c_{Q_i}) \Big ) \, \hat a_n^{u} \,, \\
    \hat a_n^{U^\prime} & = {\rm diag} \; \Big ( F(-c_{{\cal T}_{2
        i}})
    \hspace{0.5mm} F^{-1}(-c_{{\cal T}_{1 i}}) \Big ) \, \hat a_n^{U} \,,\\
    \hat a_n^U & = \frac{x_n}{\sqrt 2}\, {\rm diag} \, \left (
      F^{-1}(-c_{{{\cal T}}_{2i}}) \right ) \, \bm{Y}_d^\dagger \,
    \big [ \bm{Y}_u^{\dagger} \big ]^{-1} \, {\rm diag} \left (
      F^{-1}(c_{u^c_i}) \right ) \, \hat a_n^{u^c} \,.
  \end{split}
\end{equation}
It is also easy to show that the eigenvectors satisfy the sum rules
\begin{equation}
  \begin{split}
    \hat a_n^{u^c \hspace{0.5mm} \dagger} \, \hat a_n^{u^c} + \hat
    a_n^{u^\prime \hspace{0.5mm} \dagger} \, \hat a_n^{u^\prime} = 1
    \,, \qquad \hat a_n^{u \hspace{0.5mm} \dagger} \, \hat a_n^{u} +
    \hat a_n^{U \hspace{0.5mm} \dagger} \, \hat a_n^U + \hat
    a_n^{U^\prime \hspace{0.5mm} \dagger} \, \hat a_n^{U^\prime} = 1
    \, .
  \end{split}
\end{equation}

With these relations at hand, it is straightforward to derive analytic
expressions for the ${\cal O} (v^2/\Mkk^2)$ corrections to ${\bm
\Phi}_{q}$, ${\bm \Phi}_{Q}$, and $\bm{\Delta \tilde{g}}_h^q$. Studying first the {\it custodial model}, 
we find in the case of down-type quarks,
\begin{equation} \label{eq:PhiD}
  \begin{split} 
    & {\bm \Phi}_d = {\bm x}_d \, {\bm U}_d^\dagger \, \, {\rm diag }
    \left [ \frac{1}{1 - 2 \hspace{0.25mm}c_{Q_i}} \left (
        \frac{1}{F^2(c_{Q_i})} - 1 + \frac{F^2(c_{Q_i})}{3 + 2
          \hspace{0.25mm}
          c_{Q_i}} \right ) \right ] {\bm U}_d \, {\bm x}_d \,, \\
    & {\bm \Phi}_D = {\bm x}_d \, {\bm W}_d^\dagger \, \, {\rm diag }
    \left [ \frac{1}{1 - 2 \hspace{0.25mm}c_{{{\cal T}}_{2 i}}} \left
        ( \frac{1}{F^2(c_{{{\cal T}}_{2 i}})} \left [ 1 + \frac{1 - 2
            \hspace{0.25mm} c_{{{\cal T}}_{2 i}}}{F^2(-c_{{{\cal
                  T}}_{1 i}})} \right ] - 1 + \frac{F^2(c_{{{\cal
                T}}_{2 i}})}{3 + 2 \hspace{0.25mm} c_{{{\cal T}}_{2
              i}}} \right ) \right ]
    {\bm W}_d \, {\bm x}_d \,, \\
    & {\bm{\Delta \tilde{g}}}_h^d  = \frac {\sqrt{2} \, v^2 }{3 M_{\rm
        KK}^2} \; {\bm U}_d^\dagger \; {\rm diag} \left[ F(c_{Q_i})
    \right] \hspace{0.25mm} \bm{Y}_d \hspace{0.25mm} \bm{Y}_d^\dagger
    \hspace{0.25mm} \bm{Y}_d \; {\rm diag } \left[ F(c_{{\cal T}_{2
          i}}) \right] {\bm W}_d\,,
  \end{split}
\end{equation}
while for up-type quarks we obtain 
\beq \label{eq:PhiU}
  \begin{split} 
    & {\bm \Phi}_u = {\bm x}_u \, {\bm U}_u^\dagger \, \, {\rm diag }
    \left [ \frac{1}{1 - 2 \hspace{0.25mm}c_{Q_i}} \left (
        \frac{1}{F^2(c_{Q_i})} \left [ 1 + \frac{1 - 2 \hspace{0.25mm}
            c_{Q_i}}{F^2(-c_{Q_i})} \right ] - 1 +
        \frac{F^2(c_{Q_i})}{3 + 2 \hspace{0.25mm} c_{Q_i}} \right )
    \right ] \, {\bm U}_u \, {\bm x}_u \,,\\
    & {\bm \Phi}_U = {\bm x}_u \, {\bm W}_u^\dagger \, \, \Bigg \{ \,
    {\rm diag} \left [ \frac{1}{1 - 2 \hspace{0.25mm}c_{u^c_i}} \left
        ( \frac{1}{F^2(c_{u^c_i})} - 1 + \frac{F^2(c_{u^c_i})}{3 + 2
          \hspace{0.25mm} c_{u^c_i}} \right ) \right ] + \frac{1}{2}
    \, {\rm diag} \left ( F^{-1}(c_{u^c_i}) \right ) \!
    \hspace{0.25mm} \bm{Y}_u^{-1} \,
    \bm{Y}_d \hspace{8mm} \\
    & \mbox{} \phantom{xxxxxxxxxi} \times \hspace{0.5mm} {\rm diag}
    \left(\frac{1}{F^2(-c_{{\cal T}_{2i}})} + \frac{1}{F^2(-c_{{\cal
            T}_{1i}})} \right) \bm{Y}_d^\dagger \,\big [
    \bm{Y}_u^{\dagger} \big ]^{-1} \, {\rm diag} \left (
      F^{-1}(c_{u^c_i}) \right ) \Bigg \} \,
    {\bm W}_u \, {\bm x}_u\,, \\
    & {\bm{\Delta \tilde{g}}}_h^u = \frac {\sqrt{2} \, v^2 }{3 M_{\rm
        KK}^2} \; {\bm U}_u^\dagger \; {\rm diag} \left[ F(c_{Q_i})
    \right] \hspace{0.25mm} \bm{Y}_u \hspace{0.25mm} \bm{Y}_u^\dagger
    \hspace{0.25mm} \bm{Y}_u \; {\rm diag } \left[ F(c_{u^c_i})
    \right] {\bm W}_u\,,
  \end{split}
\eeq
with ${\bm x}_u \equiv {\rm diag} (m_u, m_c, m_t)/M_{\rm KK}$. Remember that
$\bm U_u$ ($\bm W_u$) are the left- (right-)handed rotation matrices
diagonalizing the effective up-type Yukawa coupling. 

At this point some comments are in order. First, notice that the ZMA expressions
in the {\it minimal model} can be obtained from the expressions above by 
dropping all terms involving the zero-mode profiles $F(-c_{{{\cal T}}_{1 i}})$, $F(-c_{{{\cal
T}}_{2 i}})$, and $F(-c_{Q_i})$ and performing the replacements
$c_{{{\cal T}}_{2 i}} \to c_{d_i}$, $c_{u^c_i}\to c_{u_i}$. Note that $\bm \Phi_d$ is the same in both models.
The corrections ${\bm{\Delta \tilde{g}}}_h^{d,u}$ have to be divided by a factor of 2 in 
the minimal model. 
The new terms in (\ref{eq:PhiD}) and (\ref{eq:PhiU}) arise from the admixture
of the ${\bm S}^{{\cal T}_1 (-)}_n (t)$, ${\bm S}^{{\cal T}_2 (-)}_n
(t)$, and ${\bm S}^{Q (-)}_n (t)$ profiles in the corresponding
zero-mode wave functions. In each case, the suppression by $v/M_{\rm
  KK}$ due to the admixture is offset by the ${\cal O} (\Mkk/v)$
enhancement of the $Z_2$-odd $(-)$ profile relative to its $(+)$
counterpart. For $c_{Q_i}<1/2$, the leading contribution in ${\bm
  \Phi}_u$ is numerically enhanced by a factor of 2 with respect to
the minimal model. If the $Z d_L^i \bar d_L^j$ vertices are protected
from fermion mixing by (\ref{eq:cbcb}) and $c_{{\cal T}_{1i}} < 1/2$,
then the same is true for ${\bm \Phi}_D$.  Depending on the structure
of the Yukawa matrices and bulk masses, a similar enhancement is
possible in ${\bm \Phi}_U$.  Notice that the extra suppression by
factors of $m^q_n/v$ in ${\bm \Phi}_{d,D,u,U}$ imply that for light
quark flavors the Higgs-boson FCNCs arising from the latter terms are
parametrically suppressed relative to those mediated by the exchange
of a $Z$ boson. This makes the chirally unsuppressed contributions
$\bm{\Delta \tilde{g}}_h^{d,u}$, that arise from the $Z_2$-odd Yukawa
couplings, the dominant sources of flavor violation in the Higgs
sector. This has been pointed out in \cite{Azatov:2009na}.
As far as the ${\cal O} (v^2/M_{\rm KK}^2)$ corrections to $\bm{\Delta \tilde{g}}_h^{d,u}$,
as given in (\ref{eq:PhiD}) and (\ref{eq:PhiU}) are concerned, we find perfect
agreement with the results presented in the latter article for the
case of a brane-localized Higgs sector. The results presented here 
generalize these findings to our exact treatment of KK profiles and 
thus to all orders in $v/M_{\rm KK}$, for both the minimal as well as 
the custodial RS model. Notice also that the factor of $1/3$ arising in 
the ZMA expressions for $\bm{\Delta\tilde{g}}_h^{d,u}$ follows immediately 
if one applies (\ref{eq:delthe}) to a composite operator containing two $Z_2$-odd
fermion fields.

A model-independent analysis of the flavor misalignment of the SM fermion 
masses and the Yukawa couplings has been presented in \cite{Agashe:2009di}. 
There it has been shown that, in models where the Higgs is a bound state of a new
strongly-interacting theory, at the level of dimension six, chirally unsuppressed
contributions to flavor-changing Higgs-boson vertices generically will
arise from composite operators like $\bar q_L^{\hspace{0.5mm} i}
\hspace{0.25mm} H \hspace{0.25mm} q_R^j \hspace{0.5mm} (H^\dagger H)$.
If present, the latter terms will
dominate over the chirally suppressed contributions originating from
operators of the form $\bar q_L^{\hspace{0.5mm} i} \hspace{0.25mm} D
\!\!\!\!/\, \hspace{0.25mm} q_L^j \hspace{0.5mm} (H^\dagger H)$,
because the couplings $y_{q \ast}$ of the composite Higgs to the other
strong interacting states can be as large as $y_{q \ast}^2/(16
\pi^2) \gg m_q/v$. Notice that in our concrete model, considering all
relevant dimension-six operators in lowest-order of the mass insertion
approximation, leads to the ZMA results (\ref{eq:PhiD}) and
(\ref{eq:PhiU}) quantifying the misalignment between the Yukawa
couplings and the zero-mode masses (see \cite{Azatov:2009na} for a
illuminating discussion). We emphasize that in our exact solution
(\ref{eq:mis1}), (\ref{eq:Higgscorrection}), (\ref{eq:Phi}), and
(\ref{eq:gtil2}), all new-physics effects induced by the mass
insertions are resummed to all orders in $v^2/\Mkk^2$ at tree level. 
Note that in the case of $\Delta F = 2$ processes, the importance of Higgs FCNCs
turns out to be limited. The most pronounced effects occur in the case
of the CP-violating parameter $\epsilon_K$, but are still
typically smaller than the corrections due to KK gluon exchange
\cite{Duling:2009pj}. In Chapter~\ref{sec:Pheno} we will apply the results derived
here to analyze $\Delta F = 1$ Higgs-couplings as well as 
the impact of the RS model on Higgs-boson production and decay at 
hadron colliders.

\chapter{5D Propagators}
\label{sec:5Dprop}
\vspace{-6mm} 

We have seen that particles in the presence of a compactified extra dimension can be described by towers of
KK modes. Often it turns out to be easier to avoid the presence of these KK excitations as virtual particles
in amplitudes and to work directly with the 5D fields. This evades the necessity of performing infinite sums over 
profiles (that just appeared due to the decomposition), weighted by 4D propagators, which can easily become impractical.
In this section we will derive propagators of 5D gauge bosons, including their scalar components, and of fermions.
While such propagators have been studied in the literature \cite{Randall:2001gb,Puchwein:2003jq,Carena:2004zn,Falkowski:2007hz,Csaki:2010aj},
we will include effects of EWSB, mediated by a boundary Higgs-scalar, as well as present for the first time analytic
solutions that include the full flavor structure of the SM. The flavor mixing will be included in a completely general
way. Our results will make calculations of loop-mediated flavor-changing processes in RS models, which in the KK picture
involve multiple sums over fermion profiles, more feasible. The following formulae hold for the minimal RS variant, however, 
the generalization to the custodial model is straightforward. 

\section{Massive Gauge Bosons}
The quadratic part of the gauge-boson action, which is the starting point to derive the corresponding 5D propagators, 
has already been given in \ref{Sgauge2new}, employing the gauge fixing (\ref{Sgf}). It is possible to retain this gauge-fixing Lagrangian
in the 5D approach due to relations between the scalar fields, as discussed below (\ref{Sgf}). This will lead to a propagator
that mixes the scalar components of the gauge fields with the brane-localized Goldstone bosons.
However, for the following applications of the 5D propagators it will be sufficient and more convenient to follow \cite{Csaki:2005vy}
and to use the gauge fixing
\begin{equation}\label{LGF}
\begin{split}
   {\cal L}_{\rm GF2}
   &= - \frac{1}{2\xi} \left( \partial^\mu A_\mu - \xi \left[ 
    \frac{\partial_\phi\,e^{-2\sigma(\phi)} A_\phi}{r^2} \right] 
    \right)^2 - \frac{1}{2\xi} 
    \left( \partial^\mu Z_\mu - \xi \left[ 
    \frac{\partial_\phi\,e^{-2\sigma(\phi)} Z_\phi}{r^2} \right] 
    \right)^2 \\
   &\quad\mbox{}- \frac{1}{\xi} 
    \left( \partial^\mu W_\mu^+ - \xi \left[ 
  \frac{\partial_\phi\,e^{-2\sigma(\phi)} W_\phi^+}{r^2} \right] 
    \right)
    \left( \partial^\mu W_\mu^- - \xi \left[ 
   \frac{\partial_\phi\,e^{-2\sigma(\phi)} W_\phi^-}{r^2} \right] 
    \right) \\
   +& \delta(|\phi|-\pi)
     \left[- \frac{1}{2\xi} 
      \left( \partial^\mu Z_\mu - \frac{\xi}{r} M_Z\,\varphi^3 
      \right)^2 
      - \frac{1}{\xi} 
      \left( \partial^\mu W_\mu^+ - \frac{\xi}{r} M_W\,\varphi^+ 
      \right)
      \left( \partial^\mu W_\mu^- - \frac{\xi}{r} M_W\,\varphi^-
      \right)
     \right].
\end{split}
\end{equation} 
This still eliminates mixed terms containing vector bosons and scalar fields. The quadratic terms in the resulting 
action, denoted by $\tilde S_{\rm gauge,2}$, follow from (\ref{Sgauge2new}) by a proper replacement of the gauge 
fixing
\beq
\label{eq:S2g}
\tilde S_{\rm gauge,2}=\int d^4x\,r\int_{-\pi}^\pi\!d\phi\left({\cal L}_{\rm gauge,2}-{\cal L}_{\rm GF}+{\cal L}_{\rm GF2}\right).
\eeq
When deriving the propagators corresponding to the fields present in the action (\ref{eq:S2g}), 
we have to be careful to take into account the non-trivial determinant of the RS metric
in the right way in order to arrive at the correct covariant expressions in curved space-time.

\subsection{Derivation of the Green's Functions}
 
We start with the propagator for the vector components of the massive $Z$ bosons, \ie, the two point Green's function $D_{\mu\nu}(x_1,x_2,\phi_1,\phi_2)\equiv
\langle Z_\mu(x_1,\phi_1) Z_\nu(x_2,\phi_2) \rangle$. Afterwards, we will turn to the scalar component.
The propagators for the other gauge bosons can be derived in a similar way.

We perform a Fourier transformation to momentum space for the non-compactified 4D space-time but work in position space 
for the compactified direction. In this way we can easily impose the correct BCs at the orbifold fixed 
points and account for geometric overlaps with KK-decomposed fields at vertices. The sought position/momentum-space propagator, 
$\tilde D_{\nu\rho}(p,\phi,\phi^\prime)$, corresponding to a four-momentum transfer $p^\mu$, is given by the solution to the 
differential equation
\beq
\begin{split}
    &\left[ - p^2 \eta^{\mu \nu} +(1-\frac 1 \xi) p^\mu p^\nu
      -\partial_\phi\,\frac{e^{-2\sigma(\phi)}}{r^2} \partial_\phi \eta^{\mu \nu} \right.\\
   &\ \mbox{}\left.\quad\mbox{} +\delta(|\phi|-\pi) \left(-\frac 1 \xi p^\mu p^\nu
      +\frac{M_Z^2}{r}\eta^{\mu \nu}\right)\right]\,\tilde D_{\nu\rho}(p,\phi,\phi^\prime)=\frac{i}{2r}\, \delta^\mu_\rho\, 
      \delta(\phi-\phi^\prime).
\end{split}
\eeq 
Note that the factor of 1/2 on the right hand side above is necessary for the correct normalization on the 
orbifold.
The ansatz \cite{Randall:2001gb}
\begin{equation}
\tilde D_{\mu\nu}(p,\phi,\phi^\prime)=-i G_p(\phi, \phi^\prime)\, \left( \eta_{\mu\nu}-\frac{p_\mu p_\nu}{p^2}\right)-i G^\prime_p(\phi, \phi^\prime)\, \frac{p_\mu p_\nu}{p^2}
\end{equation}
leads to
\begin{equation}
\begin{split}
    &\delta^\mu_\rho\left( p^2 + \partial_\phi\,\frac{e^{-2\sigma(\phi)}}{r^2} \partial_\phi - \frac{M_Z^2}{r} 
    \delta(|\phi|-\pi) \right) G_p(\phi, \phi^\prime)\\
    &-\frac{p^\mu p_\rho}{p^2}\left[\left( p^2 + \partial_\phi\,\frac{e^{-2\sigma(\phi)}}{r^2} \partial_\phi 
    - \frac{M_Z^2}{r} \delta(|\phi|-\pi) \right) G_p(\phi, \phi^\prime)\right.\\
&\left. \quad - \left( \frac{p^2}{\xi} + \partial_\phi\,\frac{e^{-2\sigma(\phi)}}{r^2} \partial_\phi - \left( \frac{M_Z^2}{r}-\frac{p^2}{\xi}\right) \delta(|\phi|-\pi)\right)G_p^\prime(\phi, \phi^\prime) \right]=\frac{1}{2r}\,\delta^\mu_\rho\, \delta(\phi-\phi^\prime).
\end{split}
\end{equation} 
It follows that $G_p^\prime=\tilde G_{\frac{p}{\sqrt \xi}}$, where $\tilde G_p$ has exactly the same form as $G_p$ in the bulk, 
but different BCs on the IR brane.

The shape of the solution is now determined by
\begin{equation}\label{eq:gaugepde}
\left( p^2 + \partial_\phi\,\frac{e^{-2\sigma(\phi)}}{r^2} \partial_\phi - \frac{M_Z^2}{r} \delta(|\phi|-\pi) \right) G_p(\phi, \phi^\prime) 
=\frac{1}{2r}\, \delta(\phi-\phi^\prime).
\end{equation}
Switching to $t$ coordinates, we arrive at
\beq
\label{eq:vecsol}
\begin{split}
G_p(t, t^\prime)=N t_< t_> &\left(A\, J_1(p/\Mkk\, t_<)+B\, Y_1(p/\Mkk\, t_<) \right) \\
&\times\left(C\, J_1(p/\Mkk\, t_>)+D\, Y_1(p/\Mkk\, t_>)\right)\,,
\end{split}
\eeq
where, as before $t_<={\rm min}(t,t^\prime)$ and $t_>={\rm max}(t,t^\prime)$, and $p\equiv\sqrt{p^2}$.
The correct normalization is obtained by matching the result to the $\delta$-distribution
on the right-hand side of (\ref{eq:gaugepde}).
Integrating this equation over $\phi$ in an infinitesimal interval around $\phi=\phi^\prime$ leads to
\beq
N=\frac{L}{4 r (A D-B C) \Mkk^2}\,.
\eeq

The remaining coefficients are dictated by the BCs, which can also be determined from (\ref{eq:gaugepde}) and read
(remember that the vector components are chosen to be even under the $Z_2$ parity)
\begin{equation}
\begin{split}
\partial_{\phi_<} G_p\big{|}_{\phi_<=0}=&\, 0\,,\\
\partial_{\phi_>} G_p\big{|}_{\phi_>=\pi^-}=&-\frac{r M_Z^2}{2 \epsilon^2}  G_p\big{|}_{\phi_>=\pi}\,.
\end{split}
\end{equation}
This leads to
\beq
\begin{split}
A=&Y_0\left(p\epsilon/\Mkk \right)\,,\quad B=-J_0\left(p\epsilon/\Mkk\right)\,,\\
C=&Y_0\left(p/\Mkk\right)+\frac{M_Z^2}{2 p\, \epsilon} Y_1\left(p/\Mkk\right)\,,
\quad D=-J_0\left(p/\Mkk\right)-\frac{M_Z^2}{2 p\, \epsilon} J_1\left(p/\Mkk\right)\,.
\end{split}
\eeq
For $\tilde G_p$ the same BCs hold with the replacement $M_Z^2\rightarrow M_Z^2-r p^2$ which translates into the 
same replacement in the coefficients $C$ and $D$.

We now turn to the derivation of the propagator for the scalar components of the 5D gauge fields. From the action (\ref{eq:S2g}) we deduce
\begin{equation} 
     \left( p^2+\frac{\xi}{r^2} \partial_\phi^2 e^{-2\sigma(\phi)} \right) \tilde D_{\phi\phi}(p,\phi,\phi^\prime) 
     = i \frac{r}{2}\, e^{2\sigma(\phi)} \delta(\phi-\phi^\prime)\,.
\end{equation} 
Defining $\tilde D_{\phi\phi}(p,\phi,\phi^\prime)  \equiv \frac i \xi \tilde D_{5\frac{p}{\sqrt\xi}}^\prime(\phi,\phi^\prime)$ 
leads to
\begin{equation} 
     \left( p^2+\frac{1}{r^2} \partial_\phi^2 e^{-2\sigma(\phi)} \right) \tilde D_{5p}^\prime (\phi,\phi^\prime)
     =\frac{r}{2}\,e^{2\sigma(\phi)} \delta(\phi-\phi^\prime)\,.
\end{equation}
The fully normalized solution can finally be written as
\begin{equation}
\label{eq:scasol}
\begin{split}
        \tilde D_{5p}^\prime(t,t^\prime)= 
\frac{L^2 k\, t_<^2 t_>^2}{4\pi (A D-B C) \Mkk^4}&\left(A\, J_0(p/\Mkk\, t_<)+
B\, Y_0(p/\Mkk\, t_<) \right)\\ &\times\left(C\, J_0(p/\Mkk\, t_>)+D\, Y_0(p/\Mkk\, t_>) \right),
\end{split}
\end{equation}
where the coefficients are derived from the (Dirichlet) BCs and read
\begin{equation}
\begin{split}
A=&Y_0\left(p\epsilon/\Mkk\right),\quad B=-J_0\left(p\epsilon/\Mkk\right)\,,\\
C=&Y_0\left(p/\Mkk\right),\quad D=-J_0\left(p/\Mkk\right).
\end{split}
\end{equation}

\subsection{Limits of the Propagators}

In the following we discuss several limits in the momentum of the 5D propagators. 
For this purpose, we switch to the euclidean momentum, \ie, 
we substitute $p^0 = i p^0_E$ and evaluate scalar products in euclidean space. 
The solutions (\ref{eq:vecsol}) and (\ref{eq:scasol}) become
\begin{equation}
\begin{split}
        G_p(t,t^\prime)= 
\frac{k\, t_< t_>}{2(B C-A D) \Mkk^2}&\left(A\, I_1(p_E /\Mkk\, t_<)+B\, K_1(p_E /\Mkk\, t_<) \right) \\
&\times\left(C\, I_1(p_E /\Mkk\, t_>)+D\, 
K_1(p_E /\Mkk\, t_>) \right)\,,\\
\end{split}
\eeq
\beq
\begin{split}
A=&K_0\left(p_E\, \epsilon/\Mkk\right)\,,\quad B=I_0\left(p_E\, \epsilon/\Mkk\right)\,,\\
C=&K_0\left(p_E /\Mkk\right)-\frac{M_Z^2}{2 {p_E} \epsilon} K_1\left(p_E /\Mkk\right),\quad D=I_0
\left(p_E /\Mkk\right)+\frac{M_Z^2}{2 {p_E} \epsilon} I_1\left(p_E /\Mkk\right)\,,
\end{split}
\end{equation}
where $p_E\equiv\sqrt{p_E^2}$ and $M_Z^2\rightarrow M_Z^2-r p^2$ for $\tilde G_p$, as well as
\begin{equation}
\begin{split}
        \tilde D_{5p}^\prime(t,t^\prime)=
\frac{L^2 k\, t_<^2 t_>^2}{2 \pi^2(B C-A D) \Mkk^4}&\left(A\, I_0(p_E/\Mkk\, t_<)+B\, K_0(p_E/\Mkk\, t_<) \right)\\
&\times\left(C\, I_0(p_E/\Mkk\, t_>)+D\, K_0(p_E/\Mkk\, t_>) \right),\\
\end{split}
\end{equation}
\begin{equation}
\begin{split}
&A=K_0\left(p_E\,\epsilon/\Mkk\right),\quad B=-I_0\left(p_E\,\epsilon/\Mkk\right)\,,\\
&C=K_0\left(p_E/\Mkk\right),\quad D=-I_0\left(p_E/\Mkk\right).
\end{split}
\end{equation}

\subsubsection*{Small Momenta}
\underline{Vectors:}
The limit of $p_E\ll\Mkk$ is particularly interesting, since in that case the 5D gauge-boson propagator 
\beq
   \langle Z_\mu(x,\phi) Z_\nu(x^\prime,\phi^\prime) \rangle = \frac1r\sum_n\,\chi_n^Z(\phi)\,\chi_n^Z(\phi^\prime)
   \langle Z_\mu^{(n)}(x) Z_\nu^{(n)}(x^\prime) \rangle 
\eeq
is related to the KK sums derived in Section~\ref{sec:KKsum}.
The successive expansion of the expression $(-r\, G_p(\phi, \phi^\prime))$ in $p_E$ corresponds to the series of KK 
sums on the right hand side of (\ref{expandedprop}).
In particular, the lowest order KK sum (\ref{eq:minsum}) will be given by
the first term in the expansion
\beq
\sum_n\,\frac{\chi_n^Z(t)\,\chi_n^Z(t^\prime)}{m_n^2}= \Sigma_Z^{(1)\,{\rm min}} = \lim_{p_E\rightarrow 0} \left(- r G_p\right)\,.
\eeq
We arrive at
\begin{equation}
G_p(t_<,t_>) \xrightarrow[p_E\ll \Mkk]{} -\frac{1}{M_Z^2} + \frac{L(t_>^2-1)}{4\pi r \Mkk^2}\,.
\end{equation}
We emphasize that this is the {\it exact} result for the corresponding KK sum,
\ie, no higher order terms are missing, {\it c.f.} (\ref{eq:Sigmafinal}).
For the massless case ($M_Z\rightarrow 0$) we get \cite{Randall:2001gb}
\begin{equation}
G_p(t_<,t_>) \xrightarrow[p_E\ll \Mkk]{} -\frac{1}{2 \pi r p_E^2},
\end{equation}
and the next term in the $p_E^2/\Mkk^2$ expansion reads
\begin{equation}\label{eq:4Ferm1}
-\frac{1}{8 \pi r \Mkk^2 L}+ \frac{\left(t_<^2+t_>^2-2t_<^2\,  {\rm ln}t_<-2t_>^2\,  {\rm ln}t_>-2L\, t_<^2\right)}{8 \pi r \Mkk^2}.
\end{equation}
This coincides with the KK sum (\ref{important2}).
Note that, again, we neglect terms of $\ord(\epsilon)$.\\
\\
\underline{Scalars:}
The scalar propagator behaves like
\begin{equation}
 \tilde D_{5p}^\prime(t_<,t_>) \xrightarrow[p_E\ll \Mkk]{} \frac{L^2 t_<^2 t_>^2 {\rm ln}t_>\, (L+{\rm ln}t_<)}{2 \pi r\, \pi^2 \Mkk^4}.
\end{equation}

\subsubsection*{Large Momenta}

\underline{Vectors:} We first consider the limit $p_E \gg \Mkk$, but $p_E\, t_</\Mkk \ll 1$ and $p_E\, t_>/\Mkk \ll 1$, which means that the momenta are below the position-dependent cutoff, see (\ref{eq:poscut}). We arrive at
\begin{equation}\label{eq:lp1}
G_p(t_<,t_>) \rightarrow -\frac{L}{2 \pi r p_E^2\, ({\rm ln}(2k/p_E)-\gamma_e)}.
\end{equation}
Note that the regions of momenta that fulfill the hierarchies quoted above get smaller and finally vanish, the closer the propagator is evaluated to the TeV brane. Therefore the expression is just applicable for large momenta near the Planck brane.
On the TeV brane, the large-momentum limit $p_E \gg \Mkk$ (but still $p_E<k$) leads to
\begin{equation}
G_p(1,1) \rightarrow -\frac{L}{2 \pi r\, p_E\, \Mkk}\,.
\end{equation}
The propagator changes to a $1/p_E$ behavior and thus cannot be trusted for $p_E\gg \Mkk$ on the TeV brane. Here, 
the theory has to be cut off in the TeV region, as discussed before.
Finally, for $p_E \gg \Mkk$ and $p_E\, t_</\Mkk > 1$,\, $p_E\, t_>/\Mkk > 1$,
the propagator becomes
\begin{equation}
\label{eq:lpprop}
G_p(t_<,t_>) \rightarrow - e^{-p_E \frac{(t_>-t_<)}{\Mkk}}\, \frac{L \sqrt{t_>\, t_<}}{4 \pi\, r p_E\, \Mkk},
\end{equation}
and vanishes if $t_<$ and $t_>$ are well separated from each other.\\
\\
\underline{Scalars:}
In the following we consider the scalar propagator in the limit of large momenta. For $p_E \gg \Mkk$, but $p_E\, t_</\Mkk \ll 1$ and 
$p_E\, t_>/\Mkk \ll 1$, the Green's function becomes
\begin{equation}
 \tilde D_{5p}^\prime(t_<,t_>) \rightarrow -\frac{L^3 t_>^2 t_<^2 (L+{\rm ln}(t_<)) \left({\rm ln} \left(\frac{p_E\, t_>}{2 \Mkk}\right)+\gamma_e \right)}
 { 2 \pi r\, \pi^2 \Mkk^4 \left({\rm ln} \left(\frac{p_E}{2 k}\right)+\gamma_e \right)}\,.
\end{equation}
On the TeV brane it vanishes due to the BC, while for $p_E \gg \Mkk$ and $p_E\, t_</\Mkk > 1$,\, $p_E\, t_>/\Mkk > 1$ we arrive at
\begin{equation}
 \tilde D_{5p}^\prime(t_<,t_>) \rightarrow -e^{-p_E \frac{t_>-t_<}{\Mkk}}\, \frac{L^3\, t_>^{3/2} t_<^{3/2}}{4 \pi r \pi^2 p_E\, \Mkk^3}\,.
\end{equation}
The results derived here will be applied in Section~\ref{sec:AMM} to calculate the anomalous magnetic moment of the muon in the RS setup.
\vspace{-6mm}
\section{Massive Fermions}

We consider the 5D action for quarks as given in (\ref{eq:Sferm2}). The generalization to the lepton sector as well as to
the custodial model is straightforward. First we introduce our notation. We merge the $SU(2)_L$ doublets $Q$ and singlets 
$q^c$ into 2-component vectors 
\begin{equation}
  \Omega_q\equiv
\left(\begin{matrix}
Q\\
q^c
\end{matrix}\right),\quad
  \bar\Omega_q\equiv
\left(\begin{matrix}
\bar Q & \bar q^c
\end{matrix}\right)\,,
\end{equation}
and introduce the projection operators
\begin{equation} 
P^Q\,\Omega_q = \left(\begin{matrix} Q\\ 0 \end{matrix}\right), \qquad
P^q\,\Omega_q = \left(\begin{matrix} 0\\ q^c \end{matrix}\right)\,,
\end{equation}
acting on states in the doublet/singlet (representation) space.
In matrix notation, corresponding to the basis introduced above, they read
\beq
\label{eq:PQ}
  P^Q = 
\left(\begin{matrix}
1 & 0\\
0 & 0
\end{matrix}\right),\quad
  P^q =
\left(\begin{matrix}
0 & 0\\
0 & 1
\end{matrix}\right).
\eeq
Furthermore we introduce the off-diagonal operators 
\beq
\label{eq:PQq}
  P^{Qq} = 
\left(\begin{matrix}
0 & 1\\
0 & 0
\end{matrix}\right),\quad
   P^{qQ} = 
\left(\begin{matrix}
0 & 0\\
1 & 0
\end{matrix}\right),
\eeq
which pick out the singlet and doublet component of $\Omega_q$, respectively, and raise/lower it to the other position in the vector.
The bilinear terms in the action (\ref{eq:Sferm2}) can now be written as
\begin{equation}
\label{eq:5Dact}
\begin{split}
   S_{\rm ferm,2} 
   =\sum_{q^c=u^c,d^c} & \int d^4x\, r\int_{-\pi}^\pi\!d\phi\,\bigg\{ e^{-3\sigma(\phi)}
    \bar \Omega_q\,i\delslash\,\Omega_q - \frac{e^{-2\sigma(\phi)}}{r}  
    \bar \Omega_q\,\partial_\phi\,e^{-2\sigma(\phi)}\gamma^5\,\Omega_q \\
   & - e^{-4\sigma(\phi)}\, \sgn(\phi) \bar \Omega_q\,\left(
    P^Q \bm{M}_Q\,+ P^q \bm{M}_{q^c}\right)\Omega_q \\
   &- \delta^\eta(|\phi|-\pi)\,e^{-3\sigma(\phi)} 
    \frac{\sqrt2 \pi v}{L} \Big[ \bar \Omega_u\,\left(\bm{Y}_u^C\, P^{Qq}+\bm{Y}_u^{S\dagger}\, P^{qQ} \right) P_R\,
   \Omega_u\\
   &+ \bar \Omega_d\,\left(\bm{Y}_d^C\, P^{Qq} +\bm{Y}_d^{S\dagger}\, P^{qQ} \right)  P_R\,\Omega_d + 
    \mbox{h.c.} \Big] \bigg\} \,,
\end{split}
\end{equation}
where $\Omega_{u,d}\equiv(u,u^c)^T,\,(d,d^c)^T$ and $\bm{M}_{q^c}\equiv \bm{M}_{q}$, defined before. The 4D Yukawa matrices $\bm{Y}_q^{C,S}$ describe the couplings of the Higgs
doublet to $Z_2$-even and -odd fermions, respectively, normalized according to (\ref{eq:Y4Ddef}).
Note that we will understand $Q\equiv (Q,0)^T$ and $q\equiv (0,q)^T$ as two-component vectors in the 
following derivations.

\subsection{Derivation of the Green's Functions}

Before calculating the 5D propagator $\bm{G}_q(p,\phi,\phi^\prime)\equiv\langle \Omega_q(p,\phi)\bar \Omega_q(-p,\phi^\prime)
\rangle$ with $q=u,d$ explicitly, let us have a look at the matrix structure of this object. As the following derivations will be similar
for up-type and down-type quarks, we will drop the subscript $q$ of the propagator from now on. 
First, $\bm{G}$ is a 2 $\times$ 2 matrix in representation space with projections $\bm{G}^{Qq}\equiv P^Q\bm{G}P^q = \langle Q \bar q \rangle$, 
\etc\ Second, it is a 2 $\times$ 2 matrix in Dirac space with $\bm{G}_{LR}^{Qq}\equiv P_L\bm{G}^{Qq}P_L = \langle Q_L \bar q_R \rangle$, 
\etc\ Furthermore, it is also a 3 $\times$ 3 matrix in flavor space. It is determined from 
\beq \label{eq:fermeq}
\begin{split}
\bigg[e^{\sigma(\phi)} \pslash - \frac{e^{2\sigma(\phi)}}{r} \partial_\phi\,e^{-2\sigma(\phi)}\gamma^5 - \sgn(\phi) 
   \bigg( P^Q \bm{M}_Q\,+ P^q \bm{M}_q\bigg)& \\
   \quad\mbox{}- \delta^\eta(|\phi|-\pi)\,e^{\sigma(\phi)}  \frac{\sqrt2 \pi v}{L} \left( \left(\bm{Y}_q^C\, P^{Qq}
   +\bm{Y}_q^{S\dagger}\, P^{qQ} \right) P_R\right.&\\       
   \left.+\left(\bm{Y}_q^{C\dagger}\, P^{qQ}+\bm{Y}_q^S\, P^{Qq} \right) 
   P_L\right)\bigg]\, \bm{G}(\phi,\phi^\prime) &= i \frac{e^{4\sigma(\phi)}}{2 r} \delta(\phi-\phi^\prime) {\bf 1}\,,
\end{split}
\eeq
where we have suppressed the dependence of $\bm{G}(\phi,\phi^\prime)$ on the four-momentum transfer and the index
$q$ now corresponds to up type singlets. 
Applying all possible combinations of projection operators introduced above from the left and the 
right on eq.~(\ref{eq:fermeq}), we arrive at four blocks of differential equations. Each of them contains four coupled 
equations for certain components of the propagator, while the four blocks are decoupled from each other. Note that the 
matrix equation above mixes different components in representation space, due to the Yukawa couplings. For example, 
$P^QP^{Qq}\bm{G}P^Q=P^{Qq}\bm{G}P^Q$ belongs to the non-vanishing entry of $\bm{G}^{qQ}$, mediating off-diagonal transitions in 
representation space. However, it features the corresponding entry in the upper-left corner. After 
employing the ansatz
\beq
\label{eq:ansatzprop1}
\begin{split}
\bm{G}_{LL}^{QQ}(t,t^\prime)&\equiv i P^Q P_L \pslash\,  \bm{G}_{++}^{QQ}(t,t^\prime)\,,\qquad
\bm{G}_{RL}^{QQ}(t,t^\prime)\equiv i P^Q P_R  \bm{G}_{-+}^{QQ}(t,t^\prime),\\
\bm{G}_{LL}^{qQ}(t,t^\prime)&\equiv i P^{qQ} P_L \pslash\,  \bm{G}_{-+}^{qQ}(t,t^\prime)\,,\qquad
\bm{G}_{RL}^{qQ}(t,t^\prime)\equiv i P^{qQ} P_R  \bm{G}_{++}^{qQ}(t,t^\prime),
\end{split}
\eeq
the first block of equations becomes
\beq
\begin{split}
\label{eq:block1}
  \frac t \Mkk \bm{G}_{-+}^{QQ}(t,t^\prime)+(t \partial_t-2-\bm{c}_Q) \bm{G}_{++}^{QQ}(t,t^\prime) =&     
  {\delta^\eta(t-1)}\frac{v}{\sqrt2\Mkk}\bm{Y}_q^S\bm{G}_{-+}^{qQ}(t,t^\prime)\,,\\
  \frac{tp^2}{\Mkk} \bm{G}_{++}^{QQ}(t,t^\prime)-(t \partial_t-2+\bm{c}_Q) \bm{G}_{-+}^{QQ}(t,t^\prime) =&     
  {\delta^\eta(t-1)}\frac{v}{\sqrt2\Mkk}\bm{Y}_q^C\bm{G}_{++}^{qQ}(t,t^\prime)
  +\frac{t^4}{\epsilon^4}\delta(t-t^\prime)t^\prime\,,\\
  \frac t \Mkk \bm{G}_{++}^{qQ}(t,t^\prime)+(t \partial_t-2+\bm{c}_q) \bm{G}_{-+}^{qQ}(t,t^\prime) =&     
  {\delta^\eta(t-1)}\frac{v}{\sqrt2\Mkk}\bm{Y}_q^{C\dagger} \bm{G}_{++}^{QQ}(t,t^\prime)\,,\\
  \frac{tp^2} \Mkk \bm{G}_{-+}^{qQ}(t,t^\prime)-(t \partial_t-2-\bm{c}_q) \bm{G}_{++}^{qQ}(t,t^\prime) =&     
  {\delta^\eta(t-1)}\frac{v}{\sqrt2\Mkk}\bm{Y}_q^{S\dagger}\bm{G}_{-+}^{QQ}(t,t^\prime)\,,\\
\end{split}
\eeq
where we have performed a transformation to the coordinate $t$. Remember the definition of the regularized $\delta$-distribution
(\ref{eq:deltadis}) and that $\bm{c}_{Q,q}\equiv\pm \bm{M}_{Q,q}/k$.
The first propagator in (\ref{eq:ansatzprop1}) mediates diagonal transitions in both representation space, as well as in chirality,
the second is diagonal in representation space but off-diagonal in chirality, the third vice versa, and the fourth propagator is 
off-diagonal with respect to both spaces. The subscripts $+(-)$ denote even (odd) functions in $t,t^\prime$ with respect to the $Z_2$ parity on the 
orbifold. The corresponding propagator functions are still $3\times 3$ matrices in flavor space, while the further matrix 
structure has been pulled out in terms of chirality projectors and the matrices defined in (\ref{eq:PQ}) and (\ref{eq:PQq}). The 
other three blocks of equations can be obtained from (\ref{eq:block1}) by the replacements
\beq
\label{eq:repl1}
Z_2: +\leftrightarrow -\,,\quad (t \partial_t-2)\rightarrow -(t \partial_t-2)\,,\quad \bm{Y}_q^C\leftrightarrow\bm{Y}_q^S\,,
\eeq
\beq
\label{eq:repl2} Z_2: +\leftrightarrow -\,,\quad Q\leftrightarrow q\,,\quad \bm{c}_Q \leftrightarrow - \bm{c}_q \,,\quad \bm{Y}_q^C
\leftrightarrow\bm{Y}_q^{S\dagger}\,,
\eeq
\beq
\label{eq:repl3}
Q\leftrightarrow q\,,\quad (t \partial_t-2)\rightarrow -(t \partial_t-2)\,,\quad \bm{c}_Q \leftrightarrow - \bm{c}_q \,,\quad 
\bm{Y}_q^{C,S}\rightarrow\bm{Y}_q^{C,S\dagger}\,,
\eeq
where the first replacements (first two replacements in (\ref{eq:repl2})) correspond to the indices of the ansatz, respectively, given by
\beq
\label{eq:others1}
\begin{split}
\bm{G}_{RR}^{QQ}(t,t^\prime)&\equiv i P^Q P_R \pslash\,  \bm{G}_{--}^{QQ}(t,t^\prime)\,,\qquad
\bm{G}_{LR}^{QQ}(t,t^\prime)\equiv i P^Q P_L  \bm{G}_{+-}^{QQ}(t,t^\prime)\,,\\
\bm{G}_{RR}^{qQ}(t,t^\prime)&\equiv i P^{qQ} P_R \pslash\,  \bm{G}_{+-}^{qQ}(t,t^\prime)\,,\qquad
\bm{G}_{LR}^{qQ}(t,t^\prime)\equiv i P^{qQ} P_L  \bm{G}_{--}^{qQ}(t,t^\prime)\,,
\end{split}
\eeq
\beq
\label{eq:others2}
\begin{split}
\bm{G}_{LL}^{qq}(t,t^\prime)&\equiv i P^q P_L \pslash\,  \bm{G}_{--}^{qq}(t,t^\prime) \,,\qquad
\bm{G}_{RL}^{qq}(t,t^\prime)\equiv i P^q P_R  \bm{G}_{+-}^{qq}(t,t^\prime)\,,\\
\bm{G}_{LL}^{Qq}(t,t^\prime)&\equiv i P^{Qq} P_L \pslash\,  \bm{G}_{+-}^{Qq}(t,t^\prime)\,,\qquad
\bm{G}_{RL}^{Qq}(t,t^\prime)\equiv i P^{Qq} P_R  \bm{G}_{--}^{Qq}(t,t^\prime)\,,
\end{split}
\eeq
\beq
\label{eq:others3}
\begin{split}
\bm{G}_{RR}^{qq}(t,t^\prime)&\equiv i P^q P_R \pslash\,  \bm{G}_{++}^{qq}(t,t^\prime)\,,\qquad
\bm{G}_{LR}^{qq}(t,t^\prime)\equiv i P^q P_L  \bm{G}_{-+}^{qq}(t,t^\prime)\,,\\
\bm{G}_{RR}^{Qq}(t,t^\prime)&\equiv i P^{Qq} P_R \pslash\,  \bm{G}_{-+}^{Qq}(t,t^\prime)\,,\qquad
\bm{G}_{LR}^{Qq}(t,t^\prime)\equiv i P^{Qq} P_L  \bm{G}_{++}^{Qq}(t,t^\prime).
\end{split}
\eeq

In the following we will show explicitly how to solve the first block. The other components of
the propagator can be obtained with the help of the replacements (\ref{eq:repl1})-(\ref{eq:repl3}),
taking into account the correct boundary conditions.
First note that the presence of the $\delta$-distributions gives rise to discontinuities of some components of the propagator (or its derivative) 
at the boundaries (in the limit $\eta\rightarrow0$) and at $t=t^\prime$. For example $\bm{G}_{RL}^{QQ}(t,t^\prime)$ will jump at $t=t^\prime$, 
whereas $\bm{G}_{LL}^{QQ}(t,t^\prime)$ will be continuous at that point while its derivative $\partial_t \bm{G}_{LL}^{QQ}(t,t^\prime)$ will jump. 
The behavior at the boundaries is similar to that of the KK modes in the decomposed theory and has already been discussed in Section~\ref{sec:fermions}. 
It is straightforwardly generalized to the 5D propagators.
Combining the first two equations of block (\ref{eq:block1}) decouples the system and leads to a second order differential equation 
in $t$ for $\bm{G}_{++}^{QQ}(t,t^\prime)$, valid for $t\in(\epsilon,1)$, $t\neq t^\prime$, which reads
\beq
\left(t^2\partial_t^2-4t\partial_t+t^2\frac{p^2}{\Mkk^2}-\bm{c}_Q^2+\bm{c}_Q+6\right) \bm{G}_{++}^{QQ}=0.
\eeq
The solution to this equation is given by
\beq
\label{eq:5Dfermsol1}
   \bm{G}_{++}^{QQ}(t,t^\prime) = t^{5/2}\bigg({\rm dg}\left(J_{c_{Q_i}-\frac12}(p/\Mkk\,t)\right)\bm{a}_{++}^{QQ{>\atop<}}(t^\prime) 
   +{\rm dg}\left(J_{-c_{Q_i}+\frac12}(p/\Mkk\, t)\right) \bm{b}_{++}^{QQ{>\atop<}}(t^\prime)\bigg)\,(t
   {\genfrac{}{}{0pt}{2}{>}{<}} t^\prime)\,.
\eeq
Note that, here and in the following dg $\equiv$ diag. For the case of $c_{Q_i}=1/2+z,\ z\in \mathbb{Z}$, the solution has to be obtained
from (\ref{eq:5Dfermsol1}) by a limiting procedure.
The propagator (\ref{eq:5Dfermsol1}) depends on $t^\prime$ through the 3 $\times$ 3 coefficient matrices ${\bm{a}_{++}^{QQ}}^{>\atop<}
(t^\prime)$ and ${\bm{b}_{++}^{QQ}}^{>\atop<}(t^\prime)$. These are to be determined from the BCs at the branes and the jump 
conditions at $t=t^\prime$, which introduce the dependence on the second coordinate. They will be discussed in detail below. 
Similarly, we can derive second order equations for the other components present in (\ref{eq:block1}), which leads to
\beq
\begin{split}
   \bm{G}_{-+}^{QQ}(t,t^\prime) =& t^{5/2}\bigg({\rm dg}\left(J_{-c_{Q_i}-\frac12}(p/\Mkk\,t)\right)\bm{a}_{-+}^{QQ{>\atop<}}(t^\prime) 
   +{\rm dg}\left(J_{c_{Q_i}+\frac12}(p/\Mkk\, t)\right) \bm{b}_{-+}^{QQ{>\atop<}}(t^\prime)\bigg),\, (t \genfrac{}{}{0pt}{2}{>}{<} t^\prime)\\
   \bm{G}_{-+}^{qQ}(t,t^\prime) =& t^{5/2}\bigg({\rm dg}\left(J_{-c_{q_i}-\frac12}(p/\Mkk\,t)\right)\bm{a}_{-+}^{qQ}(t^\prime) 
   +{\rm dg}\left(J_{c_{q_i}+\frac12}(p/\Mkk\, t)\right) \bm{b}_{-+}^{qQ}(t^\prime)\bigg)\,,\\
   \bm{G}_{++}^{qQ}(t,t^\prime) =& t^{5/2}\bigg({\rm dg}\left(J_{c_{q_i}-\frac12}(p/\Mkk\,t)\right)\bm{a}_{++}^{qQ}(t^\prime) 
   +{\rm dg}\left(J_{-c_{q_i}+\frac12}(p/\Mkk\, t)\right) \bm{b}_{++}^{qQ}(t^\prime)\bigg)\,.
\end{split}
\eeq

Up to now, we have only determined the general form of the solutions following from the second order differential equations in $t$.
These feature 12 coefficient functions which still have to be derived. In the following we will again work in the limit 
$\bm{Y}_q^S=\bm{Y}_q^C\equiv\bm{Y}_q$. On the UV brane, we impose standard Dirichlet BCs for the propagators that are $Z_2$ odd in $t$
\beq
\label{eq:5DUVBC}
\bm{G}_{-+}^{QQ}(\epsilon,t^\prime)=0\,,\quad \bm{G}_{-+}^{qQ}(\epsilon,t^\prime)=0.
\eeq 
The IR BCs follow from a regularization of the delta distributions at $t=1$, in analogy to the derivation within the KK decomposed
theory presented before, and read
\beq
\label{eq:5DIRBC}
\begin{split}
-\bm{G}_{-+}^{qQ}(1^-,t^\prime)=\frac{v}{\sqrt{2}\Mkk}\bm{\tilde Y}_q^\dagger \bm{G}_{++}^{QQ}(1^-,t^\prime)\,,\\
\bm{G}_{-+}^{QQ}(1^-,t^\prime)=\frac{v}{\sqrt{2}\Mkk}\bm{\tilde Y}_q^\dagger \bm{G}_{++}^{qQ}(1^-,t^\prime)\,,
\end{split}
\eeq 
with $\bm{\tilde Y}_q$ as defined in (\ref{eq:Yukresc}).
By integrating the second equation in (\ref{eq:block1}) over an infinitesimal interval around 
$t=t^\prime$, we obtain in addition the jump condition 
\beq
\label{eq:5j1}
\lim_{\eta\rightarrow0}\left[\bm{G}_{-+}^{QQ}(t,t^\prime)\right]_{t=t^\prime-\eta}^{t=t^\prime+\eta}= -\frac{t^{\prime 4}}{\epsilon^4}\bm{1}\,.
\eeq
Furthermore, from the second order equation for $\bm{G}_{++}^{QQ}$ we obtain
\beq
\label{eq:5j2}
\lim_{\eta\rightarrow0}\left[\partial_t \bm{G}_{++}^{QQ}(t,t^\prime)\right]_{t=t^\prime-\eta}^{t=t^\prime+\eta}= 
\frac{t^{\prime4}}{\epsilon^4\Mkk}\bm{1}\,,
\eeq
as well as from continuity of $\bm{G}_{++}^{QQ}$ at $t=t^\prime$
\beq
\label{eq:5j3}
\lim_{\eta\rightarrow0}\left[\bm{G}_{++}^{QQ}(t,t^\prime)\right]_{t=t^\prime-\eta}^{t=t^\prime+\eta}=0\,.
\eeq
So far we have collected 7 conditions for the 12 unknowns functions of $t^\prime$ which is not sufficient for determining 
the solutions uniquely. Imposing just the BCs above is not enough, we have to extract additional information 
from the system (\ref{eq:block1}). This is indeed possible. In the course of deriving the decoupled equations, we lost the 
information about the relative normalization of the components of the propagator with respect to each other. This can be implemented by 
evaluating (\ref{eq:block1}) at $t=\epsilon$, taking into account (\ref{eq:5DUVBC}), which leads to
\beq
\label{eq:5Dateps}
\begin{split}
   \left.(t \partial_t - 2 - \bm{c}_Q) \bm{G}_{++}^{QQ}(t,t^\prime)\right|_{t=\epsilon}=&\ 0\,,\\
   \left.(t \partial_t - 2 + \bm{c}_Q) \bm{G}_{-+}^{QQ}(t,t^\prime)\right|_{t=\epsilon}=&\ \frac{\epsilon\, p^2}{\Mkk} 
   \bm{G}_{++}^{QQ}(\epsilon,t^\prime)\,,\\
   \left.(t \partial_t - 2 + \bm{c}_q) \bm{G}_{-+}^{qQ}(t,t^\prime)\right|_{t=\epsilon}=&\, -\frac{\epsilon}{\Mkk} 
   \bm{G}_{++}^{qQ}(\epsilon,t^\prime)\,,\\
   \left.(t \partial_t - 2 - \bm{c}_q) \bm{G}_{++}^{qQ}(t,t^\prime)\right|_{t=\epsilon}=&\ 0\,.
\end{split}
\eeq
The last missing piece of information is provided by evaluating the second equation in (\ref{eq:block1}) at 
$t=t^\prime\pm\eta$
\beq
\label{eq:5j4}
   \lim_{\eta\rightarrow0} \left[(t \partial_t - 2 + \bm{c}_Q) \bm{G}_{-+}^{QQ}(t,t^\prime)
   \right]_{t=t^\prime-\eta}^{t=t^\prime+\eta}=\lim_{\eta\rightarrow0}\left[\frac{t p^2}{\Mkk}   
   \bm{G}_{++}^{QQ}(t,t^\prime)\right]_{t=t^\prime-\eta}^{t=t^\prime+\eta}\,.
\eeq
It is now possible to determine the coefficient functions by solving the equations (\ref{eq:5DUVBC})-(\ref{eq:5j4}) 
for them. We obtain
\beq
\label{eq:5Dcoefs}
\begin{split}
   {\bm{a}_{++}^{QQ}}^<(t^\prime)=&\,{\rm dg}\left(\frac{J_{-c_{Q_i}-\frac12}(p/\Mkk\,\epsilon)}{J_{c_{Q_i}+\frac12}
   (p/\Mkk\,\epsilon)}\right){\bm{b}_{++}^{QQ}}^<(t^\prime) \,,\\
   {\bm{a}_{++}^{QQ}}^>(t^\prime)=& {\bm{a}_{++}^{QQ}}^<(t^\prime)- \frac{\pi}{2 \Mkk\, \epsilon^4}\, t^{\prime 5/2}\, {\rm dg}
   \left(J_{-c_{Q_i}+\frac12}(p/\Mkk\,t^\prime)\sec(c_{Q_i} \pi)\right)\,,\\ 
   {\bm{b}_{++}^{QQ}}^>(t^\prime)=&\,{\bm{b}_{++}^{QQ}}^<(t^\prime)+\frac{\pi}{2 \Mkk\, \epsilon^4}\, t^{\prime 5/2}\, 
   {\rm dg}\left(J_{c_{Q_i}-\frac12}(p/\Mkk\,t^\prime)\sec(c_{Q_i} \pi)\right) \,,\\
   {\bm{a}_{-+}^{QQ}}^<(t^\prime)=&\,-p\, {\bm{b}_{++}^{QQ}}^<(t^\prime) \,,\quad
   {\bm{b}_{-+}^{QQ}}^<(t^\prime)=\, p\, {\bm{a}_{++}^{QQ}}^<(t^\prime)
   \,,\\
   {\bm{a}_{-+}^{QQ}}^>(t^\prime)=&\,-p\,{\bm{b}_{++}^{QQ}}^>(t^\prime)
   \,,\quad 
   {\bm{b}_{-+}^{QQ}}^>(t^\prime)=\,p\,{\bm{a}_{++}^{QQ}}^>(t^\prime) 
   \,,\\
   \bm{a}_{++}^{qQ}(t^\prime)=&\,{\rm dg}\left(\frac{J_{-c_{q_i}-\frac12}(p/\Mkk\,\epsilon)}{J_{c_{q_i}+\frac12}
   (p/\Mkk\,\epsilon)}\right){\bm{b}_{++}^{qQ}}(t^\prime)\,,\\
   \bm{a}_{-+}^{qQ}(t^\prime)=&\,p^{-1}\,\bm{b}_{++}^{qQ}(t^\prime)\,,\quad
   \bm{b}_{-+}^{qQ}(t^\prime)=\,-p^{-1} \bm{a}_{++}^{qQ}(t^\prime)
\end{split}
\eeq
\nolinebreak where $\bm{b}_{++}^{QQ<}(t^\prime)$ and $\bm{b}_{++}^{qQ}(t^\prime)$ can be determined from the IR BCs (\ref{eq:5DIRBC}).
Finally, we arrive at
\beq
\label{eq:5DcoefIR1}
\begin{split}
\small
   &\bm{b}_{++}^{QQ<}(t^\prime)=\frac{1}{\Mkk}\frac{\pi}{2 \epsilon^4}t^{\prime 5/2}\\
   &\Bigg{(}{\rm dg}\left(\frac{1}
   {J_{c_{Q_i}+\frac12}(p/\Mkk\,\epsilon)}
   \left(J_{-c_{Q_i}-\frac12}(p/\Mkk\,\epsilon)J_{c_{Q_i}+\frac12}(p/\Mkk)-J_{c_{Q_i}+\frac12}(p/\Mkk\,\epsilon)
   J_{-c_{Q_i}-\frac12}(p/\Mkk)\right)\right)\\
   &-\frac{v^2}{2\Mkk^2} \bm{\tilde Y}_q\, {\rm dg}\left(\frac{J_{-c_{q_i}-\frac12}(p/\Mkk\,\epsilon)
   J_{c_{q_i}-\frac12}(p/\Mkk)+J_{c_{q_i}+\frac12}(p/\Mkk\,\epsilon)J_{-c_{q_i}+\frac12}(p/\Mkk)}
   {J_{-c_{q_i}-\frac12}(p/\Mkk\,\epsilon)J_{c_{q_i}+\frac12}(p/\Mkk)-J_{c_{q_i}+\frac12}(p/\Mkk\,\epsilon)
   J_{-c_{q_i}-\frac12}(p/\Mkk)}\right)\bm{\tilde Y}_q^{\dagger}\\ 
   &{\rm dg}\left(\frac{1}{J_{c_{Q_i}+\frac12}(p/\Mkk\,\epsilon)}
   \left(J_{-c_{Q_i}-\frac12}(p/\Mkk\,\epsilon)J_{c_{Q_i}-\frac12}(p/\Mkk)+J_{c_{Q_i}+\frac12}(p/\Mkk\,\epsilon)
   J_{-c_{Q_i}+\frac12}(p/\Mkk)\right)\right)\Bigg{)}^{-1}\\
   &\Bigg{(}\frac{v^2}{2\Mkk^2} \bm{\tilde Y}_q\, {\rm dg}\left(\frac{J_{-c_{q_i}-\frac12}(p/\Mkk\,\epsilon)
   J_{c_{q_i}-\frac12}(p/\Mkk)+J_{c_{q_i}+\frac12}(p/\Mkk\,\epsilon)J_{-c_{q_i}+\frac12}(p/\Mkk)}
   {J_{-c_{q_i}-\frac12}(p/\Mkk\,\epsilon)J_{c_{q_i}+\frac12}(p/\Mkk)-J_{c_{q_i}+\frac12}(p/\Mkk\,\epsilon)
   J_{-c_{q_i}-\frac12}(p/\Mkk)}\right)\bm{\tilde Y}_q^{\dagger}\\
   &{\rm dg}\left(J_{c_{Q_i}-\frac12}(p/\Mkk\,t^\prime)J_{-c_{Q_i}+\frac12}(p/\Mkk)-J_{-c_{Q_i}+\frac12}(p/\Mkk\,t^\prime)
   J_{c_{Q_i}-\frac12}(p/\Mkk)\right)\\
   &+ {\rm dg}\left(J_{c_{Q_i}-\frac12}(p/\Mkk\,t^\prime)J_{-c_{Q_i}-\frac12}(p/\Mkk)+J_{-c_{Q_i}+\frac12}(p/\Mkk\,t^\prime)
   J_{c_{Q_i}+\frac12}(p/\Mkk)\right)\Bigg{)}{\rm dg}\left(\sec(c_{Q_i}\pi)\right)\,,\\
   \end{split}
\eeq
\\
\vspace{-3.5mm}
\beq
\begin{split}
   &\bm{b}_{++}^{qQ}(t^\prime)=\frac{p}{\Mkk} \frac{v}{\sqrt 2\Mkk}\frac{\pi}{2 \epsilon^4}t^{\prime 5/2}\\
   &{\rm dg}\left(J_{c_{q_i}+\frac12}(p\Mkk\,\epsilon)\left(J_{-c_{q_i}-\frac12}(p\Mkk\,\epsilon)J_{c_{q_i}
   +\frac12}(p\Mkk)-J_{c_{q_i}+\frac12}(p\Mkk\,\epsilon)
   J_{-c_{q_i}-\frac12}(p\Mkk)\right)^{-1}\right)
   \bm{\tilde Y}_q^\dagger\\
   &\Bigg\{\Bigg{(}{\rm dg}\left(\frac{J_{-c_{Q_i}-\frac12}(p/\Mkk\,\epsilon)J_{c_{Q_i}+\frac12}(p/\Mkk)-J_{c_{Q_i}
   +\frac12}(p/\Mkk\,\epsilon)J_{-c_{Q_i}-\frac12}(p/\Mkk)}
   {J_{-c_{Q_i}-\frac12}(p/\Mkk\,\epsilon)J_{c_{Q_i}-\frac12}(p/\Mkk)+J_{c_{Q_i}+\frac12}(p/\Mkk\,\epsilon)
   J_{-c_{Q_i}+\frac12}(p/\Mkk)}\right)\\
   &- \frac{v^2}{2 \Mkk^2} \bm{\tilde Y}_q\, {\rm dg}\left(\frac{J_{-c_{q_i}-\frac12}(p/\Mkk\,\epsilon)
   J_{c_{q_i}-\frac12}(p/\Mkk)+J_{c_{q_i}+\frac12}(p/\Mkk\,\epsilon)J_{-c_{q_i}+\frac12}(p/\Mkk)}
   {J_{-c_{q_i}-\frac12}(p/\Mkk\,\epsilon)J_{c_{q_i}+\frac12}(p/\Mkk)-J_{c_{q_i}+\frac12}(p/\Mkk\,\epsilon)
   J_{-c_{q_i}-\frac12}(p/\Mkk)}\right) \bm{\tilde Y}_q^\dagger\Bigg{)}^{-1}\\
   &\Bigg{(}\frac{v^2}{2 \Mkk^2} \bm{\tilde Y}_q\, {\rm dg}\left(\frac{J_{-c_{q_i}-\frac12}(p/\Mkk\,\epsilon)
   J_{c_{q_i}-\frac12}(p/\Mkk)+J_{c_{q_i}+\frac12}(p/\Mkk\,\epsilon)J_{-c_{q_i}+\frac12}(p/\Mkk)}
   {J_{-c_{q_i}-\frac12}(p/\Mkk\,\epsilon)J_{c_{q_i}+\frac12}(p/\Mkk)-J_{c_{q_i}+\frac12}(p/\Mkk\,\epsilon)
   J_{-c_{q_i}-\frac12}(p/\Mkk)}\right) \bm{\tilde Y}_q^\dagger \\
   &{\rm dg}\left(J_{c_{Q_i}-\frac12}(p/\Mkk\,t^\prime)J_{-c_{Q_i}+\frac12}(p/\Mkk)-J_{-c_{Q_i}+\frac12}
   (p/\Mkk\,t^\prime)J_{c_{Q_i}-\frac12}(p/\Mkk)\right)\\ 
   &+  {\rm dg}\left(J_{c_{Q_i}-\frac12}(p/\Mkk\,t^\prime)
   J_{-c_{Q_i}-\frac12}(p/\Mkk)+J_{-c_{Q_i}+\frac12}(p/\Mkk\,t^\prime)J_{c_{Q_i}+\frac12}(p/\Mkk)\right)
   \Bigg{)}\\
   & + {\rm dg}\left(J_{c_{Q_i}-\frac12}(p/\Mkk\,t^\prime)
   J_{-c_{Q_i}+\frac12}(p/\Mkk)-J_{-c_{Q_i}+\frac12}(p/\Mkk\,t^\prime)J_{c_{Q_i}-\frac12}(p/\Mkk)
   \right) \Bigg\}{\rm dg}\left(\sec(c_{Q_i} \pi)\right)\,.
\end{split}
\eeq
In the case of a vanishing Higgs VEV, the coefficient $\bm{b}_{++}^{qQ}(t^\prime)$ is zero. This 
makes the propagators mediating singlet-doublet transitions vanish, which could be expected from (\ref{eq:5Dact}).
In this limit, the other coefficient becomes
\beq
\label{eq:5DcoefIR2s}
\begin{split}
\small
   &\bm{b}_{++}^{QQ<}(t^\prime)\xrightarrow[v\rightarrow0]{}
   \frac{1}{\Mkk}\frac{\pi}{2\epsilon^4}t^{\prime 5/2}\,{\rm dg}(\sec(c_{Q_i}\pi))\\
   &{\rm dg}\left(J_{c_{Q_i}+\frac12}(p/\Mkk\,\epsilon)
   \frac{J_{c_{Q_i}-\frac12}(p/\Mkk\,t^\prime)
   J_{-c_{Q_i}-\frac12}(p/\Mkk)+J_{-c_{Q_i}+\frac12}(p/\Mkk\,t^\prime)J_{c_{Q_i}+\frac12}(p/\Mkk)}
   {J_{-c_{Q_i}-\frac12}(p/\Mkk\,\epsilon)J_{c_{Q_i}+\frac12}(p/\Mkk)-J_{c_{Q_i}+\frac12}(p/\Mkk\,\epsilon)
   J_{-c_{Q_i}-\frac12}(p/\Mkk)}\right)\,.
\end{split}
\eeq

\subsection{Limit of Small Momentum Transfer}
Let us finally give the expressions for the Green's functions in the limit of a negligible momentum transfer.
For $p\to 0$ the propagators reduce to
\beq
\begin{split}
\bm{G}_{LL}^{QQ}(t,t^\prime)&\to 0\,,\\[3mm]
\bm{G}_{LL}^{qQ}(t,t^\prime)&\to 0\,,
\end{split}\quad
\begin{split}
\bm{G}_{RL}^{QQ}(t,t^\prime)&\to -i P^Q P_R\,\frac{t^2 t^{\prime 2}}{2\,\epsilon^4}\,{\rm dg}\left((t^\prime/t)^{c_{Q_i}}\right) \theta(t-t^\prime)\,,\\
\bm{G}_{RL}^{qQ}(t,t^\prime)&\to -i P^{qQ} P_R\, \frac{\Mkk}{\sqrt 2 v} \frac{t^2 t^{\prime 2}}{\epsilon^4}\, {\rm dg}\left(t^{c_{q_i}}\right)
\tilde{\bm{Y}}^{-1}{\rm dg}\left( t^{\prime{c_{Q_i}}} \right)\,.
\end{split}
\eeq
The zeroes reflect the fact that the chirality-diagonal propagators are induced only by a non-vanishing momentum-flow.

\chapter{Warped Extra Dimensions: Phenomenology}
\label{sec:Pheno}
\vspace{-6mm} 

After studying in detail the setup, the interactions, as well as KK sums in both the minimal as well as the custodial RS model, we will now
apply these results to analyze the phenomenology of these models. First, we will give an overview of the status of RS models,
in particular concerning the compatibility of a low KK scale with experimental constraints.
In Section~\ref{sec:pheno1}, we have already explored electroweak precision observables and discussed possibilities to end up
with KK scales in the range of a few TeV, which can \eg be achieved by employing a custodial symmetry. 
We have also argued that the RS-GIM mechanism will suppress the majority of flavor-changing transitions in the quark sector below 
their experimental limits. Thus, at first sight the (custodial) RS model seems to offer a reasonable chance to be directly discovered at the LHC, when
running at $\sqrt{s}=14$\,TeV.

In the last decade, various studies of the flavor structure of RS models have been performed. Properties of the (generalized) 
CKM matrix, neutral-meson mixing, and CP violation were studied in \cite{Huber:2003tu}. $Z$-mediated FCNCs in the Kaon system 
were examined in \cite{Burdman:2002gr} and effects of KK gauge bosons on CP asymmetries in rare hadronic $B$-meson decays, induced 
by $b\to s$ transitions, were explored in \cite{Burdman:2003nt}. A first analysis of $\Delta F=2$ and $\Delta F=1$ processes in 
the RS framework was presented in \cite{Agashe:2004ay,Agashe:2004cp}, where the second paper treats a variety of possible effects 
and analyzes several rare decay processes. The branching ratios for the flavor-changing top-quark decays $t\to c Z (\gamma,g)$  
were discussed in \cite{Agashe:2006wa}. Model-independent approaches for analyzing NP contributions to $\Delta F=2$ 
and $\Delta F=1$ operators were developed in \cite{Agashe:2005hk,Davidson:2007si}. The first complete examination of all operators 
relevant to $K$--$\bar K$ mixing was presented in \cite{Csaki:2008zd}. A comprehensive analysis of $B_{d,s}$--$\bar B_{d,s}$ mixing 
has been performed in \cite{Blanke:2008zb}. Moreover, rare leptonic $K$- and $B$-meson decays \cite{Chang:2007uz,Blanke:2008yr} as well as
the radiative $B \to X_s \gamma$ decay \cite{Agashe:2008uz} have been studied quite recently. A comprehensive
analysis of $\Delta F=1$ and $\Delta F=2$ processes, taking into account the exchange of the whole towers 
of KK excitations has been presented in \cite{Bauer:2009cf}. Higgs- \cite{Azatov:2009na,Agashe:2009di}
and radion-mediated \cite{Azatov:2008vm, Davoudiasl:2009xz} FCNCs have been investigated, too.
It has been recognized that the only observables where a considerable fine-tuning of parameters seems to be unavoidable are 
the neutron electric dipole moment \cite{Agashe:2004ay,Agashe:2004cp} and CP-violating effects in the neutral Kaon system \cite{Csaki:2008zd,Davidson:2007si,Bona:2007vi,Gedalia:2009ws}. For generic parameters,
these turn out to be too large unless the scale of the first KK-gauge bosons is raised to $\ord(10)$\,TeV,
beyond the reach of the LHC. Due to these problems,
several modifications of the quark flavor sector of warped models have been proposed. Many of them try to implement the idea of MFV into 
the RS setup by means of a bulk flavor symmetry \cite{Santiago:2008vq,Fitzpatrick:2007sa,Csaki:2008eh,Csaki:2009wc},
see also \cite{Chen:2009gy}. Others explore the possibility of textures of the 5D Yukawa matrices \cite{Chang:2008zx} or use a bulk-Higgs  
\cite{Agashe:2008uz,Gedalia:2009ws} to weaken the constraints. The idea of spontaneous 
CP violation in the context of RS models has been used to address the problem of too large electric dipole moments \cite{Cheung:2007bu}.

Recently, a model of ``flavor triviality'' has been proposed \cite{Delaunay:2010dw}, where the solution to the puzzle of fermion 
hierarchies is given up (see also \cite{Cacciapaglia:2007fw,Csaki:2009bb}). 
This model assumes that both the bulk and the IR brane are invariant under the (gauged) SM flavor symmetries. 
The Yukawa hierarchy is assumed to be generated by some unknown physics on the UV brane. The model belongs to the MFV class and can have a 
KK scale as low as 2\,TeV, while being in agreement with flavor as well as electroweak precision constraints. Beyond that, it offers an 
interesting phenomenology \cite{Delaunay:2011vv}. However, we do not want to give up the anarchic approach to flavor and thus assume other alternatives of
flavor protection to be at work.

What concerns the lepton sector, flavor violation also leads to generically quite stringent bounds on the masses of the first KK gauge bosons in the range of
$\gtrsim$5\,TeV \cite{Csaki:2010aj,Agashe:2006iy}. These bounds, however, depend quite strongly on the concrete model parameters.
They can be lowered by an appropriate choice of the lepton representations under the custodial gauge group \cite{Agashe:2009tu}
or by flavor symmetries \cite{Chen:2008qg,Perez:2008ee,Csaki:2008qq,delAguila:2010vg}.

After this general survey of the status of RS models with respect to (precision) measurements, let us review to what extend
the setup can answer some of the additional open questions in the SM, besides solving the flavor puzzle and the gauge hierarchy problem.
First of all, it has been shown that warped extra dimensions offer the possibility of high-scale gauge-coupling unification \cite{Randall:2001gc,Goldberger:2002hb,Agashe:2002pr}.
Moreover, viable dark matter scenarios arise naturally in some variants of RS setups, typically involving (sterile) neutrinos,
GUTs, fields with special boundary conditions, or discrete symmetries like KK parity \cite{Carena:2009yt,Agashe:2004ci,Agashe:2004bm,Kadota:2007mv,Agashe:2007jb,Panico:2008bx,Haba:2009xu}.
Brane fluctuations, so-called branons, could furnish WIMP dark matter, see \eg \cite{Cembranos:2003mr}.
In addition, the strong CP problem can be addressed in the context of RS models via
spontaneous CP violation \cite{Cheung:2007bu}.
Note finally that there are interesting possibilities to generate cosmic inflation in (warped) extra dimensions \cite{Dvali:1998pa,Kachru:2003sx,Sundrum:2009ii}.

Subsequent to this overview we will analyze in detail various observables, which
are expected to show interesting signatures of the RS setup.
The first part of this chapter is devoted to precision tests. We start by discussing
modifications of SM input parameters due to warped extra dimensions. We will go into a tension
between different determinations of the $W^\pm$-boson mass and elaborate on RS modifications 
concerning the Higgs VEV. Next, we will continue our examination of the Peskin-Takeuchi parameters.
A careful analysis of the bottom-quark pseudo observables at the $Z$-pole follows,
where we will study in detail the parameter dependence and the level
of custodial protection for the RS corrections. A possibility to improve the agreement of
the forward-backward asymmetry in bottom-quark pair production at the $Z$-pole with experiment will be examined.
Randall-Sundrum models have been used to address the enhancement in the $t\bar t$ 
forward-backward asymmetry, measured at Tevatron, with respect to the theoretical prediction. 
We will investigate if it is possible to account for this anomaly in an anarchic approach to flavor.
After that, we apply the formalism derived in Chapter~\ref{sec:5Dprop} to examine RS corrections to the anomalous magnetic moment of the muon.
Then we will turn to flavor physics and CP violation. First we will study 
numerically the unitarity violating corrections to the matrix which would be identified with the CKM matrix of the SM. 
After an analysis of rare top-quark
decays, the phenomenological survey will be continued with a discussion of $B_s^0$--$\bar B_s^0$ mixing.
A calculation of the absorptive part of the mixing amplitude in the presence of new physics
is presented and the impact of the Randall-Sundrum setup on several CP-violating
observables is investigated. 
Finally, we will perform a detailed survey on Higgs Physics in warped extra dimensions. We will analyze
the Higgs-boson production cross sections at Tevatron and at the LHC, as well as the most important branching fractions.
For the first time, a complete one-loop calculation for the (custodial) RS model is presented, incorporating the effects stemming 
from the extended electroweak gauge-boson and fermion sectors.
Although deviations are generically expected for an IR-brane localized Higgs sector, the RS modifications will actually exceed
the expectations. This observation could have an significant impact on the LHC physics program.

In order to present predictions for different observables in the RS setup we need to specify
a large number of model parameters. These are, besides the NP scale $\Mkk$ and the logarithm of the inverse warp factor $L$, the 
bulk mass parameters of the 5D fermion fields as well as the 5D Yukawa matrices. 
As explained in Section~\ref{sec:hierarchies}, their choice is restricted by the fact that the known 
values of the quark masses and CKM matrix elements should be reproduced within errors, but this information still leaves some freedom.

In parts of our following analysis, we will perform a scan over the parameter space of the RS model. For this purpose, we generate 10000 randomly
chosen parameter sets using uniform initial distributions for the input parameters. The parameter ranges are $\Mkk\in [1,10]$\,TeV for the KK scale 
and $|(Y_{u,d})_{ij}|\in [0.1,3]$ for the Yukawa couplings. Moreover, we require that the Wolfenstein parameters $\bar\rho$ and $\bar\eta$ agree 
with experiment within $1\sigma$. 
The bulk-mass parameters are then determined using the warped-space Froggatt-Nielsen formulae given in Section~\ref{sec:hierarchies},
such that all points reproduce the quark masses and CKM parameters with a global $\chi^2/{\rm dof}$ of better than 11.5/10 (corresponding to 68\% CL),
subject to the constraints $|c_{Q_i,q_i}|<1$ and $c_{u_3}\in\,]-1/2,1]$. This large set of points provides a reasonable range of predictions that can 
be obtained for a given observable.\footnote{For more details concerning the algorithm that generates the parameter points 
of the RS model, the interested reader is referred to \cite{Bauer:2009cf}.} Only for a subset of the 10000 scatter points the 
predictions for the $Z b\bar b$ couplings in the minimal RS model are consistent with experiment at 99\% CL. We will consider these 
constraints by discarding points 
that are in conflict with them. We use $\Mkk=1.5$\,TeV as a default KK scale for the minimal RS model, corresponding to masses of 
about 3.7\,TeV for the first KK excitations of the gauge bosons. We sometimes use twice that value, $\Mkk=3$\,TeV, as a second reference 
point. For the custodial model, being less constraint by electroweak precision measurements, we will use the scales 
$\Mkk=1$\,TeV and $\Mkk=2$\,TeV. Unless noted otherwise, we set $L=\ln (10^{16})$ for the logarithm of the inverse warp factor.
It will be instructive to have default sets of input parameters, which are consistent with all experimental constraints concerning the 
quark masses and CKM parameters. Such sets will be given in Appendix~\ref{app:points}.

The following is based on \cite{Casagrande:2008hr,Casagrande:2010si,Bauer:2010iq,Goertz:2011nx}. 
Note that the numerical analysis of \cite{Casagrande:2008hr} has been redone with the same parameter sets as used for the 
analysis of the custodial model. Moreover, the Yukawa couplings involving $Z_2$-odd fields, that have been neglected in the latter 
article, are included. The same holds true for the $\ord(v^2/\Mkk^2)$ shift in the Higgs VEV.

\section{Precision Tests}

\subsection{Modifications of SM Parameters}
\label{sec:mod}

Let us first have a look on how values of SM-input parameters will change if the RS setup is realized.
Recall that the zero modes of the photon and gluon in RS models, interpreted as the SM photon and gluon, have flat profiles in the 
extra dimension. Thus their tree-level couplings to the zero mode fermions are universal and flavor diagonal, just like in 
the SM, see (\ref{eq115}). As a consequence, the low energy extraction of the fine structure constant $\alpha$, defined in the Thompson limit, 
as well as the strong coupling $\alpha_s$ are to excellent approximation unaffected by the presence of the extra dimension.
Moreover, the definition of the weak mixing angle $\theta_w$ in RS, in terms of (4D) weak gauge couplings, agrees with the one 
of the SM, see (\ref{eq:weinbRS}), which results in the relation
\beq\label{SU2gdef}
   g^2 = \frac{4\pi\alpha}{s_w^2} \,.
\eeq 
Defined in this way, $\theta_w$ can be extracted from measuring the left-right polarization asymmetries of light fermions at
the $Z$-pole. 
Due to the RS-GIM mechanism, the corrections to the
$Z$ couplings are given to very good approximation by the universal prefactor in (\ref{eq:Zff}) and (\ref{eq:Zffcus}), 
which cancels in the asymmetry
\beq\begin{split}
   A_f 
   &= \frac{\Gamma(Z\to f_L\bar f_R) - \Gamma(Z\to f_R\bar f_L)}%
           {\Gamma(Z\to f_L\bar f_R) + \Gamma(Z\to f_R\bar f_L)}
   = \frac{(g_L^f)^2 - (g_R^f)^2}{(g_L^f)^2 + (g_R^f)^2} \\
   &\approx 
    \frac{(1/2-|Q_f|s_w^2)^2 - (Q_fs_w^2)^2}%
         {(1/2-|Q_f|s_w^2)^2 + (Q_fs_w^2)^2}  \,.
\end{split}
\eeq
From measuring $A_f $ one can thus directly determine the weak mixing angle, which will have the same value
in the SM and in RS models. Note that this discussion holds for the  minimal as well as for the custodial RS model (where $g\to g_L$).

\begin{figure}[!t]
\begin{center}
\mbox{\includegraphics[height=3cm]{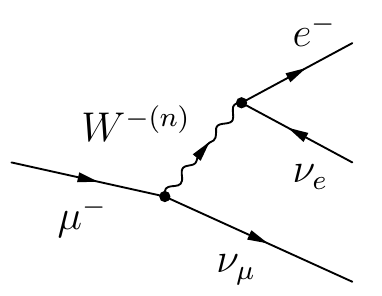}}
\vspace{-2mm}
\parbox{15.5cm}{\caption{\label{fig:muondecay}
Tree-level contributions to $\mu^-\to e^-\nu_\mu\bar\nu_e$ arising from the exchange of a $W^-$ boson and its KK partners. See \cite{Casagrande:2008hr} and text for details.}}
\end{center}
\end{figure}

We now turn to phenomenological consequences of the RS corrections to the $W^\pm$ mass.
They have already played a role in determining the $T$ parameter. However, will give here a more detailed analysis, including a survey 
about different possibilities to determine the $W^\pm$-boson mass and RS predictions in this context.

The mass of the $W^\pm$-bosons can be measured in several ways, see Section~\ref{sec:Higgs}. First, it can be obtained directly
from $W^\pm$-decays. Just from this single experiment it is not possible to differentiate between the SM and the RS model, since the 
RS corrections to the $W^\pm$ mass, as given in (\ref{eq:m02}) and (\ref{eq:mwmz}), can be absorbed into a redefinition
of the Higgs VEV in (\ref{eq:MWSM}), which will then differ from the SM value of $v_{SM}=246\,$GeV \cite{Bouchart:2009vq}.  
From (\ref{eq:m02}) one deduces that within the minimal RS model one should use the VEV
\beq
	v_{\rm RS}=v_{\rm SM}\left(1 +\frac{v_{\rm SM}^2}{2 \Mkk^2}\frac{g^2}{8} \left(L-1+\frac{1}{2L}\right) + \ord\left( \frac{v_{\rm SM}^4}{\Mkk^4} \right)
	\right) \,.
\eeq
Here, we have neglected small loop corrections, which are subleading compared to the L-enhanced RS corrections (for low KK scales). 
For the custodial model we derive from (\ref{eq:mwmz}) the VEV shift
\beq
\label{eq:VEVshift}
	v_{\rm RS_C}=v_{\rm SM}\left(1 +\left(\frac{v_{\rm SM}^2}{2 \Mkk^2}\left(\frac{g_L^2}{8} \left(L-1+\frac{1}{2L}\right)+\frac{g_R^2}{8}\,L\right)\right) + \ord\left( \frac{v_{\rm SM}^4}{\Mkk^4} \right)
	\right) \,.
\eeq
For the case of $g_L=g_R$ the effect is thus approximately twice as big as in the minimal RS model and can become quite significant for low KK scales.
We will always employ $v_{\rm RS}$ $(v_{\rm RS_C})$ for the minimal (custodial) RS model, but drop the subscript, as it will be clear from the context. 
Note that in some cases, the VEV shift will correspond to a higher order correction in $v^2/\Mkk^2$ that will not be considered,
however \eg in Higgs couplings of the $W^\pm$- and $Z$-bosons or of SM quarks it will be a LO correction and needs to be taken into account. 

The correction to the $W^\pm$-boson mass can become observable if several quantities are measured. As explained in \ref{sec:Higgs},
one can also determine the $W^\pm$ mass indirectly by measuring the electromagnetic
coupling, the weak mixing angle and the Fermi constant, making use of the SM relation (\ref{eq:Mindr}). The former quantities are expected 
to feature negligible RS corrections, as discussed above. The Fermi constant $G_F$, obtained from muon decay however, will receive sizable 
corrections in the RS setup. At tree level, this decay receives contributions from the whole towers of the $W^\pm$ bosons and their KK excitations, 
as depicted in Figure~\ref{fig:muondecay}. The corresponding KK sum over intermediate states is given in (\ref{eq:Sigmafinal})
and below for the minimal model. 
As muon decay involves only light leptons we can ignore the terms proportional to $t^2$ and ${t^\prime}^2$ due to the RS-GIM mechanism. 
Thus only the constant term in 
(\ref{eq:Sigmafinal}) contributes. We arrive at the RS prediction
\beq\label{GFcor}
   \frac{G_F}{\sqrt2} = \frac{g^2}{8m_W^2} 
   \left[ 1 + \frac{m_W^2}{2\Mkk^2} \left( 1 - \frac{1}{2L} \right) 
      + \ord\left( \frac{m_W^4}{\Mkk^4} \right) \right],
\eeq
valid both for the minimal as well as for the custodial RS model.
Here, the RS-correction term receives a contribution $(1-1/L)$ from the modification of the $W^\pm$-boson ground state as well as a subleading 
term $1/(2L)$ from the tower of KK excitations.

Thus when applying the SM relation (\ref{eq:Mindr}) to determine the $W^\pm$-boson mass, we will actually not measure its real mass, but 
rather 
\beq\label{eq:RSMWrelation}
   (m_W)_{\rm indirect}^{RS} 
   = m_W \left[ 1 - \frac{m_W^2}{4\Mkk^2} 
   \left( 1 - \frac{1}{2L} \right) 
   + \ord\left( \frac{m_W^4}{\Mkk^4} \right) \right] .
\eeq
\begin{figure}[!t]
\begin{center}
\mbox{\includegraphics[height=2.85in]{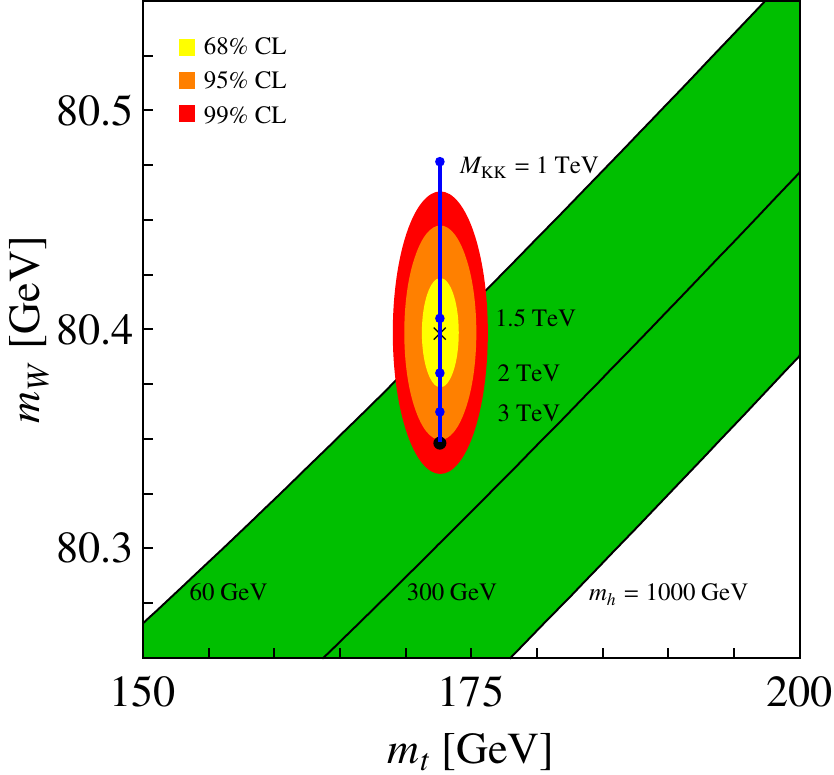}}
\vspace{-2mm}
\parbox{15.5cm}{\caption{\label{fig:MWplot}
Regions of 68\%, 95\%, and 99\% probability in the $m_t$--$m_W$ plane due to the direct measurements of $m_W$ and $m_t$ at LEP2 and the Tevatron. 
The black dot shows the SM prediction based on $G_F$ for our reference input values, whereas the green (medium gray) shaded band corresponds to 
the SM expectation for varying values of the Higgs-boson mass of $m_h\in [60,1000]$\,GeV. The blue (dark gray) line and the corresponding
points represent the RS prediction for $\Mkk\in [1,10]$\,TeV, valid for both RS variants studied in this thesis. See \cite{Casagrande:2008hr} 
and text for details.}}
\end{center}
\end{figure}
In Figure~\ref{fig:MWplot} we show again the comparison between the indirect constraint and the direct measurements of $m_W$ and $m_t$ 
in the SM (see Section~\ref{sec:Higgs}), this time overlaid with the RS prediction for $m_W$, following from (\ref{eq:RSMWrelation}) for $\Mkk\in 
[1,10]$\,TeV and depicted by the blue (dark gray) line and points. The SM limit ($\Mkk \to \infty$) is shown by the black dot, where we 
assume the SM reference point from Appendix~\ref{app:ref} with $m_h=150\,$ GeV. 
Recall that the ellipses show the regions of 68\%, 95\%, and 99\% probability, following from the direct measurements at LEP2 
and the Tevatron. The SM prediction based on (\ref{eq:Mindr}), for Higgs-boson masses in the range of 
$m_h\in [60,1000]$\,GeV, is depicted by the shaded band. 
As noted in Section~\ref{sec:Higgs}, the direct and indirect $W^\pm$-boson mass determinations, assuming the SM to be valid, deviate by about 
$2 \sigma$ (\ref{eq:mwmasses}). However, such a shift of about $50$\,MeV can be explained in the RS model. Here, we do not expect the 
two determinations, using SM formulae, to match, see (\ref{eq:RSMWrelation}). Using this relation to determine the real $W^\pm$ mass 
from the indirect measurement we see from the blue (dark gray) line, that a perfect match can be 
achieved for KK scales slightly above $\Mkk=1.5$\,TeV.  Note that also for a heavy Higgs boson, 
agreement between $m_W$ and $(m_W)_{\rm indirect}$ can be reached at the 99\% CL level
for KK scales above 1\,TeV. Taking for example $m_h=400$\,GeV ($m_h=1000$\,GeV) would indicate KK scales of
1.5\,TeV (1\,TeV).

Due to the RS corrections to the $W^\pm$- and $Z$-boson masses, the SM relation between the weak mixing angle and the masses of the weak 
gauge bosons, encoded in the $\varrho$ parameter (\ref{eq:rhopar}), will deviate from its SM value $\varrho_{SM}=1$ (at the tree level).
If the $W^\pm$-boson mass is measured directly, we derive from (\ref{eq:m02}) the result
\beq\label{eq:rho1}
   \varrho=\left[
   1 + s_w^2\,\frac{m_Z^2}{2\Mkk^2} 
   \left( L - 1 + \frac{1}{2L} \right) 
   + \ord\left( \frac{m_Z^4}{\Mkk^4} \right) \right] .
\eeq
Note that the correction term features a contribution enhanced by a factor of $L$. This $L$-enhanced term vanishes, if the extended gauge group
of the custodial RS model is present. It is canceled by the additional corrections appearing in the contributions to 
the mass formulae (\ref{eq:mwmz}).

Alternatively, if $m_W$ is determined indirectly, we arrive via 
(\ref{eq:RSMWrelation}) at the RS prediction
\beq\label{varrho}
   \varrho = 1 + \frac{m_Z^2}{2\Mkk^2}
   \left( L s_w^2 - 1 + \frac{1}{2L} \right)  
   + \ord\left( \frac{m_Z^4}{\Mkk^4} \right) .
\eeq
Remember that the $T$ parameter, measuring differences in the corrections to the $W^\pm$ and $Z$ vacuum polarization amplitudes,
which contain differences in the corresponding mass corrections, is (trivially) related to the $\varrho$ parameter. 
As we have already analyzed RS corrections to the $T$ parameter numerically in Section~\ref{sec:pheno1} (which will be continued below),
we will stop our discussion of the $\varrho$ parameter here.

\subsection[$S$, $T$ and $U$ Parameters]{\boldmath $S$, $T$ and $U$ \unboldmath Parameters}

We will now compute the $S$, $T$, and $U$ parameters in the custodial RS model for completeness and compare the findings to the minimal model. 
The set of oblique corrections, defined in (\ref{eq:STUdef}), 
can be calculated as described in Section~\ref{sec:pheno1}, where we have already performed the corresponding calculation for the minimal RS model. 
The non-zero tree-level correlators $\Pi_{aa} (0)$ with $a = W, Z$ are
calculated from the corrections to the zero-mode masses
(\ref{eq:mwmz}) and profiles (\ref{eq:expprof}), where the latter also
give rise to non-zero derivatives $\Pi_{aa}^{\hspace{0.25mm} \prime}
(0)$ of the correlators at zero momentum. We find to LO
\begin{equation} \label{eq:cor}
  \begin{split}
    \Pi_{WW}(0)&=-\frac{g_L^2 v^4}{32 \Mkk^2}
    \left[g_L^2\left(L-\frac 1{2L}\right)+g_R^2 L\right] ,\\
    \Pi^{\hspace{0.25mm} \prime}_{WW}(0)&=\frac{g_L^2v^2}{8 \Mkk^2}
    \left(1-\frac 1 L\right) ,\\
    \Pi_{ZZ}(0)&=-\frac{(g_L^2+g_Y^2) \, v^4}{32
      \Mkk^2}\left[\left(g_L^2+g_Y^2\right)\left(L-\frac
        1{2L}\right)+\left(g_R^2-g_Y^2\right) L\right] ,\\
    \Pi^{\hspace{0.25mm} \prime}_{ZZ}(0)&=\frac{(g_L^2+g_Y^2) \,
      v^2}{8 \Mkk^2} \left(1-\frac 1 L\right) .
  \end{split}
\end{equation}
Inserting these expressions into (\ref{eq:STUdef}) yields the tree-level prediction
\begin{equation} \label{eq:STUcu}
  S=\frac{2\pi v^2}{\Mkk^2}\left(1-\frac 1L\right) , \qquad
  T=-\frac{\pi v^2}{4 \, c^2_w \Mkk^2}\, \frac1L\,, \qquad U=0 \,,
\end{equation}
in agreement with \cite{Agashe:2003zs, Delgado:2007ne}. In contrast
to the results in the minimal model, given in (\ref{eq:STURS}), there is no $L$-enhanced term in the $T$
parameter. Like in the $\varrho$ parameter, it has been canceled by the additional corrections
appearing in the mass formulae (\ref{eq:mwmz}),
which introduce extra terms in the correlators $\Pi_{WW}(0)$ and
$\Pi_{ZZ}(0)$, reflecting the underlying (gauged) custodial symmetry. The $S$ and $U$ parameters however remain unaffected.

\begin{figure}[!t]
\begin{center} 
\hspace{-2mm}
\mbox{\includegraphics[height=2.85in]{ST1.pdf}} 
\qquad 
\mbox{\includegraphics[height=2.85in]{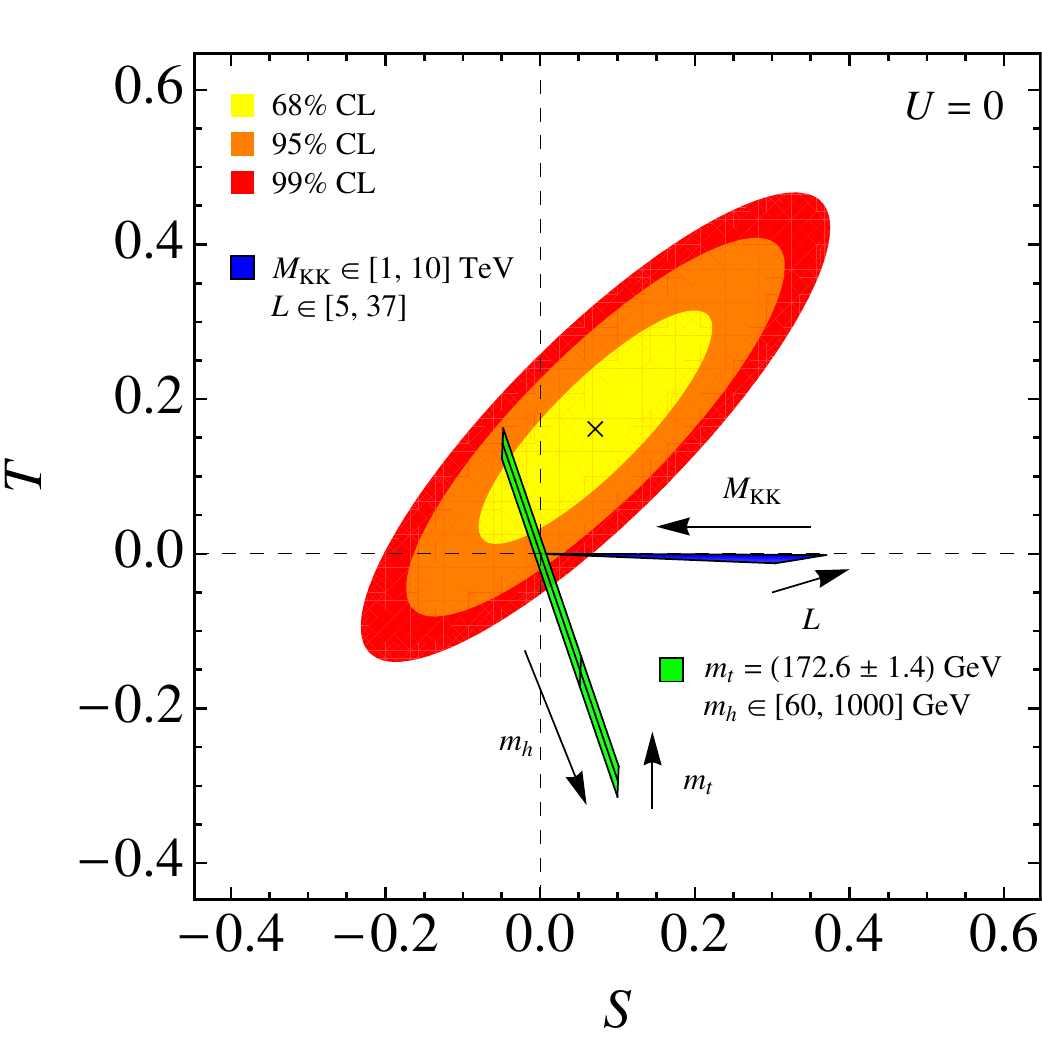}}
\vspace{-2mm}
\parbox{15.5cm}{\caption{\label{fig:STmincus}
Right: Corrections in the $S$--$T$ plane due to the custodial RS model for $\Mkk\in [1,10]$\,TeV and $L\in [5,37]$, 
depicted by the blue (dark-shaded) shaded area. The experimental regions of 68\%, 95\%, and 99\% probability are shown by 
the colored ellipses. The green (light-shaded) stripe shows the SM predictions for $m_t=(172.6\pm 1.4)$\,GeV and 
$m_h\in [60,1000]$\,GeV. Left: corresponding plot in the minimal RS proposal for comparison.
See \cite{Casagrande:2008hr} and text for details.}}
\end{center}
\end{figure}

The contributions to the $S$ and $T$ parameters in the custodial RS model are shown in the right panel of Figure~\ref{fig:STmincus}.
The experimental regions of 68\%, 95\%, and 99\% probability in the $S$--$T$ plane are again depicted by the colored ellipses, while 
the SM predictions for different values of $m_h$ and $m_t$ are shown by the green (light-shaded) stripe. The blue 
(dark-shaded) area represents the corrections of the custodial RS model for different values of the KK scale and the volume factor $L$.
The corresponding contributions in the minimal RS model, already given in Section~\ref{sec:pheno1} are shown in the left panel for comparison.
Requiring the corrections of the custodial model to satisfy the experimental bounds from $S$ and $T$ leads, for the SM reference point of
Appendix~\ref{app:ref} and $L=37$, to the constraint of $\Mkk>2.4$\,TeV $@\, 95\%\, {\rm CL}$ which we quoted in (\ref{eq:MKKboundscustodial}). This 
is now driven by the $S$ parameter, see Figure~\ref{fig:STmincus}, and is slightly lower than the corresponding constraint in the minimal RS model
with a heavy Higgs boson $m_h \lesssim 1$\, TeV, given in (\ref{eq:MKKboundsheavyhiggs}).

Quantum corrections to the $T$ parameter have been studied in \cite{Agashe:2003zs}, in the context of the gauged custodial symmetry
without an additional protection for  $Z b_L \bar b_L$ couplings. It has been shown that, due to this symmetry, the loop contributions 
to $T$ are UV-finite. The dominant one-loop corrections, coming from fermions in the loop, are found to be positive and allow to lower the KK 
scale for a sufficiently light Higgs boson.
For the model including quark bi-doublets under $SU(2)_L\times SU(2)_R$ to protect the $Z b_L\bar b_L$ coupling, the 
one-loop corrections to $T$ have been studied in \cite{Carena:2006bn, Carena:2007ua}. It has been observed that contributions due to KK 
excitations that couple to the Higgs via the top Yukawa coupling generically induce a negative shift in the $T$ parameter. It is clear from Figure \ref{fig:STmincus} that a total negative correction to $T$, together with a positive value of $S$ would be rather 
problematic. However, the authors state that positive corrections to $T$ are possible in some regions of parameter space.
The one-loop corrections to the $S$ parameter in the custodial RS model arising from (4D) Higgs loops have been calculated in
\cite{Burdman:2008gm} and found to be logarithmically UV divergent. This could result in a large and positive $S$ parameter,
which is clearly rather problematic in points of the consistency of the global fit of the oblique electroweak precision observables. 
A more detailed analysis of RS loop corrections to the Peskin-Takeuchi parameters would certainly be interesting.

\subsection{Bottom-Quark Pseudo Observables}
\label{sec:bpseudo}

We now turn to a more detailed analysis of the bottom quark pseudo observables,
which we already have started in Section~\ref{sec:pheno1} to motivate the custodial extension of the RS model.
In order to arrive at simple analytic expressions for $g_L^b \equiv \left( g_L^d
\right)_{33}$ and $g_R^b \equiv \left( g_R^d \right)_{33}$,
we employ the ZMA to our closed and exact formulae of Section~\ref{sec:custodialprotection}.
This offers the possibility to clearly interpret the RS corrections in terms of 
model parameters.
In the numerical studies, we will however use the exact expressions.
First, we need formulae for the leading contributions to the overlap
integrals $\big ( \Delta^{(\prime)}_{D,d} \big)_{33}$. For the minimal model they
have been given in (\ref{ZMA1}). For the custodial RS variant, the corresponding 
expressions turn out to be identical to those of the minimal model (with $c_{f_i}\to c_{{{\cal T}}_{2 i}}$
and $c_{F_i}\to c_{Q_i}$).
Remember that, while $\bm{\varepsilon}_{D,d}^{(\prime)}$ vanish at leading
order in the ZMA and can be neglected, the matrices $\bm\delta_{D,d}$ have to be considered
as they enter (\ref{eq:gLR}) and (\ref{gLR}) with an unsuppressed coefficient. 
The ZMA expressions for $\bm \delta_D$ have been given in (\ref{eq:ZMA2}) and (\ref{eq:deltaD2}), 
and the last missing ingredient ${\bm \delta}_d$ takes the form as given in (\ref{eq:ZMA2}) 
for both RS variants.

After a Taylor expansion of the mixing matrices ${\bm U}_d$ and ${\bm
W}_d$ in powers of the Cabibbo angle $\lambda$, as done in Section~\ref{sec:hierarchies},\footnote{
Note that these LO formulae still hold for the custodial model.} 
we finally arrive at
\begin{align} \label{eq:ZbbRS} 
  g_L^b & = \left( - \frac12 + \frac{s_w^2}{3} \right) \Bigg [ 1 -
  \frac{m_Z^2}{2\Mkk^2}\, \frac{F^2(c_{b_L})}{3 + 2 \mkern+2mu
    c_{b_L}} \left( \omega_Z^{b_L} L - \frac{5+2c_{b_L}}{2 (3 + 2
      \mkern+2mu c_{b_L})} \right)
  \Bigg ] \nonumber \\
  &\mkern+20mu + \frac{m_b^2}{2\Mkk^2}\, \Bigg \{ \frac{1}{1 - 2
    \mkern+2mu c_{b_R}} \left( \frac{1}{F^2(c_{b_R})} \left [ 1 -
     \omega_Z^c \frac{1 - 2 \mkern+2mu c_{b_R}}{1 - 2 \mkern+2mu
        c_{b^{\prime}_R}} \right ] - 1 + \frac{F^2(c_{b_R})}{3 + 2
      \mkern+2mu c_{b_R}} \right)
  \notag \\
  &\mkern+105mu + \sum_{i=1}^2 \frac{|(Y_d)_{3i}|^2}{|(Y_d)_{33}|^2}
  \frac{1}{1 - 2 \mkern+2mu c_{{\cal T}_{2 i}}} \frac{1}{F^2(c_{b_R})}
  \left [ 1 - \omega_Z^c \frac{1 - 2 \mkern+2mu c_{{{\cal T}}_{2 i}}}{1 - 2
      \mkern+2mu c_{{{\cal T}}_{1 i}}} \right ]
  \Bigg \} \, , \\[10pt]
  g_R^b & = \frac{s_w^2}{3} \, \Bigg [ 1 - \frac{m_Z^2}{2\Mkk^2}\,
  \frac{F^2(c_{b_R})}{3 + 2 \mkern+2mu c_{b_R}} \left( \omega^{b_R}_Z
    L - \frac{5 + 2 \mkern+2mu c_{b_R}}{2 (3 + 2
      \mkern+2mu c_{b_R})} \right) \Bigg ] \notag \\
  &\mkern+20mu - \frac{m_b^2}{2\Mkk^2}\, \Bigg \{ \frac{1}{1 - 2
    \mkern+2mu c_{b_L}} \left( \frac{1}{F^2(c_{b_L})} - 1 +
    \frac{F^2(c_{b_L})}{3 + 2 \mkern+2mu c_{b_L}} \right) +
  \sum_{i=1}^2 \frac{|(Y_d)_{i3}|^2}{|(Y_d)_{33}|^2} \frac{1}{1 - 2
    c_{Q_i}} \frac{1}{F^2(c_{b_L})} \Bigg \} \, , \hspace{6mm} \notag
\end{align}
where in the {\it custodial RS model} $c_{b_L} \equiv c_{Q_3}$, $c_{b_R} \equiv c_{{{\cal
T}}_{2\hspace{0.05mm}3}}$, $c_{b^\prime_R} \equiv c_{{{\cal
T}}_{1\hspace{0.05mm}3}}$, and $\omega_Z^c=1$. Furthermore, $m_b \equiv m_b (\Mkk)$
denotes the bottom-quark $\overline{\rm MS}$ mass evaluated at the KK
scale and we have demanded $c_{b_R^\prime} ,\hspace{0.5mm} c_{{{\cal T}}_{1 i}} < 1/2$ . 
Notice that we kept $c_{{{\cal T}}_{1 i}} \neq c_{{{\cal
T}}_{2i}}$, thereby allowing the $P_{LR}$ symmetry to be broken by the
triplet bulk masses. We also retained the parameters $\omega_Z^{b_L}$
and $\omega_Z^{b_R}$. In the custodial model one has
\begin{equation} \label{eq:omegabR}
\omega_Z^{b_L} = 0\,,\  \omega^{b_R}_Z = \frac{3 c_w^2}{s_w^2} \approx 10.0 \,,
\end{equation}
where in order to arrive at the numerical values, we have employed
$s_w^2 \approx 0.23$, corresponding to the value of the weak mixing
angle at the $Z$-pole.\footnote{The electromagnetic coupling 
and the weak mixing angle are running parameters in the low-energy effective
theory obtained after integrating out the RS contributions at the scale
$\Mkk$. We included the associated large logarithms effectively by
replacing $s_w^2(\Mkk)$ by $s_w^2(m_Z)$ in the couplings
$g_{L,R}^b$. However, the value of the bottom-quark mass
entering the matching is frozen at the high scale and does not evolve
in the effective theory.}

In the {\it minimal RS model}, (\ref{eq:ZbbRS}) holds with 
\beq
\omega_Z^c=0,\,\ \omega_Z^{b_L} = 
\omega_Z^{b_R} = 1\,,\ c_{b_R} \equiv c_{d_3}\,,\ 
c_{{{\cal T}}_{2i}} \to c_{di}.
\eeq
In that case, the non-universal corrections always reduce the couplings with respect to their SM values in magnitude. 
With the freedom of reparametrization invariance, see (\ref{RPI1}) and (\ref{RPI2a}), one can
redistribute contributions between the left-handed and right-handed couplings, without changing the zero-mode
quark masses and CKM parameters. However, the value of $F(c_{b_L})$ cannot be made too small due to the need
of reproducing the large top-quark mass with $\ord(1)$ Yukawa couplings.
    
In the custodial model, we observe that the non-universal
corrections to the $Z b \bar b$ couplings reduce both $g_L^b$ and
$g_R^b$ if the extended $P_{LR}$ symmetry (\ref{eq:extendedPLR}) is at
work. If one allows the $P_{LR}$ symmetry to be broken by $c_{{\cal
T}_{1i}} \neq c_{{\cal T}_{2i}}$, then the shift in $g_L^b$ can also
be positive as a result of fermion mixing. As we will see,
this always worsens the quality of the $Z \to b \bar b$ fit. Due to the custodial
protection $\omega_Z^{b_L} = 0$, the constraints arising from the
bottom-quark pseudo observables are naively much less stringent
in the custodial model with respect to the minimal model, where the shift
$\delta g_L^b$ is large and positive while $\delta g_R^b$ is small and
negative. In order to gauge the improvement 
and to fully understand the parameter dependence, in particular on 
the bulk mass parameters $c_{{\cal T}_{1i}}$, one has to perform a detailed
numerical analysis. This exercise is the subject of the next pages.

Consider the ratio of the width of the $Z$-boson decay into bottom
quarks and the total hadronic width, $R_b^0$, the bottom-quark
left-right asymmetry, $A_b$, and the forward-backward asymmetry for
bottom quarks, $A_{\rm FB}^{0,b}$. The dependences of these quantities
on the left- and right-handed bottom-quark couplings are given by
\cite{Field:1997gz}
\begin{equation}\label{eq:bPOtheory}
  \begin{split}
    R_b^0 &= \left [ 1 + \frac{4 \; {\displaystyle \sum}_{q=u,d}
        \left[ (g_L^q)^2 + (g_R^q)^2\right]}%
      {\eta_{\rm QCD}\,\eta_{\rm QED} \left[ (1-6z_b) (g_L^b-g_R^b)^2
          + (g_L^b+g_R^b)^2 \right]}
    \right]^{-1}\! , \\
    A_b &= \frac{2\sqrt{1-4z_b}\,\, {\displaystyle
        \frac{g_L^b+g_R^b}{g_L^b-g_R^b}}}%
    {1-4z_b+(1+2z_b) {\displaystyle \left(
          \frac{g_L^b+g_R^b}{g_L^b-g_R^b} \right)^2}} \,, \qquad
    A_{\rm FB}^{0,b} = \frac34\,A_e\hspace{0.25mm}A_b \,.
  \end{split}
\end{equation}
Radiative QCD and QED corrections are encoded by the factors
$\eta_{\rm QCD}=0.9954$ and $\eta_{\rm QED}=0.9997$, while the
parameter $z_b\equiv m_b^2(m_Z)/m_Z^2=0.997\cdot 10^{-3}$ describes
the effects of the non-zero bottom-quark mass. Since to an excellent
approximation one can neglect the RS contributions to the left- and
right-handed couplings of the light quarks, $g_{L,R}^q$, and to the
asymmetry parameter of the electron, $A_e$, we will fix these
quantities to their SM values $(g_L^u)_{\rm SM}=0.34674$,
$(g_R^u)_{\rm SM} =-0.15470$, $(g_L^d)_{\rm SM}=-0.42434$,
$(g_R^d)_{\rm SM}=0.077345$ \cite{LEPEWWG:2005ema}, and $(A_e)_{\rm
SM} =0.1462$ \cite{Arbuzov:2005ma}. The quoted numbers correspond to
the SM input parameters given in Appendix~\ref{app:ref}.

Evaluating the relations (\ref{eq:bPOtheory}) using $\big ( g_L^b \big
)_{\rm SM}=-0.42114$ and $\big ( g_R^b \big)_{\rm SM}=0.077420$
\cite{LEPEWWG:2005ema}, we obtain for the central values of the
bottom-quark pseudo observables
\begin{equation}\label{eq:bPOSM}
  \big (R_b^0 \big )_{\rm SM} = 0.21578 \,, \qquad 
  \big ( A_b \big )_{\rm SM} = 0.935 \,, \qquad 
  \big ( A_{\rm FB}^{0,b} \big )_{\rm SM} = 0.1025 \,.
\end{equation}
One should compare these numbers with the experimental results
\cite{LEPEWWG:2005ema}
\begin{equation}\label{eq:bPOsexp}
  \begin{array}{l}
    \big ( R_b^0 \big)_{\rm exp} = 0.21629\pm 0.00066 \,, \\[0.25mm] 
    \big ( A_b \big)_{\rm exp} = 0.923\pm 0.020 \,, \\[1mm]
    \big ( A_{\rm FB}^{0,b} \big )_{\rm exp} = 0.0992\pm 0.0016 \,, 
  \end{array}
  \qquad 
  \rho = \begin{pmatrix}
    1.00 \, & \, -0.08 & \, -0.10 \\ 
    \, -0.08 & \, 1.00 & \, 0.06 \\ 
    -0.10 \, & \, 0.06 & \, 1.00 
  \end{pmatrix} ,
\end{equation}
where $\rho$ is the correlation matrix. While the $R_b^0$ and $A_b$
measurements agree within $+0.8\sigma$ and $-0.6\sigma$ with their SM
predictions for $m_h = 150 \, {\rm GeV}$, the $A_{\rm FB}^{0,b}$
measurement quoted in (\ref{eq:bPOsexp}) is almost $-2.1\sigma$ away from its SM
expectation.\footnote{For $m_h=115$\,GeV the discrepancy in $A_{\rm
FB}^{0,b}$ would amount to around $-2.5\sigma$.} Shifts of order
$+20\%$ and $-0.5\%$ in the right- and left-handed bottom-quark
couplings relative to the SM could explain the observed
discrepancy. Such a pronounced correction in $g_R^b$ would affect
$A_b$ and $A_{\rm FB}^{0,b}$, which both depend linearly on the ratio
$g_R^b/g_L^b$, in a significant way, while it would not spoil the good
agreement in $R_b^0\propto (g_L^b)^2 + (g_R^b)^2$.

\begin{figure}[!t]
\begin{center}
\vspace{-0.5cm}
\mbox{\includegraphics[height=2.8in]{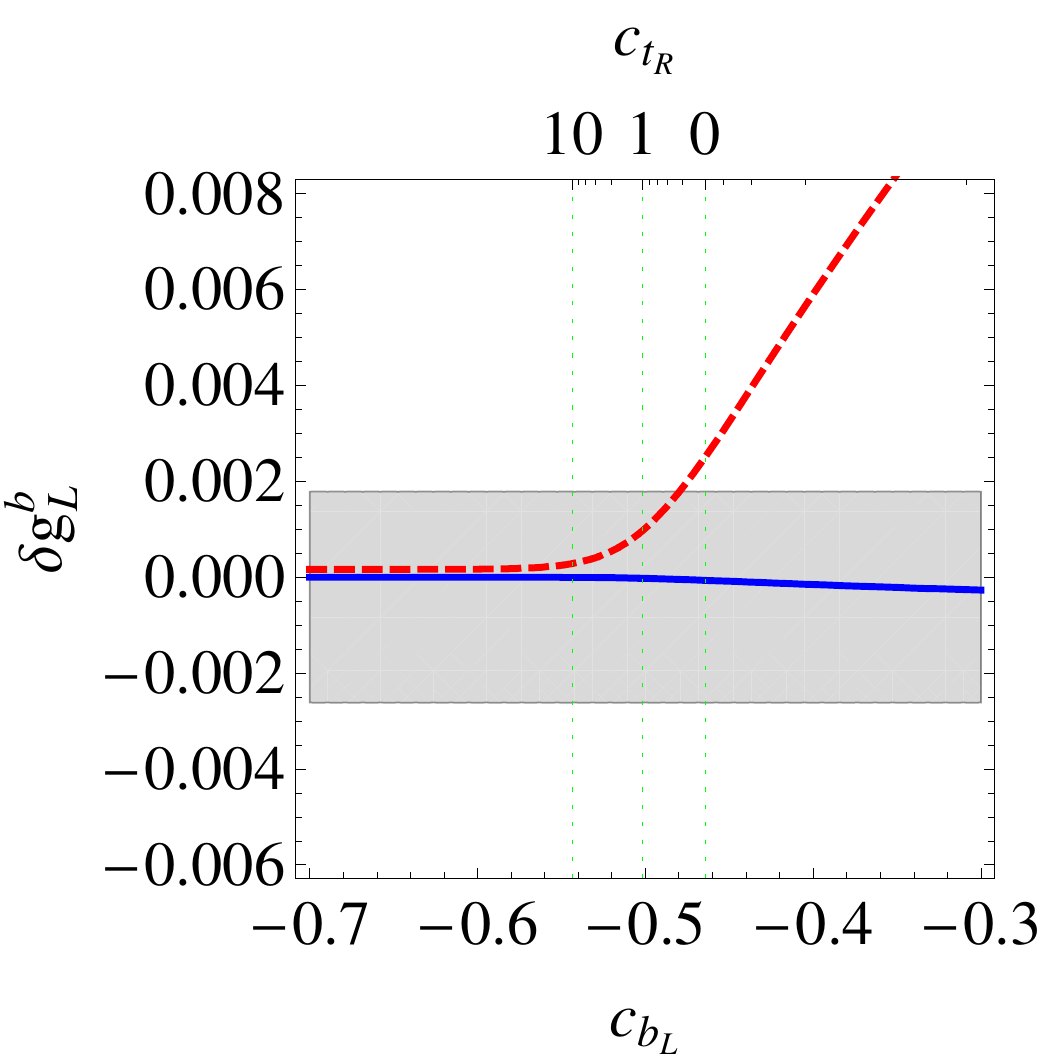}}
\qquad 
\mbox{\includegraphics[height=2.8in]{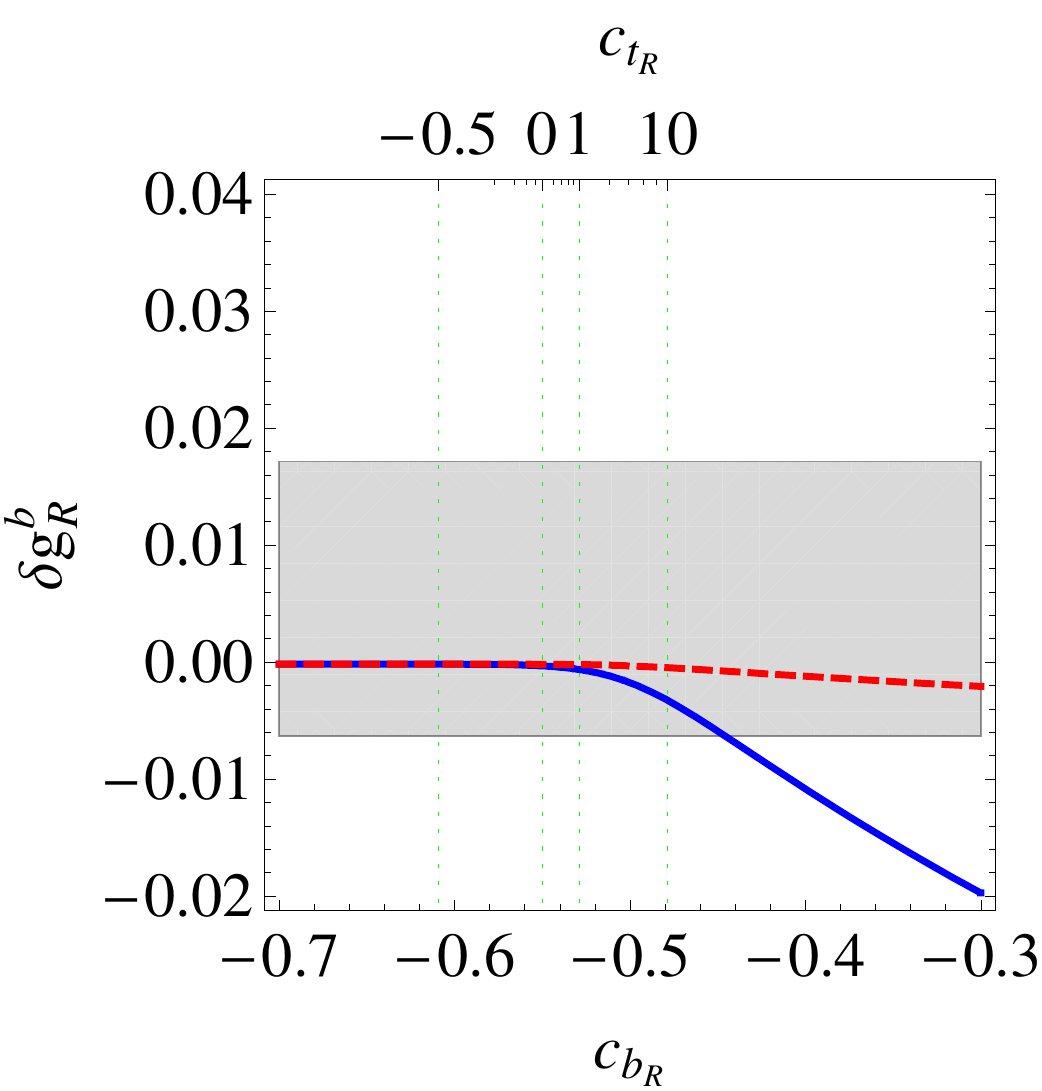}}
\vspace{-2mm}
\parbox{15.5cm}{\caption{\label{fig:dgLRb} Anomalous couplings $\delta
    g_L^b$ (left) and $\delta g_R^b$ (right) in the RS setup 
    as functions of $c_{b_L}$ and $c_{b_R}$. The blue solid (red dashed) lines correspond to the
    predictions obtained in the RS model with extended $P_{LR}$
    symmetry (minimal RS model). Bulk mass parameters not explicitly
    shown are set to $-1/2$, and all elements of the down-type Yukawa
    matrix entering the prediction are taken to be equal to 1 in magnitude. The light gray
    bands indicate the experimentally allowed 99\%~CL ranges. See \cite{Casagrande:2010si} 
    and text for details.}}
\end{center}
\end{figure}
Our predictions for the anomalous
couplings $\delta g_{L,R}^b \equiv g_{L,R}^b - \big ( g_{L,R}^b \big
)_{\rm SM}$\,, following from (\ref{eq:ZbbRS}), are shown
in Figure~\ref{fig:dgLRb} as functions of the bulk mass parameters
$c_{b_{L,R}}$. These are the most important parameters as they enter (\ref{eq:ZbbRS}) through 
their zero-mode profiles $F(c_{b_{L,R}})$. Similar plots have been presented in
\cite{Carena:2006bn}. Correlations between these bulk masses will be taken into account further
below. The shown curves correspond to $c_{Q_i} =
c_{{\cal T}_{1i}} = c_{{\cal T}_{2i}} (= c_{d_i}) = -1/2$ and $|(Y_d)_{3i}| =
|(Y_d)_{i3}| = |(Y_d)_{33}| = 1$ with $i = 1,2$. First we
identify the large and always positive corrections $\delta g_{L}^b$ in the minimal
RS model (red dashed line, {\it c.f.} Figure \ref{fig:STgLR}) for $c_{b_L}>-1/2$, which is the generic range in order to
reproduce the large third-generation masses, see Figure~\ref{fig:cs}.
In contrast, the prediction for $\delta
g_L^b$ in the RS model with extended $P_{LR}$ symmetry (blue solid
line) is, owing to (\ref{eq:custodialQ}), essentially independent of
$c_{b_L}$.\footnote{In order not to induce unacceptably large
corrections to $\delta g_R^b$ due to fermion
mixing, one has to require $c_{b_L} \gtrsim -0.55$.} It is thus generically 
within the experimental 99\% CL bound (light gray band), which
gives a strong motivation to protect the $Z b_L \bar b_L$ vertex
through the mechanism of \cite{Agashe:2006at}. Notice that in the
case of the minimal RS model, $\delta g_L^b$ can be suppressed by
pushing the right-handed top quark very close to the IR brane. This
feature is illustrated by the ticks on the upper border of the frame
in the left panel. The given values of $c_{t_R} \equiv c_{{u^{(c)}_3}}$ have
been obtained by solving $m_t = v/\sqrt{2} \, \left |(Y_u)_{33}
\right| \left |F(c_{b_L}) \hspace{0.25mm} F(c_{t_R}) \right |$ for the
bulk mass parameter $c_{t_R}$, see (\ref{eq:quarkmasses}), evaluating the 
top-quark $\overline{\rm MS}$ mass at $\Mkk = 1 \, {\rm TeV}$ and setting $|(Y_u)_{33}| =
3$. For smaller (larger) values of $|(Y_u)_{33}|$ the ticks are
shifted to the right (left).

In the case of $\delta g_R^b$ we observe instead, that as a result of
(\ref{eq:omegabR}), the corrections to the anomalous coupling are
always larger in the RS model with extended $P_{LR}$ symmetry (blue
solid line) than in the minimal formulation (red dashed
line).\footnote{Notice that in order to reproduce the large top-quark
mass with Yukawa couplings of $\ord (1)$ one has to require $c_{t_R} >
-1/2$, corresponding to $c_{b_R} \gtrsim -0.6$.} It is however
important to remark that even in the former case the $Z b_R \bar b_R$
coupling is predicted to be SM-like, since shifts in $\delta g_R^b$
outside the experimental 99\%~CL range (light gray band) would require
the bulk mass parameter of the right-handed top quark to be
significantly larger than 1. Such a choice appears unnatural, since
$c_{t_R} > 1$ implies that the corresponding bulk mass exceeds the
curvature scale, in which case the right-handed top quark should be
rather treated as a brane-localized field and not considered as a bulk fermion. The latter
feature can be inferred from the ticks on the upper border of the
frame in the right panel. They have been obtained by combining the
equality $m_b = v/\sqrt{2} \, \left |(Y_d)_{33} \right| \left
|F(c_{b_L}) \hspace{0.25mm} F(c_{b_R}) \right |$ with the one for
$m_t$ given earlier, solving again for $c_{t_R}$. The Yukawa
parameters have been fixed to $|(Y_d)_{33}| = 1$ and $|(Y_u)_{33}| =
3$. For smaller (larger) values of $|(Y_d)_{33}|$ the ticks move to
the right (left). Rescaling $|(Y_u)_{33}|$ has the opposite
effect. This observation brings us to the conclusion that,
irrespectively of the RS bulk gauge group, naturalness in combination
with the requirement to reproduce the observed top- and bottom-quark
masses excludes large corrections to $\delta g_R^b$ in models of
warped extra dimensions in which the left-handed bottom and top quark
reside in the same multiplet. 

\begin{figure}[!t]
\begin{center} 
\hspace{-2mm}
\mbox{\includegraphics[height=2.85in]{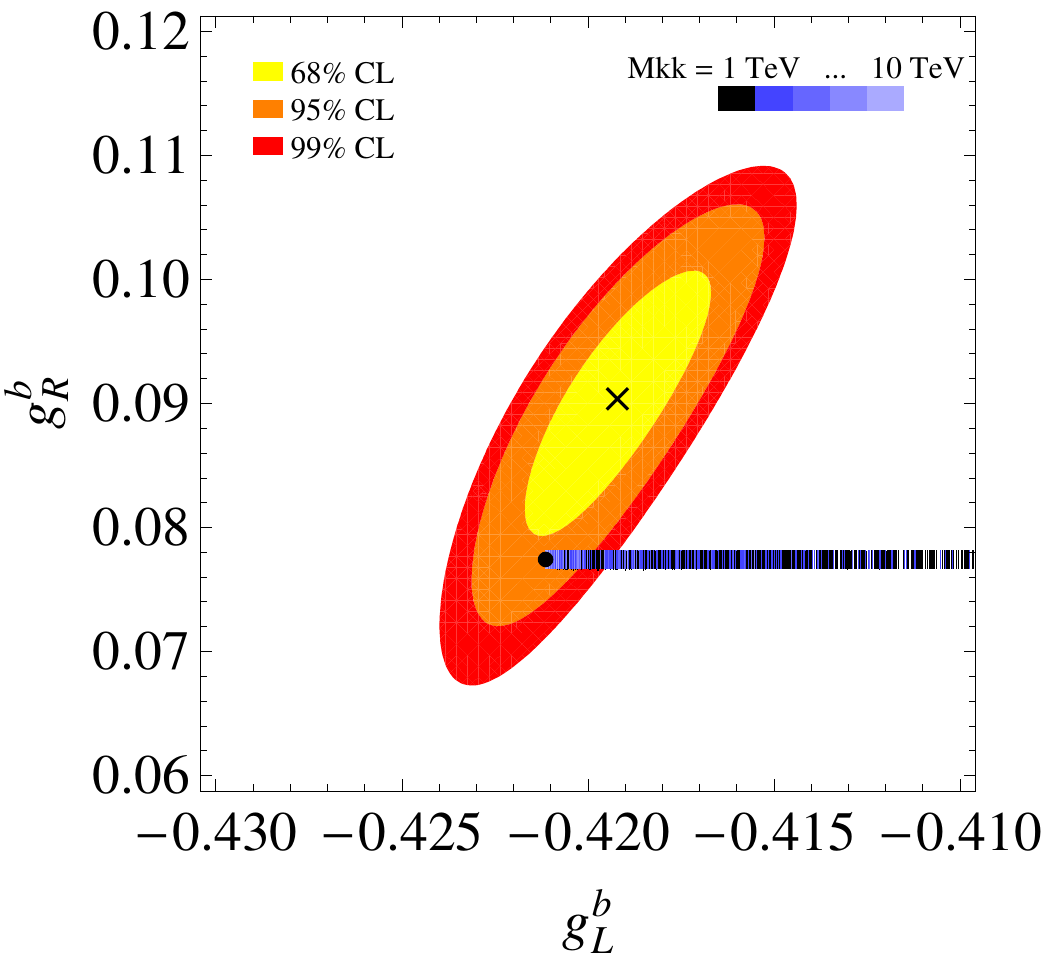}} 
\quad 
\mbox{\includegraphics[height=2.85in]{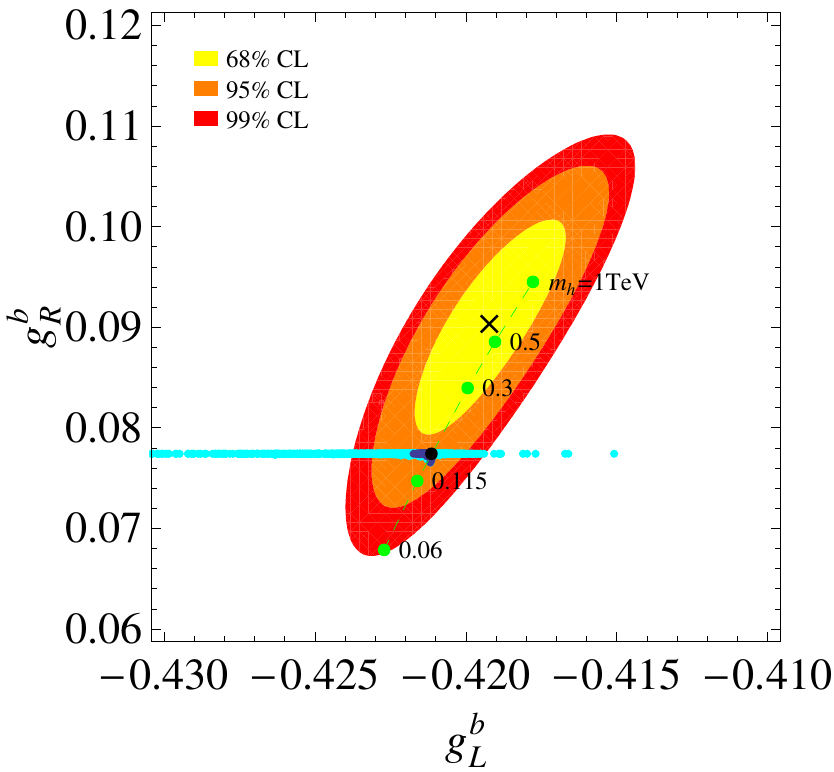}} 
\vspace{-2mm}
\parbox{15.5cm}{\caption{\label{fig:gLgRplots} 
    Regions of 68\%, 95\%, and 99\%
    probability in the $g_L^b$--$g_R^b$ plane. The horizontal stripes
    consist of a large number of points in parameter space for the minimal RS model (left)
    and the custodial model (right). The blue
    (cyan) points in the right panel represent the result for $c_{b^\prime_R} = c_{b_R}$
    ($c_{b^\prime_R} \neq c_{b_R}$). The black dot is the SM
    expectation for the reference point, and the green dashed line
    indicates the SM predictions for $m_h\in [0.06,1]$\,TeV. See \cite{Casagrande:2008hr,Casagrande:2010si} and text
    for details.}}
\end{center}
\end{figure}

The predictions in the $g_L^b$--$g_R^b$ plane for a scan over 10000 randomly generated points
in the RS parameter space, as described at the beginning of this chapter, are shown in Figure~\ref{fig:gLgRplots}. The results, derived by
applying the exact formulae for the $Zb\bar b$ couplings, will confirm our findings from above and allow to study the potential size 
of $P_{LR}$ breaking corrections in the custodial RS setup. The regions of $68\%$, $95\%$, 
and $99\%$~CL, obtained from a global fit to the $Z \to b\bar b$ pseudo observables (\ref{eq:bPOsexp}), are indicated by
the colored ellipses. The RS predictions for the minimal (custodial) RS model are superimposed on the left (right) panel.
For the latter, the predictions with (without) extended $P_{LR}$ symmetry are depicted as blue (cyan) scatter points. 
As discussed before, it is evident that the prediction for $g_L^b$ in the minimal model is always larger than the SM reference value indicated by the black dot, 
while $g_R^b$ is essentially unaffected. The corresponding negative shift is tiny.
In turn, the values $g_{L,R}^b$ are necessarily shifted further away from the best fit $g_L^b=-0.41918$ and $g_R^b=0.090677$ (black cross). 
For the custodial model featuring the extended $P_{LR}$ symmetry, the prediction is more or less SM-like.
The potential size of $P_{LR}$ symmetry-breaking corrections is shown
by the cyan points in the right panel. They have been obtained by allowing the bulk mass
parameters $c_{{\cal T}_{1i}}$\,, with $i = 1, 2, 3$\,, to take any value in
the range $[-1,0]$.\footnote{Note that here we still keep $g_L=g_R$, as we will
do for the remainder of this thesis.} While in the case of the extended symmetry, the small RS
contributions always drive $g_L^b$ to smaller values with respect to
the SM reference point, in the latter case moderate
positive and large negative corrections in $g_L^b$ are possible,
leading further away from the best fit values. In both cases $g_R^b$ remains
essentially unaffected, confirming the discussion in the context of 
Figure \ref{fig:dgLRb}.\footnote{Also in the custodial model the corrections to $g_R^b$ are
always negative but tiny and hence hardly visible in the figure.} 
In conclusion, in the custodial RS model, as well as in the model 
featuring a SM gauge group, the corrections (\ref{eq:ZbbRS}) alone
cannot account for the positive shift in $g_R^b$ needed to explain the
anomaly in $A_{\rm FB}^{0,b}$.
This should be contrasted with the analysis of \cite{Djouadi:2006rk}, 
which finds sizable corrections in $\delta g_R^b$. However, this is due to 
chosen bulk mass parameters $c_{b_{L,R}}$ and $c_{t_R}$ that lead to bottom- and 
top-quark masses of $m_b \approx 40 \, {\rm GeV}$ and $m_t \approx 75 \, {\rm GeV}$, 
which are in conflict with observation. Let us finally mention that, if the left-handed
bottom and top quarks arise as an admixture of the zero-mode fields of
two $SU(2)_L$ doublets, then the bottom- and top-quark masses are
determined by two independent sets of bulk mass parameters, so that it
is possible to account simultaneously for the quark masses and mixings
as well as for the $A_{\rm FB}^{0,b}$ anomaly \cite{Bouchart:2008vp}.

\begin{figure}[!t]
\begin{center} 
\hspace{-2mm}
\mbox{\includegraphics[height=2.85in]{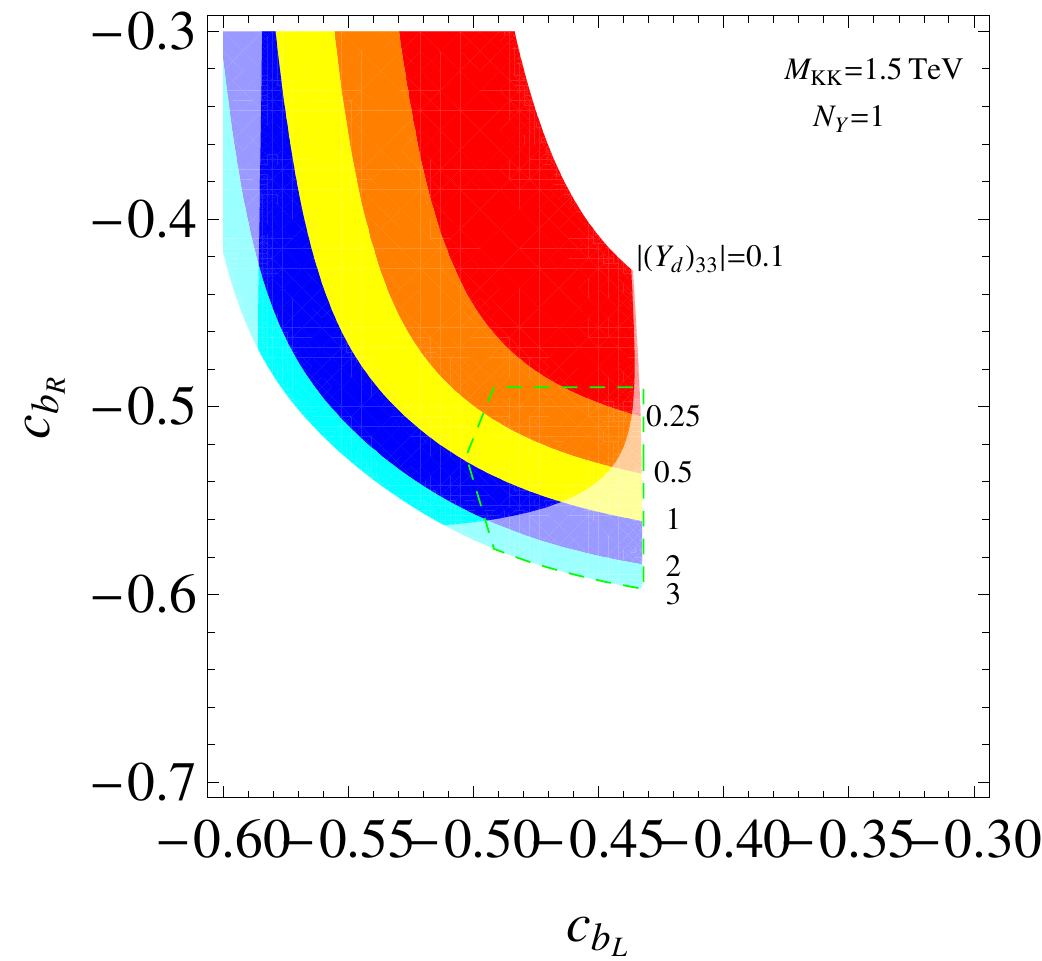}} 
\quad 
\mbox{\includegraphics[height=2.85in]{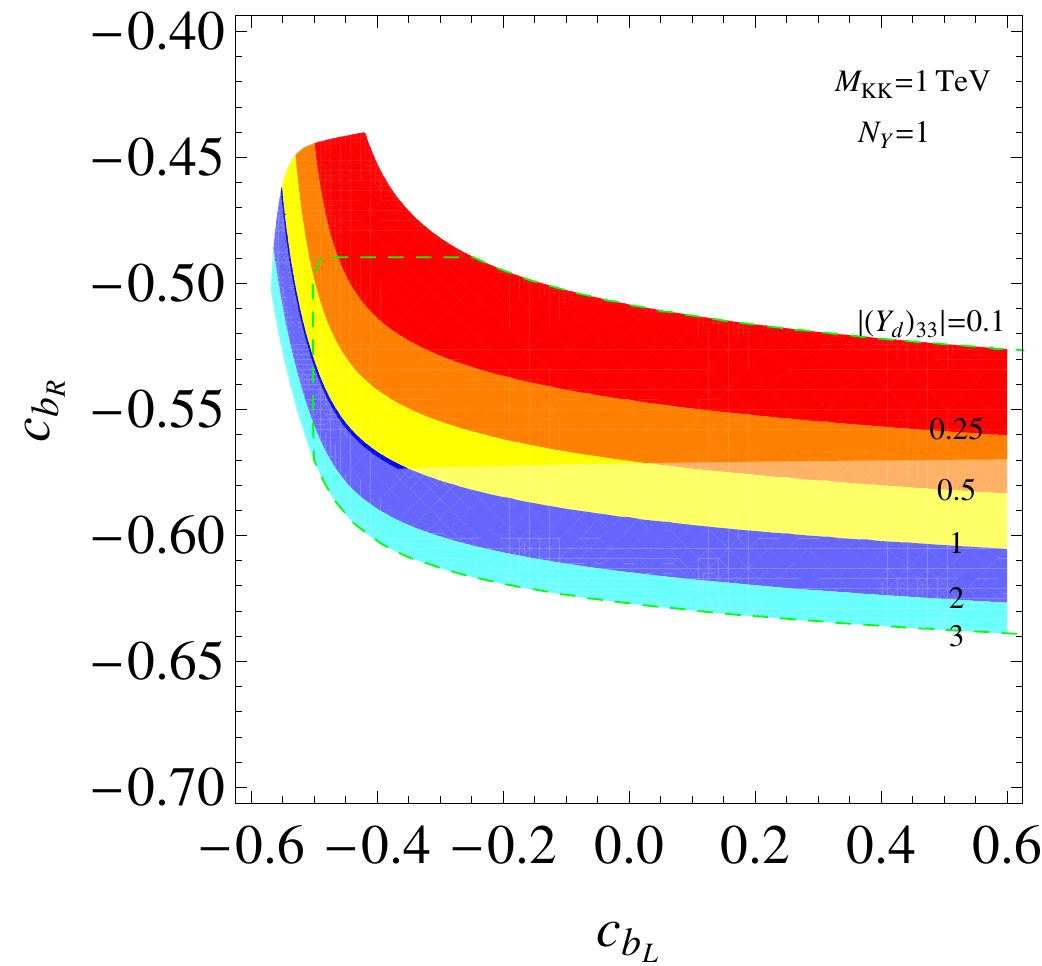}} 
\vspace{-2mm}
\parbox{15.5cm}{\caption{\label{fig:cplots} 
    Region of 99\%
    probability in the $c_{b_L}$--$c_{b_R}$ plane for the minimal RS 
    model for $\Mkk=1.5$\,TeV (left) and the custodial extension (right)
    for $\Mkk=1$\,TeV. We set $c_{Q_i} = c_{{\cal T}_{1i}} =c_{{\cal
    T}_{2i}} (= c_{d_i}) = -1/2$ for $i,j = 1,2$ and $N_Y \equiv |(Y_d)_{ij}|/|(Y_d)_{33}|=1$. 
    The colored contours indicate the value of
    $|(Y_d)_{33}|$ necessary to reproduce the
    bottom-quark mass. For the minimal RS model the results with (without) the $m_b^2/\Mkk^2$
    terms are indicated by bright (faint) colors.
    For the custodial model the whole colored region corresponds to
    the RS results with $c_{b^\prime_R} = c_{b_R}$, while only the parameter
    space indicated by bright colors is accessible for $c_{b^\prime_R}
    = 0$. The green dashed, fin-shaped regions contains 99\% of the
    parameter points that lead to consistent values of the quark
    masses and mixings. See \cite{Casagrande:2008hr,Casagrande:2010si} and text for details.}}
\end{center}
\end{figure}

The apparent large positive corrections to $g_L^b$ in the minimal RS variant imply that 
the $R_b^0$, $A_b$, and $A_{\rm FB}^{0,b}$ measurements impose stringent constraints on the parameter space 
of this RS scenario. As discussed before, the distribution of points depends strongly on the bulk mass parameters $c_{b_L}$ 
and $c_{b_R}$, while the exact values of the elements of the down-type Yukawa matrices, as well as the bulk masses of the first two 
generations and the KK scale have only a minor impact on the overall picture. 
The last fact is illustrated by the blue coloring of the RS predictions in the left panel of Figure~\ref{fig:gLgRplots}.
Also for low KK scales, indicated in black, the RS predictions enter the 2\,$\sigma$ range.
This explains the rather low bound on the KK scale quoted in Section~\ref{sec:pheno1}.

In summary, the allowed values of $c_{b_{L,R}}$ in
the minimal model are strongly constrained by the $Z\to b\bar b$ pseudo observables, whereas the other parameters are only weakly 
bounded. The former feature is illustrated in the left panel of Figure~\ref{fig:cplots}, which shows the regions of 99\% probability in 
the $c_{b_L}$--$c_{b_R}$ plane for the minimal RS model for $\Mkk=1.5$\,TeV. The colored contours indicate the magnitude $|(Y_d)_{33}|$ 
necessary to achieve the correct value of the bottom-quark mass. This requirement correlates $c_{b_L}$ and $c_{b_R}$. 
We find that under the restriction $0.1<|(Y_d)_{33}|<3$ the allowed $c_{b_{L,R}}$ parameters all lie in the 
intervals $c_{b_L}\in [-0.59,-0.44]$  and $c_{b_R}>-0.56$  
for $\Mkk=1.5$\,TeV. For $\Mkk=3$\,TeV the corresponding ranges 
read $c_{b_L}\in [-0.61,-0.16]$ and $c_{b_R}>-0.59$. 
Note that the RS results with (without) the $m_b^2/\Mkk^2$ terms in (\ref{eq:ZbbRS}) are indicated by bright (faint) colors. 
These terms result from the matrix elements $(\delta_D)_{33}$ and $(\delta_d)_{33}$ in (\ref{gLR}), arising due to fermion mixing, 
They have usually been neglected in the literature. Comparing the allowed regions makes clear that neglecting these contributions is 
in general not justified. 

Increasing the magnitude $|(Y_d)_{33}|$ decreases the available parameter space. 
This can be understood from the fact that for a larger Yukawa coupling the profiles have to be localized closer to the UV brane in order to reproduce
the correct bottom-quark mass. {\it Naively} one might expect that this feature diminishes the RS corrections. However, the corrections
due to fermion mixing increase in that case, as it is reflected by the dependence on the {\it inverse} zero-mode profile $F^{-1}(c_{b_{L,R}})$.
Notice also that the terms in (\ref{eq:ZbbRS}) proportional to $m_Z^2/\Mkk^2$ depend linearly on $L$, whereas the terms proportional to $m_b^2/\Mkk^2$ are independent of the RS volume factor. In consequence these latter terms cannot be removed by truncating the volume of the RS background and 
are typically dominant for moderate and small values of $L$.
The requirement of obtaining consistent values for all the quark masses and CKM parameters restricts the parameter space further. 
This constraint is indicated by the green dashed regions in Figure~\ref{fig:cplots}, which contains 99\% of the allowed
parameter points. 

The impact of a possible breaking of the $P_{LR}$ symmetry of the custodial
setup by the bulk-mass parameters $c_{{\cal T}_{1i}}$ is shown in the right 
panel of Figure~\ref{fig:cplots}. The plot is analogous to the one shown in 
the left panel with however $\Mkk=1$\,TeV and the  $m_b^2/\Mkk^2$ terms included
for the whole colored region. This region corresponds to the case of the extended 
$P_{LR}$ symmetry. Allowing for $c_{b_R^\prime} \neq c_{b_R}$ potentially cuts 
away a sizable part of parameter space. This is demonstrated by the bright 
colored region, which shows the allowed range for the choice 
$c_{b_R^\prime} = 0$. The $P_{LR}$-breaking correction to $g_{b_L}$ in 
(\ref{eq:ZbbRS}) arising from $c_{b_R^\prime} \neq c_{b_R}$ scales like $-v^2/\Mkk^2 \, 
| (Y_d)_{33}|^2 \, F^2 (c_{b_L})$. This explains, why values $|(Y_d)_{33}| \gtrsim
1$ are not compatible with the $Z \to b \bar b$ data in the case of
$c_{b_R^\prime} = 0$.

In the light of the insufficient corrections to $g_R^b$ to explain the 
$A_{\rm FB}^{0,b}$ anomaly with the help of RS contributions alone, 
note finally that a perfect fit can however be achieved by allowing for 
a heavy Higgs boson. The corrections
\begin{equation} \label{eq:Zbbmh}
  \Delta g_L^b = 1.77 \cdot 10^{-3} \, \ln \frac{m_h}{m_h^{\rm ref}} \,, \qquad 
  \Delta g_R^b = 0.92 \cdot 10^{-2} \, \ln \frac{m_h}{m_h^{\rm ref}}
\end{equation}
in $g_{L,R}^b$ due to a Higgs-boson mass different from the reference
value $m_h^{\rm ref} = 150 \, {\rm GeV}$ are both positive for $m_h >
m_h^{\rm ref}$. The relations (\ref{eq:Zbbmh}) parametrize the leading
logarithmic Higgs-mass dependences of $g_{L,R}^b$. They have been
derived with the help of {\tt ZFITTER}.\footnote{The default flags of {\tt ZFITTER}
version 6.42 are used, except for setting {\tt ALEM=2} to take into
account the externally supplied value of $\Delta\alpha^{(5)}_{\rm
had}(m_Z)$.} The shifts in the $Z b \bar b$ couplings for $m_h
\in [0.06, 1] \, {\rm TeV}$ are indicated by the green dashed line in
the right panel of Figure~\ref{fig:gLgRplots}. One observes that a
Higgs-boson mass in the ballpark of $m_h=0.5$\,TeV would bring the
predictions of $g_{L,R}^b$ so close to the best fit values that
already the small corrections in the custodial RS model with extended $P_{LR}$
symmetry are sufficient to reach the minimum of the $\chi^2$
distribution. Warped models with the Higgs field localized in the IR
might thus indirectly allow for an explanation of the $A_{\rm
FB}^{0,b}$ anomaly, since, as we have argued in Section~\ref{sec:SMinB},
the Higgs boson is naturally expected to be heavy in these set-ups. However, remember that
in RS models with custodial protection of the $T$ parameter 
a large Higgs-boson mass is not unproblematic and can potentially spoil the 
global electroweak fit.

\subsection{Top-Quark Forward-Backward Asymmetry}
\label{sec:afbt}

As explained in Section~\ref{sec:hierarchies}, the top quark plays a special role
in RS models with flavor anarchy. Being mostly composite (from the holographic point of view)
it interacts significantly with the KK excitations. As a consequence, non-negligible effects of the RS setup are expected
in top-quark observables.
  
The CDF and D{\O} experiments at the Tevatron have
collected thousands of $t\bar t$ events, allowing to
measure the top-quark mass, $m_t$, and its total inclusive cross
section, $\sigtot$, with an accuracy of below $1\%$ \cite{top:2009ec}
and $10\%$ \cite{CDFnotetot, D0notetot}, respectively. 
However, for the search of BSM physics, kinematic
distributions and charge asymmetries in $t \bar t$ production are more
interesting. These observables are particularly sensitive to
non-standard dynamics. Studies in that direction
have been performed at the Tevatron \cite{Aaltonen:2007dz, Aaltonen:2007dia, Abazov:2008ny}, and
a result for the $t \bar t$ invariant mass spectrum, $\dsig$, obtained 
from CDF data has been presented \cite{Bridgeman:2008zz,Aaltonen:2009iz}. 

Interestingly, also the $t \bar t$  forward-backward asymmetry, $\AFBt$, has 
been measured \cite{CDFbrandnew,Aaltonen:2011kc,Schwarz:2006ud,Abazov:2007qb,Aaltonen:2008hc,
publicCDF} and persistently found to be larger than
expected. 
In the laboratory ($p \bar p$) frame, the most recent CDF
result, obtained from semileptonic $t \bar t$ events, reads
\beq \label{eq:AFBexp}
\left ( \AFBt \right )_{\rm exp}^{p\bar p} = (15.0 \pm
5.0_{\rm{\hspace{0.5mm} stat.}} \pm 2.4_{\rm{\hspace{0.5mm} syst.}} )
\, \% \,.
\eeq

In the SM the asymmetry $\AFBt$ vanishes at LO in QCD.
This suppression makes it a well suited observable for the potential discovery of NP.
At NLO or ${\cal O} (\alpha_s^3)$ it receives non-vanishing contributions. These arise 
from the interference of one-loop QCD box diagrams with tree-level gluon exchange and the
interference of initial- and final-state radiation. Including NLO QCD and
electroweak corrections \cite{Kuhn:1998jr, Kuhn:1998kw}, the
SM prediction in the $p \bar p$ frame for the inclusive asymmetry is
\cite{Antunano:2007da}
\beq \label{eq:AFBSM}
\left ( \AFBt \right )_{\rm SM}^{p \bar p} = (5.1 \pm 0.6) \, \%
\,.
\eeq
The total error includes the uncertainties due to
different choices of the parton distribution functions (PDFs) and the
factorization and renormalization scales, as well as due to a variation of $m_t$
within its experimental error. General arguments suggest
that the prediction (\ref{eq:AFBSM}) is robust with respect to higher-order
QCD corrections \cite{Melnikov:2010iu}. This is supported by explicit
calculations of $(\AFBt)_{\rm SM}$, which include the resummation of logarithmically
enhanced threshold effects at NLO \cite{Almeida:2008ug} and
NNLO \cite{Ahrens:2010zv} and which are in substantial agreement with the latter number,
making it a firm SM prediction.

The CDF result (\ref{eq:AFBexp}) deviates from this prediction by 
nearly $2\,\sigma$. 
A recent combination of this measurement with a new result obtained in the di-lepton channel
\cite{note10398} slightly increases the discrepancy \cite{note10584}.
In addition, the very recent D{\O} measurement \cite{Abazov:2011rq} finds a similar deviation from the SM prediction.\footnote{Beyond that, 
in an approach based only on the rapidity of the final state lepton,
the asymmetry $A_{FB}^l=(15.2 \pm 4.0)\,\%$ is obtained, to be compared with the theory prediction
by {\tt MC@NLO} of $(2.1 \pm 0.1)\,\%$, which amounts to a $\gtrsim 3 \sigma$ deviation.}
An effect of high statistical significance ($3.4\,\sigma$) has been observed
in the distribution of the asymmetry at high invariant masses $M_{t \bar t}>0.45$\,TeV \cite{Aaltonen:2011kc}.
The sharp growth of the excess with $M_{t \bar t}$ suggests the exchange of a new heavy particle in the production
of top quarks.

The persistently large values of the observed asymmetry over the last years 
have triggered a lot of activity in the theory community
\cite{Delaunay:2011vv,Djouadi:2009nb, Ferrario:2009bz, Jung:2009jz, Cheung:2009ch,
  Frampton:2009rk, Shu:2009xf, Arhrib:2009hu, Dorsner:2009mq,
  Jung:2009pi, Cao:2009uz, Barger:2010mw, Cao:2010zb, Xiao:2010hm,
  Chivukula:2010fk,Grinstein:2011yv,Djouadi:2011aj}.\footnote{For an overview including additional references, see \cite{Gresham:2011fx}.} 
  An apparent option is to generate $\AFBt$ already at LO by tree-level exchange of 
  new heavy particles with axial-vector couplings to fermions. However, a viable model 
  must simultaneously avoid generating too large corrections
  to $\sigtot$ or $\dsig$, which are both in good agreement with the SM prediction.
  This makes it non-trivial to explain the large experimental value.
  A first class of models features new physics in the $t$ channel or $u$ channel with large
  flavor-violating couplings due to vector-boson exchange,
  namely $Z^\prime$ \cite{Jung:2009jz, Jung:2009pi, Cao:2009uz,
  Barger:2010mw, Cao:2010zb, Xiao:2010hm}, or
  $W^\prime$ bosons \cite{Cheung:2009ch, Barger:2010mw, Cao:2010zb}, or due to the 
  exchange of color singlet, triplet or sextet scalars \cite{Shu:2009xf, Arhrib:2009hu,
  Dorsner:2009mq, Jung:2009pi, Cao:2009uz, Cao:2010zb,Grinstein:2011yv}. In this context,
  note that it is not always clear, how the necessary flavor-changing
  couplings can be generated without {\it ad hoc}
  assumptions. A second class of models involves tree-level
  exchange of new vector states in the $s$-channel, which have to exhibit sizable axial-vector couplings to
  both the light quarks, $g_A^q$, and the top quark, $g_A^t$  \cite{Delaunay:2011vv,Djouadi:2009nb, Ferrario:2009bz,
  Frampton:2009rk, Jung:2009pi, Cao:2009uz, Cao:2010zb,
  Chivukula:2010fk,Grinstein:2011yv,Djouadi:2011aj}. Suitable candidates are color octets, which maximize the
  interference with QCD.\footnote{For a thorough analysis of direct and indirect constraints on massive spin-one color
  octets see \cite{Haisch:2011up}.}
  The new states need to have axial-vector couplings
  to the first and the third generation of quarks with opposite
  sign \cite{Ferrario:2008wm}, implying $g_A^q \hspace{0.25mm} g_A^t < 0$,
  in order to achieve a positive shift in $\AFBt$.
  Examples of theories that turn out to lead to a positive shift in the charge 
  asymmetry are flavor non-universal chiral color models \cite{Frampton:2009rk, Chivukula:2010fk},
  and warped extra dimension \cite{Delaunay:2011vv,Djouadi:2009nb,Djouadi:2011aj}, which both contain heavy partners of the SM gluon.
    Indeed the RS model seems to be a prototype for models that address such an anomaly. 
    However, a thorough analysis is necessary to answer the question if the RS setup can really naturally 
    account for the enhancement in the asymmetry.

  In the following we will show that, in the wide class of BSM scenarios that 
  rely on the virtual exchange of vector bosons in the $s$ channel, the NLO corrections to $\AFBt$ can
  exceed the LO corrections if the axial-vector couplings to light
  quarks are suppressed.\footnote{Note that the importance of NLO corrections has
  been briefly mentioned in \cite{Ferrario:2009ee} in the context of
  the charge asymmetry in the exclusive channel $p\bar p\rightarrow
  t\bar t X$.} This applies in particular to NP models that explain the hierarchical
  structures observed in the masses and mixing of the SM fermions
  geometrically, see Section \ref{sec:hierarchies}. Since this sequestering of flavor
  allows for an effective suppression of FCNCs, it is likely to be an integral part of 
  any setup that addresses the problems of the SM via a low NP scale
  in the reach of the LHC. Although our discussion will be inspired by warped extra dimensions,
  as a prototype of the models described above, the use of EFT methods will make the considerations 
  rather model-independent.
 
  We will work out in detail the relevant LO {\it and NLO} corrections to $\AFBt$ in the RS setup. 
  To this end, the heavy KK modes of the model are integrated out explicitly. At Born level, contributions due to the exchange of KK gluons and photons, the
  $Z$ boson and its KK partners, as well as the Higgs boson are included.
  Beyond LO, we consider the interference of tree-level KK-gluon exchange 
  with one-loop QCD box diagrams and corresponding bremsstrahlungs corrections. 
  We will then answer the question about the potential of the RS setup
  to explain the observed $\AFBt$ anomaly. 
  As we stick to the EFT approach, our final results will be applicable to a wide class of scenarios 
  with non-standard dynamics above the electroweak scale. 
  
\subsubsection{Top-Antitop Production}
\label{sec:ttSM}

At the Tevatron, $t \bar t$ pairs are produced in proton-antiproton
collisions, $p\bar{p}\to t\bar{t}X$. In the SM, this hadronic process 
receives partonic tree-level contributions from quark-antiquark annihilation and gluon
fusion.
\beq \label{eq:SMtreeprocesses}
\begin{split}
  q(p_1) + \bar{q}(p_2) &\to t(p_3) + \bar{t}(p_4) \,, \\
  g(p_1) + g(p_2) &\to t(p_3) + \bar{t}(p_4) \,.
\end{split}
\eeq
Here, $p_{1,2}$ denote the four-momenta of the initial state partons, which
can be expressed as fractions $x_{1,2}$ of the four-momenta $P_{1,2}$ of
the colliding hadrons, $p_{1,2}=x_{1,2} \hspace{0.25mm} P_{1,2}$. The hadronic 
center-of-mass (CM) energy squared is given by $s = (P_1+P_2)^2$ . The partonic 
cross section is a function of the kinematic invariants
\beq \label{eq:invariants}
  \hat{s} = (p_1 + p_2)^2 \,, 
  \ \ t_1 = (p_1 - p_3)^2 - m_t^2 = -\frac{\hat s}{2} ( 1 - \beta \cos\theta )\,,
  \ \ u_1 = (p_2 - p_3)^2 - m_t^2 = -\frac{\hat s}{2} ( 1 + \beta \cos\theta ) \,,
\eeq
which for $m_t\to 0$ reduce to the Mandelstam variables.
In the following we will be interested in the differential cross section,
with respect to the angle $\theta$ between $\vec{p}_1$ and
$\vec{p}_3$ in the partonic CM frame, and with respect to the 
invariant mass $\Mtt = \sqrt{(p_3 + p_4)^2}$ of the $t
\bar t$ pair. Thus, we have expressed $t_1$ and $u_1$ in
terms of $\theta$ and the top-quark velocity $\beta = \sqrt{1-\rho} \,,\ 
\rho = 4m_t^2/\hat s$.
Momentum conservation at Born level implies that $\hat{s} + t_1 + u_1 = 0$.

We write the hadronic differential cross section as
\beq \label{eq:dsdc}
\frac{d\sigma^{p\bar{p}\rightarrow t\bar t X}}{d\cos\theta} =
\frac{\alpha_s}{m_t^2} \sum_{i,j} \int_{4m_t^2}^s \frac{d \hat s}{s}
\, \ff_{ij}\big(\hat s/s,\mu_f\big) \, K_{ij} \left (
  \frac{4m_t^2}{\hat s},\cos\theta,\mu_f \right ) \,,
\eeq
where $\mu_f$ denotes the factorization scale. The corresponding parton luminosity functions
are given by
\beq \label{eq:luminosities}
\ff_{ij}(y,\mu_f) = \int_y^1 \frac{dx}{x} \, f_{i/p}(x,\mu_f) \,
f_{j/\bar{p}}(y/x,\mu_f) \,,
\eeq
and, for $ij = q \bar q, \bar q q$, are understood to be summed over all light 
species of quarks. Finally, the functions $f_{i/p}(x,\mu_f)$ ($f_{i/\bar{p}}(x,\mu_f)$) are
the universal (non-perturbative) PDFs, which describe the probability of finding the
parton $i$ in the proton (antiproton) with longitudinal momentum
fraction $x$. The hard-scattering kernels $K_{ij}
(\rho,\cos\theta,\mu_f)$ are related to the partonic cross sections
and have a perturbative expansion in $\alpha_s$
\beq \label{eq:Cijexp} 
K_{ij} (\rho,\cos\theta,\mu_f) = \sum_{n = 0}^\infty \left (
  \frac{\alpha_s}{4\pi} \right )^n \, K_{ij}^{(n)}
(\rho,\cos\theta,\mu_f) \,.
\eeq
At LO in $\alpha_s$ only the kernels with $ij = q \bar q, \bar q
q, gg$ are non-zero within the SM. For the amplitudes
corresponding to $s$-channel gluon exchange we obtain
\begin{eqnarray} \label{eq:SMLO}
\begin{split}
  K_{q\bar{q}}^{(0)} &= \alpha_s\,\frac{\pi \beta
    \rho}{8}\,\frac{\CF}{\Nc}\,\left( \frac{t_1^2+u_1^2}{\hat{s}^2} +
    \frac{2m_t^2}{\hat s} \right) , \\[1mm]
K_{gg}^{(0)} &= \alpha_s\,\frac{\pi \beta \rho}{8(\Nc^2-1)} \left( \CF
  \, \frac{\hat{s}^2}{t_1u_1} - \Nc \right) \left[
  \frac{t_1^2+u_1^2}{\hat{s}^2} + \frac{4m_t^2}{\hat{s}} -
  \frac{4m_t^4}{t_1u_1} \right]\,.
\end{split}
\end{eqnarray} 
Note that the coefficient $K_{\bar{q} q}^{(0)}$ is obtained from $K_{q
  \bar{q}}^{(0)}$ by replacing $\cos\theta$ with $-\cos\theta$. The
color factors of $SU(3)_c$ are given by $\Nc = 3$ and $\CF = 4/3$.

The forward-backward asymmetry in top-pair production is defined as
\beq\label{eq:fbas}
\AFBt \equiv \frac{ \displaystyle \int_0^1 d \cos
  \theta \; \frac{d \sigma^{p\bar{p} \to t \bar{t} X}}{d \cos{\theta}}
  - \int_{-1}^0 d \cos \theta \; \frac{d \sigma^{p\bar{p} \to t
      \bar{t} X}}{d \cos{\theta}}} {\displaystyle \int_0^1 d \cos
  \theta \; \frac{d\sigma^{p\bar{p} \to t \bar{t} X}} {d \cos{\theta}}
  + \int_{-1}^0 d \cos \theta \; \frac{d\sigma^{p\bar{p} \to t\bar{t}
      X}}{d \cos{\theta}}}\, .
\eeq
 Our notation is such that in a process labeled by the superscript
$p\bar{p} \to t\bar{t} X$ ($p\bar{p} \to \bar{t} t X$) the angle
$\theta$ corresponds to the scattering angle of the top (antitop)
quark in the partonic CM frame.
Since QCD is symmetric under charge conjugation, which implies that
\beq \label{eq:charge}
\left. \frac{d \sigma^{p\bar{p} \to  t \bar{t} X}}{d \cos{\theta}}
\right|_{\cos \theta = c} = \left. \frac{d \sigma^{p\bar{p} \to \bar{t} t
      X}}{d \cos{\theta}} \right|_{\cos \theta = - c} \,,
\eeq
for any fixed value of $c$, the forward-backward asymmetry can also be understood as a charge asymmetry
\beq
\label{eq:Afbc}
\AFBt = \Atc \equiv \frac{ \displaystyle \int_0^1 d \cos \theta \, \frac{d
    \sigma_a}{d \cos{\theta}}} { \displaystyle \int_0^1 d \cos \theta
  \, \frac{d\sigma_s}{d \cos{\theta}}} = \frac{\sigma_a}{\sigma_s}\,,
\eeq
where the charge-asymmetric ($a$) and -symmetric ($s$) averaged differential cross sections are defined as \cite{Almeida:2008ug} 
\beq \label{eq:symasym}
  \frac{d \sigma_{a,s}}{d \cos{\theta}} \equiv \frac{1}{2}
  \left[\frac{d \sigma^{p\bar{p} \to t\bar{t} X}}{d \cos{\theta}}
 \mp \frac{d \sigma^{p\bar{p} \to \bar{t} t X}}{d \cos{\theta}}
  \right] \,.
\eeq
This notation, with $d \sigma^{p\bar{p} \to t\bar{t} X}/d \cos{\theta}$ given in
(\ref{eq:dsdc}), will turn out to be convenient for the following discussion.
The total hadronic cross section is now given by
\beq \label{eq:cs}
\sigtot =\int_{-1}^1 d\cos\theta \, \frac{d\sigma_s}{d\cos\theta} \,.
\eeq
We write the asymmetric contribution to the cross section as 
\beq \label{eq:sigmatotRSLO}
\sigma_a = \frac{\alpha_s}{m_t^2} \, \sum_{i,j} \int_{4m_t^2}^s
\frac{d \hat s}{s} \, \ff_{ij}\big( \hat s/s,\mu_f\big) \,
A_{ij} \left ( \frac{4m_t^2}{\hat s} \right ) \,,
\eeq
and similar for the case of the symmetric contribution $\sigma_s$, where the hard-scattering charge-asymmetric
coefficient $A_{ij} (4 m_t^2/\hat s)$ is to be replaced by its symmetric counterpart $S_{ij} (4 m_t^2/\hat s)$.
Note that the perturbative expansions of the symmetric and asymmetric kernels are defined in analogy to (\ref{eq:Cijexp}).

In the SM, the LO coefficients of the symmetric contribution to the cross section read
\beq \label{eq:SMsymLO}
\begin{split}
  S^{(0)}_{q\bar{q}} &= \alpha_s \, \frac{\pi\beta\rho}{27}\,(2+\rho)\,,\\
  S^{(0)}_{gg} &= \alpha_s\, \frac{\pi\beta\rho}{192} \left [
    \frac{1}{\beta} \, \ln \left ( \frac{1+\beta}{1-\beta} \right )
    \left ( 16 + 16 \rho + \rho^2\right ) - 28 -31\rho \right ] ,
\end{split}
\eeq
while the asymmetric contributions $A_{q\bar{q}}^{(0)}$ and
$A_{gg}^{(0)}$ both vanish.

Below, we will discuss in detail how at NLO a non-zero coefficient
$A_{q \bar q}^{(1)}$ is generated in the SM. This will lead to a charge 
asymmetric cross section which is suppressed by $\alpha_s/(4\pi)$ with 
respect to the symmetric one.

\subsubsection{Cross Section and Asymmetry in Warped Models}
\label{sec:AFBtRS}

For the following discussion, recall from Section~\ref{sec:hierarchies}  
the RS mechanism of generating fermion hierarchies geometrically, via wave-function 
overlaps. An important consequence, following from the structure of the overlap integrals is, 
that the effective coupling strength of KK gluons to
heavy quarks is enhanced relatively to the SM couplings by a factor
$\sqrt{L}\,$, see Section~\ref{sec:gaugecouplings}. Since left- and right-handed fermions are
localized differently in the bulk\footnote{Remember that although the bulk-masses are associated
to the representations under the gauge group and not to the chiralities, a $SU(2)_L$-doublet
zero-mode will be mostly left-handed, \etc, so that for zero-modes 
the both can be identified to first approximation.}, the KK-gluon couplings to
quarks are in general not purely vector-like, but receive
non-vanishing axial-vector components. These couplings lead
to a charge asymmetric cross section $\sigma_a$ already at LO
due to quark-antiquark annihilation $q \bar q \to t \bar t$ mediated
by tree-level exchange of KK gluons in the $s$ channel. 

\begin{figure}
\vspace{-0.2cm}
\begin{center}
\includegraphics[height=2.3in]{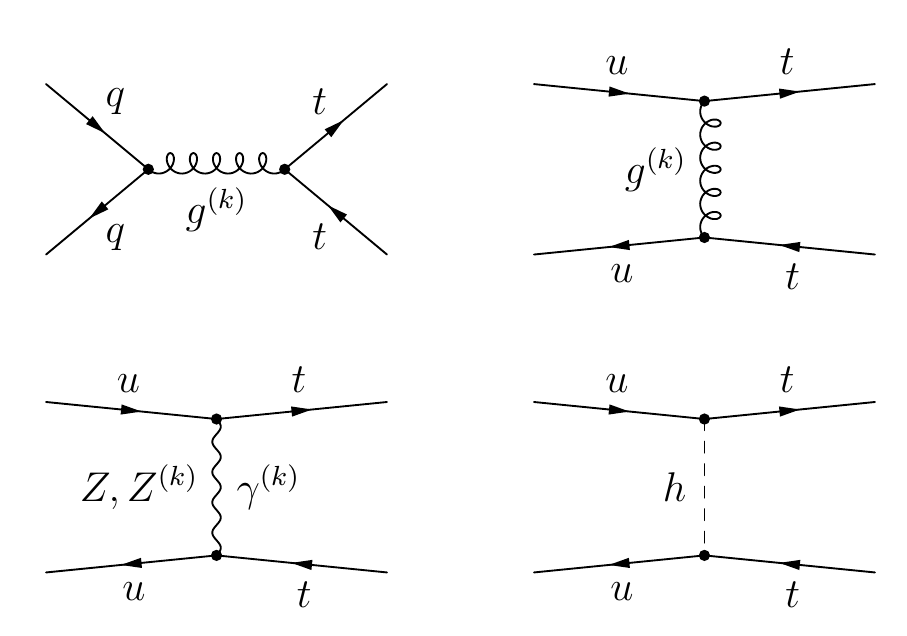}
\end{center}
\vspace{-0.5cm}
\begin{center}
  \parbox{15.5cm}{\caption{\label{fig:RSBorn} Upper row: Tree-level
      contributions 
      to $q \bar q \to t \bar t$ (left) and $u
      \bar u \to t \bar t$ (right) transitions, mediated by $s$- and
      $t$-channel exchange of KK gluons. Lower row: Tree-level
      contributions to the $u \bar u \to t \bar t$ transition arising
      from $t$-channel exchange of the $Z$ boson and its KK partners, 
      of KK photons, as well as of the Higgs boson. The $s$-channel
      ($t$-channel) amplitudes involve all light up-
      and down-type (up-type) quarks. See \cite{Bauer:2010iq} and text for details.}}
\end{center}
\end{figure}

Further tree-level contributions to $\AFBt$ arise in RS models 
due to the fact that KK gluons and photons, the $Z$ boson
and its KK excitations, as well as the Higgs boson feature flavor
non-diagonal couplings, which have been discussed in
Chapter~\ref{sec:WED}. These couplings lead to flavor-changing $u \bar u 
\to t \bar t$ transitions, affecting the $t$ channel. In principle, also $d \bar d 
\to t \bar t$ transitions receive corrections due to the $t$-channel exchange of the 
$W^\pm$ bosons and their KK excitations. 
However, it turns out that these effects are negligibly small for
realistic input parameters and can be ignored in the following.
In contrast to processes with quarks in the initial state,
the gluon-fusion channel $g g \to t \bar t$ does not receive a correction at
the Born level. Due to the orthonormality of gauge-boson profiles, the coupling of two
massless gluons to a KK gluon is zero. The Feynman diagrams that need to be considered 
at the tree-level are shown in Figure~\ref{fig:RSBorn}. 

\paragraph{Calculation of LO effects}

As we expect the NP scale (here $M_{\rm NP}=\Mkk$) to be
at least of the order of several times the electroweak scale $\Mkk>>M_{\rm EW}$ and also well
above the energies directly probable at Tevatron, virtual effects 
appearing in RS models can be described by an effective low-energy theory consisting out of
dimension-six operators, see Appendix~\ref{app:EFT}. In the case at hand, 
the appropriate effective Lagrangian which accounts for the effects of intermediate vector and scalar
states reads
\beq \label{eq:Leff}
{\cal L}_{\rm eff} = \sum_{q,u} \sum_{A, B = L, R} \Big [
\hspace{0.25mm} C_{q \bar q, AB}^{(V, 8)} \hspace{0.25mm} Q_{q \bar q,
  AB}^{(V, 8)} + C_{t \bar u, AB}^{(V, 8)} \hspace{0.25mm} Q_{t
  \bar u, AB}^{(V, 8)}  + C_{t \bar u, AB}^{(V, 1)} \hspace{0.25mm} Q_{t \bar
  u, AB}^{(V, 1)} + C_{t \bar u, AB}^{(S, 1)} \hspace{0.25mm}
Q_{t \bar u, AB}^{(S, 1)} \hspace{0.25mm} \Big ] \,,
\eeq
where
\beq \label{eq:operators}
\begin{split}
  Q_{q \bar q, AB}^{(V, 8)} & = (\bar q \hspace{0.25mm} \gamma_\mu
  \hspace{0.25mm} T^a \hspace{0.25mm} P_A \hspace{0.25mm} q) (\bar t
  \hspace{0.25mm} \gamma^\mu \hspace{0.25mm} T^a \hspace{0.25mm}
  P_B \hspace{0.25mm} t) \,, \\
  Q_{t \bar u, AB}^{(V, 8)} & = (\bar u \hspace{0.25mm} \gamma_\mu
  \hspace{0.25mm} T^a \hspace{0.25mm} P_A \hspace{0.25mm} t) (\bar t
  \hspace{0.25mm} \gamma^\mu \hspace{0.25mm} T^a \hspace{0.25mm} P_B
  \hspace{0.25mm} u) \,, \\
   Q_{t \bar u, AB}^{(V, 1)} & = (\bar u \hspace{0.25mm} \gamma_\mu
  \hspace{0.25mm} P_A \hspace{0.25mm} t) (\bar t \hspace{0.25mm}
  \gamma^\mu \hspace{0.25mm} P_B \hspace{0.25mm} u) \,, \\
  Q_{t \bar u, AB}^{(S, 1)} & = (\bar u \hspace{0.25mm} P_A
  \hspace{0.25mm} t) (\bar t \hspace{0.25mm} P_B \hspace{0.25mm} u)
  \,.
\end{split} 
\eeq 
The sum over $q$ ($u$) involves all light (up-type) quark
flavors and the superscripts $V$ and $S$ ($8$ and $1$) label vector and scalar (color-octet
and -singlet) contributions, respectively.
The $SU(3)_c$ generators $T^a$ are normalized such 
that ${\rm Tr} \left ( T^a T^b
\right ) = \TF \, \delta_{ab}$ with $\TF = 1/2$.

With the help of the effective Lagrangian (\ref{eq:Leff}) we can now easily calculate
the interference between the tree-level matrix element
describing $s$-channel gluon exchange in the SM and the $s$- and $t$-channel
new-physics contributions arising from the diagrams displayed in Figure
\ref{fig:RSBorn}. 
We arrive at the hard-scattering kernels
\begin{eqnarray} \label{eq:RSLO}
\begin{split}
  K_{q\bar{q}, \rm RS}^{(0)} &= \frac{\beta\rho}{32} \,
  \frac{\CF}{\Nc} \left [ \frac{t_1^2}{\hat s} \, C_{q \bar q,
      \hspace{0.25mm} \perp}^{(V, 8)} + \frac{u_1^2}{\hat s} \, C_{q
      \bar q, \hspace{0.25mm} \parallel}^{(V, 8)}+ m_t^2 \left ( C_{q
        \bar q, \hspace{0.25mm} \parallel}^{(V, 8)} + C_{q \bar q,
        \hspace{0.25mm} \perp}^{(V, 8)} \right )
  \right ] , \\[1mm]
  K_{t\bar{u}, \rm RS}^{(0)} &= \frac{\beta\rho}{32} \,
  \frac{\CF}{\Nc} \left [ \left ( \frac{u_1^2}{\hat s} + m_t^2 \right
    ) \left ( \frac{1}{\Nc} \hspace{0.25mm} C_{t \bar u,
        \hspace{0.25mm} \parallel}^{(V, 8)} - 2 \hspace{0.25mm} C_{t
        \bar u, \hspace{0.25mm} \parallel}^{(V, 1)} \right ) + \left (
      \frac{t_1^2}{\hat s} + m_t^2 \right ) C_{t \bar u,
      \hspace{0.25mm} \perp}^{(S, 1)} \right ]\,, \hspace{6mm}
\end{split}
\end{eqnarray}
where we have defined appropriate combinations of Wilson coefficients,
\beq \label{eq:Cparallelperp}
C_{ij, \hspace{0.25mm} \parallel}^{(P, a)} = {\rm Re} \left [
  C_{ij,LL}^{(P, a)} + C_{ij,RR}^{(P, a)} \right ] \,, \qquad \quad
C_{ij, \hspace{0.25mm} \perp}^{(P, a)} = {\rm Re} \left [
  C_{ij,LR}^{(P, a)} + C_{ij,RL}^{(P, a)} \right ] \,.
\eeq
Just like in the SM, the coefficient $K_{\bar{q} q , \rm
  RS}^{(0)}$ $\big ( K_{\bar{t} u ,\rm RS}^{(0)} \big )$ is obtained
from $K_{q \bar{q}, \rm RS}^{(0)}$ $\big ( K_{t \bar{u}, \rm RS}^{(0)}
\big )$ by replacing $\cos\theta$ with $-\cos\theta$.

The LO corrections to the symmetric and asymmetric parts of the cross section, as defined in
(\ref{eq:sigmatotRSLO}), are now obtained by integrating over $\cos\theta$. 
We find in the partonic CM system
\beq \label{eq:SLONP}
\begin{split} 
  S_{u\bar{u},\rm RS}^{(0)} & = \frac{\beta\rho}{216} \, (2 + \rho) \,
  \hat s \, \left [ C_{u\bar u, \parallel}^{(V,8)} + C_{u\bar u,
      \perp}^{(V,8)} + \frac{1}{3} \hspace{0.25mm} C_{t \bar
      u, \parallel}^{(V,8)} - 2 \hspace{0.25mm} C_{t \bar
      u, \parallel}^{(V,1)} \right ] + f_S (z) \,
    \tilde C_{t \bar u}^{S}  \,, \\[1mm]
  S_{d\bar{d},\rm RS}^{(0)} & = \frac{\beta \rho}{216} \, (2 + \rho)
  \, \hat s \, \left [ C_{d\bar d, \parallel}^{(V,8)} + C_{d\bar d,
      \perp}^{(V,8)} \right ]\,,
\end{split}
\eeq 
as well as
\beq \label{eq:ALONP}
\begin{split} 
  A_{u\bar{u},\rm RS}^{(0)} & = \frac{\beta^2\rho}{144} \, \hat s \,
  \left [ C_{u\bar u, \parallel}^{(V,8)} - C_{u\bar u, \perp}^{(V,8)}
    + \frac{1}{3} \hspace{0.25mm} C_{t \bar u, \parallel}^{(V,8)} - 2
    \hspace{0.25mm} C_{t \bar u, \parallel}^{(V,1)} \right ]
   + f_A (z) \, \tilde C_{t \bar
      u}^{S}  \,, \\[1mm]
  A_{d\bar{d},\rm RS}^{(0)} & = \frac{\beta^2\rho}{144} \, \hat s \,
  \left [ C_{d\bar d, \parallel}^{(V,8)} - C_{d\bar d, \perp}^{(V,8)}
  \right ]\,.
\end{split}
\eeq
Obviously, there is no flavor-changing $t$-channel contribution to
the coefficients involving down-type quarks. For the following
considerations it is important to note that the 
coefficients $C_{q \bar q, \parallel}^{(V,8)}$
and $C_{q \bar q, \perp}^{(V,8)}$ enter (\ref{eq:SLONP}) in the combination $C_{q \bar
  q}^V \equiv \big ( C_{q \bar q, \parallel}^{(V,8)} + C_{q \bar q,
  \perp}^{(V,8)} \big )$, while in (\ref{eq:ALONP}) they always appear in
the form $C_{q \bar q}^A \equiv \big ( C_{q \bar q, \parallel}^{(V,8)}
- C_{q \bar q, \perp}^{(V,8)} \big )$. This reflects the fact
that the symmetric (asymmetric) LO cross section $\sigma_s$
($\sigma_a$) measures the product $g_V^q \hspace{0.25mm} g_V^t$ $\big
(g_A^q \hspace{0.25mm} g_A^t \big )$ of the vector (axial-vector)
parts of the couplings of the KK gluons to light quarks and top
quarks. 
As we want to include a potentially light Higgs boson with
$m_h \ll M_{\rm KK}$ into our analysis, we have kept the 
full Higgs-boson mass dependence arising from the $t$-channel
propagator.
  This dependence is described by the phase-space factors
  $f_{S,A} (z)$ with $z \equiv m_h^2/m_t^2$, which are given explicitly 
  in Appendix~\ref{app:phasespace}. The new coefficient $\tilde C_{t \bar u}^{S}$ 
  is the dimensionless counterpart of $C_{t \bar u, \perp}^{(S,1)}$.\footnote{Note that we do not have to introduce form factors 
  for the $t$-channel contribution arising from $Z$-boson zero-mode exchange,
  since these corrections are of $\ord (v^4/\Mkk^4)$, which we
  will neglect in the following. The $t$-channel exchange 
  requires a flavor changing transition at {\it both} vertices, which at $\ord (v^2/\Mkk^2)$ 
  is only mediated by the non-factorizable $t_<$ contribution in the KK sum. 
  However, this is only induced by the KK excitations, see (\ref{eq:Sigmafinal}) 
  and the discussion below.} 

  So far, the expressions (\ref{eq:SLONP}) and (\ref{eq:ALONP}) contain in a
  model-independent way possible new-physics contributions to
  $\sigma_{s,a}$ that arise from the operators given in (\ref{eq:operators}).
  Thus they are appropriate for general models that feature 
  tree-level exchange of color-octet vectors in the $s$ and $t$ channels, 
  as well as $t$-channel corrections due to both new color-singlet, vector, and scalar
  states, making them useful beyond the context of RS models.
  However, we now want to apply the results to models with a warped extra dimension.
  
  The Wilson coefficients appearing in $S_{ij, \rm RS}^{(0)}$
  and $A_{ij, \rm RS}^{(0)}$ can be worked out with the help of the formulae derived in Section~\ref{sec:KKsum}. 
  In the minimal RS formulation we find\footnote{See 
  \cite{Bauer:2009cf} for the effective $D=6$ Hamiltonian of the minimal RS model.} 
  \begin{eqnarray} \label{eq:wilsonexplicit}
\begin{split}
  C^{(V, 8)}_{q \bar q, \parallel} & = -\frac{2 \pi \alpha_s}{\Mkk^2}
  \left \{ \frac{1}{L}- \sum_{a=Q,q} \left[
      (\Delta^\prime_a)_{11}+(\Delta^\prime_a)_{33}-2 L \,
      ({\widetilde \Delta}_a)_{11} \otimes({\widetilde
        \Delta}_a)_{33} \right] \right \}\,,\\
  C^{(V, 8)}_{q \bar q, \perp} & = -\frac{2 \pi \alpha_s}{\Mkk^2}
  \left \{ \frac{1}{L}- \sum_{a=Q,q} \Big [
    (\Delta^\prime_a)_{11}+(\Delta^\prime_a)_{33} \Big ] + 2L \left [
      ({\widetilde \Delta}_Q)_{11} \otimes({\widetilde \Delta}_q)_{33}
      + ({\widetilde \Delta}_q)_{11} \otimes({\widetilde
        \Delta}_Q)_{33}
    \right ] \right \}\,,\\
  C^{(V, 8)}_{t \bar u, \parallel} & = -\frac{4 \pi
    \alpha_s}{\Mkk^2}\, L\sum_{a=U,u}\Big[ ({\widetilde
    \Delta}_a)_{13}\otimes({\widetilde
    \Delta}_a)_{31}\Big]\,, \\
  C^{(V, 1)}_{t \bar u, \parallel} & = -\frac{4 \pi \alpha_e}{\Mkk^2}
  \, \frac{L}{s^2_w c^2_w} \left [ \left ( T_3^u-s^2_w Q_u \right )^2
    ({\widetilde \Delta}_U)_{13}\otimes({\widetilde \Delta}_U)_{31} +
    \left(s^2_w Q_u\right)^2 ({\widetilde \Delta}_u)_{13} \otimes
    ({\widetilde \Delta}_u)_{31} \right ] \\ & \phantom{xx} - \frac{4
    \pi \alpha_e}{\Mkk^2} \, L\, Q_u^2\sum_{a=U,u}\Big[ ({\widetilde
    \Delta}_a)_{13}\otimes({\widetilde
    \Delta}_a)_{31}\Big]\,, 
\end{split}
\end{eqnarray}
for $q = u, d$ and $Q = U, D$. Because the coefficient $C^{(S, 1)}_{t
  \bar u, \perp}$ is of ${\cal O} (v^4/\Mkk^4)$, we do not present its 
  explicit form. Similar expressions with the index $1$
  replaced by $2$ hold if the initial-state quarks belong to the
  second generation. The effective couplings $(\Delta^{(\prime)}_{Q,q})_{ij}$, encoding
the overlap between KK gauge bosons and $SU(2)_L$ doublet and singlet 
quarks of generations $i$ and $j$, can be found in (\ref{eq:overlapints1}).
The non-factorizable products $({\widetilde \Delta}_Q)_{ij} \otimes({\widetilde
\Delta}_q)_{kl}\,$, \etc\,, arising from the $t_<$ contribution
in the KK sums of Section~\ref{sec:KKsum} are given in Appendix~\ref{app:wil}
for completeness. The Wilson coefficients (\ref{eq:wilsonexplicit}) are understood to be 
evaluated at the scale $\Mkk$. The inclusion of RG effects (see Appendix~\ref{app:EFT})
from the evolution down to the top-quark mass scale has only a subleading impact on the results. Details on this evolution
can be found in Appendix~\ref{app:RGEafb}. While the expressions for
$C^{(V, 8)}_{q \bar q, \parallel}$, $C^{(V, 8)}_{q \bar q, \perp}$,
and $C^{(V, 8)}_{t \bar u, \parallel}$ are exact, in the coefficient
$C^{(V, 1)}_{t \bar u, \parallel}$ containing the exchange of towers with a 
massive zero mode, we have only kept the leading terms in
$v^2/\Mkk^2$.  The complete expression for $C^{(V, 1)}_{t \bar
  u, \parallel}$, including the subleading ${\cal O} (v^4/\Mkk^4)$ 
  effects from the corrections due to the mixing of fermion zero modes with their KK
excitations, can be obtained from the exact formulae given in Chapter~\ref{sec:WED}. 

The expressions (\ref{eq:wilsonexplicit}) can be easily generalized to 
the custodial RS model based on the $SU(2)_L \times SU(2)_R \times U(1)_X \times P_{LR}$ bulk
gauge group with the help of the expressions presented in sections \ref{sec:custo}
and \ref{sec:KKsum}. One finds that the left-handed part of the $Z$-boson contribution 
to $C_{t\bar u,\parallel}^{(V,1)}$ is enhanced by a factor of around 3. The
right-handed contribution, however, is protected by the custodial symmetry and thus
smaller by a factor of roughly $1/L \approx 1/37$.
Importantly, the KK-gluon contributions $C_{q \bar q,\parallel}^{(V,8)}$,
$C_{q \bar q,\perp}^{(V,8)}$, and $C_{t \bar u,\parallel}^{(V,8)}$,
remain unchanged at LO in ${\cal O} (v^2/\Mkk^2)$. As these corrections
will turn out to be the most significant ones, this implies
that the predictions for the $t \bar t$ observables
are rather model-independent.

Explicit analytic expressions for the Wilson coefficients
(\ref{eq:wilsonexplicit}) in the ZMA are given in Appendix~\ref{app:wil}. 
Considering just the $\alpha_s$ contributions,
and suppressing relative $\ord (1)$ factors and numerically
subleading terms, one finds from the expressions given in
(\ref{eq:wilsonuuZMA}) the scaling relations
\beq \label{eq:SALOscaling1}
\begin{split}
  S_{u\bar u, {\rm RS}}^{(0)} & \, \sim \, \frac{4 \pi
    \alpha_s}{\Mkk^2} \sum_{A = L, R }F^2 (c_{t_A}) \,, \\[1mm]
  A_{u\bar u, {\rm RS}}^{(0)} & \, \sim \, -\frac{4 \pi
    \alpha_s}{\Mkk^2} \, L \, \left \{ \prod_{q = t, u} \Big [
    F^2(c_{q_R}) - F^2(c_{q_L}) \Big ] + \frac{1}{3} \sum_{A = L, R}
    F^2(c_{t_A}) F^2(c_{u_A}) \right \} ,
\end{split}
\eeq
for the (up-quark) coefficient functions introduced in (\ref{eq:SLONP}) and
(\ref{eq:ALONP}).
Here, $c_{t_L}\equiv c_{Q_3}$, $c_{t_R}\equiv c_{u_3}$, $c_{u_L}\equiv
c_{Q_1}$, and $c_{u_R}\equiv c_{u_1}$.

Given that the bulk-mass parameters of the top and up quarks satisfy $c_{t_A} > -1/2$ and $c_{u_A} <
-1/2$, as required to reproduce their masses in an anarchic approach to flavor
(see Section~\ref{sec:hierarchies}, Figure \ref{fig:cs}), the zero-mode profile-factors can be
approximated by
\beq
F^2(c_{t_A}) \approx 1 + 2 c_{t_A} \,, \qquad F^2(c_{u_A}) \approx (-1
- 2 c_{u_A}) \, e^{L \hspace{0.25mm} (2 c_{u_A} + 1)} \,, 
\eeq
where $A = L,R$. Note that the difference of bulk mass parameters for light
quarks $(c_{u_L}-c_{u_R})$ is typically small and positive, while
$(c_{t_L}-c_{t_R})$ can be of $\mathcal{O}(1)$ and is usually negative. 
Applying the approximations given above and expanding in
powers of $(c_{u_L} - c_{u_R})$, we finally find
\begin{eqnarray} \label{eq:SALOscaling2}
\begin{split}
  S_{u\bar u, {\rm RS}}^{(0)} \, & \sim \, \frac{4 \pi
    \alpha_s}{\Mkk^2} \, 2 \left ( 1+ c_{t_L} + c_{t_R} \right ) \,, \\[1mm]
  A_{u\bar u, {\rm RS}}^{(0)} \, & \sim \, \frac{4 \pi
    \alpha_s}{\Mkk^2} \, 2 \hspace{0.25mm} L \, e^{L (1 + c_{u_L} +
    c_{u_R} )} \left ( 1 + c_{u_L} + c_{u_R} \right ) \\ & \quad \,
  \times \left \{ \left ( 2 +\frac{1}{3} \right ) L \left ( c_{t_L} -
      c_{t_R} \right ) \left ( c_{u_L} - c_{u_R} \right ) +
    \frac{1}{3} \left ( 1 + c_{t_L} + c_{t_R} \right ) \right \}.
\end{split}
\end{eqnarray}
The symmetric function $S_{u\bar u, {\rm RS}}^{(0)}$ is entirely
induced by $s$-channel KK-gluon exchange, whereas the contributions to the
asymmetric coefficient $A_{u\bar u, {\rm RS}}^{(0)}$ arise from
$s$ channel as well as $t$ channel exchange, corresponding to the term(s) with
coefficient $2$ and $1/3$ in the curly bracket, respectively.

From the relations (\ref{eq:SALOscaling2}) one can read off a couple 
of interesting consequences. First, the symmetric contribution
$S_{u\bar u, {\rm RS}}^{(0)}$, entering the RS prediction for $\sigma_s$ 
in (\ref{eq:sigmatotRSLO}), is in
this approximation independent of the localization of the up-quark
fields and strictly positive (as long as $c_{t_A} > -1/2$). This
leads to an enhancement of the inclusive $t \bar t$ production
cross section, which gets the more pronounced the closer the right-
and left-handed top-quark profiles are localized towards the IR brane.

On the other hand, the asymmetric function $A_{u\bar u, {\rm RS}}^{(0)}$ is exponentially 
suppressed for UV-localized up quarks, \ie, $c_{u_A} < -1/2$. 
For typical values of the bulk-mass parameters of $c_{t_L} = -0.34$, $c_{t_R} = 0.57$, 
$c_{u_L} = -0.63$, and $c_{u_R} = -0.68$ \cite{Bauer:2009cf}, 
one finds numerically that the first term in the curly bracket of (\ref{eq:SALOscaling2}), which is
suppressed by the small difference $(c_{u_L} - c_{u_R})$ of bulk mass parameters,
but enhanced by the volume factor $L$, is larger in magnitude
than the second one by about a factor of 10. As a consequence, to
first order the charge asymmetry can be described by only keeping
the effects from $s$-channel KK-gluon exchange. Since generically $(1
+ c_{u_L} + c_{u_R}) \hspace{0.25mm} (c_{u_L} - c_{u_R}) < 0$, we
find that a positive LO contribution to $A_{u \bar u,
  \rm RS}^{(0)}$ needs $(c_{t_L} - c_{t_R})$ to be negative. This
can be achieved in a natural way by localizing the right-handed top quark sufficiently
far in the IR. Employing the formulae derived in Section~\ref{sec:hierarchies}, 
we obtain, to leading powers in hierarchies, the condition
\beq\label{eq:ctR}
c_{t_R}
\, \gtrsim \, \frac{m_t}{\sqrt{2} \hspace{0.25mm} v \left | Y_t
  \right |} - \frac{1}{2} \,.
\eeq 
The top-quark mass is understood to be normalized at the KK
scale and $Y_t \equiv (Y_u)_{33}$. Numerically, we find that for
$m_t(1\,{\rm TeV}) = 144 \, {\rm GeV}$ and $|Y_t| =1$, values of
$c_{t_R}$ bigger than 0 lead to $A_{u\bar u ,\rm RS}^{(0)} >0$ and
in consequence to a positive shift in $\sigma_a$.  

However, the exponential suppression of $A_{u\bar u, {\rm RS}}^{(0)}$ due to the UV-localization
of up quarks, as well as the small difference in the bulk-masses for
their chiral components (leading to very suppressed axial-vector couplings)
render the tree-level contribution to the charge asymmetry in the RS framework tiny.\footnote{See also the
  statements made in \cite{Agashe:2006hk} concerning the mostly
  vector-like couplings of light quarks.}  As we will explicitly
  verify in our numerical analysis, the inclusion of electroweak corrections
due to Born-level exchange of the $Z$ boson and its KK excitations, 
KK excitations of the photon, and of the Higgs boson, do
not change this conclusion.

\paragraph{Calculation of NLO effects}

We have seen that in models with small axial-vector couplings to light 
quarks and no significant FCNC effects in the $t$ channel, the charge-asymmetric
cross section $\sigma_a$ is suppressed at LO. In the following we want to study
if this suppression can be evaded by going to NLO, after
paying the price of an additional factor of $\alpha_s/(4 \pi)$. 
Therefore, we first recall how the charge asymmetry arises in the SM (in QCD).
Since QCD has only vector couplings, the lowest-order
processes $q \bar q \to t \bar t$ and $gg \to t \bar t$, appearing at
$\ord (\alpha_s^2)$, do not contribute to $\AFBt$. Starting at
$\ord (\alpha_s^3)$, quark-antiquark annihilation $q\bar{q} \to t \bar
t \hspace{0.5mm} (g)$, as well as flavor excitation $qg \to q t \bar
t$ receive charge-asymmetric contributions \cite{Kuhn:1998jr,
  Kuhn:1998kw}. Gluon fusion $gg \to t \bar t \hspace{0.5mm}
(g)$, remains charge-symmetric to all orders in perturbation theory. 

Using charge conjugation invariance of QCD, one can show that, as far as the
virtual corrections to $q\bar{q}\rightarrow t\bar{t}$ are concerned,
only the interference between the lowest-order and the QCD box graphs
generates an asymmetry at NLO. For the real bremsstrahlungs contributions, 
along the same lines only the interference between amplitudes that are odd under the 
exchange of $t$ and $\bar t$ contribute to the asymmetric cross section. Since the axial-vector 
current is even under this exchange, the NLO contribution to the asymmetry is completely
due to vector-current contributions. 

\begin{figure}[!t]
\begin{center}
\includegraphics[height=1.29in]{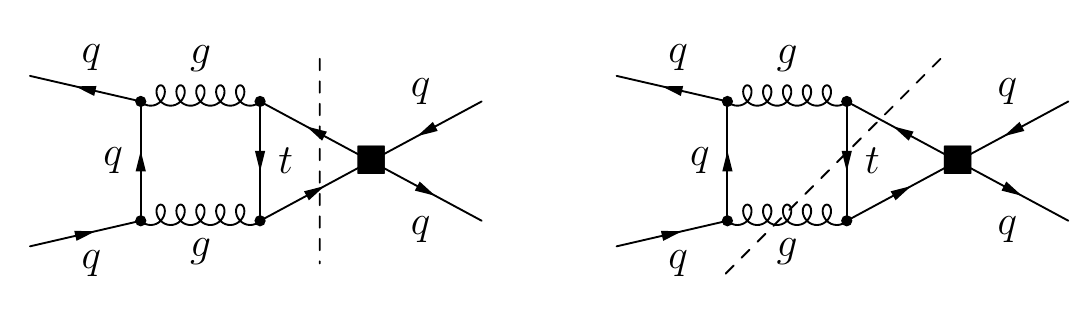}
\end{center}
\vspace{-1.2cm}
\begin{center}
  \parbox{15.5cm}{\caption{\label{fig:RSLoop} Representative diagrams
      contributing to $\AFBt$ at NLO. The black square indicates the insertion of 
      an effective operator. The two-particle (three-particle) cut,
      represented by a dashed line, corresponds to the interference
      of $q \bar q \to t \bar t$ ($q \bar q \to t \bar t
      \hspace{0.25mm} g$) with $Q_{q \bar q, AB}^{(V, 8)}$. 
      The SM contribution can be obtained by replacing the
      operator insertion by $s$-channel gluon exchange. See \cite{Bauer:2010iq} and text for details}}
\end{center}
\end{figure}

This implies that at NLO the charge-asymmetric cross section is proportional to the $d_{abc}^2
= \left (2 \hspace{0.25mm} {\rm Tr} \left ( \{ T^a, T^b \} T^c \right
  ) \right )^2$ terms that result
from the interference of both the one-loop box and the $t \bar t g$ final
state diagrams with the tree-level quark-antiquark annihilation diagram
\cite{Kuhn:1998jr, Kuhn:1998kw}. The
relevant Feynman diagrams are obtained from the ones shown in
Figure~\ref{fig:RSLoop} by replacing the operator insertions by $s$-channel
gluon exchange. 
The QCD expression for $\sigma_a$ can be derived from
generalizing the result from the electromagnetic process $e^+ e^- \to
\mu^+ \mu^-$ \cite{Berends:1973fd, Berends:1982dy} by an appropriate
replacement of the QED coupling and the electromagnetic
charges. Explicit expressions for the asymmetric contributions to the
$t\bar{t}$ production cross section in QCD are presented in
\cite{Kuhn:1998kw}. Because contributions from flavor excitation are
negligibly small at the Tevatron, they will not be taken into account in
the following.

From the considerations above we learn that, beyond LO, vector couplings alone are 
sufficient to generate non-vanishing values of $\AFBt$. 
In the case of the EFT
(\ref{eq:Leff}) this means that cut diagrams like the ones shown in
Figure~\ref{fig:RSLoop}, can give a sizable contribution to the charge
asymmetry, if the symmetric combination $C_{q \bar q}^V = \big ( C_{q \bar
  q, \parallel}^{(V,8)} + C_{q \bar q, \perp}^{(V,8)} \big )$ of
Wilson coefficients is large enough. Interestingly, this combination is
not suppressed by quark localizations in the RS setup
but can become sizable due to the large overlap of the third-generation 
up-type quark wave functions with those of the KK gluons. From (\ref{eq:SLONP}),
(\ref{eq:ALONP}), and (\ref{eq:SALOscaling2}) it is not difficult to
show that for the RS model the NLO corrections to
$\sigma_a$ should dominate over the LO corrections, if the
condition
\beq \label{eq:naive} 
\frac{\alpha_s}{4\pi} \, (1+c_{t_L}+c_{t_R}) \, \gtrsim \, L \, e^{L
  (1+ c_{u_L}+c_{u_R}) }
\eeq 
is satisfied.\footnote{This relation only corresponds to a
  crude approximation, valid up to $\ord (1)$ factors.} 
  Employing the values $c_{t_L} = -0.34$, $c_{u_L} =
-0.63$, and $c_{u_R} = -0.68$, the condition (\ref{eq:naive}) 
tells us that for
$c_{t_R} = 0.57$ the NLO contributions exceed the LO
corrections by a factor of $\sim$ 25. This first look suggests that it might be
possible to lift the LO suppression of the asymmetry and reach contributions to $\AFBt$ 
at the per cent level at NLO with typical and completely natural choices 
of parameters. 

Unlike in QCD, the RS model features further diagrams that generate a charge-asymmetric cross section
at NLO, besides those shown in Figure~\ref{fig:RSLoop}. Self-energy, vertex, and
counterterm diagrams will also contribute to the asymmetry. However, just as the Born-level 
contribution, these corrections are all exponentially suppressed by the UV
localization of the light-quark fields (and the small axial-vector
coupling of the light quarks for what concerns the contributions from
the operators $Q_{q\bar q, AB}^{(V,8)}$). Compared to the tree-level
corrections, these contributions are thus suppressed by an additional
factor of $\alpha_s/(4 \pi)$, so that they can be safely ignored. Moreover, we will not consider corrections due to box diagrams 
involving the virtual exchange of KK gluons.

Encouraged by these considerations we perform a
calculation of $\AFBt$ in the RS model beyond LO. The Feynman
graphs that we consider are those displayed in Figure \ref{fig:RSLoop}. The most important 
contributions stem from the interference
of $q \bar q \to t \bar t$ and $q \bar q \to t \bar t \hspace{0.25mm} g$ with $Q_{q \bar q, AB}^{(V, 8)}$.
After integrating over $\cos
\theta$, we obtain in the partonic CM frame ($q \bar q = u \bar u,
d\bar d$) the asymmetric function\footnote{Note that the quark luminosities $\ff_{ij}(\hat
s/s,\mu_f)$ fall off strongly with $\hat s$, behaving roughly like
$1/\hat s^2$. This compensates the factor of $\sqrt{s}$ in (\ref{eq:NPNLO}) 
so that the integrated asymmetric cross section in (\ref{eq:sigmatotRSLO}) is
saturated well before the upper integration limit $s$ is reached \cite{Bauer:2010iq}.}
\beq\label{eq:NPNLO}
A^{(1)}_{q\bar{q},\rm RS} = \frac{\hat{s}}{16\pi\alpha_s}
\hspace{0.5mm} C_{q\bar q}^V \hspace{0.5mm} A^{(1)}_{q\bar{q}}\,.
\eeq 
Here, $A^{(1)}_{q\bar{q}}$ is the NLO asymmetric SM coefficient,
normalized according to (\ref{eq:Cijexp}), which can be parametrized
by the function
\beq 
\label{eq:afun}
  A^{(1)}_{q\bar{q}} = \frac{\alpha_s\,d_{abc}^2}{16N_c^2} \;
  5.994\,\beta\rho\,\Big [1 + 17.948\,\beta - 20.391\,\beta^2 +
  6.291\,\beta^3 + 0.253\,\ln\left (1-\beta \right )\Big ]\,, 
\eeq 
which is accurate to the permille level. 
Here,
$N_c = 3$ and $d_{abc}^2 = \left (\Nc^2-1 \right ) \left
  (\Nc^2-4 \right )/\Nc = 40/3$. 
The function (\ref{eq:afun}) has been obtained
by integrating the expressions for the charge-asymmetric contributions to the
differential $t \bar t$ production cross section given in
\cite{Kuhn:1998kw} over the relevant phase space. For more details see \cite{Bauer:2010iq}. 
To judge the quantitative impact of the RS contribution (\ref{eq:NPNLO}) we will now perform a numerical
analysis.

\subsubsection{Numerical Analysis and Discussion}
\label{sec:numerics}

For the numerical analysis we have to take into account that
the Wilson coefficients appearing in the effective Lagrangian
(\ref{eq:Leff}) are not only constrained by the measurements of the
forward-backward asymmetry $\AFBt$, but also by the total cross 
section $\sigtot$, and the $t\bar t$ invariant mass spectrum $\dsig$. 
The Tevatron results ($\sqrt{s} = 1.96 \, {\rm TeV}$) for
these quantities read \cite{CDFnotetot,
  Bridgeman:2008zz, Aaltonen:2009iz}
\beq
\begin{split} 
  (\sigtot)_{\rm exp} =& (7.50 \pm 0.31_{\rm stat.} \pm 0.34_{\rm
    syst.} \pm 0.15_{\rm lumi.})\,\rm pb\,,
  \\ \label{eq:EXPss}
  \left (\frac{d \sigtot}{d \Mtt} \right )_{\rm exp}^{\Mtt \, \in \,
    [800, 1400]\,{\rm GeV}} =& (0.068 \pm 0.032_{\rm stat.} \pm 0.015_{\rm
    syst.}  \pm 0.004_{\rm lumi.})\,\frac{\rm fb}{\,\rm GeV}\,, 
\end{split}
\eeq
where the quoted individual errors are of statistical and systematic
origin, and due to the luminosity uncertainty, respectively. 
In the case of the $t\bar t$ invariant mass spectrum, we consider only 
the last bin of the CDF measurement, \ie, $\Mtt \in \, [800, 1400]$\,GeV, 
as this is most sensitive to the presence of new degrees of freedom with 
masses in the TeV range.

These results are to be compared to the predictions obtained in
the SM, supplemented by the dimension-six Lagrangian
(\ref{eq:Leff}). In the following we will ignore tiny contributions 
due to the (anti)strange-, (anti)charm-, and (anti)bottom-quark content 
of the proton (antiproton). By convoluting the kernels
(\ref{eq:SLONP}) with the parton luminosities $\ff_{ij}(\hat s/s,\mu_f)$,
by means of the charge-symmetric analogon of formula
(\ref{eq:sigmatotRSLO}), one finds in terms of the dimensionless
coefficients $\tilde C_{q \bar q}^V \equiv 1\, {\rm TeV}^2 \, C_{q
  \bar q}^V$ and $\tilde C_{t \bar u}^V \equiv 1\, {\rm TeV}^2 \, \big
( 1/3 \hspace{0.5mm} C_{t \bar u, \parallel}^{(V,8)} - 2
\hspace{0.25mm} C_{t \bar u, \parallel}^{(V,1)} \big )$ the
RS results \cite{Bauer:2010iq}
\beq
\begin{split} 
\label{eq:RSresig}
  (\sigtot)_{\rm RS} =& \left [ 1 + 0.053 \hspace{0.5mm} \big ( \tilde C_{u \bar
      u}^V + \tilde C_{t \bar u}^{V} \big ) - 0.612\, \tilde C_{t \bar
      u}^{S} + 0.008 \, \tilde C_{d \bar d}^V \hspace{0.25mm} \right ]
  \left ( 6.73^{+0.52}_{-0.80} \right ) {\rm pb} \,, \\ 
  \left (\frac{d \sigtot}{d \Mtt} \right )^{\Mtt \, \in \, [800, 1400]
    \,GeV}_{ \rm RS} =& \left [ 1 + 0.33 \hspace{0.5mm} \big ( \tilde C_{u \bar
      u}^V + \tilde C_{t \bar u}^{V} \big ) - 0.81 \, \tilde C_{t \bar
      u}^{S} + 0.02 \, \tilde C_{d \bar d}^V \hspace{0.25mm} \right ]
  \left ( 0.061^{+0.012}_{-0.006} \right ) \frac{\rm fb}{\rm GeV} \,.
  \hspace{4mm} 
\end{split}
\eeq
The numerical factors multiplying $\tilde C_{t \bar u}^{S}$
correspond to a Higgs mass of $m_h = 115 \, {\rm GeV}$,
which we take as the reference value for the present analysis. The Wilson coefficients are 
understood to be evaluated at $m_t$. The RG evolution
from $\Mkk$ down to $m_t$ is performed with the formulae given in
Appendix~\ref{app:RGEafb}. 
The result (\ref{eq:RSresig}) corresponds to the {\tt MSTW2008LO} PDFs~\cite{Martin:2009iq}
with renormalization and factorization scales fixed to the reference point $\mu_r = \mu_f=m_t=173.1$
\,GeV. The strong coupling constant reads $\alpha_s(m_Z)=0.139$, which evolves
to $\alpha_s(m_t)=0.126$ using one-loop RG running.

We now present our NLO prediction for the $t \bar t$ forward-backward asymmetry. Inserting (\ref{eq:ALONP}),(\ref{eq:NPNLO}),(\ref{eq:afun})
into (\ref{eq:sigmatotRSLO}) and (\ref{eq:Afbc}), performing the convolution with the {\tt MSTW2008LO} PDFs with the unphysical scales fixed to
$m_t$ and transforming to the $p\bar p$ frame one obtains \cite{Bauer:2010iq}
\beq \label{eq:AFBEFT} 
(\AFBt)^{p \bar p}_{\rm RS} = \left [ \frac{1 + 0.22
    \hspace{0.5mm} \big ( \tilde C_{u \bar u}^A + \tilde C_{t \bar
      u}^V \big ) +0.72 \hspace{0.25mm} \tilde C_{t \bar u}^S + 0.03
    \hspace{0.25mm} \tilde C_{d \bar d}^A + 0.034 \hspace{0.25mm}
    \tilde C_{u \bar u}^V + 0.005 \hspace{0.25mm} \tilde C_{d \bar
      d}^V}{1 + 0.053 \hspace{0.5mm} \big ( \tilde C_{u \bar u}^V +
    \tilde C_{t \bar u}^V \big ) -0.612 \hspace{0.25mm} \tilde C_{t
      \bar u}^S + 0.008 \hspace{0.25mm} \tilde C_{d \bar d}^V } \right
] \! \left ( 5.6^{+0.8}_{-1.0} \right ) \% \,,
\eeq  
Here, the NLO result for $\sigma_s$ has been used
for the normalization of the asymmetric cross section,
which has been calculated with the help of {\tt MCFM} \cite{MCFM}. 
All coefficient functions should be evaluated at the scale
$m_t$. The SM prediction for the asymmetry has been obtained by
integrating the formulae given in \cite{Kuhn:1998kw} over the relevant
phase space, weighted with {\tt MSTW2008LO} PDFs with the unphysical scales fixed to
$m_t$. It is in good agreement with (\ref{eq:AFBSM}) as well as with the
results of \cite{Almeida:2008ug, Ahrens:2010zv}. 

In the central value of (\ref{eq:AFBEFT}), we have decided not to include electroweak 
corrections to the asymmetric cross section. 
These have been studied in \cite{Antunano:2007da, Bernreuther:2010ny}
and found to enhance the $t \bar t$
forward-backward asymmetry by around $9\%$ to $4\%$, depending on
whether only mixed electroweak-QCD contributions or also purely
electroweak corrections are included. In order to account for the 
additional uncertainty in neglecting these effects we have added in
quadrature an error of $5\%$ to the combined scale and PDF
uncertainties.

\begin{table}
\begin{center}
\begin{tabular}{|c|c|c|c|c|c|c|c|}
  \hline
  $c_{t_L}$ & $c_{t_R}$ & $\tilde C_{u\bar u}^V/\alpha_s$ & 
  $\tilde C_{u\bar u}^A/\alpha_s$ & 
  $\tilde C_{d\bar d}^V/\alpha_s$ & 
  $\tilde C_{d\bar d}^A/\alpha_s$ & 
  $\tilde C_{t\bar u}^V/\alpha_s$ & 
  $\tilde C_{t \bar u}^S$ \\
  \hline
  $-0.41$ & $0.09$ & $4.50$ & $0.71 \cdot 10^{-2}$ & $0.68$ & 
   $-1.40 \cdot 10^{-3}$ & $-1.35 \cdot 10^{-4}$ & $8.2 \cdot 10^{-7}$ \\
  $-0.47$ & $0.48$ & $4.95$ & $0.22 \cdot 10^{-2}$ & $0.27$ & 
  $-0.03 \cdot 10^{-3}$ & $-0.70 \cdot 10^{-4}$ & $4.1 \cdot 10^{-7}$ \\
  $-0.49$ & $0.90$ & $5.31$ & $1.79 \cdot 10^{-2}$ & $0.08$ & 
  $-0.64 \cdot 10^{-3}$ & $-2.45 \cdot 10^{-4}$ & $122 \cdot 10^{-7}$ \\
  \hline
\end{tabular}
\end{center}
\caption{\label{tab:WC} 
  Results for the Wilson coefficients corresponding to three different 
  parameter points of the RS setup with $SU(2)_L \times U(1)_Y$ bulk gauge symmetry 
  and brane-localized Higgs sector. The coefficients in the first five columns (last column) scale as 
  $(1 \, {\rm TeV}/\Mkk)^2$ ($(1 \, {\rm TeV}/\Mkk)^4$).
}
\end{table}

To get a feeling for the importance of the different contributions that enter
the RS predictions (\ref{eq:RSresig}) and (\ref{eq:AFBEFT}) for
the $t \bar t$ observables, we give in Table \ref{tab:WC} numerical results
for the relevant Wilson coefficients at the KK scale.
The quoted numbers correspond to $\Mkk=1$\, TeV and can be easily translated to
other scales via the given scaling relations.
We present result for three different typical sets of parameters
that reproduce the observed quark masses as
well as the angles and the CP-violating phase in the quark mixing
matrix within errors (68\% CL). Out of these parameters we show in the table 
just the values of the left- and right-handed top-quark bulk
mass parameters $c_{t_L}$ and $c_{t_R}$ in order to keep the presentation simple.
The complete parameter points, including numerical values for the remaining bulk mass parameters and for the Yukawa matrices,
are spelled out in Appendix~\ref{app:points}.
It is important to emphasize that the magnitudes of the shown results are generic 
predictions in the allowed parameter space and do not reflect a specific choice or tuning
of model parameters.

The coefficients in the table exhibit significant hierarchies, which are given approximately by
$|\tilde C_{q \bar q}^A|/|\tilde C_{q \bar
  q}^V| = {\cal O} (10^{-3})$, $|\tilde C_{t \bar u}^V|/|\tilde C_{u
  \bar u}^V| = {\cal O} (10^{-5})$, and $|\tilde C_{t \bar
  u}^S|/|\tilde C_{u \bar u}^V| = {\cal O} (10^{-6})$. 
The contributions due to flavor changing interactions in the $t$-channel,
encoded in $\tilde C_{t \bar u}^V$ and $\tilde C_{t \bar u}^S$, are strongly
suppressed in the RS model. This is due to the fact that the light up quark is involved, for which
the RS-GIM mechanism is very effective. In the minimal RS setup,
the ratio of neutral electroweak gauge boson (Higgs-boson) to KK-gluon effects in the $t$-channel is 
roughly $1/3$ (on average $1/50$). In custodial extensions one finds
approximately the same suppressions. These factors imply that the predictions
for the $t \bar t$ observables considered here are to very good approximation
model-independent, as they do not depend sensitively on the exact
realizations of neither the electroweak gauge, nor the fermionic, nor
the Higgs sector. 
\begin{figure}[!t]
\begin{center}
\includegraphics[width=7.5cm]{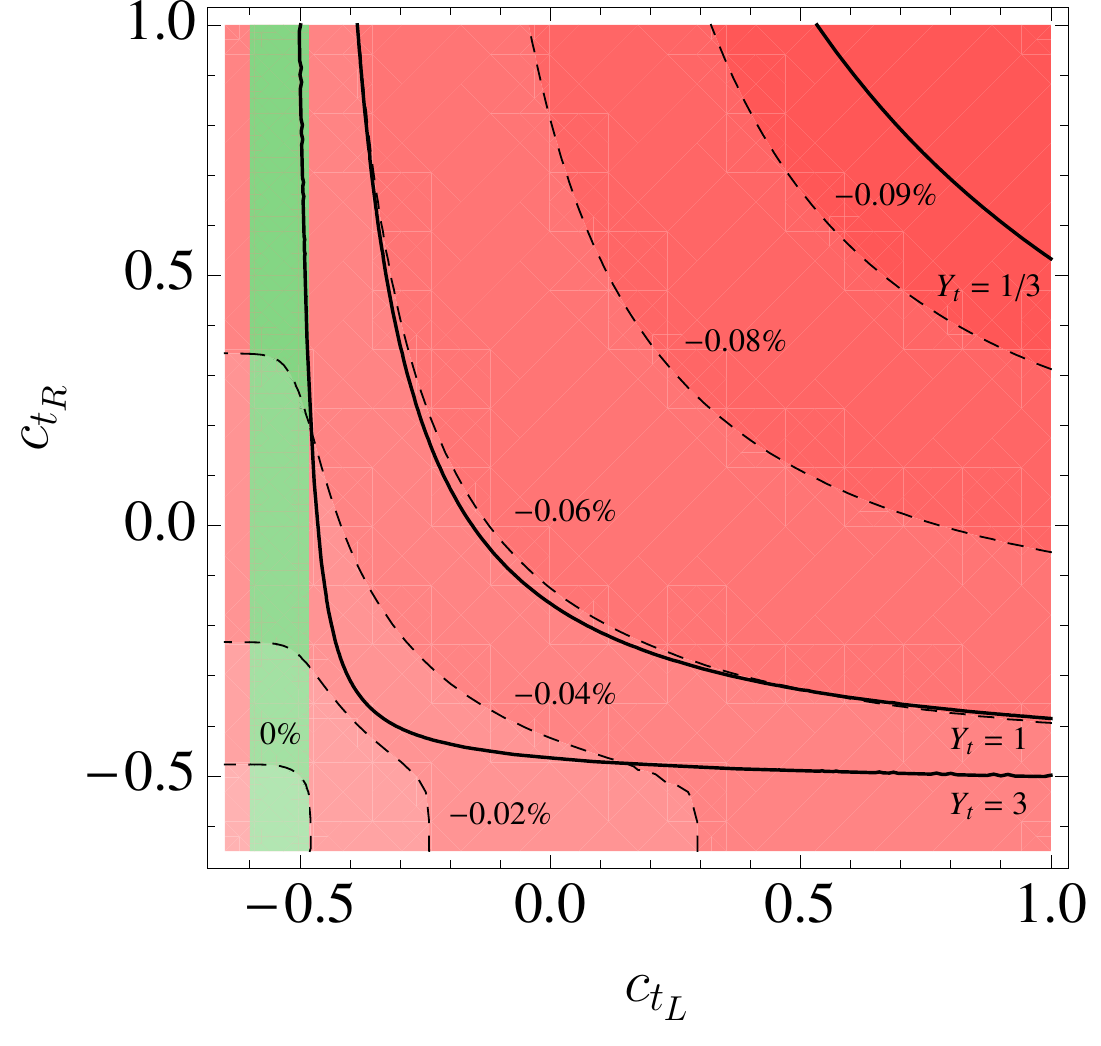}
\qquad 
\includegraphics[width=7.5cm]{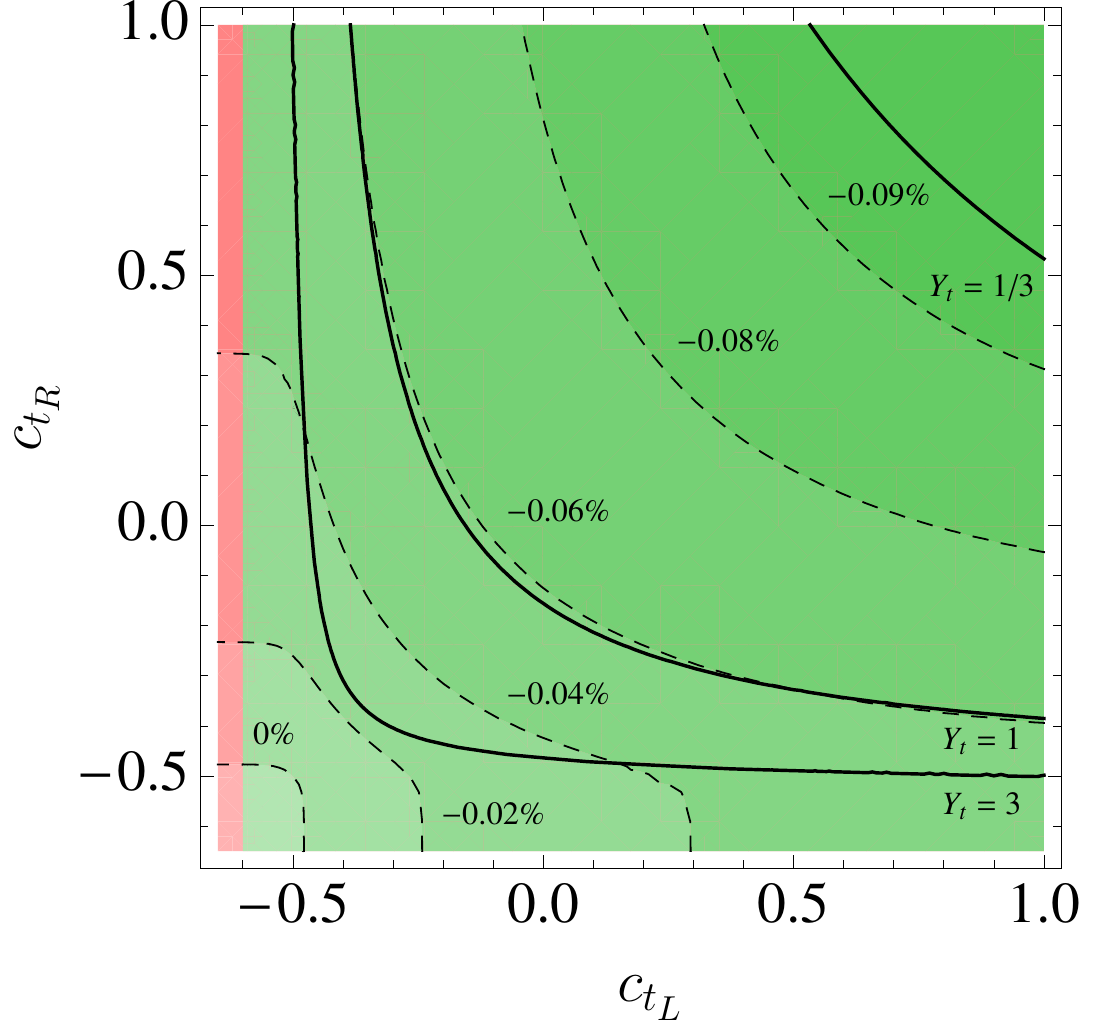}
\end{center}
\vspace{-7.5mm}
\begin{center}
  \parbox{15.5cm}{\caption{\label{fig:RSpredictions} 
      Absolute corrections to $(\AFBt)_{\rm RS}^{p \bar p}$ in the
      $c_{t_L}$--$c_{t_R}$ plane for a KK scale of $1 \, {\rm
        TeV}$. The solid lines indicate the value of $Y_t$ necessary
      to reproduce the mass of the top quark. In the left
      (right) panel the parameter region which satisfies the constraints
      from the $Z \to b \bar b$ pseudo observables for the minimal
      RS setup (custodial RS setup with extended $P_{LR}$-symmetry)
      is displayed in green. See \cite{Bauer:2010iq} and text for details.}}
\end{center}
\end{figure}
The numerical dominant corrections arise from 
$s$-channel KK-gluon exchange. Here, the contributions
from up quarks  $\tilde C_{u \bar u}^V$ and $\tilde C_{u \bar u}^A$ are a factor of a few
larger in magnitude than their counterparts involving down
quarks. This feature is amplified by the suppression
of the down-quark luminosities $\ff_{d\bar d}$ relative to their
counterparts for up-quarks entering our predictions, rendering
the contributions of the former negligible, not exceeding the per cent 
level. In the rest of this section we will thus restrict our attention to the
coefficients $\tilde C_{u \bar u}^{V, A}$ that furnish by far the
largest contributions to the $t \bar t$ observables in RS
models. 

From the first three columns of Table \ref{tab:WC}, we observe that $\tilde C_{u \bar
  u}^{V}$ grows with $(c_{t_L} + c_{t_R})$, \ie, with
  increasing localization of the top-quark in the IR. This behavior has been expected from
  (\ref{eq:SALOscaling2}) and (\ref{eq:ctR}). A similar (expected) trend in terms of $c_{t_R}$
  can be seen in $\tilde C_{u \bar u}^{A}$, however being less pronounced.
  Our main conclusion from the last Section~of strongly suppressed
  axial-vector couplings, $|\tilde C_{u\bar u}^A|/|\tilde C_{u\bar u}^V| \ll 1$ 
  is also confirmed by the numerical analysis. Inserting the corresponding values 
  into the numerator of (\ref{eq:AFBEFT}), we observe that also our third expectation
  holds true. Indeed, in RS models the NLO corrections to the asymmetric cross section, 
  arising from $\tilde C_{u \bar u}^{V}$, are significantly bigger than the LO contributions, 
  stemming from $\tilde C_{u \bar u}^{A}$. Numerically, it turns out that the vector-current 
  contributions are, despite their loop suppression, typically larger by about a factor of 100 compared to the 
  corrections due to the axial-vector current. In the light of the experimentally observed enhancement 
  of the asymmetry with respect to the SM prediction, this strong enhancement looks promising at first sight.
  
  However, a closer look at (\ref{eq:AFBEFT}) shows that in the ratio of the asymmetric and symmetric cross 
  sections the effects of $\tilde C_{u \bar u}^{V}$ tend to cancel. Because both $\sigma_a^{p \bar p}$ and
   $\sigma_s$ are enhanced for $\tilde C_{u \bar u}^V > 0$, but the dependence of $\sigma_a^{p \bar p}$
   on $\tilde C_{u \bar u}^V$ is weaker than the one of $\sigma_s$, the found positive values of $\tilde C_{u \bar
  u}^V$ will effectively lead to a {\it reduction} and not to an enhancement of the $t \bar t$ forward-backward asymmetry.
  Here, $\sigma_a^{p \bar p}$ denotes the asymmetric contribution to the cross section in the $p\bar p$ frame.
  Due to the fact that $\tilde C_{u \bar u}^V > 0$ is a robust prediction of the RS framework, following from the 
  IR localization of the top quark, we conclude that the RS corrections to $\AFBt$ are necessarily negative. 
  
  Due to the aforementioned cancellation, however, the impact of the RS setup on the $t \bar t$ forward-backward asymmetry is very small.
  This is confirmed by the numerical results presented in Figure \ref{fig:RSpredictions}. Here, we show our predictions for the absolute
  RS corrections to the forward-backward asymmetry in the $p \bar p$ frame as a function of $c_{t_L}$ and $c_{t_R}$. 
  We have employed $\Mkk = 1 \, {\rm TeV}$ and the typical bulk-mass parameters of $c_{u_L} =c_{d_L} =
 -0.63$, $c_{u_R} = -0.68$, $c_{d_R} = -0.66$, $c_{c_L} =c_{s_L} =
 -0.56$, $c_{c_R} = -0.53$, $c_{s_R} = -0.63$. Moreover, we have set
  all minors of $Y_{u,d}$ equal, however $Y_t = (Y_u)_{33}$ is allowed to vary in order to 
  reproduce the correct top-quark mass. Note that only the dominant KK-gluon corrections 
  to $\tilde C_{u \bar u}^{V,A}$ have been considered. In the left
  (right) panel the parameter region which satisfies the 99\%\,CL constraints
  from the $Z \to b \bar b$ pseudo observables for the minimal
  RS setup (custodial RS setup with extended $P_{LR}$-symmetry)
  is displayed in green. These constraints have been discussed in detail
  in Section~\ref{sec:bpseudo}.
   Both panels exhibit that in the whole $c_{t_L}$--$c_{t_R}$ plane the RS 
   corrections to $(\AFBt)_{\rm RS}^{p \bar p}$ interfere destructively
   with the SM contributions. On the other hand, even for the optimistic value of $\Mkk = 1 \,
   {\rm TeV}$, the maximal possible effect after imposing the $Z \to b \bar b$
   constraints amounts to $-0.10\%$ ($-0.05\%$ ) for the extended 
   (minimal) RS model. While the $Z \to b \bar b$ constraint is very restrictive in the minimal model, 
   cutting the allowed parameter space to a thin stripe of $c_{t_L} \in [-0.60, -0.49]$, it does not
   affect the extended scenario significantly.
    Note that the inclusion of the NLO RS contributions to the asymmetric cross section
   is important to arrive at a consistent result. They contribute at the same order to 
   (\ref{eq:AFBEFT}) as the RS tree-level corrections to the symmetric cross section.
   Including all RS corrections, we obtain for the three 
   parameter points considered before the absolute shifts of $-0.04\%$, $-0.05\%$, 
   and $-0.05\%$ with respect to the SM value.
   In conclusion, despite the strong IR localization of top quarks,
   the RS impact on $\AFBt$ is deemed to be far too small to be able to explain the
   observed discrepancy between experiment and the SM expectation.

The results presented here should be contrasted with the analysis performed in
\cite{Djouadi:2009nb}, which finds positive corrections to the $t
\bar t$ forward-backward asymmetry of up to $5.6\%$ ($7\%$) arising
from KK gluons ($Z^\prime$-boson exchange) at LO. These large
corrections are generated by localizing the left- and right-handed components 
of the light-quark fields at different ends of the extra dimension by choosing
$c_{u_L} = c_{d_L} \in [ -0.4, 0.4]$ (IR-localized) and $c_{u_R} =
c_{d_R} = -0.8$ (UV-localized), which leads to large axial-vector couplings and 
thus sizable corrections to $\tilde C_{ q \bar q}^A$.\footnote{Notice that the convention 
for the bulk-mass parameters used in \cite{Djouadi:2009nb} differs from
ours by an overall sign.} However, in an anarchic approach to flavor such a
choice is in conflict with observation, as it fails to reproduce
the hierarchies of light-quark masses and mixings.
Also the more recent analyses of \cite{Delaunay:2011vv} (\cite{Djouadi:2011aj})
abandon the anarchic approach to flavor completely (to some extend).

\begin{figure}[t!]
\begin{center}
\includegraphics[width=7.5cm]{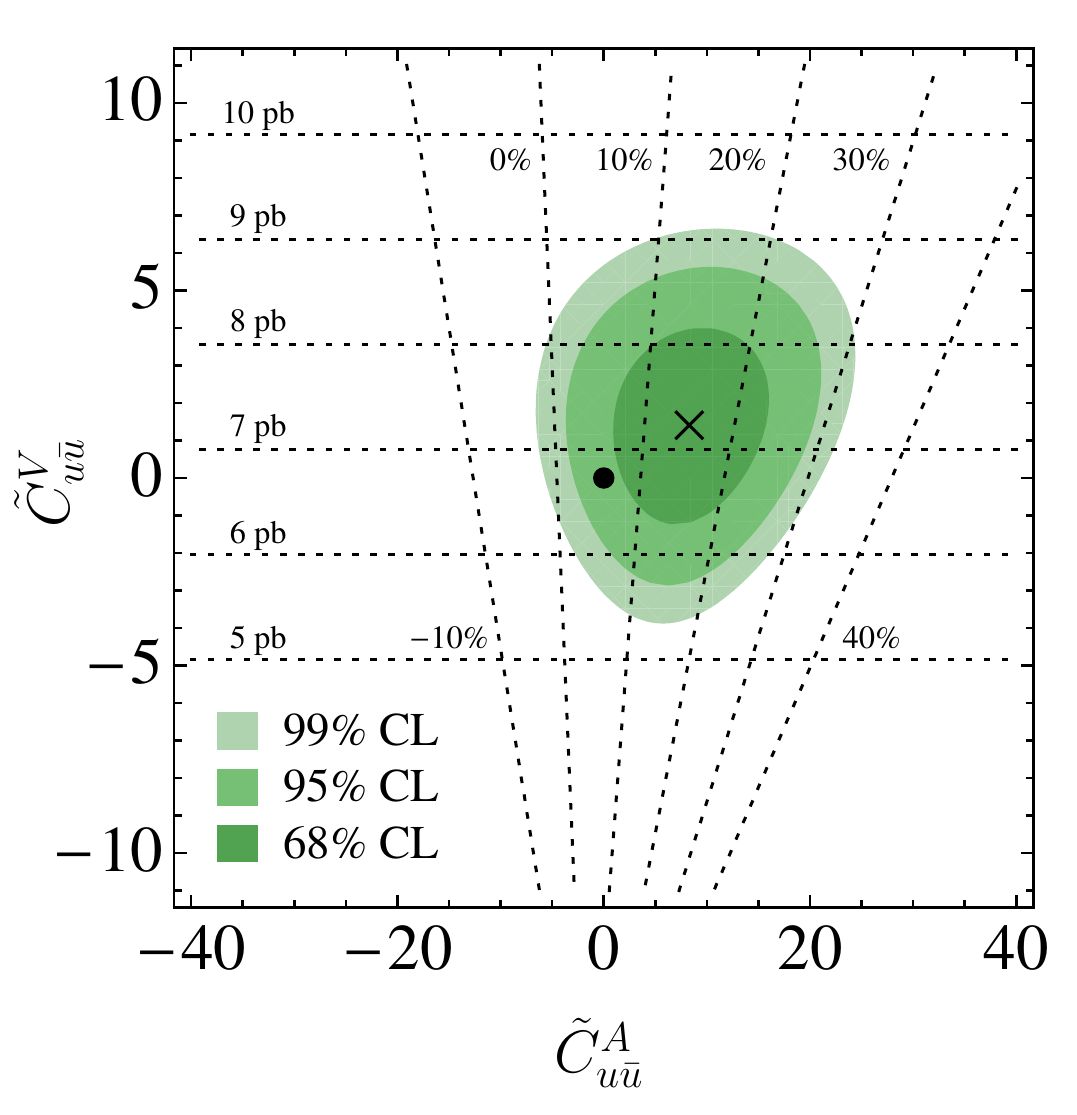}
\end{center}
\vspace{-7.5mm}
\begin{center}
  \parbox{15.5cm}{\caption{\label{fig:cvca} Results of a combined fit
      to $\sigtot$, the last bin of $\dsig$, and the value of
      $(\AFBt)^{p \bar p}$ allowing for NP in $s$-channel
      exchange. The green contours indicate the
      experimentally allowed regions of 68\%, 95\%, and 99\%
      probability in the $\tilde C_{u \bar u}^A$--$\tilde C_{u \bar
        u}^V$ plane. The horizontal (almost vertical) dashed lines
      correspond to the values of the total $t \bar t$ cross section
      (forward-backward asymmetry in the $p \bar p$ frame). Further
      details can be found in the text. Figure from \cite{Bauer:2010iq}.}}
\end{center}
\end{figure}

Our studies have shown that the sensitivity of $\AFBt= \sigma_a/\sigma_s$ to vector 
currents is not very pronounced. Our arguments 
do not only apply to the RS setup but to the broader class of models with new heavy
vector states that have suppressed axial-vector couplings to light
quarks. In such new-physics scenarios large contributions to the $t \bar t$ forward-backward asymmetry
are essentially impossible to achieve, once the experimental
information on $\sigtot$ and the high-energy tail of the $t
\bar t$ invariant mass spectrum $\dsig$ is taken into account. 
This feature is illustrated in Figure~\ref{fig:cvca}, which shows the results of a
fit to the $t \bar t$ data (\ref{eq:AFBexp}) and
(\ref{eq:EXPss}), in the presence of NP in the $s$
channel. The green contours display the experimentally allowed
regions of 68\%, 95\%, and 99\% probability in the 
$\tilde C_{u \bar u}^{A}$--$\tilde C_{u \bar u}^{V}$
plane. They makes evident, that in order to achieve a significant improvement 
in the quality of the fit, large corrections to the axial-vector coefficient $\tilde C_{u
\bar u}^A$ are needed. Vector contributions $\tilde C_{u \bar u}^V$
alone are not sufficient to get from the SM point (black dot) at
$(0,0)$ to the best-fit value (black cross) at $(8.3, 1.4)$. 
If we require the $t \bar t$ predictions to be within the combined
95\% (99\%) CL region, the maximal possible values for $(\AFBt)^{p \bar p}$
from vector contributions alone are $5.8\%$ ($6.0\%$).
A possibly large correction to the $t \bar t$ forward-backward asymmetry inevitably 
has to arise from tree-level effects involving either axial-vector currents in the $s$ channel with
flavor-specific couplings of opposite sign to light quarks and top
quarks or large flavor-changing currents in the $t$ channel. 
Both of these options are not easy to realize in explicit BSM
models, without {\it ad hoc} assumptions about the structure of the light-quark sector. 
In consequence, there seems to be a tension
between generating large effects in $\AFBt$ and achieving a natural solution to
the flavor problem.

We have seen that it is not very likely to find hints for warped extra dimensions in measuring the 
$t \bar{t}$ forward-backward asymmetry. Also the current experimental and theoretical errors on the total
and the differential cross sections $\sigtot$ and $\dsig$ are still too large to see a possible
impact of the RS setup (see (\ref{eq:EXPss}) and (\ref{eq:RSresig})). However, in the much cleaner environment of a possible International Linear Collider
(ILC), avoiding the complications of colored particles in the initial state, corresponding observables allow to probe very high KK scales.
For example a sensitivity to KK masses of (10-30)\,TeV is possible by a measurement of $\sigtot$ to better than $1\%$ or of the
top-quark left-right asymmetry to the same accuracy \cite{DePree:2006ah}.
Another promising sector where we expect sizable deviations from SM expectations and sensitivities to large KK scales, without having
to wait for a new linear collider is Higgs physics, as we will demonstrate in Section~\ref{sec:RSHiggs}.

\subsection{The Anomalous Magnetic Moment of the Muon}
\label{sec:AMM}

In the following, we will apply some of the results derived in Chapter~\ref{sec:5Dprop} to study 
the anomalous magnetic moment of the muon in the background of a 
warped extra dimension. In particular, we will show that the corresponding one-loop diagram
leads to a finite result.
As the muon is a light fermion, we may assume it to be localized on the Planck brane
to good approximation. Thus, we will neglect possible contributions from KK fermions in the following.
It will turn out that, due to the special structure of the emerging sum,
the calculation is also possible in the decomposed theory. We will present this 
derivation further below. 
Although the RS EFT is defined with a cutoff, showing the convergence is useful,
as this means that it will be possible to sum up the complete tower in closed form.

\begin{figure}[!t]
\begin{center}
		\includegraphics[width=5.5cm]{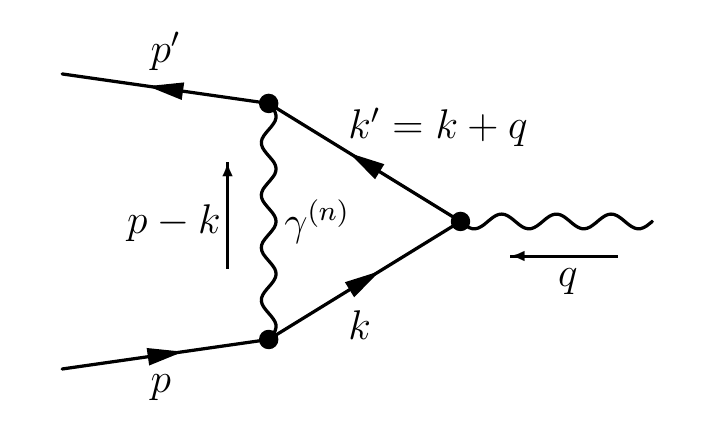}
\caption{\label{fig:vertex} NLO contribution to the 
magnetic moment of a fermion. A straight line corresponds to a fermion while
a wiggly line represents a photon. In the RS model, heavy KK excitations of 
the photon propagate in the loop. See text for details.}
\end{center}
\end{figure}

The magnetic moment  $\bm{\mu}_f$ of a fermion of mass $m_f$ 
and electromagnetic charge $e$ is proportional to its spin $\bm{S}$ via
\begin{equation}
\bm{\mu}_f= g_f \ \frac{e}{2m_f} \bm{S}.
\end{equation} 
The constant $g_f$ is called the {\it Land\'{e} $g$-factor}. 
In terms of the form factors $F_1(q^2)$ and $F_2(q^2)$ it is given as
\begin{equation}
g_f = 2\left[F_1(0)+F_2(0)\right].
\end{equation}
These form factors are defined as coefficients 
of the possible Lorentz structures in the vertex function
\begin{equation}
\Gamma^\mu(p^\prime,p) = \gamma^\mu F_1(q^2) 
+ \frac{i \sigma^{\mu\nu}q_\nu}{2m_f} F_2(q^2)\,,
\end{equation}
where $p$ ($p^\prime$) is the momentum of the incoming (outgoing) 
fermion and $q=p^\prime-p$ is the momentum of the external photon,
see \eg \cite{Peskin:1995ev}. 
Note that $F_1(0)=1$ to all orders in perturbation theory and $F_2(0)=0$ to leading order. At the 
one-loop order, QED predicts a deviation from the Dirac value for an elementary 
fermion $g_f=2$, which is determined by the vertex-correction diagram 
depicted in Figure~\ref{fig:vertex} (with $\gamma^{(n)}\to \gamma$). In the case of the electron, 
this contribution was first calculated by Schwinger \cite{Schwinger:1948iu} 
who determined the {\it anomalous magnetic moment} to be 
\begin{equation}
a_e\equiv\frac{g_e-2}{2}=\frac{\alpha}{2\pi}.
\end{equation}

In the RS model, corrections to $a_f$ arise 
at the one-loop level due to the exchange of a photon that propagates into 
the extra dimension, as indicated by $\gamma^{(n)}$ in Figure~\ref{fig:vertex}. 
In the following, we will show analytically that these corrections are finite,
working with the five dimensional mixed momentum/position-space propagators 
derived in Chapter~\ref{sec:5Dprop}. 
To this end, we look at the region where the loop-momentum becomes large
compared to the KK scale, as this region is critical for the convergence of
the result.

\subsubsection{5D Calculation}

As we study the case of a UV-brane localized fermion,
we use the photon propagator for $p\gg\Mkk$, given in (\ref{eq:lpprop}),
at $t=t^\prime=\epsilon$.

Writing the vertex function as $\Gamma^\mu=\gamma^\mu+\delta\Gamma^\mu$, we
obtain at $\ord(\alpha)$
\beq
\label{eq:inti}
\bar u(p^\prime) \delta\Gamma^\mu u(p)=
2 i e^2 \frac{i L \epsilon}{2 \Mkk}
\int \frac{d^4k}{(2\pi)^4}
\frac{\bar u(p^\prime) \left[\kslash\gamma^\mu\kslash^\prime+m_\mu^2\gamma^\mu-2m_\mu(k+k^\prime)^\mu\right]  u(p)}
{\sqrt{(k-p)^2+i\epsilon}\, ({k^\prime}^2-m_\mu^2+i\epsilon)
(k^2-m_\mu^2+i\epsilon)}\,.
\eeq
Here, $u(p)$ and $\bar u(p^\prime)$ are Dirac spinors, belonging to the in- and out-going muon,
respectively (see Figure~\ref{fig:vertex}), and the $i\epsilon$ prescription shifts the poles of the expression above
infinitesimally away from the real axis. 
Note that we have already performed the trivial integral over the extra dimension.
While the numerator algebra is in analogy to the SM calculation \cite{Peskin:1995ev},
after Feynman parametrization, the denominator becomes 
\begin{equation}
\int_0^1 dx dy dz\, \delta(x+y+z-1) z^{-1/2}\frac{3}{4 D^{5/2}}\,,
\end{equation}
with
\begin{equation}
D=l^2-\Delta+i \epsilon\,.
\end{equation}
Here, as in the SM calculation,
\begin{equation}
\Delta=-xy q^2+(1-z)^2 m_\mu^2\,,\quad l\equiv k+y q -z p\,.
\end{equation}
Note that (\ref{eq:inti}) is just correct for $(k-p)_E \gg\Mkk$, which is the region we are interested in.
In consequence, after Wick rotation, we evaluate the integral
\begin{equation}\label{eq:hint}
\int_{\Mkk}^\infty dl_E \frac{l_E^3}{(l_E^2+\Delta)^{5/2}}=\frac{2 \Delta + 3 \Mkk^2}{3(\Delta+\Mkk^2)^{3/2}}\,.
\end{equation}
The remaining Feynman-parameter integral can easily be solved after performing an expansion in $m_\mu/\Mkk$.
We finally arrive at the UV contribution to the form factor
\begin{equation}
F_2^{{\rm RS}_{\rm UV}}(0)=\frac{\alpha}{2 \pi}\, 
L \epsilon\left(-\frac{8}{35} \frac{m_\mu^2}{\Mkk^2} 
+ \ord\left(\frac{m_{\mu}^4}{\Mkk^4}\right)\right)\,,
\end{equation}
which shows the finiteness of the RS corrections.

\subsubsection{4D Calculation}

In the following we calculate the RS corrections to the anomalous magnetic 
moment of the muon in the KK decomposed theory. To this end, we 
have to compute the contributions to $F_2(0)$ stemming from the 
infinite sum over vertex-correction diagrams in Figure \ref{fig:vertex}, 
with massive KK photons running in the loop. After performing the loop integral,
we arrive at
\begin{equation}\label{F24D1}
F_2(0)= \frac{\alpha}{2 \pi} 
\left(1+2 \sum_{n=1}^\infty \int_0^1 dz \int_0^{1-z} dy\,  
\frac{m_\mu^2\, z(1-z)}{m_\mu^2(1-z)^2+m_n^2z}\, I_n\right),
\end{equation} 
where $m_n$ $(n>0)$ is the mass of the $n^{th}$ KK excitation of the photon. The interactions of the fermions 
with the photons in the extra dimension are described by 
\begin{equation}
I_n=2 \pi \chi^\gamma_n(0)\chi^\gamma_n(0), 
\end{equation}
where $\chi^\gamma_n(\phi)$ is the profile of the $n^{th}$ KK photon. 
Expanding (\ref{F24D1}) in the small ratio $m_\mu/m_n$ and performing 
the Feynman parameter integrals, we get
\begin{equation}\label{F24D2}
F_2(0)= \frac{\alpha}{2 \pi} 
\left(1+\frac{2 m_\mu^2}{3}  \sum_{n=1}^\infty \frac{I_n}{m_n^2} 
\left(1+\ord\left(\frac{m_{\mu}^2}{m_n^2}\right)\right)\right).
\end{equation} 
We are lucky because we can actually perform the infinite sum over boson profiles, 
weighted by inverse powers of KK masses, appearing in (\ref{F24D2}). 
Neglecting subleading terms in $\epsilon$ it reads (see (\ref{important2}))
\begin{equation}
\sum_{n=1}^\infty \frac{I_n}{m_n^2}=\frac{1}{2 \Mkk^2}
\frac{1}{2L}+\ord\left(\frac{v^2}{\Mkk^4}\right),
\end{equation}
which finally leads to
\begin{equation}
F_2(0)=\frac{\alpha}{2 \pi} 
\left(1+ \frac{m_\mu^2}{3 \Mkk^2}\frac{1}{2L}\right) 
+ \ord\left(\frac{m_{\mu}^2 v^2}{\Mkk^4}\right).
\end{equation} 

The RS correction to the anomalous magnetic moment of the muon $a_\mu$
scales like\\ $\alpha/(2\pi)\, m_\mu^2/(L \Mkk^2)$. It is orders of magnitude to small to 
explain the observed discrepancy in $a_\mu$, mentioned in Section~\ref{sec:SMProblems}.
While the derivations presented here were mainly for illustration purposes,
there are more complicated situations, where the approach of using 5D propagators
will make calculations more feasible, see Chapter~\ref{sec:concl}.

\section{Flavor Physics}
\subsection{The CKM Matrix}
\label{sec:CKMPheno}

Remember that our definition of the CKM matrix $\bm{V}_{\! L}$ via effective four-fermion interactions includes the 
exchange of the whole tower of $W^\pm$ bosons and their KK excitations. As discussed in Section~\ref{sec:4Fint}, this matrix 
is not unitary (which would also hold true for the mixing matrix $\bm{\tilde V}_L$ defined via the $W u^i_{L} d^j_{L}$ vertex ).

As a measure of unitarity violation, one can consider the deviation of the sum of the 
squares of the elements in the first
row of ${\bm V}_{\!  L}$ from unity \cite{Bauer:2009cf},
\begin{equation}
  \Delta_1^{\rm non} = 1 - (|V_{ud}|^2 + |V_{us}|^2 + |V_{ub}|^2) = \left (\bm{1}
    - \bm{V}_{\! L}\bm{V}_{\! L}^\dagger \right )_{11} \,.
\end{equation}
After expanding the mixing matrices ${\bm U}_{u,d}$ in powers of the
Cabibbo angle $\lambda$, using the warped-space Froggatt-Nielsen
formulae given in Section~\ref{sec:hierarchies}, 
and normalizing the result to the typical value of the bulk mass parameter $c_{Q_1} \approx
-0.63$, we obtain
\begin{equation} \label{eq:delta1nonapproximate}
  \begin{aligned}
    \Delta_1^{\rm non} &\approx 2 \cdot 10^{-6} \left(
      \frac{F(c_{Q_1})}{F(-0.63)} \right)^2 \left( \frac{\Mkk}{\rm TeV}
    \right)^{-2} \\
    &\quad \times \left[ \, \left| \, {\rm diag}\left(
          \sqrt{\frac{2}{3+2c_{Q_i}}}\; \right) \vec{u} \; \right|^2 -
      \frac{1}{4} \left| {\rm diag}\left( \sqrt{\frac{2}{1-2c_{{\cal
                  T}_{1i}}}}\; \right) \hspace{0.5mm} \bm{Y}_{\!d}^T
        \vec{u} \hspace{0.5mm} \right|^2 \right] ,
  \end{aligned}
\end{equation} 
where the vector $\vec{u}$ is given by minors of $\bm{Y}_u$
\begin{equation}
  \vec{u} = \big(1, -(M_u)_{21}/(M_u)_{11}, (M_u)_{31}/(M_u)_{11}\big) \,.
\end{equation}

The first contribution in the square brackets in
\eqref{eq:delta1nonapproximate} stems from the exchange of the whole
tower of $W^\pm$ bosons and is also present in the {\it minimal RS model}. It
provides a strictly positive contribution to $\Delta_1^{\rm non}$, which
is typically well below the current experimental uncertainty of 
$6.5 \cdot 10^{-4}$ \cite{Bauer:2009cf}. However, the effects due to 
the admixture of $U^\prime$ and $D^\prime$ quarks add to
\eqref{eq:delta1nonapproximate} with opposite sign and can in
principle lead to negative values of $\Delta_1^{\rm non}$. This is not
possible in the minimal RS variant. A detailed discussion of the
breakdown of the unitarity of the quark mixing matrix in the minimal
RS model has been presented in \cite{Bauer:2009cf}. A similar analysis
for the custodial setup has been performed in \cite{Buras:2009ka}. 
However, in that paper the CKM matrix is defined via the $W u^i_{L} d^j_{L}$ 
vertex and not via the effective four-fermion interactions as discussed above. 
This prevents us from an easy comparison of the results 
in \cite{Buras:2009ka} with ours.
The non-unitarity of the CKM matrix has also been touched on previously in \cite{Huber:2003tu,Cheung:2007bu}. 
Yet, a thorough discussion of all relevant effects has not been given in these articles. 
Note that, while the misalignment between the mass and flavor eigenbases 
in the sector of quark zero modes is small - as reflected by the Froggatt-Nielsen like
mechanism of generating fermion hierarchies in the RS setup - this is not the case for the higher KK levels.

The mixings between the states of the first level of KK excitations are encoded in the flavor vectors $\vec a_{4-9}^{(U,D)}$ and 
$\vec a_{4-9}^{(u,d)}$ for the minimal RS variant (and the respective expressions for the custodial model).
Naively, one would expect the mixings between KK fermions to be suppressed by the Higgs VEV over the KK scale.
On the contrary, one finds very large effects, especially for down type quarks, due to the near degeneracy of the corresponding
bulk masses. The mass splittings of the undisturbed KK states, before EWSB, are typical of 
the order of 100\,GeV. Since this is not large compared to $v$, the Yukawa couplings generically induce $\ord(1)$ mixings among the KK excitations of the same KK level (see \cite{Casagrande:2008hr}).
These mixings give rise to unsuppressed flavor changing transitions through KK modes within loop diagrams.
Numerical values for the neutral-current as well as charged-current mixing-matrices for a default parameter point can be found in \cite{Casagrande:2008hr}.
In that article we also presented a numerical analysis of the left- and right-handed CKM matrices, defined via the $W u^i d^j$ vertex.
\subsection{Rare Decays}
\label{sec:rare}

As they are suppressed within the SM, FCNCs offer a promising possibility to discover BSM physics.
Due to the large mass of the top quark, that leads to its IR localization, 
one naturally expects sizable effects of RS models in processes involving flavor-changing top-quark
couplings. Since FCNCs in the up-type quark sector are less
constrained by $K$- and $B$-meson physics than those in the down-type
quark sector, the presence of such anomalous couplings of non-negligible size 
is not ruled out experimentally. In consequence, radiative and rare $\Delta F=1$ processes involving the top quark 
offer a high potential to test the RS setup. In the following we study the rare decays 
$t\rightarrow cZ$ and $t\to c h$.

\subsubsection[Rare Decay $t\rightarrow cZ$]{Rare Decay \boldmath$t\rightarrow cZ$\unboldmath}
\label{sec:tcZ}

From (\ref{eq:Zff}) we can derive the branching ratio for the decay $t\rightarrow cZ$,
which is given to excellent approximation by 
\begin{align}
    {\cal B}(t\to c Z) &= \frac{2\left( 1-r_Z^2 \right)^2 \left(1+2r_Z^2
      \right)}%
    {\left( 1-r_W^2 \right)^2 \left( 1+2r_W^2 \right)} \nonumber\\
    \pagebreak
    &\quad\times \left\{ \left| \left( g_L^u \right)_{23} \right|^2 +
      \left| \left( g_R^u \right)_{23} \right|^2 - \frac{12r_c r_Z^2}
      {\left( 1-r_Z^2 \right) \left( 1+2r_Z^2 \right)}\, \mbox{Re}\big[
      \left( g_L^u \right)_{23}^\ast
      \left( g_R^u \right)_{23} \big] \right\} \nonumber\\
    &\approx 1.842\,\Big[ \left| \left( g_L^u \right)_{23} \right|^2 +
    \left| \left( g_R^u \right)_{23} \right|^2 \Big] -
    0.048\,\mbox{Re}\big[ \left( g_L^u \right)_{23}^\ast \left( g_R^u
    \right)_{23} \big] \,, 
\end{align}
where $r_i\equiv m_i^{\rm pole}/m_t^{\rm pole}$. For simplicity we
keep only terms up to first order in $v^2/\Mkk^2$ and the
charm-quark mass ratio $r_c\approx 8.7\cdot 10^{-3}$. 
As given above, this formula holds for both RS variants studied in this thesis.

The flavor-changing couplings in the {\it custodial model} are given by
\begin{equation}
\label{eq:gtcZ}
  \begin{split}
    \left( g_L^u \right)_{23} &= - \frac{m_Z^2}{2\Mkk^2} \left(
      \frac12 - \frac23 s_w^2 \right) \Big[ \omega_Z^{u_L} L \left(
      \Delta_U \right)_{23} - \left( \Delta'_U \right)_{23} \Big]
    - \left( \delta_U \right)_{23} \,, \\ 
    \left( g_R^u \right)_{23} 
   &= \frac{m_Z^2}{2\Mkk^2}\,\frac23 s_w^2\,
    \Big[  \omega_Z^{u_R} L \left( \Delta_u \right)_{23} 
    - \left( \Delta'_u \right)_{23} \Big] 
    + \left( \delta_u \right)_{23}  \, .
  \end{split}
\end{equation}
The ZMA expressions for the matrices $\bm{\Delta}_U$,
$\bm{\Delta}'_U$, and $\bm{\Delta}'_u$ are obtained from (\ref{ZMA1})
by the replacements $c_{F_i} \to c_{Q_i}$, $c_{f_i} \to
c_{u^c_i}$. In
the same approximation one has $\bm{\delta}_U = 1/2 \, \bm{\Phi}_U$
with $\bm{\Phi}_U$ introduced in (\ref{eq:PhiU}) and
\begin{equation} \label{eq:deltau}
  {\bm \delta}_u = \frac{1}{2} \; {\bm x}_u \, {\bm U}_u^\dagger \, \,
  {\rm diag }\left [ \frac{1}{1 - 2 \hspace{0.25mm}c_{Q_i}} \left (
      \frac{1}{F^2(c_{Q_i})} \left [ 1 - \frac{1 - 2 \hspace{0.25mm}
          c_{Q_i}}{F^2(-c_{Q_i})} \right ] - 1 + \frac{F^2(c_{Q_i})}{3 + 2
        \hspace{0.25mm} c_{Q_i}} \right ) \right ] \, {\bm U}_u \, {\bm x}_u
  \,.
  \vspace{0.4cm}
\end{equation}
In the {\it minimal model}, (\ref{eq:gtcZ}) holds true with
$\omega_Z^{u_L}=\omega_Z^{u_R}=1$ and an additional factor of 1/2 in front of $\left( \delta_{U,u} \right)_{23}$.
The corresponding ZMA expressions are given in (\ref{ZMA1}) and (\ref{eq:ZMA2}).
Notice that, compared to the ZMA result in the minimal RS model, the mixing matrix $\bm{\delta}_u$ above contains an
additional term involving the zero-mode profile $F(-c_{Q_i})$.

\begin{figure}[!t]
\begin{center} 
\hspace{-2mm}
\mbox{\includegraphics[height=2.85in]{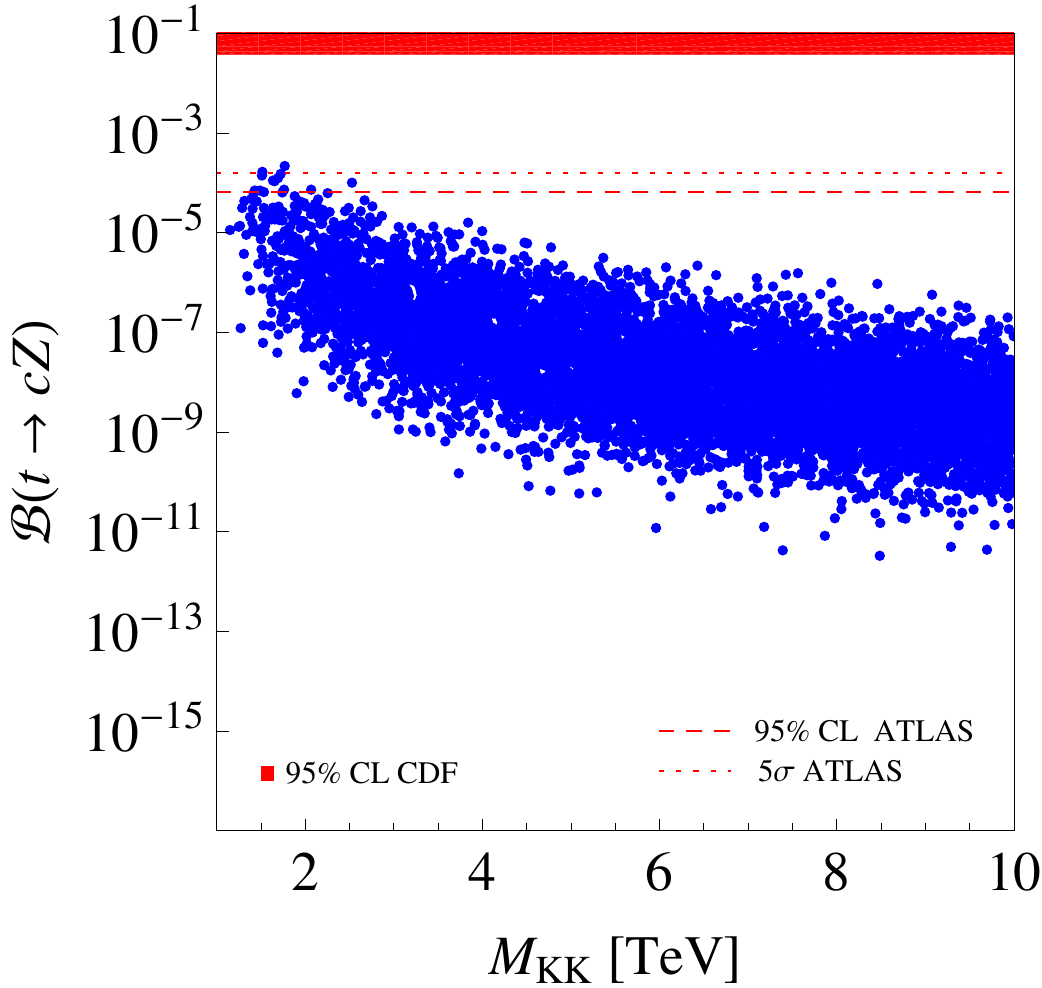}} 
\qquad 
\mbox{\includegraphics[height=2.85in]{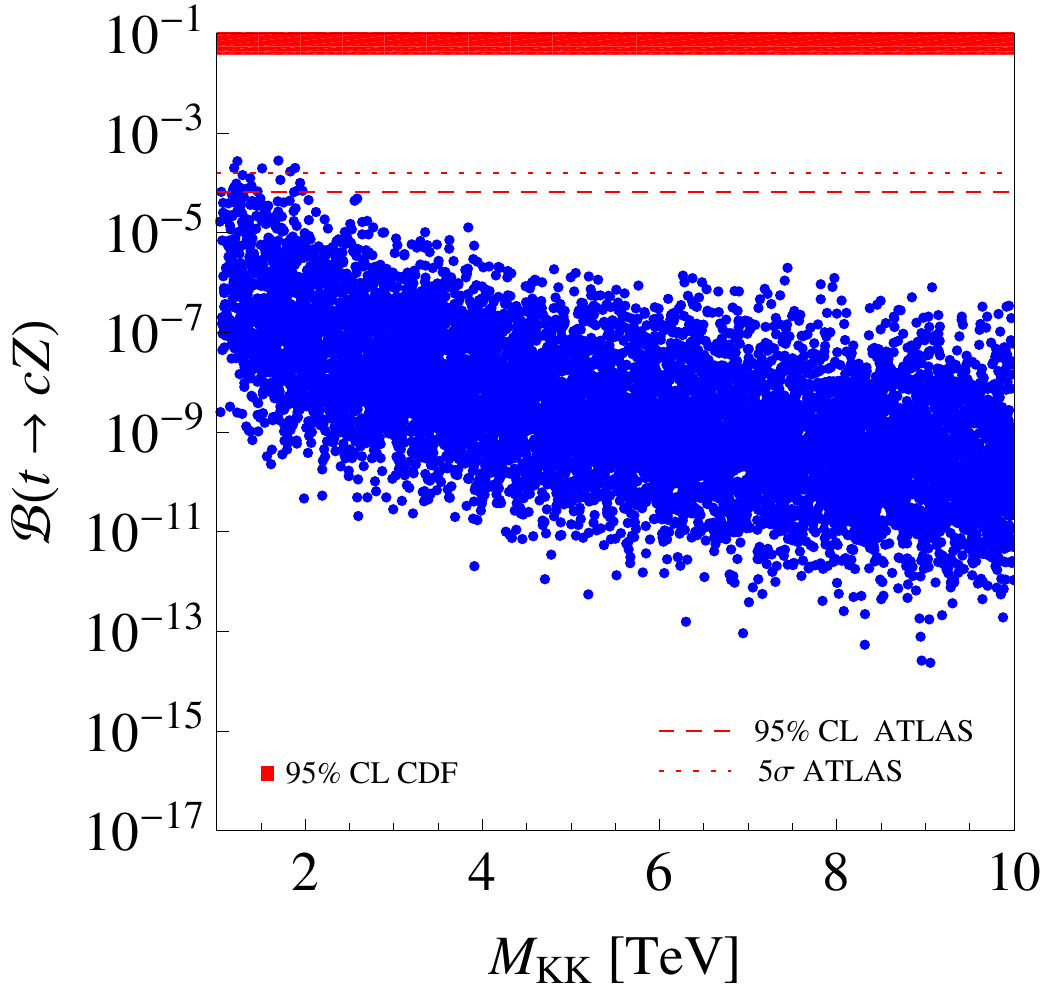}}
\vspace{-1mm}
\parbox{15.5cm}{\caption{\label{fig:tcplots}
    Branching ratio of the
    rare decay $t\to c Z$ as a function
    of $\Mkk$ in the minimal RS model (left) and in the variant with extended 
    custodial protection $c_{{{\cal T}}_{1 i}} = c_{{{\cal T}}_{2 i}}$ (right).
    Points that do not satisfy the constraints form the $Z\to b\bar b$ pseudo observables 
    are rejected. The red band
    is excluded at 95\%~CL by the CDF search for $t\to u(c)
    Z$. The red dotted and dashed lines
    indicate the expected discovery and exclusion sensitivities of
    ATLAS for 100\,fb$^{-1}$ integrated luminosity. See \cite{Casagrande:2008hr,Casagrande:2010si} and text for details.}}
\end{center}
\end{figure}

Inserting the quantum numbers of the representation
(\ref{eq:multiplets}) into (\ref{eq:omegaZ}), we see that the leading
contribution to $\left( g_L^u \right)_{23}$ in the custodial model
is enhanced by a factor
\begin{equation} 
\omega_Z^{u_L} = \frac{2 c_w^2}{1 - \frac{4}{3} s_w^2}
\approx 2.2 \,.
\end{equation} 
In contrast to the minimal model, the
right-handed coupling does not receive an $L$-enhanced contribution,
because 
\beq
\omega_Z^{u_R}=0\,.
\eeq 
Moreover, the contribution that is inversely
proportional to $F^2(c_{Q_i})$ in $\bm{\delta}_u$ is highly suppressed,
if $c_{Q_i} < 1/2$, since $F^2(-c_{Q_i}) \approx 1 - 2 \hspace{0.25mm}
c_{Q_i}$ in such a case. 
The leading corrections to the $Z u_R^i \bar
u_R^j$ vertices due to quark mixing are therefore protected by the
custodial symmetry. While these features remove a possible source of
large effects associated with the composite nature of the right-handed
top quark, they also imply that the chirality of the $Z tc$ interactions in
the model under consideration is predicted to be
left-handed. Of course, other
choices of the quantum numbers of the right-handed up-type quarks than
those in Table~\ref{tab:charges} are possible, so that the RS
framework does not lead to a firm prediction of the chirality of
$Z tc$ interactions.
In the minimal RS formulation a slight preference for left-handed couplings 
is given due to the prefactors in (\ref{eq:gtcZ}), however, there is also
a large part of the parameter space in which the chirality of $Z tc$ interactions
is predominantly right handed. For more details, also on correlations with
$B\to X_s l^+l^-$ decays, see \cite{Casagrande:2008hr}.

The predictions for ${\cal B}(t\to c Z)$ in the minimal as well as in the 
custodial RS model with extended $P_{LR}$ symmetry, as a function of $\Mkk$, are shown in
the left and right panel of
Figure~\ref{fig:tcplots}, respectively. The experimental upper
bound on FCNC $t\to u(c) Z$ from the CDF experiment amounts to ${\cal
  B}(t\to u(c) Z)<3.7\%$ at 95\%~CL \cite{tcZ:2008aaa} and is shown as
a band. At the LHC, one can search for rare FCNC top-quark transitions
in top-quark production and decays. The ATLAS \cite{Carvalho:2007yi}
and CMS \cite{Ball:2007zza} collaborations have examined this
possibility in simulation studies. The minimal branching ratio ${\cal
  B}(t\to c Z)$ allowing for a $5\sigma$ signal discovery with
100\,fb$^{-1}$ integrated luminosity is expected to be $1.6\cdot
10^{-4}$ at ATLAS. In the absence of a signal, the expected upper
bound at 95\%~CL is $6.5\cdot 10^{-5}$. These values are visualized by
the red dotted and dashed lines in the plots. Our numerical studies
show that in the custodial setup, for low KK mass scales in the ballpark of 2
TeV,\footnote{Corresponding to masses of the lightest KK gauge bosons
  of around 5 TeV.} the branching ratio ${\cal B}(t\to c Z)$ can
come close to the region which can be probed at the LHC.\footnote{As a
  result of $|F(c_{Q_1})|/|F(c_{Q_2})| \sim \lambda$ the branching
  ratio of $t \to uZ$ is typically suppressed by two orders of
  magnitude compared to $t \to cZ$, rendering the former mode
  unobservable at the LHC. Similar statements apply to the branching
  ratio of $t \to uh$.} 
Such a low KK scale is a realistic possibility in RS models with
custodial protection. In the minimal RS model the possible branching
ratios are smaller. To a large extend this is due to the rejection of points
which fail to satisfy the constraints form the $Z\to b\bar b$ pseudo observables. 
The custodial protection of the $Z b_L \bar b_L$ vertex thus leads indirectly to
improved prospects of a detection of the decay $t \to c Z$ at the LHC.
Note that there is a strong correlation between the left-handed $Ztc$ and $Zb \bar b$ couplings
 due to the fact that the left-handed top quark resides in the same 
$SU(2)_L$ doublet as the corresponding bottom quark. 
On the other hand, the range of predictions in the custodial model also reaches lower values
compared to the minimal model. This is due to the custodial protection of the right-handed
couplings. These are anti-correlated with the left-handed ones due to the
mass relations (\ref{eq:FNm}). Points that show a pronounced right-handed contribution in the minimal model
feature a strongly suppressed $Ztc$ coupling in the custodial model.

\subsubsection[Rare Decay $t\rightarrow ch$]{Rare Decay \boldmath$t\rightarrow ch$\unboldmath}
\label{sec:tch}

\begin{figure}[!t]
\begin{center} 
\hspace{-2mm}
\mbox{\includegraphics[height=2.85in]{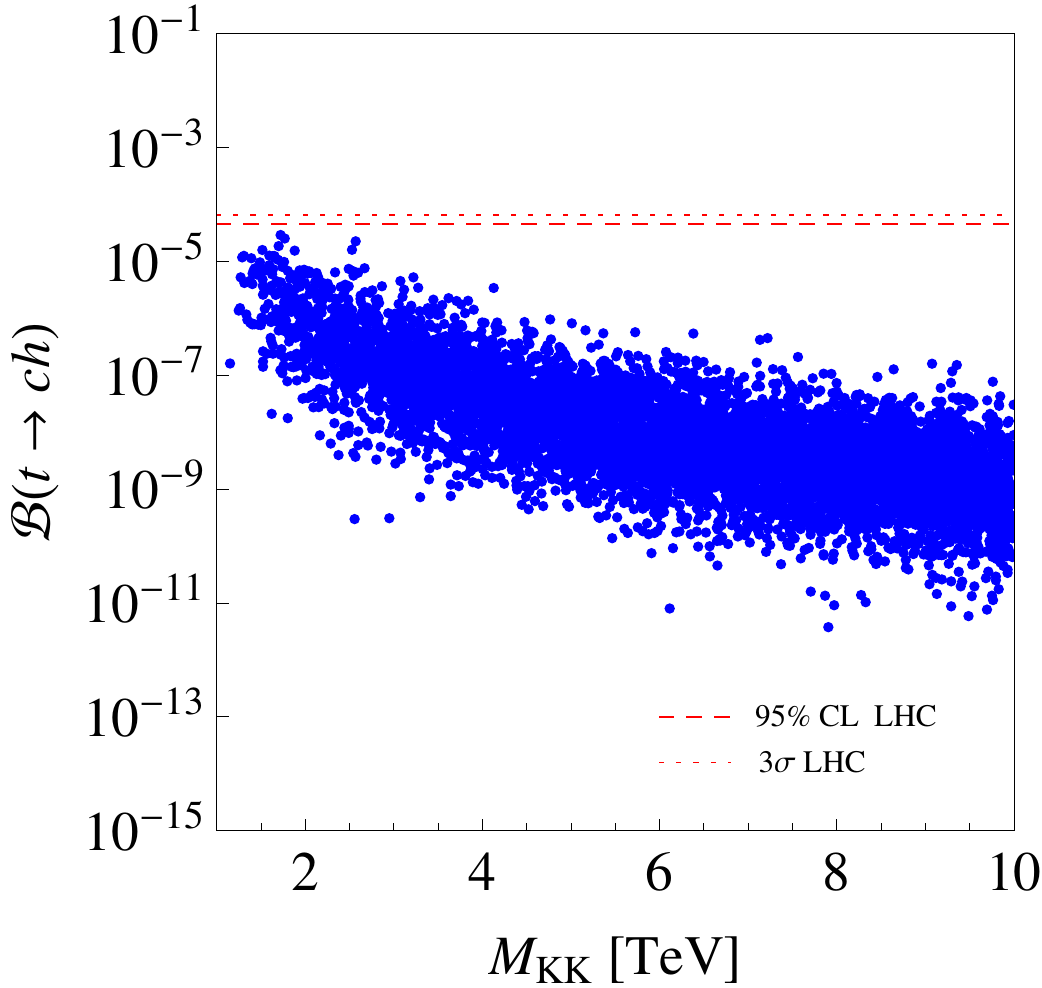}} 
\qquad 
\mbox{\includegraphics[height=2.85in]{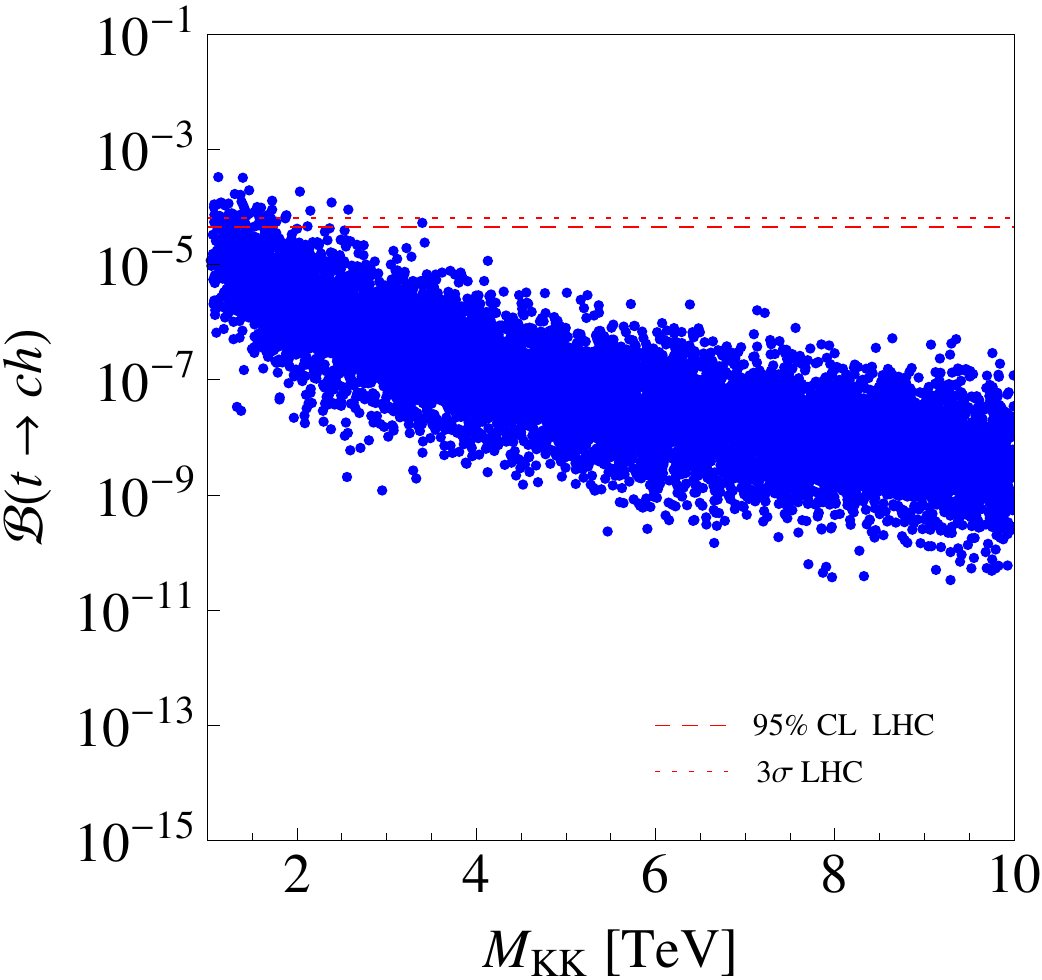}}
\vspace{-1mm}
\parbox{15.5cm}{\caption{\label{fig:tchplots}
    Branching ratio of the
    rare decay $t \to c h$ as a function
    of $\Mkk$ in the minimal RS model (left) and the RS model with extended 
    custodial protection (right). The red dotted and dashed lines 
    indicate the expected discovery and exclusion sensitivities of
    the LHC for 100\,fb$^{-1}$ integrated luminosity. See \cite{Casagrande:2010si} and text for details.}}
\end{center}
\end{figure}

The general form of the interactions of fermions with the Higgs boson
has been given in (\ref{eq:hff}). These couplings allow for the
flavor-changing decay $t\to c h$ (if kinematically accessible) with a branching ratio
\begin{eqnarray} \label{BRthc} 
  {\cal{B}}(t\to c h) = \frac{2 \left( 1-r_h^2 \right)^2 r_W^2}{\left(
      1-r_W^2 \right)^2 \left( 1+2r_W^2 \right) g^2} \left\{ \left| \left(
        g_h^u \right)_{23} \right|^2 + \left| \left( g_h^u \right)_{32}
    \right|^2 + \frac{4r_c}{1-r_h^2}\, \mbox{Re}\big[ \left( g_h^u
    \right)_{23} \left( g_h^u \right)_{32} \big] \right\} , \hspace{8mm}
\end{eqnarray}
where as before $r_i\equiv m_i^{\rm pole}/m_t^{\rm pole}$, and $g$ is
the $SU(2)_L$ gauge coupling. Again, we have included terms up to
first order in the charm-quark mass. In our numerical analysis we will
use $r_h=0.87$, corresponding to a Higgs-boson mass of $m_h=150$\,GeV.

The predictions for ${\cal B}(t\to c h)$ in the minimal RS model
(custodial RS model with extended $P_{LR}$ symmetry) as a function 
of $\Mkk$ are shown in the left (right) panel of Figure~\ref{fig:tchplots}. The LHC is expected to
provide a $3\sigma$ evidence for ${\cal B}(t\to ch)$ larger than
$6.5\cdot 10^{-5}$ or set an upper bound of $4.5\cdot 10^{-5}$ with
95\%~CL if the decay is not observed \cite{AguilarSaavedra:2000aj}.
These limits are indicated by the red dotted and dashed lines in the
plot. We see that for low KK mass scales, the predicted values for the 
branching ratio in the custodial scenario can even exceed the LHC reach, so 
that a detection of a possible RS signal with $t \to ch$ could become 
reality. Let us add that without the inclusion of the Yukawa couplings 
involving $Z_2$-odd fermion profiles, the obtained branching fractions 
would be typically smaller by almost two orders of magnitude, see \cite{Casagrande:2008hr}. 
In the minimal RS model the prospects for an observation of $t \to c h$ turn 
out to be less favorable, since the constraints from $Z \to b \bar b$ 
typically eliminate those points in parameter space that would show 
pronounced effects.

\subsection[CP Violation in $B_s^0$-Meson Decays]{CP Violation in $\bm{B_s^0}$-Meson Decays}
\label{sec:asl}

We have seen in Section~\ref{sec:hierarchies} that the RS setup features nine new phases with respect to
the SM, providing new sources of CP violation. In this section, we will study 
CP violating observables in the framework of warped extra dimensions. We will focus on
$B^0_s$-mesons, since due to the hierarchy of CKM matrix elements, CP violation in $B^0_s$ mixing should be
tiny within the SM (and models of MFV). This offers a nice prerequisite to find signs for NP
like warped extra dimensions. Moreover, $B^0_s$ mesons are promising since they are the only mesons that are expected to 
mix but do not contain first-generation quarks. This lets us expect potentially more significant NP corrections. 
Beyond that, some anomalies have been reported in the $B^0_s$-meson sector, recently (see Section~\ref{sec:SMProblems} 
and the remainder of this section).
We will analyze corrections to the width difference $\Delta\Gamma_s$, to the 
semileptonic CP asymmetry $A_{\rm SL}^s$, as well as to the CP asymmetry $S_{\psi\phi}$.
We will perform a thorough calculation of the decay amplitude $\Gamma_{12}^s$ in the framework 
of an EFT, based on operator product expansion. As a by-product, we identify a SM contribution missing in the literature.
However, although it is of the same size as terms that have been considered, it belongs to a class of numerically suppressed terms
and its omission has no significant impact on the numerical predictions. 
Due to the EFT approach, our results can be used for many new physics models.

\subsubsection[Theory of $B$-Meson Mixing and CP Violation]{Theory of $\bm{B}$-Meson Mixing and CP Violation}

We start with a breif review of the theory of CP violation in $B_q^0$\,-meson $(q=d,s)$ decays. 
For a comprehensive introduction see \eg \cite{Nir:2001ge,Grossman:2010gw,Buras:2005xt,Nir:cargese}.
One can distinguish between three types of CP violation in meson decays: \\[-5mm]
\begin{enumerate}[(a)]
\item{CP violation in mixing}
\item{CP violation in decay}
\item{CP violation in the interference of decays with and without mixing.}
\end{enumerate}
Before discussing the origin of these different classes, we have to introduce the theoretical 
framework to describe $B_q^0$--$\bar{B}_q^0$ mixing. 

For times $t$ that are much larger than the strong 
interaction scale, the time evolution of the neutral $B_q^0$--$\bar{B}_q^0$ system can be 
described by a time-dependent Schr\"odinger equation \cite{Weisskopf:1930au,Weisskopf:1930ps,Sachs}.
Writing an arbitrary neutral $B$-meson state at time $t$ as a superposition of
the strong interaction eigenstates $B_q^0$ and $\bar{B}_q^0$
\beq
 |B_q^{0{\rm\, phys}}(t)\rangle= b(t) |B_q^0\rangle+\bar b(t) |\bar B_q^0\rangle\,,
\eeq
its time evolution is given by
\beq
    i \frac{d}{dt}
		\left( \begin{array}{c}
    b(t) \\
    \bar{b}(t)
   \end{array} \right) 
   ={\bm{H}}^q
   \left( \begin{array}{c}
    b(t) \\
    \bar{b}(t)
   \end{array} \right).
\eeq
The effective $2 \times 2$ Hamiltonian 
\beq
{\bm{H}}^q=\left(\bm{M}^q-\frac{i}{2}\, \bm{\Gamma}^q \right)
\eeq
governs the oscillation and 
decay of $B$-mesons and therefore contains an anti-hermitian part proportional to the decay matrix $\bm{\Gamma}^q$. 
This matrix and the mass matrix $\bm{M}^q$ are hermitian and obtained in QFT as the absorptive and dispersive part of the amplitude governing $\left(B_q^0,\bar{B}_q^0\right) \leftrightarrow \left(B_q^0,\bar{B}_q^0\right)$ transitions, respectively. For example, $M^q_{12}$
corresponds to the dispersive part of the $B_q^0 \rightarrow \bar{B}_q^0$ transition.
CPT invariance guarantees that $M^q_{11}=M^q_{22}$ and $\Gamma^q_{11}=\Gamma^q_{22}$ and hermiticity leads to
$M^q_{21}=M^{q \ast}_{12}$ and $\Gamma^q_{21}=\Gamma^{q \ast}_{12}$.
The eigenstates of ${\bm{H}}^q$ correspond to
\beq
|B_q^L\rangle=p|B_q^0\rangle+q|\bar{B}_q^0\rangle,\,\quad |B_q^H\rangle=p|B_q^0\rangle-q|\bar{B}_q^0\rangle\,,
\eeq
with eigenvalues 
\beq
\mu_{L,H}^q=M_{L,H}^q+i \Gamma_{L,H}^q\,,
\eeq
where $|p|^2+|q|^2=1$.
Here, $M_{L,H}^q$ and $\Gamma_{L,H}^q$ denote the masses and decay widths of the light and heavy mass eigenstates $B_q^{L,H}$, respectively.
We define the mass difference as well as the width difference between these states as
\beq
\Delta m_{B_q}\equiv M_H^q-M_L^q,\,\quad \Delta\Gamma_q\equiv\Gamma^q_L-\Gamma^q_H\,.
\eeq
Solving the eigenvalue equation leads to
\beq
\begin{split}
\Delta\Gamma_q=& -\,\frac{4\,{\rm Re}(M_{12}^q\Gamma_{12}^{q\ast})}{\Delta m_{B_q}}\,,\quad  
(\Delta m_{B_q})^2-\frac14(\Delta\Gamma_q)^2 = (4 |M^q_{12}|^2-|\Gamma^q_{12}|^2)\,,\\
\frac{q}{p}=&-\frac{2 M_{12}^{q\ast}-i\Gamma_{12}^{q\ast}}{\Delta m_{B_q}+\frac{i}{2}\Delta\Gamma_q}\,.
\end{split}
\eeq
It is experimentally known that for $B$-mesons $\Delta m_{B_q}\gg\Delta\Gamma_q$
and in consequence also $|M_{12}^q|\gg|\Gamma_{12}^q|$ \cite{Nir:2001ge,Buskulic:1995yf}.
It follows that to zeroth order in $|\Gamma_{12}^q|/|M_{12}^q|$, which is a very good approximation,
\beq
\label{eq:DMM}
\Delta m_{B_q}= 2 |M_{12}^q|\,.
\eeq 
This results in \cite{Buras:1984pq,Dunietz:1987yt}
\beq\label{eq:DeltaG}
\Delta\Gamma_q= -\,\frac{2\,\text{Re}(M_{12}^q\Gamma_{12}^{q\ast})}{|M_{12}^q|} 
=2\,|\Gamma_{12}^q|\,\cos \phi_q\,,
\eeq
where we have defined the relative phase $\phi_q$ between the mixing and the decay amplitude according to
the convention
\beq\label{eq:phase}
\displaystyle
\frac{M_{12}^q}{\Gamma_{12}^q}= -\,\frac{|M_{12}^q|}{|\Gamma_{12}^q|}\;e^{i\phi_q}\,,\qquad
\phi_q=\arg(-M_{12}^q\Gamma_{12}^{q\ast})\,.
\eeq
Note that, including corrections to first order in $|\Gamma_{12}^q|/|M_{12}^q|$ 
\beq
\frac{q}{p}=-\frac{|M_{12}^{q\ast}|}{|M_{12}^q|}\left[1-\frac12{\rm Im}\left(\frac{\Gamma_{12}^q}{M_{12}^q}\right)\right]\,.
\eeq

Moreover, we define the decay amplitudes of $B$-mesons, decaying into a final state $f$ via a decay Hamiltonian ${\cal H}_d$, as
\beq
A_f=\langle f|{\cal H}_d|B\rangle\,, \quad \bar{A}_f=\langle f|{\cal H}_d|\bar{B}\rangle\,,
\eeq
where CP relates $A_f$ and $\bar A_f$.\footnote{Note that we will not always use an index to denote the decaying meson type, 
which however should be clear from the context.} 
Now we can identify conditions for the different types of CP violation mentioned above to be present (see \eg \cite{Nir:2001ge}).
\begin{enumerate}
\item{{\bf CP violation in mixing}: $|q/p|\neq 1$\\
This type of CP violation is induced, if the mass eigenstates of the mesons differ from the CP eigenstates. It requires a relative
phase between the dispersive and the absorptive part of the $B_q^0 \rightarrow \bar{B}_q^0$ transition, \ie, $\phi_q\neq0,\pi$\,.
CP violation in mixing can be observed through a non-vanishing {\it semileptonic CP asymmetry}
\beq
A_{\rm SL}^q=\frac{\Gamma(\bar{B}_q^{0{\rm\, phys}}(t)\to l^+ \nu_l X)-\Gamma(B_q^{0{\rm\, phys}}(t)\to l^- \bar \nu_l X)}
{\Gamma(\bar{B}_q^{0{\rm\, phys}}(t)\to l^+ \nu_l X)+\Gamma(B_q^{0{\rm\, phys}}(t)\to l^- \bar\nu_l X)}\,,
\eeq 
where 
\beq\label{eq:ASL}
A_{\rm SL}^q=\frac{1-|q/p|^4}{1+|q/p|^4}={\rm Im}(\Gamma_{12}^q/M_{12}^q)\,.
\eeq
Note that, here and in the following, we denote with $B_q^{0{\rm\, phys}}(t)$ ($\bar{B}_q^{0{\rm\, phys}}(t)$) a
time evolved, initially pure $B_q^0$-meson ($\bar{B}_q^0$-meson).
}
\item{{\bf CP violation in decay}: $|\bar A_{\bar f}/A_f|\neq1$\\
CP violation in the decay of a meson requires at least two terms in the decay amplitude to have different weak phases, \ie, phases due to complex
parameters in the Lagrangian, and different strong phases, \ie, phases due to intermediate on-shell states. 
For example this type of CP violation is the only source of the {\it CP asymmetry in charged $B$-meson decays}, given by
\beq
a_{f^\pm}=\frac{\Gamma(B^+\to f^+)-\Gamma(B^-\to f^-)}
{\Gamma(B^+\to f^+)+\Gamma(B^-\to f^-)}\,.
\eeq
This asymmetry can be calculated from the decay amplitudes as
\beq
\frac{1-|\bar{A}_{f^-}/A_{f^+}|^2}{1+|\bar{A}_{f^-}/A_{f^+}|^2}.
\eeq
}
\item{{\bf CP violation in the interference of decays with and without mixing}: Im$(\lambda_f)\neq~0$\\
This type of CP violation is an effect of the interference between an amplitude for a meson to first mix
and then decay and a direct decay amplitude into the same final state. It is induced by a non-vanishing value for the imaginary part of the quantity
\beq
\lambda_f=\frac{q}{p}\frac{\bar A_f}{A_f}\,.
\eeq
It enters the {\it time dependent CP asymmetry} in the decay into a CP eigenstate $f=\bar f=f_{\rm CP}$
\beq
\label{eq:tdcp}
{\cal A}_{f_{\rm CP}}^q(t)=\frac{\Gamma(\bar{B}_q^{0{\rm\, phys}}(t)\to f_{\rm CP})-\Gamma(B_q^{0{\rm\, phys}}(t)\to f_{\rm CP})}
{\Gamma(\bar{B}_q^{0{\rm\, phys}}(t)\to f_{\rm CP})+\Gamma(B_q^{0{\rm\, phys}}(t)\to f_{\rm CP})}\,,
\eeq
which is given by \cite{Nir:2001ge}
\beq
\begin{split}
{\cal A}_{f}^q(t)=S_f \sin(\Delta m_{B_q}t)
-C_f\cos(\Delta m_{B_q}t)\,,\\ 
S_f=\frac{2{\rm Im}\lambda_f}{1+|\lambda_f|^2}\,,\qquad
C_f=\frac{1-|\lambda_f|^2}{1+|\lambda_f|^2}\,.
\end{split}
\eeq
In the case that CP violation from mixing as well as from the direct decay are negligible, 
one has $|\lambda_{f_{\rm CP}}|=1$ and thus $C_{f_{\rm CP}}=0$. In consequence Im$(\lambda_{f_{CP}})\neq0$ 
is the only contributing effect.}
\end{enumerate}
Note that all the three types of CP violation have been experimentally observed in Kaon physics,
namely by measuring non-zero values for the quantities ${\rm Re}(\epsilon_K)$, ${\rm Re}(\epsilon_K^\prime)$, and ${\rm Im}(\epsilon_K)$, 
respectively (see \eg \cite{Nir:2001ge}).

\subsubsection[$B_s^0$-Meson Observables in the SM and Beyond]{$\bm{B_s^0}$-Meson Observables in the SM and Beyond} 

Many of the quantities introduced above have been measured in the
$B_s^0$-system, which allows to confront theoretical predictions with experiment.

For example, CDF and D{\O} presented combined results in the $(\beta_s^{J/\psi\phi},\Delta\Gamma_s)$--plane \cite{pubn} which
differ from the SM prediction by about $2\sigma$.\footnote{However, the
latest individual CDF results disagree by $1\sigma$ only \cite{pubn2}.}
Here, $\beta_s^{J/\psi\phi}\in[-\pi/2,\pi/2]$ is the CP-violating phase 
in the interference of mixing and decay (see below), obtained from the time-dependent angular analysis 
of flavor-tagged $B_s^0\rightarrow J/\psi\phi\,$ decays.
In the SM it is given by \cite{Lenz:2006hd,Ligeti:2006pm}
\beq
\displaystyle
\beta_s^{J/\psi\phi}=-\arg\left(- \frac {\lambda_{t}^{bs}}{\lambda_{c}^{bs}}\right)
= 0.020\pm0.005\,,
\eeq
with $\lambda_{q}^{bs}=V_{qb}V_{qs}^*\,$.
The SM value for the width difference reads \cite{Lenz:2011ti}
\beq
\Delta\Gamma_s^{\rm SM}=(0.087\pm 0.021)\, {\rm ps}^{-1}\,.
\eeq
 
The SM prediction for the semileptonic CP asymmetry ${(A_{\rm SL}^s)}_{\rm SM}=(1.9\pm0.3)\cdot 10^{-5}$ \cite{Lenz:2011ti}, which is 
often denoted by $a_{\rm sl}^s$ or $a_{\rm fs}^s$ in the literature, agrees with the direct measurement 
${(A_{\rm SL}^s)}_{\rm exp}=-0.0017\pm0.0092$ \cite{Asner:2010qj} within the (large) error. 
However, recent measurements of the like-sign dimuon charge asymmetry $A_{\rm SL}^b$ \cite{Abazov:2011yk}, which 
measures a combination of $A_{\rm SL}^s$ and its counterpart $A_{\rm SL}^d$ of the $B_d^0$-meson sector 
\cite{Grossman:2006ce}, imply a deviation in $A_{\rm SL}^s$ of almost $2\sigma$.

In the presence of NP contributions to $M_{12}^s$ and $\Gamma_{12}^s$, 
the observables in $B_s^0$-physics will receive modifications. 
We follow \cite{Dighe:2010nj} and extend the SM expressions according to
\begin{align}
\label{eq:MGNP}
\begin{split}
M_{12}^s&=M_{12}^{s\,\rm{SM}}+M_{12}^{s\,\rm{NP}} = M_{12}^{s\,\rm{SM}} R_M\, e^{i\phi_M}\,,\\
\Gamma_{12}^s&=\Gamma_{12}^{s\,\rm{SM}}+\Gamma_{12}^{s\,\rm{NP}} = 
\Gamma_{12}^{s\,\rm{SM}} R_\Gamma\, e^{i\phi_\Gamma}\,.
\end{split}
\end{align}
Applying these relations to (\ref{eq:DeltaG}) we arrive at the width difference in the presence of NP \cite{Grossman:1996era,Dunietz:2000cr}
\beq\label{eq:DGammaapprox}
\Delta\Gamma_s=2\,|\Gamma_{12}^{s\,\rm{SM}}|\,R_\Gamma\,\cos(\phi_s^{\rm SM}+\phi_M-\phi_\Gamma)\,,
\eeq
whereas the semileptonic CP asymmetry (\ref{eq:ASL}) becomes
\beq
A_{\rm SL}^s=\,\frac{|\Gamma_{12}^{s\,{\rm SM}}|}{|M_{12}^{s\,{\rm SM}}|}\,
\frac{R_\Gamma}{R_M}\;\sin (\phi_s^{\rm SM}+\phi_M-\phi_\Gamma)\,.
\eeq

In order to obtain predictions for these observables in RS models, a calculation of the 
corresponding corrections $R_{M,\Gamma}$ and $\Phi_{M,\Gamma}$ is necessary.
Within the SM, the leading contribution to the dispersive part of the $B_s^0$--$\bar B_s^0$ 
mixing amplitude appears at the one-loop level. If NP involves FCNCs at the tree level, 
these give rise to sizable corrections to $M_{12}^s$ and thus also to the mass 
difference $\Delta m_{B_s}$, see (\ref{eq:DMM}).
In the context of RS scenarios, the corrections to $M_{12}^s$ have been
calculated in \cite{Bauer:2009cf,Blanke:2008zb} (see also \cite{Agashe:2004ay,Agashe:2004cp} for
a first estimate).

Moreover, the presence of tree-level FCNCs and right-handed 
charged-current interactions gives rise to new decay diagrams.
However, NP corrections to the absorptive part of the amplitude are generically suppressed by 
$m_W^2/M_{\rm NP}^2$ with respect to the SM contribution, where $M_{\rm NP}$ is the NP mass scale. 
Yet, as a concrete calculation of such corrections in RS models has not been presented before, we will provide 
it in the following, for the sake of obtaining a more quantitative prediction. We will give the results both for the minimal as well as the custodial
RS variant. Beyond that, since we use an EFT approach, our results will be applicable to a more general
class of NP models. Recently, model-independent estimates on $A_{\rm SL}^s$ in the presence 
of heavy gluons have been presented in \cite{Datta:2010yq}, taking into account modifications in $\Gamma_{12}^s$. 
NP contributions from electroweak (EW) penguin operators as well as right-handed charged 
currents have not been considered. However, in RS models the former can compete with or even 
dominate contributions from QCD penguins. Moreover, part of the latter tend to 
give the dominant contribution to $\Gamma_{12}^{s\,{\rm RS}}$ for the most natural choice of 
input parameters.

\subsubsection[Calculation of $\Gamma_{12}^s$ in the Presence of New Physics]{Calculation of $\bm \Gamma_{12}^s$ in the Presence of New Physics}
\vspace{-0.4cm}
Within the SM, $\Gamma_{12}^s$ has been calculated to NLO in QCD  
\cite{Lenz:2006hd,Beneke:1996gn,Beneke:1998sy,Dighe:2001gc,Beneke:2003az,Ciuchini:2003ww,Badin:2007bv}.
It is given by the hadronic matrix element of the transition operator, 
which converts $\bar B^0_s$~into~$B^0_s$
\begin{align}
\begin{split}
\displaystyle
\Gamma_{12}^s=&\;\frac 1{2m_{B_s}}\, \langle B_s^0| {\mathcal T} | \bar B_s^0 \rangle\,,\\
{\mathcal T}=&\;\text{Disc} \int d^4x\,\frac i2\,T\,\left[
{\mathcal H}_{\rm eff}^{\Delta B=1}(x) {\mathcal H}_{\rm eff}^{\Delta B=1}(0)\right]\,.
\end{split}
\end{align}
Taking the discontinuity in the expression above projects
out those intermediate states, that are on-shell. 
A systematic evaluation of the matrix element in powers of $1/m_b$ is 
possible in the framework of the heavy-quark expansion (HQE),
assuming local quark-hadron duality (for a review and references see \cite{Neubert:1997gu}).
At zeroth order, the momentum of the $B$-meson in its rest frame is equal to the 
momentum of the bottom quark, while the strange-quark momentum is set to zero. 
Long-lived intermediate states (with respect to hadronic scales) would normally jeopardize a short-distance treatment
of the transition amplitude. For $B_q^0$-mesons, however, the
bottom quark mass corresponds to an additional short-distance scale,
leading to a large energy transfer into the intermediate states.
Thus, at typical hadronic distances $x>1/m_b$, the transition of
$\bar B^0_s$ into $B^0_s$ is again a local process \cite{Beneke:1996gn}
and the matrix element can be expanded in terms of local $\Delta B=2$ operators.

We derive the leading contribution to $\Gamma_{12}^s$ in the 
presence of NP. The most important corrections to the SM result are given by the interference 
of SM and NP insertions. QCD corrections are implemented by evolving the Wilson coefficients of the 
$\Delta B=1$ operators from the matching scale down to $m_b$.
The leading SM contributions can be written as matrix elements of the $\Delta B=2$~operators\\[-3mm]
\begin{align}\label{eq:Q1Q2}
\begin{split}
\Q_1=&\,(\bar s_i b_i)_{\small V-A}(\bar s_j b_j)_{\small V-A}\,,\\
\Q_2=&\,(\bar s_i b_i)_{\small S+P}(\bar s_j b_j)_{\small S+P}\,,
\end{split}
\end{align}
where $i$ and $j$ denote color indices (that are summed over). The notation $V\pm A$ denotes 
the Dirac structure $\gamma^\mu(1 \pm \gamma^5)$ in between the spinors, whereas $S\pm P$ corresponds to 
$(1\pm \gamma^5)$.  
Right-handed charged currents, which occur in RS models, bring about the necessity of
further $\Delta B=2$ operators, \vspace{-4mm}
\begin{align}\label{eq:Q3Q4Q5}
\begin{split}
\Q_3=&\,(\bar s_i b_j)_{\small S+P}(\bar s_j b_i)_{\small S+P}\,,\\
\Q_4=&\,(\bar s_i b_i)_{\small S-P}(\bar s_j b_j)_{\small S+P}\,,\\
\Q_5=&\,(\bar s_i b_j)_{\small S-P}(\bar s_j b_i)_{\small S+P}\,,
\end{split}
\end{align}
due to the interference of SM with NP insertions.
The appropriate $\Delta B=1$ Hamiltonian, allowing for new right-handed charged currents
as well as FCNCs, is given by  
\beq\label{eq:Heff}
\begin{split}
{\mathcal H}_{\rm eff}^{\Delta B=1}=\,\frac {G_F}{\sqrt 2}\,\lambda_{c}^{bs}
&\,\Big{[}\sum_{i=1,2}\Big{(}C_iQ_i  \,+ \,C_i^{LL}Q_i\,
+\,C_i^{LR}Q_i^{LR}\,+\, C_i^{RL}Q_i^{RL}\Big{)}
+\sum_{i=3}^{10} C_i Q_i \Big{]}\\
&+\sum_{i=3}^{10}\left(C_i^{\rm NP} Q_i + \widetilde C_i^{\rm NP}\widetilde Q_i\right).
\end{split}
\eeq
The operators
\begin{align}\label{eq:Qcharge}
\begin{split}
Q_1&=(\bar s_i c_j)_{\small V-A}(\bar c_j b_i)_{\small V-A}\,,\\
Q_2&=(\bar s_i c_i)_{\small V-A}(\bar c_j b_j)_{\small V-A}\,,\\
Q_1^{LR}&=(\bar s_i c_j)_{\small V-A}(\bar c_j b_i)_{\small V+A}\,,\\
Q_2^{LR}&=(\bar s_i c_i)_{\small V-A}(\bar c_j b_j)_{\small V+A}\,,\\
\end{split}
\end{align}
arise from (KK) $W^\pm$-boson exchange ($Q_{1,2}$), as well as from
right handed charged currents ($Q_{1,2}^{LR}$) which correspond to a new structure - compared to the SM -
see Section~\ref{sec:4Fint}. 
The operators $Q_i^{RL}$ are chirality flipped with respect to $Q_i^{LR}\,$. 
We do not include operators of the type $RR$ as their coefficients scale like 
$v^4/\Mkk^4$ in the models at hand. Due to the hierarchies in the CKM matrix and
the RS-GIM mechanism, it is sufficient to restrict ourselves on $c$ quarks 
as intermediate states, when calculating RS corrections involving the charged current-sector. 
For the SM contribution however, we will include all combinations $uc$, $cu$, and $uu$ in our analysis, 
in addition to the operators given above. Concerning the NP corrections $LL,LR,RL$, we factor out the CKM factor 
$\lambda_{c}^{bs}$ only for conven\-ience. 

As explained in Section~\ref{sec:4Fint}, the experimentally determined values for $V_{cb}$ and $V_{cs}$
from semileptonic $B$ and $D$ decays should be identified with the exchange of all $SU(2)_L$ gauge bosons.
In turn, the NP coefficients $C_{1,2}^{LL}$ arise only due to the non-factorizable corrections in (\ref{eq:LL}), 
which can not be absorbed into $\lambda_{c}^{bs}$. In addition to the operators discussed so far, there are also
QCD penguin operators
\begin{align}\label{eq:QCDpeng}
\begin{split}
Q_3=&\, (\bar s_i b_i)_{\small V-A} {\sum}_q\,(\bar q_j q_j)_{\small V-A}\,,\\
Q_4=&\, (\bar s_i b_j)_{\small V-A} {\sum}_q\,(\bar q_j q_i)_{\small V-A}\,,\\ 
Q_5=&\, (\bar s_i b_i)_{\small V-A} {\sum}_q\,(\bar q_j q_j)_{\small V+A}\,,\\
Q_6=&\, (\bar s_i b_j)_{\small V-A} {\sum}_q\,(\bar q_j q_i)_{\small V+A}\,,
\end{split}
\end{align}
and EW penguin operators
\begin{align}\label{eq:EWpeng}
\begin{split}
\displaystyle
Q_7=&\,\frac 32\,(\bar s_i b_i)_{\small V-A} {\sum}_q Q_q\,(\bar q_j q_j)_{\small V+A}\,,\\
Q_8=&\,\frac 32\,(\bar s_i b_j)_{\small V-A} {\sum}_q Q_q\,(\bar q_j q_i)_{\small V+A}\,,\\
Q_9=&\,\frac 32\,(\bar s_i b_i)_{\small V-A} {\sum}_q Q_q\,(\bar q_j q_j)_{\small V-A}\,,\\
Q_{10}=&\,\frac 32\,(\bar s_i b_j)_{\small V-A} {\sum}_q Q_q\,(\bar q_j q_i)_{\small V-A}\,,
\end{split}
\end{align}
with $q=u,c,d,s$.

Here, no CKM factors are involved and all light quarks have to be kept as intermediate states,
when considering neutral-current insertions only.
Finally, there exist also chirality-flipped operators with respect to (\ref{eq:QCDpeng}) and 
(\ref{eq:EWpeng}), denoted by $\widetilde Q_{3..10}$.
Note that the possibility of a flavor change on both vertices for NP penguins,
as well as the dependence of the Wilson coefficients on the quark flavor $q$,
can be safely neglected due to an additional RS-GIM suppression.
For the same reason the chirality flipped penguins $\widetilde C_{3..10}^{\rm RS}$ can be neglected
compared to $C_{3..10}^{\rm RS}$ for $bs$ transitions \cite{Bauer:2009cf}.
 
Despite of the $\alpha/\alpha_s$-suppression, the EW penguin operators in the minimal RS model
can dominate over the gluon penguins \cite{Agashe:2004ay,Agashe:2004cp,Bauer:2009cf}.
This is due to the enhancement of the leading correction to the left-handed $Z$-coupling
by the RS-volume $L$, see Section~\ref{sec:KKsum}. Note that this is not the case in the custodial 
RS variant featuring a protection for the $Zb_L\bar b_L$ vertex. 
The RS Wilson coefficients for the penguin operators can be obtained from Section~\ref{sec:KKsum} and have been worked out in 
\cite{Bauer:2009cf}. We give them in Appendix~\ref{app:WilsonsP} for completeness.   
There further is the possibility of flavor-changing Higgs couplings which,
however, can be neglected compared to the contributions of flavor-changing heavy gauge bosons in RS models.

In double-penguin insertions, we include all light quarks with masses set to zero, besides for~$m_c$.
Double penguins also allow for leptons within the cut-diagram. However, as the related SM coefficient is suppressed 
by $\alpha/\alpha_s$, it is not possible to obtain large effects from $\bar sb\rightarrow\bar \tau \tau$
transitions, which are less constrained by experiment \cite{Bauer:2010dga}. 
Note that this conclusion can be evaded, if this transition is mediated 
at tree level by light NP particles in the range of $\sim 100\,$GeV. In that case, the double NP insertion can
become comparable to the SM diagrams \cite{Alok:2010ij}.
Possible candidates are scalar leptoquarks \cite{Dighe:2010nj,Dighe:2007gt}.
Given the loose bounds imposed by existing tree- and loop-level mediated $B^0_{d,s}$-meson decays, 
it has been shown recently \cite{Bobeth:2011st}, that the presence of a single
$(\bar s b)(\bar \tau \tau)$ operator can lead model-independently to an enhancement of $\Gamma_{12}^s$ 
of maximally 40\% compared to the SM value.

Neglecting intermediate leptons, we find to LO in the HQE
\begin{align}
\label{eq:gamma12}
\Gamma_{12}^s=-&\,\frac{m_b^2}{12\pi (2 M_{B_s})} \,G_F^2{(\lambda_{c}^{bs})}^2\,\sqrt{1-4z}\nonumber \\
\Bigg{\lbrace}&\Big{[}(1-z)(\Sigma_1+\Sigma_1^{LL})
  +\frac 12(1-4z)(\Sigma_2+\Sigma_2^{LL})\, +3z \,(\Sigma_3\,+K_3^{' LL})\nonumber \\
 &\ -\,\frac 32\,\sqrt z\, (\Sigma_1^{LR} +\Sigma_2^{LR}+K_3^{'LR}+K_4^{'LR})\nonumber \\
&\  +\frac 1{\sqrt{1-4z}}\Big{(}(3\bar K''_1 +K''_{s1} +\frac 32 \bar K''_2 + \frac 12 K''_{s2})
  \, +\, \frac{\lambda_u^{bs}}{\lambda_c^{bs}}\,(1-z)^2\big{(}(2+z)K_1+(1-z)K_2\big{)}\nonumber \\
   &\  +\frac 12 \frac{{(\lambda_u^{bs})}^2}{{(\lambda_c^{bs})}^2}\,(2K_1+K_2) \Big{)}\Big{]}\,
           \langle{\Q_1}\rangle\nonumber \\
 +&\, \Big{[}(1+2z)(\,\Sigma_1+\Sigma_1^{LL}-\,\Sigma_2-\Sigma_2^{LL})
  \,-\,3\,\sqrt z\,(2\Sigma_1^{LR} +\Sigma_2^{LR} -K_4^{'LR}) \nonumber \\
&\  +\frac 1{\sqrt{1-4z}}\Big{(}(3\bar K''_1 +K''_{s1}-3\bar K''_2-K''_{s2})\nonumber \\
  &\  +\,2\, \frac{\lambda_u^{bs}}{\lambda_c^{bs}}\,(1-z)^2(1+2z)(K_1-K_2)
           +\,\frac{{(\lambda_u^{bs})}^2}{{(\lambda_c^{bs})}^2}(K_1-K_2)\Big{)}  \Big{]}\,
   \langle{\Q_2}\rangle \nonumber \\
 -&\;3\,\sqrt z\,(\Sigma_1^{LR} +2\Sigma_2^{LR}+K_3^{'LR})\,\langle \Q_3 \rangle \;
   +\, 3\sqrt z\,(\Sigma_1^{RL}-K_3^{'RL})\,\langle\Q_4\rangle
\; +\,3\sqrt z\,(\Sigma_2^{RL}-K_4^{'RL})\,\langle\Q_5\rangle \,\Bigg{\rbrace} \nonumber \\
-&\frac{m_b^2}{12\pi (2 M_{B_s})}\,\sqrt 2\,G_F\lambda_{c}^{bs}\,\sqrt{1-4z}\nonumber \\
\Bigg{\lbrace}&\Big{[}(1-z)\,\Sigma_1^{\rm NP}
+\frac 12(1-4z)\,\Sigma_2^{\rm NP} +\,3z \,\Sigma_3^{\rm NP}\nonumber \\
&\ + \frac 1{\sqrt{1-4z}}(3\bar K_1^{''\rm NP} +K_{s1}^{''\rm NP}
+ \frac 32 \bar K_2^{''\rm NP} + \frac 12 K_{s2}^{''\rm NP})\Big{]}\,\langle{\Q_1}\rangle\nonumber \\
&\,+\Big{[}(1+2z)(\,\Sigma_1^{\rm NP}-\,\Sigma_2^{\rm NP})\nonumber \\
&\ + \frac 1{\sqrt{1-4z}}(3\bar K_1^{''\rm NP}+K_{s1}^{''\rm NP}
 -3\bar K_2^{''\rm NP}-K_{s2}^{''\rm NP})\Big{]}\,\langle{\Q_2}\rangle
\,+\;\ord\left(\frac 1{m_b}\right)\Bigg{\rbrace}\,,
\end{align}

where $z=m_c^2/m_b^2\,$ and 
$\langle\Q\rangle\,\equiv\,\langle B_s^0|\, \Q\,|\bar B_s^0\rangle$.
In order to get a compact result, we have defined the linear combinations 
($A,B \in \lbrace L,R \rbrace$)
\begin{align}\label{eq:coefsum}
\begin{split}
\Sigma_i=&\,K_i+K'_i+K''_i\, , \\
\Sigma_i^{AB}=&\,K_i^{AB}+K_i^{' AB},\ \ i=1,2,\\
\Sigma_3=&\,K'_3+K''_3\,,\\
\Sigma_i^{\rm NP}=&\,K_i^{'\rm NP}+K_i^{''\rm NP}\,\ \ i=1,2,3\,,
\end{split}
\end{align}
where the coefficients on the right-hand side of (\ref{eq:coefsum}) are again
linear combinations of Wilson coefficients. 
In agreement with \cite{Beneke:1996gn} we get ($C_{i+j}\equiv C_i+C_j$)
\beq
\begin{split}
K_1=&\,N_cC_1^2+2\,C_1C_2\,,\quad K_2=\,C_2^2\,,\\
K'_1=&\,2\,(N_cC_1C_{3+9}+C_1C_{4+10}+C_2C_{3+9})\,,\\
K'_2=&\,2\,C_2C_{4+10}\,,\\
K'_3=&\,2\,(N_cC_1C_{5+7}+C_1C_{6+8}+C_2C_{5+7}+C_2C_{6+8})\,,\\
K''_1=&\,N_cC_{3+9}^2+2\,C_{3+9}C_{4+10}
      +N_cC_{5+7}^2+2\,C_{5+7}C_{6+8}\,,\\
K''_2=&\,C_{4+10}^2+C_{6+8}^2\,,\\ 
K''_3=&\,2(N_c C_{3+9}C_{5+7}+C_{3+9}C_{6+8}
      +C_{4+10}C_{5+7}+C_{4+10}C_{6+8})\,.
\end{split}
\eeq 
The combinations $K_i$ are due to insertions of charged-current operators
and are responsible for the dominant contribution in the SM.
The coefficients $K'_i$ and $K''_i$ correspond to the interference of charged-current operators
with penguin operators and penguin-penguin insertions, respectively.
As we consider light quarks ($q=u,d,s\,$) in
the limit $m_q=0$, there is a cancellation in the EW penguin sector
due to the electric charges. The coefficients $\bar K_i^{''}$ therefore resemble 
the $K_i^{''}$, with $C_{7..10}$ set to zero.
For strange quarks as intermediate states, there is a second possibility for
the penguin insertion. In the limit $m_s=0$, there are additional contributions from
\beq
\begin{split}
K''_{s1}=&\,(2+N_c)(C_4- C_{10}/2)^2 
     + 2\,(N_c+1)(C_3-C_9/2)(C_4-C_{10}/2) 
     + 2\,(C_3-C_9/2)^2\,,\\
K''_{s2}=&\, 2\,(C_3-C_9/2)(C_4-C_{10}/2) + (C_3-C_9/2)^2 \,.
\end{split}
\eeq
Note that these terms have not been taken into account in \cite{Beneke:1996gn}.
On the other hand, as all double-penguin insertions are numerically suppressed, this omission 
has no significant effect.

Next we come to the interference of SM diagrams with NP penguins, which is 
collected in
\beq
\begin{split}
K_1^{'\rm NP}=&\,2\,(N_cC_1C_{3+9}^{\rm NP}+C_1C_{4+10}^{\rm NP}+C_2C_{3+9}^{\rm NP})\,,\\
K_2^{'\rm NP}=&\,2\,C_2C_{4+10}^{\rm NP}\,,\\
K_3^{'\rm NP}=&\,2\,(N_cC_1C_{5+7}^{\rm NP}+C_1C_{6+8}^{\rm NP}+C_2C_{5+7}^{\rm NP}+C_2C_{6+8}^{\rm NP})\,,\\
K_{s1}^{''\rm NP}=&\,2\,\big{(}(N_c+2) C_4(C_4^{\rm NP}-C_{10}^{\rm NP}/2)
                  + (N_c+1) C_4(C_3^{\rm NP}-C_9^{\rm NP}/2)\\
                 &\ + (N_c+1) C_3(C_4^{\rm NP}-C_{10}^{\rm NP}/2)
		   + 2 C_3(C_3^{\rm NP}-C_9^{\rm NP}/2) \big{)}\,,\\
K_{s2}^{''\rm NP}=&\,2\,\big{(}C_3(C_3^{\rm NP}-C_9^{\rm NP}/2)+C_3(C_4^{\rm NP}-C_{10}^{\rm NP}/2)
			   +C_4(C_3^{\rm NP}-C_9^{\rm NP}/2)\big{)}\,
\end{split}
\eeq 
and 
\beq
\begin{split}
K_1^{''\rm NP}=&\,2\,(N_cC_3C_{3+9}^{\rm NP}+C_3C_{4+10}^{\rm NP}+C_4C_{3+9}^{\rm NP}
     +N_cC_5C_{5+7}^{\rm NP}+C_5C_{6+8}^{\rm NP}+C_6C_{5+7}^{\rm NP})\,,\\
K_2^{''\rm NP}=&\,2\,(C_4C_{4+10}^{\rm NP}+C_6C_{6+8}^{\rm NP})\,,\\
K_3^{''\rm NP}=&\,2\,(N_c C_3C_{5+7}^{\rm NP}
               +C_3C_{6+8}^{\rm NP}+C_4C_{5+7}^{\rm NP}+C_4C_{6+8}^{\rm NP}\\
               &\,+N_c C_5C_{3+9}^{\rm NP}+C_5C_{4+10}^{\rm NP}+C_6C_{3+9}^{\rm NP}+C_6C_{4+10}^{\rm NP})\,.
\end{split}
\eeq
Here, we have neglected the tiny contributions from the interference of SM EW penguins with NP graphs. 
In addition, there are contributions from the interference of NP charged currents with SM penguins
\beq
\begin{split}
K_1^{'LL}=&\,2\,(N_c C_3C_1^{LL}+C_3C_2^{LL}+C_4C_1^{LL})\,,\\
K_2^{'LL}=&\,2\,C_4C_2^{LL} \,,\\
K_3^{'LL}=&\,2\,(N_c C_5C_1^{LL}+C_5C_2^{LL}+C_6C_1^{LL}+C_6C_2^{LL})\,,\\
K_1^{'LR}=&\,2\,(N_c C_3C_1^{LR}+C_3C_2^{LR}+C_4C_1^{LR})\,,\\
K_2^{'LR}=&\,2\,C_4C_2^{LR}\,,\\
K_3^{'LR}=&\,2\,(N_c C_5C_1^{LR}+C_5C_2^{LR}+C_6C_1^{LR})\,,\\
K_4^{'LR}=&\,2\,C_6C_2^{LR}\,.
\end{split}
\eeq
The corrections to the purely charged-current interactions are given by
\beq
\begin{split}
K_1^{LL}=&\,2\,\big{(}N_cC_1C_1^{LL}+C_1C_2^{LL}+C_2C_1^{LL}\big{)}\,,\\\
K_2^{LL}=&\,2\,C_2C_2^{LL}\,,\\
K_1^{LR}=&\,2\,(N_cC_1C_1^{LR}+C_1C_2^{LR}+C_2C_1^{LR})\,,\\
K_2^{LR}=&\,2\,C_2C_2^{LR}\,.
\end{split}
\eeq
The coefficients $K_i^{(')RL}$ resemble $K_i^{(')LR}$, with $C_i^{LR}$
replaced by $C_i^{RL}$.
In oder to arrive at the form (\ref{eq:gamma12}) we have used several chiral Fierz identities \cite{fierz}.
A nice method to obtain them in an easy way is presented in \cite{Nishi:2004st}.
All NP coefficients should by calculated at the NP mass scale and then evolved
down to $m_b$. Explicit expressions for the Wilson coefficients of the minimal as well as the custodial RS model
are given in the appendices \ref{app:WilsonsP} and \ref{app:charged}. 

The dispersive part of the mixing amplitude has been calculated for RS models in the literature \cite{Bauer:2009cf,Blanke:2008zb}.
The required effective Hamiltonian reads
\beq
 {\mathcal H}_{\rm eff}^{\Delta B=2}=\sum_{i=1}^5 C_i \Q_i+ \sum_{i=1}^3 \widetilde C_i \widetilde\Q_i\,.
\eeq
Note that there are no tree-level contributions to $C_{2,3}$ and $\widetilde C_{2,3}$ in the RS model. 
The RS correction to 
\beq\label{eq:M12me}
2\, m_{B_s} M_{12}^s=\, \langle B_s^0|\, {\mathcal H}_{\rm eff}^{\Delta B=2}\, | \bar B_s^0 \rangle\,
\eeq
is given by \cite{Bauer:2009cf,Blanke:2008zb}
\beq\label{eq:M12RS}
M_{12}^{s\,{\rm RS}}=\, \frac 43\, m_{B_s}  f_{B_s}^2 \Big{[} 
\left(C_1^{\rm RS}(\bar m_b)+\widetilde C_1^{\rm RS}(\bar m_b)\right) B_1 
+\frac 34\, R(\bar m_b)\,C_4^{\rm RS}(\bar m_b)B_4
+\frac 14\,R(\bar m_b)\,C_5^{\rm RS}(\bar m_b)B_5\Big{]}\,.
\eeq
The bag parameters $B_{1,4,5}$, related to hadronic matrix elements evaluated on the lattice, 
are listed in (\ref{eq:bagfac}). The $\Delta B=2$ Wilson coefficients 
can be found in Appendix~\ref{app:mixing}.
Compared to $C_1^{\rm RS}(\bar m_b)$, the coefficient $C_4^{\rm RS}(\bar m_b)$ is suppressed by 
about two orders of magnitude due to a stronger RS-GIM mechanism. The coefficients 
$\widetilde C_1^{\rm RS}(\bar m_b)$ and $C_5^{\rm RS}(\bar m_b)$ are even further suppressed.  
The SM mixing amplitude can be obtained from \cite{Buras:1998raa,Lenz:2006hd,Buras:1990fn}
and reads
\beq\label{eq:M12SM}
\displaystyle
M_{12}^{s\,{\rm SM}}=\frac{G_F^2}{12\,\pi^2}\,
{(\lambda^{bs}_t)}^2 m_W^2 m_{B_s} \eta_B f_{B_s}^2 B_1 S_0(x_t)\,.
\eeq
Here $\eta_B=0.837$ includes NLO QCD corrections in naive dimensional reduction (NDR).
$S_0(x_t)$ is the Inami-Lim function and $x_t=\bar m_t(\bar m_t)^2/m_W^2$ with $\bar m_t(\bar m_t)=(163.8\pm2.0)\,$GeV. 
We use $m_{B_s}=5.366(1)\,$GeV \cite{Nakamura:2010zzi} and 
$f_{B_s}=(238.8\pm9.5)\,$MeV \cite{Laiho:2009eu} for the $B_s^0$ meson mass and decay constant, respectively.
If not stated otherwise, all other experimental input for this section is taken from \cite{Nakamura:2010zzi}.

\subsubsection{Numerical Analysis for RS Models and Discussion}\label{sec:num}

We now present our numerical predictions for the RS setup, based on the anarchic parameter sets
described at the beginning of this Chapter.
In Table \ref{tab:coef}, we give the contributions of selected individual ingredients of $\Gamma_{12}^s$ 
(\ref{eq:gamma12}). The SM coefficients are taken from \cite{Buchalla:1995vs}. 
As they are not supplemented with a CKM factor in (\ref{eq:Heff}),
we rescale the RS penguin coefficients - for instance 
$\tilde K_2^{'\rm RS}\equiv \sqrt 2\,{(G_F \lambda_c^{bs})}^{-1} K_2^{'\rm RS}$ 
(SM: $\tilde K'_2=K'_2$) - to allow for an easier comparison.
The mean absolute values of our RS predictions should be compared to the corresponding 
SM results, where the numbers have to be multiplied by the order of magnitude 
given in the last column. 
The maximum values exceed the given numbers by at least one order of magnitude, 
as suggested by the large standard deviations.
We have set the NP scale to $\Mkk=2\,$TeV and have discarded all points, which are in conflict
with the $Z\rightarrow b\bar b$ pseudo observables. 
For the purpose of this analysis, we reject all points lying outside the $95\%$ confidence region 
in the $g_L^b-g_R^b$ plane.
For the minimal RS model this sets a quite stringent upper limit on $c_{b_L}$, 
see Section~\ref{sec:bpseudo}. Within the custodial RS variant with a protection for the $Z b_L\bar b_L$-vertex, the respective 
bound vanishes.

\begin{table}
\begin{center}
\begin{tabular}{|c|c|c|c|c|c|c|}
  \hline
  Model\slash Coef.  & $\, |\tilde K'_2|\,$ & $\, |\tilde K''_2|\,$  
                     & $|K_2^{(LL)}|$ & $|K_2^{LR}|$ & $|K_2^{RL}|$ & $\times$ \\
  \hline
  SM  & $0.543$ & $0.016$ & $12.656$ & - & - & $10^{-1}$ \\
  \hline
  mean (min RS)  & $0.16$ & $0.03$ & $0.01$ & $4.40$ & $0.04$ & $10^{-3}$\\
  stand. dev.  & $0.17$ & $0.03$ & $0.05$ & $7.41$ & $0.06$ & $10^{-3}$\\
  \hline
  mean (cust)  & $0.94$ & $0.06$ & $0.23$ & $2.22$ & $0.03$ & $10^{-3}$ \\
  stand. dev.  & $1.39$ & $0.09$ & $1.38$ & $4.98$ & $0.05$ & $10^{-3}$ \\
  \hline
\end{tabular}
\end{center}
\caption{\label{tab:coef}  Selected SM-penguin and charged-current 
  coefficients contributing to $\Gamma_{12}^s$ compared to 
  the mean absolute values of the corresponding RS coefficients for 
  $\Mkk=2\,$TeV and $\mu=\bar m_b\,$. See text for details. }
\end{table}

Neglecting experimental constraints, there is no difference between the minimal and the custodial 
RS variant at LO in $v^2/\Mkk^2$ in the charged current sector (see Appendix~\ref{app:charged}). 
For the natural assumption of $c_{Q_2}<-1/2\,$, see Figure~\ref{fig:cs}, the biggest correction stems from the operator $Q_2^{LR}$. 
This is easy to understand if we apply the Froggatt-Nielsen analysis of Section~\ref{sec:hierarchies}
to (\ref{eq:CLL}) and (\ref{eq:CLRCRL}). 
Setting all Yukawa factors to one and performing an expansion in the Wolfenstein parameter
$\lambda$, we find as a crude approximation
\begin{align}
\displaystyle
C_2^{LL}\propto\;& \frac{m_W^2}{2\Mkk^2}\, L\;F(c_{Q_2})^2 F(c_{Q_3})^2\,,\\
C_2^{LR}\propto\;& \frac{v^2}{2\Mkk^2}\,\frac{F(c_{Q_3})}{F(c_{Q_2})}\, F(c_{u_2}) F(c_{d_3})
\propto \frac{m_c m_b}{\Mkk^2}\,\frac 1{F(c_{Q_2})^2} \,,\nonumber\\
C_2^{RL}\propto\;&\frac{v^2}{\Mkk^2}\,F(c_{u_2}) F(c_{d_2})\propto \frac{2\,m_c m_s}{\Mkk^2}
\frac 1{F(c_{Q_2})^2} \nonumber\,.
\end{align}
Note that the importance of $C_2^{LR}$ grows with increasing {\it UV-localization} of the $(c_L,s_L)^T$ doublet,
which explains its rather large size. This effect is caused by fermion mixing. Technically, the chiral suppression
in small masses is lifted by the inverse zero-mode profiles (which are strongly suppressed) entering $C_2^{LR}$.
The same holds true for $C_2^{RL}$, which is however smaller by a factor of $m_s/m_b$.
The coefficients $C_1^{AB}$ with $A,B \in \lbrace L,R \rbrace$ are zero at the matching scale, but generated
through operator mixing in the evolution down to $\mu=\bar m_b$.
At $\mu=\bar m_b$, the values of $|K_1^{AB}|$ amount to about a third of the corresponding values of $|K_2^{AB}|$ .
In the RS model the contributions from the coefficients $C_i^{LL}$ and $C_i^{RL}$ can be safely neglected, just as those
of the chirality flipped penguins.

The coefficients $K_i^{'{\rm RS}}$ and $K_i^{''{\rm RS}}$ grow with an increasing IR localization value of 
$c_{b_L}$ and $ c_{s_L} \equiv c_{Q_2}$.
The reason is, that the RS corrections due to penguin operators are dominated by overlap 
integrals of left-handed fermions with intermediate
KK-gauge bosons and mixing effects of the latter with the $Z$-boson.
The corresponding RS expressions are given in (\ref{eq:Cpenguin}). As KK modes are 
peaked towards the IR brane, overlap integrals with UV localized fermions 
are exponentially suppressed and the RS-GIM mechanism is at work. 
As mentioned before, the leading correction due to $Z$ exchange is enhanced by a factor $L$ within the minimal RS variant.
Nevertheless, because of the stringent bounds from $Zb\bar b$, the total penguin contributions remain 
smaller than in the custodial model.
In both RS setups, it is sufficient to consider only the contributions stemming from
the coefficients $K_i^{'{\rm NP}}$ in the neutral-current sector.
The impact of double penguin contributions amounts typically to 
about $1\%$ of the leading correction due to charged currents. 

In order to arrive at a global picture, we have to evaluate the whole expressions (\ref{eq:gamma12})
and (\ref{eq:M12RS}).
In terms of
\begin{equation}
R(\mu)\equiv {\left(\frac{M_{B_s}}{\bar{m}_b(\mu)+\bar{m}_s(\mu)}\right)}^2\,,
\end{equation}
the matrix elements are given by
\begin{align}\label{eq:bag}
\begin{split}
\langle\Q_1\rangle&\,=\,\frac 83\, M_{B_s}^2 f_{B_s}^2 B_1(\mu)\,,\\
\langle\Q_2\rangle&\,=-\frac 53\, M_{B_s}^2 f_{B_s}^2 \,R(\mu)\, B_2(\mu)\,,\\
\langle\Q_3\rangle&\,=\,\frac 13\, M_{B_s}^2 f_{B_s}^2 R(\mu) B_3(\mu)\,,\\
\langle\Q_4\rangle&\,=\,2\, M_{B_s}^2 f_{B_s}^2 R(\mu) B_4(\mu)\,,\\
\langle\Q_5\rangle&\,=\,\frac 23\, M_{B_s}^2 f_{B_s}^2 R(\mu) B_5(\mu)\,.
\end{split}
\end{align}
The bag parameters $B_i$ are obtained from the lattice. We take the values of
\cite{Becirevic:2001xt} in the NDR-$\overline{\rm MS}$ scheme of \cite{Beneke:1998sy}.
They read
\begin{align}\label{eq:bagfac}
& B_1=0.87(2)\left(^{+5}_{-4}\right),\ \ B_2=0.84(2)(4),\ \ B_3=0.91(3)(8),\nonumber \\
& B_4=1.16(2)\left(^{+5}_{-7}\right),\ \ B_5=\,1.75(3)\left(^{+21}_{-6}\right),
\end{align}
where the first (second) number in brackets corresponds to the statistical (systematic) error. 
In order to resum large logarithms we employ $\bar z=\bar m_c^2(\bar m_b)/\bar m_b^2(\bar m_b)=0.048(4)$ 
\cite{Lenz:2006hd} in our numerical analysis. We use $\bar m_b(\bar m_b) = (4.22 \pm 0.08)$\,GeV 
and $\bar m_s(\bar m_b) = (0.085 \pm 0.017)$\,GeV for the quark masses in the $\overline{\rm MS}$ scheme.

\begin{figure}[t!]
\begin{center}
\includegraphics[width=6.5cm]{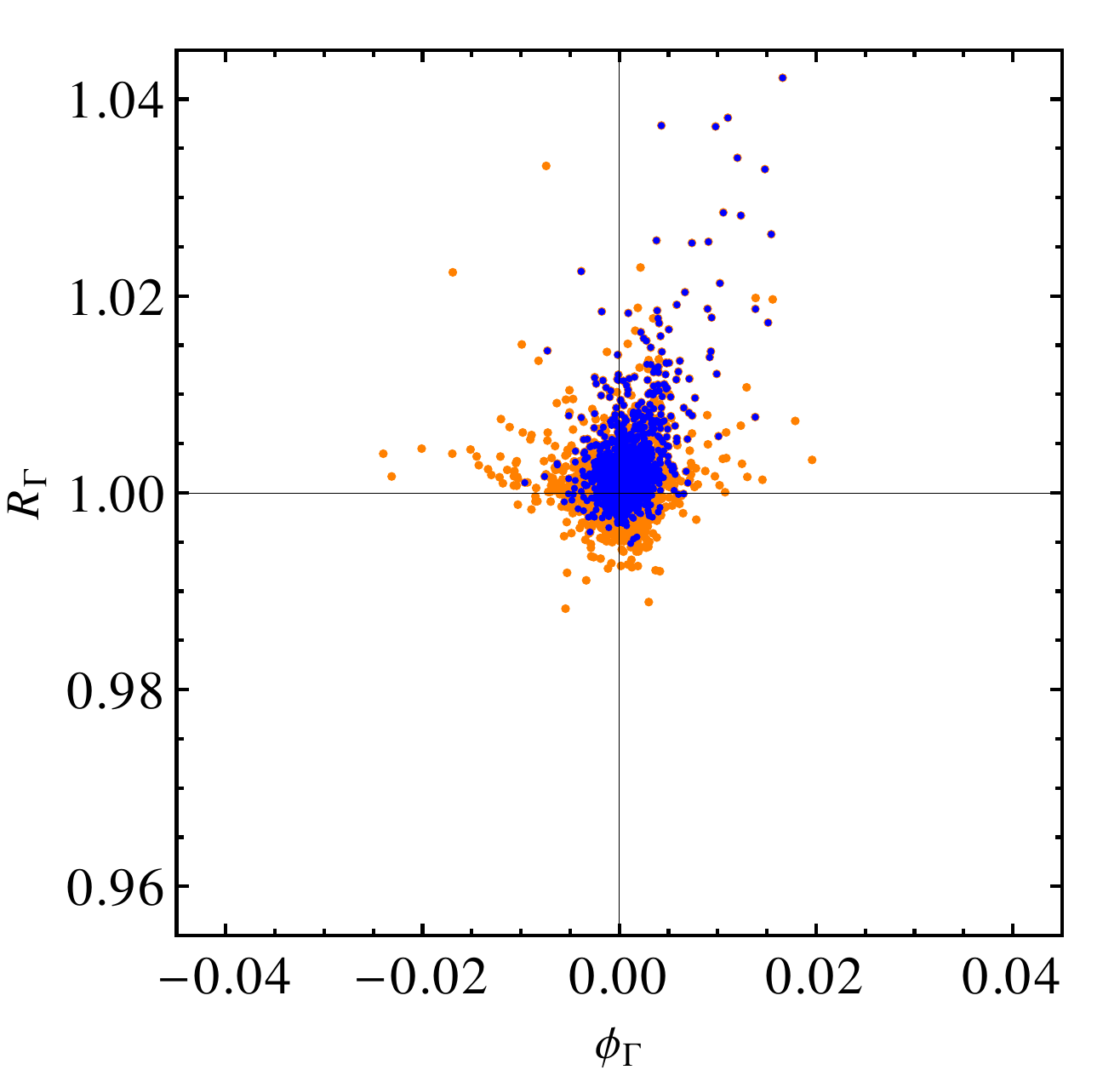}\quad
\includegraphics[width=6.5cm]{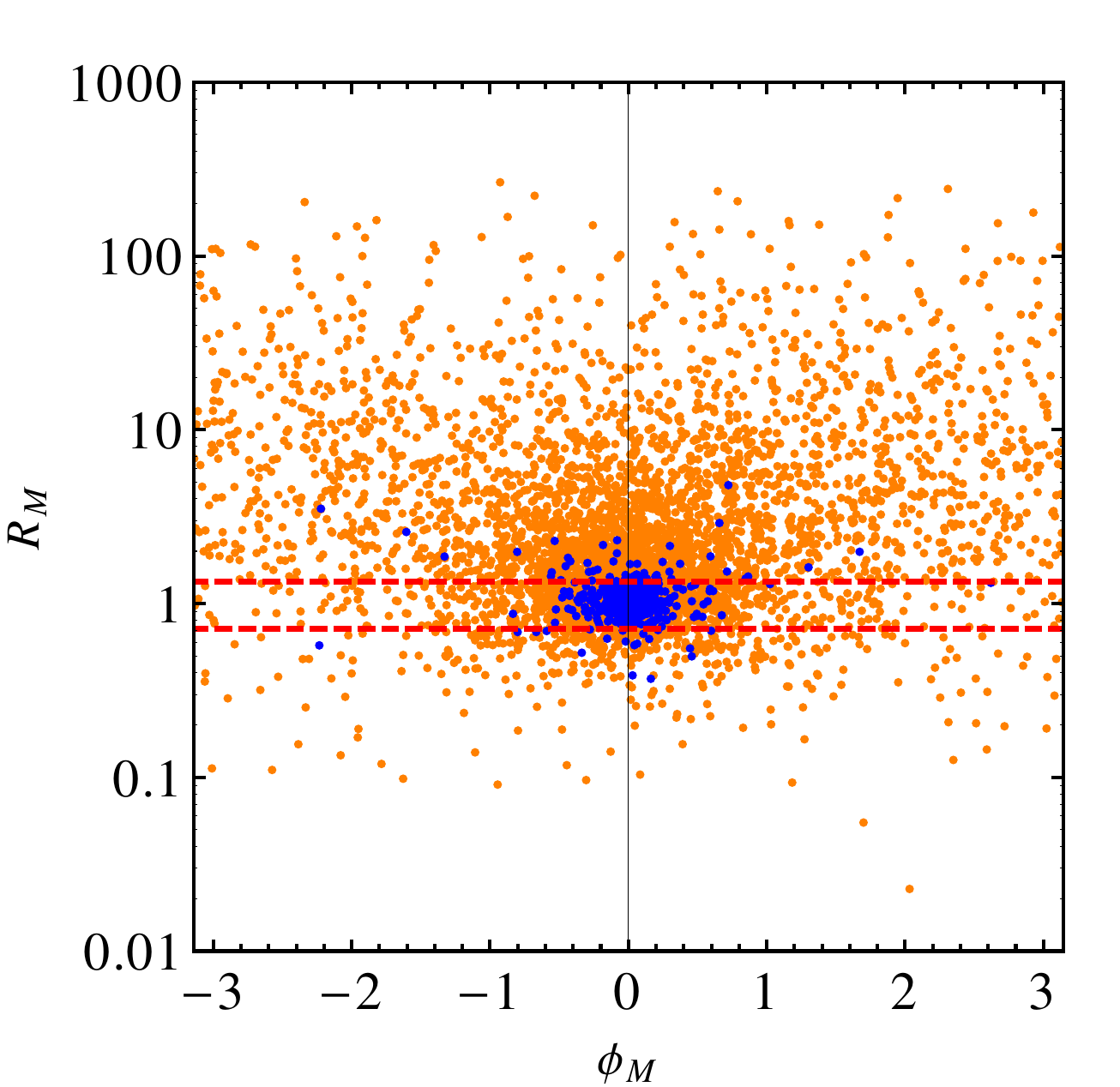}
\end{center}
\begin{center}
  \parbox{15.5cm}{\caption{\label{fig:RS} 
  RS corrections to the magnitude and CP-violating phase of the $\bar B_s^0$--$B_s^0$ decay amplitude, 
  $R_\Gamma$ and $\phi_\Gamma$ (left), as well as of the dispersive part of the mixing amplitude, 
  $R_M$ and $\phi_M$ (right). Blue (dark gray) points correspond to the minimal, orange (light gray) to the 
  custodial RS model. The dashed lines indicate the $95\%$ confidence region
  from the measurement of $\Delta m_{B_s}$. See \cite{Goertz:2011nx} and text for details.}}
\end{center}
\end{figure}

The left panel of Figure~\ref{fig:RS} shows the RS corrections to the magnitude 
and CP-violating phase of the $\bar B_s^0$--$B_s^0$ decay width, $R_\Gamma$ and $\phi_\Gamma$, for the 
10000 parameter sets, generated as described before. For the analysis of this section we do not vary the 
KK scale but rather chose $\Mkk=2\,$TeV. The blue (dark gray) points correspond 
to the minimal RS model, where only points that are in agreement with the 
$Z\rightarrow b\bar b$ pseudo observables are shown. 
The orange (light gray) points correspond to the custodial extension, where the latter constraint
effectively vanishes. 
As we are just interested in the approximate size of RS corrections, we work with the LO SM expressions. 
However, for precise predictions for a certain parameter point, one should include the full NLO corrections 
to $\Gamma_{12}^s$ and $M_{12}^s$.
As expected, the RS corrections to $|\Gamma_{12}^s|$ are rather small, typically not exceeding $\pm 4\%$.
The right panel displays the corrections to the magnitude and phase of the dispersive part of the mixing amplitude, 
$R_M$ and $\phi_M$. For this analysis, important constraints arise from the measurement of the $\bar B_s^0$--$B_s^0$ 
oscillation frequency. The corresponding result for the mass difference reads \cite{Abulencia:2006ze}
\beq\label{eq:msexp}
\Delta m_{B_s}^{\rm exp}= (17.77\pm 0.10\, ({\rm stat})\pm0.07\,({\rm syst}))\,{\rm ps}^{-1}\, ,
\eeq
and is in good agreement with the SM prediction of $(17.3\pm2.6)\,{\rm ps}^{-1}$ \cite{Lenz:2011ti}.
As a consequence, we exclude all points with $R_M \not\in [0.718,1.336]$, which corresponds to the $95\%$ CL region of (\ref{eq:msexp}).
This is indicated by the dashed lines. Note that for a sufficient amount of scatter points, the phase correction $\phi_M$  
can take any value in $[-\pi,\pi]$ within the custodial RS model. 
Comparing the results for the new phase $\phi_\Gamma$ to those for $\phi_M$, we see that the former can be neglected
to very good approximation, what we will do from now on. 

\begin{figure}[t!]
\begin{center}
\includegraphics[width=6.5cm]{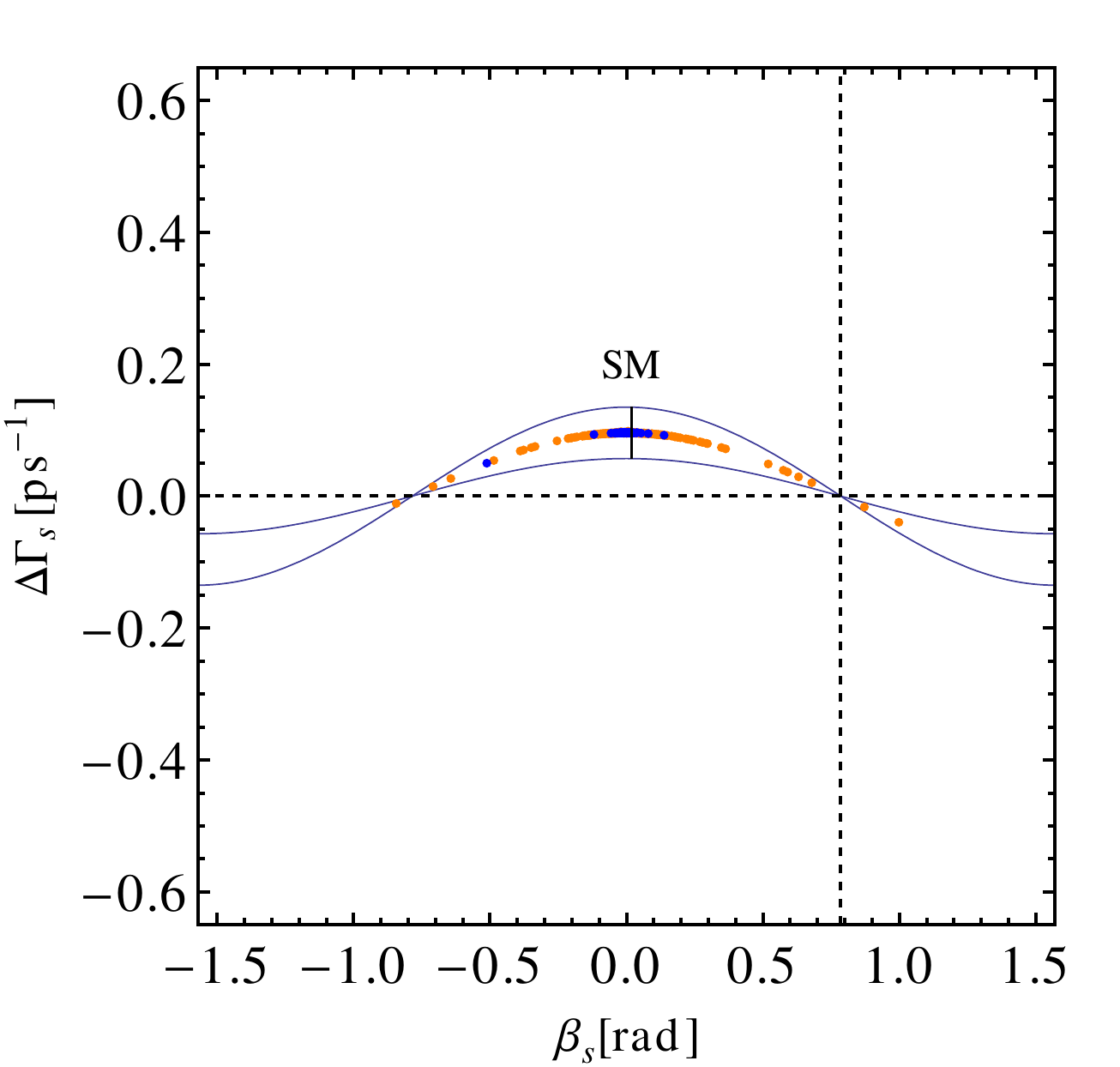}\quad
\includegraphics[width=6.4cm]{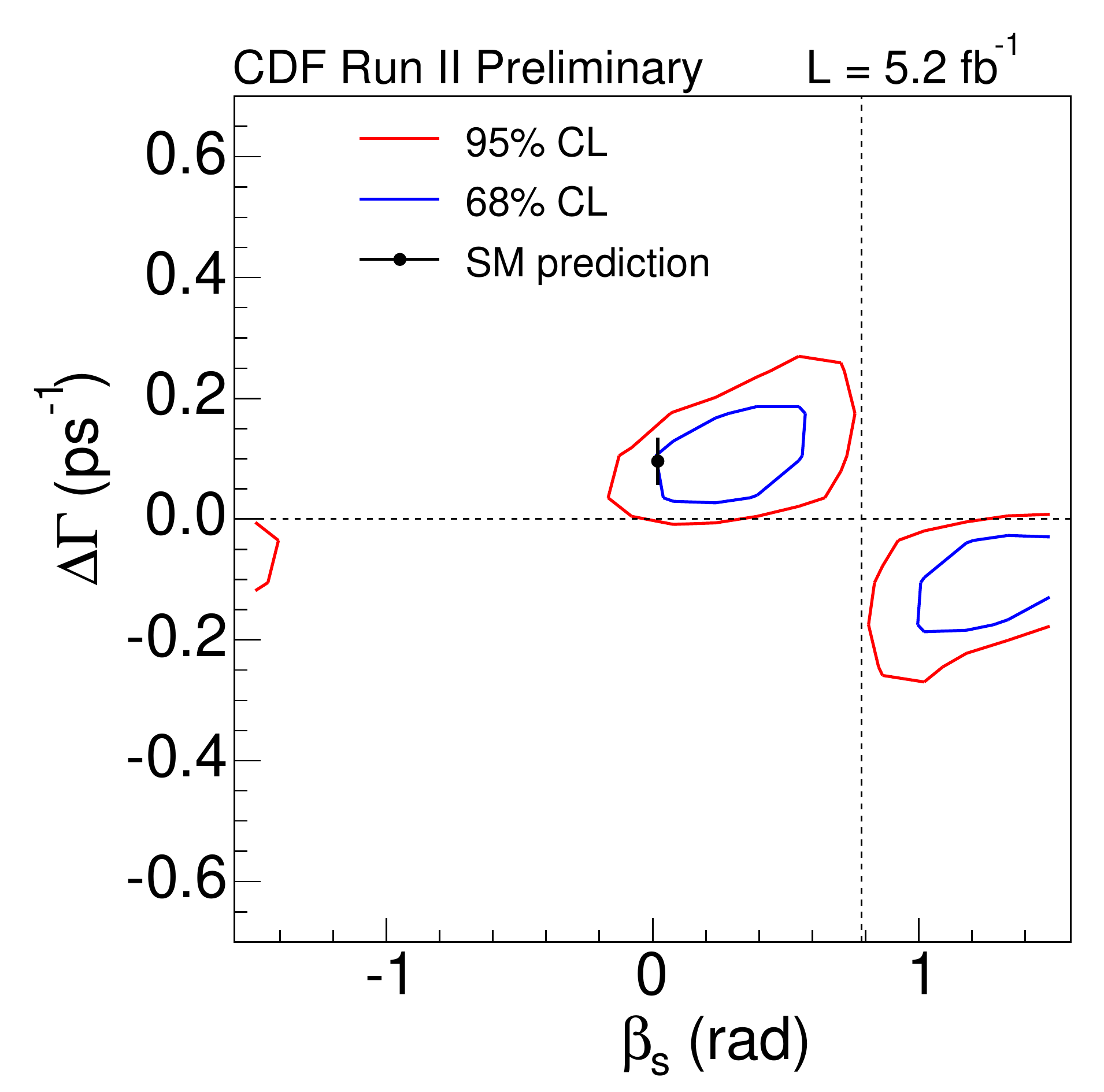}
\end{center}
\vspace{-2.5mm}
\begin{center}
  \parbox{15.5cm}{\caption{\label{fig:gammabeta} Left: Corrections 
   in the $\Delta\Gamma_s^{\rm SM}/\beta_s\,$-plane for
   the minimal (blue/dark gray) and custodial (orange/light gray) RS model.
   Bounds from $Z b\bar b$, $\Delta m_{B_s}$, and $\epsilon_K$ are satisfied. See \cite{Goertz:2011nx} and text for details.
   Right: Experimental constraints from flavor-tagged
   $B_s^0 \rightarrow J/\psi\phi\,$ decays. Figure from \cite{pubn2}. }}
\end{center}
\end{figure}
 
\begin{figure}[t!]
\begin{center}
\includegraphics[width=6.7cm]{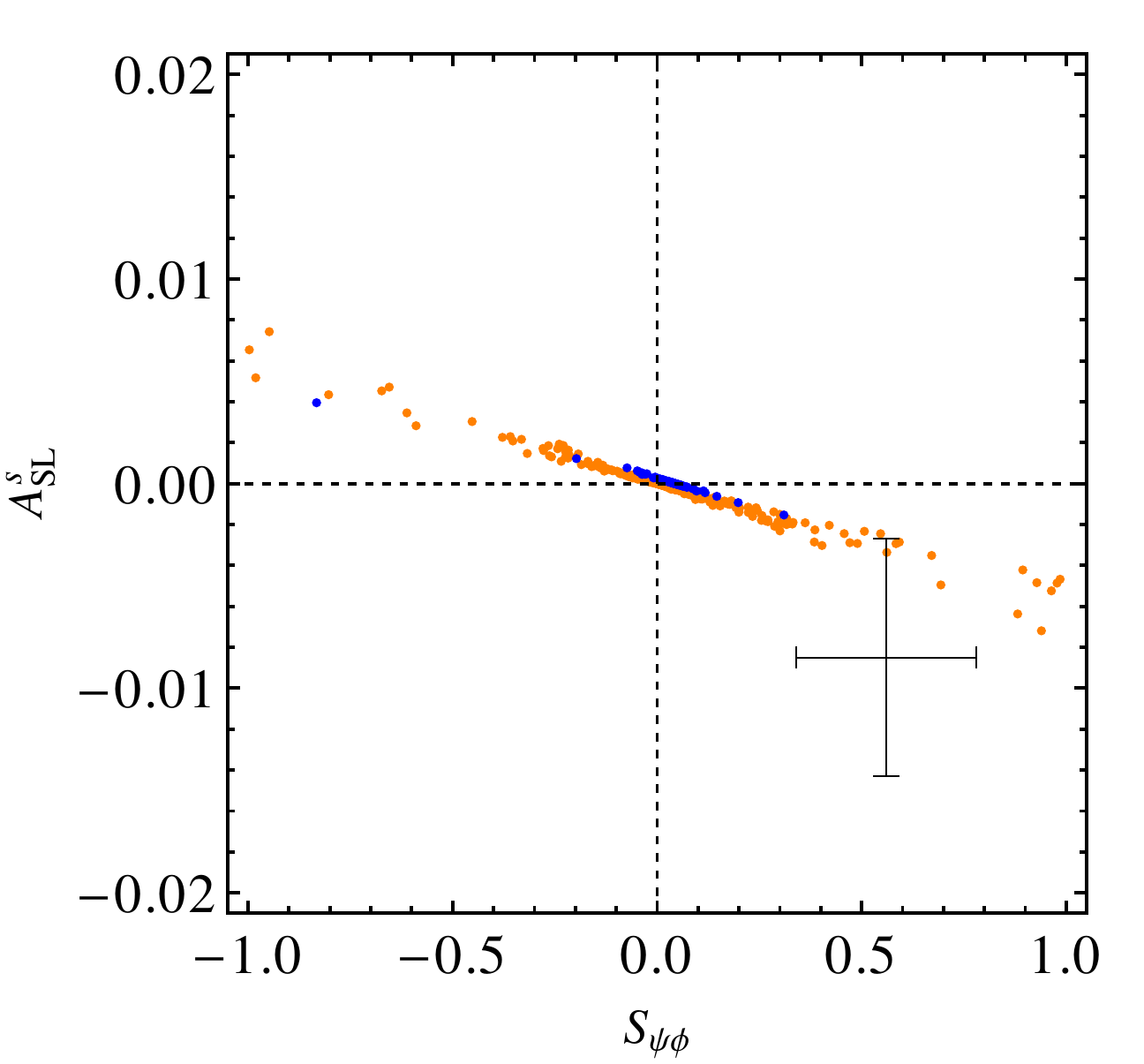}
\end{center}
\vspace{-2.5mm}
\begin{center}
  \parbox{15.5cm}{\caption{\label{fig:ASL} Corrections 
   in the $A_{\rm SL}^s/S_{\psi\phi}\,$-plane for
   the minimal (blue/dark gray) and custodial (orange/light gray) RS model.
   Bounds from $Z b\bar b$, $\Delta m_{B_s}$, and $\epsilon_K$ are satisfied. See \cite{Goertz:2011nx} and text for details. }}
\end{center}
\end{figure}

Another important constraint for our analysis comes from the CP-violating observable 
$\epsilon_K=\epsilon_K^{\rm SM}+\epsilon_K^{\rm RS}$ \cite{Bauer:2008xb,Bauer:2009cf,Csaki:2008zd,Blanke:2008zb}. 
Explicitly, we demand $|\epsilon_K|\in [1.2,3.2]\cdot 10^{-3}$ \cite{Bauer:2009cf}, 
where 
\beq
\epsilon_K=\frac{\kappa_\epsilon\, e^{i\varphi_\epsilon}}{\sqrt 2\,{(\Delta m_K)}_{\rm exp}}\,
{\rm Im}({M_{12}^{K{\,\rm SM}}}+{M_{12}^{K{\,\rm RS}}})\,,
\eeq
with $\varphi_{\epsilon} = (43.51\pm0.05)^{\circ}$ \cite{Nakamura:2010zzi} and $\kappa_{\epsilon} = 0.92\pm0.02$ 
\cite{Buras:2008nn}. The neutral Kaon mixing amplitude is defined in analogy to (\ref{eq:M12me}). 
The input data needed for the calculation can be found in Appendix~B of \cite{Bauer:2009cf}. 
As it turns out, without some tuning, the prediction for $\epsilon_K$ is generically too large.
Note that the dangerous contributions from the operators $Q_{4,5}^{sd}$ \cite{Csaki:2008zd}, which can become comparable to those
of $Q_1^{sd}$ due to $R_K=(M_K/(\bar m_d+\bar m_s))^2\approx 20\,$ for $\mu=2\,$GeV and a more pronounced RG running, could be suppressed 
by imposing a $U(3)$ flavor symmetry in the right-handed down-quark sector \cite{Santiago:2008vq}, as mentioned in 
Section~\ref{sec:hierarchies}. Although this symmetry will inevitably be broken by the Yukawa couplings,
the possibility to choose all down-type bulk masses equally allows to forbid tree-level 
down-type FCNCs in the ZMA. This can be seen from (\ref{eq:Cmix}), since ${(\bm{W}_d^\dagger)}_{mj}{(\bm{W}_d)}_{jn}=0\,$ 
for $m\neq n$ due to the unitarity of $\bm{W}_d\,$. Non-vanishing contributions from the exchange of KK gauge bosons arise 
only at $\ord(v^4/\Mkk^4)$ from the mixing of the right-handed fermion zero modes with their KK excitations. 
For $\Mkk=2\,$TeV, it would thus be possible to reduce $C_{4,5}^{sd}$ by a factor of about $100$, by imposing such a symmetry. 
The same suppression factor also applies to the B-meson sector. For the coefficient $C_1^{\rm RS}$ however, such a protection
is not possible. 
In our analysis, we do not impose an additional flavor symmetry on the 
bulk masses, but rather use the bound from $\epsilon_K$ as an additional constraint on our scatter points and discard
those that are in conflict with it.

Setting the tiny SM phases to zero, the width difference (\ref{eq:DGammaapprox}) can be written as
\beq\label{eq:Dgamma}
\Delta\Gamma_s= \Delta\Gamma_s^{\rm SM}\,R_\Gamma\,\cos 2\beta_s\,,
\eeq
where $2\beta_s\approx -\phi_M$ \cite{pubn}. 
Thus, in the absence of large corrections to the magnitude of $\Gamma_{12}^s$,
NP contributions always lead to a negative shift in $\Delta\Gamma_s$ \cite{Grossman:1996era}.
Note that the preliminary CDF analysis \cite{pubn2} uses the older SM prediction 
$\Delta\Gamma_s^{\rm SM}=(0.096\pm0.039)\text{ps}^{-1}$ \cite{Lenz:2006hd},
which we will also take as central value for our calculation.
Taking the more recent value will not change our conclusions.
In the left panel of Figure \ref{fig:gammabeta} we plot our predictions 
for $\Delta\Gamma_s$ against $\beta_s$ in the RS model.
A comparison to the CDF results in the right panel leads to the conclusion
that both the minimal (blue/dark gray) as well as the custodial RS model (orange/light gray) can enter the $68\%$ confidence region and come close to
the best fit value. They stay below the desired value for $\Delta\Gamma_s$, 
as there are no sizable positive corrections to $|\Gamma_{12}^s|$. 

Neglecting the tiny SM phases and the NP phase corrections related to decay, 
the semileptonic CP asymmetry $A_{\rm SL}^s$ is 
proportional to $S_{\psi\phi}$ \cite{Ligeti:2006pm}. The latter quantity is given by the amplitude 
of the time-dependent asymmetry in $B_s^0\rightarrow J/\psi\phi$ decays, 
${\cal A}_{\rm CP}^s(t)=S_{\psi\phi}\sin(\Delta m_{B_s} t)$, see (\ref{eq:tdcp}).
Setting just the NP phase in decay to zero, one obtains the well known expression 
$S_{\psi\phi}=\sin(2\beta_s^{J/\psi \phi}-\phi_M)$ \cite{Blanke:2006ig}.
Thus one has to good approximation
\beq
\label{eq:lin}
A_{\rm SL}^s\approx-\,\frac{|\Gamma_{12}^{s\,{\rm SM}}|}{|M_{12}^{s\,{\rm SM}}|}\,
\frac{R_\Gamma}{R_M}\,S_{\psi\phi}\,.
\eeq
The results for $A_{\rm SL}$ and $S_{\psi\phi}$ in the RS model are shown in Figure \ref{fig:ASL}.
They confirm the approximately linear dependence (\ref{eq:lin}) between the plotted quantities. The experimentally favored regions 
$S_{\psi\phi}=0.56\pm0.22$ \cite{Bona:2008jn} and $A_{\rm SL}^s=-0.0085\pm0.0058$ \cite{Asner:2010qj}
are marked by the black cross. 
The latter has been obtained by a combination of the direct measurement with the results derived from 
$A_{\rm SL}^b$ together with the average 
$A_{\rm SL}^d=-0.0047\pm0.0046$ from $B$-factories.
It is evident from the plot that the best fit value of $S_{\psi\phi}$ can be reproduced 
(however, some amount of tuning will be necessary in the minimal RS variant), which has already been noted in \cite{Bauer:2009cf,Blanke:2008zb}.
Furthermore, the RS setup allows to enter the $1\sigma$ range of the measured value of $A_{\rm SL}^s$.
The same conclusion has been drawn in \cite{Datta:2010yq} recently, using a different approach.
The authors did not use any concrete sets of input parameters, but rather scanned
the predictions for FCNC vertices in a range constrained by bounds from $\Delta \Gamma_s$ 
and $\Delta m_{B_s}$.
Note that due to $S_{\psi\phi}\approx\sin 2\beta_s$, the corrections in the $\Delta\Gamma_s^{\rm SM}/\beta_s\,$-plane
and the $A_{\rm SL}^s/S_{\psi\phi}\,$-plane are correlated. An improvement with respect to experiment 
in the former leads to an improvement in the latter.

In conclusion, in particular the custodial RS model allows for a better agreement of the theoretical values
for $A_{\rm SL}^s$, $S_{\psi\phi}$, and $\Delta\Gamma_s$ with experiment. However the concrete predictions for these
observables still depend sensitively on the RS parameters (and not only on the scale $\Mkk$).
Essentially this is due to the fact that the range for the new phase $\Phi_M$ is not very limited by the constraints 
that are imposed on the parameters.

\section{Higgs Physics}
\label{sec:RSHiggs}

\begin{figure}[t!]
\begin{center}
\mbox{\includegraphics[width=6.5cm]{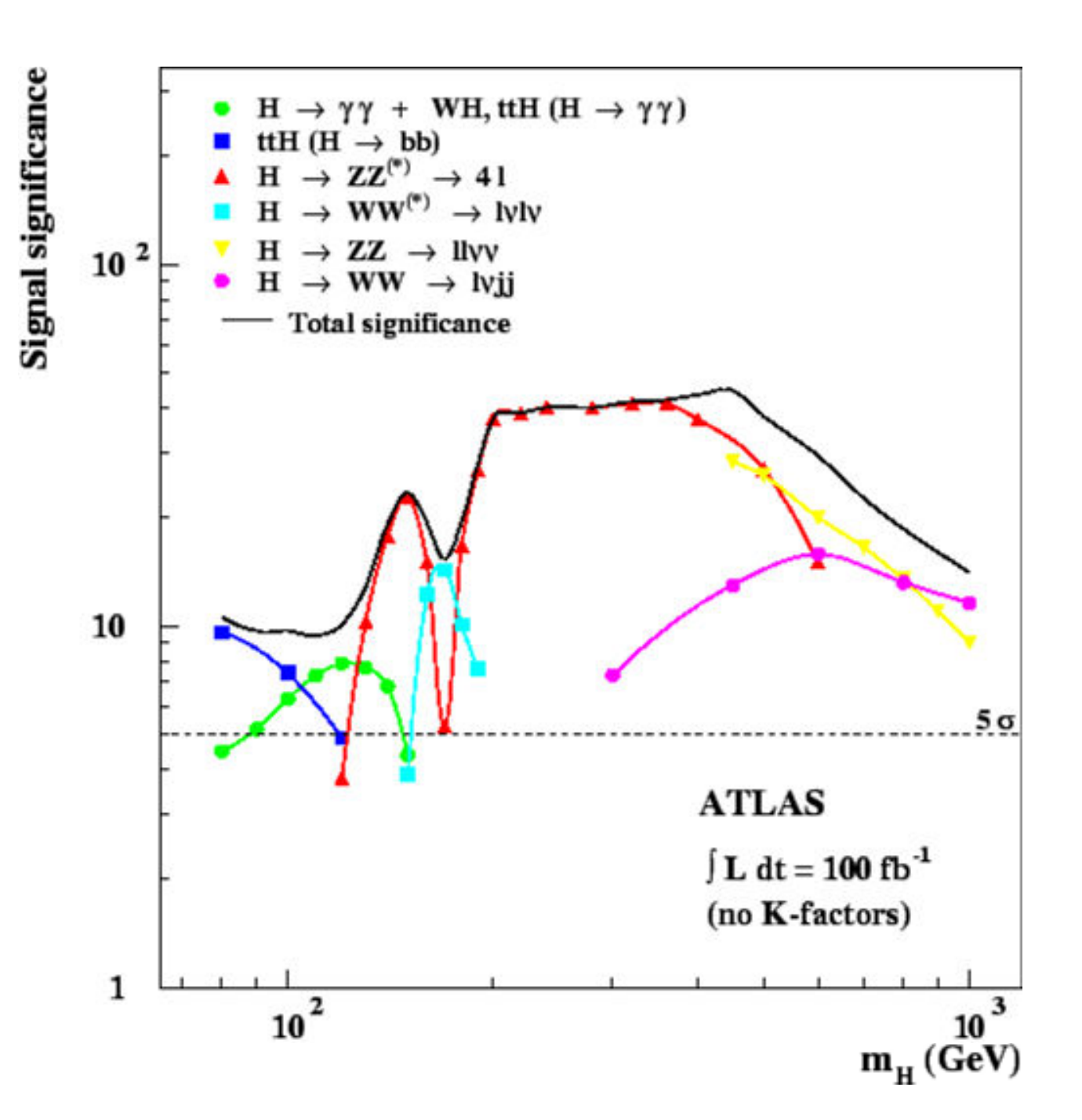}} 
\parbox{15.5cm}{\caption{\label{fig:sens} Expected statistical significance for a discovery of the SM Higgs boson at ATLAS
as a function of the Higgs-boson mass for an integrated luminosity of 100\,fb$^{-1}$. Plot from \cite{TDR} (with permission).}}
\end{center}
\end{figure}

As we have seen in Chapter~\ref{sec:IntroSM}, a Higgs boson with a mass not exceeding the TeV scale 
is an important ingredient of the SM and also of many of its extensions. However, such a boson has not 
been observed experimentally yet. 

The SM Higgs boson is expected to be found with a statistical
significance of $\gtrsim 10\,\sigma$ over the full mass range with 100\,fb$^{-1}$ integrated luminosity, 
see Figure~\ref{fig:sens}.
Imagine that we do not discover this particle at the LHC in the first years of running. Does this
already mean that we have to abandon the corresponding mechanism of EWSB? 
The answer to this question is certainly no. BSM physics could feature a standard Higgs
mechanism that however could be much harder to detect than the one of the SM,
even for a Higgs-boson mass easily accessible at the LHC. It is thus important to study Higgs physics 
in various models to be prepared for different possible scenarios. In warped extra dimensions, large effects
are expected due to the localization of this sector on the IR brane, where also
the KK modes as well as heavy SM quarks are peaked. In the following we will study Higgs-boson production 
and decay within the custodial RS model.
Although Higgs physics has been looked at in related models
\cite{Bouchart:2009vq,Azatov:2009na,Agashe:2009di,Falkowski:2007hz,Lillie:2005pt,Azatov:2010pf,Frank:2011rw,Cacciapaglia:2009ky,Espinosa:2010vn,Djouadi:2007fm},
there has not been a complete analysis of the subject in the context of the models studied in this thesis, taking into account all important effects induced by the KK excitations at the one-loop order.
This will be provided in the following.

\subsection{Higgs-Boson Production}
\label{sec:higgsproduction}

\begin{figure}[t!]
\begin{center}
\mbox{\includegraphics[width=15cm]{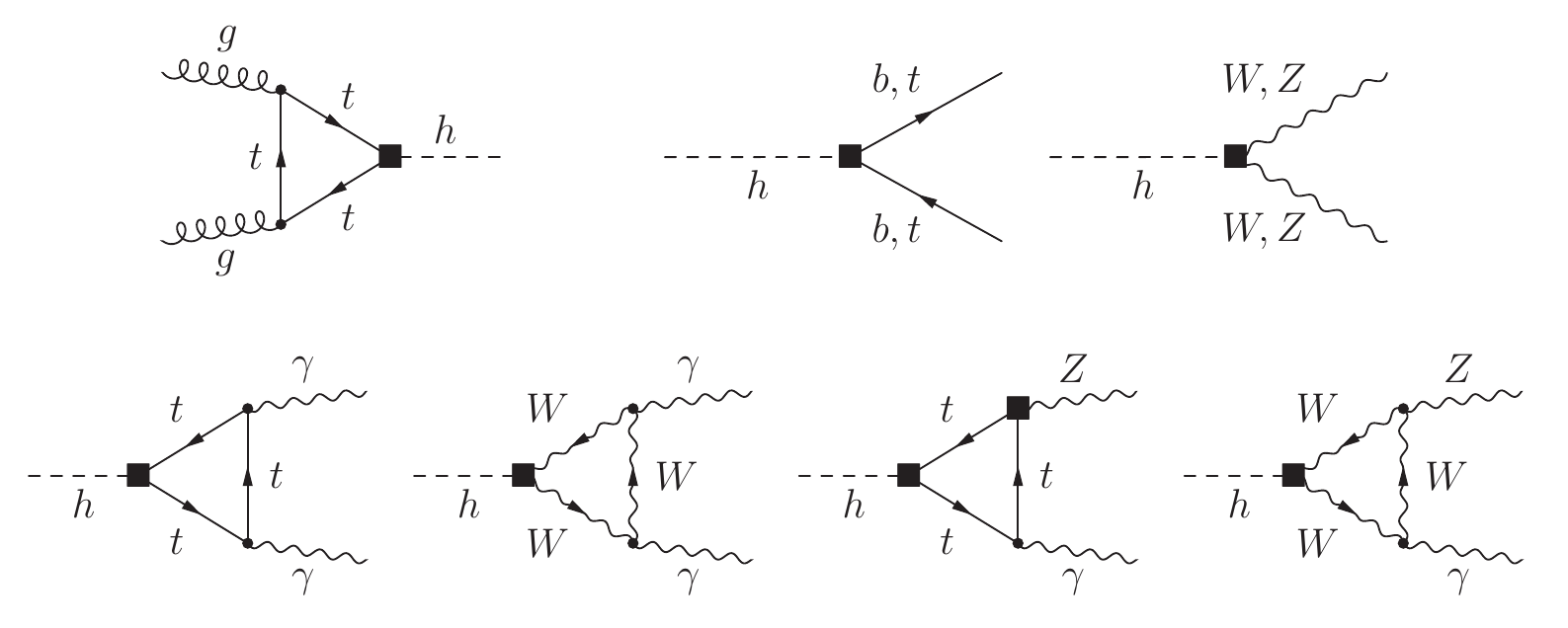}} 
\parbox{15.5cm}{\caption{\label{fig:hprodec} Examples of
    zero mode contributions to the
    production and the decay of the Higgs boson at leading order of
    perturbation theory. Vertices indicated by a black square can
    receive sizable shifts in the RS model relative to the SM
    couplings. See \cite{Casagrande:2010si} and text for details.}}
\end{center}
\end{figure}

\begin{figure}[t!]
\begin{center}
\mbox{\includegraphics[width=12.5cm]{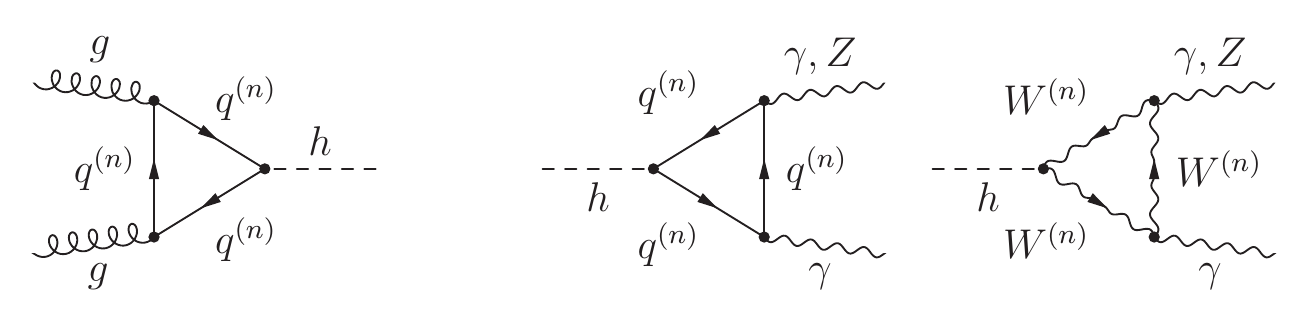}} 
\parbox{15cm}{\caption{\label{fig:hkkcontr} Examples of one-loop
    diagrams involving KK excitations that contribute to the
    production and the decay of the Higgs boson at leading order of
    perturbation theory. See \cite{Casagrande:2010si} and text for details.}}
\end{center}
\vspace{-4mm}
\end{figure}

We will start with the leading (SM) production mechanism for the Higgs
boson at hadron colliders, which is gluon-gluon fusion.
In the SM, this mechanism, which receives its dominant contribution 
from a top-quark triangle loop, has been introduced in Section~\ref{sec:Higgs}.
Within the RS framework, one has to take into account in addition the KK
tower of the top quark as well as of all the other quark flavors, since all these
modes contribute to the $gg \to h$ amplitude at $\ord
(v^2/\Mkk^2)$. The relevant Feynman diagrams are shown on the very
left in the top row of Figure~\ref{fig:hprodec} and on the left-hand
side of Figure~\ref{fig:hkkcontr}.

In order to calculate the $gg \to h$ production cross section in the
RS model, we rescale the SM prediction according to
\begin{equation} \label{eq:rescale1}
\sigma (gg \to h)_{\rm RS} = \left | \kappa_{g} \right |^2 \,
\sigma (gg \to h)_{\rm SM} \,, 
\end{equation}
where
\begin{equation} \label{eq:kappagg} 
  \kappa_{g} = \frac{{\displaystyle
      \sum}_{i = t, b} \, \kappa_i \hspace{0.25mm} A_{q}^h
    (\tau_i)\, + \hspace{0.25mm}{\displaystyle \sum}_{j = u,d,\lambda}  
    \, \nu_j }{ {\displaystyle \sum}_{i = t,
      b} \; A_{q}^h (\tau_i)} \,,
\end{equation}
and $\tau_i \equiv 4 \hspace{0,25mm} m_i^2/m_h^2$. 
The first term in the numerator corresponds to top- and bottom-quark zero modes
running in the loop, with Higgs couplings (normalized to the SM) given by~\cite{Goertz:2011ti}
\vspace{-2mm}
\beq
\label{tbratios}
\kappa_t={\rm Re}[(g_h^u)_{33}]/\left(\frac{m_t}{v_{\rm SM}}\right)\,,\quad \kappa_b={\rm Re}[(g_h^d)_{33}]/\left(\frac{m_b}{v_{\rm SM}}\right).
\vspace{-2mm}
\eeq
These couplings differ from those of the SM, $\kappa_{t,b}\neq 1$, due to contributions to the fermion masses
from compactification, see Section~\ref{sec:HcouplingsRS}, and due to $v_{\rm RS_C}\neq v_{\rm SM}$.
The form factor $A_{q}^h (\tau_i)$, needed for a correct weight of the different contributions
in (\ref{eq:kappagg}), approaches $1$ for $\tau_i \to \infty$ and vanishes proportional to
$\tau_i$ for $\tau_i \to 0$. Its analytic form is given in
Appendix~\ref{app:formfactors}. 

Due to power suppression, the only
phenomenologically relevant correction in $\sigma (gg \to h)_{\rm SM}$
from lighter fermions stems from the bottom quark. We include the
interference term of the bottom- and the
top-quark amplitude approximately, by multiplying
the cross section $\sigma (gg \to h)_{\rm SM}$ by $\big (1 + 2 \, {\rm Re} \hspace{0.25mm}
A_{q}^h (\tau_b) \big )$. Numerically, this approximate treatment
decreases the SM cross section by about $9\%$, $2\%$, and below $1\%$
for $m_h = 100 \, {\rm GeV}, 300 \, {\rm GeV}$, and $600 \, {\rm
  GeV}$, which is in good agreement with the NLO calculation
including the exact mass dependence~\cite{Spira:1995mt}. 

\begin{figure}[!t]
\begin{center} 
\hspace{-2mm}
\mbox{\includegraphics[height=2.85in]{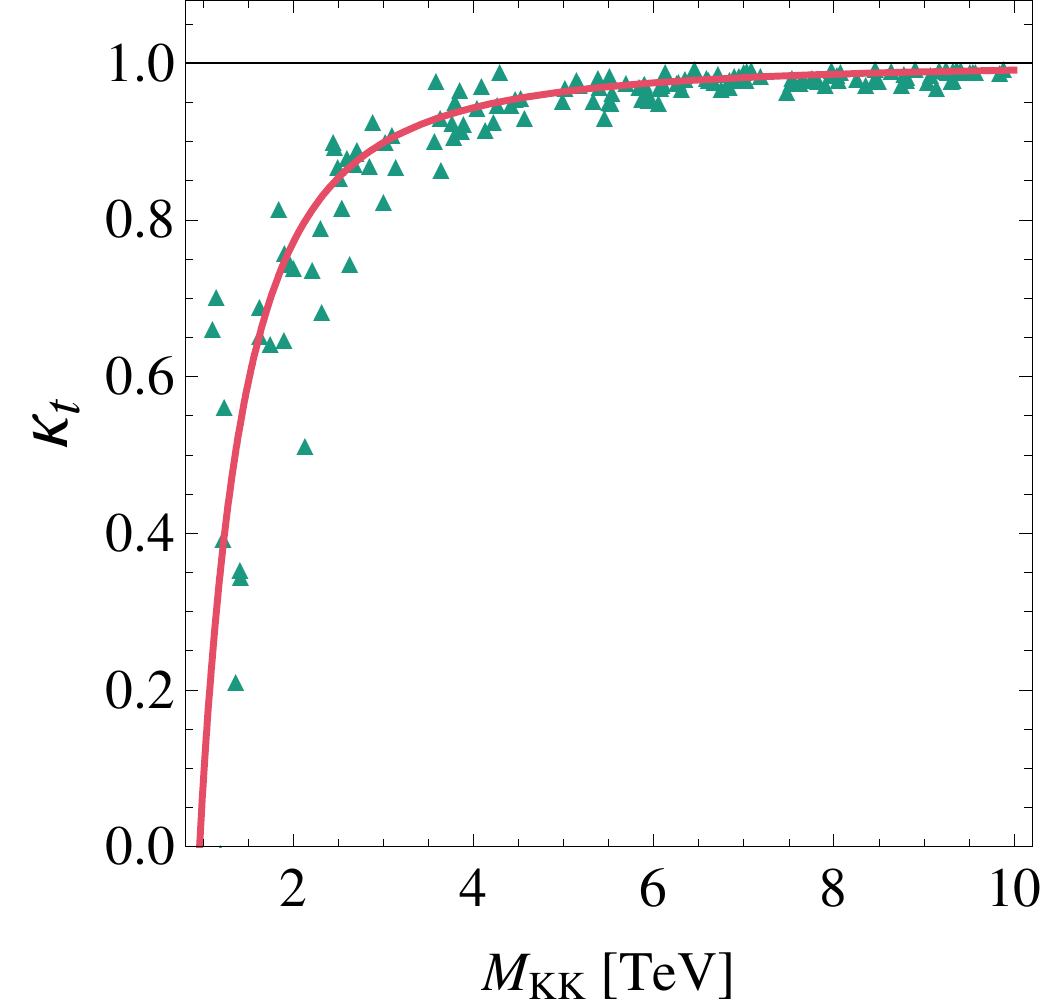}} 
\hspace{4mm}
\mbox{\includegraphics[height=2.85in]{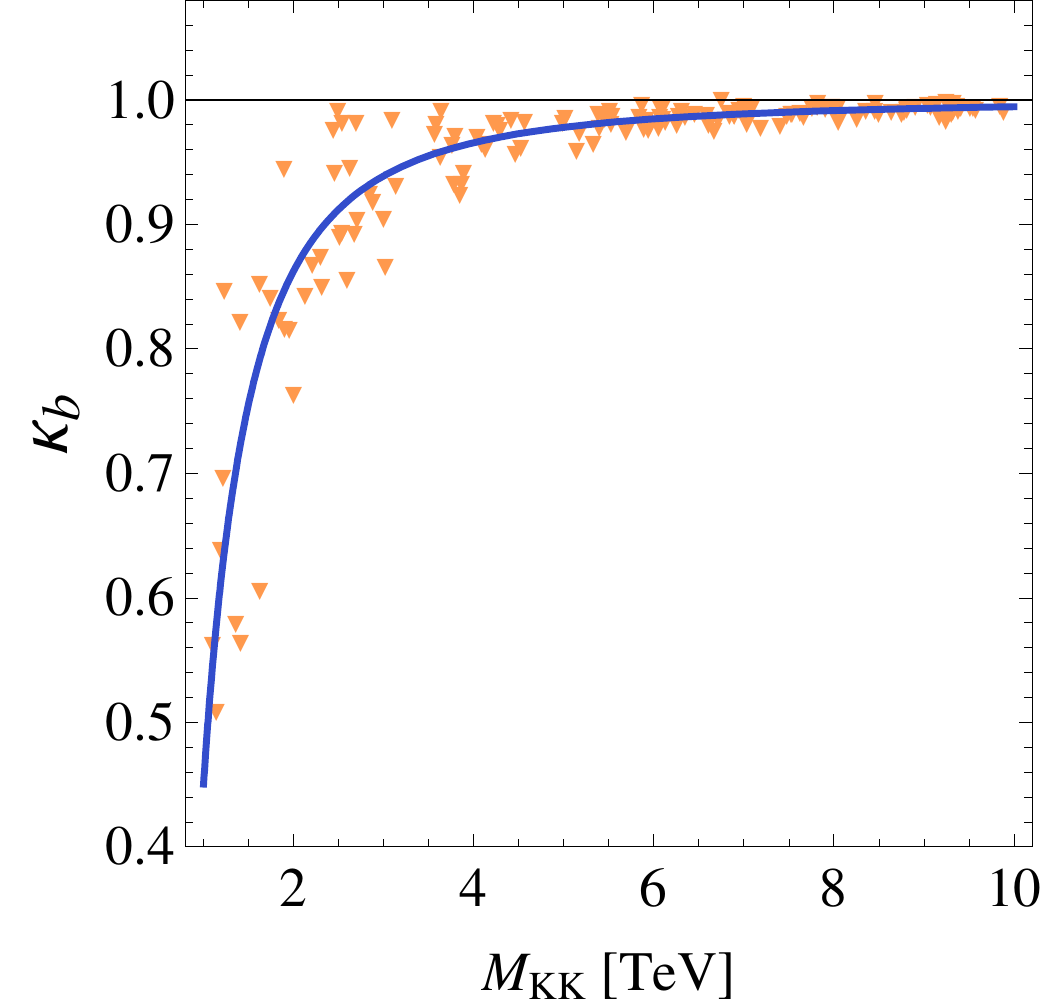}}
\parbox{15.5cm}{\caption{\label{fig:kappa} Predictions for the ratios 
    of the $ht\bar t$ (left) and $hb\bar b$ (right) coupling in the custodial RS model relative to the SM
    value. The solid lines show fits to the samples of parameter
    points. See \cite{Casagrande:2010si} and text for details.
  }}
\end{center}
\end{figure}

In Figure~\ref{fig:kappa} we show the ratios (\ref{tbratios}) as functions 
of $\Mkk$ for a set of 150 random parameter points. These correspond to the custodial 
model with extended $P_{LR}$ symmetry (\ref{eq:extendedPLR}), which we will always 
employ in the following analysis. They reproduce the quark masses
as well as the CKM parameters at 68\% CL. The same sample of model parameter points will be
used in the remainder of this section. We observe that
both the $ht \bar t$ and the $h b \bar b$ coupling are reduced in the custodial
RS scenario with respect to the SM, resulting in $\kappa_{t, b} \leq 1$. 
The same conclusion has been drawn in \cite{Azatov:2009na} for the minimal RS model. 
Numerically, we find
that for $\Mkk = 2 \, {\rm TeV}$ ($\Mkk = 3 \, {\rm TeV}$) the average
corrections amount to around $-25$\% and $-15$\% ($-10$\% and $-5$\%)
in the top- and bottom-quark sectors, respectively. Since the RS
corrections to the Higgs couplings scale as $v^2/M_{\rm KK}^2$, the
average value of the ratios $\kappa_{t, b}$ can be parametrized by
$1-a_{t, b} \, v^2/\Mkk^2$. The resulting coefficients $a_{t,b}$ are 
given in Table~\ref{tab:kappas}. The quoted values of $a_{t,b}$ have been
obtained from the best fits to the shown sample of scatter points.
Note that the RS predictions depend most strongly on the KK scale, while
the additional dependence on the explicit parameter points is less important.
Moreover, as expected from the arguments of Section~\ref{sec:hierarchies},
the correction to the top-quark coupling is more pronounced than this
to the bottom-quark coupling.

It is not difficult to understand the suppression of the Yukawa couplings 
of the third-generation quarks, $\kappa_{t,b} \leq 1$,
analytically by analyzing the structure of (\ref{eq:Higgscorrection}).
First, one has $m_3^q/v \, \big ( (\Phi_q)_{33} +
(\Phi_Q)_{33} \big) \geq 0$ since the diagonal elements of the
matrices $\bm{\Phi}_{q,Q}$ introduced in (\ref{eq:Phi}) are absolute
squares. Second, the third term in (\ref{eq:Higgscorrection}) can be
written in the ZMA as
\begin{equation} \label{eq:gtildehZMA}
\begin{split}
  (\Delta \tilde g^u_h)_{33} & = \frac{4 m_t^2}{3 v \Mkk^2} \,
  \sum_{j=1}^3 \, m_j^u \; \Big ( \bm{U}_u^\dagger \; {\rm diag} \left
    [ F^{-2}(c_{Q_i}) \right ] {\bm{U}_u} \Big )_{j3} \, \Big (
  \bm{W}_u^\dagger \; {\rm diag} \left [ F^{-2}(c_{u_i^c}) \right ]
  {\bm{W}_u} \Big )_{3j} \,
\end{split}
\end{equation}
and a similar formula applies to the case of $(\Delta \tilde g^d_h)_{33}$.
Because the diagonal elements of the matrices $\bm{U}_u^\dagger \;
{\rm diag} \left [ F^{-2}(c_{Q_i}) \right ] {\bm{U}_u}$ and
$\bm{W}_u^\dagger \; {\rm diag} \left [ F^{-2}(c_{u_i^c}) \right ]
{\bm{W}_u}$ are absolute squares, the term with $j = 3$ is
positive semi-definite. The terms with $j=1,2$, on the other hand, can
have an arbitrary complex phase. Yet, due to the strong chiral
suppression, $m_c/m_t \approx 1/275$ and $m_u/m_t \approx 10^{-5}$,
these terms cannot drive the real part of $\Delta (\tilde g^u_h)_{33}$
negative. The same
holds true for $(\Delta \tilde g^d_h)_{33}$, although the chiral
suppression is weaker in this case, $m_s/m_b \approx 1/50$ and
$m_d/m_b \approx 1/800$. Recalling that $(\Delta g_h^q)_{33} = m_3^q/v
\, \big ( (\Phi_q)_{33} + (\Phi_Q)_{33} \big) + (\Delta \tilde
g^q_h)_{33} \geq 0$ enters (\ref{eq:mis1}) with a minus sign, we
conclude that the $ht\bar t$ and $h b \bar b$ couplings are predicted
to be suppressed relative to their SM values in both the minimal and
the extended RS models. This result seems to be independent of the
particular realization of the setup and to hold for a large class
of RS setups. The same conclusion has been drawn in the context of models 
where the Higgs-boson arises as a pseudo Nambu-Goldstone boson 
\cite{Falkowski:2007hz,Low:2009di}.

The second term in the numerator of (\ref{eq:kappagg}) represents the
contribution arising from the virtual
exchange of KK quarks. Here, all quark flavors have to be taken into account,
since a possible UV localization of a zero mode does not apply for the KK 
excitations. The corresponding Feynman diagram is shown on the
very left in Figure \ref{fig:hkkcontr}. In the up-type quark sector
the associated coefficient takes the form
\begin{equation}\label{eq:nug}
  \begin{split}
    \nu_u & = \,v_{\rm SM} \;\sum_{n = 4}^\infty \; \frac{{\rm Re}\left[(g_h^u)_{nn}\right]}{m_n^u}
    \hspace{0.5mm} A_{q}^h (\tau_n^u) \\ & = \frac{2 \pi}{\epsilon L}\, \frac{v_{\rm SM}}{v_{\rm RS_C}}
    \, \sum_{n = 4}^\infty \; \frac{{\rm Re}\left[\vec a_{n}^{\hspace{0.25mm}
        U\dagger} \,\bm{C}_{n}^{U} (\pi^-) \left(\displaystyle
        \bm{1}-\frac{v_{\rm RS_C}^2}{3\hspace{0.25mm}\Mkk^2} \, \bm{\tilde
          Y}_{\vec u}\bm{\bar Y}_{\vec u}^\dagger\right)
      \bm{S}_{n}^U(\pi^-) \, \vec a_{n}^{\hspace{0.25mm} U}\right] }{x_n^u}
    \hspace{0.5mm} A_{q}^h (\tau_n^u) \,.
  \end{split}
\end{equation}
Similar relations hold in the sector of down-type and $\lambda$
quarks. Note that the VEV shift $v_{\rm SM}/v_{\rm RS_C} \neq 1$ does not contribute to $\nu_u$ at $\ord(v^2/\Mkk^2)$.
Since the mass of the first KK up-type quark is already much 
larger than the mass of the Higgs-boson, $m_4^u/\Mkk = \ord({\rm a\ few}) \hspace{0.25mm} 
\gg m_h/\Mkk$, it is an excellent approximation to replace the function $A_{q}^h
(\tau_n^u)$ by its asymptotic value of 1 obtained for $\tau_n^u \equiv
4 \left (m_n^u \right )^2/m_h^2  \, \to \, \infty$.

Before presenting our numerical results for these contributions,
some comments about the infinite sum in (\ref{eq:nug}) are in order. 
First of all, since the RS model is an EFT defined with a 
cutoff $\Lambda_{\rm UV}(\pi)\ll M_{Pl}$, the KK modes above a certain level will not be part of the
EFT anymore. Thus the potentially divergent sum is effectively cut off after a certain KK level.
However, as we will see below, there are in addition cancellations within the sum in (\ref{eq:nug}) 
that cause the terms to decrease quadratically with the mass of the KK fermions in the loop,
which renders the sum convergent. At the same time it becomes IR dominated, \ie, the lowest lying modes become most important, 
whereas the difference between truncating the sum at a finite level $n_{\rm max}$ and {\it extrapolating} till infinity
scales like $1/n_{\rm max}$. This corresponds to a subleading effect if at least a few modes are present below the cutoff.
Importantly, this line of reasoning with a cutoff prevents UV physics above this RS cutoff to modify the behavior of the KK sum.
For the following analysis we will assume the fundamental parameters of the RS model to be such that
 $\Lambda_{\rm UV}(\pi)$, which cuts off the KK sums, is above the mass scale of $\ord(4)$ complete KK levels. 
 
In the SM, the top-quark contribution to the $gg \to
h$ amplitude is proportional to $y_t/m_t$ in the decoupling limit.
As mentioned in Section~\ref{sec:Higgs}, it can be described by the effective operator
$h/v \, G^a_{\mu \nu} G^{a \, \mu \nu}$.
Naively, also the contributions of the KK levels in (\ref{eq:nug}) scale like $1/m_n$,
which would lead to a divergent sum.
However, due to low-energy theorems appropriate to external Higgs bosons with
vanishing momentum \cite{Low:2009di, Ellis:1975ap, Shifman:1979eb,
Kniehl:1995tn}, which apply to any quantum field theory,
the Wilson coefficient of the $D=5$ operator above (encoding the whole KK-towers) 
is related to the QCD $\beta$-function.
In the context of the RS framework this implies that the sum in (\ref{eq:nug})
must be convergent, because the running of $\alpha_s$ can be shown to
be logarithmic in warped extra-dimension models \cite{Randall:2001gb,
  Goldberger:2002hb,Pomarol:2000hp, Goldberger:2002cz, Agashe:2002bx, 
  Contino:2002kc, Choi:2002ps, Goldberger:2003mi}. While the
finiteness of the effective $hgg$ coupling is thus guaranteed on
general grounds, an explicit calculation of the complete sum (\ref{eq:nug}) in the KK
decomposed 4D theory for the full flavor structure turns out to 
be non-trivial. This is due to the
fact that the Higgs VEV induces $\ord(1)$ mixings between the various
modes of a single KK level, see Section~\ref{sec:CKMPheno}. For example, in
the up-type quark sector there are five types of fields, namely $u$,
$u^\prime$, $u^c$, $U^\prime$, and $U$. Each of them exists in three
different flavors, so that there are altogether 15 KK modes of similar
mass in each level. Since the mixing effects among the states of the 
same KK level are large, they cannot be
treated perturbatively, and one has to resort to numerical methods 
or 5D propagators (see Section~\ref{sec:5Dprop}) as long as one is interested in the case of three families. However,
in the toy example of a single generation, it turns out to be possible to derive 
an expression for (\ref{eq:nug}) in the KK decomposed theory, revealing analytically
the dependence on the fermion localization.

\begin{figure}[!t]
\begin{center}
\mbox{\includegraphics[height=2.7in]{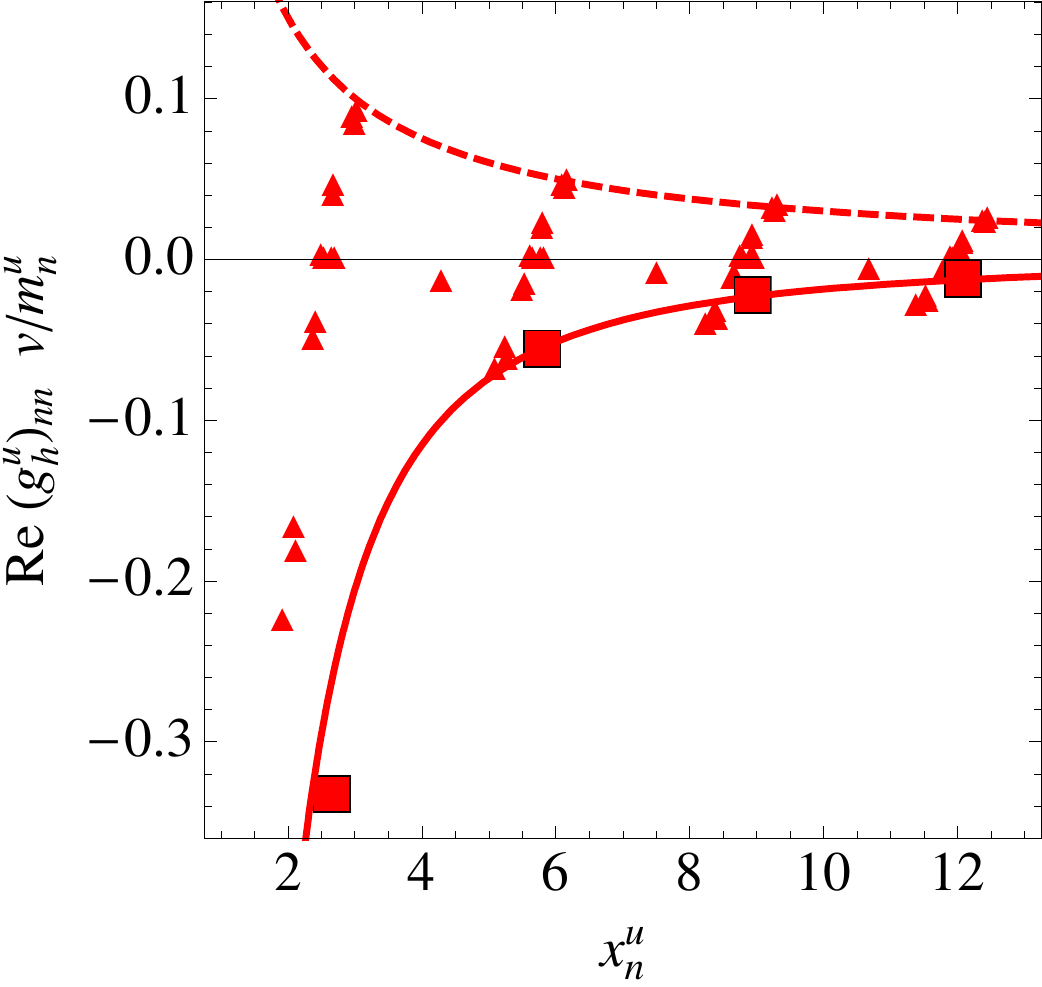}}
\qquad
\mbox{\includegraphics[height=2.7in]{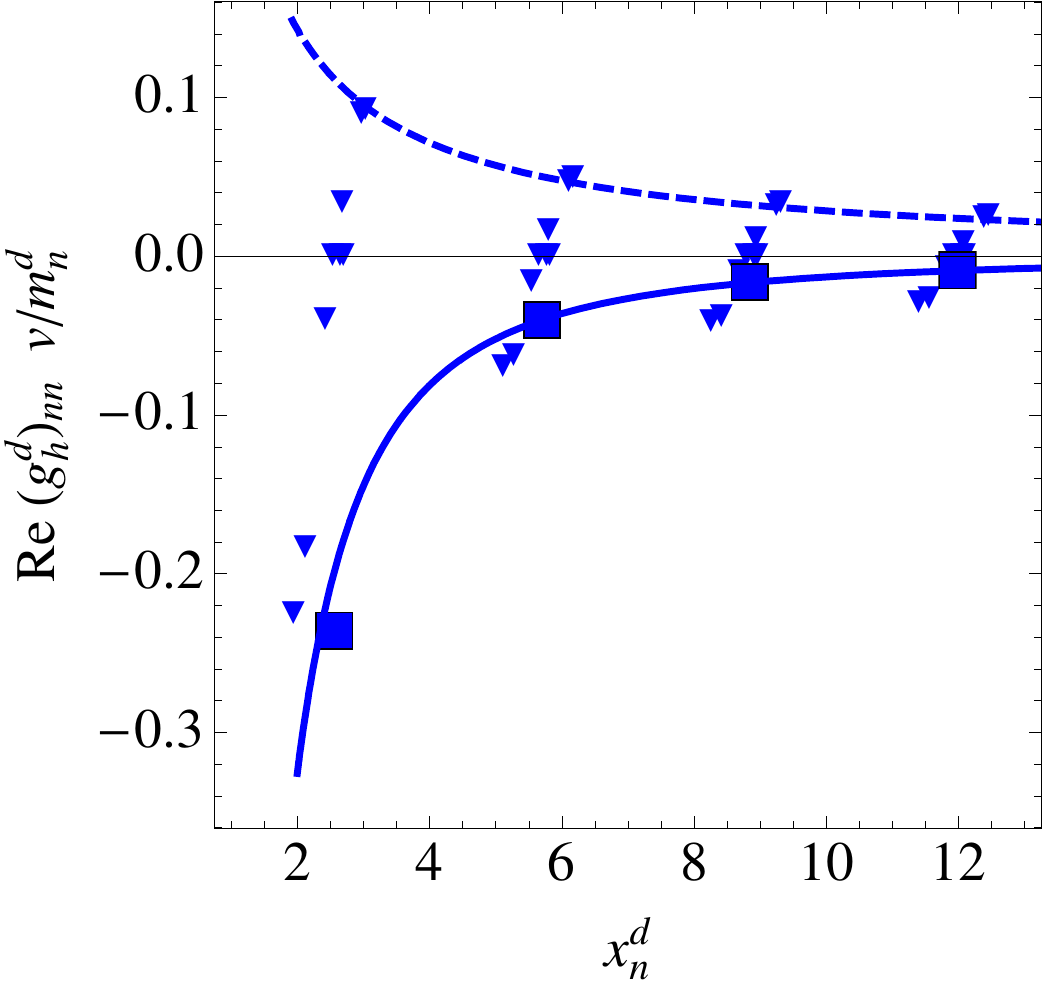}}
\parbox{15.5cm}{\caption{\label{fig:nug} Numerical results for the
    coefficients $\nu_{u,d}$ corresponding to a
    specific parameter point with $M_{\rm KK} = \, 2 \, {\rm
      TeV}$. The red (blue) dots in the left (right) panel display the
    first 60 (36) terms in the KK sum for up- (down-type) quarks,
    while the red (blue) filled boxes indicate the sums over
    complete KK levels. See \cite{Casagrande:2010si} and text for details.  }}
\end{center}
\end{figure}

In order to calculate the KK sum numerically, one first has to find
the solutions to the eigenvalue equation (\ref{eq:fermeigenvals}). In the case of the up-type quark
sector, this requires determining the roots of a $6\times 6$
determinant, which in practice turns out to be intricate, because one
needs to find suitable starting points to search for the roots
which feature tiny splittings.  We obtain these starting values by diagonalizing 
a truncated mass matrix obtained in the perturbative approach \cite{Goertz:2008vr,Huber:2003tu,
delAguila:2000kb}. In Figure~\ref{fig:nug} we display
the results of our numerical calculations for one parameter point with
$M_{\rm KK} = 2 \, {\rm TeV}$. The dots correspond to the individual terms in the sum (\ref{eq:nug}) for up- and down-type
quarks, while the filled boxes indicate the values obtained by summing
up the contributions of one KK level. Results for the exotic
$\lambda$-type quarks are not shown, since they resemble those found
in the down-type quark sector. By inspection of the two panels one
immediately notices two important features of the KK
contributions. First, even though the contribution of an individual
mode can be positive and negative, the sum over an entire KK level is
strictly negative. Second, the importance of higher-level KK sums
decreases quadratically, ensuring that (\ref{eq:nug}) converges to a
finite value. This feature is indicated by the solid lines, which
represent the best fits to $1/x_{n}^2$ including the results of the
second, third, and fourth KK-level sums. The dashed lines depict
the $1/x_{n}$ behavior of the sum over a single fermion tower. 
The convergence of the total sum is guaranteed by cancellations 
between different modes of the same KK level.

\begin{figure}[!t]
\begin{center} 
\hspace{-2mm}
\mbox{\includegraphics[height=2.85in]{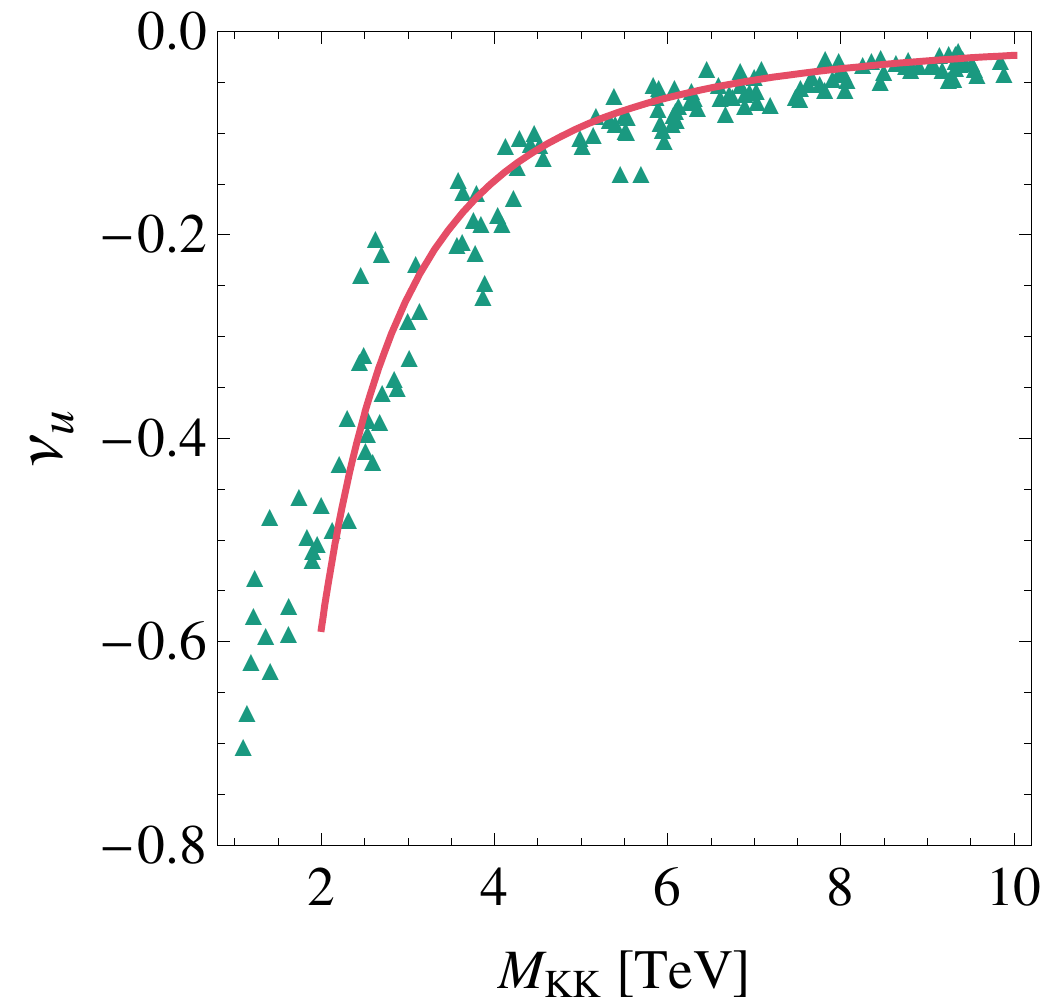}} 
\hspace{4mm}
\mbox{\includegraphics[height=2.85in]{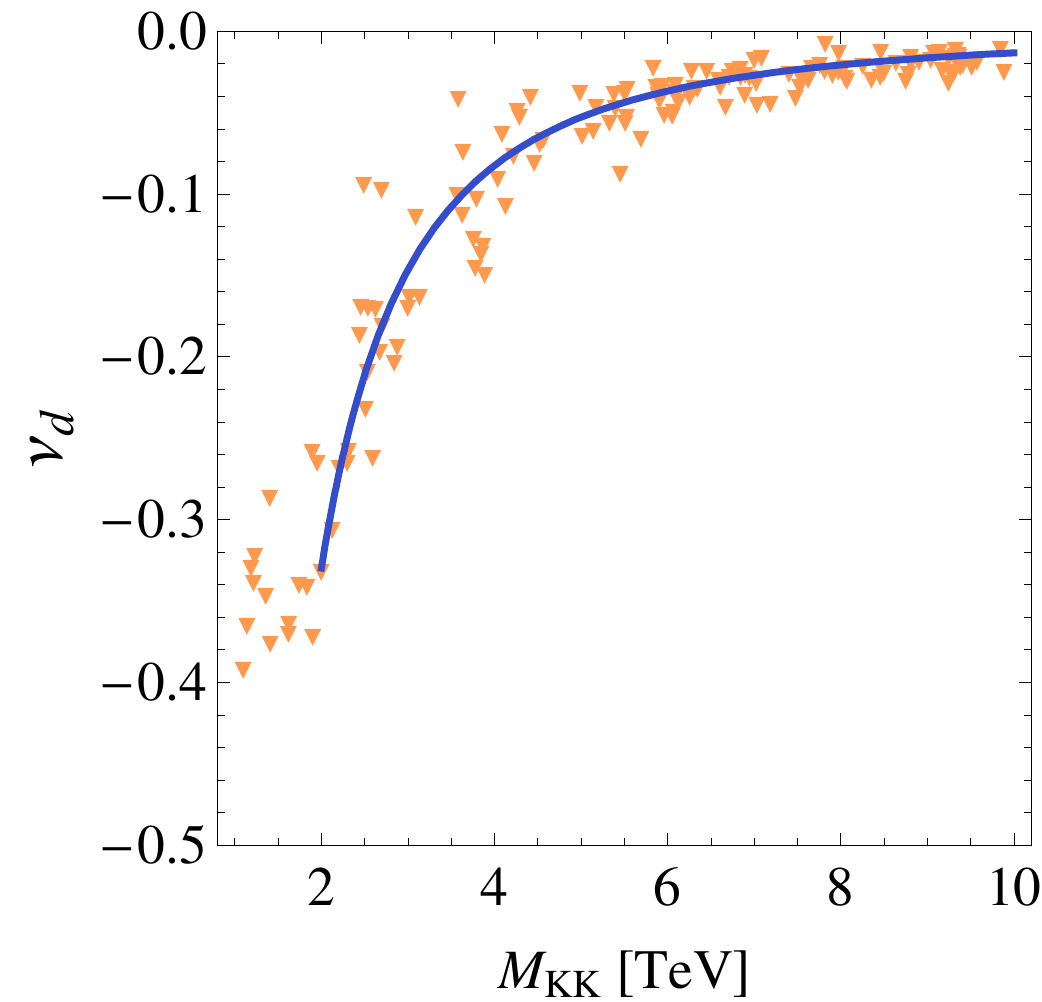}}
\parbox{15.5cm}{\caption{\label{fig:nu} Predictions for the 
    coefficients $\nu_{u}$ and $\nu_{d}$ in the custodial RS
    model. The solid lines indicate the best fits to the shown sample
    of parameter points lying in the range $\Mkk=[2,10]$\,TeV. See \cite{Casagrande:2010si} and text for details.}}
\end{center}
\end{figure}

The results for the KK sums $\nu_u$ and $\nu_d$ are shown in Figure~\ref{fig:nu}
as functions of the KK scale for our set of 150
randomly chosen parameter points. The corresponding results for 
the coefficient $\nu_\lambda$ are almost identically to those of 
$\nu_d$, and we do not show them explicitly. We see that the corrections to the
effective $hgg$ coupling that arise from triangle diagrams involving KK
quarks are all strictly negative. In the up-type quark sector the
corrections are almost a factor of 2 larger than those appearing in
the down- and $\lambda$-type quark sectors. This feature can be traced
back to the higher multiplicity of states in the former relative to
the later sectors, which suggests that $\nu_u/\nu_{d,\lambda} = 15/9
\approx 1.7$. Numerically, we find that for $\Mkk = 2 \, {\rm TeV}$
($\Mkk = 3 \, {\rm TeV}$) the average value of 
$\nu_u$ and $\nu_{d,\lambda}$ amounts to about $-0.59$ and $-0.34$
($-0.26$ and $-0.15$) with the ratio of the values being quite close
to the naive estimate. Since the leading KK-quark corrections
to the effective $hgg$ vertex decouple again as $v^2/\Mkk^2$, we parametrize
the average values of $\nu_{u, d, \lambda}$ as $a_{u, d, \lambda} \,
v^2/\Mkk^2$ and determine $a_{u, d, \lambda}$ from the
best fit to the shown sample of points restricted to the range $\Mkk =
[2, 10] \, {\rm TeV}$. The resulting numbers for the coefficients
$a_{u, d, \lambda}$ are shown in Table~\ref{tab:kappas}. 
Points with a KK scale below $2 \, {\rm TeV}$ have 
been excluded in the fit, since they depend sensitively on higher-order 
terms in $v/\Mkk$. This feature is noticeable in the plots, which show 
that for very low KK
scale the exact results for $\nu_{u,d}$ are typically above the solid
lines indicating our fits. This should be kept in mind when using the
parametrizations $a_{u,d,\lambda} \, v^2/\Mkk^2$ to calculate
$\nu_{u,d,\lambda}$ for KK scales below 2 TeV.

\begin{figure}[!t]
\begin{center} 
\hspace{-2mm}
\mbox{\includegraphics[height=2.85in]{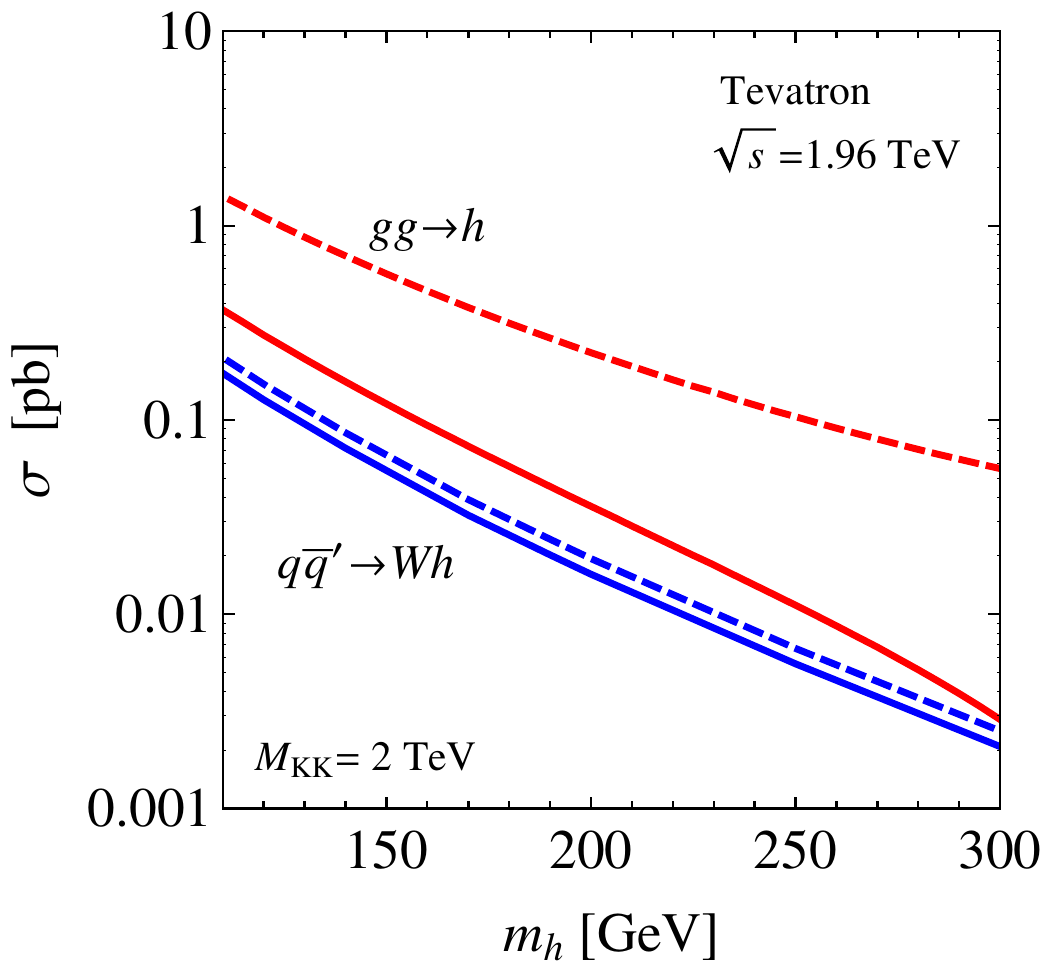}} 
\hspace{4mm}
\mbox{\includegraphics[height=2.85in]{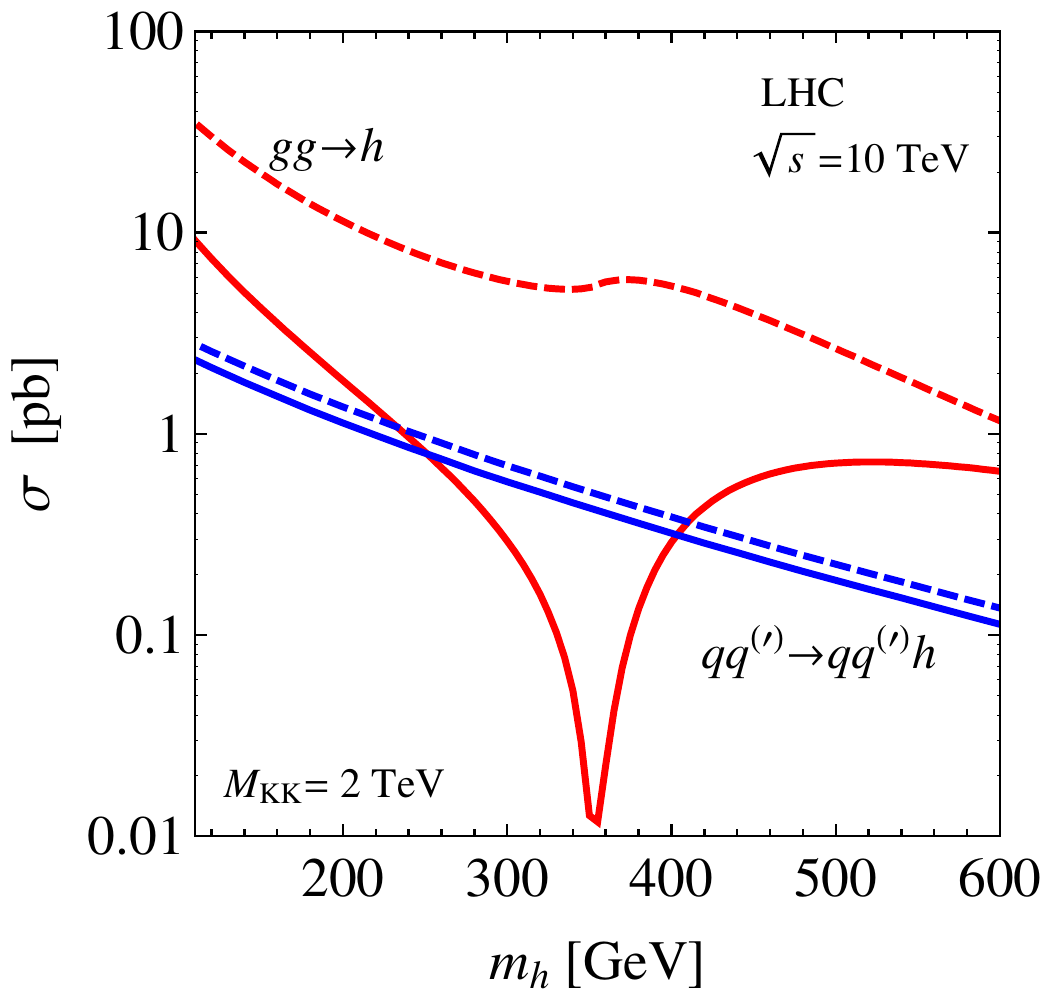}}
\vspace{2mm}
\hspace{-2mm}
\mbox{\includegraphics[height=2.85in]{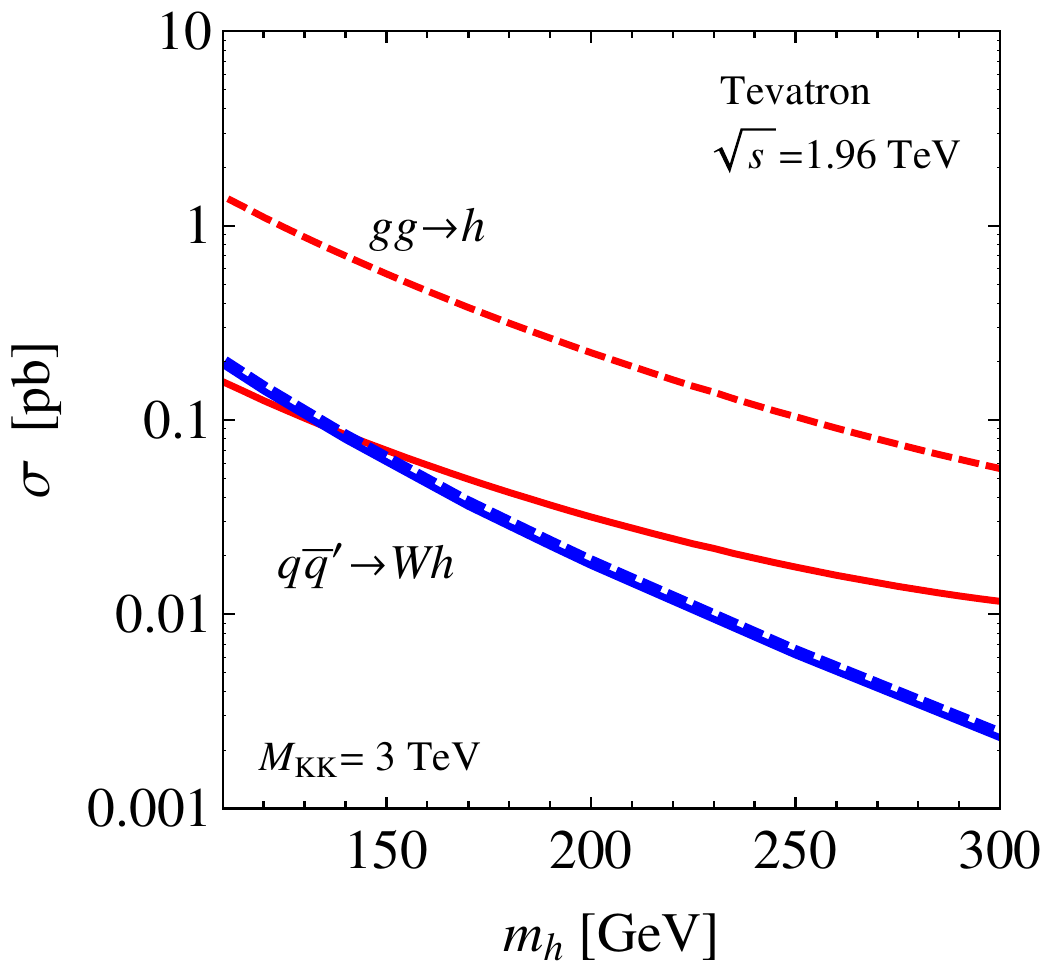}} 
\hspace{4mm}
\mbox{\includegraphics[height=2.85in]{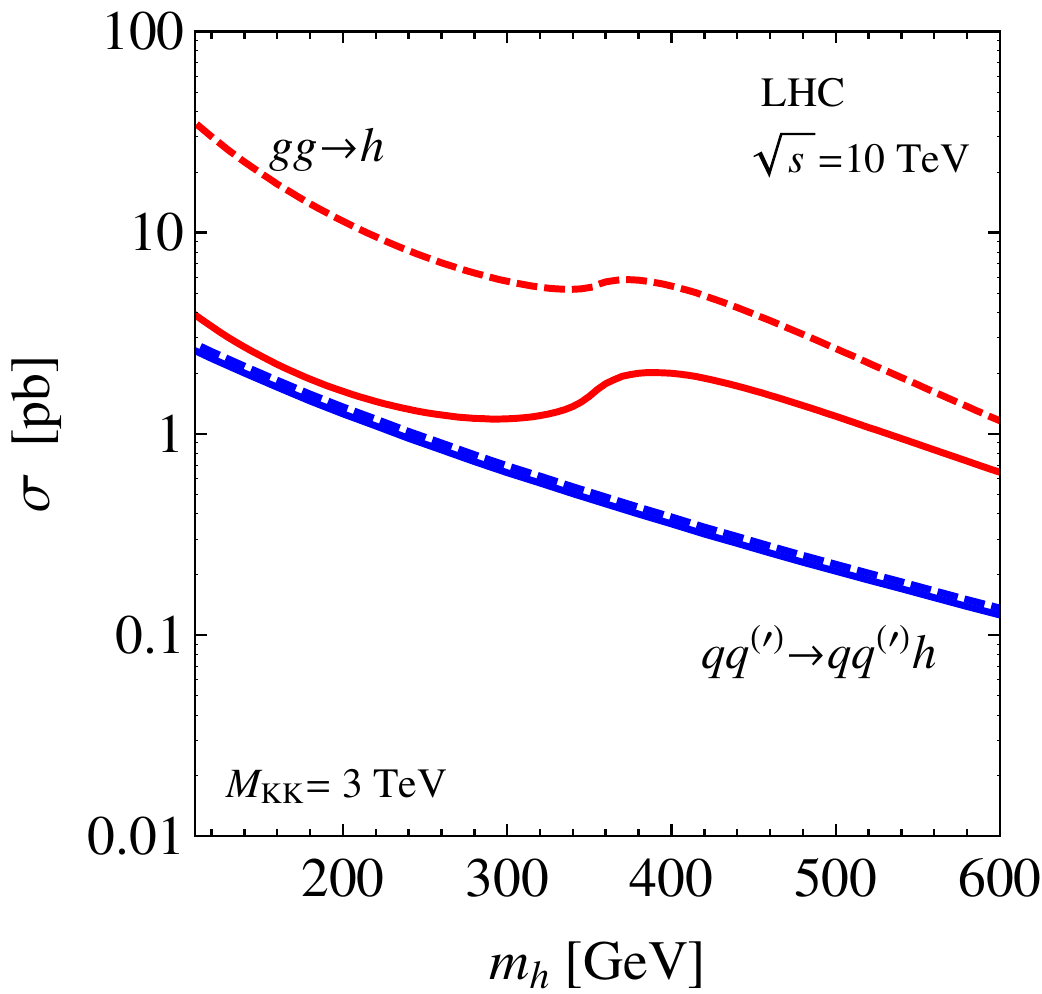}}
\parbox{15.5cm}{\caption{\label{fig:prodplots} Main Higgs-boson
    production cross sections at the Tevatron (left) and the LHC
    (right) for center-of-mass energies of $\sqrt{s} = 1.96 \, {\rm
      TeV}$ and $\sqrt{s} = 10 \, {\rm TeV}$, employing $\Mkk=2$ TeV
    (upper row) and $\Mkk=3$ TeV (lower row). The dashed lines represent the SM predictions, 
    while the solid lines correspond to the custodial RS model. In the case of the
    Tevatron the panels show gluon-gluon fusion (red) and associated
    $W^\pm$-boson production (blue), while for the LHC gluon-gluon (red) 
    and weak gauge-boson fusion (blue) are presented. See \cite{Casagrande:2010si} and text for details.}}
\end{center}
\end{figure}

Our results for the Higgs-boson production cross sections at
the Tevatron and the LHC for center-of-mass energies $\sqrt{s} = 1.96 \,
{\rm TeV}$ and $\sqrt{s} = 10 \, {\rm TeV}$ are shown in
Figure~\ref{fig:prodplots}. The SM results, on which also our
RS predictions are based, correspond to those presented in 
Figure~\ref{fig:HprodSM} and are depicted by red dashed lines. 
The solid red lines depict the RS results. The latter have
been obtained by employing (\ref{eq:rescale1}) and (\ref{eq:kappagg})
using the fit formulae for $\kappa_{t,b}$ and $\nu_{u, d, \lambda}$
discussed before. The relevant values for $a_{t,b,u,d,\lambda}$ can be
found in Table~\ref{tab:kappas}. All four panels show clearly that the
Higgs production cross sections in gluon-gluon fusion experience a
significant reduction in the custodial RS model. 
For the considered Higgs-boson masses, we
find in the case of $\Mkk = 2 \, {\rm TeV}$ ($\Mkk = 3 \, {\rm TeV}$)
suppressions that range between $-65\%$ and $-95\%$ ($-80\%$ and
$-90\%$) at the Tevatron and from $-45\%$ to almost $-100\%$ ($-45\%$ to $-90\%$)
at the LHC, see also Figure~\ref{fig:kappas}. The found depletions 
survive even at $\Mkk = 5 \, {\rm TeV}$, still reaching up to $-40 \%$ at both
colliders. Such a sensitivity to high scales is very interesting for the indirect search
for new physics. Since both the theoretical accuracy \cite{Ahrens:2008nc,
  Harlander:2002wh, Anastasiou:2002yz, Ravindran:2003um} and the
expected experimental precision \cite{Ball:2007zza, Aad:2009wy} are
at the level of 10\%, the pronounced reductions in Higgs events from
gluon-gluon fusion should be clearly visible at the LHC (for reasonably low KK scales). 
The non-trivial dependence of the RS corrections on the Higgs mass
results form an interference of zero- and KK-mode contributions.
The real part of the zero-mode amplitude increases until the $t \bar t$
threshold is reached. Above, it decreases quadratically with $m_h$,
modulo logarithmic effects. It is positive for all values of the
Higgs-boson mass. On the other hand, the real part of the amplitude
associated to the virtual exchange of KK quarks is negative and a
constant in the heavy-mass limit. Since for $\Mkk \lesssim 2 \,{\rm
TeV}$ the latter contribution is always dominant, the correction
arising from KK-quark triangle diagrams effectively flips the sign of
the real part of the total $gg \to h$ amplitude with respect to the SM
expectation for small and high Higgs masses. However, in the threshold 
region where $m_h \approx 2 \hspace{0.25mm} m_t$, the destructive 
interference between the individual contributions can become
almost perfect, leading to a strong suppression of Higgs production
via gluon-gluon fusion. This feature can be observed clearly in the
upper right panel of Figure~\ref{fig:prodplots}. Because the RS
contributions decouple rapidly for increasing KK scale, a complete
extinction of the sum of individual amplitudes is not possible for
$\Mkk \gtrsim 2 \,{\rm TeV}$. In this case, the zero-mode contribution
to $gg \to h$ dominates, and the Higgs-mass dependence of the RS prediction 
is similar to the one of the SM result. 
It is important to stress that, in spite of the in principle many parameters in the fermion 
sector of the custodial RS model, the shown results for the Higgs-boson production 
cross section depend to first order only on the overall KK-mass scale. This claim 
is supported by the narrow spread of scatter points in Figure~\ref{fig:nu}.
 
\begin{table}[!t]
\begin{center}
\begin{tabular}{|c|c|c|} 
  \hline
  $a_t$ & 
  $a_b$ & 
  $a_t^V$ \\ \hline
  $15.08$ & 
  $9.08$ & 
  $3.63$ \\
  \hline 
\end{tabular}

\vspace{4mm}

\begin{tabular}{|c|c|c|c|c|c|c|} 
  \hline
  $a_u$ & 
  $a_d$ &
  $a_\lambda$ &
  $a_{\gamma Z}^u$ & 
  $a_{\gamma Z}^d$ &
  $a_{\gamma Z}^\lambda$ &
  $a_{\gamma Z}^W$\\ \hline
  $-38.80$ & 
  $-21.80$ & 
  $-22.58$ & 
  $-46.46$ & 
  $17.98$ & 
  $-6.38$ & 
  $10.76$ \\
  \hline 
\end{tabular}
\end{center}
\parbox{15.5cm}{\caption{\label{tab:kappas} Fit coefficients entering the various 
    contributions to Higgs-boson production and decay in units of 
    $v^2/\Mkk^2$. Corrections due to zero and KK modes are 
    displayed in the upper and lower table, respectively.}
}
\end{table}

Let us now have a look on the subleading production mechanisms
at the main hadron colliders. Here, the important LHC channel
of weak gauge-boson fusion, $q q^{(\prime)} \to qq^{(\prime)} V^{\ast}V^{\ast} \to qq^{(\prime)} h$ 
with $V = W,Z$, receives only moderate
corrections of around $-10\%$ ($-5 \%$) for $\Mkk = 2 \, {\rm TeV}$
($\Mkk = 3 \, {\rm TeV}$). The same reduction will affect associated
$W^\pm$-boson production, $q \bar q^{\hspace{0.25mm} \prime} \to W^\ast
\to Wh$ at the Tevatron. The RS predictions for the corresponding 
production cross sections are illustrated by the
solid blue lines in the right and left panels of
Figure~\ref{fig:prodplots}, respectively. The corresponding SM
predictions, depicted by the blue dashed lines, are those of Figure 
\ref{fig:HprodSM}. Finally, the cross section of associated
top-quark pair production, $q \bar q \to t \bar t^\ast \to t \bar t
h$, will also experience a reduction. For values of the KK scale in
the ballpark of $2 \, {\rm TeV}$, this suppression can amount up to
$-40\%$. Since all these processes happen at tree level, the RS 
predictions have all been obtained by a simple rescaling of the 
SM results. In that context, one has to consider the VEV shift (\ref{eq:VEVshift}) which partially
cancels the depletion due to the mixing of the weak gauge bosons with their KK excitations.
This effect has been included in deriving the numerical results above.

In summary, we find that the main Higgs-boson production modes at
hadron colliders are suppressed in the custodial RS model relative to
the SM. Clearly this affects Higgs searches, as we will
detail below. Suppression effects in $gg \to h$ were also reported in
\cite{Azatov:2009na,Falkowski:2007hz,Cacciapaglia:2009ky}.
In this context, see also \cite{Espinosa:2010vn} for a detailed analysis 
of Higgs-boson production cross sections and decay rates in a related 
general setup. A direct numerical comparison of the results of these 
publications with our findings is however not possible, since
\cite{Azatov:2009na} only included zero-mode corrections, while
\cite{Falkowski:2007hz, Cacciapaglia:2009ky,Espinosa:2010vn} studied 
RS variants that differ from the set-up considered here. 
In \cite{Azatov:2010pf}, an enhancement of $gg\to h$ in a custodial RS variant has 
been found. However for a sufficiently IR-localized Higgs sector, this effect is 
caused by UV physics, above the cutoff $\Lambda_{\rm UV}$ of the RS EFT.
Considering a Higgs profile with a finite width $\sim \eta$, the effect is
driven by KK excitations with a mass of $m_n\sim 1/\eta$. It vanishes in the limit
of a small width, if the KK sum is cutoff at $\Lambda_{\rm UV}<\infty$.

In \cite{Djouadi:2007fm} the authors studied corrections to gluon-gluon
fusion arising from virtual exchange of very light fermionic KK
modes. It has been found that for a heavy bottom-quark partner
with a mass $m_{b^\prime}$ of a few hundred GeV the Higgs-boson
production cross section via $gg \to h$ can be significantly enhanced.
However, in order to achieve $m_{b^\prime} \ll \Mkk$ with the embedding 
of quarks as chosen in (\ref{eq:multiplets}), the $P_{LR}$ symmetry has 
to be broken strongly via the bulk mass parameters of the ${\cal T}_{1}$
multiplets by choosing $c_{{\cal T}_{1i}}$ rather far away from $c_{{\cal T}_{2i}}$.
While for $c_{{\cal T}_{1i}} > 1/2$ it is possible to achieve $\nu_d > 0$ and thus an enhancement of the $gg
\to h$ cross section, such choices of parameters do not naturally
reproduce the measured mass spectrum of the SM quarks
for anarchic Yukawa couplings. If on the other hand
$c_{{\cal T}_{1i}} < 1/2$, the correction $\nu_d$ remains strictly negative, and as a result the
$gg \to h$ channel experiences a reduction. Moreover,
choices for $c_{{\cal T}_{1i}}$, corresponding to a strong breaking of
the $P_{LR}$ symmetry, lead generically to a
sizable negative shift in the $Z b_L \bar b_L$ coupling through
(\ref{eq:deltaD2}), which is problematic in view of the stringent
constraints arising from the $Z \to b \bar b$ pseudo observables.

\subsection{Higgs-Boson Decay}
\label{sec:higgsdecay}

We now move on to study the decay modes of the Higgs boson. In this
context, we will consider all processes with quarks and gauge bosons
in the final state that can receive important RS corrections and have
a branching fraction larger than $10^{-4}$. As we have not explicitly
specified the embedding of the fermions in the lepton sector, we
ignore decays into taus and muons, which we however expect to be SM-like,
due to the rather UV localized profiles. Furthermore, we will not include
loop contributions of KK leptons in our analysis of the $h \to
\gamma\gamma$ and $h \to \gamma Z$ decay channels but rather estimate
their impact.

In order to be able to calculate the decay rates of the Higgs boson
into massive gauge bosons, we still need to give the RS
corrections to the $WWh$, $ZZh$, and $WWZ$ tree-level vertices. Due to
the unbroken $U(1)_{\rm EM}$ gauge group, the $WW\gamma$ coupling is
unchanged with respect to the SM to all orders in $v^2/\Mkk^2$.  The
weak couplings involving the Higgs boson are derived from the cubic
and quartic interactions given by (\ref{eq:D}). In unitary gauge, the
relevant terms in the Lagrangian read
\begin{equation} 
  {\cal L}_{\rm 4D} \ni \left(h^2+2\,h\,v_{\rm RS_C}\right)
  \left [ \, \frac{g_L^2}4 \left (1 - \Delta g_h^W \right ) W^+_\mu 
    \hspace{0.25mm} W^{-\mu}+\frac{g_L^2+g_Y^2}{8} \left (1 - 
      \Delta g_h^Z \right ) Z_\mu \hspace{0.25mm} Z^{\mu} \, \right ] ,
\end{equation}
where 
\begin{align}
  \begin{split}\label{eq:Hbosoncoupl}
    \Delta g_h^V =x_V^2\left[L\left(1+\frac{s_V^2}{c_V^2}\right)
      -1+\frac 1{2L}\right]+\ord\left(x_V^4\right) ,
  \end{split}
\end{align}
and $x_V \equiv m_V/\Mkk$ for $V = W,Z$. Due to the $P_{LR}$
symmetry (\ref{PLR}), one has $s_W^2/c_W^2 = 1$ and $s_Z^2/c_Z^2 = 1 -
2 \hspace{0.25mm} s_w^2\,$, which implies that the leading correction
due to $\Delta g_h^{W,Z}$ takes the form $-2 \hspace{0.5mm}
m_W^2/\Mkk^2 \hspace{0.5mm} L$. For $\Mkk = 2 \, {\rm TeV}$ ($\Mkk = 3
\, {\rm TeV}$) these terms amount to about $-10\%$ ($-5\%$). 
However, when comparing the size of the $WWh$ coupling to the value in the SM, one also has to take
into account the fact that the VEV $v_{\rm RS_C}$ differs from $v_{\rm SM}$.

The partial decay widths $\Gamma (h \to f)$ of the Higgs boson
decaying to a final state $f$ are again obtained by rescaling the SM
decay widths. We use
\begin{equation} \label{eq:Gammahtof}
  \Gamma (h \to f)_{\rm RS} = \left | \kappa_f \right |^2 \,
  \Gamma (h \to f)_{\rm SM} \,, 
\end{equation}
where the VEV shift enters the expressions
\begin{equation} 
  \kappa_{W} = \frac{v_{\rm RS_C}}{v_{\rm SM}}\left(1 - \Delta g_h^W\right) \,, \qquad 
  \kappa_{Z} = \frac{v_{\rm RS_C}}{v_{\rm SM}}\left(1 - \Delta g_h^Z\right) \,,
\end{equation}
for the decay of the Higgs boson into a pair of $W^\pm$ and $Z$
bosons via the ratio $v_{\rm RS_C}/v_{\rm SM}$. This ratio can be obtained from (\ref{eq:VEVshift}).
While the corresponding corrections coming with the subleading terms $\propto \Delta g_h^{W,Z}$ in the equation above are of $\ord\left(x_V^4\right)$
and will be neglected in the following,
the corrections to the leading terms have to be considered at $\ord\left(x_V^2\right)$.
We arrive at
\begin{equation} 
  \kappa_{W} = 1 - \frac{3}{4}x_W^2\left[2L
      -1+\frac 1{2L}\right]+\ord\left(x_W^4\right) \,, \quad 
  \kappa_{Z} = 1  - \frac{3}{4} x_W^2\left[2L
      -\frac{4-c_w^2}{3c_w^2}(1-\frac{1}{2L})\right]+\ord\left(x_Z^4\right) \,,
\end{equation}
where we have used $g_L=g_R$. In consequence, the depletion due to the overlap 
of profiles is weakened by the VEV shift to the few per cent level.

Note that in the minimal RS model the expressions (\ref{eq:Hbosoncoupl})
hold in the limit $s_{W,Z} \rightarrow 0$.  Our finding that the couplings $WWh$
and $ZZh$ experience a reduction from their SM expectations confirms
the model-independent statements made in \cite{Low:2009di}.
The parameters $\kappa_{g, t, b}$ for
decays into two gluons, top or bottom quarks have already been given
in (\ref{eq:kappagg}) and (\ref{tbratios}).
 
In Figure~\ref{fig:hprodec} the diagrams inducing the decay into a pair
of heavy quarks and massive gauge bosons are shown on the right in the
top row. Apart from the change in the $ht \bar t$ coupling, we neglect
RS corrections to the three-body decay $h \to tt^\ast \hspace{0.5mm}
(WW^\ast) \to tbW$. Relative to the two-body mode $h \to t \bar
t$ this amounts to a correction of (far below) 1\% in the SM. Given the
smallness of this effect, the omission of possible new-physics effects
in the $Wtb$ coupling that would affect the $h \to tbW$ channel is well justified.

The RS correction entering the Higgs decay into two photons can be accounted for by
employing
\begin{equation} \label{kappagamma} 
\kappa_{\gamma} =
  \frac{{\displaystyle \sum}_{i = t, b} \; N_c \hspace{0.5mm} Q_i^2
    \hspace{0.5mm} \kappa_i \hspace{0.25mm} A_{q}^h (\tau_i) +
    \kappa_{W} \hspace{0.25mm} A_W^h (\tau_W) + {\displaystyle
      \sum}_{j = u, d, \lambda} \; N_c \hspace{0.5mm} Q_j^2
    \hspace{0.5mm} \nu_j + \nu_\gamma^W}{{\displaystyle \sum}_{i = t,
      b} \; N_c \hspace{0.5mm} Q_i^2 \hspace{0.5mm} A_{q}^h (\tau_i) +
    A_W^h (\tau_W)}
\end{equation}
in (\ref{eq:Gammahtof}), where $N_c =3$, $Q_{t,u} = 2/3$, $Q_{b,d} =
-1/3$, $Q_\lambda = 5/3$, and $\tau_W \equiv 4 \hspace{0.25mm} m_W^2/
m_h^2$. The explicit expression for the form factor $A_{W}^h
(\tau_W)$, encoding the $W^\pm$-boson contribution, can be found in
Appendix~\ref{app:formfactors}. The first, second, and third terms in
the numerator describe the effects of virtual heavy-quark, $W^\pm$-boson,
and KK-quark exchange, respectively. The corresponding one-loop graphs
are shown on the left in the bottom row of Figure~\ref{fig:hprodec}
and in the center of Figure~\ref{fig:hkkcontr}. Note that the amplitude
$A_{W}^h (\tau_W)$ interferes destructively with the quark
contribution $A_{q}^h (\tau_i)$, falling from $-21/4$ for $\tau_W \to
\infty$ to $-15/4 - 9 \pi ^2/16$ at the $WW$ threshold $\tau_W = 1$
and finally approaching $-3/2$ in the limit $\tau_W \to 0$. 
Within the SM, the $W^\pm$-boson contribution to the $h\to
\gamma \gamma$ decay amplitude is always dominant below threshold.

Let us estimate the size of the leptonic
contributions which have not been included in (\ref{kappagamma}).
While the precise impact of these effects depends on the exact realization 
of the lepton sector, which we have not specified, it is possible to predict 
their relative sign as well as their rough magnitude.
Generalizing the result (\ref{kappagamma}) to
include contributions from triangle diagrams with KK leptons only
requires to perform the replacement
\begin{equation} \label{kkleptonsinkappaA}
  {\displaystyle \sum}_{j = u, d, \lambda} \; N_c
  \hspace{0.5mm} Q_j^2 \hspace{0.5mm} \nu_j \; \to \; {\displaystyle
    \sum}_{j = u, d, \lambda} \; N_c \hspace{0.5mm} Q_j^2
  \hspace{0.5mm} \nu_j + Q_l^2 \nu_l \, = \, \frac{4 \hspace{0.25mm} \nu_u}{3}  +  
  \frac{\nu_d}{3} +\frac{25 \hspace{0.25mm} \nu_\lambda}{3} + \nu_l \,,
\end{equation}
where $\nu_u \approx 2 \hspace{0.25mm} \nu_d \approx 2 \hspace{0.25mm}
\nu_\lambda$ and the parameter $\nu_l$ encodes the effects due to 
KK-lepton loops. Under the reasonable assumption that $\nu_l \approx
\nu_u/2$, we conclude from (\ref{kkleptonsinkappaA}) that the KK
lepton contribution to the $h \to \gamma \gamma$ amplitude amounts to
approximately $10\%$ of the KK quark corrections and interferes
constructively with the latter. In consequence, an omission of KK lepton 
effects in the calculation of $\kappa_\gamma$ has only a minor numerical
impact.

The sum entering the quantity $\nu_\gamma^W$, representing the one-loop contribution due
to $W^\pm$-boson KK modes, can be performed analytically in the
decoupling limit. The corresponding Feynman diagram is displayed on
the very right in Figure \ref{fig:hkkcontr}. Employing the results
for the KK sums derived in Section~\ref{sec:KKsum}, we obtain
\begin{equation}\label{eq:nugammaW}
\begin{split}
  \nu_\gamma^W & = \frac{2 \pi x_W^2 \left ( g_L^2 + g_R^2 \right
    )}{g_L^2} \, \sum_{n = 1}^\infty \, \frac{\vec{d}_W^{\, T}
    \hspace{0.5mm} \vec{\chi}_n^{\, W} (1) \, \vec{\chi}_n^{\, W \, T}
    (1) \hspace{0.5mm} \vec{d}_W}{\big (x_n^W \big)^2} \,
  A_W^h (\tau_n^W) +\ord(x_W^4)\\
  & = \frac{2 \pi x_W^2 \left ( g_L^2 + g_R^2 \right )}{g_L^2} \;
  \vec{d}_W^{\, T} \hspace{0.5mm} \big [ {\bm \Sigma}_W^{(1)} (1,1)- {\bm
    \Pi}_W (1,1)\big ] \hspace{0.5mm} \vec{d}_W \, \left (
    -\frac{21}{4} + {\cal O} \left (1/\tau_n^W \right ) \right ) \\
  & =-\,\frac{21}8\, \Delta g_h^W \left(1+{\cal O} \left (1/\tau_n^W
    \right )\right) ,
\end{split}
\end{equation}
where $\vec{d}_W = (c_W, -s_W )^T$ and $\tau_n^W \equiv 4
\hspace{0.25mm} \big ( m_n^{W} \big)^2/m_h^2$. Since already $m_1^W
\approx 2.5 \hspace{0.25mm} {\Mkk} \gg m_h$, the terms suppressed by
powers of $\tau_n^W$ in (\ref{eq:nugammaW}) can be ignored in
practice. Note that in $\nu_\gamma^W $,
the VEV shift contributes only at $\ord(x_W^4)$.
The result for $\Delta g_h^W$ has been given in (\ref{eq:Hbosoncoupl}).

The last missing ingredient that is needed to compute the branching fractions for
all sought decay channels in the RS model is the rescaling factor for a decay into a photon and a $Z$ boson.
It reads
\begin{equation} \label{kappagammaZ} 
  \kappa_{\gamma Z} = \frac{{\displaystyle \sum}_{i = t, b} \; N_c \,
    \displaystyle \frac{2 \hspace{0.5mm} Q_i \hspace{0.25mm} v_i}{c_w}
    \hspace{0.5mm} \kappa_i \hspace{0.5mm} \kappa_i^V \hspace{0.25mm}
    A_{q}^h (\tau_i, \lambda_i) + \kappa_{W} \hspace{0.25mm} A_W^h
    (\tau_W, \lambda_W) + {\displaystyle \sum}_{j = u, d,\lambda} \; N_c
    \, \displaystyle \frac{2 \hspace{0.5mm} Q_j \hspace{0.25mm} v_j}{c_w}
    \hspace{0.5mm} \nu^j_{\gamma Z} + \nu_{\gamma Z}^W}{{\displaystyle
      \sum}_{i = t, b} \; N_c \, \displaystyle \frac{2 \hspace{0.5mm} Q_i
      \hspace{0.25mm} v_i}{c_w} \, A_{q}^h (\tau_i, \lambda_i) + A_W^h
    (\tau_W, \lambda_W)} \,.
\end{equation}
Here $v_i \equiv T_L^{3\, i} - 2
\hspace{0.25mm} s_w^2 \hspace{0.25mm} Q_i$, and $\lambda_i \equiv 4
\hspace{0.25mm} m_i^2/m_Z^2$ for $i = t, b, W$. The amplitudes
$A_{q,W}^h (\tau_i, \lambda_i)$ encoding the effects of virtual quarks
and $W^\pm$ bosons in $h \to \gamma Z$ are collected in
Appendix~\ref{app:formfactors}. The corresponding Feynman diagrams are
shown on the right in the bottom row of Figure~\ref{fig:hprodec}. Like
in the case of $h \to \gamma \gamma$, the SM decay rate for $h \to
\gamma Z$ is in large parts of the parameter space dominated by the
$W^\pm$-boson loop contribution. The function $A_W^h (\tau_W, \lambda_W)$ 
rises from around $4.6$ to $9.8$ between $\tau_W \to \infty$ and 
$\tau_W = 1$, and then falls to approximately $0.6$ in the limit $\tau_W \to 0$.
On the other hand, one has $A_q^h (\tau_i, \lambda_i) =
-1/3$ for $\tau_i , \lambda_i \to \infty$ and $A_q^h (\tau_i,
\lambda_i) = 0$ for $\tau_i , \lambda_i \to 0$. 
 
\begin{figure}[!t]
\begin{center} 
\hspace{-2mm}
\mbox{\includegraphics[height=2.85in]{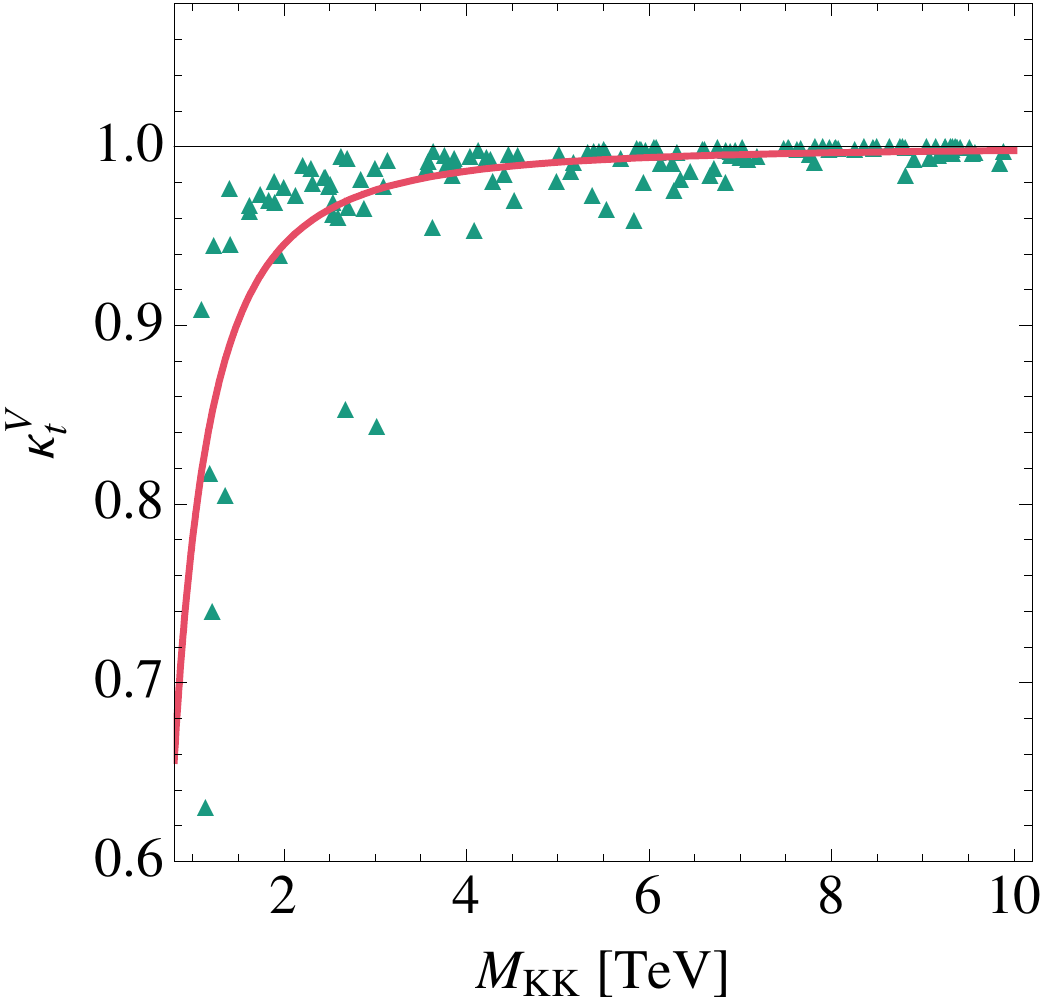}} 
\hspace{4mm}
\mbox{\includegraphics[height=2.85in]{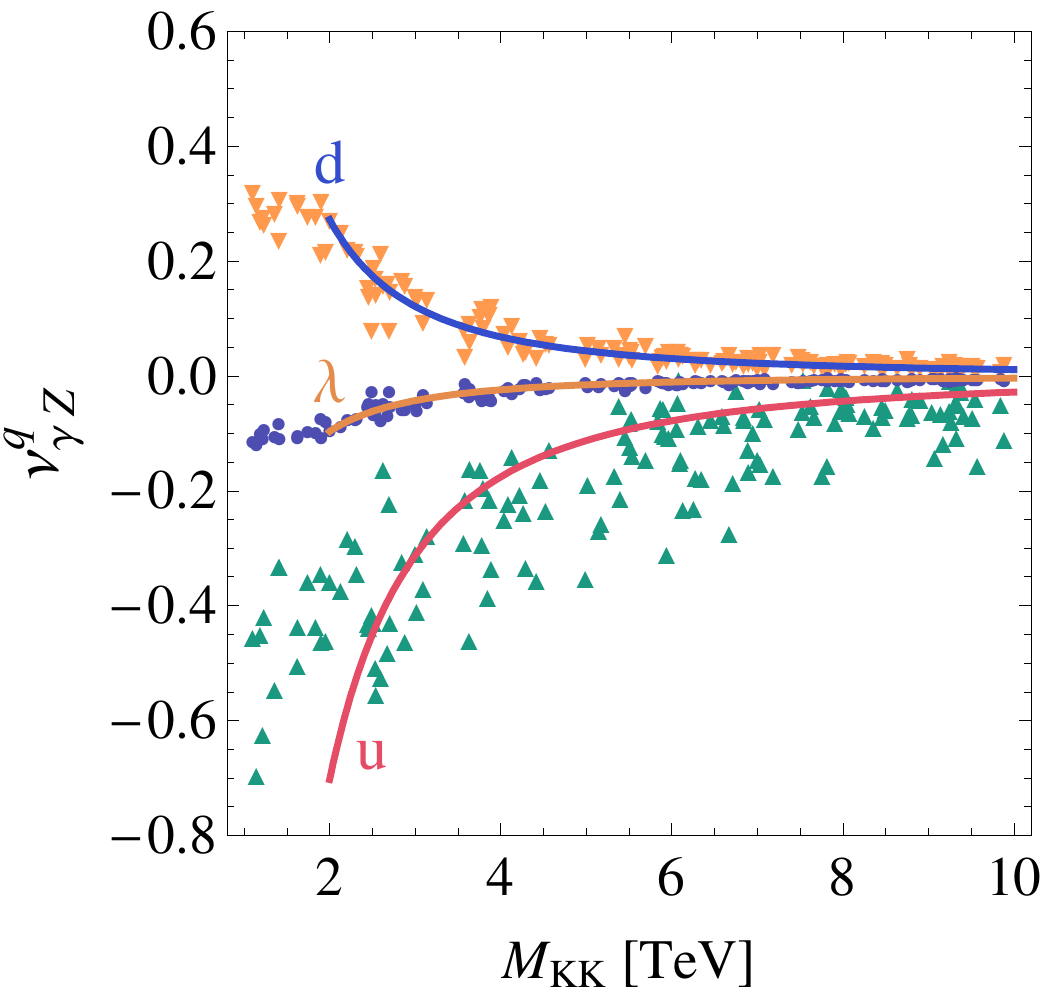}}
\parbox{15.5cm}{\caption{\label{fig:vectorZplot} Left: Predictions for the
    vector couplings of the $Z$ boson to top and bottom quarks
    in the custodial RS model. 
    Right: Different types of KK-quark contributions to the
    effective $h\gamma Z$ coupling. The solid lines show fits to the 
    scatter points. See \cite{Casagrande:2010si} and text for details.}}
\end{center}
\end{figure}

The first term in the numerator of (\ref{kappagammaZ}) depends on the
ratios
\begin{equation} \label{eq:kappaiV}
  \kappa_{t}^V = \frac{\big ( g_L^u \big )_{33} + \big ( g_R^u \big )
    _{33}}{v_t} \,, \qquad \kappa_{b}^V = \frac{\big ( g_L^d \big )_{33} +
    \big ( g_R^d \big ) _{33}}{v_b} \,,
\end{equation}
which quantify the relative shift in the vector coupling of the $Z$
boson to top and bottom quarks. In the left panel of
Figure~\ref{fig:vectorZplot} we show the predictions for
$\kappa^V_{t}$ versus $\Mkk$ for our set of 150 parameter
points. It is evident from the plot that the vector coupling of the
$Z$ boson to top quarks is always reduced in the custodial RS model
relative to the SM. Numerically, the suppression amounts to a moderate
effect of $-5\%$ ($-2.5\%$) for $\Mkk = 2 \, {\rm TeV}$ ($\Mkk = 3 \,
{\rm TeV}$). In contrast, the $Z$-boson coupling to bottom-quark pairs
is larger than its SM value, but numerically the resulting
effects turn out to be negligibly small due to the custodial
protection mechanism. Consequently, we will set $\kappa_b^V$ to 1 in
our numerical analysis. Parametrizing the average value of the
relative shift $\kappa_{t}^V$ by $(1-a_{t}^V \, v^2/\Mkk^2)$ the
coefficient $a_{t}^V$ can again be determined through a fit. Employing
the shown set of parameter points, we obtain the value for $a_{t}^V$
given in Table~\ref{tab:kappas}.

The second term in the numerator of (\ref{kappagammaZ}) encodes the
contribution arising from the $W^\pm$-boson triangle graph. The calculation 
of this zero-mode contribution is greatly simplified by the fact that at ${\cal O} (v^2/\Mkk^2)$ 
the triple gauge-boson vertex involving two $W^\pm$- and one $Z$-boson fields does not 
receive corrections in the RS model (regardless of the specific gauge
group). This can be easily seen by employing that
\begin{equation} \label{eq:one}
  \frac{2 \pi}{L} \int_{\epsilon}^1 \!  \frac{dt}{t} \, \chi^{(+)}_0
  (t) \, = \sqrt{2 \pi} + {\cal O} \left ( \frac{v^4}{\Mkk^4} \right
  )\,,
\end{equation} 
and $\big [(\vec{A}_0^a)_2 \hspace{0.5mm} \chi_0^{(-)} (t) \big
]^2 = {\cal O} (v^4/\Mkk^4)$. The expressions for $\chi_0^{(\pm)} (t)$
and $\vec{A}_0^a$ necessary to derive these results can be found in
(\ref{eq:expprof}) and (\ref{eq:vecA0a}).
By the same line of reasoning, it is also readily seen that all
quartic gauge-boson vertices first differ at order $v^4/\Mkk^4$ from
the corresponding SM expressions. In view of this extra suppression,
we will set the triple gauge-boson couplings of the zero modes to
their SM values when evaluating the Higgs-boson branching
fractions. In this approximation the effect of virtual $W^\pm$-boson
exchange to (\ref{kappagammaZ}) is simply given by the combination
$\kappa_W A_W^h (\tau_W, \lambda_W)$, which up to the different form
factor resembles the form of the corresponding term in
(\ref{kappagamma}).

The third term in the numerator of (\ref{kappagammaZ}) describes the
contribution to the $h \to \gamma Z$ amplitude stemming from the
virtual exchange of KK quarks. The corresponding one-loop diagram is
displayed in the middle of Figure \ref{fig:hkkcontr}. In the up-type
quark sector we find
\begin{equation} \label{eq:nugammaZu}
\vspace{-4.5mm}
\begin{split}
  \nu_{\gamma Z}^u =& \,v_{\rm SM} \;\sum_{n = 4}^\infty \;
  \frac{{\rm Re}\left[(g_h^u)_{nn}\right]}{m_n^u} \, \kappa_n^{u, V} A_{q}^h (\tau_n^u,
  \lambda_n^u) \\ 
  =& \, \frac{2 \pi}{\epsilon L}\, \frac{v_{\rm SM}}{v_{\rm RS_C}}\sum_{n = 4}^\infty \; \frac{{\rm Re}\left[\vec
    a_{n}^{U\dagger}\, \bm{C}_{n}^{U} (\pi^-) \left(\displaystyle
      \bm{1}-\frac{v_{\rm RS_C}^2}{3\,\Mkk^2} \bm{\tilde Y}_{\vec u}\bm{\bar
        Y}_{\vec u}^\dagger\right) \bm{S}_{n}^U(\pi^-)\, \vec
    a_{n}^{\hspace{0.25mm} U} \right]}{x_n^u}\hspace{0.5mm} \,\kappa_n^{u, V}
  A_{q}^h (\tau_n^u, \lambda_n^u) \,, \hspace{6mm}
\end{split}
\vspace{-2.5mm}
\end{equation}
where $\kappa_n^{u, V}$ denotes the relative strength of vector
coupling of the $Z$ boson to the $n^{\rm th}$ up-type quark KK mode
defined in analogy to (\ref{eq:kappaiV}), and $\lambda_n^u \equiv 4
\hspace{0.25mm} \big(m_n^u\big)^2/m_Z^2$. Similar expressions apply in
the case of down- and $\lambda$-type quark KK modes. 
Note again that the VEV shift $v_{\rm SM}/v_{\rm RS_C} \neq 1$ does not contribute to $\nu_{\gamma Z}^u$ 
at $\ord(v^2/\Mkk^2)$.
The numerical results for
$\nu_{\gamma Z}^{u}$, $\nu_{\gamma Z}^{d}$, and
$\nu_{\gamma Z}^{\lambda}$, corresponding to our set of 150 random model
parameter points, are depicted in the right panel of
Figure~\ref{fig:vectorZplot}. The solid lines indicate
the best fit of the form $a_{\gamma Z}^{u, d, \lambda} \hspace{0.5mm}
v^2/\Mkk^2$ to the sample of points with KK scales in the range $[2,
10] \, {\rm TeV}$.  As before, points with $\Mkk < 2 \, {\rm TeV}$
have been excluded in the fit, since they are subject to significant
higher-order corrections.  The corresponding coefficients $a_{\gamma
  Z}^{u, d, \lambda}$ can be found in Table~\ref{tab:kappas}. The
average values of $\nu_{\gamma Z}^{u}$, $\nu_{\gamma
  Z}^{d}$, and $\nu_{\gamma Z}^{\lambda}$ obtained from the fit
formulae are $-0.70$ ($-0.31$), $0.27$ ($0.12$), and $-0.10$ ($-0.04$)
for $\Mkk = 2 \, {\rm TeV}$ ($\Mkk = 3 \, {\rm TeV}$), respectively.

\begin{figure}[!t]
\begin{center}
\mbox{\includegraphics[height=2.85in]{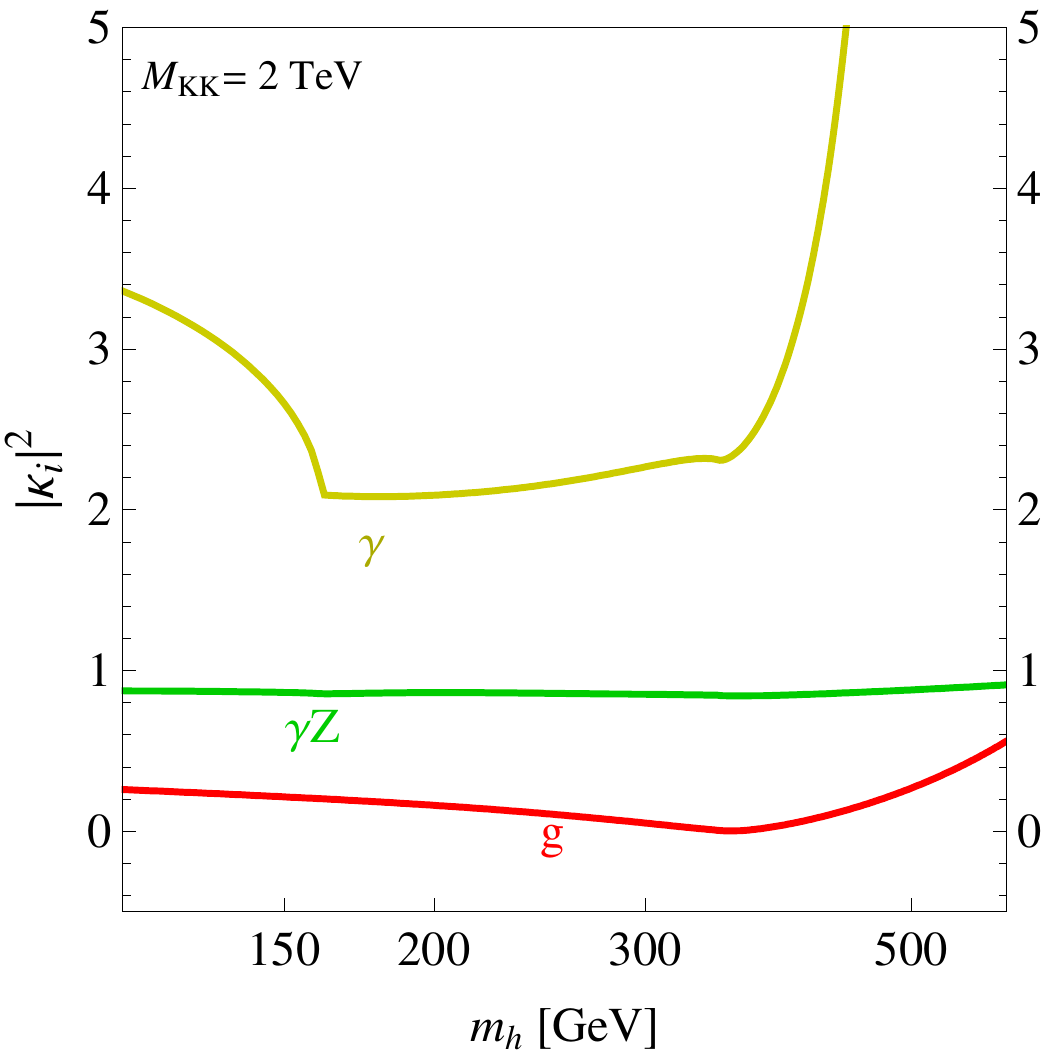}}
\hspace{6mm}
\mbox{\includegraphics[height=2.85in]{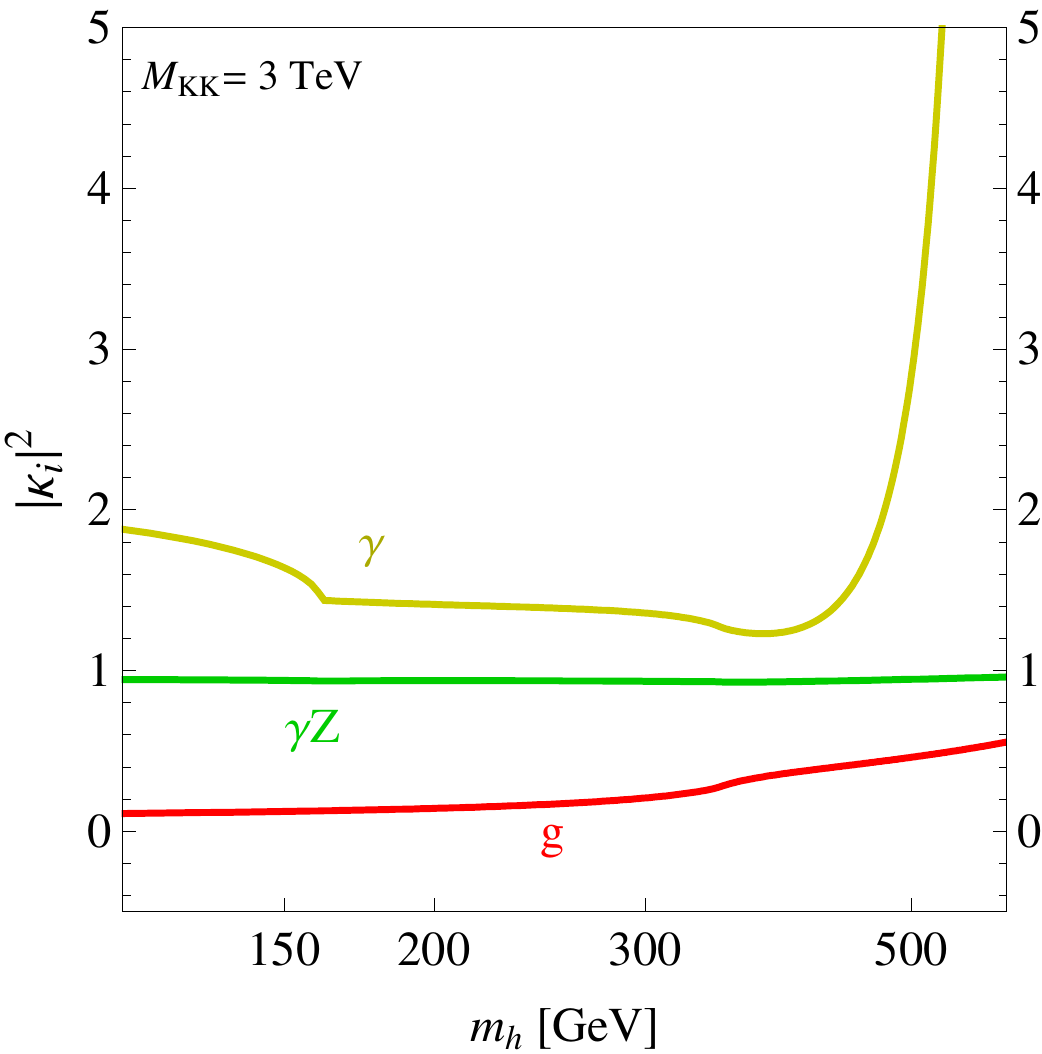}}
\parbox{15.5cm}{\caption{\label{fig:kappas} RS-correction factors
    $|\kappa_g|^2$ (red line), $|\kappa_\gamma|^2$ (yellow line), and
    $|\kappa_{\gamma Z}|^2$ (green line) as functions of the
    Higgs-boson mass, employing $\Mkk = 2 \, {\rm TeV}$ (left panel)
    and $\Mkk = 3 \, {\rm TeV}$ (right panel). See \cite{Casagrande:2010si} and text for details.
  }}
\end{center}
\end{figure}

Denoting contributions from KK-lepton triangle graphs by
$\nu_{\gamma Z}^l$, they can be included in (\ref{kappagammaZ}) via the simple replacement
\begin{equation} \label{kkleptonsinkappaAZ} 
      {\displaystyle \sum}_{j = u, d,\lambda} \; N_c \, \displaystyle
      \frac{2 \hspace{0.5mm} Q_j \hspace{0.25mm} v_j}{c_w} \hspace{0.5mm}
      \nu^j_{\gamma Z} \; \to \; \,  {\displaystyle \sum}_{j = u,
        d,\lambda} \; N_c \, \displaystyle \frac{2 \hspace{0.5mm} Q_j
        \hspace{0.25mm} v_j}{c_w} \hspace{0.5mm} \nu^j_{\gamma Z} +
      \displaystyle \frac{2 \hspace{0.5mm} Q_l \hspace{0.25mm} v_l}{c_w}
      \hspace{0.5mm} \nu^l_{\gamma Z} \,.
\end{equation}
In order to estimate the typical size of $\nu_{\gamma Z}^l$ we need an
analytic formula for the relative strength of the vector coupling
between the $Z$ boson and fermionic KK modes appearing in
(\ref{eq:nugammaZu}). We obtain
\begin{equation}
  \kappa_n^{f,V} = 1 - \frac{\left({\bm \delta}_F\right)_{nn} - 
  \left({\bm \delta}_f\right)_{nn}}{v_f} +  
  {\cal O}\left(\frac{m_Z^2}{\Mkk^2}\right) , 
\end{equation}
where the expressions for ${\bm \delta}_{F,f}$ can be found in
(\ref{eq:delta2}).  In the case of the extended $P_{LR}$ symmetry
(\ref{eq:extendedPLR}), it turns out that for down- and $\lambda$-type
KK quarks the result for $\kappa_n^{f,V}$ can be expressed in terms of
the electric charge and the third component of the weak isospin of the
involved fermion, while no such formula can be derived for up-type
quark KK modes. We get to excellent approximation ($f =d, \lambda$)
\begin{equation}
\kappa_n^{f,V} = 1 + \frac{T_L^{3 \, f_L}}{v_f} \,,
\end{equation}
which implies that all down-type ($\lambda$-type) KK-quark modes
couple with universal strength to the vector part of the $Z$-boson
coupling. It follows that in the decoupling limit, $\tau_n^f,
\lambda_n^f \to \infty$
\begin{equation}
  \left ( 1 + \frac{T_L^{3 \, f_L}}{v_f}  \right ) 
  \frac{A^h_f(\tau_n^f, \lambda_n^f)}{A^h_f(\tau_n^f)} =
  \frac{a^f_{\gamma Z}}{a_f} \,.
\end{equation}
From the numbers of the fit coefficients given in
Table~\ref{tab:kappas}, we see that this relation is satisfied
to an accuracy of around $1\%$. The KK-fermion effects in the down-
and $\lambda$-type quark sectors that contribute to $h \to gg, \gamma
\gamma$, $\gamma Z$ are thus universal, in the sense that they
can be simply obtained from each other by an appropriate replacement
of the vector couplings of the external fields.

Making now the plausible assumption that in the decoupling limit the
sums $\nu_{\gamma Z}^{d}$ and $\nu_{\gamma Z}^{l}$ differ only by the
presence of the vector couplings $\kappa_n^{d,V}$ and
$\kappa_n^{l,V}$, we obtain the following estimate for
the contribution to (\ref{kkleptonsinkappaAZ}) from leptonic relative
to down-type quark KK modes
\begin{equation}
  \frac{Q_l \hspace{0.25mm} v_l \hspace{0.25mm} \nu_{\gamma Z}^l}
  {N_c \hspace{0.25mm} Q_d \hspace{0.25mm} v_d \hspace{0.25mm} 
    \nu_{\gamma Z}^d} \approx  \frac{Q_l \hspace{0.25mm} v_l \hspace{0.25mm} 
    \kappa_n^{l,V}} {N_c \hspace{0.25mm} Q_d \hspace{0.25mm} v_d \hspace{0.25mm} 
    \kappa_n^{d,V}} = \frac{3- 6 s_w^2}{3-2 s_w^2} \approx 0.64 \,.
\end{equation}
In consequence, the sum (\ref{kkleptonsinkappaAZ}) can be approximated as
\begin{equation} \label{eq:sumlKK}
{\displaystyle \sum}_{j = u,
        d,\lambda} \; N_c \, \displaystyle \frac{2 \hspace{0.5mm} Q_j
        \hspace{0.25mm} v_j}{c_w} \hspace{0.5mm} \nu^j_{\gamma Z} +
      \displaystyle \frac{2 \hspace{0.5mm} Q_l \hspace{0.25mm} v_l}{c_w}
      \hspace{0.5mm} \nu^l_{\gamma Z} \approx 0.88 \,
      \nu_{\gamma Z}^u + 0.79 \, \nu_{\gamma Z}^d - 3.04 \, \nu_{\gamma
        Z}^\lambda +  0.50 \, \nu_{\gamma Z}^d \,.
\end{equation}
Note that the last term on the right-hand side encodes the effects due to
KK leptons, and in order to obtain the numerical values we have
inserted the relevant electroweak quantum numbers and used $s_w^2
\approx 0.23$. For $\Mkk = 2\, {\rm TeV}$, the sum above evaluates 
to $-0.11$ ($0.03$) if effects due to KK leptons are excluded (included). 
While these numbers imply that an omission of KK lepton effects can change the numerical value
of the KK fermion contribution to $h \to \gamma Z$ notably, it is not difficult to see
that the impact on (\ref{kappagammaZ}) itself is limited, since the
coefficient $\kappa_{\gamma Z}$ is dominated by the $W^\pm$-boson triangle
contribution. We thus conclude that the absence of KK-lepton
contributions in our prediction for $h \to \gamma Z$ will not change any of our
conclusions.

\begin{figure}[!t]
\begin{center}
\mbox{\includegraphics[width=13cm]{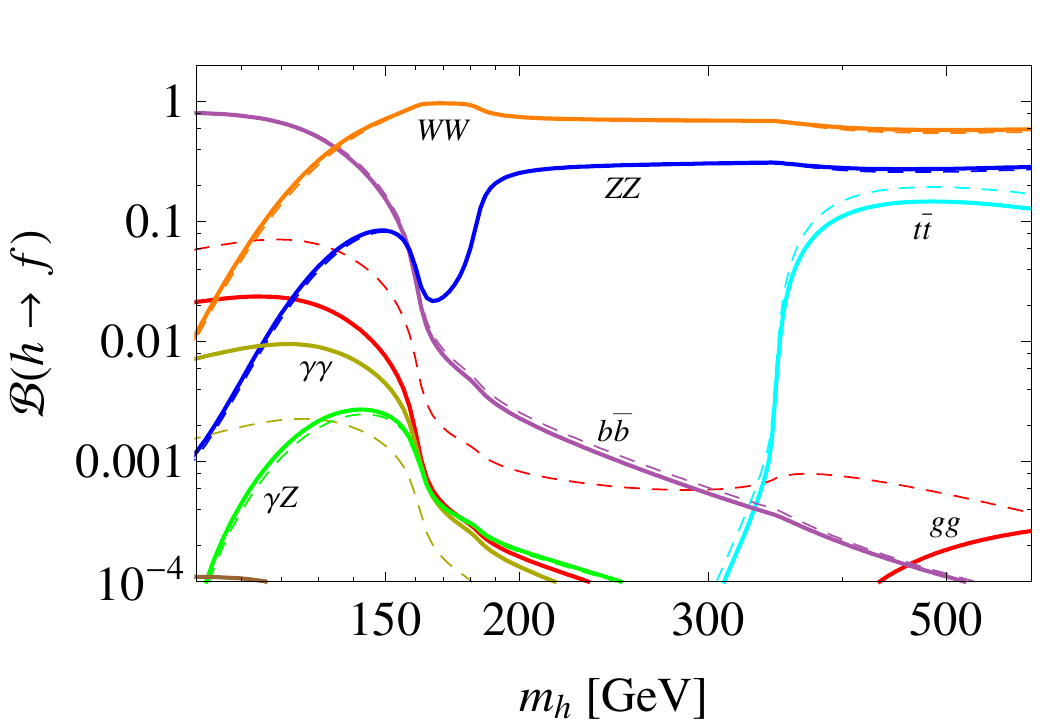}}

\mbox{\includegraphics[width=13cm]{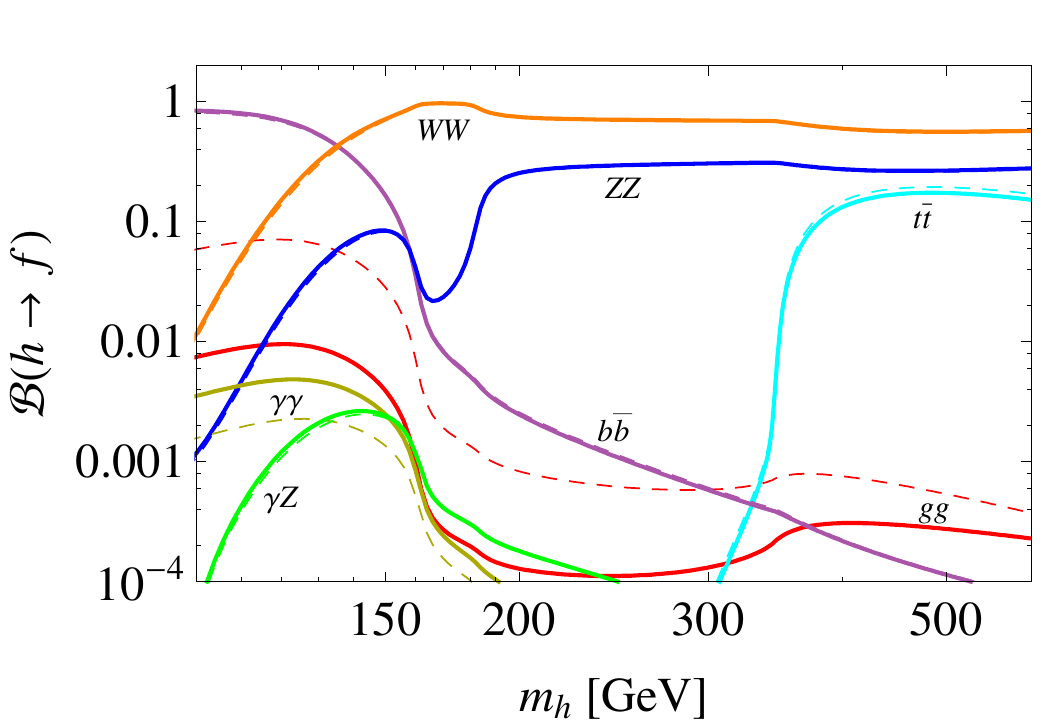}}

\vspace{4mm}

\parbox{15.5cm}{\caption{\label{fig:hXX} Branching ratios for the decays
    $h \to f$ as functions of the Higgs-boson mass for $\Mkk =2 \, {\rm TeV}$
    (upper panel) and $\Mkk = 3 \, {\rm TeV}$ (lower panel). The
    SM predictions are indicated by dashed lines, while the solid lines
    show the corresponding RS expectations. Branching fractions of
    less than $10^{-4}$ and decay channels into final states with
    muon, tau, charm-, and strange-quark pairs, which are all expected
    to remain SM-like, are not shown. See \cite{Casagrande:2010si} and text for details.}}
\end{center}
\end{figure}

The coefficient $\nu_{\gamma Z}^W$ in (\ref{kappagammaZ}) incorporates
the effects due to charged KK-boson excitations in the loop. 
The corresponding Feynman graph is displayed on the very right
in Figure~\ref{fig:hkkcontr}. This contribution reads explicitly
\begin{equation} \label{eq:nugammaZ} 
  \nu_{\gamma Z}^W = \frac{2 \pi
    x_W^2 \left ( g_L^2 + g_R^2 \right )}{g_L^2} \, \sum_{n =
    1}^\infty \, \frac{ \vec{d}_W^{\, T} \hspace{0.5mm}
    \vec{\chi}_n^{\, W} (1) \, \vec{\chi}_n^{\, W \, T} (1)
    \hspace{0.5mm} \vec{d}_W}{\big (x_n^W \big)^2} \; {\mathcal
    I}_{nn0}^{WWZ}\hspace{0.25mm} A_W^h(\tau_n^W,\lambda_n^W)\,,
\end{equation}
where
\begin{equation} \label{eq:IWWZ}
\begin{split}
  {\mathcal I}_{nn0}^{WWZ}=\,\frac{(2 \pi)^{3/2}}L
  \int_\epsilon^1\frac{dt}{t}\,\Big{[}\, &\chi_0^{(+)Z}{(\vec
    A_0^Z)}_1 \left( {\chi_n^{(+)W}}^2{(\vec A_n^W)}_1^{\,2} +
    \frac{g_Y^2}{g_L^2}\,
    {\chi_n^{(-)W}}^2{(\vec A_n^W)}_2^{\,2} \right)\\
  &-\,\sqrt{1-g_Y^4/g_L^4}\;\chi_0^{(-)Z}{(\vec
    A_0^Z)}_2\,{\chi_n^{(-)W}}^2{(\vec A_n^W)}_2^{\,2}\,\Big{]}\,,
\end{split}
\end{equation}
and $\lambda_n^W\equiv 4 \hspace{0.25mm}
\big(m_n^W\big)^2/m_Z^2$. Note that in the prefactor in the second line
of the above formula we have set $g_L=g_R$.  Since the
first term in the sum of (\ref{eq:nugammaZ}) is already suppressed by
a factor of $v^2/\Mkk^2$, the computation of $\nu_{\gamma Z}^W$ to this 
order only requires the knowledge of the overlap integral
(\ref{eq:IWWZ}) to zeroth order in the ratio of the weak over the KK
scale. We obtain
\begin{equation} \label{eq:IWWZapprox}
\begin{split}
  {\mathcal I}_{nn0}^{WWZ}=\,\frac{2 \pi}{L}
  \int_\epsilon^1\frac{dt}{t}\, \left( {\chi_n^{(+)W}}^2{(\vec
      A_n^W)}_1^{\,2} + \frac{g_Y^2}{g_L^2}\, {\chi_n^{(-)W}}^2{(\vec
      A_n^W)}_2^{\,2} \right)+\ord\left(\frac{v^2}{\Mkk^2} \right) \,.
\end{split}
\end{equation}
It is again an excellent approximation to evaluate the loop function
$A_W^h(\tau_n^W,\lambda_n^W)$ in the infinite mass limit $\tau_W^h,
\lambda_n^W \to \infty$, in which the form factor approaches $31 \,
c_w/6 - 11 \, s_w^2/(6 \, c_w) \approx 4.0$.  We perform the sum in
(\ref{eq:nugammaZ}) numerically. In this way, we find $\nu_{\gamma
  Z}^W=0.16$ $(\nu_{\gamma Z}^W = 0.07)$ for $\Mkk=2 \, {\rm TeV}$
($\Mkk=3 \, {\rm TeV}$).  Values for $\nu_{\gamma Z}^W$ corresponding
to different KK scales can be obtained by means of the fit formula
$a_{\gamma Z}^W \hspace{0.5mm} v^2/\Mkk^2$. The coefficient
$a_{\gamma Z}^W$ is given in Table~\ref{tab:kappas}.

In Figure~\ref{fig:kappas} we display the relative
corrections $|\kappa_g|^2$, $|\kappa_\gamma|^2$, and $|\kappa_{\gamma
  Z}|^2$ for $\Mkk = 2 \, {\rm TeV}$ (left) and $\Mkk = 3 \, {\rm
  TeV}$ (right). The depicted curves represent the RS results obtained
from (\ref{eq:Hbosoncoupl}) and (\ref{eq:nugammaW}) as well as from the
relevant fit formulae with the values of the coefficients collected in
Table \ref{tab:kappas}. While the behavior of $|\kappa_g|^2$ has
already been analyzed in Section~\ref{sec:higgsproduction}, we see
that $|\kappa_{\gamma Z}|^2$ is close to 1 and independent of the
value of the Higgs-boson mass.  This implies that the partial decay
width $\Gamma (h \to \gamma Z)$ in the custodial RS model is
essentially unchanged with respect to the SM. The relative correction
$|\kappa_\gamma|^2$ is, on the other hand, a non-trivial function of
$m_h$. Below the $WW$ threshold, the $W^\pm$-boson amplitude dominates the
SM $h \to \gamma \gamma$ decay rate and the contributions due to KK
quarks and $W^\pm$ bosons both interfere constructively with the SM
gauge-boson triangle graph. For $m_h = 130 \, {\rm GeV}$, the
new-physics contributions amount to around $70\%$ ($30\%$) of the
total SM amplitude for $\Mkk = 2\, {\rm TeV}$ ($\Mkk = 3\, {\rm
  TeV}$), resulting in values $|\kappa_\gamma|^2 \approx 3$
($|\kappa_\gamma|^2 \approx 1.7$).  For $m_h \gtrsim 160 \, {\rm
  GeV}$, the Higgs-mass dependence of the SM amplitude becomes less
pronounced and the RS prediction stays almost constant. The strong
rise of $|\kappa_\gamma|^2$, visible at higher values of the Higgs
mass, is due to the fact that for $m_h \approx 650 \, {\rm GeV}$
the top-quark loop nearly cancels the $W^\pm$-boson contribution in the
SM. As a result, for $m_h \gtrsim 500 \, {\rm GeV}$ the partial
width $\Gamma (h \to \gamma \gamma)$ is almost entirely due to loops
involving heavy KK modes, with the dominant contribution stemming 
from KK quarks.

Let us finally analyze the emerging picture of Higgs-boson decays, 
obtained using the above results. 
The various branching fractions in the custodial RS model are displayed
by the solid lines in Figure~\ref{fig:hXX}, whereas 
the dashed lines illustrate the SM expectations calculated with the help of {\tt HDECAY}
\cite{Djouadi:1997yw}. The RS predictions are based on the results for $\kappa_{t,b,W,Z}$ quoted above
and the curves for $|\kappa_{g, \gamma, \gamma Z}|^2$ displayed in
Figure~\ref{fig:kappas}. It is evident that in the custodial RS model
the branching ratios $h \to b \bar b$, $h \to WW$, and $h \to ZZ$
receive only insignificant corrections, not exceeding the level of
$\pm 5\%$. For $m_h\gtrsim 180 \, {\rm GeV}$ the experimentally
cleanest signature for the discovery of the Higgs boson at the LHC is
the ``golden'' channel of a decay to four leptons, $h \to Z^{(\ast)}Z^{(\ast)} \to
l^+ l^- l^+ l^-$. Since the $h \to ZZ$ branching fraction is essential
SM-like, the reduction in the $gg \to h$ production cross section will
make an observation of the Higgs boson in this channel more
difficult. 

Moderate effects occur in the non-discovery channels $h \to
\gamma Z$ and $h \to t \bar t$. In the relevant ranges for the Higgs
mass, the modifications in the branching ratios amount to around $+10
\%$ ($+10\%$) and $-25 \%$ ($-10\%$) for $\Mkk = 2 \, {\rm TeV}$
($\Mkk = 3 \, {\rm TeV}$). The most pronounced effects are found for
$h \to gg$ and $h \to \gamma \gamma$. For Higgs masses below the $WW$
threshold, the branching fraction of the former mode is reduced by a
factor of almost 4 (8), while the branching ratio of the latter
transition is enhanced by a factor of around 4 (2). The corresponding
maximal values of ${\cal B} (h \to \gamma \gamma)$ are $9.3 \cdot
10^{-3}$ ($4.8 \cdot 10^{-3}$) for $\Mkk = 2 \, {\rm TeV}$ ($\Mkk = 3
\, {\rm TeV}$) and arise at $m_h \approx 120 \, {\rm
  GeV}$. Calculating the rescaling factor $\varkappa = \left
  (\sigma_{\rm RS} (gg \to h) \hspace{1mm} {\cal B} (h \to \gamma
  \gamma)_{\rm RS} \right )/\left (\sigma_{\rm SM} (gg \to h)
  \hspace{1mm} {\cal B} (h \to \gamma \gamma)_{\rm SM} \right )$ for
$\sqrt s = 10 \, {\rm TeV}$ and the quoted maximal branching
fractions, we obtain the values $1.03$ ($0.24$). These numbers suggest
that the statistical significance for a LHC discovery of the Higgs
boson in $h \to \gamma \gamma$ can be slightly enhanced in the custodial RS
model for low KK scales. Note that if the KK scale is lowered to
$1 \, {\rm TeV}$, the branching ratio of $h \to tc$ can reach values
above $10^{-4}$ for Higgs masses above $m_h \approx 180 \, {\rm
  GeV}$. In the limit of a vanishing charm-quark mass, $r_c =
  0$, the corresponding decay rate is simply obtained from
  (\ref{BRthc}) by multiplying the branching fraction for $t \to ch$
  with $g^2 ( 1 - r_W^2 )^2 (1 + 2 \hspace{0.25mm} r_W^2)/(2
  \hspace{0.25mm} r_W^2) \, m_h/(16 \pi)$ and replacing $r_h$ through
  $r_t$. For such a low KK scale, also the decay channel $h\rightarrow
bs$ can open up below the $WW$ threshold (whose rate obeys a similar formula), but typically stays below
the level of $10^{-3}$. Like for the case of the production cross sections,
the value of the KK scale fixes the results for the Higgs-boson branching fractions to first approximation
and leaves no space for large variations. RS predictions for the various branching fractions
of the Higgs boson have been presented previously in
\cite{Azatov:2009na}. Yet, a direct comparison with our results is
difficult, as the latter work only includes RS corrections affecting
the tree-level couplings of the Higgs boson to fermions.

The analysis performed in this section demonstrates that Higgs physics at the LHC 
could be very sensitive to warped extra dimensions, even for KK scales that are 
significantly too high for a direct production of KK modes and for suppressed FCNCs. 
It provides an example for a theory that features a SM Higgs sector which, depending 
on $m_h$, could be quite hard to discover with early LHC data, despite the absence 
of NP directly at $\sim 1$\, TeV. This is mainly due to the depletion in the production 
cross section. Moreover, it could be possible that we just find a single spin-0 
Higgs-doublet with hypercharge 1/2 at the LHC, with however couplings that differ 
from the SM expectations. These modified couplings could be our only clear signal and 
hint for BSM physics. The analysis presented here demonstrates what to expect if warped 
extra dimensions are realized in nature.

\chapter{Conclusions and Outlook}
\label{sec:concl}

\vspace{-6mm} 

Some pressing unresolved questions in particle physics might be 
addressed, if the world around us features more than four dimensions.
The shortcomings, but also the successful features of the Standard Model of
particle physics have finally lead to the construction of the custodial Randall-Sundrum model with bulk fields,
an enlarged gauge symmetry group 
$SU(2)_L\times SU(2)_R\times U(1)_X\times P_{LR}$ and extended fermion representations.
This model thus follows various directions to extend the SM which have been mentioned in
the introduction of this thesis.
We have discussed in detail the way to the custodial RS model, the solution
to the gauge hierarchy problem and to the puzzle of fermion hierarchies.

A comprehensive discussion of the KK decomposition 
of RS models within the mass basis has been presented and
exact expressions to all orders in $v^2/\Mkk^2$ for the masses of the SM fields and 
their KK excitations, as well as for the profiles of these fields in the
extra dimension, have been derived. Special attention has been paid to the correct inclusion of Yukawa couplings involving
$Z_2$-odd fermion fields, which would be naively lost in a perturbative approach.
By demonstrating the analogy to the Froggatt-Nielsen mechanism,
formulae for the hierarchical quark spectrum as well as for the CKM matrix
in terms of $\ord(1)$ localization parameters of the fermion fields have been derived. 
This anarchic approach to flavor improves the predictivity of the RS setup.
We have performed a complete analysis of flavor changing effects
at tree level, pointing out the importance of fermion mixing for neutral current interactions
with the $Z$ boson.
We have also demonstrated how to perform infinite sums over KK towers in RS models.
Beyond that, analytic expressions for
five dimensional propagators for massive gauge bosons as well as for fermions have been presented.
In the latter case, we have considered the complete Yukawa structure for the first time, 
mediating flavor off-diagonal transitions.

We have studied in detail the phenomenology of the minimal as well as of the custodial Randall-Sundrum proposal.
Predictions for various observables have been presented. 
After an analysis of electroweak precision tests, we have shown that the minimal RS model, 
in combination with a heavy Higgs boson, could still furnish a viable setup.
In fact, due to the RS scale of $\Lambda_{UV}(\pi)=\ord$(TeV) at the IR brane, the natural assumption for 
the Higgs-boson mass in the RS framework would be $m_h\lesssim1\,$TeV.
Expanding our exact formulae in powers of $v^2/\Mkk^2$
has allowed to expose the dependence of observables on the input parameters of the model clearly. 
In particular, the exact approach has offered the possibility
to decipher, which contributions to the $Z\to b\bar b$
pseudo observables can be removed in the custodial setup
and which correspond to irreducible sources of $P_{LR}$-symmetry breaking.

Randall-Sundrum models have been used in the literature to address the enhancement in the $t\bar t$ 
forward-backward asymmetry measured at Tevatron. We have shown that it is not possible to
explain this anomaly in RS models with an anarchic approach to flavor.
The leading order corrections to the charge-asymmetric cross section are strongly suppressed due 
to the UV localization and mostly vector-like couplings
of light quarks. In this context we have taken into account the dominant next-to-leading order 
corrections for the first time. We have shown that they exceed the formally leading order corrections, and argued that
this observation holds true for a broad class of new physics models. However, we still found a generic tension between
generating large effects in the asymmetry and achieving a natural solution to the flavor hierarchy problem.
We have investigated the rare decays $t \to cZ$ and $t \to ch$ as well as the non-unitarity of the quark mixing
matrix. We demonstrated that among these, the most promising option to see signatures of the RS model is
due to Higgs couplings. Moreover, we have calculated the anomalous magnetic moment of the muon
in the RS setup, which was found to be orders of magnitude below the experimental bounds.
The phenomenological survey has been continued with an analysis of $B_s^0$--$\bar B_s^0$ mixing.
A calculation of the absorptive part of the mixing amplitude in the presence of new physics
has been presented and the impact of the Randall-Sundrum setup on several CP-violating
observables has been investigated. We have shown that an improved agreement of the theoretical values
for $A_{\rm SL}^s$, $S_{\psi\phi}$, and $\Delta\Gamma_s$ with experiment is possible. 

A central purpose of the Large Hadron Collider is to explore the sector of
electroweak symmetry breaking and to discover the Higgs boson. 
For the first time, we have presented a complete one-loop calculation
of all relevant Higgs-boson production and decay channels in the custodial RS setup, 
incorporating the effects stemming from the extended electroweak gauge-boson and fermion sectors.
We have shown that Higgs physics offers the possibility to test large scales at the LHC, that are 
significantly beyond the direct reach. Thus, RS models might show up first in modified
Higgs-production cross-sections or branching fractions.
Depending on its mass, the impact of the RS model could, on the other hand, make a discovery of 
the Higgs boson more difficult.
It would be worthwhile studying the impact of the RS setup on Higgs searches and exclusion limits
at the LHC and the Tevatron more detailed.

An ongoing issue is to understand RS models at the quantum level. A complete analytical calculation of loop mediated 
flavor-changing neutral currents such as $B\rightarrow X_s \gamma$ and of quantum corrections to flavor 
diagonal transitions like for $a_\mu$, including bulk fermions, is still a desideratum.
Since performing multiple infinite sums, accounting for the full RS flavor structure is not always feasible,
the 5D propagators presented in this thesis will be useful to tackle these problems. 
Moreover, employing a proper regularization of the brane-Higgs sector, 5D fermion propagators at vanishing momentum
transfer can be used to study loop mediated Higgs-production and -decay channels analytically.

\newpage

\subsubsection*{Acknowledgments}

First, I would like to owe my gratitude to my supervisor, Prof. Dr. Matthias Neubert 
who always supported me with his knowledge, guidance and encouragement.
I had the opportunity to work on a highly interesting topic. 
Additionally, I would like to thank Prof. Dr. Hubert Spiesberger for his advice in many situations.

I was glad to write my thesis in an 
inspiring, encouraging environment and I want to thank the whole THEP.
Especially, my officemates Torsten Pfoh, Michael Benzke and Alessandro Broggio for 
three cooperative years and Kaustubh Agashe, Valentin Ahrens, Volker B\"uscher, Marcela Carena, Sandro Casagrande, Hooman Davoudiasl, Uli Haisch, Nima Arkani-Hamed, Rainer H\"au\ss ling, Gabriele Honecker, Barbara J\"ager, Florian Jung, Daniel Litim, Lucia Massetti, Ben Pecjak, Martin Reuter, Christoph Schmell, Rainer Wanke, Stefan Weinzierl, Susanne Westhoff, Li Lin Yang and Omar Zanusso for helpful discussions.
Special thanks goes to Martin Bauer and Raoul Malm for checking results,
as well as to Martin Bauer, Stefan Berge, Alessandro Broggio, Tobias Hurth and Julia Seng for proof-reading! 
I also would like to thank Martin Bauer, Sandro Casagrande, Uli Haisch, Matthias Neubert, Torsten Pfoh and Susanne Westhoff 
for fruitful collaborations.

I owe my deepest gratitude to my family and my wife Julia. 
\\[2mm]
\noindent Florian Goertz
\\[3mm]
The Feynman diagrams shown in this thesis have been produced with the help of 
{\tt FeynArts} \cite{Hahn:2000kx} and {\tt Jaxodraw} \cite{Binosi:2003yf}.

\begin{appendix}

\chapter{Appendices Chapter~\ref{sec:IntroSM}}
\chaptermark{Appendices Chapter~1}

\section{Pauli, Dirac, and Gell-Mann Matrices}

\label{app:PDG}
The Pauli matrices are defined as
\beq
  \sigma^1 = \begin{pmatrix}
    0 & \,1 \\ 
    1 & \,0
  \end{pmatrix}\,,\quad
  \sigma^2 = \begin{pmatrix}
    0 & -i \\ 
    i & \,0 
  \end{pmatrix}\,,\quad
    \sigma^3 = \begin{pmatrix}
    1 & \,0 \\ 
    0 & -1 
  \end{pmatrix}\,.
\eeq

The Dirac-gamma matrices fulfill the Clifford algebra
\beq
\{\gamma^\mu,\gamma^\nu\}=2\, \eta^{\mu\nu} \bm{1}\,.
\eeq
In the {\it Weyl representation} they read ($i=1,2,3$)
\beq
  \gamma^0 = \begin{pmatrix}
    0 & \bm{1}_{2\times2} \\ 
    \bm{1}_{2\times2} & 0
  \end{pmatrix}\,,\quad
  \gamma^i = \begin{pmatrix}
    0 & \,\sigma^i \\ 
    -\sigma^i & 0
  \end{pmatrix}\,.
\eeq
The chirality matrix is defined as
\beq
\gamma^5\equiv i\gamma^0\gamma^1\gamma^2\gamma^3=\begin{pmatrix}
    -\bm{1}_{2\times2} & 0 \\ 
    0 & \bm{1}_{2\times2} 
  \end{pmatrix}\,.
\eeq

Finally, the Gell-Mann color matrices are
\beq
\begin{split}
  \lambda^1 = \begin{pmatrix}
    0 & 1 & 0\\ 
    1 & 0 & 0\\
    0 & 0 & 0
  \end{pmatrix}\,,\quad
  \lambda^2 = \begin{pmatrix}
    0 & -i & 0\\ 
    i & 0 & 0\\
    0 & 0 & 0
  \end{pmatrix}\,,\quad
    \lambda^3 = \begin{pmatrix}
    1 & 0 & 0\\ 
    0 & -1 & 0\\
    0 & 0 & 0
  \end{pmatrix}\,,\quad
  \lambda^4 = \begin{pmatrix}
    0 & 0 & 1\\ 
    0 & 0 & 0\\
    1 & 0 & 0
  \end{pmatrix}\,,\\
  \lambda^5 = \begin{pmatrix}
    0 & 0 & -i\\ 
    0 & 0 & 0\\
    i & 0 & 0
  \end{pmatrix}\,,\quad
    \lambda^6 = \begin{pmatrix}
    0 & 0 & 0\\ 
    0 & 0 & 1\\
    0 & 1 & 0
  \end{pmatrix}\,,\quad
  \lambda^7 = \begin{pmatrix}
    0 & 0 & 0\\ 
    0 & 0 & -i\\
    0 & i & 0
  \end{pmatrix}\,,\quad
  \lambda^8 = \frac{1}{\sqrt3}\begin{pmatrix}
    1 & 0 & 0\\ 
    0 & 1 & 0\\
    0 & 0 & -2
  \end{pmatrix}\,.
\end{split}
\eeq

\section{Custodial Symmetry}
\label{app:cus}

Starting from the general squared mass matrix (\ref{eq:custoM}) and demanding that one eigenvalue of the lower right block 
has to be zero (corresponding to the photon), while denoting the other one by $m_Z^2$, 
leads to the conditions
\beq
\label{eq:zeroEV}
   \rm{Det}[\left( \begin{array}{cc}
      m_3^2 & m^2 \\
       m^2 & m_0^2 \\
   \end{array} \right)]=0 \,,\quad
   \rm{Tr}[\left( \begin{array}{cc}
      m_3^2 & m^2 \\
       m^2 & m_0^2 \\
   \end{array} \right)]=m_Z^2\,.
\eeq
Identifying the eigenvector belonging to the vanishing eigenvalue as $(s_w,c_w)^T$ (forming a row of the matrix
that rotates the vector fields to the mass basis) we get
\beq
\label{eq:swcw}
\frac{s_w}{c_w}=-\frac{m^2}{m_3^2}=-\left|\frac{m_0}{m_3}\right|\,,
\eeq
where we have used the first condition of (\ref{eq:zeroEV}). From the second 
relation we now derive
\beq
m_Z^2=m_3^2\left(1+\frac{s_w^2}{c_w^2}\right)\,,
\eeq
which results in
\beq
\label{eq:mwmzgen}
\left|m_3\right|=m_Z\, c_w\,.
\eeq
We can use (\ref{eq:swcw}) to trade $m_3$ for $m_0$ which leads to $\left|m_0\right|= - m_Z\, s_w$.
Applying the last relations we arrive at the most general (squared) mass matrix, compatible with the
breaking pattern (\ref{eq:break}) of electroweak gauge symmetry
\beq
   M^2 =
   \left( \begin{array}{cccc}
    m_W^2 & 0 & 0 & 0 \\
     0 & m_W^2  & 0 & 0 \\
      0 & 0 & m_z^2\,c_w^2 & -m_Z^2\,c_w s_w \\
       0 & 0 & -m_Z^2\,c_w s_w & m_Z^2\,s_w^2 \\
   \end{array} \right)\,.
\eeq
In order to fulfill the sought mass relation (\ref{eq:mWmZ}), one needs $m_3^2=m_W^2$, as can be read off from 
(\ref{eq:mwmzgen}). Thus, the mass matrix to start with (\ref{eq:custoM}), has to fulfill
\beq
\label{eq:custom}
m_1^2=m_2^2=m_3^2=m_W^2
\eeq
in addition to the requirements following from the residual $U(1)_{EM}$ gauge invariance.

For the SM Higgs mechanism, this relation is fulfilled due to an accidental global symmetry in the Higgs Lagrangian. 
Writing the Higgs doublet as 
\beq
\Phi=\left( \begin{array}{c}
      \phi_1+i\, \phi_2 \\
      \phi_3+i\, \phi_4
   \end{array} \right)\,,
\eeq
we see that terms of the form 
\beq
\label{eq:O4}
\Phi^\dagger\Phi=\left( \begin{array}{c}
      \phi_1\\
      \phi_2\\
      \phi_3\\
      \phi_4
   \end{array} \right)^T
   \left( \begin{array}{c}
      \phi_1\\
      \phi_2\\
      \phi_3\\
      \phi_4
   \end{array} \right)
\eeq
possess an $O(4)$ symmetry which is isomorphic to $SU(2)_L\times SU(2)_R$. As it is more convenient to work with the second form of the symmetry,
we write the Higgs field as a $2 \times 2$ matrix, composed of $\Phi^c \equiv i \sigma_2 \Phi^*$  and $\Phi$ (see (\ref{eq:Hdoub+}))
\beq
\label{eq:HiggsLR}
\Phi=\left( \begin{array}{cc}
      \phi^{0*} & \phi^+ \\
      -\phi^- & \phi^0
   \end{array} \right)\,,
\eeq
which transforms under $SU(2)_L\times SU(2)_R$ as
\beq
\label{eq:LRtraf}
\Phi \to U_L\,\Phi\,U_R^\dagger\,. 
\eeq
Here, $U_{L,R}$ correspond to unitary $SU(2)_{L,R}$ transformation matrices. 
In Section~\ref{sec:custo} we also use such a form of the Higgs field, given a gauged version of the $SU(2)_L\times SU(2)_R$ symmetry.
The Higgs potential (see (\ref{eq:LHiggs})) now becomes
\beq
V(\Phi)=-\frac{\mu^2}{2}{\rm Tr}[\Phi^\dagger\Phi]+\frac{\lambda}{4}{\rm Tr}[\Phi^\dagger\Phi\,\Phi^\dagger\Phi]\,.
\eeq
It is obviously invariant under the transformation (\ref{eq:LRtraf}). The same is true for the part containing the covariant derivatives, only
if $g^\prime=0$, otherwise the coupling with $B_\mu$ breaks the global $SU(2)_R$ symmetry.
Setting $g^\prime=0$, the Higgs Lagrangian (\ref{eq:LHiggs}) has the full global $SU(2)_L\times SU(2)_R$ symmetry, given that $W_\mu^i$ transforms
as a triplet under $SU(2)_L$ and is a singlet under $SU(2)_R$. The vacuum expectation value of the Higgs field 
\beq
\langle\Phi\rangle=\frac{v}{\sqrt 2}
\left(
   \begin{array}{cc}
      1&0\\
      0&1
   \end{array} \right)
\eeq
now breaks this global symmetry down according to
\beq
\label{eq:LRbreak}
\begin{split}
SU(2)_L\times SU(2)_R & \longrightarrow SU(2)_V \,,\qquad \langle\Phi\rangle\to U_V\,\langle\Phi\rangle\,U_V^\dagger\,,\\
\mathrel{\widehat{=}}\quad O(4) & \longrightarrow  O(3)\,.
\end{split}
\eeq
In the parametrization of (\ref{eq:O4}), the residual $O(3)$ symmetry corresponds to a rotation among the components of 
$\Phi$, which do not acquire a VEV. After EWSB, this symmetry requires the mass terms of the components of $W_\mu^i$, 
transforming as a vector, to have the same coefficients (note that still $U(1)_{EM}$ is left unbroken)
\beq
m_1=m_2=m_3\,.
\eeq
This relation remains valid (at the tree level) if we allow again for $g^\prime\neq 0$ and leads to the mass relation (\ref{eq:mWmZ}). Thus we have 
identified a residual $O(3)\mathrel{\widehat{=}}SU(2)_V$ symmetry, guaranteeing that this 
mass relation is fulfilled. 

\section{Unitarity, Triviality and Vacuum-Stability Bounds}
\label{app:HB}

\paragraph{Unitarity bound:} 
To derive the unitarity bound on the Higgs-boson mass, it is useful to perform a partial wave analysis of the scattering amplitude 
\beq
A(s,t)= 16\pi\sum_{l=0}^\infty(2l+1)a_l(s)P_l(\cos\theta)\,,
\eeq
where $P_l(x)$ are Legendre Polynomials, $\cos\theta=1+2t/s$ at high energies, and $t$ is a Mandelstam variable, corresponding to the 
squared difference of an initial state momentum and a final state one, see Section~\ref{sec:afbt}.
The total cross section for a $2\to2$ process becomes
\beq
\sigma=\frac{16\pi}{s}\sum_{l=0}^\infty(2l+1)|a_l(s)|^2\,.
\eeq
Applying the optical theorem
\beq
\sigma=\frac{1}{s}\,{\rm Im}A(s,0) \,,
\eeq
which follows from unitarity, it is easy to show that
\beq\label{eq:pwcons}
\left|{\rm Re}\,a_l(s)\right|\leq\frac12
\eeq
has to hold.
Using the result for $W^\pm$-boson scattering at high energies $s\gg m_W^2$ (and for a heavy Higgs boson) \cite{Djouadi:2005gi} 
\beq
A(s,t)=-\left[2\frac{m_h^2}{v^2}+\left(\frac{m_h^2}{v}\right)^2 \frac{1}{s-m_h^2}+\left(\frac{m_h^2}{v}\right)^2 \frac{1}{t-m_h^2}\right]\,,
\eeq
we obtain from (\ref{eq:pwcons}) (in the limit $s\gg m_h^2$) the s-wave (l=0) constraint
\beq
\label{eq:HBap}
\left|{\rm Re}\,a_0\right|=\frac{m_h^2}{8 \pi v^2}\leq \frac12\ \Rightarrow m_h\lesssim870\,{\rm GeV}\,.
\eeq

\paragraph{Triviality and vacuum-stability bounds:}
In the SM, the leading dependence of the Higgs-quartic coupling on the energy scale $\mu$ is given by the renormalization 
group equation \cite{Djouadi:2005gi}
\beq
\label{eq:lambdaRGE}
\frac{d\lambda}{d\,{\rm ln}\mu^2}\simeq\frac{1}{16\pi^2}\left[6 \lambda^2+6 \lambda y_t^2-6 y_t^4-\frac{3}{2}\lambda(3g^2+g^{\prime\,2})
+\frac{3}{8}(2 g^4+(g^2+g^{\prime\,2})^2)\right]\,,
\eeq 
where $y_t=\sqrt{2}/v\,m_t$ is the top-quark Yukawa coupling. The lighter-fermion contributions can be neglected
to good approximation.
For large $\lambda$ only the first term on the right hand side of (\ref{eq:lambdaRGE}) contributes and the 
solution to this equation becomes
\beq
\lambda(\mu^2)=\lambda(\mu_0^2)\left(1-\frac{3}{8\pi^2}\lambda(\mu_0^2)\,{\rm ln}\frac{\mu^2}{\mu_0^2}\right)^{-1}\,.
\eeq
Looking at this equation we see that the Higgs quartic coupling blows up and hits a Landau pole at
\beq
\mu_c=v\, \exp\left[\frac{4\pi^2}{3 \lambda(v^2)}\right]=v \exp\left[\frac{4\pi^2 v^2}{3 m_h^2}\right]\,,
\eeq
where we have chosen the electroweak symmetry breaking scale as the reference scale, $\mu_0=v$, and used $\lambda(v^2)\,v^2=m_h^2$.
Requiring this pole not to appear until the Planck scale $\mu_c=M_{\rm Pl}$, which means that the Higgs sector of the SM
is well behaved for the whole possible energy range of the theory, demands the Higgs boson to be light. 
However, a low Landau pole at around a TeV allows for a heavy Higgs boson 
\beq
\begin{split}
\mu_c\gtrsim10^{19}\, {\rm GeV}\ \Rightarrow m_h \lesssim 145\,{\rm GeV}\\
\mu_c\gtrsim1\,{\rm TeV}\ \Rightarrow m_h \lesssim 750\,{\rm GeV}\,.
\end{split}
\eeq
Note that, including the top Yukawa as well as the gauge couplings moderately modifies these results.
A more exact perturbative constraint will be given below.

The vacuum stability bound on the Higgs-boson mass can be derived from requiring the Higgs potential to be bounded from below. 
This translates into $\lambda>0$, see (\ref{eq:LHiggs}). Thus we have to
study under which conditions (\ref{eq:lambdaRGE}) can drive the quartic coupling negative. To this end, we consider the
renormalization group equation for small $\lambda$, which leads to the solution ($\mu_0=v$)
\beq
\lambda(\mu^2)=\lambda(v^2)+\frac{1}{16\pi^2}\left[-24 \frac{m_t^4}{v^4}+\frac{3}{8}(2 g^4+(g^2+g^{\prime\,2})^2)\right]{\rm ln}\frac{\mu^2}{v^2}\,.
\eeq
Note that only the top Yukawa contribution can drive $\lambda$ negative.
If we do not want $\lambda$ to become negative before the scale $\mu_c$, we arrive at the condition
\beq
m_h^2>\frac{v^2}{16\pi^2}\left[24 \frac{m_t^4}{v^4}-\frac{3}{8}(2 g^4+(g^2+g^{\prime\,2})^2)\right]{\rm ln}\frac{\mu_c^2}{v^2}\,.
\eeq
This leads to the bounds
\beq
\begin{split}
\mu_c\gtrsim10^{19}\, {\rm GeV}\ \Rightarrow m_h \gtrsim 120\,{\rm GeV}\\
\mu_c\gtrsim1\,{\rm TeV}\ \Rightarrow m_h \gtrsim 70\,{\rm GeV}\,.
\end{split}
\eeq
Below the scale $\mu_c$ our theory is stable, whereas above, it would become unstable and has to be replaced by something new.
The values quoted here all correspond to leading order results. The running of the top-quark Yukawa
coupling is included. A proper two loop calculation, requiring perturbation theory to be valid and the 
Higgs potential to be stable below the cutoff $\mu_c$ leads to \cite{Djouadi:2005gi,Hambye:1996wb}
\beq
\label{eq:theomhba}
\begin{split}
\mu_c\gtrsim10^{16}\, {\rm GeV}\ \Rightarrow 130\,{\rm GeV} \lesssim m_h \lesssim 180\,{\rm GeV}\\
\mu_c\gtrsim1\,{\rm TeV}\ \Rightarrow 50\,{\rm GeV} \lesssim m_h \lesssim 800\,{\rm GeV}\,.
\end{split}
\eeq

\section{Effective Field Theories}
\label{app:EFT}
In this appendix we present the main concepts of effective field theories, which are used 
throughout this thesis. The fact that short-distance effects can be separated from long-distance effects,
the concept on which EFTs are based, is crucial for a step-by-step progress in physics. Excellent introductions to the 
subject, on which part of the following is based, 
can be found in \cite{Buras:1998raa,Neubert:2005mu,Polchinski:1992ed}.
By making use of the renormalization group, EFTs are well suited to deal with multi scale problems which 
often appear in nature and they provide a modern notion of renormalization. Furthermore they allow for model-independent analyses. 
We will first describe the concept formally and then from a practical point of view with the help of an example. 

Basically, the idea is the following. Consider a QFT with a certain fundamental scale $M$, which could be the mass of 
a heavy particle or some 
characteristic (euclidean) momentum transfer, and suppose we are only interested in predictions for energies
$E \ll M$.\footnote{If a problem involves several scales, we can usually consider one at a time.} We can then 
construct an EFT, valid for this (low) energy range and to be UV completed by the 
full theory, in three steps:
\begin{enumerate}
\item{
Fix a cutoff $\Lambda<M$ and divide the fields $\phi$ of the QFT into low-energy Fourier modes $\phi_L$ and 
high-energy modes $\phi_H$. One could for example use the absolute euclidean momentum $|k|$ after Wick 
rotation to distinguish these modes, such that 
\beq
\begin{split}
	\phi_L(k)= \begin{cases} \phi(k)\,,& |k|<\Lambda \\ \quad\ 0\,,& |k|\geq\Lambda \end{cases}\,,  \qquad
	\phi_H(k)= \begin{cases} \quad\ 0\,,& \quad |k|<\Lambda \\ \phi(k)\,,& \quad |k|\geq\Lambda \end{cases}\,, \qquad
	\phi=\phi_L+\phi_H\,. 
\end{split}
\eeq
In the following we will leave the concrete method of dividing the fields into low- and high energy modes open,
as it does not have to be specified for the upcoming discussion.
S-matrix elements in our EFT, which we want to be the correct theory for low 
energies $E\ll\Lambda$, should now just depend on the fields $\phi_L$. In the path 
integral formalism, they should be derived from vacuum correlation functions of the type
\beq
	\langle0|T \{ \phi_L(x_1) \dots \phi_L(x_n) \} |0\rangle = \frac{(-i)^n }{Z[0]}
		\left. \frac{\delta}{\delta J_L(x_1)} \cdots \frac{\delta}{\delta J_L(x_n)} Z[J_L]\right|_{J_L=0}\,.
\eeq
We thus need source terms only for the low energy fields in the generating functional 
\beq
Z[J_L]=\int {\cal D}\phi_L {\cal D}\phi_H\, e^{i S\left(\phi_L,\phi_H\right)+ i \int d^d x J_L(x) \phi_L(x)}\,,
\eeq
where $S\left(\phi_L,\phi_H\right)=\int d^d x {\cal L}(x)$ is the action of the QFT in $d$ space-time dimensions.
}
\item{Now we {\it integrate out} the high energy modes, which cannot be produced directly in experiments at energies 
$E<\Lambda$, by performing explicitly the corresponding path integral
\beq
e^{i S_\Lambda(\phi_L)}\equiv\int {\cal D}\phi_H\, e^{i S\left(\phi_L, \phi_H\right)}\,.
\eeq
This leads to the new generating functional
\beq
Z[J_L]=\int {\cal D}\phi_L\, e^{i S_\Lambda(\phi_L)+ i \int d^dx J_L(x) \phi_L(x)}\,,
\eeq
where the high energy modes are removed as dynamical degrees of freedom from our theory. $S_\Lambda(\phi_L)$ is called 
the {\it Wilsonian effective action}. Note that, as modes with energies above $\Lambda$ have been integrated out, this 
action is non local above this scale.}
\item{Finally we can expand this non-local action into a sum of local operators containing only low energy fields, by a 
so-called {\it Operator Product Expansion} (OPE), such that
\beq
S_\Lambda(\phi_L)=\int d^dx\,{\cal L}_{eff}^\Lambda(x)\,,
\eeq
where the (local) effective Lagrangian  
\beq
\label{eq:OPE}
{\cal L}_{eff}^\Lambda(x)=\sum_{D,i}{\frac{C_i^{(D)}}{M^{D-d}} {\cal Q}_i^{(D)}(\phi_L(x))}
\eeq
consists of an infinite series of operators with mass dimensions $D$. The corresponding dimensionless coefficients $C_i^{(D)}$ 
have to appear with appropriate powers of $M$, the only scale present in the theory, in order to make the action dimensionless.
The combined quantities $g_i^{(D)}\equiv C_i^{(D)} M^{d-D}$ are called {\it Wilson coefficients}. Note that for fixed $D$, 
just a finite number of operators can emerge ($i<\infty$), given that all fields have a strictly positive mass dimension. 
This is the case for the SM fields in $d\geq3$. 
Furthermore, it is possible that for some values of $D$, the corresponding set of operators is empty. Since operators of 
mass dimension $D$ are expected to scale like $E^{D-d}$, the expansion (\ref{eq:OPE}) features the order parameter $E/M\ll1$.}
\end{enumerate}

Having obtained the local effective action, we have to address the question how to make use of a theory that contains
apparently infinitely many parameters. The answer has already been sketched in Section~\ref{sec:SM1} and is contained in the 
considerations before. As operators with increasing mass dimension are suppressed by bigger and bigger powers of $M$ one can cut 
off the series at a certain level of precision (corresponding to a certain $D$), 
knowing that higher terms will at most give a contribution suppressed by $E/M$ with respect to the contributions considered.
In that sense, EFTs are automatically renormalizable and predictive (up to suppressed terms), 
since per definition only a finite number of parameters is present in 
the theory, after neglecting power suppressed corrections in (\ref{eq:OPE}) (assuming that $d \geq 3$). 
The argument of naturalness suggests that $C_i^{(D)} \sim \ord(1)$, since besides $M$ there is no other scale in the theory. 
Note that those coefficients now contain the information on
the high energy modes which have been removed from the action. Thus, they will 
in principle depend on the scale $\Lambda$, providing a new paradigm of renormalization. From the dimensional analysis above, 
it is evident that operators with $D>4$ will become less and 
less important at low energies. Thus they are named {\it irrelevant} (or non-renormalizable) operators, although they are very 
interesting, since they arise from integrating out heavy physics and tell us something about physics at the cutoff scale. Operators 
with $D<4$ become more important at low energies and are called {\it relevant} (or super-renormalizable), whereas those with $D=4$ are 
called {\it marginal} (or renormalizable) operators. Relevant operators are problematic since they lead to a UV sensitivity, see Section~\ref{sec:HP}.

In practice, it turns out that it is not always feasible to integrate out heavy modes formally as introduced before. It is often more
practical to follow a procedure called {\it matching} which works in the following way. We know that (\ref{eq:OPE}) can only contain 
operators compatible with the symmetries of the theory. So we can turn the procedure around and write down all operators
that are consistent with the given symmetries and depend just on the fields $\phi_L$ with so far {\it unknown} coefficients $C_i^{(D)}$, 
up to a certain mass dimension. This is now our EFT and we have already used such a procedure while 
constructing the SM Lagrangian in Section~\ref{sec:SM1} (there with the complete field content observed so far). Now we know that, 
for energies below the cutoff, the amplitudes in the EFT, after integrating out certain modes, have to agree with those obtained 
in the full theory
\beq
\label{eq:matching}
A_m \equiv \langle f_m|{\cal L}_{\rm full}| i_m \rangle \stackrel{!}{=} \sum_{D,i}\frac{C_i^{(D)}(\mu)}{M^{D-d}}
\langle f_m|{\cal Q}_i^{(D)}(\mu)| i_m \rangle\,.
\eeq
On the right hand side all insertions of operators contributing to a given matrix element up to a certain level of precision
(mass dimension $D$) have to be considered. Note that in this equation, the full Lagrangian as well as the effective Lagrangian 
have been expanded up to first order in perturbation theory. However, it is possible to extend this matching to higher orders in
the coupling constants. While not written out explicitly, at least the important higher order corrections in $\alpha_s$ are to be 
included implicitly on both sides. 
For processes mediated at leading order by $D=6$ operators and for
the case of the SM being the ``full'' theory we get at LO
\beq
\label{eq:matching2}
A_m \equiv \langle f_m|{\cal L}_{\rm SM}| i_m\rangle \stackrel{!}{=} \sum_{i}\frac{C_i^{(6)}(\mu)}{M^2}\langle f_m|{\cal Q}_i^{(6)}(\mu)| i_m \rangle + {\rm 
higher}\ D\ {\rm operators} .
\eeq
Here, we have used the scale $\mu$ instead of the hard cutoff $\Lambda$. This scale, being closer related to the more common approach of dimensional regularization, now takes the role of the cutoff.
By calculating enough matrix elements $A_m$ in the full theory, as well as in the EFT, and equating those with equal final and initial states, 
we can determine all Wilson coefficients present in the EFT, up to a sought level of precision. Note that in the following, we will use the
term Wilson coefficient also for the dimensionless coefficients $C_i^{(D)}$.
These coefficients have the important feature of being process independent, \ie, they do not depend on the external states in
(\ref{eq:matching}). Thus they can be determined using a certain class of processes and then used for predictions in other processes via the
EFT. If a process involves QCD at low scales, one usually has to resort to non-perturbative methods like lattice-QCD to calculate the 
corresponding matrix elements. However, note that the procedure of integrating out {\it high energy modes} via matching is independent 
of infrared (IR) physics. Thus, if one matches the theory at a scale where QCD is perturbative, it is possible to 
calculate the matrix elements in perturbation theory for the purpose of obtaining the corresponding Wilson coefficients.

\begin{figure}[!t]
\begin{center}
\mbox{\includegraphics[width=12cm]{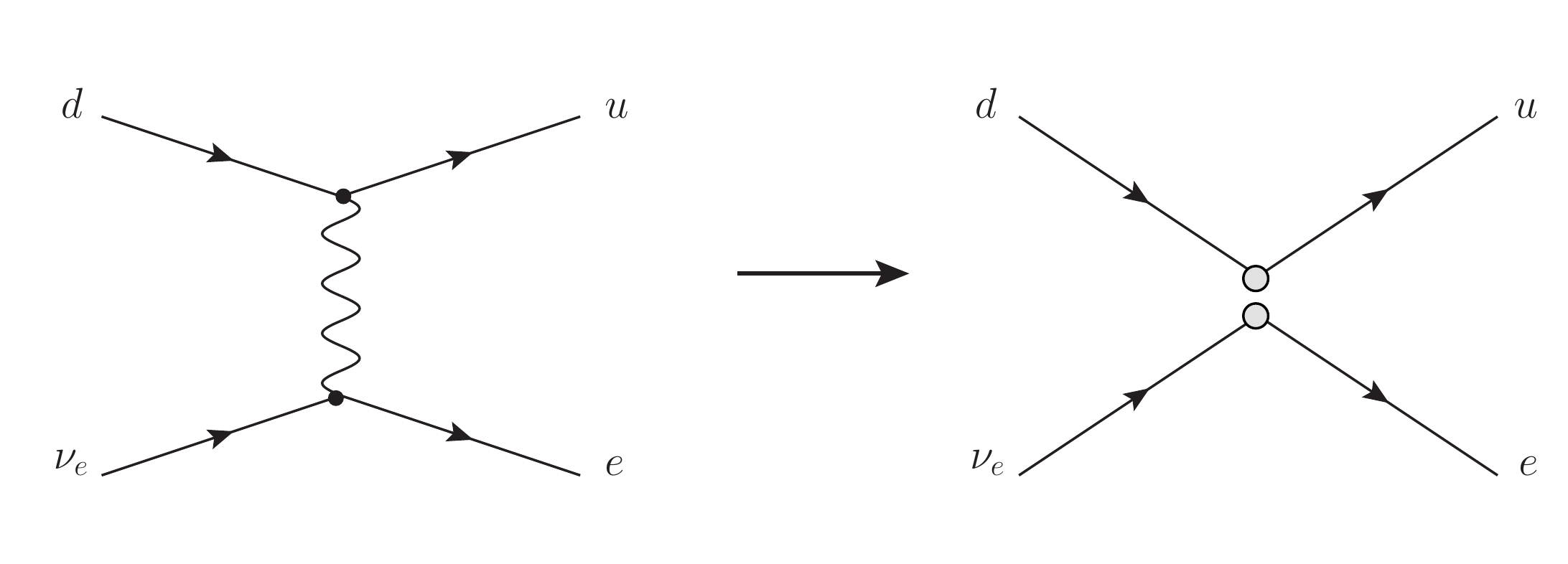}}
\parbox{15.5cm}{\caption{\label{fig:4fermi} Effective four-fermion interaction mediating $\beta$-decay, obtained by integrating out the 
heavy $W^\pm$ bosons of SM. The local four-quark vertex in the effective theory is denoted by the two gray circles
in the second graph.}}
\end{center}
\end{figure}

We will now give an example of constructing an EFT from a ``full'' theory by studying Fermi's theory of weak interactions, valid for 
$E\ll m_W$. It is obtained by integrating out the heavy $W^\pm$-bosons of the SM, resulting in local effective $D=6$ four-fermion 
vertices like the one mediating $\beta$-decay on the right hand side of Figure \ref{fig:4fermi}. The lack of the exchange of the massive 
gauge bosons leads to a violation of unitarity within the Fermi theory above the scale of electroweak interactions $M_{\rm EW}\sim m_W$. However,
as an effective theory with a cutoff at that scale, it is perfectly fine. The full SM gauge theory, including those fields, UV completes the 
Fermi theory at the electroweak scale. 
The diagram in the full theory (Figure \ref{fig:4fermi}, left) leads to the amplitude
\beq
\label{eq:4fermi1}
\begin{split}
   A&=-\frac{g^2}{8} V_{ud}\, \frac{1}{p^2-m_W^2} (\bar e\, \nu_e)_{V-A}(\bar u\, d)_{V-A}\\
   &=-\frac{g^2}{8 m_W^2} V_{ud}\, (\bar e\, \nu_e)_{V-A}(\bar u\, d)_{V-A} +\ord(\frac{p^2}{m_W^2}),
\end{split}
\eeq
where $(\bar e\, \nu_e)_{V-A} \equiv \bar e\, (1-\gamma^5) \nu_e$, \etc, and  $p^2$ is the squared momentum transfer.  
Thus we immediately see that the corresponding effective Lagrangian, which gives the same result on the EFT side, has to take the form
\beq
   {\cal L}_{eff}=- \frac{G_F}{\sqrt 2} V_{ud}\, C_1(\mu) (\bar e\, \nu_e)_{V-A}(\bar u\, d)_{V-A} + {\rm 
   higher}\ D\ {\rm operators},
\eeq
with $C_1=1$, working at tree level. Note that here we have pulled out a numerical factor times the weak coupling constant $g^2$, divided by 
the scale $m_W^2$, which are collected into the Fermi coupling constant of weak interactions $G_F/\sqrt{2}=g^2/(8 m_W^2)$, as well
as the CKM matrix element $V_{ud}$. This element is extracted exactly from the process discussed here. Furthermore we
have not made explicit the color indices which are to be contracted between the adjacent quark fields. The higher $D$ operators correspond to the 
higher powers of $\ord(p^2/m_W^2)$ in the 
expansion (\ref{eq:4fermi1}). In the same way the Wilson coefficients for four-fermion operators involving other external states can be obtained.
The result demonstrates that the reason for the apparent weakness of weak interactions compared to electromagnetism
is not a substantially smaller coupling constant but merely the large mass of the $W^\pm$ bosons which suppresses the corresponding interaction 
at low energies, as mentioned in Section~\ref{sec:Higgs}. Technically speaking, already the leading contribution corresponds to an 
irrelevant $D=6$ operator. It was another triumph
of particle physics, that the $W^\pm$ bosons were discovered with a mass close to the scale of $(G_F/\sqrt{2})^{-1/2}\approx 110$ GeV that 
was indicated indirectly by low energy experiments. By expanding the 
$W^\pm$ propagator to zeroth order in the momentum transfer, we were able to match it on a $D=6$ operator at the tree level. The same 
result can also be obtained by integrating out the $W^\pm$-bosons via the path integral formalism, as discussed before, since 
the corresponding integral is of Gaussian type. However, in more complicated situations, \eg involving QCD, this is no more true. 
For QCD, another complication arises, because it is not perturbative at low energies, so in most cases one relies on the procedure 
of matching.

While for the operator studied above, QCD corrections do not alter the Wilson coefficient (they are the same in the full theory and 
in the EFT), this is not true for generic four-quark operators like $(\bar s^{\,a} c^{\,a})_{V-A}(\bar u^{\,b} b^{\,b})_{V-A}$ obtained by 
integrating out $W^\pm$ bosons and mediating for example the decay $b_L\rightarrow  u_L\bar c_L s_L$. This partonic process is relevant for 
the hadronic decay $\bar B^0 \rightarrow \pi^+ D_s^-$ (similar processes will become relevant in Section~\ref{sec:asl}). 
Here, QCD corrections generate a second operator $(\bar s^a c^b)_{V-A}(\bar u^b b^a)_{V-A}$ with interchanged color indices at 
$\ord(\alpha_s)$.
The corresponding Wilson coefficients at that order can be obtained by matching the matrix elements at the one-loop level, including 
gluon-corrections at $\ord(\alpha_s)$ in the full theory as well as in the EFT. The coefficients now depend non-trivially on $\mu$ at 
that order, see \cite{Buras:1998raa,Neubert:2005mu}. By lowering the cutoff of the EFT, one successively integrates out high energy 
modes (virtual gluon corrections). Their effects are absorbed into the Wilson coefficients which thus, as mentioned before, 
generically depend on the scale $\mu$. Changing the scale corresponds to reshuffling contributions between the Wilson coefficient 
and the operator matrix element. We will now discuss the evolution of these coefficients with $\mu$.
 
For this purpose we consider a generic operator basis of a given dimension $D$, \ie, a complete set of operators of that mass
dimension, allowed by the symmetries of a problem $\{{\cal Q}_i(\mu)\},\, i=1,\dots,n$. Now we know that observables should not
depend on the scale $\mu$, separating high energy and low energy physics residing in the Wilson coefficients and matrix elements,
respectively. Thus we conclude
\beq
\label{eq:run1}
\frac{d}{d\, \rm{ln}\mu} \sum_{i=1}^n{C_i(\mu)\langle{\cal Q}_i(\mu)\rangle}=0.
\eeq
As we started from a complete set of operators, it must be possible to expand the derivatives of the matrix elements on the left 
hand side of the equation above in terms of the operators contained in the basis
\beq
\frac{d}{d\, \rm{ln}\mu} \langle{\cal Q}_i(\mu)\rangle\equiv \sum_{j=1}^n{ -\gamma_{ij}(\mu) \langle{\cal Q}_j(\mu)\rangle}.
\eeq
The coefficients $\gamma_{ij}$ measure the changes in the operator matrix elements under infinitesimal scale variations. They
are called {\it anomalous dimensions} as they feature deviations from the naive scaling dimensions $(D-d)$ of the 
operators, due to quantum corrections. With this definition, we derive from (\ref{eq:run1}) the famous {\it renormalization group 
equation}
\beq
\frac{d}{d\, \rm{ln}\mu}\, C_j(\mu)-\sum_{i=1}^n C_i(\mu) \gamma_{ij}(\mu) =0,
\eeq
where we have used that the ${\cal Q}_i$ are linearly independent from each other.
In matrix notation it reads
\beq
\frac{d}{d\, \rm{ln}\mu}\, \vec C(\mu)=\hat \gamma^T(\mu)\vec C(\mu).
\eeq
Here, the anomalous dimension matrix (ADM) $\hat \gamma$ depends on the scale $\mu$ just through the running coupling $\alpha_s(\mu)$ 
in QCD. If it contains off-diagonal entries, the operators mix under renormalization. Before solving this equation we first 
switch variables from $\rm{ln}\mu$ to $\alpha_s(\mu)$ to arrive at
\beq
\frac{d}{d\, \alpha_s(\mu)}\, \vec C(\mu)=\beta^{-1}(\alpha_s(\mu))\, \hat \gamma^T(\alpha_s(\mu))\, \vec C(\mu),
\eeq
where $\beta\equiv d\alpha_s(\mu)/d\, {\rm ln} \mu$ is the $\beta$-function of QCD. 
This renormalization group equation is now solved by
\beq
\label{eq:RGEsol}
   \vec C(\mu)=\hat U(\mu,m_W)\, \vec C(m_W)\,,\quad \hat U(\mu,m_W) \equiv T_\alpha\, \exp \left[\int_{\alpha_s(m_W)}^{\alpha_s(\mu)}
   d\alpha\frac{\hat \gamma^T(\alpha)}{\beta(\alpha)}\right]\,,
\eeq
which gives the evolution of the Wilson coefficients between one scale and another. Here, the initial condition has been chosen to 
be set by the values of the Wilson coefficients at the weak scale. The ADM can be calculated from the renormalization constants, 
appearing in the course of renormalizing the (effective) theory, see \cite{Buras:1998raa}. 
The corresponding exponential function in $U(\mu,m_W)$ is defined via its Taylor expansion. The ordering prescription $T_\alpha$, 
which results in matrices $\hat \gamma^T(\alpha)$ with smaller $\alpha$ standing on the right of those with larger $\alpha$ is 
necessary since in general those matrices do not commute for different values of $\alpha$.
Clearly, if the evolution is to be performed between two scales with the threshold of a particle in between, which shall be integrated out,
it can be performed in an iterative way. First in the EFT with this particle included and then,
below the corresponding scale, in a theory without that particle as a dynamical degree of freedom. For example, an evolution could be performed 
first from a high mass scale $M>m_t$ down to $m_t$ (where the top quark is integrated out) and then in a five-flavor theory further down to 
a lower scale $\mu$ by a two-step running
\beq
\vec C(\mu)=\hat U^{(5)}(\mu,m_t)\hat U^{(6)}(m_t,M) \vec C(M).
\eeq
 
Let us stress a very important virtue of this process of integrating out high energy modes and absorbing their effects into
running Wilson coefficients, described by a RG equation (\ref{eq:RGEsol}). If a problem involves 
several scales, one will generically encounter logarithms of ratios of
these scales when calculating higher order corrections. For the time being, let us assume that we want to study a process 
involving a mass scale $M$ at a second scale $\mu \ll M$, which could \eg be set by the mass of another light particle present in the
theory ($\mu\sim m$). If these 
scales are largely separated from each other, this can 
spoil the perturbative expansion in a small coupling constant since (for the case of QCD) now formally subleading corrections of the form
\beq
\label{eq:Logs}
\frac{\alpha_s(\mu)}{4\pi}\,{\rm ln}\frac{M^2}{\mu^2},
\eeq
can easily become of $\ord(1)$. 
Even if only weak couplings are involved, such logarithm terms can become
important if one has for example $M=M_{\rm GUT}$ and $\mu=m_W$.
By integrating out physics related to the larger scale one can circumvent this problem. Matching the theory at the large scale, one can 
avoid the presence of the problematic large logarithms involving the scale $M$ (\ref{eq:Logs}). Remember that this matching procedure is independent
of the low scale $m$. One can then evolve down the corresponding Wilson 
coefficients to the low scale by using the renormalization group evolution (\ref{eq:RGEsol}). This leads at leading order 
to a resummation of terms of the form $(\alpha_s(\mu)/(4\pi)\ {\rm ln}(M^2/\mu^2))^n$ (\ref{eq:Logs}), 
to all orders. These terms are now counted as $\ord(1)$, avoiding the breakdown of the perturbative series. The matrix element
in the EFT does not know about the large scale and thus, if one evaluates it (perturbatively) at the low scale it cannot feature large logarithms. 
The separation of high energy physics from low energy physics is a very important feature of EFTs. Processes involving low-energy QCD can thus be handled
by integrating out the heavy physics and run down to a scale at which the resulting matrix element can be calculated on the lattice
(or by means of other non-perturbative methods), while the high-energy
contributions reside in the Wilson coefficients. This approach will be used in Section~\ref{sec:asl}.
Here we have assumed the existence of just two distinct scales. However, the application to problems with more 
scales is possible by an iterative procedure, as indicated before.

Finally, let us see how the resummation works schematically in the case of QCD corrections to the electroweak theory. 
For that purpose we expand the $\beta$-function as well as the ADM formally up to the LO in $\alpha_s$, respectively
\beq
   \beta(\alpha_s)=-2 \alpha_s \left[\beta_0\frac{\alpha_s}{4 \pi}+\ord(\alpha_s^2)\right]  \,,\quad 
   \hat \gamma(\alpha_s)=\hat \gamma_0 \frac{\alpha_s}{4 \pi} + \ord(\alpha_s^2).
\eeq
Note that these quantities, as well as $\vec C(m_W)$ are free of large logarithms.
The evolution matrix now takes the form
\beq
\label{eq:ev}
\hat U_0(\mu,m_W)=\left[\frac{\alpha_s(m_W)}{\alpha_s(\mu)}\right]^{\frac{\hat \gamma_0}{2\beta_0}} \left(1+\ord(\alpha_s)\right).
\eeq
The matrix exponent can be evaluated via
\beq
\hat U_0(\mu,m_W)=\hat V {\rm diag}\left(\left[\frac{\alpha_s(m_W)}{\alpha_s(\mu)}\right]^{\frac{\vec \gamma_0}{2\beta_0}}\right) \hat V^{-1}
\eeq
where 
$\hat V$ diagonalizes the ADM $\hat \gamma_0^T$
\beq
\hat V^{-1} \hat \gamma_0^T \hat V = {\rm diag}\left(\vec \gamma_0\right).
\eeq
Studying the case of a single operator for simplicity and expanding
\beq
   U_0(\mu,m_W)=\left(1+\beta_0 \frac{\alpha_s}{4\pi}\,{\rm ln}\frac{m_W^2}{\mu^2}+\cdots\right)^{-\frac{\gamma_0}{2\beta_0}}
   =1-\frac{\gamma_0}{2}\frac{\alpha_s}{4\pi}\,{\rm ln}\frac{m_W^2}{\mu^2}+\ord\left(\alpha_s^2\,{\rm ln}^2\frac{m_W^2}{\mu^2}\right)+\,\cdots\,,
\eeq
we see that all the logarithms of the form $(\alpha_s/(4\pi)\ {\rm ln}(M^2/\mu^2))^n$ (\ref{eq:Logs}) are present in $U_0(\mu,m_W)$, 
which can be calculated reliably in perturbation theory.

It is possible to generalize the concepts introduced here to more complicated processes and to include \eg (loop-mediated) FCNCs relevant 
for meson mixing and decay. These processes are interesting, because they are suppressed within the SM 
and thus very sensitive to BSM contributions. Here one integrates out internal states in \eg box and penguin diagrams (for the case of the SM) 
and ends up with an effective weak Lagrangian mediating such processes. For details, we refer the reader to the literature 
\cite{Buras:1998raa,Neubert:2005mu,Buchalla:1995vs} and to Section~\ref{sec:asl} where such effective Lagrangians are used to calculate amplitudes
related to $B$-meson mixing and decay.

Let us stress again that EFTs are also well suited to match experiments with theory, without assuming a certain model. By writing 
down an OPE containing all fields up to a certain energy, consistent with the given symmetries, 
one can try to measure the various Wilson coefficients. Thus it is also possible to use the concept of EFTs without being able to integrate out 
certain modes explicitly (without knowing the underlying renormalizable model) and to learn something about nature. With the help of 
an OPE for the SM field content, one can try to collect information about its UV completion. 
Just like the $D=6$ operators of Fermi theory arise from integrating out the heavy $W^\pm$ bosons within the SM, the SM 
probably has a UV completion that replaces it above a certain scale $\Lambda_{\rm UV}$. This can be accounted for, by not truncating at the level 
of $D=4$, see Section~\ref{sec:SMProblems}. The presence of this BSM theory leaves its imprints in higher dimensional operators containing the 
SM fields. By measuring the Wilson coefficients of these operators one can quantify possible deviations from the SM in terms of 
Lagrangian parameters. Below the threshold of NP, the SM, including $D>4$ operators is an appropriate theory to describe nature. If one 
wants to be more specific and has a certain model in mind, one can calculate the Wilson coefficients in this model explicitly and 
compare them to the experimental results. By integrating out heavy fields of the model step by step one can resum large logarithms, arising when a 
theory contains several scales. These techniques will be used at several points in the phenomenology chapter (Chapter~\ref{sec:Pheno}) of 
this thesis.

\chapter{Appendices Chapter~\ref{sec:Pheno}}
\chaptermark{Appendices Chapter~5}

\section{Higgs-Boson Phase-Space Factors for $\AFBt$}
\label{app:phasespace}

In this appendix we present the explicit analytical form of the phase-space
factors, appearing in the Higgs-boson contribution to the
charge-symmetric and -asymmetric part of the $t \bar t$ cross
section, see (\ref{eq:SLONP}) and (\ref{eq:ALONP}). They read
\beq \label{eq:fSfA}
\begin{split}
  f_S (z) & = -\frac{\beta \rho}{72} \left[ 1 +\frac{\rho \left ( 1-z
      \right )}{2}+ \frac{\rho \left(4+ \rho \left (1-z \right )^2
      \right) }{8 \beta } \, \ln \left(\frac{ 2 \left ( 1+ \beta
        \right ) - \rho \left (1 - z\right )}{2 \left ( 1- \beta
        \right ) -\rho \left ( 1 - z
        \right )}\right) \right] \,, \\
  f_A (z) & = \frac{\rho }{144} \left[ 1-\rho + \frac{\rho
      \left(4+ \rho \left (1-z \right )^2 \right)}{4 } \, \ln
    \left(\frac{\rho \left(4 z + \rho \left (1-z \right )^2
        \right)}{(2 - \rho \left (1-z \right ) )^2}\right) \right ]\,,
\end{split}
\eeq 
where $z = m_h^2/m_t^2$, $\beta = \sqrt{1-\rho}$, and $\rho = 4
m_t^2/\hat s$.

\section{Tensor Integrals and Wilson Coefficients for $\AFBt$ in the ZMA}
\label{app:wil}

\sectionmark{Tensor Integrals and Wilson Coefficients in the ZMA}

The non-factorizable products of
overlaps $\bm{\widetilde\Delta}_A \otimes
\bm{\widetilde\Delta}_B$ are defined as \cite{Bauer:2009cf}
\beq\label{Deltaotimes}
\begin{split}
  \big( \widetilde\Delta_F \big)_{mn}\otimes \big(
  \widetilde\Delta_{f'} \big)_{m'n'} &= \frac{2\pi^2}{L^2\epsilon^2}
  \int_\epsilon^1\!dt \int_\epsilon^1\!dt'\,t_<^2 \\
  &\quad\times \left[ a_m^{F\dagger}\,\bm{C}_m^{Q}(\phi)\,
    \bm{C}_n^{Q}(\phi)\,a_n^{F} +
    a_m^{f\dagger}\,\bm{S}_m^{f}(\phi)\,
    \bm{S}_n^{f}(\phi)\,a_n^{f} \right] \\
  &\quad\times \left[
    a_{m'}^{f'\dagger}\,\bm{C}_{m'}^{f'}(\phi')\,
    \bm{C}_{n'}^{f'}(\phi')\,a_{n'}^{f'} +
    a_{m'}^{F'\dagger}\,\bm{S}_{m'}^{Q}(\phi')\,
    \bm{S}_{n'}^{Q}(\phi')\,a_{n'}^{F'} \right] ,
\end{split}
\eeq
{\it etc.}, where $F=U,D$ and $f=u,d$.

For the case of the up quark ($q=u$),
the ZMA results for the Wilson coefficients appearing in (\ref{eq:wilsonexplicit}) read
\begin{align} \label{eq:wilsonuuZMA}
  C^{(V, 8)}_{u \bar u, \parallel} & = -\frac{4 \pi \alpha_s}{\Mkk^2}
  \Bigg [ \frac{1}{2L} -\frac{F^2(c_{t_R}) \left ( 2 c_{t_R} + 5
    \right )}{4 (2 c_{t_R} +3)^2} -\frac{F^2(c_{t_L}) \left ( 2
      c_{t_L} + 5 \right )}{4 (2 c_{t_L} +3)^2} \nonumber \\ & \phantom{xx}
  -\frac{F^2(c_{u_R})}{4\left |(M_u)_{11} \right |^2} \sum_{i=1,2,3}
  \frac{(2 c_{u_i} + 5) \left |(M_u)_{1i} \right |^2 }{(2 c_{u_i}
    +3)^2} -\frac{F^2(c_{u_L})}{4 \left |(M_u)_{11} \right |^2}
  \sum_{i=1,2,3} \frac{(2 c_{Q_i} + 5) \left |(M_u)_{i1} \right |^2
  }{(2 c_{Q_i} +3)^2} \nonumber \\ &\phantom{xx} + \frac{L}{2} \,
  \frac{F^2(c_{t_R}) \hspace{0.25mm} F^2(c_{u_R})}{(2 c_{t_R}+3) \left
      |(M_u)_{11} \right |^2} \sum_{i=1,2,3} \frac{(c_{u_i} + c_{t_R}
    +3) \left |(M_u)_{1i} \right |^2 }{(2 c_{u_i} +3)
    (c_{u_i} + c_{t_R} +2)} \nonumber \\
  &\phantom{xx} + \frac{L}{2} \, \frac{F^2(c_{t_L}) \hspace{0.25mm}
    F^2(c_{u_L})}{(2 c_{t_L}+3) \left |(M_u)_{11} \right |^2}
  \sum_{i=1,2,3} \frac{(c_{Q_i} + c_{t_L} +3) \left |(M_u)_{i1} \right
    |^2 }{(2 c_{Q_i} +3)
    (c_{Q_i} + c_{t_L} +2)}\Bigg ] \,, \nonumber \\[6mm]
  C^{(V, 8)}_{u \bar u, \perp} & = -\frac{4 \pi \alpha_s}{\Mkk^2}
  \Bigg[ \frac{1}{2L} -\frac{F^2(c_{t_R}) \left ( 2 c_{t_R} + 5 \right
    )}{4 (2 c_{t_R} +3)^2} -\frac{F^2(c_{t_L}) \left ( 2 c_{t_L} + 5
    \right )}{4 (2 c_{t_L} +3)^2} \nonumber \\ & \phantom{xx}
  -\frac{F^2(c_{u_R})}{4\left |(M_u)_{11} \right |^2} \sum_{i=1,2,3}
  \frac{(2 c_{u_i} + 5) \left |(M_u)_{1i} \right |^2 }{(2 c_{u_i}
    +3)^2} -\frac{F^2(c_{u_L})}{4 \left |(M_u)_{11} \right |^2}
  \sum_{i=1,2,3} \frac{(2 c_{Q_i} + 5) \left |(M_u)_{i1} \right |^2
  }{(2 c_{Q_i} + 3)^2} \nonumber \\ &\phantom{xx} + \frac{L}{2} \,
  \frac{F^2(c_{t_L}) \hspace{0.25mm} F^2(c_{u_R})}{(2 c_{t_L}+3) \left
      |(M_u)_{11} \right |^2} \sum_{i=1,2,3} \frac{(c_{u_i} + c_{t_L}
    +3) \left |(M_u)_{1i} \right |^2 }{(2 c_{u_i} +3)
    (c_{u_i} + c_{t_L} +2)} \nonumber \\
  &\phantom{xx} + \frac{L}{2} \, \frac{F^2(c_{t_R}) \hspace{0.25mm}
    F^2(c_{u_L})}{(2 c_{t_R}+3) \left |(M_u)_{11} \right |^2}
  \sum_{i=1,2,3} \frac{(c_{Q_i} + c_{t_R} +3) \left |(M_u)_{i1} \right
    |^2 }{(2 c_{Q_i} +3)(c_{Q_i} + c_{t_R} +2)} \Bigg] \,.
\end{align}
Similar relations with obvious replacements hold in the case of the 
remaining light quarks $q = d, s, c$. 
For the $t$-channel Wilson coefficients in the vector channel, we get
\begin{eqnarray} \label{eq:wilsontuZMA}
\begin{split}
  C^{(V, 8)}_{t \bar u, \parallel} & = -\frac{\pi \alpha_s}{\Mkk^2} \,
  L \Bigg [ \frac{F^2(c_{t_R}) \hspace{0.25mm} F^2(c_{u_R})\left
      |(M_u)_{13} \right |^2 }{ ( 2 c_{t_R} + 3 ) ( c_{t_R} + 1) \left
      |(M_u)_{11} \right |^2} +\frac{F^2(c_{t_L}) \hspace{0.25mm}
    F^2(c_{u_L}) \left |(M_u)_{31} \right |^2 }{(2 c_{t_L} + 3)
    (c_{t_L}+1) \left |(M_u)_{11} \right |^2} \Bigg] \,,
  \\[1mm]
  C^{(V, 1)}_{t \bar u, \parallel} & = -\frac{\pi \alpha_e}{\Mkk^2} \,
  \frac{L}{s_w^2 c_w^2} \, \Bigg [ (T_3^u-Q_u s_w^2)^2 \,
  \frac{F^2(c_{t_L}) \hspace{0.25mm} F^2(c_{u_L}) \left |(M_u)_{31}
    \right |^2 }{(2 c_{t_L} + 3) (c_{t_L}+1) \left |(M_u)_{11} \right
    |^2} \\ & \hspace{2.75cm} + \left ( s_w^2 Q_u \right )^2
  \frac{F^2(c_{t_R}) \hspace{0.25mm} F^2(c_{u_R}) \left |(M_u)_{13}
    \right |^2 }{( 2 c_{t_R} + 3 ) ( c_{t_R} + 1) \left |(M_u)_{11}
    \right |^2} \Bigg ] \\ & \phantom{xx} - \frac{\pi \alpha_e
    Q_u^2}{\Mkk^2} \, L \Bigg [ \frac{F^2(c_{t_R}) \hspace{0.25mm}
    F^2(c_{u_R})\left |(M_u)_{13} \right |^2 }{ ( 2 c_{t_R} + 3 ) (
    c_{t_R} + 1) \left |(M_u)_{11} \right |^2} +\frac{F^2(c_{t_L})
    \hspace{0.25mm} F^2(c_{u_L}) \left |(M_u)_{31} \right |^2 }{ (2
    c_{t_L} + 3) (c_{t_L}+1) \left |(M_u)_{11} \right |^2} \Bigg ] \,.
\end{split}
\end{eqnarray}
The result for the Higgs-boson contribution to the $t$ channel,
which is of $\ord(v^4/\Mkk^4)$, is given by
\begin{equation}
\tilde C^{S}_{t \bar u}  = 
    \left | (g_h^u)_{13} \right |^2 + \left | (g_h^u)_{31} \right |^2 \,,
\end{equation}
where the ZMA expressions for the flavor off-diagonal Higgs couplings
are given in (\ref{eq:mis1}).

\section{RG Evolution of the Wilson Coefficients for $\AFBt$}
\sectionmark{RG Evolution of the Wilson Coefficients}
\label{app:RGEafb}

In this appendix we present analytic formulae relating the Wilson
coefficients evaluated at the top-quark mass scale $m_t$ to their
initial conditions calculated at $\Mkk \gg m_t$,
see Appendix~\ref{app:EFT}. Since in the RS model the $t$-channel Wilson 
coefficients $\tilde C_{t\bar u}^{V}$ and
$\tilde C_{t \bar u}^{S}$ are numerically irrelevant, we
will not consider their running in the following. The RG
evolution is performed at leading-logarithmic accuracy (\ie, at one-loop order),
neglecting tiny effects that arise from the mixing with QCD penguin
operators. For the $s$-channel Wilson coefficients entering 
(\ref{eq:RSresig}) and (\ref{eq:AFBEFT}), one obtains for $P = V,
A$ ({\it c.f.} (\ref{eq:ev})) \cite{Bauer:2010iq}
\beq \label{eq:c1}
\tilde C_{q\bar q}^P(m_t) = \left ( \frac{2}{3 \eta^{4/7}} +
  \frac{\eta^{2/7}}{3} \right ) \tilde C_{q\bar q}^P(\Mkk) \,,
\eeq 
where $\eta \equiv \alpha_s (\Mkk)/\alpha_s(m_t)$ is the ratio of
strong coupling constants evaluated at the relevant scales. 

In order to judge the impact of RG effects, we
evaluate (\ref{eq:c1}) explicitly, using $\alpha_s(M_Z)=0.139$, $\Mkk=1\,$TeV, and
$m_t=173.1$\,GeV, which leads to $\eta=0.803$ at one-loop order. We
obtain
\beq \label{eq:c2}
\tilde C_{q\bar q}^P(m_t) = 1.07\,\tilde C_{q\bar q}^P(\Mkk)\,.
\eeq
Thus, the RG evolution increases the Wilson
coefficients $\tilde C_{q\bar q}^P$ by about $7\%$ with respect to the
values quoted in Table \ref{tab:WC} which means that operator mixing
represents only a numerically subdominant effect.

\section{Wilson Coefficients of Penguin Operators}
\label{app:WilsonsP}

The Wilson coefficients of the penguin operators in 
(\ref{eq:Heff}), evaluated at the matching scale $\mu=\Mkk$, read at $\ord(v^2/\Mkk^2)$ 
\cite{Bauer:2009cf} 
\begin{eqnarray}\label{eq:Cpenguin}
\begin{aligned}
   C_3^{\rm RS}(\Mkk) &=
    \frac{\pi\alpha_s}{\Mkk^2}\,\frac{(\bm{\Delta}_D')_{23}}{2N_c} -
    \frac{\pi\alpha}{6\sws\cws\,\Mkk^2} (\,\bm{\Sigma}_D)_{23} \,,\\
   C_4^{\rm RS}(\Mkk) &= C_6^{\rm RS}(\Mkk) = -
    \frac{\pi\alpha_s}{2\Mkk^2}\,(\bm{\Delta}_D')_{23} \,, \\ 
   C_5^{\rm RS}(\Mkk) &=
    \frac{\pi\alpha_s}{\Mkk^2}\,\frac{(\bm{\Delta}_D')_{23}}{2N_c} \,,\\
    C_7^{\rm RS}(\Mkk) &= \frac{2\pi\alpha}{9\Mkk^2}\,(\bm{\Delta}_D')_{23} -
     \frac{2\pi\alpha}{3\cws\,\Mkk^2} (\,\bm{\Sigma}_D)_{23} \,, \\
    C_8^{\rm RS}(\Mkk) &= C_{10}^{\rm RS}(\Mkk) = 0
\,, \\
    C_9^{\rm RS}(\Mkk) &= \frac{2\pi\alpha}{9\Mkk^2}\,(\bm{\Delta}_D')_{23} +
     \frac{2\pi\alpha}{3\sws\,\Mkk^2} (\,\bm{\Sigma}_D)_{23} \,, 
\end{aligned}
\end{eqnarray}
where
\beq\label{eq:Sigmas}
   \bm{\,\Sigma}_D \equiv  \omega_Z^{d_L} L \left( \frac12 - \frac{\sws}{3} \right)
   \bm{\Delta}_D + \frac{\Mkk^2}{m_Z^2}\,\bm{\delta}_D \,.
\eeq
These results are valid for the 
minimal RS variant for $\omega_Z^{d_L}=1$, whereas 
in the custodial RS model with $P_{LR}$-symmetry, one finds 
$\omega_Z^{d_L}= 0\,$, see (\ref{eq:omegaZ_definition}) and below.
Exact analytic expressions for $\bm{\Delta}_D$, $\bm{\Delta}_D'$, 
and $\bm{\delta}_D$, as well as the corresponding ZMA results are given in
sections \ref{sec:gaugecouplings} and \ref{sec:custodialprotection}.
The evolution of the penguin coefficients down to the mass of the bottom quark is treated in Appendix~\ref{app:running}.

\section{Wilson Coefficients of Charged-Current Operators}
\label{app:charged}
\vspace{-5mm}
In this appendix we derive ZMA expressions for the Wilson coefficients of the
charged current operators in (\ref{eq:Heff}). The starting point is the (h.c.~part~of the) effective Hamiltonian
as given in (\ref{eq:HeffW}).
Performing the overlap integrals with the corresponding fermion profiles
in the ZMA, we arrive for the minimal RS model at
\beq\label{eq:LLa}
\begin{split}
{({\bf V}_L^\dagger)}_{mn}\otimes{({\bf V}_L)}_{m'n'}
=&{(\bm{U}_d^\dagger\bm{U}_u)}_{mn}{(\bm{U}_u^\dagger\bm{U}_d)}_{m'n'}\left[1+\ord\left(\frac{v^2}{\Mkk^2}\right)\right]\\
&+\frac{m_W^2}{2\Mkk^2}\,L\,{(\bm{U}_d^\dagger)}_{mi}{(\bm{U}_u)}_{in}
{(\bm{\widetilde\Delta}_{QQ})}_{ij} {(\bm{U}_u^\dagger)}_{m'j}{(\bm{U}_d)}_{jn'}\,,
\end{split}
\eeq
with the non-factorizable correction \cite{Bauer:2008xb}
\beq\label{eq:Deltatilde}
{(\bm{\widetilde\Delta}_{QQ})}_{ij}= \frac{F^2(c_{Q_i})}{3+2c_{Q_i}}\,\frac{3+c_{Q_i}+c_{Q_j}}
{2+c_{Q_i}+c_{Q_j}}\,\frac{ F^2(c_{Q_j})}{3+2c_{Q_j}}\,.
\eeq
For $B_s^0$-meson decays, we need the element $(m=2,n=2,m'=2,n'=3)$.
Here, the leading term in (\ref{eq:LLa}), together with factorizable
corrections  of the form $v^2/\Mkk^2\, (...)_{mn}\cdot (...)_{m'n'}$, 
should be identified with $\lambda^{bs}_c$. 
In the custodial RS model, one would find additional 
factorizable terms, which also will be absorbed into CKM-matrix elements.
Thus, we find at LO in $v^2/\Mkk^2\,$
\beq \label{eq:CLL}
\begin{split}
C_2^{LL}(\Mkk)&=\,\frac{m_W^2}{2\Mkk^2}\,L
{\frac{{(\bm{U}_d^\dagger)}_{2i}{(\bm{U}_u)}_{i2}}{{(\bm{U}_d^\dagger\bm{U}_u)}_{22}}\,
{(\bm{\widetilde\Delta}_{QQ})}_{ij}\, 
\frac{{(\bm{U}_u^\dagger)}_{2j}{(\bm{U}_d)}_{j3}}{{(\bm{U}_u^\dagger\bm{U}_d)}_{23}} }\,,
\end{split}
\eeq
independent of the chosen scenario, and  $C_1^{LL}(\Mkk)=0\,$.
For the mixed-chirality currents we arrive at
\beq\label{eq:LR}
\begin{split}
{({\bf V}_L^\dagger)}_{mn}\otimes\, {({\bf V}_R)}_{m'n'}
=&{(\bm{U}_d^\dagger\bm{U}_u)}_{mn}
{(\bm{x}_u \bm{U}_u^\dagger)}_{m'j}\,
f(c_{Q_j})\, {(\bm{U}_d\,\bm{x}_d)}_{jn'}\,, \\
{({\bf V}_R^\dagger)}_{mn}\otimes\, {({\bf V}_L)}_{m'n'}
=&{(\bm{x}_d\bm{U}_d^\dagger)}_{mi}\,f(c_{Q_i})\,{(\bm{U}_u\,\bm{x}_u)}_{in}
{(\bm{U}_u^\dagger\bm{U}_d)}_{m'n'}\,,
\end{split}
\eeq
where
\beq
f(c)=\,\frac 1{F^2(c)(1-2c)}\,-\frac 1{1-2 c}\,+\frac{F^2(c)}{(1+2 c)^2}\left(\frac 1{1-2 c}-1+\frac 1{3+2 c}\right)\,.
\eeq
Modifications due to the custodial model are of higher order.
We find $C_1^{LR/RL}(\Mkk)=0\,$ and
\beq\label{eq:CLRCRL}
\begin{split}
& C_2^{LR}(\Mkk)=\,
{\frac{{(\bm{x}_u \bm{U}_u^\dagger)}_{2i}\,f(c_{Q_i})\,{(\bm{U}_d\,\bm{x}_d)}_{i3}}
{{(\bm{U}_u^\dagger \bm{U}_d)}_{23}} }\,,\\
&C_2^{RL}(\Mkk)=\,
{\frac{{(\bm{x}_d\bm{U}_d^\dagger)}_{2i}\,f(c_{Q_i})\,{(\bm{U}_u\,\bm{x}_u)}_{i2}}
{{(\bm{U}_d^\dagger\bm{U}_u)}_{22}} }\,.
\end{split}
\eeq
The running down to $m_b$ is again treated in Appendix~\ref{app:running}.

\section[Wilson Coefficients of $\Delta B=2$ Operators]{Wilson Coefficients of $\bm \Delta B=2$ Operators}\label{app:mixing}

The $\Delta B=2$ operators that contribute to the $\bar B_s^0$--$B_s^0$ mixing amplitude at 
tree-level are $\Q_1$, $\widetilde \Q_1$, $\Q_4$, and $\Q_5$. 
The operators $\Q_1$ and $\widetilde \Q_1$ do not mix under 
renormalization. The anomalous dimension for both is given by
$\gamma_0^{\rm{VLL}}=6-6/N_c$ \cite{Buras:2000if}. 
The opera\-tors $\Q_{4,5}$ mix and
the corresponding ADM can be taken from \cite{Buras:2000if,Bagger:1997gg}.
The running of the coefficients is described by the general formula (\ref{eq:rundown}).
Defining $\bm{\widetilde\Delta}_{dd}$ and $\bm{\widetilde\Delta}_{Qd}$ 
in analogy to (\ref{eq:Deltatilde}), the RS coefficients in the ZMA, evaluated at the 
KK scale, are given by \cite{Bauer:2009cf}
\begin{align}\label{eq:Cmix}
   C_1^{\rm RS}(\Mkk) &=\, \frac{\pi L}{\Mkk^2}\, 
    {(\bm{U}_d^\dagger)}_{2i}{(\bm{U}_d)}_{i3}
    {(\bm{\widetilde\Delta}_{QQ})}_{ij} {(\bm{U}_d^\dagger)}_{2j}{(\bm{U}_d)}_{j3}\\
    \times&\left[
    \frac{\alpha_s}{2} \left( 1 - \frac{1}{N_c} \right) +
    Q_d^2\,\alpha + (\omega_Z^{d_Ld_L})\frac{(T_3^d-\sws\,Q_d)^2\,\alpha}
    {\sws\cws} \right] ,  \nonumber\\[2mm]
   \widetilde C_1^{\rm RS}(\Mkk) &=\, \frac{\pi L}{\Mkk^2}\, 
    {(\bm{W}_d^\dagger)}_{2i}{(\bm{W}_d)}_{i3}
    {(\bm{\widetilde\Delta}_{dd})}_{ij} {(\bm{W}_d^\dagger)}_{2j}{(\bm{W}_d)}_{j3}\nonumber\\
    \times&\left[
    \frac{\alpha_s}{2} \left( 1 - \frac{1}{N_c} \right) +
    Q_d^2\,\alpha + (\omega_Z^{d_Rd_R})\frac{(\sws\,Q_d)^2\,\alpha}
    {\sws\cws} \right] , \nonumber\\[2mm]
   C_4^{\rm RS}(\Mkk) &=- 2\alpha_s \frac{\pi L}{\Mkk^2}\, {(\bm{U}_d^\dagger)}_{2i}{(\bm{U}_d)}_{i3}
    {(\bm{\widetilde\Delta}_{Qd})}_{ij} {(\bm{W}_d^\dagger)}_{2j}{(\bm{W}_d)}_{j3}\nonumber\\[2mm]
   C_5^{\rm RS}(\Mkk) &=\, \frac{\pi L}{\Mkk^2}\,
    {(\bm{U}_d^\dagger)}_{2i}{(\bm{U}_d)}_{i3}
    {(\bm{\widetilde\Delta}_{Qd})}_{ij} {(\bm{W}_d^\dagger)}_{2j}{(\bm{W}_d)}_{j3}\nonumber\\
    &\times\left[
    \frac{2\alpha_s}{N_c} 
    - 4 Q_d^2\,\alpha
    + \omega_Z^{d_Ld_R}
    \frac{4\sws\,Q_d\,(T_3^d-\sws\,Q_d)\,\alpha}{\sws\cws} \right] .\nonumber
\end{align}
Here we have introduced the correction factors $\omega_Z^{qq'}$, which are equal to 1
in the minimal RS model, and given by
\beq
\omega_Z^{qq'} = 1+\frac 1{c_w^2-s_w^2}\,\left(\frac{s_w^2(T_L^{3q}-Q^q)-c_w^2 T_R^{3q}}{T_L^{3q}-s_w^2 Q^q}\right)\left(\frac{s_w^2(T_L^{3q'}-Q^{q'})-c_w^2 T_R^{3q'}}{T_L^{3q'}-s_w^2 Q^{q'}}\right)
\eeq
in the custodial RS variant with $P_{LR}$ symmetry.
Numerically we find $\omega_Z^{d_Ld_L}\approx 2.9$, $\omega_Z^{d_Rd_R}\approx 150.9$, and $\omega_Z^{d_Ld_R}\approx -15.7 $.
The quantum numbers $T_{L,R}^{3q}$ can be found in Section~\ref{sec:custodialprotection}.

\section[Running of the $\Delta B=1$ Coefficients]{Running of the $\bm \Delta B=1$ Coefficients}\label{app:running}

The evolution of the RS Wilson coefficients of the $\Delta B=1$ operators can again be performed along the lines as described in Appendix~\ref{app:EFT}. 
The ADM $\hat\gamma_0$ for the operator basis $\vec Q=(Q_1,Q_2,Q_{3..10})$,
which is a function of $N_c$, $n_f$, $n_u$, and $n_d$ (number of colors, flavors, up- and down-type quarks),
can be found in \cite{Buras:1992tc,Ciuchini:1993vr}. 

The evolution from $\Mkk$ to $m_b$ is given by ({\it c.f.} (\ref{eq:RGEsol}) ff.)
\beq\label{eq:rundown}
\vec C(m_b)=\hat U^{(5)}(m_b,m_t)\,\hat U^{(6)}(m_t,\Mkk)\,\vec C(\Mkk)\,,
\eeq
where, at LO in $\alpha_s$, 
\vspace{5mm}
\beq
\displaystyle
\hat U^{(n_f)}(\mu_1,\mu_2)=
\hat V {\rm diag}\left({\left[\frac{\alpha_s^{(n_f)}(\mu_2)}
{\alpha_s^{(n_f)}(\mu_1)}\right]}^{\frac{\vec\gamma_0}{2\beta_0(n_f)}}\right) \hat V^{-1}\,,
\eeq
and $\hat V$ diagonalizes ${\hat\gamma_0}^T$ via
${\rm diag}(\vec\gamma_0)=\hat V^{-1} {\hat\gamma_0}^T \hat V$.
The QCD $\beta$-function is given by $\beta_0(n_f)=(11 N_c-2\, n_f)/3$.
The operators $Q_1$ and $Q_2$ will mix independently of $n_f$, $n_u$, and $n_d$.
The evolution in the penguin sector receives a small admixture from charged current operators,
however, the operators $Q_1^{LR/RL}$ and $Q_2^{LR/RL}$ do not mix into the penguin sector.
Their internal mixing is identical to that of the $LL$ operators. There
is no mixing between charged currents of different chiralities.
For the running of the $LR/RL$ coefficients, one consequently uses
\beq
\displaystyle
\gamma_0=\left(
\begin{array}{cc}
  -\frac 6{N_c} & 6 \\
  6 & -\frac 6{N_c}
\end{array}\right)
\eeq
in (\ref{eq:rundown}),
which also describes the internal $LL$ mixing.

\section{Form Factors for Higgs-Boson Production and Decay}
\label{app:formfactors}

The form factors $A_{q,W}^h (\tau)$ and $A_{q,W}^h (\tau, \lambda)$
which describe the effects of quark and $W^\pm$-boson loops in the production
and the decay of the Higgs boson are given by \cite{Djouadi:2005gi}
\begin{align}
\allowdisplaybreaks
    A_{q}^h (\tau) & = \frac{3 \hspace{0.25mm} \tau}{2} \left [
      \hspace{0.25mm} 1 + \left ( 1 - \tau \right ) f (\tau)
      \hspace{0.25mm} \right ] \,, \nonumber\\
    A_{W}^h (\tau) & = -\frac{3}{4} \left [ \hspace{0.25mm} 2 + 3 \tau
      + 3 \tau \left ( 2 - \tau \right ) f (\tau) \hspace{0.25mm}
    \right ]
    \,, \nonumber\\[2mm]\pagebreak
    A_{q}^h (\tau, \lambda) & = - I (\tau, \lambda) + J (\tau,
    \lambda) \,, \nonumber\\[1mm]
    A_{W}^h (\tau, \lambda) & = c_w \hspace{0.5mm} \left \{ 4 \left (
        3 - \frac{s_w^2}{c_w^2} \right ) I (\tau, \lambda) + \left [
        \left ( 1 + \frac{2}{\tau} \right ) \frac{s_w^2}{c_w^2} -
        \left ( 5 + \frac{2}{\tau} \right ) \right ] J (\tau, \lambda)
    \right \} \,.
\end{align}
The functions $I(\tau, \lambda)$ and $J(\tau, \lambda)$ take the form
\begin{equation}
  \begin{split}
    I (\tau, \lambda) & = -\frac{\tau \lambda}{2 (\tau - \lambda)} \,
    \big [ f(\tau) - f(\lambda) \big ] \,, \\
    J (\tau, \lambda) & = \frac{\tau \lambda}{2 \left (\tau - \lambda
      \right )} + \frac{\tau^2 \lambda^2}{2 \left (\tau - \lambda
      \right )^2} \, \big [ f(\tau) - f(\lambda) \big ] + \frac{\tau^2
      \lambda}{(\tau - \lambda)^2} \, \big [ g(\tau) - g(\lambda) \big
    ] \,,
  \end{split}
\end{equation}
where the functions $f(\tau)$ and $g(\tau)$ read 
\begin{align}
    f(\tau) & = \begin{cases} - \displaystyle \frac{1}{4} \left [ \,
        \ln \left ( \displaystyle \frac{1 + \sqrt{1 - \tau}}{1 -
            \sqrt{1 - \tau}} \right ) - i \pi \, \right ]^2 \,, & \tau
      \leq 1 \,, \\[4mm] \arcsin^2 \left ( \displaystyle
        \frac{1}{\sqrt{\tau}}
      \right ) \,, & \tau > 1 \,, \end{cases} \\[2mm]
    g(\tau) & = \begin{cases} \sqrt{\tau - 1} \hspace{0.5mm} \arcsin
      \left ( \displaystyle
        \frac{1}{\sqrt{\tau}} \right ) \,, & \tau \leq 1 \,,\\[6mm]
      \displaystyle \frac{1}{2} \, \sqrt{1 - \tau} \, \left [ \, \ln
        \left ( \displaystyle \frac{1 + \sqrt{1 - \tau}}{1 - \sqrt{1 -
              \tau}} \right ) - i \pi \, \right ] \,, & \tau > 1
      \,. \end{cases}
\end{align}

\chapter{RS Parameter Points}
\label{app:points}

In this appendix, we specify a set of three RS parameter points, \ie, 
the bulk mass parameters of the quark fields and the Yukawa matrices,
that are used in our numerical analysis.
All the parameter sets given below have been obtained by random choice,
as described at the beginning of Chapter~\ref{sec:Pheno}.

Our first parameter point is specified by the following bulk mass
parameters and Yukawa matrices\footnote{The results are given to at least
  three significant digits.}
\beq\label{eq:cparameter1}
\begin{aligned}
   c_{Q_1} &= -0.611 \,, \qquad
    c_{Q_2} &= -0.580 \,, \qquad
    c_{Q_3} &= -0.407 \,, \\
   c_{u_1} &= -0.688 \,, \qquad
    c_{u_2} &= -0.550 \,, \qquad
    c_{u_3} &= +0.091 \,, \\
   c_{d_1} &= -0.665 \,, \qquad
    c_{d_2} &= -0.627 \,, \qquad
    c_{d_3} &= -0.577 \,,
\end{aligned}
\eeq
\beq\label{eq:yukawas1}
\begin{split}
  \bm{Y}_u &= \left(
    \begin{array}{rrr} 
      -1.303-0.364 \hspace{0.5mm} i &
      ~~-1.215+0.089 \hspace{0.5mm} i~~ &        
      -1.121-1.679 \hspace{0.5mm} i \\
      1.857+1.199 \hspace{0.5mm} i &       
      ~~2.038+1.105 \hspace{0.5mm} i~~ &       
      -0.484-0.193 \hspace{0.5mm} i \\
      -1.052+0.546 \hspace{0.5mm} i & 
      ~~-2.833+0.191 \hspace{0.5mm}
      i~~ & -1.287-1.141 \hspace{0.5mm} i
   \end{array}
 \right) , \\
 \bm{Y}_d &= \left(
   \begin{array}{rrr}
     -0.661-1.118 \hspace{0.5mm} i &       
     ~~-0.075-0.656 \hspace{0.5mm} i~~ &       
     ~~0.141-0.465 \hspace{0.5mm} i \\
     -2.070+1.364 \hspace{0.5mm} i &      
     ~-2.518+1.435 \hspace{0.5mm} i~~ &       
     0.717-0.165 \hspace{0.5mm} i \\
     0.306+2.830 \hspace{0.5mm} i &       
     ~~0.034-0.350 \hspace{0.5mm} i~~ &       
     -0.951-0.829 \hspace{0.5mm} i
   \end{array}
   \right) .
\end{split}
\eeq
Our second parameter point features
\beq\label{eq:cparameter2}
\begin{aligned}
   c_{Q_1} &= -0.646 \,, \qquad
    c_{Q_2} &= -0.573 \,, \qquad
    c_{Q_3} &= -0.449 \,, \\
   c_{u_1} &= -0.658 \,, \qquad
    c_{u_2} &= -0.513 \,, \qquad
    c_{u_3} &= +0.480 \,, \\
   c_{d_1} &= -0.645 \,, \qquad
    c_{d_2} &= -0.626 \,, \qquad
    c_{d_3} &= -0.578 \,,
\end{aligned}
\eeq
\beq\label{eq:yukawas2}
\begin{split}
 \bm{Y}_u &= \left(
   \begin{array}{rrr}
     0.637-1.800 \hspace{0.5mm} i &
     ~~1.518-2.209 \hspace{0.5mm} i~~ &       
     0.904+0.146 \hspace{0.5mm} i \\
     0.219-0.207 \hspace{0.5mm} i &      
     ~~-0.333-0.942 \hspace{0.5mm} i~~ &     
     0.597+0.020 \hspace{0.5mm} i \\
     1.829+1.538 \hspace{0.5mm} i & 
     ~~-0.018+1.772 \hspace{0.5mm} i~~ &
     -1.258+1.265 \hspace{0.5mm} i
  \end{array}
\right) , \\
\bm{Y}_d &= \left(
  \begin{array}{rrr}
    -2.835-0.946 \hspace{0.5mm} i &      
    ~~-0.404+0.746 \hspace{0.5mm} i~~ &      
    -1.135+0.060 \hspace{0.5mm} i \\
    0.724-0.350 \hspace{0.5mm} i &      
    ~-2.214-0.555 \hspace{0.5mm} i~~ &      
    0.610-0.051 \hspace{0.5mm} i \\
    0.701-0.101 \hspace{0.5mm} i &      
    ~~-0.154+0.104 \hspace{0.5mm} i~~ &      
    1.514+0.919 \hspace{0.5mm} i
  \end{array}
  \right) .
\end{split}
\eeq
Finally, the third parameter point reads 
\beq\label{eq:cparameter3}
\begin{aligned}
   c_{Q_1} &= -0.624 \,, \qquad
    c_{Q_2} &= -0.563 \,, \qquad
    c_{Q_3} &= -0.468 \,, \\
   c_{u_1} &= -0.712 \,, \qquad
    c_{u_2} &= -0.560 \,, \qquad
    c_{u_3} &= +0.899 \,, \\
   c_{d_1} &= -0.659 \,, \qquad
    c_{d_2} &= -0.642 \,, \qquad
    c_{d_3} &= -0.571 \,,
\end{aligned}
\eeq
\beq\label{eq:yukawas3}
\begin{split}
 \bm{Y}_u &= \left(
  \begin{array}{rrr}        
    -0.541+1.517 \hspace{0.5mm} i &     
    ~~-1.083+1.857 \hspace{0.5mm} i~~ &       
    1.718-2.057 \hspace{0.5mm} i \\
    0.359-1.713 \hspace{0.5mm} i &      
    ~~-2.208+1.404 \hspace{0.5mm} i~~ &      
    -1.160+0.886 \hspace{0.5mm} i \\
    -1.172-0.543 \hspace{0.5mm} i &      
    ~~-0.116-0.238 \hspace{0.5mm} i~~ &       
    -0.669-1.688 \hspace{0.5mm} i
  \end{array}
\right) , \\
\bm{Y}_d &= \left(
  \begin{array}{rrr}
    -0.878-1.677\hspace{0.5mm} i &      
    ~~0.190+0.573 \hspace{0.5mm} i~~ &      
    -0.817+2.663 \hspace{0.5mm} i \\
    -1.792+0.861 \hspace{0.5mm} i &      
    ~-2.880+0.132 \hspace{0.5mm} i~~ &      
    -0.070-1.151 \hspace{0.5mm} i \\
    -1.679+1.588 \hspace{0.5mm} i &     
    ~~0.972+0.615 \hspace{0.5mm} i~~ &      
    1.421+0.981 \hspace{0.5mm} i
  \end{array}
  \right) .
\end{split}
\eeq
Note that for the custodial model $c_{d_i}\to c_{{\cal T}_{2i}}$ and $c_{u_i}\to c_{u_i^c}$.

\chapter{Reference Values for the SM Parameters}
\label{app:ref}

The central values and errors of the $\overline{\rm MS}$ quark masses, evaluated
at the scale $\Mkk=1$\,TeV, that we use in our analysis are 
\begin{equation}\label{eq:fitmasses}
  \begin{aligned}
    m_u &= (1.5\pm 1.0)\,\mbox{MeV} \,, & \qquad m_c &= (520\pm
    40)\,\mbox{MeV} \,, & \qquad
    m_t &= (144\pm 5)\,\mbox{GeV} \,, \\
    m_d &= (3.0\pm 2.0)\,\mbox{MeV} \,, & \qquad m_s &= (50\pm
    15)\,\mbox{MeV} \,, & \qquad m_b &= (2.4\pm 0.1)\,\mbox{GeV} \,.
  \end{aligned}
\end{equation} 
They have been obtained by using the low-energy values as compiled in
\cite{Nakamura:2010zzi}. For the Wolfenstein
parameters we use \cite{Charles:2004jd}
\begin{equation}\label{eq:fitwolf}
  \lambda = 0.2265\pm 0.0008 \,, \quad 
  A = 0.807\pm 0.018 \,, \quad
  \bar{\rho} = 0.141\,_{-0.017}^{+0.029} \,, \quad 
  \bar{\eta} = 0.343\pm 0.016 \,.
\end{equation}
If not stated otherwise, the remaining SM parameters entering our 
phenomenological analysis read \cite{Nakamura:2010zzi,LEPEWWG:2005ema,Group:2008nq}
\begin{equation}
  \begin{aligned}
    \Delta\alpha^{(5)}_{\rm had}(m_Z) &= 0.02758\pm 0.00035 \,, \qquad
     & \alpha_s(m_Z) &= 0.118\pm 0.003 \,, \\
     m_W &= (80.399\pm 0.023)\,\mbox{GeV} \,, \qquad 
     & m_Z &= (91.1875\pm 0.0021)\,\mbox{GeV} \,,\\
     m_t &= (172.6\pm 1.4)\,\mbox{GeV}\,.
  \end{aligned}
\end{equation}
We refer to the central values of these quantities as SM reference
values. Unless noted otherwise, our reference value for the
Higgs-boson mass is $m_h=150$\,GeV.

\end{appendix}

\end{document}